\documentclass[structabstract]{aa}  
\usepackage[english]{babel}  
\usepackage{graphicx}  
\usepackage[utf8]{inputenc}  
\usepackage{microtype}  
\usepackage{booktabs}  
\usepackage{rotating}  
\usepackage{lscape}
\usepackage[squaren]{SIunits}
\usepackage{natbib}
\usepackage{amsmath}
\usepackage{amssymb}
\usepackage{pstricks}
\usepackage{txfonts}
\usepackage[config, labelfont={bf}]{caption,subfig}
\usepackage{tikz}
\usepackage{hyperref}

\newcommand{\tex}{T_\mathrm{{ex}}}
\newcommand{\tmb}{T_\mathrm{{MB}}}
\newcommand{\tbg}{T_\mathrm{{BG}}}

\newcommand{\tk}{T_\mathrm{{K}}}

\newcommand{\td}{T_\mathrm{{d}}}
\newcommand{\tp}[2]{T_\mathrm{{#1}}^\mathrm{{#2}}}
\newcommand{\nhtwo}{n(\mathrm{H_2})}
\newcommand{\coldhtwo}{N(\mathrm{H_2})}

\newcommand{\pot}[1]{10^{#1}}
\newcommand{\expo}[1]{\mathrm{e^{#1}}}
\newcommand{\vlsr}{V_\mathrm{{LSR}}}

\newcommand{\cm}{\usk\centi \metre}
\newcommand{\cmtab}{\centi \metre}

\newcommand{\pc}{\usk\mathrm{pc}}
\newcommand{\pctab}{\mathrm{pc}}
\newcommand{\kpc}{\usk\mathrm{kpc}}

\newcommand{\hii}{H\textsc{ii}}
\newcommand{\msun}{\usk\mathrm{M_\odot}} 
\newcommand{\msuntab}{\mathrm{M_\odot}}

\newcommand{\yr}{\usk\mathrm{yr}}

\newcommand{\lsun}{\usk\mathrm{L_\odot}}

\newcommand{\gauss}{\mathrm{G}}
\newcommand{\kms}{\usk\kilo\metre\usk\second^{-1}}
\newcommand{\kmstab}{\kilo\metre\usk\second^{-1}}
\newcommand{\kel}{\usk\kelvin}
\newcommand{\mum}{\usk\micro\metre}

\newcommand{\asec}{\ensuremath{^{\prime\prime}}}
\newcommand{\unc}[2]{_{#1}^{#2}}
\newcommand{\mvir}{M_\mathrm{vir}}

\newcommand{\Istp}[2]{^{#1}\mathrm{C}^{#2}\mathrm{O}}
\newcommand{\nobrkdash}[1]{#1\mbox{-}}
\newcommand{\optprefix}[1]{(#1\mbox{-})}
\newcommand{\rdep}{\ensuremath{R_\mathrm{dep}}}
\newcommand{\reff}{\ensuremath{R_\mathrm{eff}}}
\newcommand{\ncrit}{\ensuremath{n_\mathrm{crit}}}

\newcommand{\dgc}{D_{GC}}
\newcommand{\dnear}{D_\mathrm{near}}
\newcommand{\dfar}{D_\mathrm{far}}
\newcommand{\znear}{z_\mathrm{near}}
\newcommand{\zfar}{z_\mathrm{far}}

\newcommand{\cooz}{\mathrm{C^{17}O(1-0)}} 
\newcommand{\cott}{\mathrm{C^{17}O(3-2)}} 
\newcommand{\ceightoto}{\mathrm{C^{18}O(2-1)}} 
 
\newcommand{\kkmstab}{\mathrm{K \usk km \usk s^{-1}}} 
\newcommand{\oratio}{$[^{18}\mathrm{O}]/[^{17}\mathrm{O}]$}
\newcommand{\cratio}{$[^{12}\mathrm{C}]/[^{13}\mathrm{C}]$}
\newcommand{\oratiom}{[^{18}\mathrm{O}]/[^{17}\mathrm{O}]}
\newcommand{\cratiom}{[^{12}\mathrm{C}]/[^{13}\mathrm{C}]}

\bibpunct{(}{)}{;}{a}{}{,}

\title{ATLASGAL-selected massive clumps in the inner Galaxy: \\ I. CO depletion and isotopic ratios}
\author{
A. Giannetti \inst{\ref{ira}, \ref{unibo}}\thanks{Marco Polo fellow.}
\and F. Wyrowski \inst{\ref{mpi}}
\and J. Brand \inst{\ref{ira}, \ref{arc}} 
\and T. Csengeri \inst{\ref{mpi}}
\and F. Fontani \inst{\ref{arcetri}}
\and C.~M. Walmsley \inst{\ref{arcetri}, \ref{dias}}
\and Q. Nguyen Luong \inst{\ref{cita}}
\and H. Beuther \inst{\ref{mpia}}
\and F. Schuller \inst{\ref{eso}}
\and R. G\"usten \inst{\ref{mpi}}
\and K. Menten \inst{\ref{mpi}}
}
\institute{
INAF-Istituto di Radioastronomia, Via Gobetti 101, 40129, Bologna, Italy \label{ira}
\and Dipartimento di Astronomia, Universit\`{a} di Bologna, Via Ranzani 1, 40127, Bologna, Italy \label{unibo}
\and Max Planck Institute for Radioastronomy, auf dem H\"ugel 69, D-53121, Bonn, Germany \label{mpi}
\and Italian ALMA Regional Centre, Via Gobetti 101, 40129, Bologna, Italy \label{arc}
\and INAF-Osservatorio Astrofisico di Arcetri, Largo E. Fermi 5, 50125, Firenze, Italy \label{arcetri}
\and Dublin Institute of Advanced Studies, 31 Fitzwilliam Place, Dublin 2, Ireland \label{dias}
\and Canadian Institute for Theoretical Astrophysics, University of Toronto, St. George Street 60, M5S 3H8, Toronto, Canada \label{cita}
\and Max-Planck-Institute for Astronomy, K\"onigstuhl 17, 69117 Heidelberg, Germany \label{mpia}
\and European Southern Observatory, Alonso de C\'ordova 3107, 19001, Santiago, Chile \label{eso}
}

\abstract{
In the low-mass regime, 
molecular cores have spatially resolved temperature and density profiles
allowing a detailed study of their chemical properties. It is found that the gas-phase abundances of C-bearing molecules in cold starless cores rapidly decrease with increasing density.
Here the molecules tend to stick to the grains, forming ice mantles.
}
{
We study CO depletion in a large sample of massive clumps, and investigate 
its correlation with evolutionary stage and with the physical parameters of the sources. Moreover, we study the gradients in \cratio\ and \oratio\ isotopic ratios across the inner Galaxy, and the virial stability of the clumps.
}
{
From the ATLASGAL $870\mum$ survey we selected 102 clumps, which have masses in the range $\sim\pot{2}-3\times\pot{4}\msun$, sampling different evolutionary stages. We use low-J emission lines of CO isotopologues and the dust continuum emission to infer the depletion factor $f_D$.
RATRAN one-dimensional models were also used to determine $f_D$ and to investigate the presence of depletion above a density threshold. The isotopic ratios and optical depth were derived with a Bayesian approach.
}
{
We find a significant number of clumps with a large CO depletion, up to $\sim 20$. Larger values are found for colder clumps, thus for earlier evolutionary phases. For massive clumps in the earliest stages of evolution we estimate the radius of the region where CO depletion is important to be a few tenths of a $\pctab$. The value of the \cratio\ ratio is found to increase with distance from the Galactic centre, with a value of $\sim 66\pm12$ for the solar neighbourhood. The \oratio\ ratio is approximately constant ($\sim4$) across the inner Galaxy between $2\kpc$ and $8\kpc$, albeit with a large range ($\sim2-6$). Clumps are found with total masses derived from dust continuum emission up to $\sim 20$ times higher than $\mvir$, especially among the less evolved sources. These large values may in part be explained by the presence of depletion: if the CO emission comes mainly from the low-density outer layers, the molecules may be subthermally excited, leading to an overestimate of the dust masses.
}
{
CO depletion in high-mass clumps seems to behave as in the low-mass regime, with less evolved clumps showing larger values for the depletion than their more evolved counterparts, and increasing for denser sources. The ratios \cratio\ and \oratio\ are consistent with previous determinations, and show a large intrinsic scatter. 
}

\offprints{A. Giannetti (agianne@ira.inaf.it)}

\begin{document}  
\maketitle

\section{Introduction}\label{sec:intro}

The fundamental role of massive stars in shaping their direct environment and the galaxies they are in, makes the understanding of the process through which they form one of the major objectives yet to be achieved by astrophysicists. Progress in this sense requires a detailed observational knowledge of high-mass star forming regions from a physical and chemical point of view. Molecular line and dust continuum emission in the submm regime are among the main tools to study the first stages of massive star formation, where the sources are still embedded in the molecular gas.

In the low-mass regime, molecules such as CO and CS tend to freeze onto the dust grains in the densest and coldest part of starless cores \citep[e.g. ][]{Caselli+99, Kramer+99, BerginTafalla07, Caselli11}. This is observed as an abundance drop for these molecules (referred to as depletion) in the central parts of the core, identified by means of dust continuum emission \citep[e.g. ][and references therein]{Tafalla+02}. The evolution of the abundance of different molecules on large scales can be reproduced with time-dependent models for gas-phase chemistry \citep[e.g. ][]{Langer+00}. Comparing observations of starless cores in the mm-continuum and in molecular lines from (among other species) CO isotopologues, \citet{Tafalla+02} find that depletion of CO can be substantial, with abundances with respect to $\mathrm{H}_2$ in the central regions that are up to $1-2$ orders of magnitude below the canonical ones. 
CO depletion is a temperature- and density-sensitive process; at low temperatures and high densities the depletion is higher, because under those conditions it is easier for the molecules to attach themselves to the grains.
When protostars are formed in the core, the temperature increases, and at temperatures $T\sim 20-25\kel$ \citep[commonly found in high-mass clumps, e.g. ][]{Wienen+12} the molecules evaporate from grains back into the gas phase, and the abundance returns to canonical levels \citep[for CO $\sim\pot{-4}$ with respect to molecular hydrogen, in the solar neighbourhood; see e.g. ][]{Fontani+06}. 
The depletion in a molecular core can thus vary substantially during the process of star formation, and can be used as an evolutionary indicator.

The Atacama Pathfinder Experiment (APEX\footnote{This publication is based on data acquired with the Atacama Pathfinder Experiment (APEX). APEX is a collaboration between the Max-Planck-Institut f\"ur Radioastronomie, the European Southern Observatory, and the Onsala Space Observatory.}) \citep{Gusten+06} is a $\nobrkdash{12}\metre$ submm telescope located on Chajnator plane in Chile.
The APEX Telescope Large Area Survey of the Galaxy \citep[ATLASGAL, ][]{Schuller+09} is the first complete survey of the inner Galactic plane, carried out in the submm continuum at $870\mum$ ($345\usk\giga\hertz$), and thus sensitive to very cold dust, potentially tracing the pristine condensations where the process of star formation still has to begin. \citet{Contreras+13} compiled a catalogue of compact sources \citep[extracted with SExtractor, ][]{BertinArnouts96} of the ATLASGAL survey, and \citet{Csengeri+14_AGGC} filtered the extended emission and decomposed the remainder into Gaussian components using the GAUSSCLUMPS algorithm \citep{StutzkiGuesten90}, optimised to extract the properties of embedded sources.
The properties of a sample of massive star forming clumps in the ATLASGAL survey were studied by \citet{Urquhart+13a} and \citet{Urquhart+13b}, searching for sources associated with methanol masers \citep[from the methanol multi-beam survey,][]{Green+09, Caswell+10, Green+10, Caswell+11, Green+12} and \optprefix{ultra}compact \hii\ regions \citep[from the CORNISH survey,][]{Hoare+12, Purcell+13}, respectively.

On the other hand, the starless-phase in the high-mass regime is elusive, due to its short duration \citep[e.g. ][]{Motte+07, Russeil+10, Tackenberg+12}. Candidates have been found looking for compact and massive molecular condensations without signs of active star formation \citep[e.g. ][]{Chambers+09, Rygl+10, ButlerTan12, Rygl+13, Giannetti+13, Beuther+13}. Recent Herschel observations have also been used to search for massive starless objects \citep{Nguyenluong+11}. These clumps show lower temperatures than their star forming counterparts ($10-15\kel$ vs. $20-40\kel$). The star forming sources are characterised 
by tracers of current star formation such as PAH emission, radio continuum and maser emission.

A number of studies, usually targeting a limited sample of sources, have been carried out to investigate depletion in high-mass clumps \citep[e.g. ][]{Zinchenko+09, Miettinen+11, Hernandez+11, Rygl+13, Liu+13}, and very few address the variation with time of this essential parameter for the chemistry of the source \citep[][]{Fontani+12}. 
Discordant evidence for depletion exists for high-mass objects, with both claims of significant CO freeze-out onto grains \citep{Hernandez+11, Fontani+12, Rygl+13} and of canonical abundances \citep{Zinchenko+09, Miettinen+11}. In this work we investigate the CO abundance in a large sample of massive clumps, selected from ATLASGAL, by means of its rarer (optically thin) isotopologues. The clumps have been selected to be in different evolutionary phases, so that we can also study changes in CO abundance in massive clumps during their evolution.

The paper is organised as follows. In Sect.~\ref{sec:sample} we describe the sample and its selection; in Sect.~\ref{sec:observations} we briefly describe the observations. In Sect.~\ref{sec:results} we derive the physical properties of the clumps, and the carbon and oxygen isotopic ratios across the inner Galaxy; Sect.~\ref{sec:discussion} is dedicated to the discussion of these results, to the derivation of the CO-depletion factor ($f_D = X^E/X^O$, where $X^O$ and $X^E$ are the observed and expected abundances, respectively), and to the study of the physical and chemical changes occurring in the sources as they evolve. Finally, in Sect.~\ref{sec:summary} we draw some general conclusions and summarise our findings.

\section{The Sample}\label{sec:sample}

\begin{figure}[tb]
  \centering
  \includegraphics[angle=-90,width=0.8\columnwidth]{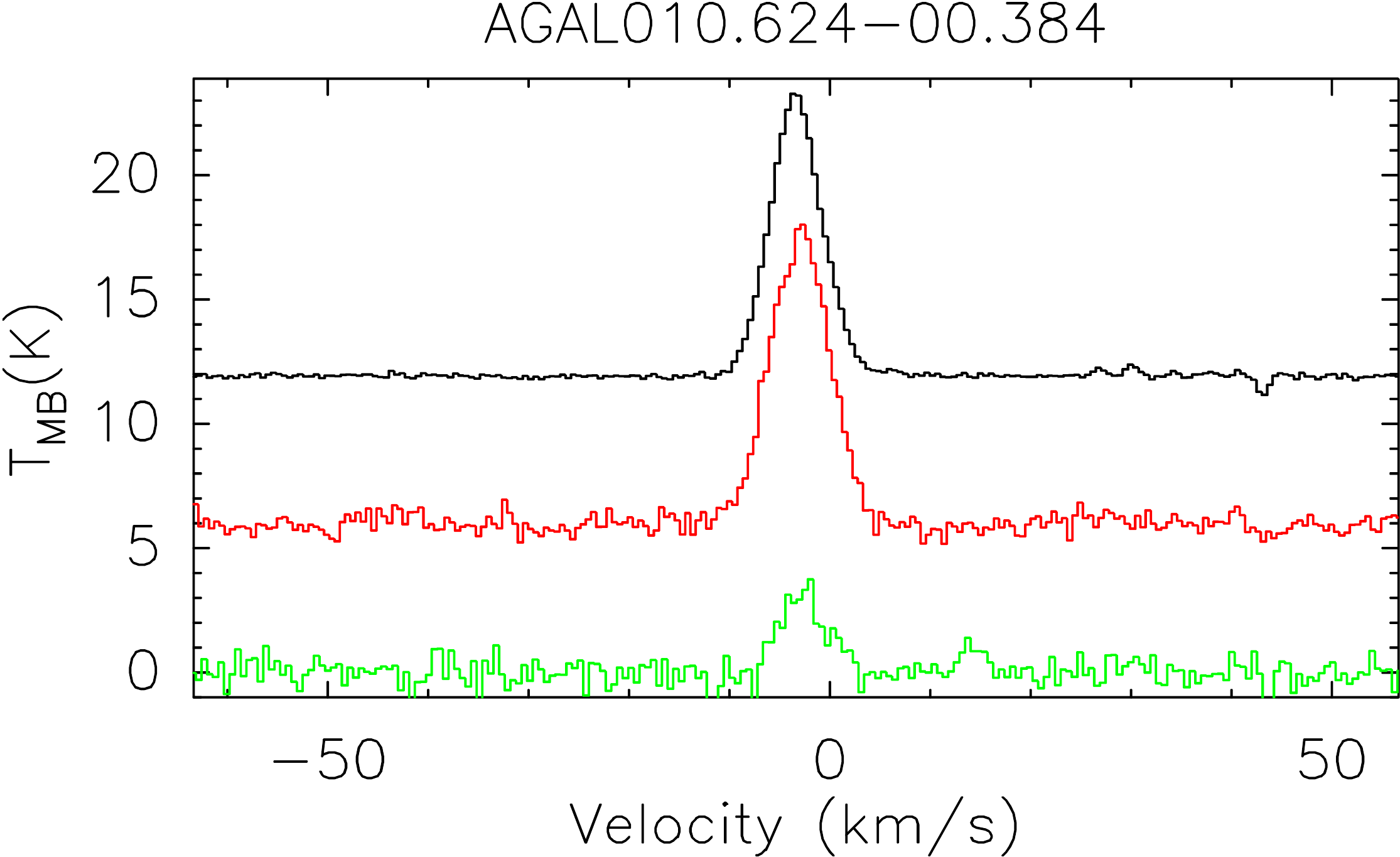} \hfill \\
  \vspace*{0.5cm}
  \includegraphics[angle=-90,width=0.8\columnwidth]{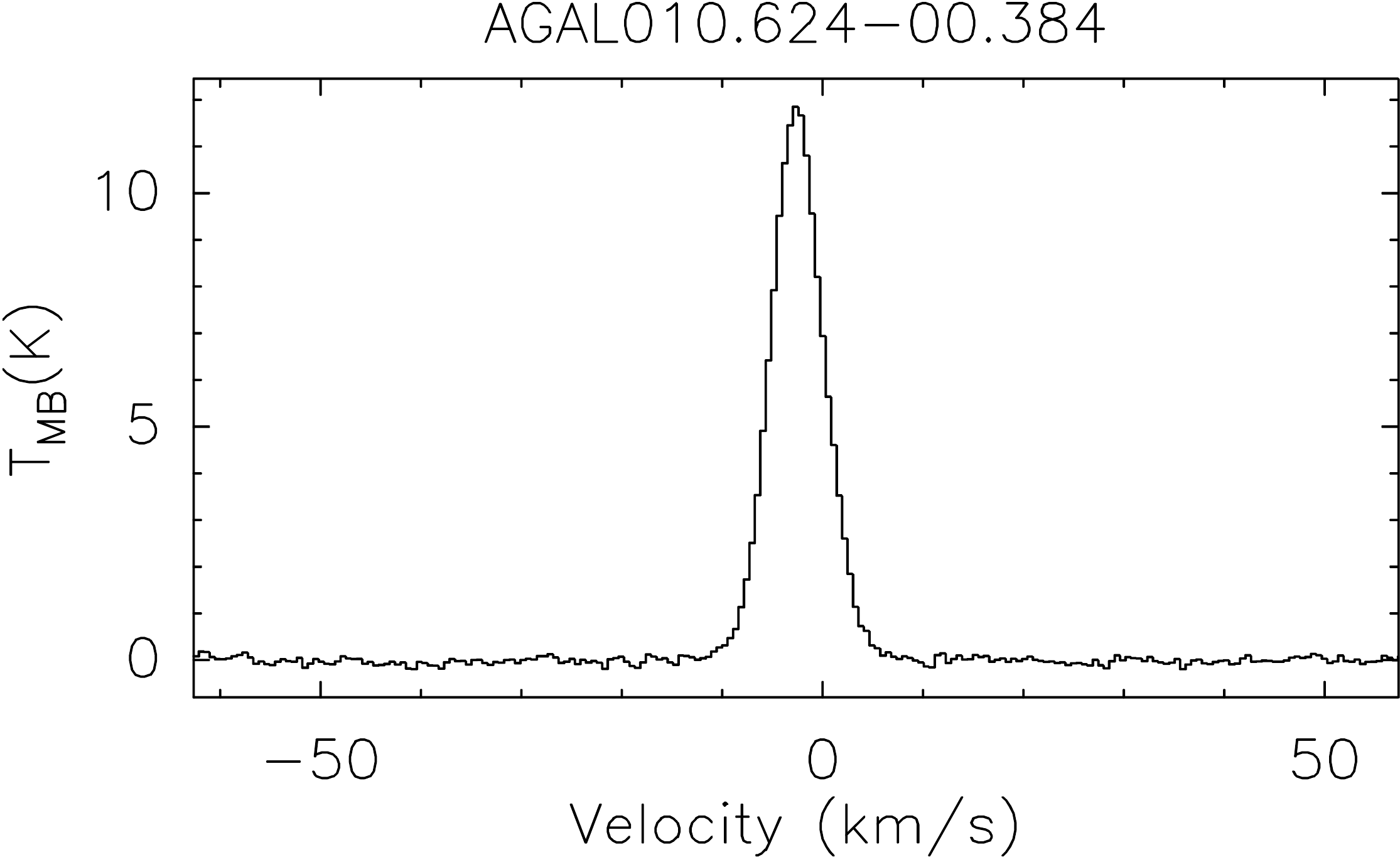}
  \caption{Example of the observed spectra. Top panel: $^{13}$C$^{18}$O$(1-0)$ (x10, green), C$^{17}$O$(1-0)$ (x3, red) and C$^{18}$O$(1-0)$ (black). The spectra are displaced for clarity. Bottom panel: C$^{17}$O$(3-2)$.}\label{fig:spectra_example}
\end{figure}

The ATLASGAL survey constitutes an excellent tool for selecting massive clumps. 
To select sources in various evolutionary phases, the clumps were extracted from the ATLASGAL database with the following criteria, each one defining a group of objects:
\begin{itemize}
 \item The 32 brightest sources of the whole survey, excluding the Central Molecular Zone (i.e. the central $\mathrm{few} \times \pot{2}\pc$) are called IRB group. These clumps are also detected in the IR;
 \item The 25 brightest sources, that are classified as a Massive Young Stellar Object (MYSO) in the Red MSX Sources survey \citep{Urquhart+08}, are hereafter referred to as the RMS group;
 \item The 23 brightest objects dark at $8\mum$ constitute the D8 group;
 \item Finally, the 22 brightest sources that are dark at $24\mum$, are hereafter called D24.
\end{itemize}
Above, ``brightest'' refers to the submm peak flux at $870\mum$. The members of each group were selected from the ATLASGAL catalogue after removing those of the preceding groups. Sources are classified as $24\mum$ or $8\mum$ dark if their average $24\mum$ or $8\mum$ flux density within the ATLASGAL beam is smaller than the average flux at the same wavelength in their vicinity.

This sample is termed the TOP100 sample of the ATLASGAL survey. Clumps in the first two groups of objects most likely contain high-mass stars and/or massive young stellar objects, while the other two are in an earlier phase of evolution \citep[see ][]{Motte+07,Nguyenluong+11}, hosting very deeply embedded (and thus heavily extincted) (proto-)stars or being still starless. Separating the sample in these four categories allows us to study possible variations of depletion as a function of evolution.

This is the first of a series of papers concerning follow-up observations of the ATLASGAL TOP100 sample.

\section{Observations} \label{sec:observations}

Single-pointing observations were carried out for all sources in the TOP100 sample with APEX/FLASH \citep[First-light APEX submillimeter heterodyne instrument; ][]{Heyminck+06} in $\cott$, on 14, 15, 24 June 2011 and 11-12 August 2011. Lower-excitation lines have been observed with APEX-1 for southern sources ($^{13}$CO, C$^{18}$O $J = 2 - 1$; observations carried out on 13-15 November 2008, and on 30 October and 1-2 November 2009), and with the EMIR (Eight mixer receiver) on the IRAM \nobrkdash{30}m telescope for northern sources (C$^{18}$O, $^{13}$CO, C$^{17}$O, $^{13}$C$^{18}$O $J = 1 - 0$; on 8-11 April 2011). The frequency of each transition and the beam size of the antenna for that molecular line are listed in Table~\ref{tab:freq_trans}. 
We have divided our sample into three subsamples, named S1, S2 and S3, according to the observational data we had available. 
Their positions and basic characteristics are listed in Tables~\ref{tab:dist_A}-\ref{tab:dist_C}.
Subsample S1 consists of 35 sources with EMIR follow-ups. In subsample S2 there are 44 sources, with APEX-1 follow-ups and subsample S3 (23 sources) were observed only with FLASH. Single Gaussians were fitted to the spectra for $^{13}$CO, C$^{18}$O and $^{13}$C$^{18}$O lines, while for $\cott$ and $\cooz$ the appropriate hyperfine splitting was taken into account. The line parameters are listed in Tables~\ref{tab:line_pars_A} to \ref{tab:line_pars_C}. Figures from \ref{fig:spectra_10_A} to \ref{fig:spectra_32_C} show the spectra for all the observed sources; an example is given in Fig.~\ref{fig:spectra_example}, showing $\cooz$, $\Istp{}{18}(1-0)$, $\Istp{13}{18}(1-0)$ and $\cott$ for AGAL010.624-00.384. The velocity resolution and the rms of the spectra are typically between $0.3-0.5\kms$ and $0.05-0.10\kel$ (in units of main beam temperature, $\tmb$), respectively.

\begin{table}
	 \centering 
	 \caption{Frequencies of the observed transitions and angular size of the beam (FWHP).} \label{tab:freq_trans}
	 \begin{tabular}{l*{6}c} 
		\toprule
		Receiver     & Transition           & $\nu$         & $\theta_\mathrm{beam}$    \\   
		             &                      &$(\giga\hertz)$& $(\arcsec)$               \\   
		\midrule
		APEX/FLASH   & $\Istp{12}{17}(3-2)$ & $337.06099$   & $19$                      \\   
		APEX/APEX-1  & $\Istp{13}{16}(2-1)$ & $220.39868$   & $28$                      \\   
		             & $\Istp{12}{18}(2-1)$ & $219.56035$   & $28$                      \\   
    IRAM\nobrkdash{30}m/EMIR & $\Istp{13}{16}(1-0)$ & $110.20135$   & $22$                      \\   
		             & $\Istp{12}{18}(1-0)$ & $109.78217$   & $22$                      \\   
		             & $\Istp{12}{17}(1-0)$ & $112.35898$   & $21$                      \\   
		             & $\Istp{13}{18}(1-0)$ & $104.71140$   & $23$                      \\   
		\bottomrule
	 \end{tabular}
\end{table}

\section{Results} \label{sec:results}

\subsection{Distance} \label{ssec:distance}

In order to derive the masses and to investigate the relative abundance of different isotopes as a function of Galactocentric distance, we searched the literature for a distance determination for each of the sources in our sample. We use direct maser parallax or spectrophotometric measurements where available ($7$ and $16$ sources respectively), otherwise ($79$ sources) we use the kinematic distance \citep[using the rotation curve of ][]{BrandBlitz93}, resolving the near-far ambiguity with H\textsc{i} self-absorption data from Wienen et al. (in prep.), or from the literature. After this process, only two sources out of $102$ still have a distance ambiguity, and for these we use the near kinematic distance. On the other hand, no such ambiguity is present for the Galactocentric distance. Our findings are summarised in Tables~\ref{tab:dist_A}-\ref{tab:dist_C}. We choose to use the \citet{BrandBlitz93} instead of the more recent one of \citet{Reid+09} because it still the best sampled in terms of Heliocentric and Galactocentric distances and Galactocentric azimuth.

\subsection{Excitation temperatures} \label{ssec:tex}

The excitation temperature ($\tex$) is derived from the ratio $R_{ij}$ of the integrated line intensities of the transitions $i \rightarrow i-1$ and $j \rightarrow j-1$,
\begin{equation}
R_{ij} \equiv \frac{\int \tp{MB,\emph{i} \rightarrow \emph{i}-1}{} dV}{\int \tp{MB,\emph{j} \rightarrow \emph{j}-1}{} dV} = \frac{I(i \rightarrow i-1)}{I(j \rightarrow j-1)},
\label{eq:ratio}
\end{equation}
where $I(i \rightarrow i-1)$ stands for the integrated intensity of the molecular transition $i \rightarrow i-1$ and the main beam temperature is
\begin{equation}
\tmb = \eta [J(\tex)-J(\tbg)] (1-\expo{-\tau}),
\label{eq:det_eq}
\end{equation}
with 
\begin{equation}
J(T) = \frac{h \nu}{k_B} \frac{1}{(\expo{\emph{h} \nu / (\emph{k}_\emph{B} \emph{T})} - 1)}.
\end{equation}
In Eq.~\ref{eq:det_eq}, $\eta$ is the source filling factor, $\tbg$ is the background temperature and $\tau$ is the optical depth.
To solve Eq.~\ref{eq:ratio} for $\tex$ we neglected the background contribution $J(\tbg)$, and assumed that the emission is optically thin, and that it traces the same volume of gas for both transitions. From the latter assumption it follows that the filling factor $\eta$ cancels out in the ratio, if the angular resolution is similar, and that the lines have similar profiles.
For subsample S1 $R_{ij}$, and thus $\tex$, is obtained from $\cott$ and $\cooz$, while for subsample S2 it is obtained from $\cott$ and $\ceightoto$, assuming an isotopic ratio $\oratiom=4$ (see Sect.~\ref{ssec:abund_dgc}) and correcting $R_{ij}$ for the optical depth of $\ceightoto$ (see Sect.~\ref{ssec:abund_dgc}). 
The values derived for $\tex$ range from $\sim5\kel$ to $\sim70\kel$ and are listed in Tables~\ref{tab:tex_a} and \ref{tab:tex_b}, with their uncertainties for individual sources. 
Since we do not have maps, we cannot smooth molecular data to a common resolution to derive the temperature, removing the filling factor from the ratio. This is not a problem for subsample S1, because $\cott$, observed with APEX, and $\cooz$, observed with the \nobrkdash{30}m, have similar angular resolutions ($19\asec-21\asec$), but it may be an issue for subsample S2 ($19\asec-28\asec$). As an example, if one assumes a source size of $70\arcsec$ and tries to account for the different angular resolutions using the correction factor $(\theta_{beam}^2+\theta_{source}^2)/\theta_{source}^2$, the ratio $R_{ij}$ varies by $\sim 10\%$. This implies a decrease in $\tex$ of $\lesssim 10\%$ for a $\tex\lesssim20\kel$ and up to $25\%$ for $\tex\sim70\kel$. For smaller sources the difference increases and reaches maximum for point sources. To this source of uncertainty, one has to add that the assumed value of the isotopic ratio $\oratiom=4$ may not be appropriate for specific sources (cf. Fig.~\ref{fig:ref_isotopic_ratios}b). The highest excitation temperatures are found in S2, some of them possibly caused by the uncertainties discussed above.
Equation~\ref{eq:ratio} is implemented in JAGS\footnote{\url{http://mcmc-jags.sourceforge.net/}} (Just Another Gibbs Sampler) to estimate $\tex$ and its uncertainty, directly from the measured quantities. In the procedure, only temperatures up to $\sim100\kel$ were considered to be physically plausible. This is also motivated by the transitions used, for which the method is increasingly insensitive above $\sim 30\kel$.

For S3 we used the average values of $\tex$ from S1 and S2, listed in Table~\ref{tab:tex_class}, for the appropriate group of sources to derive the $\Istp{12}{17}$ column density, and the mass from the submm continuum.

\subsection{Isotopic abundance variations in the Galaxy} \label{ssec:abund_dgc}

\begin{figure*} 
\centering 
\includegraphics[angle=-90,width=0.77\textwidth]{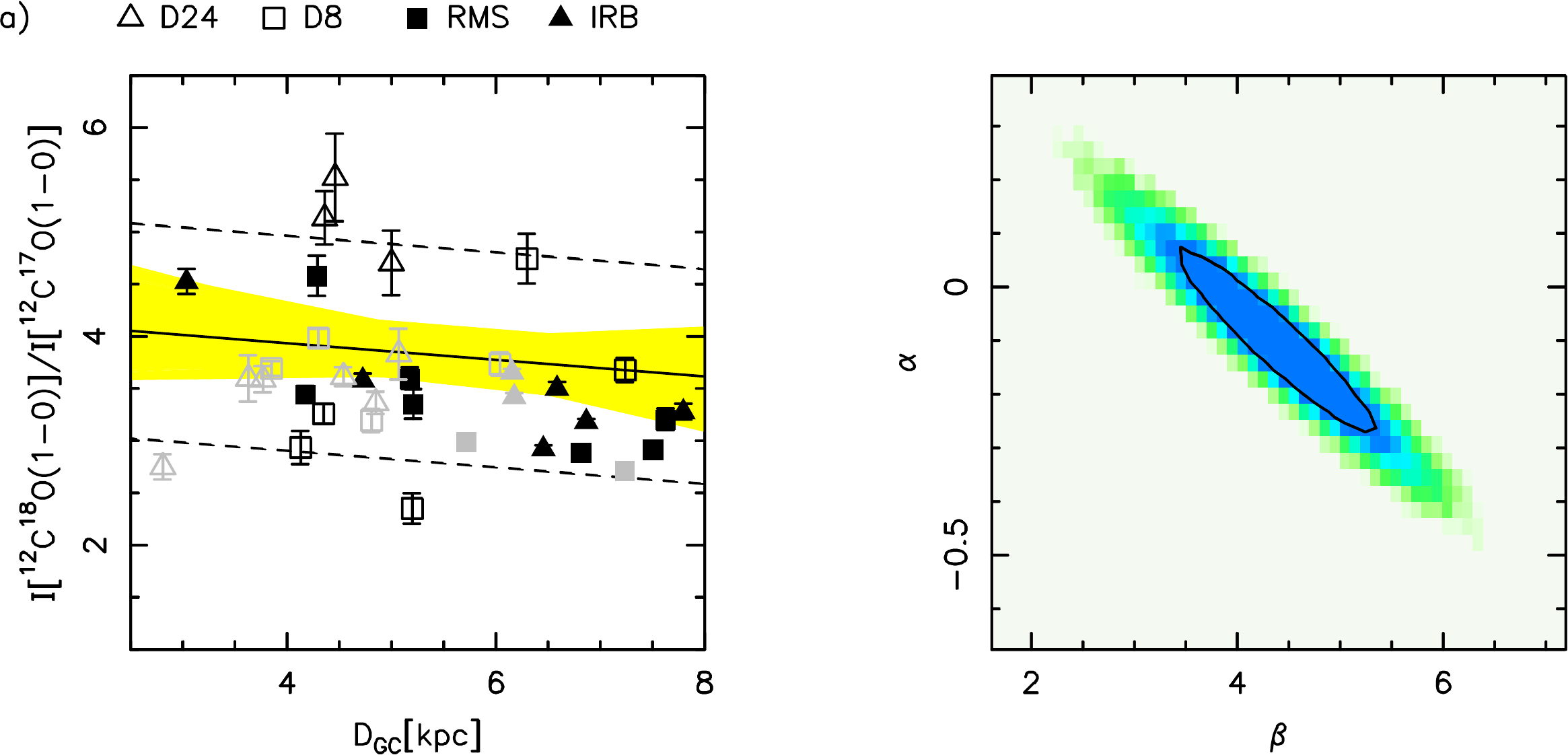} \\ \vspace{0.4cm}
\includegraphics[angle=-90,width=0.77\textwidth]{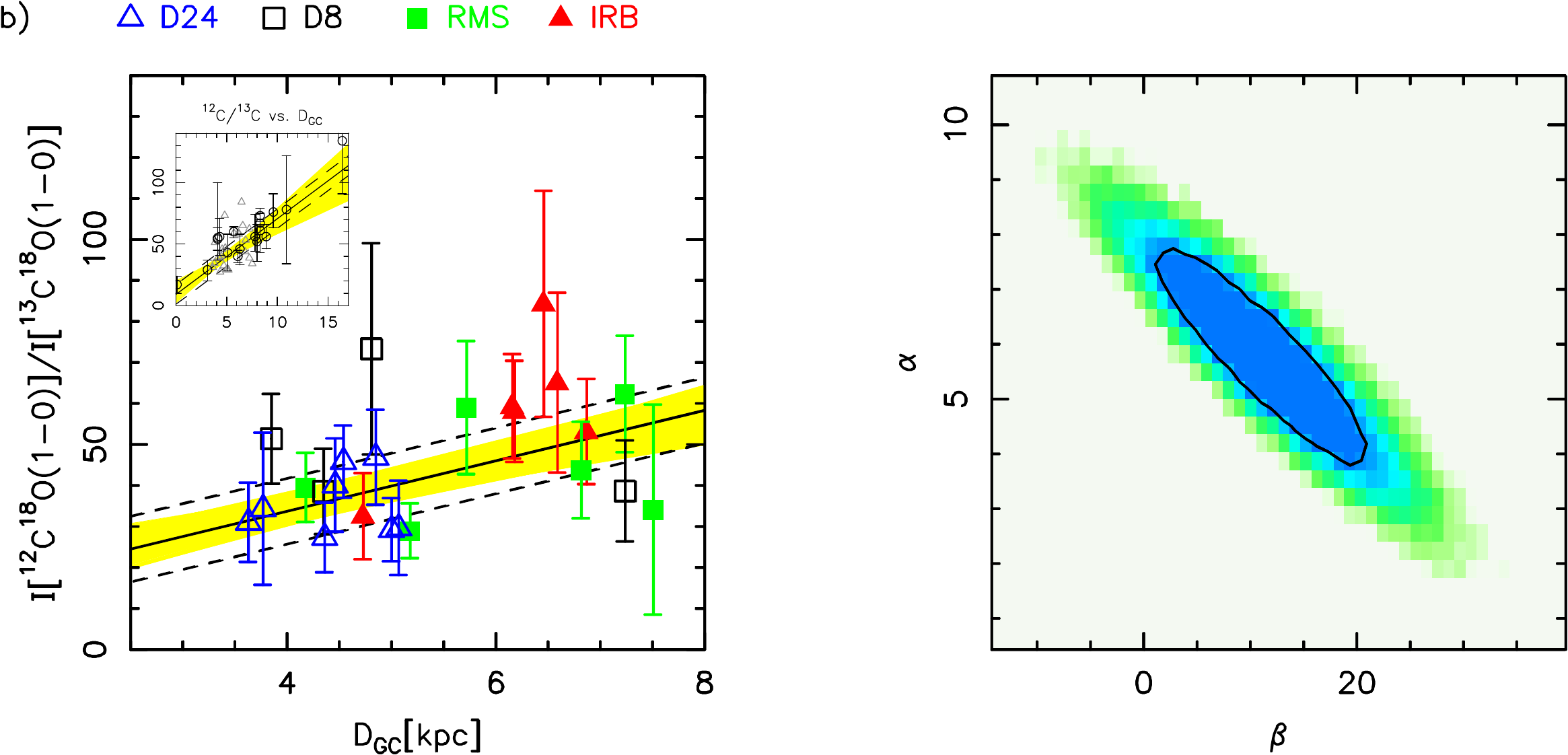} \\ \vspace{0.4cm}
\includegraphics[angle=-90,width=0.77\textwidth]{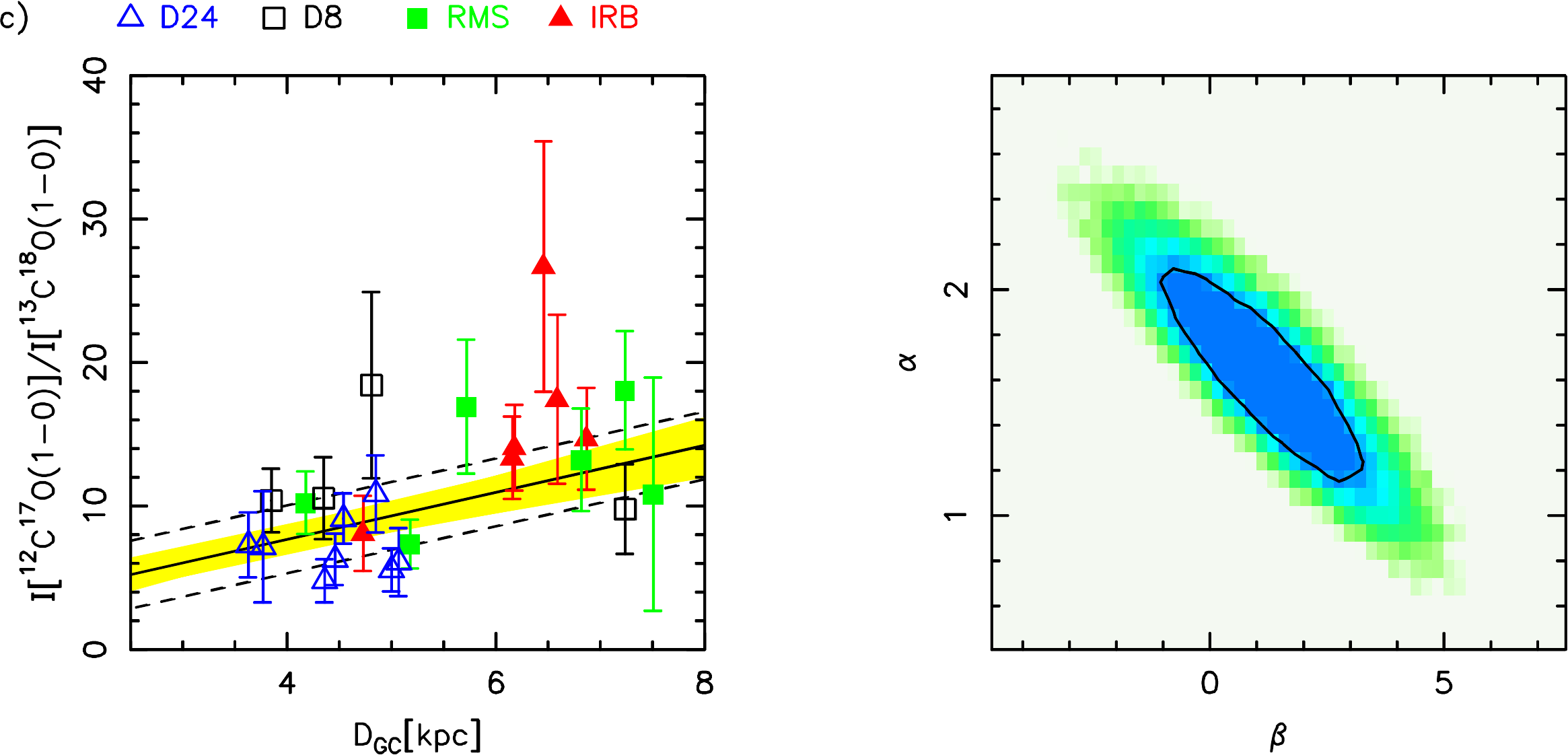} 
\caption{Ratio of integrated fluxes of different CO isotopologues as a function of Galactocentric radius. In the right panel of each row we show the joint probability distribution of the parameters of the fit ($y=\alpha x +\beta$); in black we indicate the $68\%$ contour. In the left panel of each row, the black solid line is the best fit, the yellow shaded area shows the $68\%$ uncertainty, and the dashed lines show the intrinsic scatter of the relation. Panel \textbf{(a)} shows the ratio $I[\Istp{12}{18}(1-0)]/I[\Istp{12}{17}(1-0)]$ (see Eq.~\ref{eq:ratio}), each group of sources is shown with a different symbol, as indicated above the left panel. Black points are those with an estimated $\Istp{12}{18}(1-0)$ optical depth less than $0.3$ (only from the detection equation), grey points are those with $\tau>0.3$. Panel \textbf{(b)} shows the ratio $I[\Istp{12}{18}(1-0)]/I[\Istp{13}{18}(1-0)]$; each group of sources is shown with a different symbol and colour, as indicated. The small box in the top left corner shows the points from \citet{Milam+05} as black circles, and those from this work as grey triangles. Finally, panel \textbf{(c)} is the same as (b) for the ratio $I[\Istp{12}{17}(1-0)]/I[\Istp{13}{18}(1-0)]$.} \label{fig:isotopic_ratios}
\end{figure*}

To derive more accurate optical depths in Sect.~\ref{ssec:tau_N} we start from a first estimate of the relative abundance of the different isotopes of C and O as a function of the Galactocentric distance $\dgc$ in the range $2-8\usk\kilo\pctab$. The isotopic gradients can then be used as \textit{priors} for the procedure to derive $\tau$ \citep[see ][for a description of the Bayesian approach to statistics]{BolstadIBS}.
This can be achieved by comparing the column densities of different carbon monoxide isotopologues, for which we use the $(1-0)$ transition of $\Istp{}{18}$, $\Istp{}{17}$ and $\Istp{13}{18}$. If we assume that the transitions are optically thin, the integrated flux of a transition is proportional to the column density, and we can approximate the relative abundance of C or O isotopes with the ratio of the integrated fluxes of the lines, if we are using the same rotational transition (i.e. with similar excitation conditions, and likely tracing the same volume of gas). Using the same rotational transition thus implies that we do not need to know the actual value of $\tex$, which is the same for both isotopologues.
The ratio of the fluxes was corrected for the small differences in the transition frequency \citep[see ][]{Linke+77}.
Panel (a) in Figure~\ref{fig:isotopic_ratios} shows that we find no clear trend with $\dgc$ for \oratio, and our values of the ratio show a large scatter. A constant value of this ratio in space and time is to be expected if $^{18}$O is a secondary nucleosynthesis product (those that can be produced only in the presence of pre-existing seed nuclei, generated by previous stellar generations), like $^{17}$O, and  they both come from the same primary element, $^{16}$O \citep[][]{WilsonRood94}. 
On the other hand, \citet{Wouterloot+08} find that the \oratio\ ratio tends to increase in the outer Galaxy, out to $\dgc\sim16\usk\kilo\pctab$.

We have fit the formula $N(\Istp{}{18})/N(\Istp{}{17}) = \alpha \dgc + \beta$ to the data (see Fig.~\ref{fig:isotopic_ratios}), selecting sources with low optical depth ($< 0.3$, i.e. a correction of $\sim15\%$ in column density; see Eq.~\ref{eq:corr_factor}), derived from the detection equation alone.
The fit gives a small negative slope, $\sim -0.1\unc{-0.2}{+0.1}$. The slope is still consistent with zero and it is so small that the resulting values of \oratio\ would change by only $\sim 1$ ($25\%$ of the assumed value of $4$) over the whole range of $\dgc$ of the S1 sample. This variation is of the order of intrinsic scatter of the fit ($\sim1.0\unc{-0.2}{+0.2}$). 
We note that a model with only one free parameter (the intrinsic scatter, and $\oratiom = 4$ across the inner Galaxy) is to be favoured with respect to a model with three free parameters (the slope, the intercept and the intrinsic scatter): a Bayesian model comparison shows that the \textit{odds ratio} \citep[see e.g. ][]{Gregory} is about $15$ in favour of the first model. 
This thanks to the fact that more complex models are automatically penalised in the Bayesian approach, and are to be preferred only if the data justify the added complexity (Occam's Razor).
Therefore, we assume an average constant value for $\oratiom=4$, independently of $\dgc$, with an intrinsic scatter as given by the fit; this value of the ratio is consistent with the measurements of \citet{Wouterloot+08} and \citet{WilsonRood94} for the same range of $\dgc$.

We also investigate the variation of the \cratio\ ratio as a function of $\dgc$, derived from both $I[\Istp{12}{18}(1-0)]/I[\Istp{13}{18}(1-0)]$ and $I[\Istp{12}{17}(1-0)]/I[\Istp{13}{18}(1-0)]$. The former gives a direct estimate of the \cratio\ ratio, but it may be affected by optical depth issues, while the latter involves only optically thin transitions, but the ratio \oratio, despite not being dependent on the Galactocentric distance, may vary up to $\sim50\%$ in different sources, thus increasing the scatter in the relation. The measured ratios and the results of fitting a straight line to the data points \citep[using as priors the results reported by ][ see below]{Milam+05} are shown in Fig.~\ref{fig:isotopic_ratios}b and c. From these panels, one can see that \cratio\ increases with $\dgc$, in agreement with previous works \citep{LangerPenzias90, WilsonRood94}, and expected for a primary/secondary product ratio \citep[see ][]{WilsonRood94}; in fact, primary elements are produced independently of the metallicity of the stars, while the production of secondary elements increases for higher metallicities. We find the relations $\cratiom = 6.2\unc{-2.1}{+1.1} \dgc + 9.0\unc{-6.2}{+9.9}$ and $(\cratiom)/(\oratiom) = 1.6\unc{-0.3}{+0.8} \dgc+ 1.1\unc{-1.4}{+1.9}$, respectively from $I[\Istp{12}{18}(1-0)]/I[\Istp{13}{18}(1-0)]$ and $I[\Istp{12}{17}(1-0)]/I[\Istp{13}{18}(1-0)]$. We note that the uncertainties reported in the relations are derived \textit{marginalising} \citep[see e.g. ][]{Jaynes03,BolstadIBS,Gregory} the probability distributions shown in the right panels of Fig.~\ref{fig:isotopic_ratios}b and c, but the values of the parameters of the fit $\alpha$ and $\beta$ ($y=\alpha x + \beta$) are strongly correlated and the yellow-shaded area in the left panels of the figure, representing the fit uncertainty, takes this into account. The above fit results give a \cratio\ ratio for the solar neighbourhood of $\sim62$ and $\sim60$ respectively, the latter derived with a fixed \oratio\ of 4.
In this case as well source-to-source variations are found, resulting in an intrinsic scatter of $\sim8\unc{-4}{+3}$ ($\sim15\%$ of the \cratio\ at the position of the Sun), obtained from the fit procedure. The estimated scatter in isotopic ratios for sources at a given $\dgc$ can be caused by a multiplicity of processes at work in the specific source (such as chemical fractionation and selective photodissociation) differently affecting the specific molecules, from our simple estimate of $\tau$, or can be intrinsic if the metallicity and the star formation history are different or other processes have an important role in modifying the isotopic ratios (e.g. non-efficient mixing, radial mixing, cloud mergers). 
\citet{Milam+05} use three different molecular species, CN, CO, and H$_2$CO, to investigate the \cratio\ ratio across the Galaxy, finding that all species give consistent results. The authors use this result to show that photodissociation does not have systematic and strong effects on the isotopic ratios. In addition, they compare the data with the predictions from a simple model for chemical fractionation, showing that they are at odds, thus concluding that also this process does not play a fundamental role.
In the small panel in Fig.~\ref{fig:isotopic_ratios}b we also show the average values of \cratio\ reported by \citet{Milam+05} as black open circles, with the error bars showing the range in the measured \cratio\ from the different species (or the given uncertainty where only one measurement is available), along with our points (grey open triangles). The comparison illustrates that the points are fully consistent with our measurements and fit. The fact that we find a smaller intercept is mainly due to the inclusion of Sgr B2 ($\dgc=0.1\pc$) and WB189 ($\dgc=16.4\pc$) in the work of \citet{Milam+05} and also to the unweighted fit that they use; the points with a high \cratio\ and a large uncertainty (cf. their Figure~2) could slightly increase the intercept.

\subsection{Optical depths and column densities} \label{ssec:tau_N}

Assuming Local Thermodynamic Equilibrium (LTE), we can compute the column density of the molecules using the excitation temperature, according to
\begin{equation}
N = C(\tau) f(\tex) \int \tmb dV,
\label{eqn:N}
\end{equation}
where $\tau$ is the optical depth of the transition, and $f(\tex)$ is a term grouping together all the constants and the terms depending on $\tex$ \citep[see e.g. ][]{KramerWinnewisser91}.
For a precise determination of the column density we have to estimate the optical depth $\tau$ of the lines. The correction factor $C(\tau)$ for Eq.~\ref{eqn:N} can be expressed in the form \citep[for $\tau\lesssim 2$][]{GoldsmithLanger99}:
\begin{equation}
 C(\tau)=\frac{\tau}{1-\expo{-\tau}}.
\label{eq:corr_factor}
\end{equation}
Typically C$^{18}$O and C$^{17}$O have column densities so low that their emission is optically thin, meaning that the correction for $\tau$ is small and therefore the molecular column density is proportional to the integrated flux of the lines.
However, it is worthwhile to check if the emission in environments having high column densities (as we expect for the clumps in the TOP100 sample) may still be assumed to be optically thin. The optical depth of a transition can be estimated by means of the detection equation or from the ratio of the integrated flux of the same transition, coming from different isotopologues, if the relative abundance is known \citep[e.g. ][]{Hofner+00},
\begin{equation}
R_{ij} = \frac{\int \tp{MB,\emph{i}}{} \; dV}{\int \tp{MB,\emph{j}}{} \; dV} = \frac{1-\expo{-\tau_\emph{i}}}{1-\expo{-\varphi \tau_\emph{i}}},
\end{equation}
where $\varphi$ is the relative abundance of the two isotopologues ($X_j/X_i$).
To derive the optical depth we used different methods, depending on the data available.

\textit{Subsample S1:} For these sources we have several CO isotopologues observed in the same transition, namely $\Istp{12}{18}$, $\Istp{12}{17}$ and  $\Istp{13}{18}(1-0)$. We derive and make a fit to the probability distribution of the three line ratios by means of Monte Carlo Markov chains. The results of the fits to the probabilities are used to describe the distribution out of which the measured ratios are extracted. The $\tau$ of the lines of the different isotopologues are interconnected by means of
\begin{equation}
\tau_{\Istp{12}{17}} = \tau_{\Istp{12}{18}}/A, 
\label{eq:A}
\end{equation}
\begin{equation}
\tau_{\Istp{13}{18}} = \tau_{\Istp{12}{18}}/B, 
\label{eq:B}
\end{equation}
\begin{equation}
\tau_{\Istp{13}{18}} = \tau_{\Istp{12}{17}} A/B \equiv \tau_{\Istp{12}{17}}/C,
\end{equation}
where $A$ and $B$ are the relative abundances \oratio\ and \cratio, respectively. The relations
\begin{equation}
\frac{\int \tp{MB,\Istp{12}{18}}{} dV}{\int \tp{MB,\Istp{12}{17}}{} dV} \equiv R_1 = \frac{1-\expo{-\tau_{\Istp{12}{18}}}}{1-\expo{-\tau_{\Istp{12}{18}}/\emph{A}}},
\end{equation}
\begin{equation}
\frac{\int \tp{MB,\Istp{12}{18}}{} dV}{\int \tp{MB,\Istp{13}{18}}{} dV} \equiv R_2 = \frac{1-\expo{-\tau_{\Istp{12}{18}}}}{1-\expo{-\tau_{\Istp{12}{18}}/\emph{B}}},
\end{equation}
\begin{equation}
\frac{\int \tp{MB,\Istp{12}{17}}{} dV}{\int \tp{MB,\Istp{13}{18}}{} dV} \equiv R_3 = \frac{1-\expo{-\tau_{\Istp{12}{18}}/\emph{A}}}{1-\expo{-\tau_{\Istp{12}{18}}/\emph{B}}},
\end{equation}
\begin{equation}
\tp{MB,\Istp{12}{18}}{} = \eta [J(\tex)-J(\tbg)] (1-\expo{-\tau_{\Istp{12}{18}}}), \; \mathrm{and}
\end{equation}
\begin{equation}
\tp{MB,\Istp{12}{17}}{} = \eta [J(\tex)-J(\tbg)] (1-\expo{-\tau_{\Istp{12}{18}}/\emph{A}})
\end{equation}
were simultaneously solved with Monte Carlo Markov chains in JAGS, thus combining the information from the detection equation and the line ratios for a more accurate determination of $\tau$ and of the isotopic ratios. \citet{Hofner+00} have maps in $\Istp{}{17}(2-1)$ for a sample of similar sources at a comparable angular resolution, and find them to be extended; therefore, for simplicity we assume $\eta = 1$ for these calculations. The \textit{priors} for $A$ and $B$ are Gaussian curves, centred on the value expected from the fit of the isotopic ratios as a function of $\dgc$, with a dispersion equal to the intrinsic scatter of the fit (see Sect.~\ref{ssec:abund_dgc}); the \textit{prior} for $\tau$ comes from the fit of the hyperfine structure of the $\cooz$ emission. This procedure returns $\tau_{\Istp{12}{18}}$, $\tau_{\Istp{12}{17}}$, $A$, $B$ and $C$, and their respective uncertainties for each source in the subsample S1, refining our estimate of the isotopic ratios and of the optical depth. We use these values of the \oratio\ and \cratio\ isotopic ratios in Sect.~\ref{ssec:CO_depl} to derive a more accurate $f_D$, and in Sect.~\ref{ssec:refined_estimate} to derive again the gradient of relative abundance as a function of $\dgc$ (see Fig.~\ref{fig:ref_isotopic_ratios}). The results of this procedure for the refined estimate of the \cratio\ and \oratio\ ratios are discussed in Sect.~\ref{ssec:refined_estimate} and summarised in Table~\ref{tab:tau_A}. We note that a significant fraction of the sources ($\sim55\%$) shows an optical depth $> 0.3$ for $\Istp{12}{18}(1-0)$. On the other hand, only one source has $\tau_{\Istp{12}{17}}>0.3$, while $\tau\lesssim0.2$ for the other sources; similar low values for the $\Istp{12}{17}$ opacity were obtained by \citet{Hofner+00}.

\textit{Subsample S2:} For this subsample we have observations of $\Istp{13}{16}$ and $\Istp{12}{18}(2-1)$.
Thus, we derive $\tau$ from the ratio of these two transitions. To estimate the uncertainty in the optical depth, we calculate the probability distribution of the line ratio, considering the calibration uncertainty and the noise of the spectrum. This can be translated into a distribution of $\tau$ for $\Istp{12}{18}(2-1)$, using the appropriate value for \cratio\ and $[^{18}$O$]/[^{16}$O$]$ at the $\dgc$ of the source.
The results are listed in Table~\ref{tab:line_pars_B}.

\textit{Subsample S3:} Finally, for the last subsample, the only measure for $\tau$ we have comes from the fit of the hyperfine structure of $\Istp{12}{17}(3-2)$. However, the satellites are too close to be spectrally resolved, meaning $\tau$ is highly uncertain. Thus, we assume that the emission is optically thin. 

The column densities of the CO isotopologues are listed in Tables~\ref{tab:cd_a}, \ref{tab:cd_b} and \ref{tab:cd_c}.

\subsection{Column density of molecular hydrogen}

The column density of molecular hydrogen $\coldhtwo$ was calculated from the ATLASGAL peak flux, according to \citet{Schuller+09},
\begin{equation}
\coldhtwo = \gamma \frac{F_p}{2.8 \usk m_p  \usk \Omega_B \usk \kappa_{870} \usk B_{870}(\td)},
\label{eq:cold_h2_schuller}
\end{equation}
where $F_p$ is the peak flux at $870\mum$, the factor $2.8$ accounts for the presence of helium \citep[$\sim10\%$ in number, e.g.][]{Allen73}, $\kappa_{870} = 1.8 \cm^2\usk\gram^{-1}$ is the dust opacity at $870\mum$ \citep[derived from the numbers given in ][]{OssenkopfHenning94}, $m_p$ is the proton mass, $\Omega_B$ is the beam solid angle for a beam size of $\approx 19.2\arcsec$ (i.e. $9.817\pot{-9} \steradian$), $B_{870}(\td)$ is the emission at $870\mum$ of a blackbody with a dust temperature $\td$ and $\gamma$ is the gas-to-dust ratio, assumed to be $100$.  
To derive the column density of molecular hydrogen we assume that gas and dust are coupled, and that the molecules are in LTE, thus $\td = \tk = \tex$ \citep[for a comparison of independently derived $\tk$ and $\td$ for a similar source sample ][]{Giannetti+13}.
For the subsample S3, we do not have a direct estimate of the temperature. Thus, we assign the typical temperature of the group to which the object belongs, with an uncertainty equal to the dispersion of that group (see Sect.~\ref{sec:discussion} and Table~\ref{tab:tex_class}).

\subsection{Masses}

The clump mass is derived from the integrated $870\mum$ flux as given in the latest ATLASGAL catalogue (for details, see \citealt{Csengeri+14_AGGC}), through
\begin{equation}
M = \frac{\gamma S_{870} D^2}{\kappa_{870} B_{870}(\td)},
\label{eq:mass}
\end{equation}
where $S_{870}$ is the $870\mum$ integrated flux, and $D$ is the distance.
The objects are massive, ranging from $\sim\pot{2}\msun$ to $\sim3 \times \pot{4}\msun$: we are dealing with extreme sources in the ATLASGAL survey, many of which are known sites of high-mass star formation. For subsample S3 the same assumptions for the temperature are made as for the determination of column densities. The masses are listed in Tables~\ref{tab:mass_a} to \ref{tab:mass_c}.

\begin{figure} 
\centering 
\includegraphics[angle=-90,width=0.8\columnwidth]{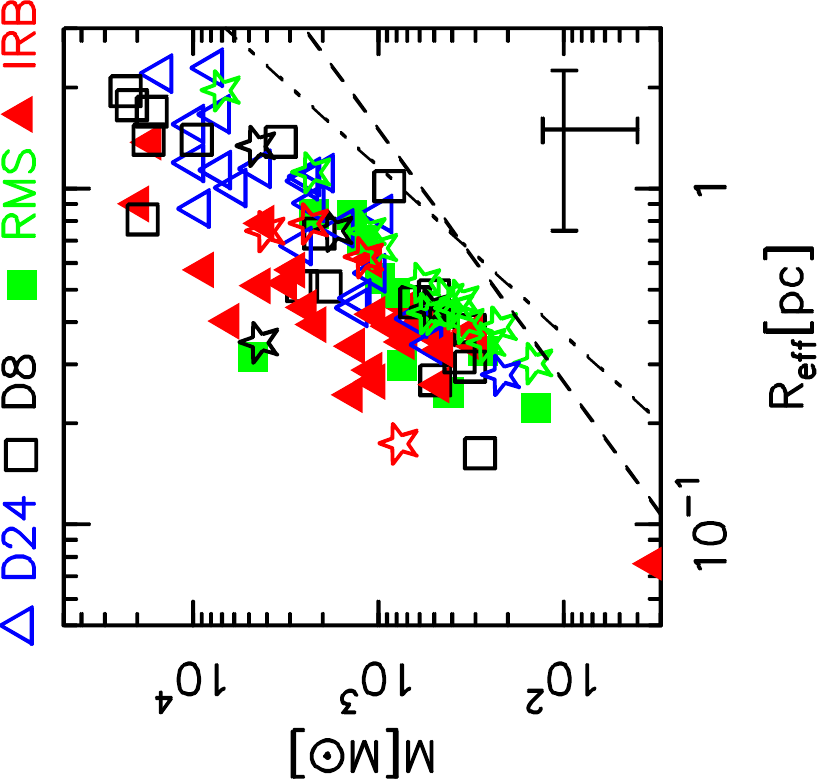} 
\caption{Mass-radius plot. $\reff$ is calculated according to \citet{Rosolowsky+10}, using the sizes from \citet{Csengeri+14_AGGC}, and it is approximately equal to the FWHM size of the clump. The \citet{KauffmannPillai10} relation (dashed line) was rescaled to match our values of the dust opacity. The dash-dotted line indicates the threshold of $\Sigma = 0.05 \usk\gram\cm^{-2}$ for massive clumps proposed by \citet{Urquhart+13a} for massive star formation to occur. All the sources are found above it, suggesting that they are potentially forming massive stars. A typical uncertainty for the mass (for subsamples S1 and S2) is shown in the bottom right corner. We also indicate the effect of a hypothetical uncertainty of $50\%$ in $\reff$. Each group of sources is shown with a different symbol and colour, as indicated. The stars refer to sources of subsample S3, while their colour still identifies the group (red = IRB, green = RMS, black = D8, and blue = D24).}  \label{fig:m_r}
\end{figure}

\citet{KauffmannPillai10} derived a relation between mass and radius, to separate clumps with the potential of forming massive stars from those without. A similar threshold for high-mass star formation, with a constant surface density $\Sigma = 0.05 \usk\gram\cm^{-2}$, was derived by \citet{Urquhart+13a} for massive star forming clumps selected cross-correlating the ATLASGAL catalogues with the catalogue of the methanol multi-beam survey and the CORNISH survey. Figure~\ref{fig:m_r} shows that all of the clumps in this sample are found above this empirical relations, implying that these clumps are potentially forming massive stars. In the figure, we used the $\reff$ definition from \citet{Rosolowsky+10} and the sizes of the two-dimensional Gaussian fitted to the emission in \citet{Csengeri+14_AGGC}. With this definition, $\reff$ is approximately equal to the FWHM size of the clump.

\section{Discussion}\label{sec:discussion}

\begin{table}
	 \centering 
	 \caption{Mean and median values for $\tex[\kelvin]$ for the four groups, with their dispersion.} \label{tab:tex_class}
	 \begin{tabular}{l*{5}c} 
		\toprule
		Group    & IRB       & RMS       & D8        & D24       \\
		\midrule
		Mean     & $47$      & $26$      & $22$      & $11$      \\
		Median   & $47$      & $24$      & $16$      & $10$      \\
		$\sigma$ & $14$      & $12$      & $13$      & $3$       \\
		\bottomrule
	 \end{tabular}
\end{table}

\subsection{Linewidths and temperatures}

\begin{figure}[t]
\centering 
\includegraphics[angle=-90,width=0.7\columnwidth]{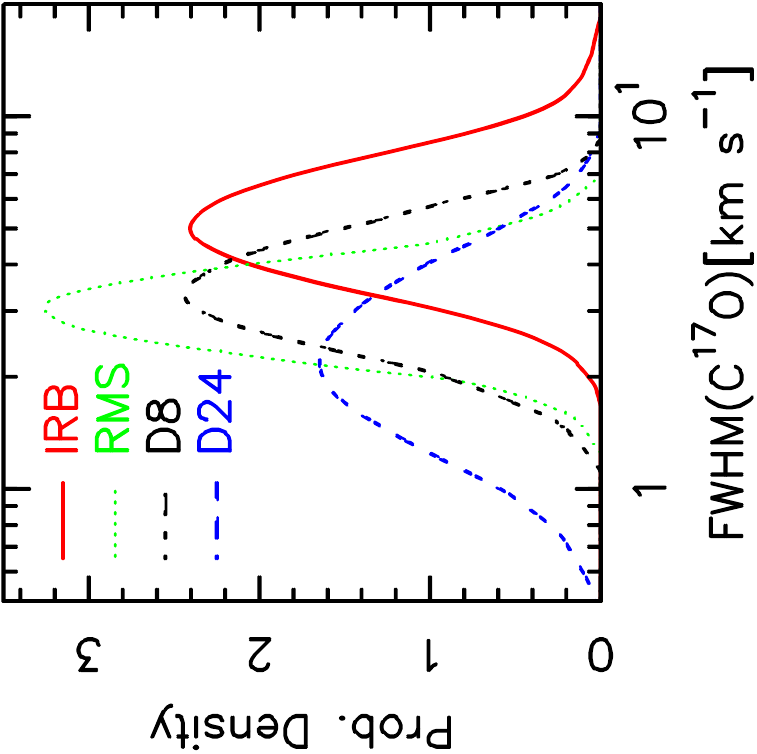} 
\caption{Kernel density estimates applied to the linewidths of $\Istp{12}{17}(3-2)$, for the four groups.} \label{fig:dens_dv_sep}
\end{figure}

Figure~\ref{fig:dens_dv_sep} shows the kernel density estimates \citep[e.g. ][]{WandJones95} applied to the linewidths of $\Istp{12}{17}(3-2)$, for the four groups of sources. We used the \textit{KernSmooth} package and an Epanechnikov kernel. The bandwidth was derived with the direct plug-in method \citep{SheatherJones91}. In practice, the procedure convolves the discrete data points with a kernel function, obtaining a smooth probability density function. The typical linewidth of the $\Istp{12}{17}$ lines for IRB is much larger than that of the other groups. The less evolved RMS and D8 sources show similar values of $\Delta V$, slightly larger than those of D24 and smaller than those typical of the IRB. This may indicate that \optprefix{proto}stars embedded in the molecular gas and dust deliver energy to the surrounding material, making it more turbulent than before their formation, which is reflected in the width of molecular lines. Alternatively, the IR-bright clumps may have been more turbulent from the start, and move much faster through the IR-dark phase, where we find very few clumps with $\Delta V \gtrsim 4\kms$.

\begin{figure}[t]
\centering 
\includegraphics[angle=-90,width=0.7\columnwidth]{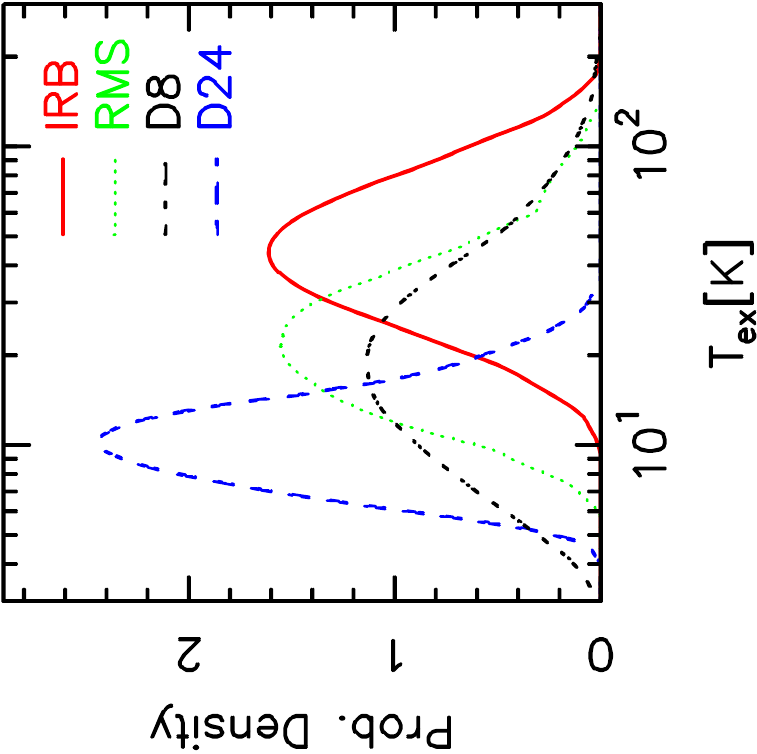} 
\caption{Kernel density estimates applied to the excitation temperatures $\tex$, for the four groups.} \label{fig:dens_tex_sep}
\end{figure}

The kernel density estimates for $\tex$, shown in Figure~\ref{fig:dens_tex_sep}, are obtained in the same way as those for the linewidths. The distributions show that the four groups of objects have a different characteristic temperature, listed in Table~\ref{tab:tex_class}. This indicates an increasing temperature as evolution proceeds. A similar result for gas and/or dust temperature is found in e.g. \citet{Rygl+10}, \citet{Wienen+12} and \citet{Giannetti+13}. 

\subsection{Refined estimate of the isotopic ratios} \label{ssec:refined_estimate}

\begin{figure*} 
\centering 
\includegraphics[angle=-90,width=0.9\textwidth]{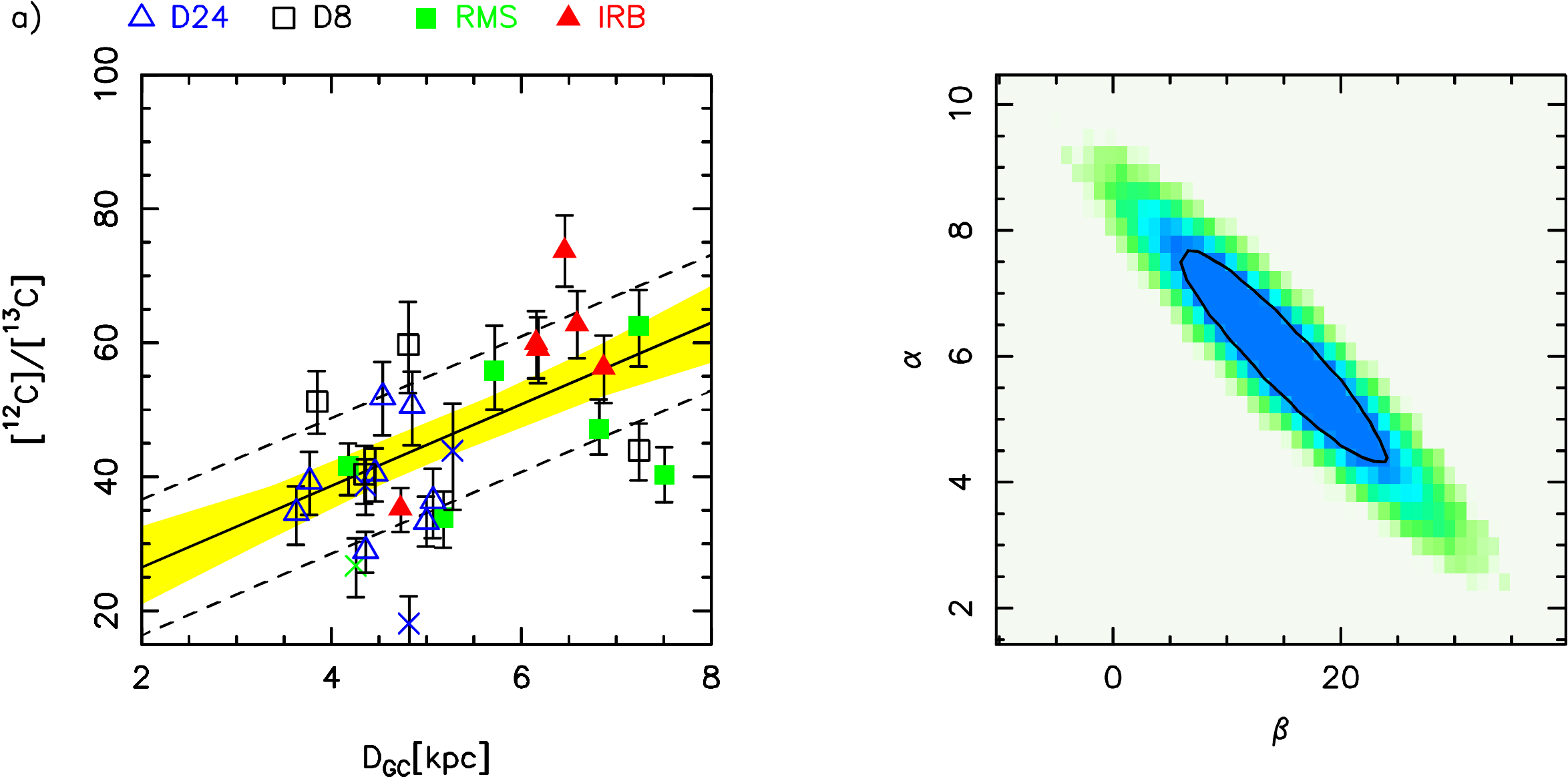} \\
\vspace{0.8cm}
\hspace{0.03\columnwidth}
\includegraphics[angle=-90,width=0.414\textwidth]{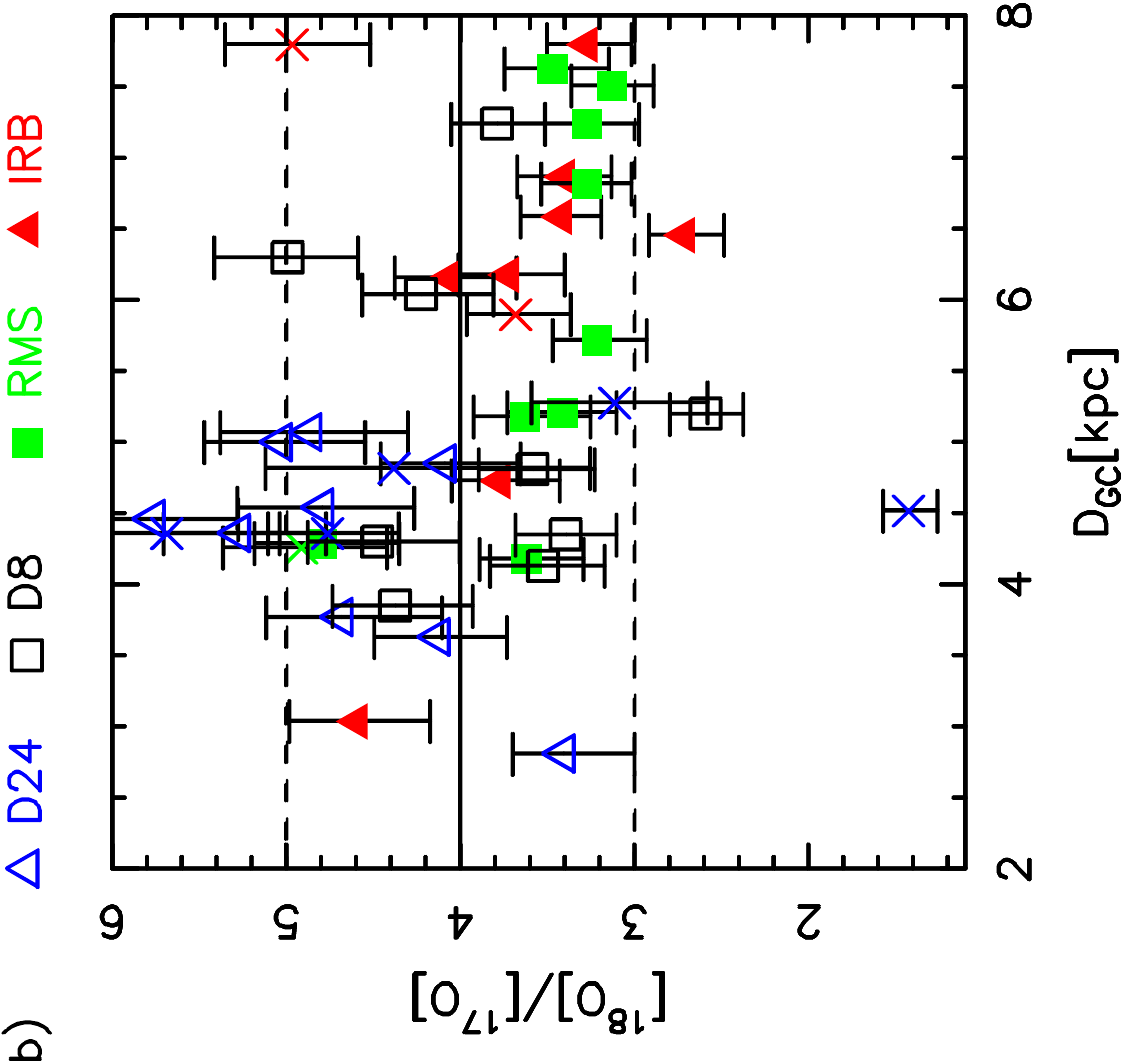} \hspace{0.135\columnwidth}
\includegraphics[angle=-90,width=0.414\textwidth]{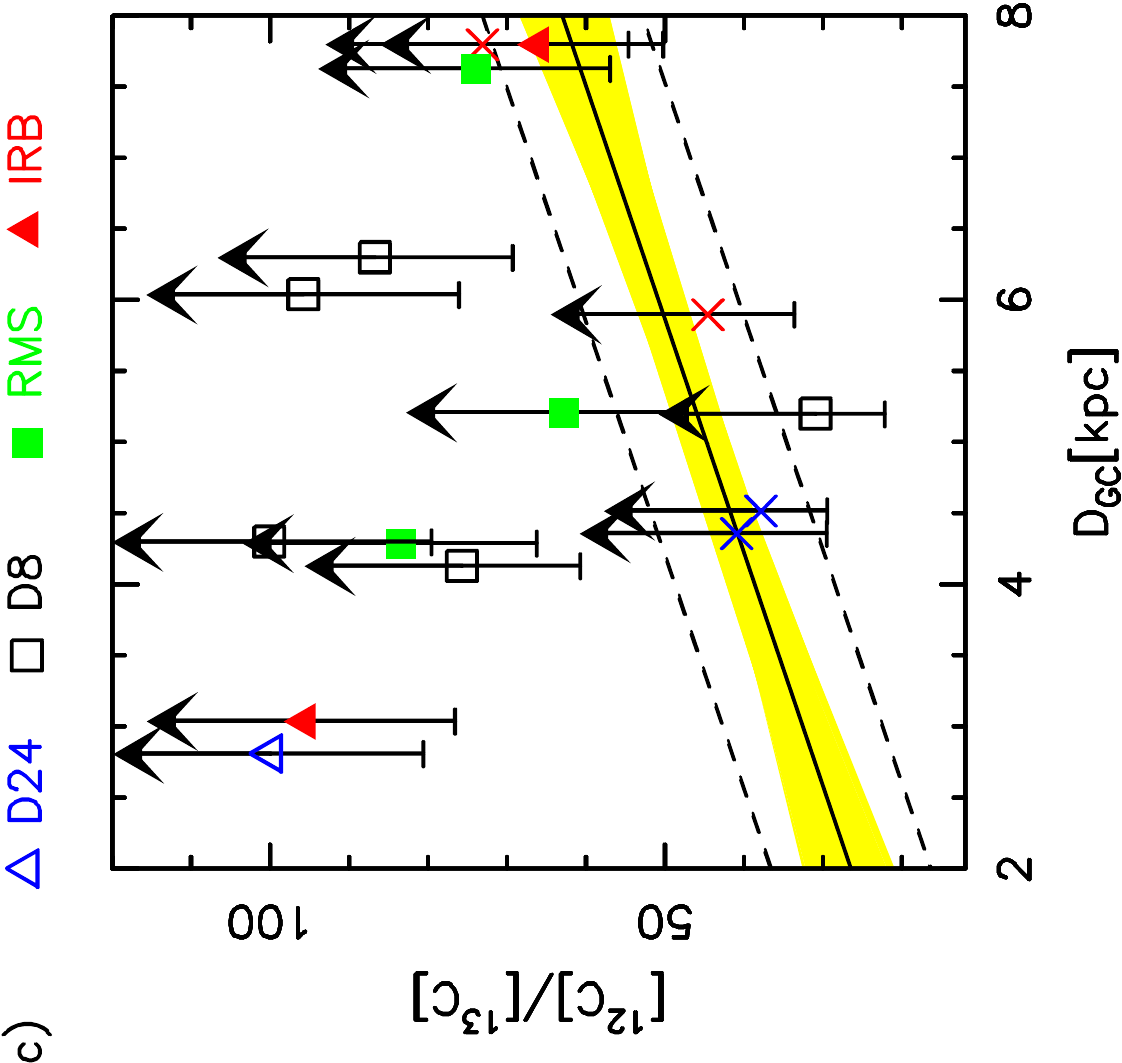} 
\caption{Ratio of different CO isotopologues as a function of Galactocentric radius, as determined from the procedure described in Sect.~\ref{ssec:tau_N}. Each group of sources is shown with a different symbol and colour, as indicated. Velocity components different from the main one are indicated with crosses; the colour of the cross refers to the main component classification. In \textbf{(a)} we show \cratio\ for the sources detected in $\Istp{13}{18}(1-0)$. The black solid line is the best fit, the yellow-shaded area in the left panel indicates the $68\%$ uncertainty, and the dashed lines show the intrinsic scatter of the relation. In the right panel we show the joint probability distribution of the parameters of the fit ($y=\alpha x +\beta$); in black we indicate the $68\%$ contour. Panel \textbf{(b)} is the same as the left panel in (a), for \oratio. Panel \textbf{(c)} shows \cratio\ for the sources undetected in $\Istp{13}{18}(1-0)$ (see text). We show again the fit from the left panel in (a) for clarity.} \label{fig:ref_isotopic_ratios}
\end{figure*}

Figure~\ref{fig:ref_isotopic_ratios} shows the refined estimate of the isotopic ratios for the sources in S1; we also included different velocity components observed in the spectra (indicated with crosses), with the appropriate estimate of the Galactocentric distance obtained with the rotation curve of \citet{BrandBlitz93}.
First of all, both the isotopic ratios \oratio\ and \cratio\ show a large scatter for sources at similar $\dgc$, confirming the existence of source-to-source variations in the relative abundance of these isotopes at the same $\dgc$.

Here we derive the \oratio\ ratio in massive clumps and find it to be consistent with the current estimate of $\sim4$ for $2 \usk\kilo\pctab \lesssim \dgc \lesssim 8 \usk\kilo\pctab$, albeit with a very large scatter that cannot be due to the uncertainty associated with the inferred values of this ratio. The \oratio\ ratio is found to range between $\sim 2$ and $\sim 6$, as can be seen in Fig.~\ref{fig:ref_isotopic_ratios}b. A few sources have $\oratiom\sim5-6$, especially objects in groups D8 and D24, values similar to that found in pre-solar grains \citep[$\oratiom\sim 5.5$, reported in e.g. ][]{Prantzos+96}. \citet{Prantzos+96} discuss how $^{18}$O poses a problem for models of chemical evolution of the Galaxy. Neither [$^{18}$O]/[$^{16}$O] nor \oratio\ can be reproduced by simple models describing the chemical evolution of the Milky Way. The former because it appears to be larger now in the solar vicinity compared to the time of the Sun's formation, whereas it is predicted to decrease with time; the latter because it appears to be substantially larger in the pre-solar cloud than in the present-day ISM, while it is predicted to remain constant \citep[e.g. ][]{Prantzos+96}.
The most probable solution to this discrepancy is the pollution of the pre-solar cloud by a previous generation of massive stars undergoing Type II Supernovae (SNe) events, as discussed in detail by \citet{Young+11} \citep[for ${[^{18}\mathrm{O}]/[^{16}\mathrm{O}]}$ a role could also have been played by the position of the Sun at the time of its formation; see ][]{WielenWilson97}.

\citet{Young+11} show that values as high as those found in pre-solar grains can be reached if intermediate-mass Type II SNe events (from B stars) of a previous stellar generation pollute the surrounding medium, arguing that more massive progenitors are not able to produce similar values of the ratio \oratio. It could be that the sources with a high \oratio\ ratio had an evolution similar to that of the solar system in its earliest formation stages. Because several of these sources are near the inner molecular ring, many massive stars were formed in their vicinity, making this pollution scenario a reasonable one. 

The procedure outlined in Sect.~\ref{ssec:tau_N} also yields values of $B$ ($=\cratiom$, see Eq.~\ref{eq:B}) for those sources without a $\Istp{13}{18}(1-0)$ detection (panel (c) of Fig.~\ref{fig:ref_isotopic_ratios}): we used a flux for the undetected $\Istp{13}{18}$ transition equal to $1\sigma$ with $\sigma$ derived from the spectrum as an input for the Monte Carlo calculations. Some of these sources have quite high values of \cratio\ given their $\dgc$, even if the uncertainties are large. Such large values of \cratio\ may be caused by simple dilution effects if the area of $\Istp{13}{18}(1-0)$ emission is substantially smaller than those of $\Istp{12}{18}(1-0)$ and $\Istp{12}{17}(1-0)$, or if not, it may be caused for instance by selective depletion of the heavier molecule. Excluding these non-detections and using the values for the carbon isotopic ratio from Sect.~\ref{ssec:tau_N} we find a gradient of $\cratiom = 6.1\unc{-1.8}{+1.1} \dgc+ 14.3\unc{-7.2}{+7.7}$, with an intrinsic scatter of $10.1\unc{-2.5}{+2.0}$; this relation gives a most probable value for \cratio\ of $\sim66\pm12$ for the solar vicinity. The relation derived is consistent with that derived from simpler estimates, but each point has reduced uncertainties (cf. Figures~\ref{fig:isotopic_ratios}b and \ref{fig:ref_isotopic_ratios}), thanks to the better estimate of $\tau$ and the combination of all the line ratios. We note that the slope of the relation does not vary much, but the intercept has increased slightly, probably because of the correction for optical depth effects. This new estimate highlights even more the large scatter of the points. 

No clear systematic difference in the isotopic ratios is found between the groups. This indicates that the isotopic ratios set by the complex interplay of the processes mentioned in Sect.~\ref{ssec:abund_dgc} do not change appreciably on a large scale over the short time spanned by the first phases of high-mass star formation.

\subsection{Depletion of CO} \label{ssec:CO_depl}

\begin{figure*} 
\centering 
\includegraphics[angle=-90,width=0.7\textwidth]{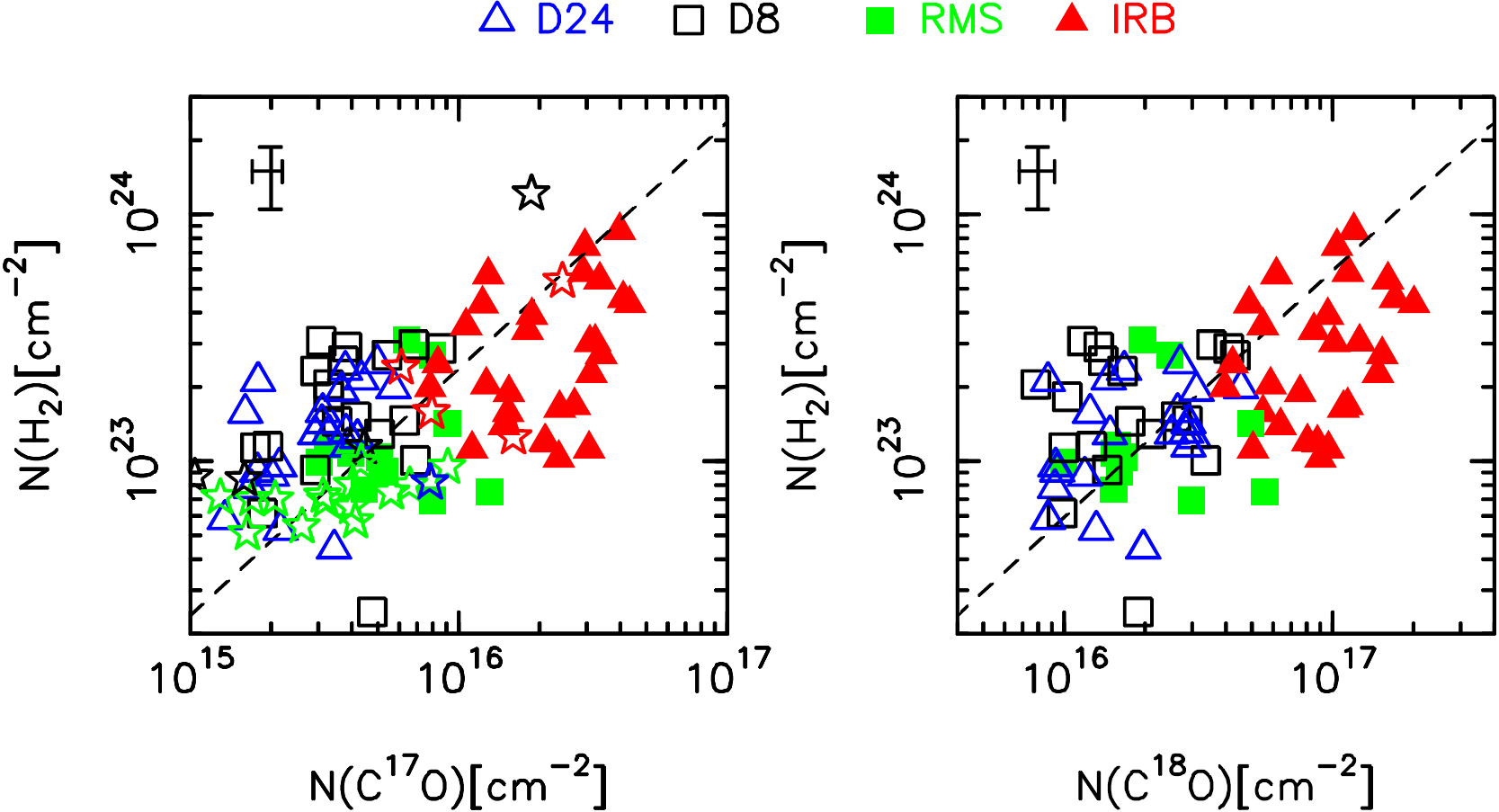} 
\caption{Comparison of $\coldhtwo$ and the CO isotopologues column density. Each group of sources is shown with a different symbol and colour, as indicated. The stars refer to sources of subsample S3, while their colour still identifies the group (red = IRB, green = RMS, black = D8, and blue = D24). The dashed lines indicate the canonical abundance assuming local isotopic ratios. A typical uncertainty (for subsamples S1 and S2) is shown in the top left corner of each panel.} \label{fig:co_col_dens}
\end{figure*}

Figure~\ref{fig:co_col_dens} shows $\coldhtwo$ as a function of $N(\Istp{}{18})$ and $N(\Istp{}{17})$ for the TOP100 sample. The dashed line indicates the canonical abundance of the CO isotopologue in the solar neighbourhood ($\sim1.7\times\pot{-7}$ for $\Istp{}{18}$ and $\sim4.2\times\pot{-8}$ for $\Istp{}{17}$). This is a first, simple approach for investigating CO depletion: in several sources there seems to be an abundance lower than the canonical one, suggesting that CO may indeed be frozen onto the surface of the dust grains. We observe that there appears to be a difference between groups D8 and D24 and groups IRB and RMS, with objects from the former two groups reaching markedly lower abundances. 

\begin{figure} 
\centering 
\includegraphics[width=0.9\columnwidth]{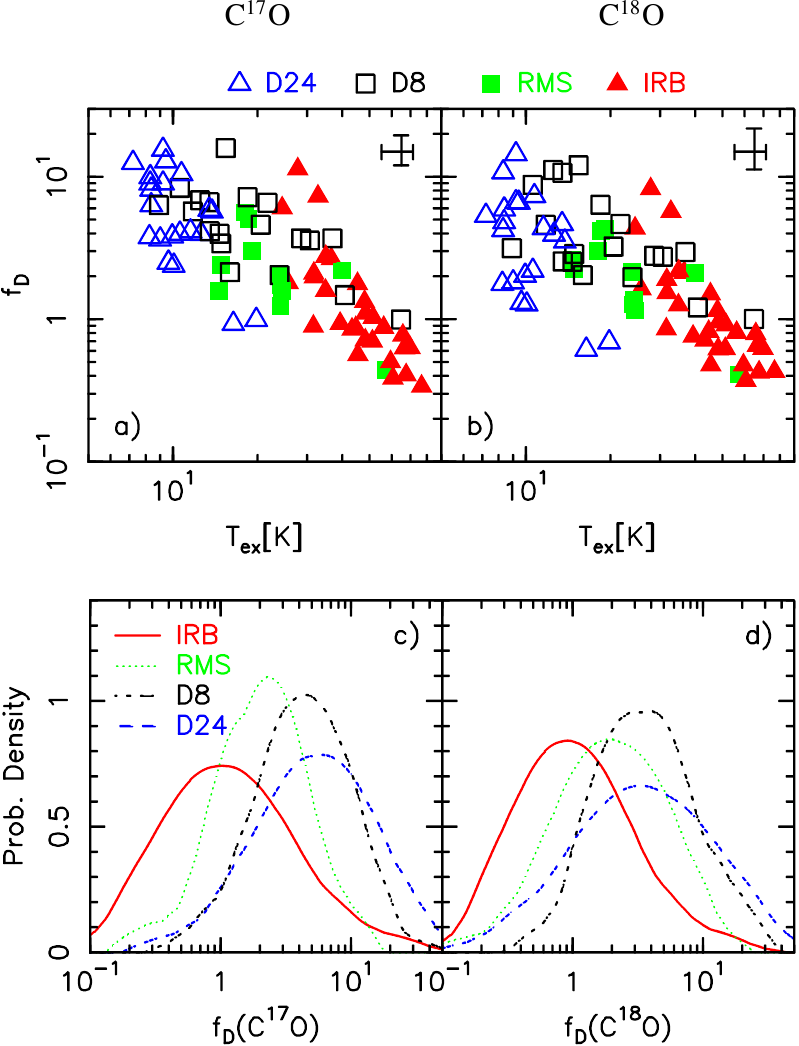} 
\caption{Panels \textbf{(a)} and \textbf{(b)} show CO depletion factors from $\Istp{12}{17}$ and $\Istp{12}{18}$ as a function of $\tex$, respectively. A typical uncertainty is shown in the top right corner. Each group of sources is shown with a different symbol and colour, as indicated. Panels \textbf{(c)} and \textbf{(d)} show the kernel density estimates for the depletion factors, from $\Istp{12}{17}$ and $\Istp{12}{18}$, respectively.} \label{fig:tex_depl}
\end{figure}

\begin{figure} 
\centering 
\includegraphics[angle=-90,width=0.9\columnwidth]{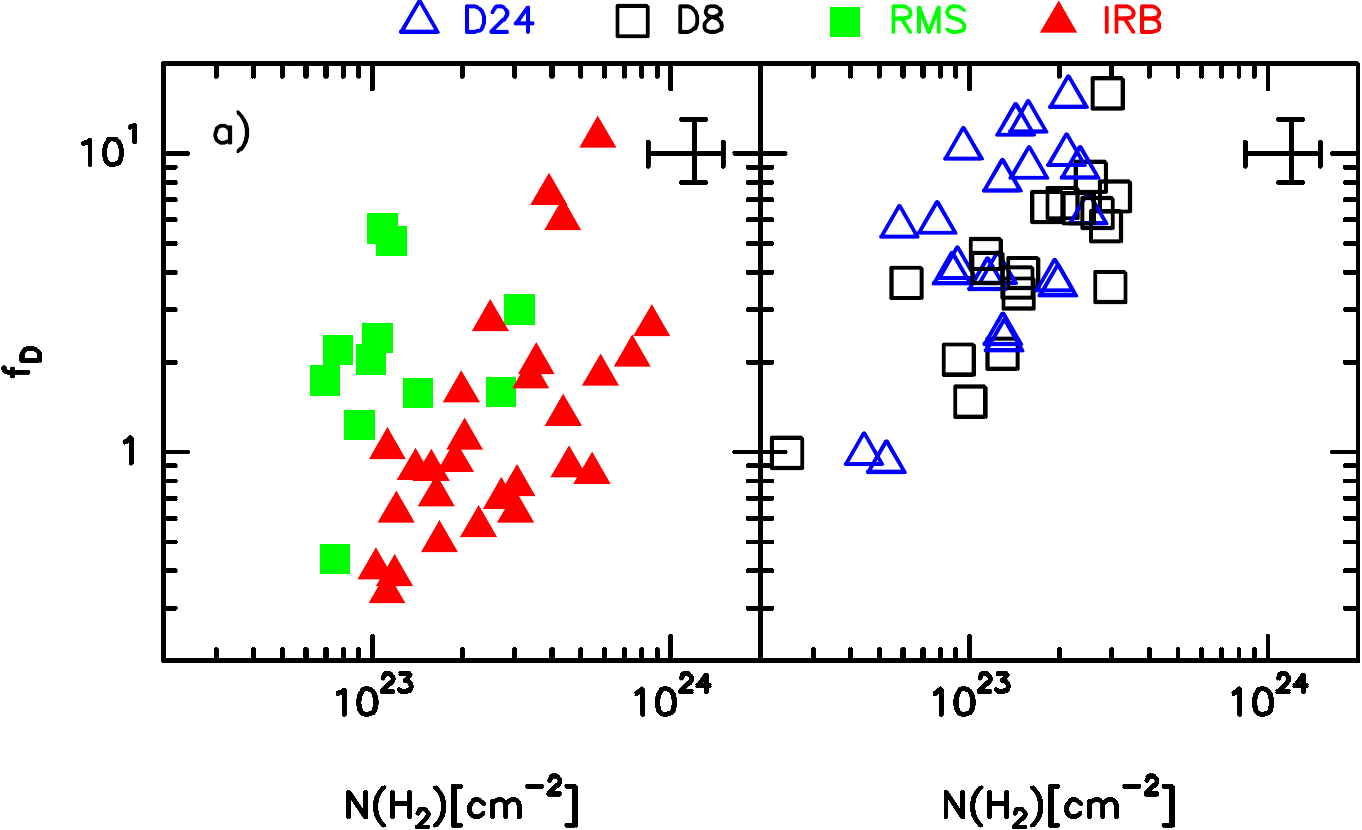} \vspace*{0.4cm} \\ 
\includegraphics[angle=-90,width=0.9\columnwidth]{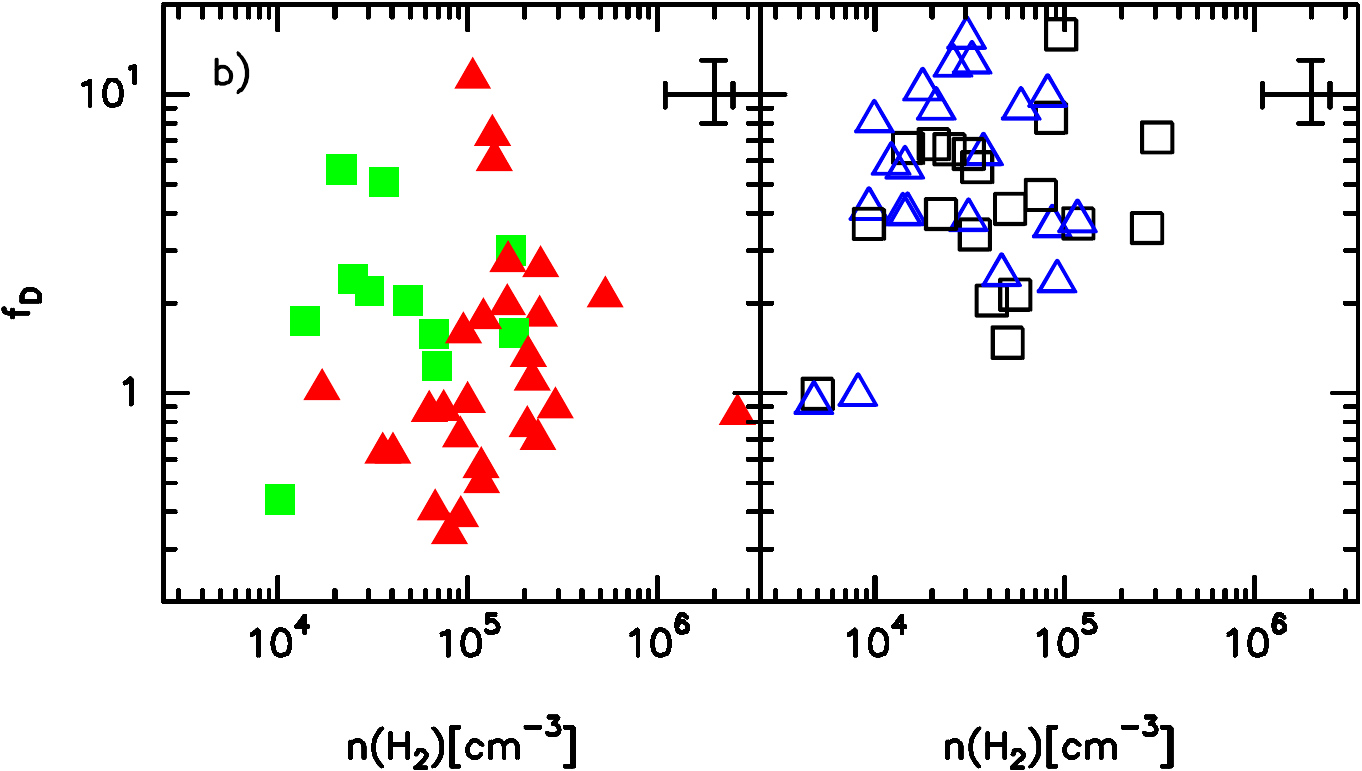} \vspace*{0.4cm} \\
\includegraphics[angle=-90,width=0.9\columnwidth]{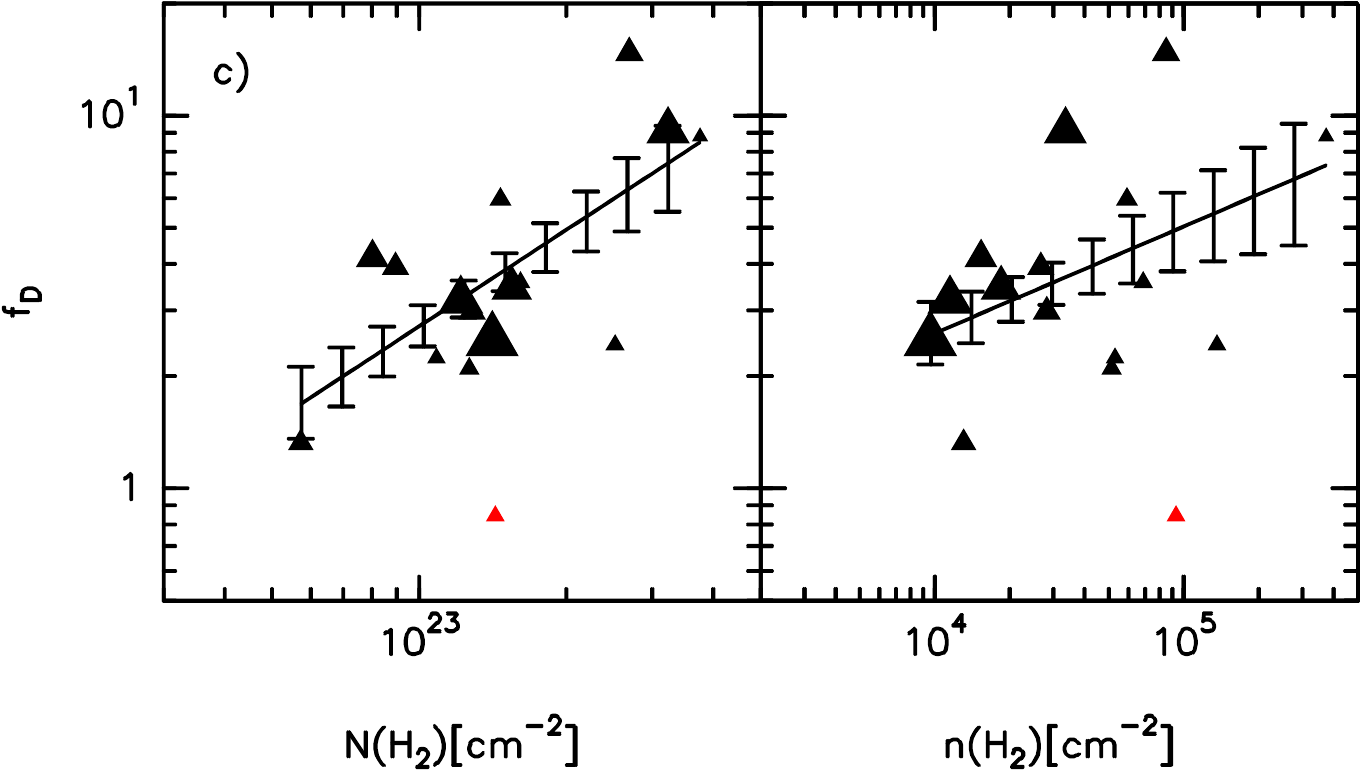}  
\caption{Panel \textbf{(a)} shows the CO depletion factor (from $\Istp{12}{17}$) as a function of $\coldhtwo$. Each group of sources is shown with a different symbol and colour, as indicated. Panel \textbf{(b)} is the same as (a), but for $f_D$ as a function of $\nhtwo$. Panel \textbf{(c)} shows that removing the $\tex$-dependence the trends are still visible (see text). 
The solid lines are simple non-parametric regressions made using the R \citep{Rsoft} \textit{np} package \citep{HayfieldRacine08} (excluding the outlier indicated in red); bootstrapped errors are indicated. 
Symbol sizes are proportional to the distance of the source.} \label{fig:dens_depl}
\end{figure}

Panels (a) and (b) of Fig.~\ref{fig:tex_depl} show the CO depletion factor as a function of the excitation temperature. Here, the depletion factor, which is defined as the ratio of the expected and observed abundance, is determined taking into account the CO abundance gradient with $\dgc$, deriving the expected $\Istp{12}{18}$ abundance according to \citet[][and references therein]{Miettinen+11},
\begin{equation}
X^E_{\Istp{12}{18}}(\dgc) = \frac{X^E_{\Istp{12}{16}}(\dgc)}{[^{16}\mathrm{O}]/[^{18}\mathrm{O}](\dgc)},
\end{equation}
where $X^E_{\Istp{12}{16}}(\dgc)$ and $[^{16}\mathrm{O}]/[^{18}\mathrm{O}](\dgc)$ are the $\Istp{12}{16}$ and $[^{16}\mathrm{O}]/[^{18}\mathrm{O}]$ Galactic gradients, respectively. In detail, the former is $X^E_{\Istp{12}{16}}(\dgc) = X_{\Istp{12}{16}}(8.5\kpc) \mathrm{e}^{1.105-0.13 \dgc [\kilo\pctab]}$, with $X_{\Istp{12}{16}}(8.5\kpc) = 9.5 \times \pot{-5}$, and the latter is $[^{16}\mathrm{O}]/[^{18}\mathrm{O}](\dgc) = 58.8 \dgc[\kilo\pctab] +37.1$.
The expected $\Istp{12}{17}$ was calculated with the \oratio\ derived as described above for the subsamples S1 and S2 (see Sect.~\ref{ssec:tau_N}). Colder sources show larger CO depletion factors, suggesting that in less evolved sources CO in the gas phase is less abundant than in more evolved ones.
Figure~\ref{fig:dens_depl} shows that $f_D$ (from $\Istp{12}{17}$, in this figure) also correlates with the peak column and volume densities \citep[as also found by ][ with a lower angular resolution]{Liu+13}, increasing for higher densities. The peak volume density was calculated according to $\nhtwo = \coldhtwo / (2 \reff)$. 
The average $f_D$ decreases from the D24- to the IRB objects as visible in panels (c) and (d) of Fig.~\ref{fig:tex_depl}; the kernel density estimates for the depletion factors were derived as for the linewidths and $\tex$ (see Sect.~\ref{ssec:tex}). Mean and median values for $f_D$ for each group are listed in Table~\ref{tab:fd_class}. 
The vertical offset between the four groups in Fig.~\ref{fig:dens_depl} may be caused by the depletion and evaporation timescales, that depend on density and temperature, which is different, on average, for each group. 
Denser objects are more prone to high-levels of molecular depletion, because the depletion timescale (i.e. the time after which depletion becomes important) decreases with increasing density, while for higher temperatures, molecules tend to evaporate more rapidly, as the evaporation timescale decreases for increasing temperatures. We note that CO abundances lower than the canonical one are observed towards warm protostellar envelopes \citep[e.g. ][ and references therein]{Fuente+12}; the reason for this is as yet unclear, and the authors suggest that conversion of CO in more complex molecules may happen on the grain surface and/or that the shocks and UV radiation illuminating the walls of the cavities produced by the protostellar outflow may be important factors in explaining the measured CO abundance. 

\begin{figure}[tb]
  \centering
  \includegraphics[width=0.9\columnwidth]{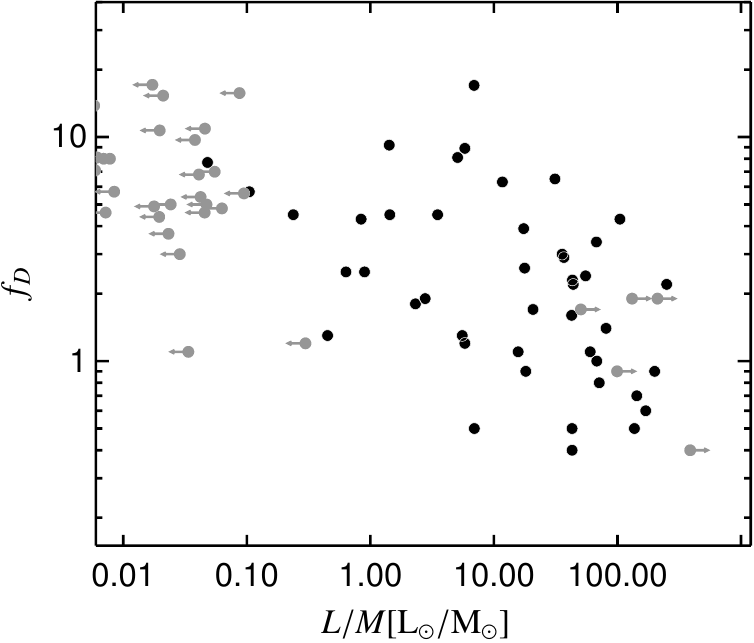}
  \caption{Depletion factor as a function of the $L/M$ ratio. Grey symbols indicate upper and lower limits, depending on the direction of the arrow.} \label{fig:fd_l_m}
\end{figure}

That $f_D$ decreases with age of the source is supported by Fig.~\ref{fig:fd_l_m}, where we find a weak anti-correlation between the depletion factor and the $L/M$ ratio, which we use as an indication of the time evolution of the source \citep[e.g. ][]{Molinari+08, LopezSepulcre+11}. We obtain monochromatic luminosities by cross-matching our sources with the MSX \citep{Egan+03} and WISE \citep{Wright+10} point-source catalogues (see Csengeri et al. in prep. for details); from these, we extrapolated the bolometric luminosities following \citet{Davies+11}, based on the results of \citet{Mottram+11}.
In Fig.~\ref{fig:fd_l_m}, using also the sources with upper and lower limits (for a total of $79$ sources) on the IR flux, we get a Pearson correlation coefficient of $-0.40$, while if we exclude them, $44$ objects remain, and we obtain a correlation coefficient of $-0.34$. This means that there is  a probability of $\sim1.1\%$ of getting similar results from an uncorrelated sample. 
The large scatter can be due to several reasons. The various distances to the targets imply that we probe different spatial scales, where the depletion factor may vary significantly, influencing the average value we derive. For sources in D8 and D24 the ratio between the depletion and the free-fall timescales may differ in different sources, and some of the less evolved ones may not have lived enough for the depletion to become important (see below), or have rather high temperatures ($\gtrsim20\kel$) making the evaporation of molecules from the grains much faster (cf. Fig.~\ref{fig:tex_depl}). In addition, the luminosity estimate is uncertain due to the extrapolation from monochromatic luminosity to bolometric luminosity.

The observed trends could be caused by the assumption of $\tex = \tk = \td$. In order to check if the trends are real, we performed two different tests. In the case of Fig.~\ref{fig:tex_depl} we also used different dust temperatures for the groups: $\td = 10\kel$ for group D24, $\td = 18 \kel$ for group D8, and $\td = 25\kel$ for groups IRB and RMS \citep[cf. e.g, ][]{Giannetti+13, Sanchez-Monge+13b}. We then obtain an average $f_D$ of $\sim 6$ for D24 and $\sim2-3$ for the IRB. Calculating the average $f_D$ in bins of $\tex$ shows that the depletion factor decreases only slightly or stays constant for $\tex \lesssim 25\kel$ ($f_D\sim7-5$), while it decreases more rapidly for $\tex \gtrsim 25 \kel$ (down to $\sim 2$). On the other hand, to test the dependence of $f_D$ on the density (Fig.~\ref{fig:dens_depl}), we selected the sources in a smaller range of $\tex$ ($14-24\kel$, to remove extreme temperatures and have still $\sim20$ sources) and recomputed the densities and depletion factors assuming a constant $\td=20\kel$. Panel (c) of Fig.~\ref{fig:dens_depl} shows that the trends are still visible. The symbol sizes are proportional to the distance of the sources, showing that the trends are not due to distance effects. 

In recent works (\citealt{Wienen+12} and Wienen et al. in prep.) the ammonia (1,1), (2,2) and (3,3) inversion transitions were observed towards a subsample of sources in the TOP100 sample. The observations were carried out with Effelsberg and Parkes telescopes, thus sampling a larger scale than our observations, though with a typical filling factor of $0.1$ \citep{Wienen+12}. The temperatures derived from ammonia are typically higher than those calculated in this work for sources in groups D8 and D24, especially for the latter, with typical values $\sim20\kel$.
We tried to reproduce the observed line intensities with RATRAN assuming a temperature of $20\kel$: the results are briefly discussed in Sect.~\ref{ssec:ratran_model}.

\begin{figure} 
\centering 
\includegraphics[width=0.8\columnwidth]{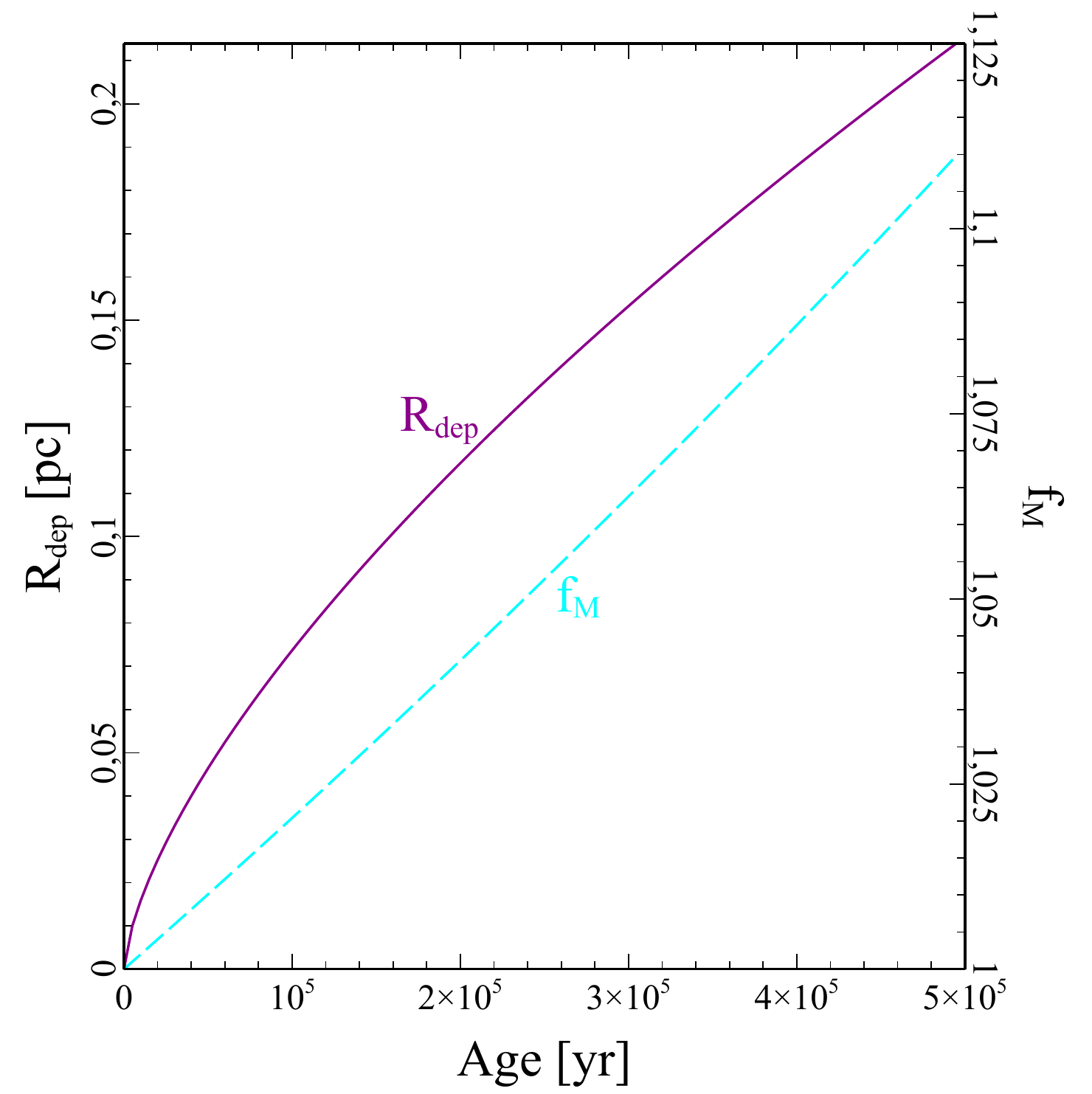}
\caption{Evolution of $\rdep$ and $f_M$ (see text) with age for a typical starless clump with $M = 550\msun$ and a radius of $1\pc$, $\nhtwo \propto r^{-1.5}$ and $T\sim10\kel$.} \label{fig:age_rdep}
\end{figure}

For the sources in groups D8 and D24 we can derive the timescale for CO depletion $\tau_{dep}$, using the expression given in \citet{BerginTafalla07} and the peak volume density of molecular hydrogen. We find that $\tau_{dep}$ is in the range $\pot{3} - \pot{5}\yr$, of the same order of magnitude as the free-fall timescale, which is a rough measure of the starless/IR-quiet clumps lifetime. It is interesting to note that three out of four sources with $f_D<2$ indeed have the largest $\tau_{dep}/\tau_{ff}$ of the sample ($\gtrsim1.5$). \citet{Tackenberg+12} and \citet{Csengeri+14_AGGC} find that lifetimes of massive starless/IR-quiet clumps are of the order of $\pot{4}-\pot{5}\yr$; we can compare this number with the timescale for depletion as a function of radius in clumps with a $\nhtwo \propto r^{-1.5}$ density profile (approximately the mean value for the clumps in the ATLASGAL survey). In practice, assuming a mass for the clump unequivocally sets the volume density of molecular hydrogen at all radii. Using again the equation of \citet{BerginTafalla07}, one can assign a characteristic timescale for CO depletion at each $r$, using the above-cited $\nhtwo$. We can thus derive an estimate of the size $\rdep$ of the central depletion hole as the radius at which $\tau_{dep}$ matches the typical lifetime of massive starless clumps. As illustrated in Fig.~\ref{fig:age_rdep}, this procedure yields values of $\rdep$ in the interval $\sim 0.02-0.1\pc$ for a clump with a mass within $1\pc$ of $\sim550\msun$ and an age of $\pot{4}\yr$ and $\pot{5}\yr$, respectively. The radius increases for larger masses. For a distance of $4\kpc$ a radius of $0.1\pc$ would be slightly in  excess of $5\arcsec$. Figure~\ref{fig:age_rdep} also shows the quantity $f_M$,
\begin{equation}
  f_M = \frac{M_{clump}}{M_{clump} - M(r < \rdep)} = \frac{1}{1 - (R_{eff}/\rdep)^{1.5}},
\end{equation}
which is the ratio between the total mass and the mass outside the depletion radius. If we assume that after $\tau_{dep}(\nhtwo)$ all CO within $\rdep$ is locked onto grains, this is also an alternative way of measuring depletion.

We note that assuming $\tau_{dep} = \tau_{ff}$, one can derive an expression for the critical density $n_{\mathrm{H},crit}$ above which all CO is depleted, that depends only on the temperature $T$ and on the mean grain cross section per H atom $\langle \sigma_\mathrm{H} \rangle$,
\begin{multline}
  n_{\mathrm{H},crit} = \left( \frac{3 \pi}{32 G \mu m_\mathrm{H}} \right)^{-1} \left( \frac{1}{\langle \sigma_\mathrm{H} \rangle S v_{th,\mathrm{CO}}} \right) \\ \approx 5 \times \pot{4} \left( \frac{T}{10\kel} \right)^{-1} \left( \frac{\langle \sigma_\mathrm{H} \rangle}{\pot{-21}\cm^2} \right)^{-2} \left( \frac{S}{1} \right)^{-2} [\cm^{-3}],
  \label{eq:crit_dens}
\end{multline}
where the mean cross section per H atom is $\langle \sigma_\mathrm{H} \rangle = \sigma_g n_g / n_\mathrm{H} \approx \pot{-21}\cm^2$, $v_{th,\mathrm{CO}}$ is the thermal velocity for CO and $S$ is the sticking probability. If coagulation proceeds maintaining constant the density of the grains material (and a spherical shape) $\langle \sigma_\mathrm{H} \rangle$ is proportional to the inverse of the grain size $a$ \citep[see e.g. ][]{Flower+05}, and the critical density ($\propto a^2$) would increase in the central regions where the grains are larger. The value for $n_{\mathrm{H_2},crit}$ (which is equal to $0.5 n_{\mathrm{H},crit}$ in a fully molecular environment) derived from this simple estimate is in good agreement with studies of low-mass starless cores \citep[see e.g. ][]{Bacmann+02}.

\begin{table}[tb]
	 \centering 
	 \caption{Mean and median values for the depletion factor ($f_D$) for the four groups, with the dispersion of the points.} \label{tab:fd_class}
	 \begin{tabular}{l*{5}c} 
		\toprule
		Group    & IRB   & RMS   & D8    & D24    \\
		\midrule
		Mean     & $2  $ & $2  $ & $5  $ & $6  $  \\
		Median   & $1  $ & $2  $ & $4  $ & $5  $  \\
		$\sigma$ & $2  $ & $1  $ & $3  $ & $3  $  \\
		\bottomrule
	 \end{tabular}
\end{table}

\subsubsection{RATRAN modelling} \label{ssec:ratran_model}

The previous analysis assumes that molecules are in LTE. In order to have more solid results for CO abundances we used RATRAN\footnote{\url{www.sron.rug.nl/~vdtak/ratran/}} \citep{HogerheijdeVanDerTak00} to build one-dimensional models of the clumps for typical parameters of the sample. To reproduce the observed CO line intensities and ratios, we build two grids of models, one set with a central heating source, the other with a constant temperature. For the former we vary the luminosity of the central source $L$ \citep[in 3 steps: $\pot{2}\lsun$, $5\times\pot{3}\lsun$, $\pot{5}\lsun$, which is then translated in a power-law temperature profile; see ][]{Rowan-Robinson80,WolfireCassinelli86} and the clump mass (from $\sim200\msun$ to $\sim45000\msun$ in 6 steps), while for the latter we vary the temperature (from $5\kel$ to $15\kel$ in 4 steps)
and the mass (as in the other grid). For the models with a central heating source we use grains with thin ice mantles, while for the isothermal clumps we use grains with a thick ice coating, as suggested by their derived $f_D$. The thick ice mantle on the dust grains increases their opacities, thus reducing the column density of dust (and thus of molecular hydrogen, for a fixed gas-to-dust ratio) needed to obtain a given flux. The clump outer radius, the linewidth and the index $p$ for the power law describing the density profile were fixed at $1\pc$, $4\kms$ and $-1.5$, respectively. We also assumed a typical distance of $4\usk\kilo\pctab$. The models were convolved with a Gaussian as large as the beam size for each wavelength, to simulate our observations.  

\begin{figure*} 
\centering 
\includegraphics[angle=-180,width=0.7\textwidth]{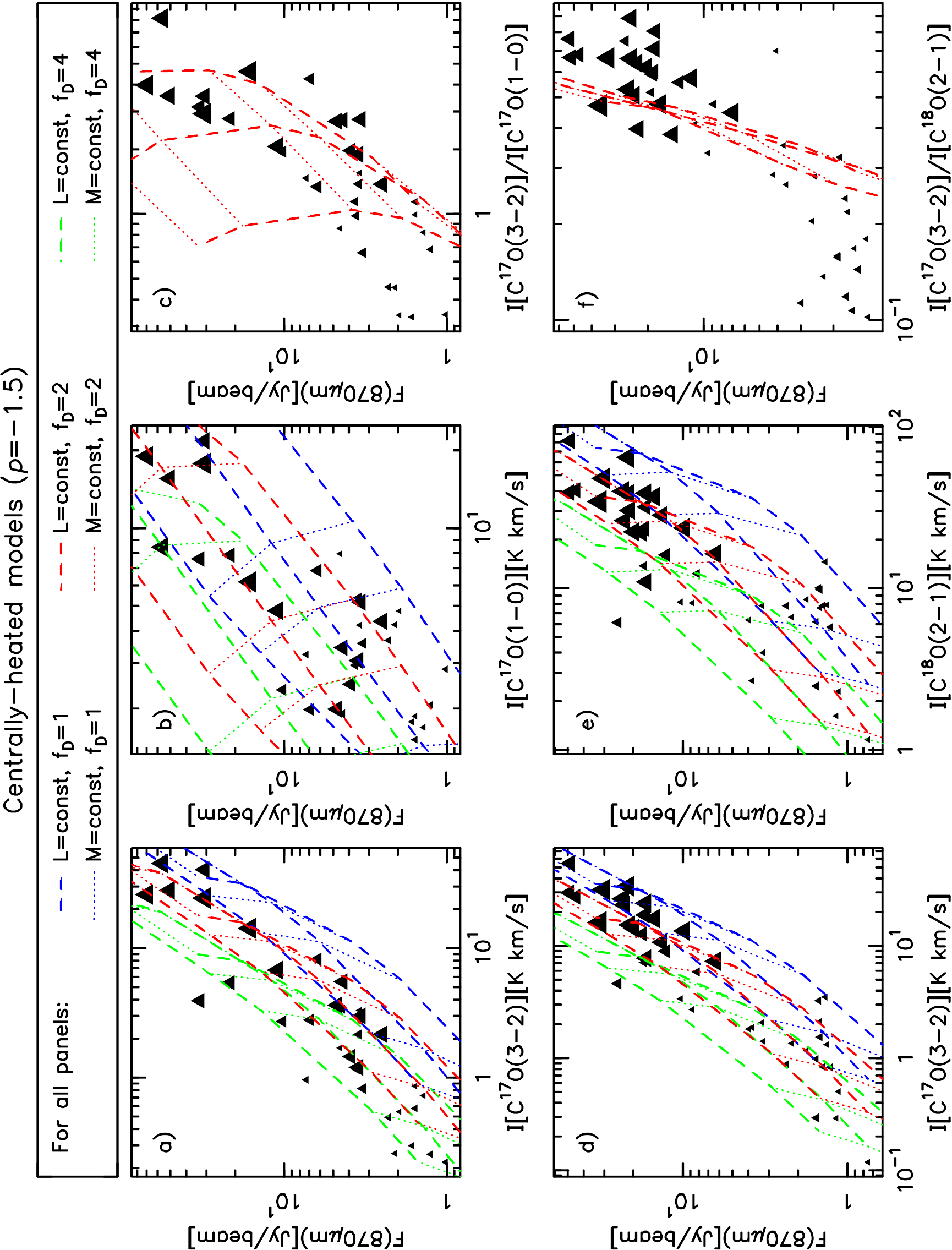} \\
\caption{Line and continuum fluxes of the sources, compared to grids of centrally heated models for typical sources parameters ($D = 4\kpc$, $\Delta V = 4\kms$, $\reff = 1 \pc$, $n \propto r^{-1.5}$), convolved with the appropriate beam. The size of the symbol is proportional to the logarithm of MSX flux in band E ($21\mum$). The grids include a central object heating the gas, and differ for depletion factors identified with different colours, indicated above the panels (blue: $f_D = 1$, red: $f_D = 2$, green: $f_D = 4$). In each panel, the dashed lines connect models with constant luminosity of the central object (three lines, from $\pot{2}\lsun$ to $\pot{5}\lsun$) and varying mass. In a similar way, the dotted lines connect models with the same mass (six lines from $\sim200\msun$ to $\sim45000\msun$) and varying luminosity of the central object. Panels (a) to (c) refer to subsample S1, panels (d) to (f) to subsample S2.} \label{fig:ratran_gen_pl}
\end{figure*}

\begin{figure*} 
\centering 
\includegraphics[angle=-180,width=0.7\textwidth]{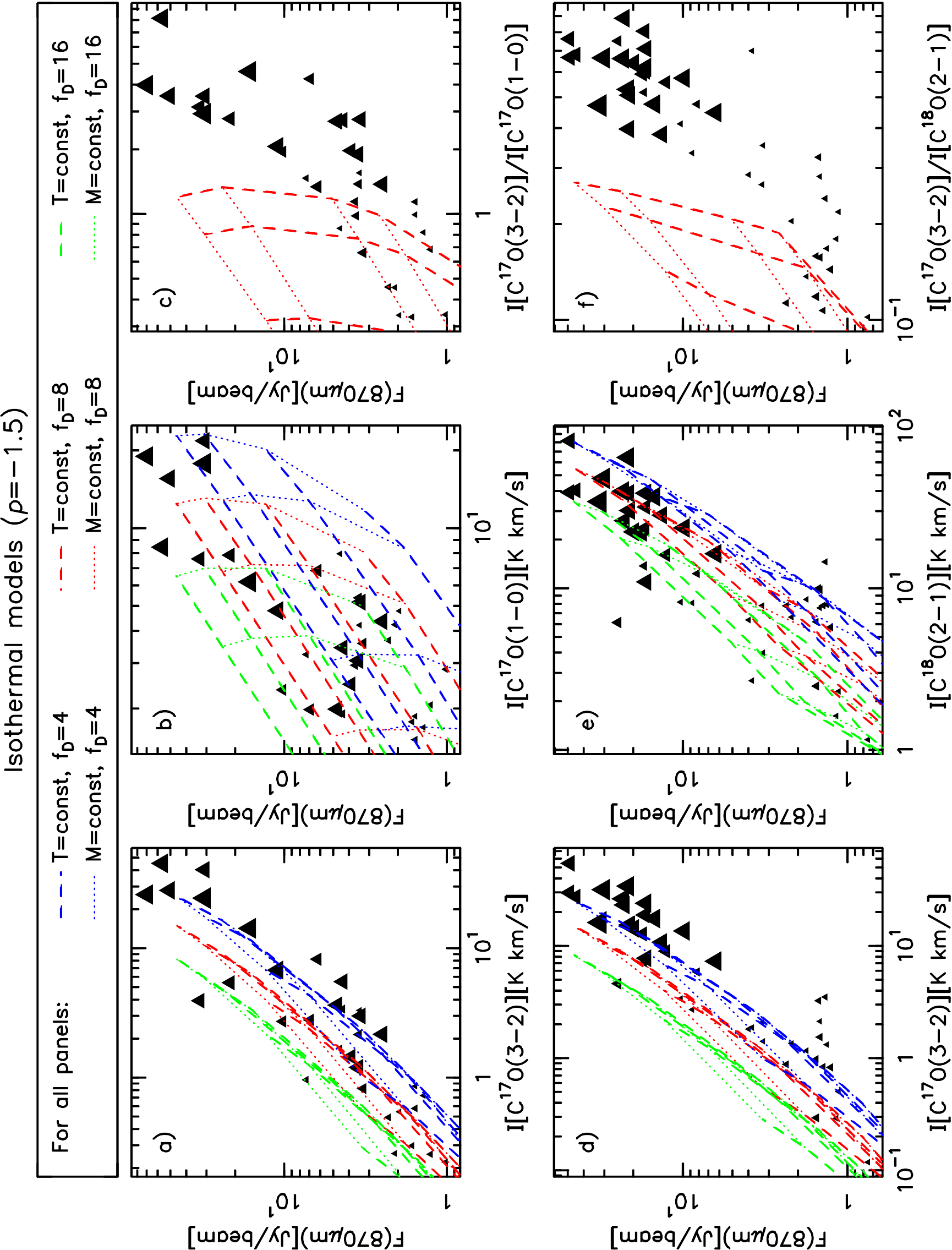}
\caption{Line and continuum fluxes of the sources, compared to grids of isothermal models for typical sources parameters ($D = 4\kpc$, $\Delta V = 4\kms$, $\reff = 1 \pc$, $n \propto r^{-1.5}$), convolved with the appropriate beam. The size of the symbol is proportional to the logarithm of MSX flux in band E ($21\mum$). 
The grids differ for depletion factors identified with different colours, indicated above the panels (blue: $f_D = 4$, red: $f_D = 8$, green: $f_D = 16$). In each panel, the dashed lines connect models with constant temperature (four lines, from $5\kel$ to $15\kel$) and varying mass. In a similar way, the dotted lines connect models with the same mass (six lines from $\sim200\msun$ to $\sim45000\msun$) and varying temperature. Panels (a) to (c) refer to subsample S1, panels (d) to (f) to subsample S2.} \label{fig:ratran_gen_iso}
\end{figure*}

Figure~\ref{fig:ratran_gen_pl} shows the results for centrally-heated models. The data are shown as black triangles, with marker sizes proportional to the logarithm of the MSX flux in the E band ($21\mum$). Panels (a), (b) and (c) refer to subsample S1, where we compare $\cott$, $\cooz$ and the submm flux; on the other hand, panels (d), (e) and (f) show the results for subsample S2, using $\cott$ and $\ceightoto$. In the figure we draw grids with three different depletion factors (coded in three different colours), as shown above the panels. The dotted lines indicate clumps with the same mass, while dashed lines indicate models with the same luminosity of the central object. 
Figure~\ref{fig:ratran_gen_iso} is the same as Fig.~\ref{fig:ratran_gen_pl}, but for isothermal models. Differently from Fig.~\ref{fig:ratran_gen_pl}, dashed lines indicate models with the same temperature, rather than luminosity of the central object. Again, three values of $f_D$ are considered, but they are higher in this case ($1-4$ vs. $4-16$).

The figures show that the sources with line ratios $\Istp{12}{17}(3-2)/\Istp{12}{17}(1-0) \gtrsim 1$ or $\Istp{12}{17}(3-2)/\Istp{12}{18}(2-1) \gtrsim 0.3$ and large submm fluxes (corresponding also to sources with high MSX fluxes) are much better reproduced by a centrally heated clump, with only moderate depletion. On the other hand, sources with very low line ratios are consistent with cold, isothermal clumps, but need a larger depletion factor to reproduce the line fluxes. The fact that for subsample S2 the points are much more scattered around the model grid may be the result of having assumed the same \oratio\ for all sources, while, as observations have different angular resolutions, the geometry of the region and/or a different size of the source than the one assumed for the grid ($1\pc$) may have a larger impact on the measured ratios.

\begin{figure*}[t]
	\centering
	\includegraphics[angle=-90,width=\textwidth]{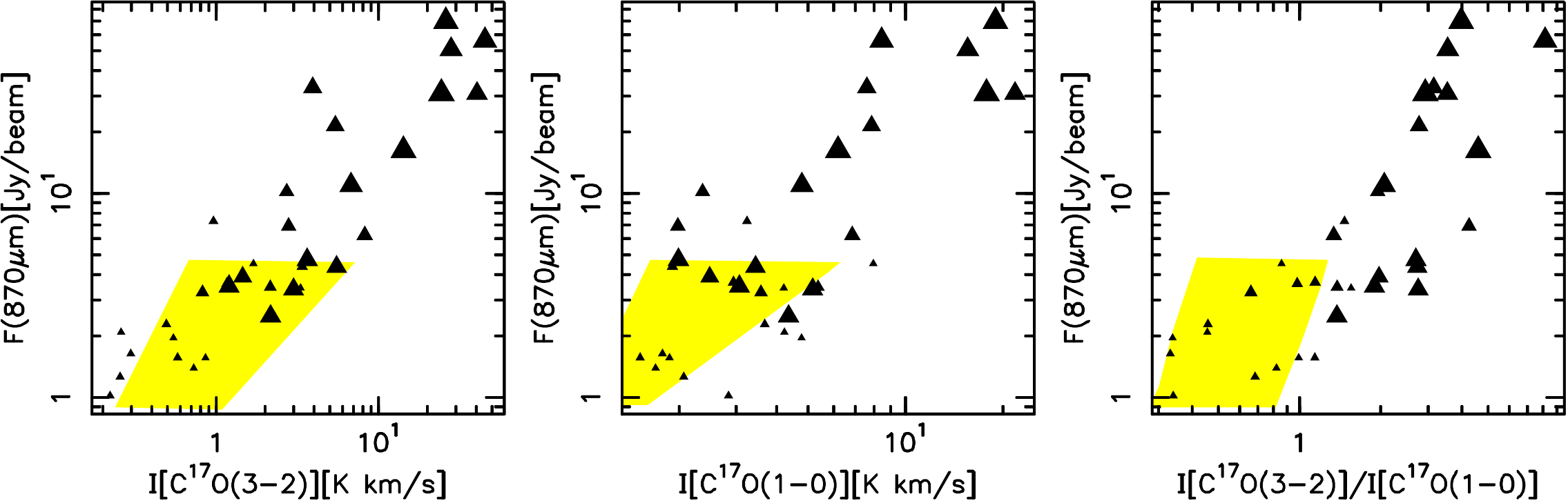}
	\caption{Same as panels (a), (b) and (c) in Fig.~\ref{fig:ratran_gen_iso}, but for models with a drop-profile in the $\Istp{12}{17}$ abundance (see text); the ranges in peak submm fluxes, line intensities and ratios spanned by the models are indicated as a yellow-shaded area.} \label{fig:ratran_gen_hole}
\end{figure*}

For sources in D8 and in D24 we tried to reproduce the observed line ratios using temperatures of the order of those suggested by ammonia. With the simple spherical models considered here, it is not possible to do so for the ratios $I[\Istp{12}{17}(3-2)]/I[\Istp{12}{17}(1-0)]\lesssim1$ with a constant CO depletion. Even limiting the volume density of the clump to $1-3\times\pot{4}\cm^{-3}$ does not yield an acceptable combination of line intensities, ratios and peak fluxes in the submm regime. 
A different approach is to use a drop profile for the CO abundance, i.e. to assume that the abundance is canonical when the density is below a given value, while all CO is locked onto grains for densities above this value. In this case, we constructed a grid of large clumps with a radius $\sim 2\pc$, to try to simulate the contribution from the external layers. We varied the mass in $10$ steps (in the range $550-2900\msun$ within $1\pc$), the critical density defining the size of the central depletion hole (in the range $\pot{4}-\pot{5}\cm^{-3}$) and the temperature of the depleted layers (in the range $8-15\kel$). The temperature of the external layer was assumed to be constant and equal to $20\kel$.  Figure~\ref{fig:ratran_gen_hole} shows the results of these models, as a yellow-shaded area, compared with the peak fluxes and line intensities of S1. The spread in line intensities is due to the different critical densities above which we assume that CO is completely depleted, while the range in submm flux is mainly caused by the different masses of the clump; the temperature of the central regions has a minor impact for the low values considered. Despite the rather high temperature in the low-density external parts of the clumps, the observed molecular-line ratios can be reproduced with this kind of model. 

In conclusion, both using a constant CO abundance throughout the clump and the temperature suggested by the CO line ratio or a drop-profile model we find indications that depletion is important for cold, massive clumps. On the one hand, we have a simple order-of-magnitude estimate of $f_D$, that can be even $10-20$ for cold sources, and on the other hand we can derive the rough size of the depletion hole, to be compared with that derived from the typical lifetimes of massive starless clumps (see Sect.~\ref{ssec:CO_depl}). Both methods suggest depletion zones with radii $\lesssim0.1\pc$, for a typical clump.

\subsubsection{RATRAN modelling of individual sources}\label{ssec:ind_sou_ratran}

A more detailed study is carried out for some individual sources in the sample, to properly take into account the source size, radial density distribution, linewidth, expected abundance and distance. The sources were selected from subsample S1 not to be extremely elongated from the $870\mum$ images. We selected more sources from among groups D8 and D24 objects to confirm and constrain better their large depletion factors.
We build a grid of models with RATRAN, with $13$ equally spaced steps in mass, and $11$ in depletion factor. Depending on whether the clump is centrally heated or isothermal, $15$ equal logarithmic steps in luminosity or $15$ equal steps in temperature. The radius of the cloud was fixed at the value of $\reff$ as given in the ATLASGAL catalogue.
For sources in the first two groups we use models with a central heating source and thin ice mantles, while for groups D8 and D24 we use models with a constant temperature and thick ice mantles for the dust.

We compare the observed values of line and peak $870\mum$ continuum fluxes with those predicted by the model, assigning a probability to each set of free parameters of the model ($M$, $L$ or $T$, $f_D$). For the centrally heated models this probability is calculated according to
\begin{equation}
\begin{split}
 P(M,L,f_D|D,model) = & \: \frac{1}{\psi} \: P(D|M,L,f_D,model) \: \times \\
                                          & \: P(M,L,f_D|model),
\end{split}
 \label{eq:bayes_TOP100}
\end{equation}
where $P(D|M,L,f_D,model)$ is the likelihood, evaluated following
\begin{equation}
  P(D|M,L,f_D,model) = \prod_{i=0}^3 \expo{\frac{-(F_i-F_{mod,i})^2}{2\sigma_{F,i}^2}}.
  \label{eq:likelihood_eval_TOP100}
\end{equation}
In Eq.~\ref{eq:likelihood_eval_TOP100} $F_i$ are the observed integrated line and $870\mum$ peak fluxes, $F_{mod,i}$ are the same fluxes, predicted by the model and convolved with the beam of the observations and $\sigma_{F,i}$ are the flux uncertainties. The normalisation constant $\psi$ is chosen so that
\begin{equation}
 \int P(M,L,f_D | D, model) \: dM \: dL \: df_D = 1.
\end{equation}
Finally, $P(M,L,f_D|model)$ is the \textit{prior}. The priors for $M$ and $f_D$ are constants on the parameter range shown in Fig.~\ref{fig:ind_sou_ratran}. The luminosity derived in this way may not be well constrained, since it is derived only by the $\Istp{12}{17}(1-0)/(3-2)$ line ratio in the central regions probed by our single-pointing observations (i.e. representing the average $\tex$ along the line-of-sight), in the hypothesis of spherical symmetry. Therefore, using luminosities reported in literature, we set loosely informative Gaussian priors on $L$ for the bright sources.
\citet{Fazal+08}, \citet{vanderTak+13} and \citet{Gaume+93} give a luminosity (scaled to the distance used here) of $\sim8\times\pot{3}\lsun$, $\sim\pot{5}\lsun$ and $\sim6\times\pot{4}\lsun$
for AGAL19.882-00.534, AGAL034.258+00.154 and AGAL049.489-00.389, respectively.   

For isothermal models, the equations for assigning probability are identical to those listed above, with temperature instead of luminosity. The prior for the temperature is taken to be constant.

\begin{figure*}[tbp]
  \centering
  \includegraphics[angle=-90,width=0.93\columnwidth]{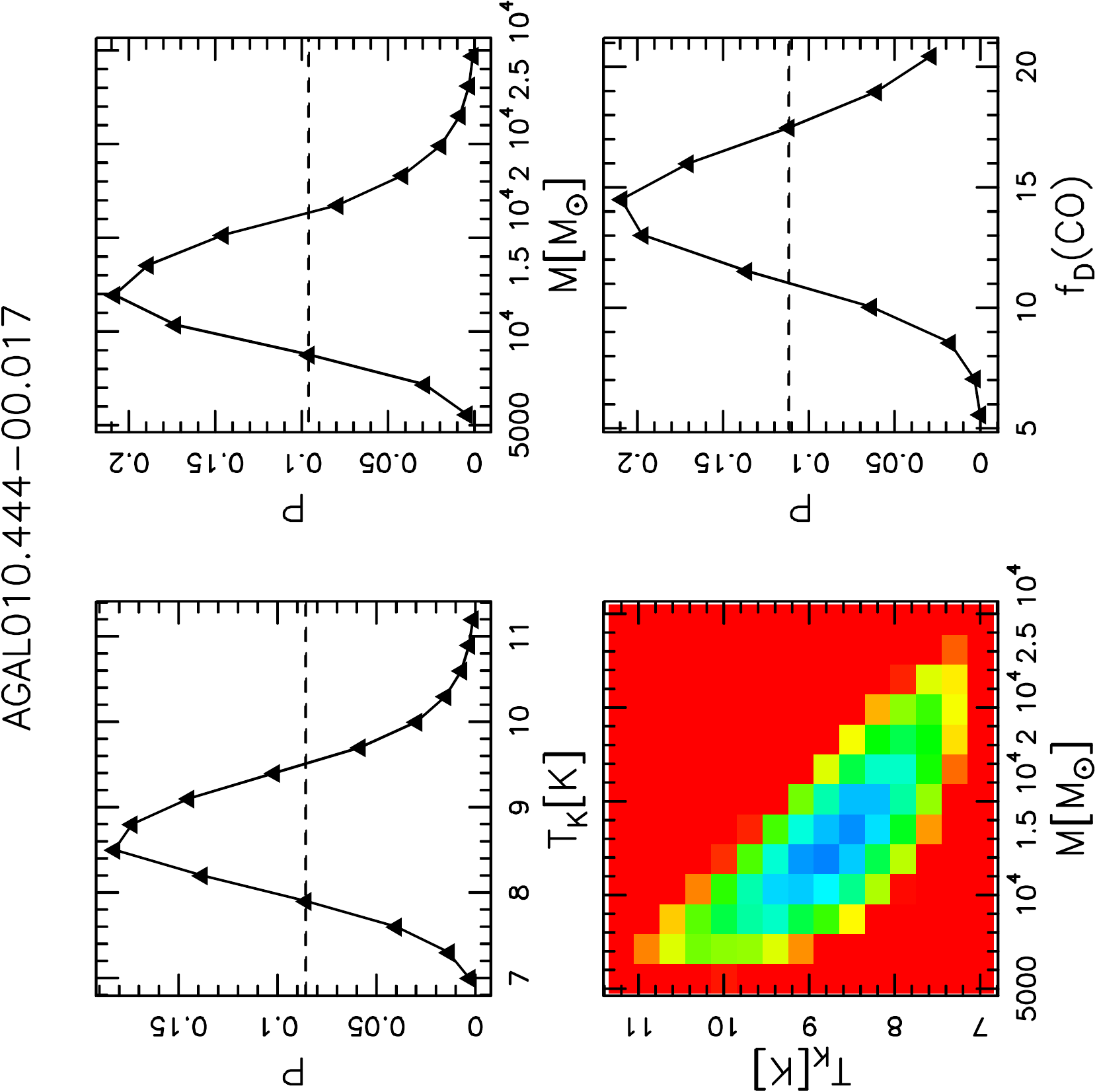} 
  \hspace*{0.06\columnwidth}
  \includegraphics[angle=-90,width=0.9\columnwidth]{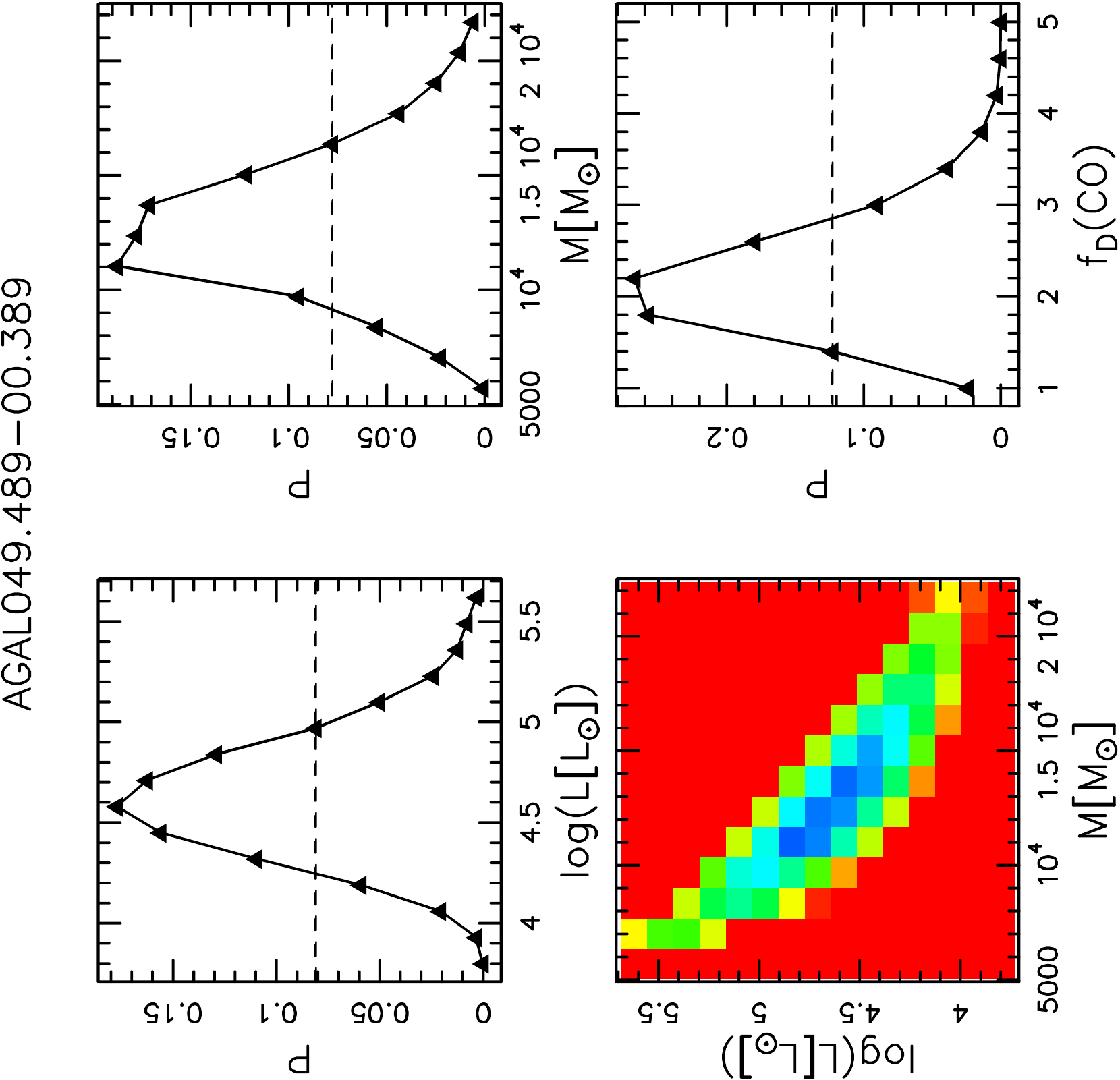}
  \hfill
  \caption{Examples of the RATRAN results for individual sources, using models with a constant abundance profile; the left source belongs to D24, whereas the right one to IRB. For each source the panels show: \textbf{(top left)} marginal probability distribution of the temperature/luminosity, depending on the group of the source, \textbf{(top right)} marginal probability distribution of the mass, \textbf{(bottom left)} joint probability distribution of mass and $T$ or $L$ (depending on whether model is centrally heated or isothermal), \textbf{(bottom right)} marginal probability distribution of the depletion factor. All sources modelled are shown in Fig.~\ref{fig:ind_sou_ratran}, available in the electronic version.} \label{fig:ind_sou_ratran_main_body}
\end{figure*}

Figure~\ref{fig:ind_sou_ratran_main_body} shows two examples of the probability distributions derived for the modelled sources; the complete figure is available in the electronic version (Fig.~\ref{fig:ind_sou_ratran}).

The probability distribution as a function of mass and luminosity for the centrally heated sources, and mass and temperature for the isothermal objects is computed by integrating the probability cube along the depletion axis. 
Tables~\ref{tab:fd_ratran_pl} and \ref{tab:fd_ratran_cold} show the parameters of the sources derived from RATRAN models and the width of their probability distribution, calculated integrating the probability cube along the other two axes. 
All regions include a large quantity of gas and dust, and the temperatures for groups D8 and D24 are very low. 

\begin{table*}[tb]
	\footnotesize
	 \centering 
	 \caption{Parameters derived from RATRAN models for individual sources in groups IRB and RMS, assuming a constant abundance profile.} \label{tab:fd_ratran_pl}
	\begin{tabular}{l*{10}c} 
		 \toprule 
		Source                        &  $L(mean)$                  &  $L(mode)$                  &  $68\%$int                  &  $M(mean)$                  &  $M(mode)$                  &  $68\%$int                  &  $f_D(mean)$                &  $f_D(mode)$                &  $68\%$int               \\

		                              &  $(\pot{3}\lsun)$           &  $(\pot{3}\lsun)$           &  $(\pot{3}\lsun)$           &  $(\pot{2}\msun)$           &  $(\pot{2}\msun)$           &  $(\pot{2}\msun)$           &                             &                             &                          \\

		\midrule
		AGAL019.882-00.534            &$ 20                        $&$ 16                        $&$ 5-43                      $&$ 5.4                       $&$ 5.3                       $&$ 3.3-7.0                   $&$ 2.4                       $&$ 2.1                       $&$ 1.3-3.3                $\\
		AGAL034.258+00.154            &$ 16                        $&$ 8.9                       $&$ 4.0-24                    $&$ 12                        $&$ 12                        $&$ 8-15                      $&$ 1.2                       $&$ 1.3                       $&$ 0.7-1.8                $\\
		AGAL049.489-00.389            &$ 56                        $&$ 38                        $&$ 17-100                    $&$ 130                       $&$ 124                       $&$ 82-167                    $&$ 2.2                       $&$ 2.0                       $&$ 1.2-3.0                $\\
		\bottomrule 
	\end{tabular}
\tablefoot{The columns show the mean, mode and the shortest $68\%$ interval of the derived luminosity, mass and depletion factor.}
\end{table*}

\begin{table*}[tb]
	\footnotesize
	 \centering 
	 \caption{Parameters derived from RATRAN models for individual sources in groups D8 and D24, assuming a constant abundance profile.} \label{tab:fd_ratran_cold}
	\begin{tabular}{l*{10}c} 
		 \toprule 
		Source                        &  $\tk(mean)$               &  $\tk(mode)$               &  $68\%$int                  &  $M(mean)$                  &  $M(mode)$                  &  $68\%$int                  &  $f_D(mean)$                &  $f_D(mode)$                &  $68\%$int               \\

		                              &  $(\kelvin)$                &  $(\kelvin)$                &  $(\kelvin)$                &  $(\pot{2}\msun)$           &  $(\pot{2}\msun)$           &  $(\pot{2}\msun)$           &                             &                             &                          \\

		\midrule
		AGAL008.684-00.367            &$ 12.7                      $&$ 12.5                      $&$ 11.0-14.0                 $&$ 215                       $&$ 205                       $&$ 151-266                   $&$ 5.0                       $&$ 5.0                       $&$ 4.0-6.0                $\\
		AGAL010.444-00.017            &$ 8.8                       $&$ 8.5                       $&$ 7.9-9.6                   $&$ 157                       $&$ 145                       $&$ 104-202                   $&$ 14.4                      $&$ 14.5                      $&$ 10.7-17.6              $\\
		AGAL013.178+00.059            &$ 13.9                      $&$ 13.6                      $&$ 11.9-15.6                 $&$ 29                        $&$ 28                        $&$ 21-36                     $&$ 11.7                      $&$ 11.1                      $&$ 8.7-14.4               $\\
		AGAL014.492-00.139            &$ 8.9                       $&$ 8.8                       $&$ 8.0-9.8                   $&$ 34                        $&$ 31                        $&$ 24-43                     $&$ 9.3                       $&$ 8.7                       $&$ 6.8-11.7               $\\ 
		AGAL018.606-00.074            &$ 16.3                      $&$ 15.0                      $&$ 11.8-18.0                 $&$ 44                        $&$ 37                        $&$ 24-57                     $&$ 3.1                       $&$ 3.0                       $&$ 1.9-4.0                $\\ 
		AGAL028.564-00.236            &$ 8.3                       $&$ 8.3                       $&$ 7.3-9.1                   $&$ 99                        $&$ 95                        $&$ 66-127                    $&$ 7.2                       $&$ 6.7                       $&$ 5.4-8.8                $\\
		AGAL034.411+00.234            &$ 16.1                      $&$ 15.5                      $&$ 14.2-17.7                 $&$ 5.4                       $&$ 5.1                       $&$ 3.9-6.6                   $&$ 9.5                       $&$ 9.1                       $&$ 7.2-11.5               $\\
		\bottomrule 
	\end{tabular} 
\tablefoot{The columns show the mean, mode and the shortest $68\%$ interval of the derived temperature, mass and depletion factor.}
\end{table*}
                    
We note that groups IRB and RMS and groups D8 and D24 indeed show different depletion factors, in the ranges between $1-3$ and $3-15$, respectively. These depletion factors are averaged along the line-of-sight and in the beam. Because groups D8 and D24 contain sources that are in an earlier stage of evolution than IRB and RMS objects, this result confirms that, as evolution proceeds, the molecules are evaporated from the dust grains in the gas phase, and thus the CO-depletion decreases in the environments around high-mass stars. 
Larger values of $f_D$ are found by \citet{Fontani+12}, for a separate sample of massive clumps. In that work the column density of molecular hydrogen is calculated with the expression of \citet{Beuther+05erratum}, yielding column densities $\sim 2.7$ times larger than the expression used in the present paper, because a different dust opacity is used. \citet{Hernandez+11} study in detail depletion in a single IR-dark cloud, showing that a large mass of gas is affected by CO depletion. 
The large depletion factors found in the present paper, together with those derived by \citet{Fontani+12} show that the freeze-out phenomenon not only occurs in high-mass clumps, but it also involves large masses of gas (at least tens of $\msuntab$), as suggested by \citet{Hernandez+11}. 

It is reasonable to expect that the abundance profile is not constant within the clump, 
and that the depletion factors in the densest and coldest regions are larger than those derived;
for example, in \citet{Zhang+09} it is claimed that depletion factors may reach values $\gtrsim1000$ in the central regions of low-mass cores. Our estimate may be, in fact, a lower limit for apparently starless objects, an average within an APEX beam along the line-of-sight, since we do not take into account such variable abundance profiles. A simple test performed with RATRAN shows that if we assume a depletion increasing with density, we find that the depletion factors in the inner regions may be much larger than those found with a constant profile, and, because the outer and less dense layers dominate the emission in this situation, one can use a larger gas and dust temperature to reproduce the observed line ratio, due to non-LTE excitation (similarly to the drop-profile models described in Sect.~\ref{ssec:ratran_model} and below), thus decreasing the mass by a factor of $\sim 3$ for ``cold'' sources. However, without a map of molecular emission we do not have enough constraints for these more sophisticated models. Mapping of the clumps in molecular lines and at higher angular-resolution may unveil the actual mass of gas suffering from depletion and constrain the abundance gradient, as well as a more accurate determination of temperature and mass. 

On the other hand, \citet{Zinchenko+09} and \citet{Miettinen+11} report determination of canonical abundance of CO towards massive clumps. In both works the sources show signs of active star formation, thus being similar to our more evolved sources. For these clumps with a central heating source the abundance may rise near to the \optprefix{proto}star, due to the energy injected in the medium.

\begin{figure*}[tbp]
  \centering
  \includegraphics[angle=-90,width=0.8\textwidth]{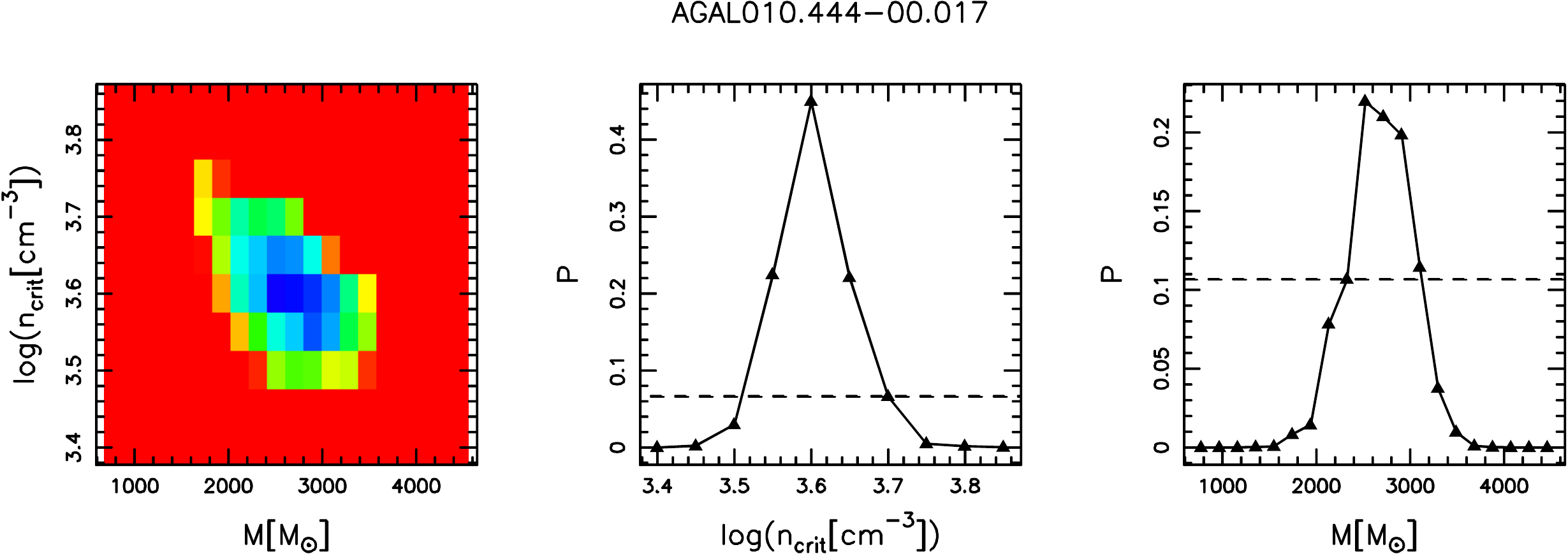} \hfill
  \caption{Example of the RATRAN results for individual sources, using models with a drop profile. The panels show: \textbf{(left)} joint probability distribution of mass and critical density of molecular hydrogen above which all CO is locked onto dust grains, \textbf{(centre)} marginal probability distribution of critical density, \textbf{(right)} marginal probability distribution of mass. All sources modelled are shown in Fig.~\ref{fig:RATRAN_ind_drop}, available in the electronic version.} \label{fig:RATRAN_ind_drop_main_body}
\end{figure*}

\medskip
We try to model the seven sources of Table~\ref{tab:fd_ratran_cold} of groups D8 and D24 also using models with a drop profile.
The model used in Sect.~\ref{ssec:ratran_model} showed that in these cases the temperature of the layers where all CO is locked onto dust grains is not very important in determining the continuum, therefore we fixed it to $20\kel$ for simplicity. The outer non-depleted layers also have $\td=20\kel$ (except for one case, as indicated in Table~\ref{tab:RATRAN_ind_drop}). The radius of the clump was fixed to twice the value of $\reff$, in order to account for lower-density layers. We varied the mass in $20$ equal steps, and the critical density of $\mathrm{H}_2$ $\ncrit$ above which CO is completely depleted, in $10$ equal logarithmic steps.
The results are summarised in Table~\ref{tab:RATRAN_ind_drop}  and shown in Fig.~\ref{fig:RATRAN_ind_drop} (available in the electronic version), of which we give an example in Fig.~\ref{fig:RATRAN_ind_drop_main_body}.

\begin{table*}[tb]
	\footnotesize
	 \centering 
	 \caption{Parameters derived from RATRAN models with a drop profile for individual sources in groups D8 and D24.} \label{tab:RATRAN_ind_drop}
	\begin{tabular}{l*{9}c} 
		 \toprule 
		Source                        &  $M(mean)$                  &  $M(mode)$                  &  $68\%$int                  &  $\ncrit(mean)$             &  $\ncrit(mode)$             &  $68\%$int               &  $\rdep(model)$             &  $\rdep(\tau_{dep})$     \\

		                              &  $(\pot{2}\msun)$           &  $(\pot{2}\msun)$           &  $(\pot{2}\msun)$           &  $(\pot{3}\cm^{-3})$        &  $(\pot{3}\cm^{-3})$        &  $(\pot{3}\cm^{-3})$     &  $(\pctab)$                 &  $(\pctab)$              \\

		\midrule
		AGAL008.684-00.367            &$ 74                        $&$ 70                        $&$ 61-85                     $&$ 13.7                      $&$ 13.8                      $&$ 10.4-18.2              $&$ \sim0.7                   $&$ \sim0.3                $\\
		AGAL010.444-00.017            &$ 27                        $&$ 25                        $&$ 22-32                     $&$ 4.0                       $&$ 4.0                       $&$ 3.2-5.3                $&$ \sim0.8                   $&$ \sim0.2                $\\
	AGAL013.178+00.059\tablefootmark{a}   &$ 9.3                       $&$ 8.9                       $&$ 7.9-11                    $&$ 21.5                      $&$ 20.4                      $&$ 15.7-29.2              $&$ \sim0.3                   $&$ \sim0.2                $\\
		AGAL014.492-00.139            &$ 6.8                       $&$ 6.9                       $&$ 5.6-7.7                   $&$ 12.2                      $&$ 12.6                      $&$ 10.0-14.7              $&$ \sim0.3                   $&$ \sim0.2                $\\ 
		AGAL018.606-00.074            &$ 19                        $&$ 19                        $&$ 16-22                     $&$ 28.7                      $&$ 25.6                      $&$ 14.4-56.9              $&$ \sim0.2                   $&$ \sim0.2                $\\ 
		AGAL028.564-00.236            &$ 16                        $&$ 16                        $&$ 14-18                     $&$ 10.0                      $&$ 10.0                      $&$ 8.0-12.7               $&$ \sim0.4                   $&$ \sim0.2                $\\
		AGAL034.411+00.234            &$ 2.8                       $&$ 2.8                       $&$ 2.4-3.3                   $&$ 78.7                      $&$ 77.5                      $&$ 71.0-87.8              $&$ \sim0.2                   $&$ \sim0.3                $\\
		\bottomrule 
	\end{tabular} 
\tablefoot{
The columns show the mean, mode and the shortest $68\%$ interval of the derived mass and critical density of molecular hydrogen above which CO is completely locked up onto grains, the radius of the central depletion zone derived from the RATRAN models ($\rdep(model)$) and from the depletion timescale as a function of radius ($\rdep(\tau_{dep})$, see text) for a lifetime of the starless clump $\sim\pot{5}\yr$, respectively. \\
\tablefoottext{a}{For this source, to obtain an acceptable fit we used $\td=30\kel$ and $f_D=2$ for the external layers.}
}
\end{table*}

Sources with low $\Istp{12}{17}(3-2)/\Istp{12}{17}(1-0)$ are usually much better reproduced than those with $\Istp{12}{17}(3-2)/\Istp{12}{17}(1-0)\sim 1$, where the models predict an excess of $\Istp{12}{17}(1-0)$. This more detailed analysis shows that the critical density of molecular hydrogen above which all CO is depleted is very similar to that estimated through Eq.~\ref{eq:crit_dens} and that the typical depletion radii derived from the source-specific drop profile models are usually larger than those derived from the depletion timescale as a function of radius (see Sect.~\ref{ssec:CO_depl}); for two sources the discrepancy is even a factor of $3-4$ (or more, for lifetimes shorter than $\pot{5}\yr$).
Given the simplicity of the models and of the estimate obtained using the expression for $\tau_{dep}$ with an age of $\pot{5}\yr$ ($\sim \tau_{ff}$; see Sect.~\ref{ssec:CO_depl}) for the starless clumps, this discrepancy is not surprising. However, it is interesting to note that the larger radii are found for two of the most distant sources (AGAL008.684-00.367 and AGAL010.444-00.017), whereas the nearest source (AGAL034.411+00.234, $1.6\kpc$) has $\rdep(model)\sim\rdep(\tau_{dep})$. This might be the result of a clumpy medium within the beam, where the superposition of smaller regions where depletion is important simulate a larger $\rdep$. 
In any case, the results obtained with these models confirm that a large mass of gas appears to be affected by CO depletion, on average $\sim20\%$ of the total mass (from $\sim 5\%$ up to $\sim 50\%$ in one case).

Another interesting thing to note is that the mass of gas within $\reff$ is up to a factor of $5$ smaller than in the case of a constant abundance profile and $\td=\tex$ (cf. Tables~\ref{tab:fd_ratran_cold} and \ref{tab:RATRAN_ind_drop}) due to the increased temperature and model clump radius ($2\reff$, see above). This difference in mass may be the dominant term in the discrepancy observed in the ratio $\alpha_{vir}=\mvir/M$ between the sources in IRB and RMS, and D8 and D24 (see Sect.~\ref{ssec:virial_stability}).
In addition, it must be kept in mind that when using CO isotopologues to investigate the properties of massive clumps, the emission may come mainly from the outer layers even though the line is optically thin, whereas the inner, dense layers contribution to the molecular emission is minor because of depletion.

\subsection{Stability of the clumps} \label{ssec:virial_stability}

\begin{figure} 
\centering 
\includegraphics[angle=-90,width=0.7\columnwidth]{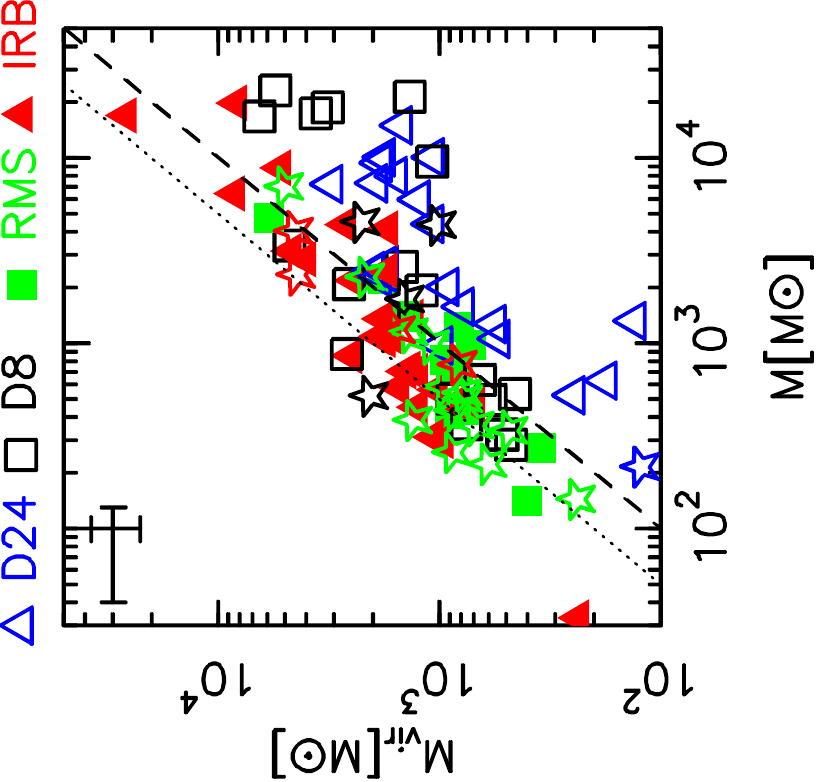} \\
\vspace*{0.5cm}
\includegraphics[angle=-90,width=0.7\columnwidth]{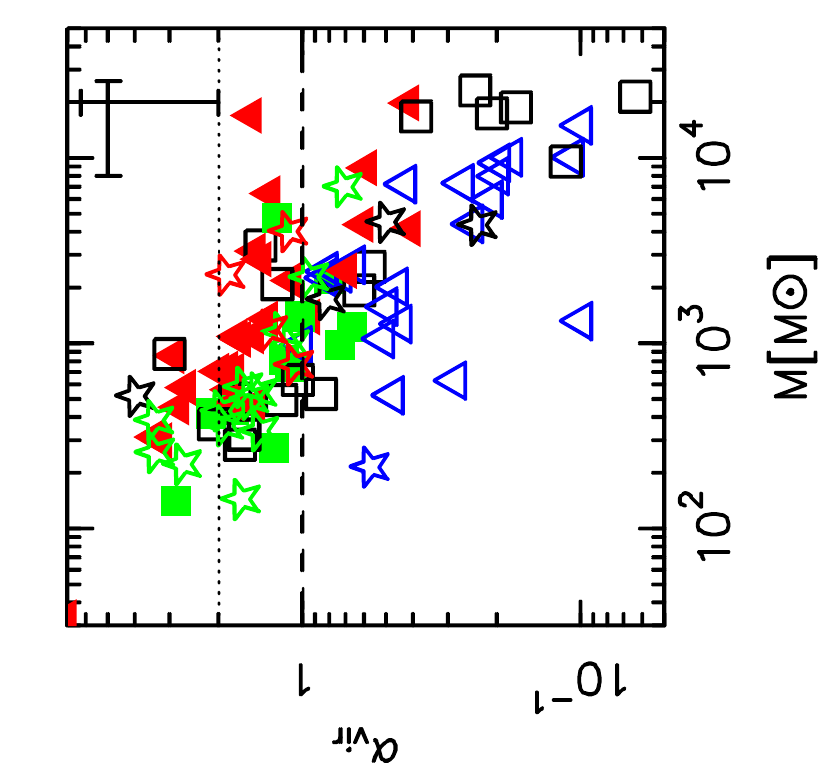} 
\caption{Comparison between the gas mass derived from ATLASGAL fluxes and the virial mass. In the upper panel we show $M \usk vs. \usk \mvir$, in the lower one $M \usk vs. \usk \alpha_{vir}$ ($= \mvir/M$). Each group of sources is shown with a different symbol and colour, as indicated. The stars refer to sources of subsample S3, while their colour still identifies the group (red = IRB, green = RMS, black = D8, and blue = D24). A typical uncertainty (for subsamples S1 and S2) is shown in the top left or right corner. The dashed and dotted lines indicate $\mvir=M$ and $\mvir=2M$, respectively (see text).} \label{fig:m_mvir}
\end{figure}

A simple analysis of the gravitational stability can be performed by comparing the virial mass with the mass calculated from the submm emission by means of Eq.~\ref{eq:mass}, assuming $\td = \tk = \tex$ and that the clump is isothermal. \citet{MacLaren88} give a simple expression to evaluate the virial mass in the case of a spherical clump of radius $r$, with a given density profile:
\begin{equation}
 \mvir[\msuntab] = k \usk r[\pctab] \Delta V^2[\kms].
\end{equation}
In our case we assumed an average radial density profile $\nhtwo \propto r^{-1.5}$ (see Sect.~\ref{ssec:ratran_model}), implying $k=170$, $r=\reff$ (approximately the FWHM of the clump), and we use the $\Istp{}{17}(3-2)$ line to derive $\Delta V$, so that we use the same line for all subsamples. Figure~\ref{fig:m_mvir} shows this comparison, with the dashed and dotted lines indicating $\mvir=M$ and $\mvir=2M$, respectively (see below): in the upper panel we show $M \usk vs. \usk \mvir$ and in the lower one the virial parameter $\alpha_{vir}=\mvir/M$ as a function of $M$. Objects in groups IRB and RMS are mostly distributed around the virial equilibrium line, while the most massive ones in groups D8 and D24 appear to be unstable, i.e. with $\mvir \ll M$. The most massive sources in groups D8 and D24 have virial masses $\sim 10$ times lower than the mass derived from dust emission.

Here we try to understand if these massive sources do really have virial masses much lower than their actual mass, or if it is due to an observational effect. There is the possibility that the temperature we derive is underestimated and that the mass is consequently overestimated. The temperature of apparently starless clumps was previously measured by e.g. \citet{Rygl+10}, \citet{Giannetti+13} and \citet{Sanchez-Monge+13b} making use of other temperature probes, typically NH$_3$. The characteristic $\tk$ was found to be $\sim10-15\kel$, slightly larger than what we measure for group D24. However it is similar to that of group D8 and it is difficult to explain such a large difference in mass as being a consequence of this difference in temperature. The ammonia temperatures measured by \citet{Wienen+12} and Wienen et al. (in prep.) suggest temperatures of $\sim20\kel$ for D8 and D24. Even allowing such temperature accounts for a change by only a factor $\lesssim3$. On the other hand, the temperature we measure in IR-bright clumps may be relatively high and not representative of the whole clump, thus leading one to underestimate $M$. Again it is difficult to explain a difference of a factor of up to $20$ in mass, because $\tk$ from ammonia observations is typically around $20-40\kel$ for such sources.
Depletion of CO may play a role in this: if the degree of depletion is very high in the centre, we might not be able to probe the physical conditions of the gas there, where the density is very high. In this case we would be tracing only the gas in a more external layer (see also Sect.~\ref{ssec:ratran_model}). If the source is starless, we do not expect the temperature in this external layer to be significantly lower than in the centre; on the contrary, in low-mass sources starless objects seem to have a small negative gradient \citep[e.g. ][]{BerginTafalla07}, being colder in the centre, as fewer photons can penetrate the dense layers. On the other hand, we do not measure a large average depletion for star forming sources (see Sect.~\ref{ssec:CO_depl}). It is possible to have a larger effect on the mass if the density in the external layers is low, and the molecule is not in LTE. A simple test of the impact of this on the mass of the clump is described in Sect.~\ref{ssec:ind_sou_ratran}, and it appears to be a major source of uncertainty. The presence of a thick ice coating on the dust grains changes the dust opacity; according to \citet{OssenkopfHenning94} it may increase by $\sim30\%$ at these frequencies, with respect to a model with thin ice mantles. Using a model with thick mantles for ``cold'' clumps in groups D8 and D24 would decrease their masses, but not enough to significantly increase the derived $\mvir/M$ ratios. Another source of uncertainty for $\alpha_{vir}$ is the clump size. The ratio between the major and minor axis of the source is typically between $1$ and $2$. Therefore, this may cause an uncertainty on the virial mass of a factor of $2$ for some clumps.
The assumed gas-to-dust ratio of $100$ may not be applicable to  all sources in the sample, but we do not know enough about its variation to estimate its influence on the derived masses.
The uncertainties described above do not seem to be able to account for the largest differences between the mass derived from dust emission and the virial mass.

A clump mass measured from the submm emission larger than the virial mass was found also by \citet{Hofner+00} and \citet{Fontani+02}. In the latter work, the authors use CH$_3$CCH as a temperature probe, which is a symmetric-top molecule. This molecule gives a reliable temperature estimate, thus reducing the uncertainties on the mass. 
\citet{Giannetti+13} also have temperature determinations both from ammonia and SED fitting, and find that the most massive clumps in their sample have $M>\mvir$. 
In a recent work \citet{Kauffmann+13} study in detail the virial parameter in molecular clouds, for a large collection of objects, from entire clouds to cores. These authors find that in high-mass sources the virial mass can be much lower than the mass derived from observations of the dust. The authors consider $\alpha_{vir}\sim2$ as a limit for the gas motions alone to prevent collapse, to take into account a wide range of shapes and density gradients. However, 
our definition of $\mvir$ already takes into account a density gradient ($n \propto r^{-1.5}$), implying values of $\alpha_{vir}$ about $20\%$ lower than theirs. 
They also discuss the observational uncertainties on $\alpha_{vir}$ and conclude that the very low values of the virial parameter are likely to be real. \citet{Kauffmann+13} fit a straight line to the $\log(M)-\log(\alpha_{vir})$ diagram for high-mass clumps, finding that the slope is $\sim -0.5$, and that $\alpha_{vir,min} \sim 0.2$ from the fit, for high-mass sources. An analogous fit yields very similar results in both parameters for the clumps in the present paper, excluding the sources in groups IRB and RMS, to have a comparable sample. As shown in \citet{Kauffmann+13}, despite the low values of the virial parameter, clumps with $M\gg\mvir$ are not likely to be undergoing gravitational collapse.
\citet{LopezSepulcre+10}, on the other hand, find masses consistent with the virial mass. Thus, discordant claims on the virial stability of massive clumps exist in the literature. 

If these objects are really unstable, the magnetic field may play a significant role in opposing the collapse. For the most massive clumps, fields of the order of a $\milli\gauss$ are needed to halt the collapse. In particular, using the expression from \citet{BertoldiMcKee92} for cold, magnetised clouds, the average critical magnetic field strength ranges between $0.3-3 \usk\milli\gauss$. Magnetic fields of this order of magnitude have been measured in regions of high-mass star formation \citep[e.g. ][]{Crutcher05,Girart+09}. If we accept that the derived mass may be overestimated by up to a factor of $5$, a smaller magnetic field strength is enough to stabilise the clumps, and the magnetic and turbulent energy density are then approximately of the same order of magnitude even for the most unstable sources

\section{Summary and conclusions} \label{sec:summary}

We studied several CO isotopologues in $\nobrkdash{870\mum}\mathrm{bright}$ clumps of the ATLASGAL survey, to investigate the depletion of carbon monoxide, and the \cratio\ and \oratio\ isotopic ratios in regions of (potential) massive star formation. This ``TOP100'' sample consists of 102 clumps.

The sample is selected to include the brightest sources in the submm in different evolutionary stages, separated in four groups \citep[from IR-bright to $24\mum$-dark sources; ][ see Sect.~\ref{sec:sample}]{Motte+07, Nguyenluong+11}. 

From the ratio of different rotational transitions of the observed CO isotopologues we estimate an excitation temperature, with which we derive the molecular column density corrected for optical depth; in order to have a consistent estimate of the optical depth and a refined estimate of the isotopic ratios, we combined all available information from line intensities, ratios and hyperfine structure with a Bayesian approach, allowing the relative isotopic abundances to vary (see Sect.~\ref{ssec:tau_N}).
Comparing the CO isotopologue-column densities with those of H$_2$, derived from the dust emission at $870\mum$, we find that a significant fraction of the sources suffer from depletion, especially the ones with a low $\tex$. 

The main result of this work is that we find that, just like for low-mass cores, depletion of CO is relevant also in massive sources during their early life, and varies with evolution.
Groups D8 (the brightest $8\mum$-dark objects) and D24 (the brightest $24\mum$-dark sources), typically show larger depletion factors and lower temperatures than the more evolved groups IRB and RMS (the brightest objects of the whole survey and the brightest among the remaining sources classified as MYSOs in the RMS survey, respectively; the clumps in both classes have mid-IR emission). 
The depletion factors $f_D$ in the less evolved sources may be as large as $\sim 20$ (see Fig.~\ref{fig:tex_depl} and Table~\ref{tab:fd_ratran_cold}; this corresponds to $f_D\sim 55$ using the dust opacities adopted in \citealt{Fontani+12}). These estimates are likely to be lower limits at least for starless sources, since they are averaged along the line-of-sight and derived with a constant abundance profile. On the other hand, the more evolved sources show a typical depletion $\sim 3-10$ times lower (see Sect.~\ref{ssec:CO_depl}). The larger depletion found in groups D8 and D24 was confirmed with one-dimensional models made with RATRAN, to take into account possible non-LTE effects. 
Therefore, massive objects seem to follow an evolution of CO depletion similar to that of low-mass objects, where carbon monoxide is frozen onto grains, before the feedback from star formation evaporates the molecules back into the gas phase. The column and volume densities are also found to correlate with the depletion factor: denser sources have a larger $f_D$, on average. However, among different groups the typical depletion at a given density decreases with increasing temperature, i.e. for more evolved sources. 
The dependence of the depletion factor on the temperature and density could be explained by the effect of these quantities on the evaporation and freeze-out timescales.

Another way to reproduce the observations is to use models with a central drop in CO abundance. We used models where all CO is locked onto grains for densities above a critical value, whereas below this density, the abundance is canonical. In this case as well the observations can be qualitatively reproduced, again suggesting that CO depletion is important in the dense layers of massive clumps. The estimated critical densities of molecular hydrogen are similar to those found in low-mass starless cores, around a $\mathrm{few} \times \pot{4}\cm^{-3}$. Both these models and the comparison of typical lifetimes of starless/IR-quiet clumps with the timescale for depletion as a function of radius suggest that the radius of the central depletion hole is roughly in the range $\sim 0.02-0.1\pc$.
However, when trying to model individual sources, the radius of this region is found to lie between $\approx0.2-0.8\pc$, even a factor of $\sim3-4$ higher with respect to that estimated from the depletion timescale (see Table~\ref{tab:RATRAN_ind_drop}), confirming that a large mass of gas appears to be affected by CO depletion. This is not surprising, considering that these are simple order-of-magnitude estimates. It is interesting to note that $\rdep$ is typically $0.2-0.3\pc$ for sources with distances $\lesssim6\kpc$, and is $0.7-0.8\pc$ for two out of three of the more distant sources. This may simply indicate that, because the medium is clumpy, more than one condensation affected by depletion exists within the beam. 

A simple test for the stability of clumps is performed comparing the total mass $M$ derived from dust continuum emission and the virial mass. The clumps are found to be near virial equilibrium for groups IRB and RMS. On the other hand, several sources of groups D8 and D24 seem to be gravitationally unstable, in the sense that $\mvir \ll M$. We consider some sources of uncertainty and conclude that this is likely to be real, at least for the sources where $\mvir / M$ is the smallest. The presence of a central region affected by strong CO depletion may be the most important factor of uncertainty in the virial parameter $\alpha_{vir}$. This is because the clumps used for such models have masses within $\reff$ $2-5$ times lower than those derived with models with $\tex=\td$ and a constant abundance profile. 

We also investigated the \oratio\ and \cratio\ isotopic ratios in  the inner Galaxy. We find no significant gradient for \oratio\ as a function of $\dgc$ for $2\kpc \lesssim \dgc \lesssim 8\kpc$, and the ratios are consistent with $\sim4$ with an intrinsic scatter of $\sim 1$. We find that a few sources with $\dgc\sim4\kpc$ have \oratio\ values of $\sim5.5$, similar to the values measured for the pre-solar cloud.
\cratio\ is found to increase with $\dgc$, as predicted by the models of chemical evolution of the Galaxy \citep[see e.g.][]{Prantzos+96}, with $\cratiom\sim66\pm12$ in the solar neighbourhood; however, the intrinsic scatter of the relation is as large as $\sim7-13$. \citet{Milam+05} show that this large scatter is not likely due to processes such as chemical fraction of selective photodissociation, therefore leaving intrinsic differences (e.g. metallicity, star formation history) between sources, or other processes unaccounted for (such as cloud mergers, non-efficient- or radial gas mixing) to explain it. 

\begin{acknowledgements}
This work was partly funded by the Marco Polo program (Universit\`a di Bologna), making it possible for AG to spend three months at the Max Planck Institute f\"ur Radioastronomie in Bonn. AG thanks the Max Planck Institute f\"ur Radioastronomie and FW for their hospitality and support. T.Cs.'s contribution was funded by the ERC Advanced investigator grant GLOSTAR (247078). This research made use of data products from the Midcourse Space Experiment. Processing of the data was funded by the Ballistic Missile Defense Organization with additional support from NASA Office of Space Science. This research has also made use of the NASA ADS, SIMBAD, CDS (Strasbourg) and NASA/IPAC Infrared Science Archive, which is operated by the Jet Propulsion Laboratory, California Institute of Technology, under contract with the National Aeronautics and Space Administration.
\end{acknowledgements}

\bibliographystyle{bibtex/aa}
\bibliography{bibtex/biblio.bib}

\onltab{1}{
\appendix
\section{Tables}

\begin{landscape}

\begin{table}[p] 
\scriptsize 
\centering 
\caption{Distances and classification for sources in subsample S1, derived from C$^{17}$O(3-2).} \label{tab:dist_A}
\medskip 
  \begin{tabular}{lccccccccccccc} 
\toprule  
        Source              & RA             & DEC            & $V_{LSR}$  & $D_{near}$ & $z_{near}$ & $D_{far}$  &  $z_{far}$ &  HI\tablefootmark{a} & $D$ liter       & Method     & Adopted $D$& $\dgc$     & Group  \\
                            & (HH:MM:SS)     & (DD:MM:SS)     & $(\kmstab)$& (kpc)      & (pc)       & (kpc)      & (pc)       &                      & (kpc)           &            & (kpc)      & (kpc)      &        \\
\midrule 
       AGAL008.684-00.367   &$ 18:06:23.35  $&$ -21:37:05.2  $&$   37.99  $&$     4.8  $&$   -30.6  $&$    12.0  $&$   -77.1  $&$                 F  $&$     4.5^{ 2}  $&$      HI  $&$    12.0  $&$     3.9  $& D8     \\
       AGAL008.706-00.414   &$ 18:06:36.72  $&$ -21:37:18.6  $&$   39.44  $&$     4.9  $&$   -35.0  $&$    11.9  $&$   -86.1  $&$                 F  $&$            -  $&$       -  $&$    11.9  $&$     3.8  $& D24    \\
       AGAL010.444-00.017   &$ 18:08:44.59  $&$ -19:54:36.4  $&$   75.87  $&$     6.0  $&$    -1.8  $&$    10.7  $&$    -3.2  $&$                 F  $&$    11.0^{ 2}  $&$      HI  $&$    10.7  $&$     2.8  $& D24    \\
       AGAL010.472+00.027   &$ 18:08:38.24  $&$ -19:51:51.5  $&$   67.56  $&$     5.7  $&$     2.8  $&$    11.0  $&$     5.4  $&$                 F  $&$    11.2^{ 2}  $&$      HI  $&$    11.0  $&$     3.0  $& IRB    \\
       AGAL010.624-00.384   &$ 18:10:28.82  $&$ -19:55:48.4  $&$   -2.86  $&$    17.2  $&$  -115.2  $&$       -  $&$       -  $&$                 N  $&$     2.4^{ 4}  $&$      SP  $&$     2.4  $&$     6.2  $& IRB    \\
       AGAL012.804-00.199   &$ 18:14:13.34  $&$ -17:55:45.3  $&$   36.20  $&$     3.9  $&$   -13.5  $&$    12.7  $&$   -43.8  $&$                 -  $&$     2.4^{ 9}  $&$      PA  $&$     2.4  $&$     6.2  $& IRB    \\
       AGAL013.178+00.059   &$ 18:14:01.32  $&$ -17:28:38.6  $&$   50.42  $&$     4.6  $&$     5.0  $&$    11.9  $&$    12.9  $&$                 N  $&$     4.8^{12}  $&$      HI  $&$     4.6  $&$     4.1  $& D8     \\
       AGAL013.658-00.599   &$ 18:17:24.37  $&$ -17:22:08.0  $&$   48.37  $&$     4.5  $&$   -46.7  $&$    12.1  $&$  -125.8  $&$                 N  $&$    12.3^{ 2}  $&$      HI  $&$     4.5  $&$     4.3  $& RMS    \\
       AGAL014.114-00.574   &$ 18:18:13.14  $&$ -16:57:20.0  $&$   20.85  $&$     2.6  $&$   -25.7  $&$    13.9  $&$  -139.2  $&$                 N  $&$            -  $&$       -  $&$     2.6  $&$     6.0  $& D8     \\
       AGAL014.194-00.194   &$ 18:16:58.77  $&$ -16:42:17.5  $&$   39.23  $&$     3.9  $&$   -13.1  $&$    12.6  $&$   -42.4  $&$                 N  $&$            -  $&$       -  $&$     3.9  $&$     4.8  $& D8     \\
       AGAL014.492-00.139   &$ 18:17:22.09  $&$ -16:24:59.4  $&$   39.46  $&$     3.9  $&$    -9.4  $&$    12.6  $&$   -30.5  $&$                 N  $&$            -  $&$       -  $&$     3.9  $&$     4.8  $& D24    \\
       AGAL014.632-00.577   &$ 18:19:15.22  $&$ -16:30:02.3  $&$   18.53  $&$     2.3  $&$   -23.1  $&$    14.2  $&$  -142.0  $&$                 N  $&$    13.7^{13}  $&$      FR  $&$     2.3  $&$     6.3  $& D8     \\
       AGAL015.029-00.669   &$ 18:20:22.37  $&$ -16:11:38.8  $&$   19.62  $&$     2.4  $&$   -27.6  $&$    14.1  $&$  -164.4  $&$                 N  $&$     2.0^{10}  $&$      SP  $&$     2.0  $&$     6.6  $& IRB    \\
       AGAL018.606-00.074   &$ 18:25:08.27  $&$ -12:45:22.8  $&$   46.63  $&$     3.9  $&$    -5.0  $&$    12.3  $&$   -15.8  $&$                 F  $&$            -  $&$       -  $&$    12.3  $&$     5.0  $& D24    \\
       AGAL018.734-00.226   &$ 18:25:56.05  $&$ -12:42:50.8  $&$   42.61  $&$     3.6  $&$   -14.3  $&$    12.5  $&$   -49.2  $&$                 F  $&$    13.0^{ 2}  $&$      HI  $&$    12.5  $&$     5.2  $& D8     \\
       AGAL018.888-00.474   &$ 18:27:07.69  $&$ -12:41:36.4  $&$   66.06  $&$     4.7  $&$   -39.2  $&$    11.3  $&$   -93.6  $&$                 N  $&$     3.4^{ 2}  $&$      HI  $&$     4.7  $&$     4.3  $& D8     \\
       AGAL019.882-00.534   &$ 18:29:14.20  $&$ -11:50:28.4  $&$   44.97  $&$     3.7  $&$   -34.2  $&$    12.3  $&$  -115.1  $&$                 N  $&$     3.3^{ 2}  $&$      HI  $&$     3.7  $&$     5.2  $& RMS    \\
       AGAL022.376+00.447   &$ 18:30:24.26  $&$ -09:10:36.0  $&$   54.01  $&$     4.0  $&$    30.9  $&$    11.8  $&$    91.7  $&$                 N  $&$            -  $&$       -  $&$     4.0  $&$     5.1  $& D24    \\
       AGAL023.206-00.377   &$ 18:34:54.99  $&$ -08:49:15.5  $&$   78.17  $&$     5.0  $&$   -33.1  $&$    10.6  $&$   -69.7  $&$                 N  $&$    10.9^{ 2}  $&$      HI  $&$     5.0  $&$     4.3  $& D8     \\
       AGAL024.629+00.172   &$ 18:35:35.61  $&$ -07:18:17.9  $&$  116.01  $&$     6.9  $&$    20.9  $&$     8.5  $&$    25.8  $&$                 -  $&$            -  $&$       -  $&$     6.9  $&$     3.6  $& D24    \\
       AGAL028.564-00.236   &$ 18:44:18.04  $&$ -03:59:40.0  $&$   87.24  $&$     5.5  $&$   -22.4  $&$     9.5  $&$   -38.9  $&$                 -  $&$            -  $&$       -  $&$     5.5  $&$     4.5  $& D24    \\
       AGAL028.861+00.066   &$ 18:43:46.06  $&$ -03:35:32.2  $&$  104.02  $&$     6.7  $&$     7.7  $&$     8.2  $&$     9.5  $&$                 N  $&$            -  $&$       -  $&$     6.7  $&$     4.2  $& RMS    \\
       AGAL030.848-00.081   &$ 18:47:55.45  $&$ -01:53:29.5  $&$   94.35  $&$     7.2  $&$   -10.1  $&$     7.2  $&$   -10.1  $&$                 N  $&$            -  $&$       -  $&$     7.2  $&$     4.4  $& D24    \\
       AGAL030.893+00.139   &$ 18:47:13.37  $&$ -01:45:04.0  $&$   97.26  $&$     6.4  $&$    15.4  $&$     8.2  $&$    19.7  $&$                 N  $&$            -  $&$       -  $&$     6.4  $&$     4.5  $& D24    \\
       AGAL031.412+00.307   &$ 18:47:34.33  $&$ -01:12:45.5  $&$   98.15  $&$     6.7  $&$    35.7  $&$     7.8  $&$    42.0  $&$                 N  $&$     5.6^{ 4}  $&$      SP  $&$     5.6  $&$     4.7  $& IRB    \\
       AGAL034.258+00.154   &$ 18:53:18.52  $&$  01:15:01.5  $&$   58.52  $&$     4.0  $&$    10.6  $&$    10.1  $&$    27.1  $&$                 N  $&$     2.1^{ 4}  $&$      SP  $&$     2.1  $&$     6.9  $& IRB    \\
       AGAL034.401+00.226   &$ 18:53:18.79  $&$  01:24:37.8  $&$   57.44  $&$     3.9  $&$    15.4  $&$    10.1  $&$    40.1  $&$                 -  $&$     1.6^{11}  $&$      PA  $&$     1.6  $&$     7.2  $& RMS    \\
       AGAL034.411+00.234   &$ 18:53:18.18  $&$  01:25:23.0  $&$   58.01  $&$     3.9  $&$    16.0  $&$    10.1  $&$    41.2  $&$                 -  $&$     1.6^{11}  $&$      PA  $&$     1.6  $&$     7.2  $& D8     \\
       AGAL034.821+00.351   &$ 18:53:38.07  $&$  01:50:28.6  $&$   58.29  $&$     4.0  $&$    24.2  $&$    10.0  $&$    61.3  $&$                 -  $&$     3.4^{ 2}  $&$      HI  $&$     4.0  $&$     5.7  $& RMS    \\
       AGAL035.197-00.742   &$ 18:58:12.79  $&$  01:40:37.6  $&$   34.66  $&$     2.5  $&$   -32.5  $&$    11.4  $&$  -147.5  $&$                 N  $&$     2.2^{18}  $&$      PA  $&$     2.2  $&$     6.8  $& RMS    \\
       AGAL037.554+00.201   &$ 18:59:09.95  $&$  04:12:16.3  $&$   86.09  $&$     6.7  $&$    23.4  $&$     6.7  $&$    23.4  $&$                 -  $&$     5.0^{ 2}  $&$      HI  $&$     6.7  $&$     5.2  $& RMS    \\
       AGAL043.166+00.011   &$ 19:10:13.40  $&$  09:06:11.2  $&$    5.26  $&$     0.4  $&$     0.1  $&$    12.0  $&$     2.3  $&$                 -  $&$    11.4^{ 7}  $&$      PA  $&$    11.4  $&$     7.8  $& IRB    \\
       AGAL049.489-00.389   &$ 19:23:44.15  $&$  14:30:29.5  $&$   57.16  $&$     5.5  $&$   -37.5  $&$     5.5  $&$   -37.5  $&$                 T  $&$     5.4^{ 6}  $&$      PA  $&$     5.4  $&$     6.5  $& IRB    \\
       AGAL053.141+00.069   &$ 19:29:17.80  $&$  17:56:17.9  $&$   22.29  $&$     1.9  $&$     2.4  $&$     8.3  $&$    10.3  $&$                 -  $&$     1.6^{ 2}  $&$      HI  $&$     1.9  $&$     7.5  $& RMS    \\
       AGAL059.782+00.066   &$ 19:43:10.83  $&$  23:44:03.3  $&$   23.12  $&$     2.6  $&$     2.9  $&$     6.0  $&$     6.9  $&$                 -  $&$     2.2^{19}  $&$      PA  $&$     2.2  $&$     7.6  $& RMS    \\
\bottomrule                                                                                                                                                                                     
\end{tabular} 

\\ 
\medskip 
\tablefoot{
\tablefoottext{a}{From Wienen et al., in prep.} \\
1: Caswell 1975, 2: Green \& McClure-Griffith 2011, 3: Reid et al. 2009, 4: Moises et al. 2011, 5: Bik et al. 2005, 6: Sato et al. 2010, 7: Gwinn et al. 1992, 8: Blum et al. 2001, 9: Immer et al. 2012, 10: Xu et al. 2011, 11: Kurayama et 
al.  2011, 12: Pandian et al. 2008, 13: Sewilo et al. 2004, 14: Bica et al. 2004, 15: Busfield et al. 2006, 16: Urquhart et al. 2011, 17: Figueredo et al. 2005, 18: Zhang et al. 2009, 19: Xu et al. 2009. \\
Methods: PA = Direct parallax measurement; SP = Spectrophotometric measurement; H\textsc{i} = H\textsc{i} self-absorption; FR = Formaldehyde self-absorption. \\
In the ``H\textsc{i}'' column T, N and F indicate the tangent point, the near and far distances, respectively.\\
The columns show the object's name, RA, DEC, $\vlsr$, the near distance $\dnear$, the height above the Galactic plane for the near distance $\znear$, the far distance $\dfar$, the height above the Galactic plane for the far distance $\zfar$, the distance ambiguity resolution according to Wienen et al., in prep., literature values for 
the distance and the method with which it was determined, the adopted distance and the resulting Galactocentric distance, and the source classification, according to Sect.~\ref{sec:sample}.
}
\end{table} 

\end{landscape} 
}
\onltab{2}{
\begin{landscape}

\begin{table}[p] 
\scriptsize 
\centering 
\caption{Distances and classification for the sources in subsample S2, derived from C$^{17}$O(3-2).} \label{tab:dist_B}
\medskip 
\begin{tabular}{lccccccccccccc} 
\toprule 
        Source              & RA             & DEC            & $V_{LSR}$  & $D_{near}$ & $z_{near}$ & $D_{far}$  &  $z_{far}$ &  HI        & $D$ liter       & Method     & Adopted $D$& $\dgc$   & Group  \\
                            & (HH:MM:SS)     & (DD:MM:SS)     & $(\kmstab)$& (kpc)      & (pc)       & (kpc)      & (pc)       &            & (kpc)           &            & (kpc)      & (kpc)      &        \\
\midrule 
       AGAL301.136-00.226   &$ 12:35:34.89  $&$ -63:02:32.8  $&$  -39.15  $&$     4.4  $&$   -17.5  $&$     4.4  $&$   -17.5  $&$       F  $&$     4.3^{ 2}  $&$      HI  $&$     4.4  $&$     7.3  $& IRB    \\
       AGAL305.209+00.206   &$ 13:11:14.09  $&$ -62:34:41.5  $&$  -42.53  $&$     4.8  $&$    17.7  $&$     4.8  $&$    17.7  $&$       F  $&$     3.5^{14}  $&$      SP  $&$     3.5  $&$     7.1  $& IRB    \\
       AGAL305.794-00.096   &$ 13:16:32.84  $&$ -62:49:41.3  $&$  -40.82  $&$     5.0  $&$    -8.7  $&$     5.0  $&$    -8.7  $&$       F  $&$     3.5^{14}  $&$      SP  $&$     3.5  $&$     7.0  $& D24    \\
       AGAL317.867-00.151   &$ 14:53:16.60  $&$ -59:26:33.9  $&$  -40.19  $&$     3.0  $&$    -7.8  $&$     9.7  $&$   -25.5  $&$       N  $&$            -  $&$       -  $&$     3.0  $&$     6.6  $& D8     \\
       AGAL318.779-00.137   &$ 14:59:33.06  $&$ -59:00:33.5  $&$  -38.14  $&$     2.8  $&$    -6.7  $&$    10.0  $&$   -24.1  $&$       N  $&$            -  $&$       -  $&$     2.8  $&$     6.7  $& D8     \\
       AGAL320.881-00.397   &$ 15:14:33.18  $&$ -58:11:28.5  $&$  -45.30  $&$     3.2  $&$   -22.4  $&$     10.0  $&$   -69.4  $&$       F  $&$            -  $&$       -  $&$    10.0  $&$     6.3  $& D24    \\
\bottomrule 
\end{tabular} 

\\ 
\medskip 
\tablefoot{
As in Table~\ref{tab:dist_A}.
}
\end{table} 

\end{landscape}
\clearpage
}
\onltab{3}{
\begin{landscape}

\begin{table}[p] 
\scriptsize 
\centering 
\caption{Distances and classification for the sources in  subsample S3 derived from C$^{17}$O(3-2).} \label{tab:dist_C}
\medskip 
  \begin{tabular}{lccccccccccccc} 
\toprule 
        Source              & RA             & DEC            & $V_{LSR}$  & $D_{near}$ & $z_{near}$ & $D_{far}$  &  $z_{far}$ &  HI        & $D$ liter       & Method     & Adopted $D$& $\dgc$     & Group  \\
                            & (HH:MM:SS)     & (DD:MM:SS)     & $(\kmstab)$& (kpc)      & (pc)       & (kpc)      & (pc)       &            & (kpc)           &            & (kpc)      & (kpc)      &        \\
\midrule 
       AGAL030.818-00.056   &$ 18:47:46.82  $&$ -01:54:24.6  $&$   97.78  $&$     6.4  $&$    -6.2  $&$     8.2  $&$    -7.8  $&$       N  $&$            -  $&$       -  $&$     6.4  $&$     4.4  $& IRB    \\
       AGAL305.192-00.006   &$ 13:11:13.73  $&$ -62:47:27.1  $&$  -34.19  $&$     3.5  $&$    -0.4  $&$     6.3  $&$    -0.7  $&$       F  $&$     3.5^{14}  $&$      SP  $&$     3.5  $&$     7.1  $& RMS    \\
       AGAL305.562+00.014   &$ 13:14:26.24  $&$ -62:44:24.9  $&$  -39.80  $&$     4.9  $&$     0.9  $&$     4.9  $&$     0.9  $&$       F  $&$     3.5^{14}  $&$      SP  $&$     3.5  $&$     7.1  $& RMS    \\
       AGAL309.384-00.134   &$ 13:47:24.03  $&$ -62:18:09.5  $&$  -51.28  $&$     5.3  $&$   -12.1  $&$     5.3  $&$   -12.1  $&$       N  $&$     3.5^{ 2}  $&$      HI  $&$     5.3  $&$     6.6  $& D8     \\
       AGAL310.014+00.387   &$ 13:51:38.36  $&$ -61:39:15.1  $&$  -41.31  $&$     3.6  $&$    24.4  $&$     7.3  $&$    49.4  $&$       N  $&$            -  $&$       -  $&$     3.6  $&$     6.8  $& RMS    \\
       AGAL313.576+00.324   &$ 14:20:08.34  $&$ -60:42:05.4  $&$  -46.94  $&$     3.8  $&$    21.4  $&$     7.9  $&$    44.9  $&$       -  $&$     3.0^{ 2}  $&$      HI  $&$     3.8  $&$     6.5  $& RMS    \\
\bottomrule 
\end{tabular} 

\\ 
\medskip 
\tablefoot{
As in Table~\ref{tab:dist_A}.
}
\end{table} 

\end{landscape}
}

\onltab{4}{
\begin{landscape} 

\begin{table}[p] 
\centering 
\scriptsize 
\caption{Line parameters for subsample S1.} \label{tab:line_pars_A}
  \begin{tabular}{lrrrrrrrrrrrrrr} 
\toprule 
Source & \multicolumn{7}{c}{$\cooz$} & \multicolumn{7}{c}{$\cott$} \\ 
\cmidrule(r){2-8}
\cmidrule(r){9-15}
& $\int \tmb dV$ & $\tmb$ & rms & $\vlsr$ & $\sigma(\vlsr)$ & $\Delta V$ & $\sigma(\Delta V)$ & $\int \tmb dV$ & $\tmb$ & rms & $\vlsr$ & $\sigma(\vlsr)$ & $\Delta V$ & $\sigma(\Delta V)$ \\ 
& $(\kkmstab)$ & $(\kelvin)$ & $(\kelvin)$ & $(\kmstab)$ & $(\kmstab)$ & $(\kmstab)$ & $(\kmstab)$ & $(\kkmstab)$ & $(\kelvin)$ & $(\kelvin)$ & $(\kmstab)$ & $(\kmstab)$ & $(\kmstab)$ & $(\kmstab)$ \\ 
\midrule 
    AGAL008.684-00.367   &$    7.69  $&$    1.32  $&$    0.08  $&$   37.19  $&$    0.07  $&$     4.1  $&$     0.1  $&$     6.9  $&$     1.8  $&$    0.07  $&$   37.26  $&$    0.04  $&$     3.6  $&$     0.1  $\\
    AGAL008.706-00.414   &$    2.75  $&$    0.66  $&$    0.07  $&$   38.64  $&$    0.12  $&$     2.2  $&$     0.3  $&$     1.0  $&$     0.3  $&$    0.07  $&$   37.62  $&$    0.15  $&$     1.8  $&$     0.2  $\\
    AGAL010.444-00.017   &$    4.08  $&$    0.64  $&$    0.07  $&$   75.07  $&$    0.52  $&$     3.8  $&$     1.7  $&$     1.9  $&$     0.7  $&$    0.07  $&$   74.84  $&$    0.10  $&$     2.8  $&$     0.2  $\\
    AGAL010.472+00.027   &$    7.34  $&$    1.11  $&$    0.07  $&$   66.76  $&$    0.10  $&$     4.4  $&$     0.5  $&$    24.1  $&$     2.9  $&$    0.10  $&$   66.33  $&$    0.12  $&$     6.9  $&$     0.3  $\\
    AGAL010.624-00.384   &$   21.05  $&$    2.96  $&$    0.08  $&$   -3.02  $&$    0.00  $&$     5.8  $&$     0.1  $&$    77.6  $&$    11.8  $&$    0.08  $&$   -2.81  $&$    0.01  $&$     5.9  $&$     0.1  $\\
\bottomrule 
\end{tabular} 

\tablefoot{
The columns show the integrated intensity, the main beam temperature, the rms of the spectrum, the $\vlsr$ and the width of the line, with their uncertainties for the specified molecular lines.
}
\end{table} 

\end{landscape}
\clearpage
}
\onltab{5}{
\begin{landscape}

\begin{table}[p] 
\centering 
\scriptsize 
\caption{Line parameters C$^{18}$O$(1-0)$ for subsample S1.}  \label{tab:line_pars_c18o_A}
  \begin{tabular}{lrrrrrrr} 
\toprule 
Source & $\int \tmb dV$ & $\tmb$ & rms & $\vlsr$ & $\sigma(\vlsr)$ & $\Delta V$ & $\sigma(\Delta V)$ \\ 
& $(\kkmstab)$ & $(\kelvin)$ & $(\kelvin)$ & $(\kmstab)$ & $(\kmstab)$ & $(\kmstab)$ & $(\kmstab)$ \\ 
\midrule 
    AGAL008.684-00.367 &$     26.42$&$      5.22$&$      0.08$&$     37.27$&$      0.02$&$       4.9$&$       0.1$\\
    AGAL008.706-00.414 &$      9.22$&$      3.12$&$      0.07$&$     38.70$&$      0.02$&$       2.9$&$       0.1$\\
    AGAL010.444-00.017 &$     10.49$&$      2.55$&$      0.06$&$     75.59$&$      0.07$&$       4.1$&$       0.2$\\
    AGAL010.472+00.027 &$     30.71$&$      4.61$&$      0.08$&$     66.80$&$      0.06$&$       6.2$&$       0.1$\\
    AGAL010.624-00.384 &$     71.13$&$     11.33$&$      0.07$&$     -2.87$&$      0.01$&$       6.0$&$       0.1$\\
    AGAL012.804-00.199 &$     54.46$&$      8.66$&$      0.06$&$     35.51$&$      0.01$&$       5.9$&$       0.1$\\
\bottomrule 
\end{tabular} 

\tablefoot{
As in Table~\ref{tab:line_pars_A}.
}
\end{table} 

\end{landscape}
\clearpage
}
\onltab{6}{
\begin{landscape}

\begin{table}[p] 
\centering 
\scriptsize 
  \begin{tabular}{lrrrrrrr} 
\toprule 
Source & $\int \tmb dV$ & $\tmb$ & rms & $\vlsr$ & $\sigma(\vlsr)$ & $\Delta V$ & $\sigma(\Delta V)$ \\ 
& $(\kkmstab)$ & $(\kelvin)$ & $(\kelvin)$ & $(\kmstab)$ & $(\kmstab)$ & $(\kmstab)$ & $(\kmstab)$ \\ 
\midrule 
    AGAL008.684-00.367 &$     81.21$&$     11.93$&$      0.07$&$     36.71$&$      0.03$&$       6.7$&$       0.1$\\
    AGAL008.706-00.414 &$     39.14$&$      8.99$&$      0.06$&$     38.50$&$      0.04$&$       3.9$&$       0.1$\\
    AGAL010.444-00.017 &$     39.31$&$      7.02$&$      0.07$&$     75.48$&$      0.10$&$       5.4$&$       0.3$\\
    AGAL010.472+00.027 &$    146.22$&$     18.55$&$      0.08$&$     66.61$&$      0.02$&$       7.2$&$       0.1$\\
    AGAL010.624-00.384 &$    258.35$&$     34.56$&$      0.07$&$     -3.03$&$      0.01$&$       7.0$&$       0.1$\\
    AGAL012.804-00.199 &$    176.77$&$     23.40$&$      0.07$&$     35.33$&$      0.02$&$       7.4$&$       0.1$\\
\bottomrule 
\end{tabular} 

\tablefoot{
As in Table~\ref{tab:line_pars_A}.
}
\end{table} 

\end{landscape}
\clearpage
}
\onltab{7}{
\begin{landscape}

\begin{table}[p] 
\centering 
\scriptsize 
\caption{Line parameters $^{13}$C$^{18}$O$(1-0)$ for subsample S1.}  \label{tab:line_pars_13c18o_A}
  \begin{tabular}{lrrrrrrr} 
\toprule 
Source & $\int\tmb dV$ & $\tmb$ & rms & $\vlsr$ & $\sigma(\vlsr)$ & $\Delta V$ & $\sigma(\Delta V)$ \\ 
& $(\kkmstab)$ & $(\kelvin)$ & $(\kelvin)$ & $(\kmstab)$ & $(\kmstab)$ & $(\kmstab)$ & $(\kmstab)$ \\ 
\midrule 
    AGAL008.684-00.367 &$      0.67 $&$      0.11 $&$      0.02 $&$      37.1 $&$       0.4 $&$       5.6 $&$       1.0 $\\
    AGAL008.706-00.414 &$      0.35 $&$      0.09 $&$      0.02 $&$      38.4 $&$       1.0 $&$       3.8 $&$       2.4 $\\
    AGAL010.444-00.017 &$     \dots $&$     \dots $&$      0.03 $&$     \dots $&$     \dots $&$     \dots $&$     \dots $\\
    AGAL010.472+00.027 &$     \dots $&$     \dots $&$      0.04 $&$     \dots $&$     \dots $&$     \dots $&$     \dots $\\
    AGAL010.624-00.384 &$      1.40 $&$      0.26 $&$      0.03 $&$      -2.7 $&$       0.4 $&$       5.1 $&$       0.9 $\\
    AGAL012.804-00.199 &$      1.09 $&$      0.19 $&$      0.02 $&$      35.3 $&$       0.4 $&$       5.4 $&$       1.0 $\\
\bottomrule 
\end{tabular} 

\tablefoot{
As in Table~\ref{tab:line_pars_A}.
}
\end{table} 

\end{landscape}
\clearpage
}
\onltab{8}{
\begin{landscape}

\begin{table}[p] 
\centering 
\scriptsize 
\caption{Line parameters for subsample S2.}  \label{tab:line_pars_B}
  \begin{tabular}{lrrrrrrrrrrrrrrrrrr} 
\toprule 
Source & \multicolumn{10}{c}{$\ceightoto$} & \multicolumn{8}{c}{$\cott$} \\ 
& $\int \tmb dV$ & $\tmb$ & rms & $\vlsr$ & $\sigma(\vlsr)$ & $\Delta V$ & $\sigma(\Delta V)$ & $\tau$ & $68\%$int.     & $\int \tmb dV$ & $\tmb$ & rms & $\vlsr$ & $\sigma(\vlsr)$ & $\Delta V$ & $\sigma(\Delta V)$ & $\tau$ & $\sigma(\tau)$ \\ 
\cmidrule(r){2-10}
\cmidrule(r){11-19}
& $(\kkmstab)$ & $(\kelvin)$ & $(\kelvin)$ & $(\kmstab)$ & $(\kmstab)$ & $(\kmstab)$ & $(\kmstab)$ & & & $(\kkmstab)$ & $(\kelvin)$ & $(\kelvin)$ & $(\kmstab)$ & $(\kmstab)$ & $(\kmstab)$ & $(\kmstab)$ & & \\ 
\midrule 
  AGAL301.136-00.226   &$  54.90  $&$  33.94  $&$   0.09  $&$ -39.15  $&$   0.01  $&$   4.7  $&$   0.1  $&$   0.4  $&$   0.3-   0.6  $&$   21.88  $&$    3.93  $&$    0.10  $&$  -39.28  $&$    0.02  $&$   4.7  $&$   0.1  $&$   0.1  $&$   0.1  $\\
  AGAL305.209+00.206   &$  42.23  $&$  21.22  $&$   0.08  $&$ -42.53  $&$   0.02  $&$   6.2  $&$   0.1  $&$   0.3  $&$   0.2-   0.3  $&$   16.19  $&$    2.40  $&$    0.08  $&$  -42.15  $&$    0.05  $&$   4.9  $&$   0.2  $&$   0.9  $&$   0.1  $\\
  AGAL305.794-00.096   &$  11.87  $&$  11.32  $&$   0.08  $&$ -40.82  $&$   0.03  $&$   3.1  $&$   0.1  $&$   1.1  $&$   0.7-   1.6  $&$    1.99  $&$    0.69  $&$    0.09  $&$  -40.91  $&$    0.14  $&$   2.3  $&$   0.4  $&$   0.7  $&$   0.6  $\\
  AGAL317.867-00.151   &$  10.73  $&$   8.21  $&$   0.11  $&$ -40.19  $&$   0.04  $&$   3.6  $&$   0.1  $&$   0.4  $&$   0.2-   0.5  $&$    3.07  $&$    0.76  $&$    0.08  $&$  -40.61  $&$    0.12  $&$   3.7  $&$   0.5  $&$   0.1  $&$   5.5  $\\
  AGAL318.779-00.137   &$  11.03  $&$   7.55  $&$   0.11  $&$ -38.14  $&$   0.04  $&$   4.1  $&$   0.1  $&$   0.5  $&$   0.3-   0.6  $&$    2.26  $&$    0.72  $&$    0.09  $&$  -39.36  $&$    0.11  $&$   2.8  $&$   0.3  $&$   0.1  $&$   0.4  $\\
\bottomrule 
\end{tabular} 

\tablefoot{
The columns show the integrated intensity, the main beam temperature, the rms of the spectrum, the $\vlsr$, the width and the opacity of the line, with their uncertainties for the specified molecular lines.
}
\end{table} 

\end{landscape}
\clearpage
}
\onltab{9}{
\begin{landscape}

\begin{table}[p] 
\centering 
\scriptsize 
\caption{Line parameters C$^{17}$O$(3-2)$ for subsample S3.}  \label{tab:line_pars_C}
  \begin{tabular}{lrrrrrrrrr} 
\toprule 
Source & $\int \tmb dV$ & $\tmb$ & rms & $\vlsr$ & $\sigma(\vlsr)$ & $\Delta V$ & $\sigma(\Delta V)$ & $\tau$ & $\sigma(\tau)$ \\ 
& $(\kkmstab)$ & $(\kelvin)$ & $(\kelvin)$ & $(\kmstab)$ & $(\kmstab)$ & $(\kmstab)$ & $(\kmstab)$ & & \\ 
\midrule 
    AGAL030.818-00.056 &$     12.15$&$      1.98$&$      0.07$&$     97.78$&$      0.05$&$       5.5$&$       0.1$&$       2.0$&$       0.1$\\
    AGAL305.192-00.006 &$      3.16$&$      0.86$&$      0.07$&$    -34.19$&$      0.09$&$       3.0$&$       0.3$&$       0.9$&$       0.1$\\
    AGAL305.562+00.014 &$     16.84$&$      4.79$&$      0.09$&$    -39.80$&$      0.02$&$       3.0$&$       0.1$&$       4.8$&$       0.1$\\
    AGAL309.384-00.134 &$      6.25$&$      1.76$&$      0.09$&$    -51.28$&$      0.06$&$       3.1$&$       0.1$&$       1.8$&$       0.1$\\
    AGAL310.014+00.387 &$      7.55$&$      2.02$&$      0.09$&$    -41.31$&$      0.04$&$       3.0$&$       0.1$&$       2.0$&$       0.1$\\
    AGAL313.576+00.324 &$      7.69$&$      1.65$&$      0.09$&$    -46.94$&$      0.05$&$       3.5$&$       0.3$&$       1.6$&$       0.2$\\
\bottomrule 
\end{tabular} 

\tablefoot{
As in Table~\ref{tab:line_pars_B}.
}
\end{table} 

\end{landscape}
\clearpage
}
\onltab{10}{
\begin{table} 
  \centering 
  \tiny 
  \caption{Excitation temperature for subsample S1.} \label{tab:tex_a}
    \begin{tabular}{l*{4}c} 
      \toprule 
    Source                        &  $\tex$                     &  $68\%$int.              & Group \\

				  &  $(\kelvin)$                &  $(\kelvin)$             &       \\

    \midrule
    AGAL008.684-00.367            &$ 11.8                      $&$ 10.6-12.6              $& D8    \\
    AGAL008.706-00.414            &$ 8.4                       $&$ 7.7-8.9                $& D24   \\
    AGAL010.444-00.017            &$ 9.2                       $&$ 8.5-9.8                $& D24   \\
    AGAL010.472+00.027            &$ 27.9                      $&$ 20.4-32.5              $& IRB   \\
    AGAL010.624-00.384            &$ 31.7                      $&$ 22.6-37.2              $& IRB   \\
    AGAL012.804-00.199            &$ 25.8                      $&$ 18.8-29.7              $& IRB   \\
\bottomrule 
  \end{tabular} 

  \tablefoot{
  In the Group column the identifiers between parentheses are the classifications of the main component.
  }
\end{table} 
}
\onltab{11}{
\begin{table} 
  \centering 
  \tiny 
  \caption{Excitation temperature for subsample S2.}  \label{tab:tex_b}
    \begin{tabular}{l*{4}c} 
    \toprule 
    Source                        &  $\tex$                     &  $68\%$int.              & Group \\

				  &  $(\kelvin)$                &  $(\kelvin)$             &       \\

    \midrule
    AGAL301.136-00.226            &$ 31.9                      $&$ 18.4-35.2              $& IRB   \\
    AGAL305.209+00.206            &$ 35.0                      $&$ 19.2-40.6              $& IRB   \\
    AGAL305.794-00.096            &$ 9.6                       $&$ 8.2-10.3               $& D24   \\
    AGAL317.867-00.151            &$ 20.5                      $&$ 14.6-22.4              $& D8    \\
    AGAL318.779-00.137            &$ 13.5                      $&$ 10.6-14.5              $& D8    \\
    AGAL320.881-00.397            &$ 19.8                      $&$ 14.7-22.2              $& D24   \\
\bottomrule 
  \end{tabular} 

  \tablefoot{
  As in Table~\ref{tab:tex_a}
  }
\end{table} 
}
\onltab{12}{
\begin{table*}[p]
  \centering 
  \scriptsize 
  \caption{Optical depths and isotopic ratios for subsample S1.} \label{tab:tau_A}
  \begin{tabular}{l*{11}c} 
	\toprule 
		Source                         & $\tau(\mathrm{C^{18}O}(1-0))$    & $68\%$int.                       & $\tau(\mathrm{C^{17}O}(1-0))$   & $68\%$int.                       & $\frac{[\mathrm{^{18}O}]}{[\mathrm{^{17}O}]}$   & $68\%$int.                       & $\frac{[\mathrm{^{12}C}]}{[\mathrm{^{13}C}]}$   & $68\%$int.                       & $\frac{[\mathrm{^{12}C^{17}O}]}{[\mathrm{^{13}C^{18}O}]}$   & $68\%$int.                       \\ 
	\midrule 
		AGAL008.684-00.367             & $0.77                          $ & $0.61          -0.91           $ &$0.17                          $ & $0.15          -0.20           $ &$4.4                           $ & $3.9           -4.7            $ &$51.2                          $ & $46.4          -55.8           $ &$11.7                          $ & $10.7          -12.6           $ \\
		AGAL008.706-00.414             & $1.00                          $ & $0.73          -1.21           $ &$0.21                          $ & $0.17          -0.25           $ &$4.7                           $ & $4.1           -5.1            $ &$39.3                          $ & $34.3          -43.7           $ &$8.4                           $ & $7.6           -9.2            $ \\
		AGAL010.444-00.017             & $0.67                          $ & $0.50          -0.80           $ &$0.20                          $ & $0.16          -0.23           $ &$3.4                           $ & $3.0           -3.7            $ &$100.0                         $ & $80.6          -116.9          $ &$29.4                          $ & $23.4          -34.1           $ \\
\bottomrule 
\end{tabular} 

  \tablefoot{
  The columns show the opacity for $\Istp{}{18}(1-0)$, $\Istp{}{17}(1-0)$, the \oratio, the \cratio\ and $(\cratiom)/(\oratiom)$, with their respective uncertainties for subsample S1. Where the derived $\tau$ is less than $0.10$, we indicate $\tau<0.10$, to show that the emission is optically thin.
  }
\end{table*} 
\clearpage
}

\onltab{13}{

\begin{landscape}
\begin{table} 
  \centering 
  \tiny 
  \caption{Molecular column densities for subsample S1.} \label{tab:cd_a}
    \begin{tabular}{l*{12}c} 
    \toprule 
    Source                        &  $N(\Istp{12}{18})$         &  $68\%$int.                 &  $N(\Istp{12}{17},1-0)$     &  $68\%$int.                 &  $N(\Istp{12}{17},3-2)$     &  $68\%$int.                 &  $N(\Istp{13}{18})$         &  $68\%$int.                 &  $\coldhtwo$                &  $68\%$int.              & Group \\

				  &$(\pot{16}\times\cmtab^{-2})$&$(\pot{16}\times\cmtab^{-2})$&$(\pot{15}\times\cmtab^{-2})$&$(\pot{15}\times\cmtab^{-2})$&$(\pot{15}\times\cmtab^{-2})$&$(\pot{15}\times\cmtab^{-2})$&$(\pot{13}\times\cmtab^{-2})$&$(\pot{13}\times\cmtab^{-2})$&$(\pot{23}\times\cmtab^{-2})$&$(\pot{23}\times\cmtab^{-2})$&    \\

    \midrule
    AGAL008.684-00.367            &$ 4.1                       $&$ 3.5-4.7                   $&$ 8.6                       $&$ 7.6-9.5                   $&$ 7.0                       $&$ 6.1-8.0                   $&$ 74.8                      $&$ 61.5-86.9                 $&$ 2.9                       $&$ 2.1-3.5                $& D8    \\
    AGAL008.706-00.414            &$ 1.5                       $&$ 1.2-1.7                   $&$ 2.9                       $&$ 2.5-3.3                   $&$ 2.2                       $&$ 1.9-2.6                   $&$ 36.5                      $&$ 29.6-42.0                 $&$ 1.3                       $&$ 0.9-1.6                $& D24   \\
    AGAL010.444-00.017            &$ 1.5                       $&$ 1.2-1.7                   $&$ 4.3                       $&$ 3.8-4.9                   $&$ 3.4                       $&$ 2.9-3.9                   $&$ <3.6                      $&$ \dots                     $&$ 2.1                       $&$ 1.6-2.6                $& D24   \\
\bottomrule 
  \end{tabular} 

  \tablefoot{
  The columns show the column density of $\Istp{12}{18}$, of $\Istp{12}{17}$, derived both from the $(1-0)$ and $(3-2)$ transitions, of $\Istp{13}{18}$ and of molecular hydrogen, with their respective uncertainties. 
  The last column indicates the group of the source; the identifiers between parentheses are the classifications of the main component.
  }
\end{table} 
\end{landscape}
}
\onltab{14}{
\begin{table*} 
  \centering 
  \tiny 
  \caption{Molecular column densities for subsample S2.} \label{tab:cd_b}
    \begin{tabular}{l*{8}c} 
    \toprule 
    Source                        &  $N(\Istp{12}{18})$         &  $68\%$int.                 &  $N(\Istp{12}{17})$         &  $68\%$int.                 &  $\coldhtwo$                &  $68\%$int.              & Group \\

				  &$(\pot{16}\times\cmtab^{-2})$&$(\pot{16}\times\cmtab^{-2})$&$(\pot{15}\times\cmtab^{-2})$&$(\pot{15}\times\cmtab^{-2})$&$(\pot{23}\times\cmtab^{-2})$&$(\pot{23}\times\cmtab^{-2})$&    \\

    \midrule
    AGAL301.136-00.226            &$ 5.5                       $&$ 4.7-6.0                   $&$ 10.7                      $&$ 9.4-11.6                  $&$ 3.6                       $&$ 1.4-4.6                $& IRB   \\
    AGAL305.209+00.206            &$ 4.0                       $&$ 3.4-4.3                   $&$ 7.9                       $&$ 6.8-8.6                   $&$ 2.0                       $&$ 0.7-2.8                $& IRB   \\
    AGAL305.794-00.096            &$ 2.5                       $&$ 1.7-3.0                   $&$ 3.3                       $&$ 2.2-4.0                   $&$ 1.3                       $&$ 0.9-1.6                $& D24   \\
    AGAL317.867-00.151            &$ 1.0                       $&$ 0.8-1.1                   $&$ 1.8                       $&$ 1.4-1.9                   $&$ 1.1                       $&$ 0.6-1.5                $& D8    \\
    AGAL318.779-00.137            &$ 1.3                       $&$ 1.0-1.5                   $&$ 1.9                       $&$ 1.4-2.3                   $&$ 1.1                       $&$ 0.7-1.5                $& D8    \\
    AGAL320.881-00.397            &$ 2.0                       $&$ 1.7-2.2                   $&$ 3.4                       $&$ 2.8-3.8                   $&$ 0.4                       $&$ 0.2-0.6                $& D24   \\
\bottomrule 
  \end{tabular} 

  \tablefoot{
  The columns show the column density of $\Istp{12}{18}$, of $\Istp{12}{17}$ and of molecular hydrogen, with their respective uncertainties. 
  The last column indicates the group of the source; the identifiers between parentheses are the classifications of the main component.
  }
\end{table*} 
}
\onltab{15}{
\begin{table*} 
  \centering 
  \tiny 
  \caption{Molecular column densities for subsample S3.} \label{tab:cd_c}
    \begin{tabular}{l*{6}c} 
    \toprule 
    Source                        &  $N(\Istp{12}{17})$         &  $68\%$int.                 &  $\coldhtwo$                &  $68\%$int.              & Group \\

				  &$(\pot{15}\times\cmtab^{-2})$&$(\pot{15}\times\cmtab^{-2})$&$(\pot{23}\times\cmtab^{-2})$&$(\pot{23}\times\cmtab^{-2})$&    \\

    \midrule
    AGAL030.818-00.056            &$ 6.1                       $&$ 4.5-7.5                   $&$ 2.4                       $&$ 0.9-3.5                $& IRB   \\
    AGAL305.192-00.006            &$ 1.7                       $&$ 1.2-1.8                   $&$ 0.7                       $&$ 0.2-0.9                $& RMS   \\
    AGAL305.562+00.014            &$ 9.1                       $&$ 6.4-10.4                  $&$ 1.0                       $&$ 0.3-1.2                $& RMS   \\
    AGAL309.384-00.134            &$ 4.4                       $&$ 2.0-4.9                   $&$ 1.1                       $&$ 0.2-1.4                $& D8    \\
    AGAL310.014+00.387            &$ 4.0                       $&$ 2.9-4.6                   $&$ 0.8                       $&$ 0.2-1.0                $& RMS   \\
    AGAL313.576+00.324            &$ 4.1                       $&$ 2.7-4.6                   $&$ 0.6                       $&$ 0.2-0.7                $& RMS   \\
\bottomrule 
  \end{tabular} 

  \tablefoot{
  The columns show the column density of $\Istp{12}{17}$ and of molecular hydrogen, with their respective uncertainties. 
  The last column indicates the group of the source.
  }
\end{table*} 
}

\onltab{16}{
\begin{table*} 
  \centering 
  \tiny 
  \caption{Masses from $870\mum$ dust emission and virial masses for subsample S1.} \label{tab:mass_a}
    \begin{tabular}{l*{7}c} 
    \toprule 
    Source                        &  $M$                        &  $68\%$int.              &  $\mvir$                    &  $68\%$int.              & Group \\
				  &  $(\pot{3}\times\msuntab)$  & $(\pot{3}\times\msuntab)$&  $(\pot{3}\times\msuntab)$  & $(\pot{3}\times\msuntab)$&       \\
    \midrule
    AGAL008.684-00.367            &$ 17.40                     $&$ 7.46-24.78             $&$ 3.62                      $&$ 2.77-4.52              $& D8    \\
    AGAL008.706-00.414            &$ 14.97                     $&$ 5.83-20.39             $&$ 1.50                      $&$ 0.96-1.96              $& D24   \\
    AGAL010.444-00.017            &$ 10.09                     $&$ 3.51-13.94             $&$ 1.80                      $&$ 1.19-2.30              $& D24   \\
    AGAL010.472+00.027            &$ 19.99                     $&$ 6.07-27.62             $&$ 8.35                      $&$ 6.02-10.36             $& IRB   \\
    AGAL010.624-00.384            &$ 1.06                      $&$ 0.37-1.52              $&$ 1.78                      $&$ 1.32-2.21              $& IRB   \\
    AGAL012.804-00.199            &$ 2.21                      $&$ 1.29-3.02              $&$ 2.44                      $&$ 2.19-2.67              $& IRB   \\
\bottomrule 
  \end{tabular} 

  \tablefoot{
  The columns show the mass derived from the dust continuum emission and the virial mass, with their respective uncertainties.
  The last column indicates the group of the source.
  }
\end{table*} 
}
\onltab{17}{
\begin{table*} 
  \centering 
  \tiny 
  \caption{Masses from $870\mum$ dust emission and virial masses for subsample S2.} \label{tab:mass_b}
    \begin{tabular}{l*{6}c} 
    \toprule 
    Source                        &  $M$                        &  $68\%$int.              &  $\mvir$                    &  $68\%$int.              & Group  \\
				  &  $(\pot{3}\times\msuntab)$  & $(\pot{3}\times\msuntab)$&  $(\pot{3}\times\msuntab)$  & $(\pot{3}\times\msuntab)$&        \\
    \midrule
    AGAL301.136-00.226            &$ 2.49                      $&$ 0.58-3.22              $&$ 1.74                      $&$ 1.34-2.10              $& IRB    \\
    AGAL305.209+00.206            &$ 1.06                      $&$ 0.32-1.48              $&$ 1.78                      $&$ 1.45-2.07              $& IRB    \\
    AGAL305.794-00.096            &$ 1.06                      $&$ 0.53-1.35              $&$ 0.54                      $&$ 0.30-0.70              $& D24    \\
    AGAL317.867-00.151            &$ 0.36                      $&$ 0.09-0.47              $&$ 0.76                      $&$ 0.43-0.96              $& D8     \\
    AGAL318.779-00.137            &$ 0.63                      $&$ 0.21-0.85              $&$ 0.66                      $&$ 0.42-0.80              $& D8     \\
    AGAL320.881-00.397            &$ 2.01                      $&$ 0.67-2.72              $&$ 0.93                      $&$ 0.71-1.14              $& D24    \\
\bottomrule 
  \end{tabular} 

  \tablefoot{
  As in Table~\ref{tab:mass_a}.
  }
\end{table*} 
}
\onltab{18}{
\begin{table*} 
  \centering 
  \tiny 
  \caption{Masses from $870\mum$ dust emission and virial masses for subsample S3.} \label{tab:mass_c}
    \begin{tabular}{l*{6}c} 
    \toprule 
    Source                        &  $M$                        &  $68\%$int.              &  $\mvir$                    &  $68\%$int.              & Group  \\
				  &  $(\pot{3}\times\msuntab)$  & $(\pot{3}\times\msuntab)$&  $(\pot{3}\times\msuntab)$  & $(\pot{3}\times\msuntab)$&        \\
    \midrule
    AGAL030.818-00.056            &$ 4.04                      $&$ 0.93-6.78              $&$ 4.51                      $&$ 3.51-5.63              $& IRB    \\
    AGAL305.192-00.006            &$ 0.43                      $&$ 0.08-0.58              $&$ 0.85                      $&$ 0.59-1.06              $& RMS    \\
    AGAL305.562+00.014            &$ 0.50                      $&$ 0.08-0.65              $&$ 0.77                      $&$ 0.64-0.91              $& RMS    \\
    AGAL309.384-00.134            &$ 1.74                      $&$ 0.18-2.45              $&$ 1.42                      $&$ 1.06-1.80              $& D8     \\
    AGAL310.014+00.387            &$ 0.58                      $&$ 0.10-0.89              $&$ 0.95                      $&$ 0.69-1.17              $& RMS    \\
    AGAL313.576+00.324            &$ 0.26                      $&$ 0.03-0.36              $&$ 0.87                      $&$ 0.60-1.12              $& RMS    \\
\bottomrule 
  \end{tabular} 

  \tablefoot{
  As in Table~\ref{tab:mass_a}.
  }
\end{table*} 

\clearpage
}

\onlfig{1}{
\section{Spectra}

\begin{figure*} 
\centering 
\includegraphics[angle=-90,width=0.3\textwidth]{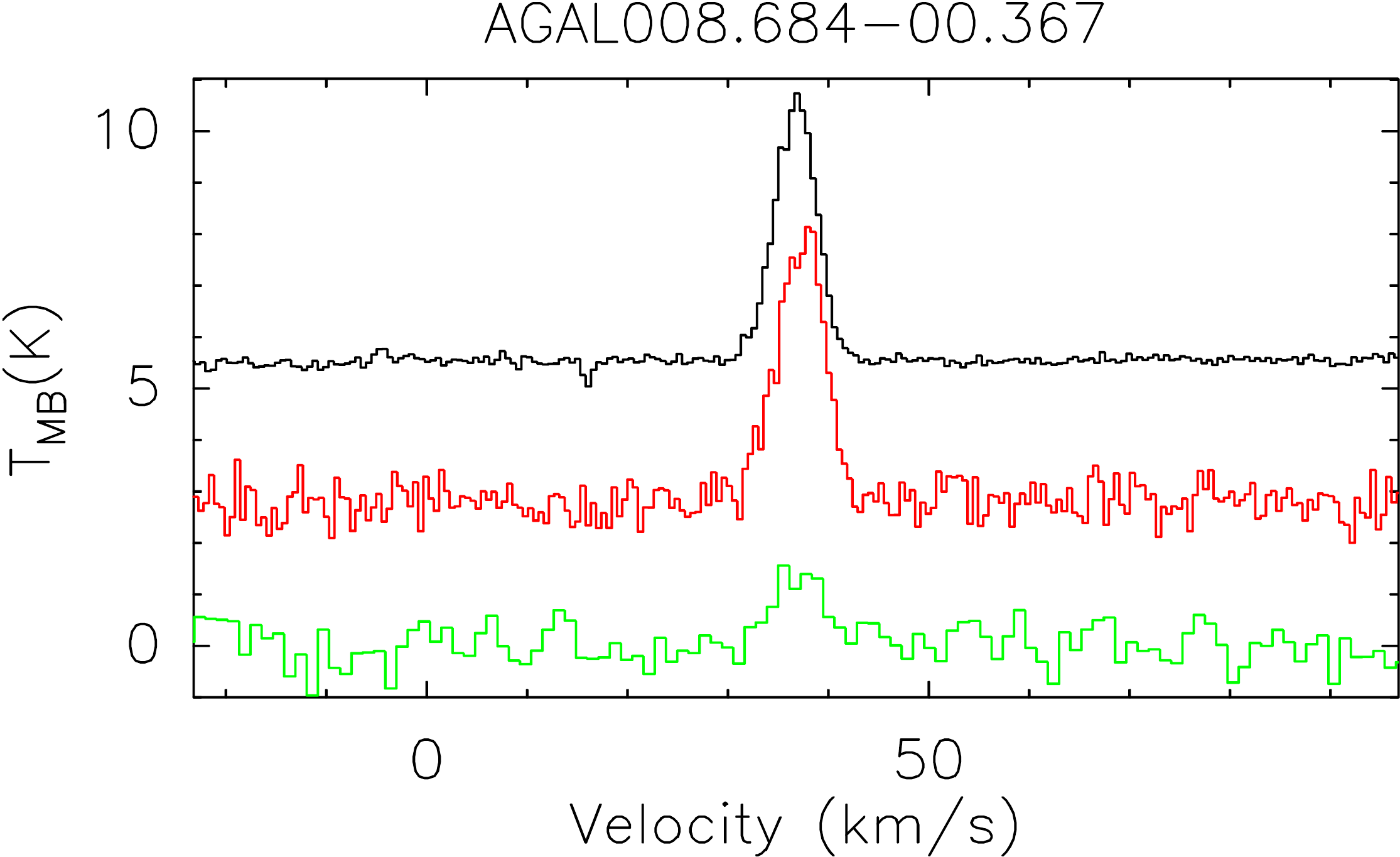}
\includegraphics[angle=-90,width=0.3\textwidth]{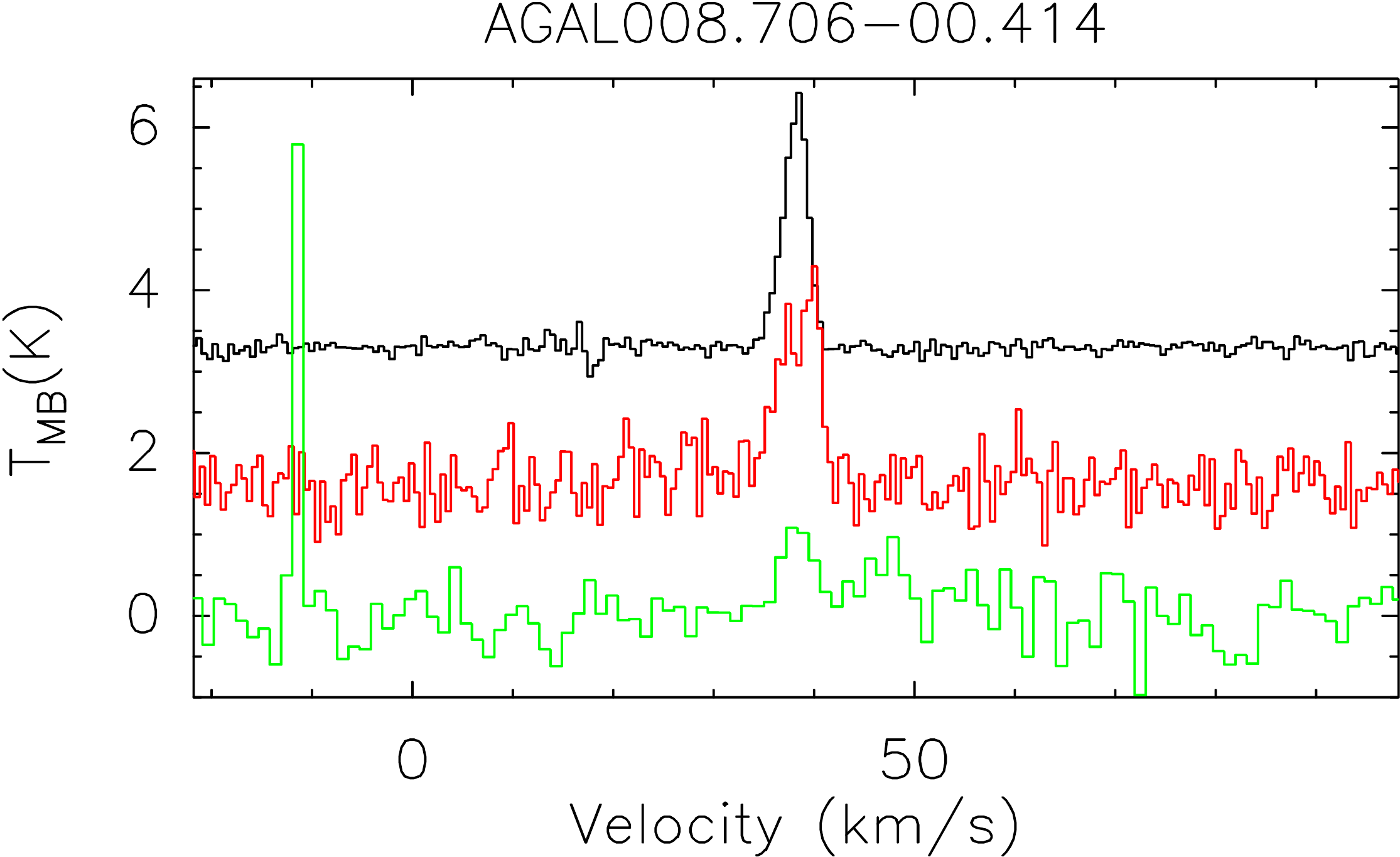}
\includegraphics[angle=-90,width=0.3\textwidth]{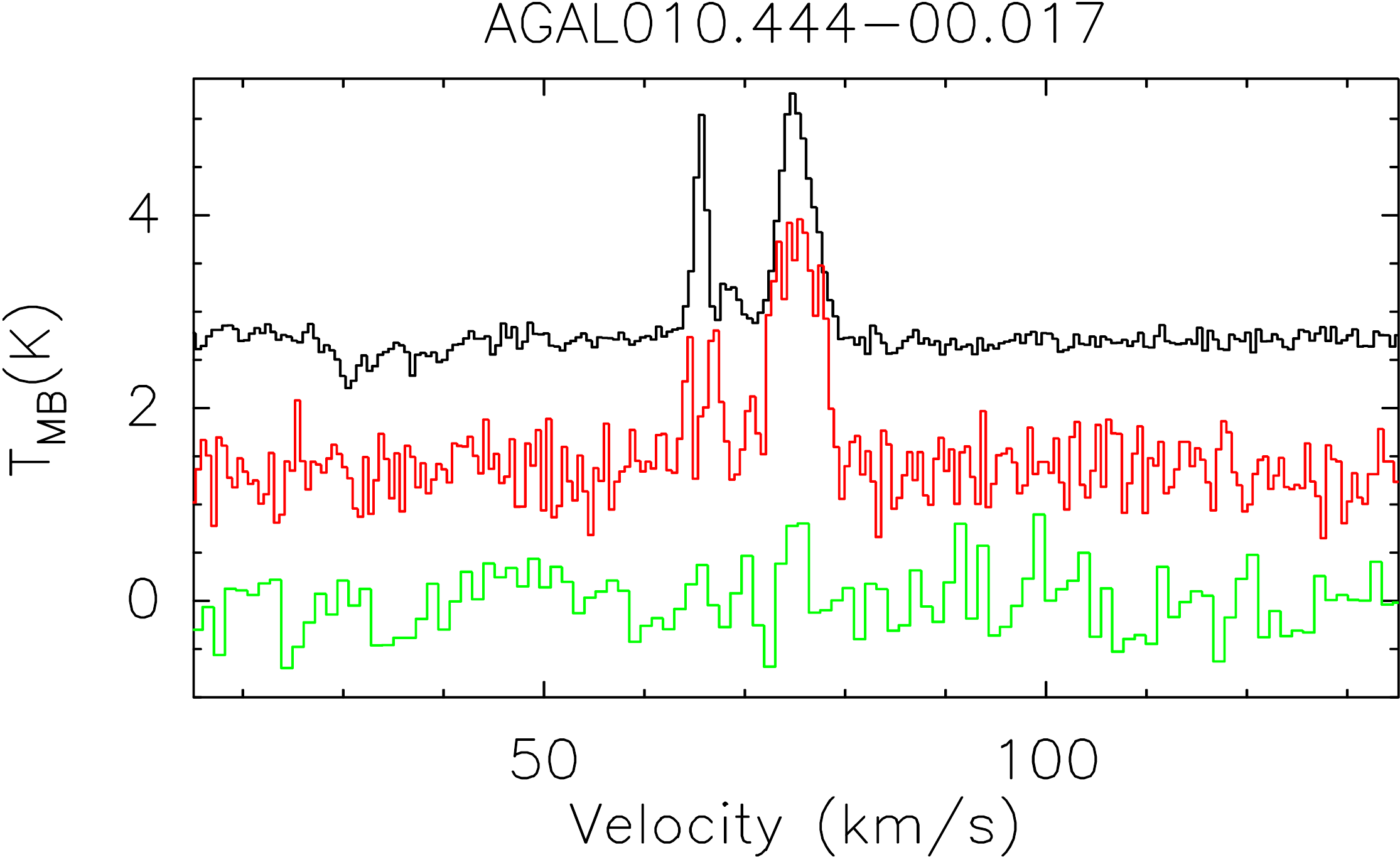} \\
\includegraphics[angle=-90,width=0.3\textwidth]{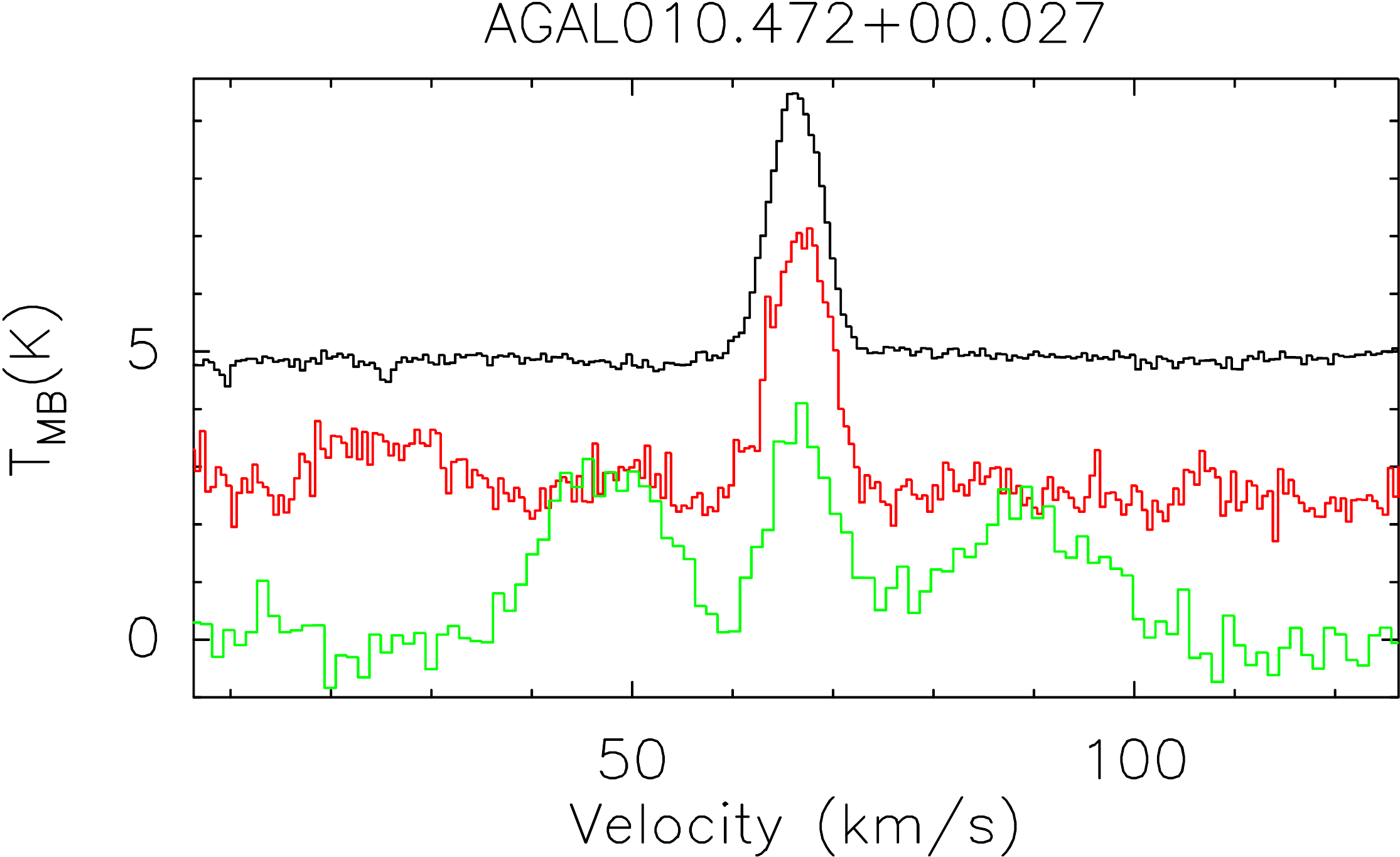}
\includegraphics[angle=-90,width=0.3\textwidth]{fig_clean/{G10.62-0.38}.pdf}
\includegraphics[angle=-90,width=0.3\textwidth]{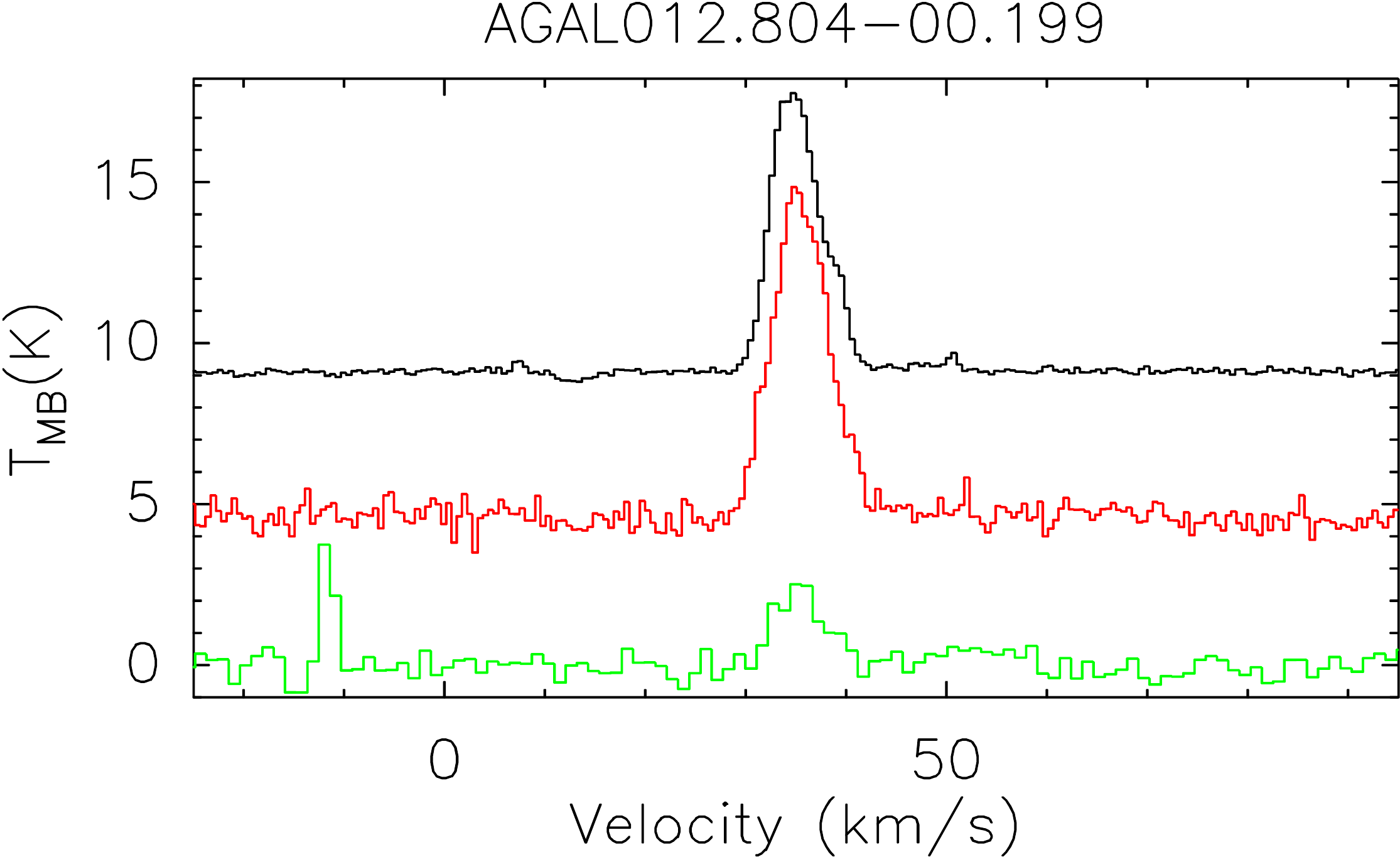} \\
\includegraphics[angle=-90,width=0.3\textwidth]{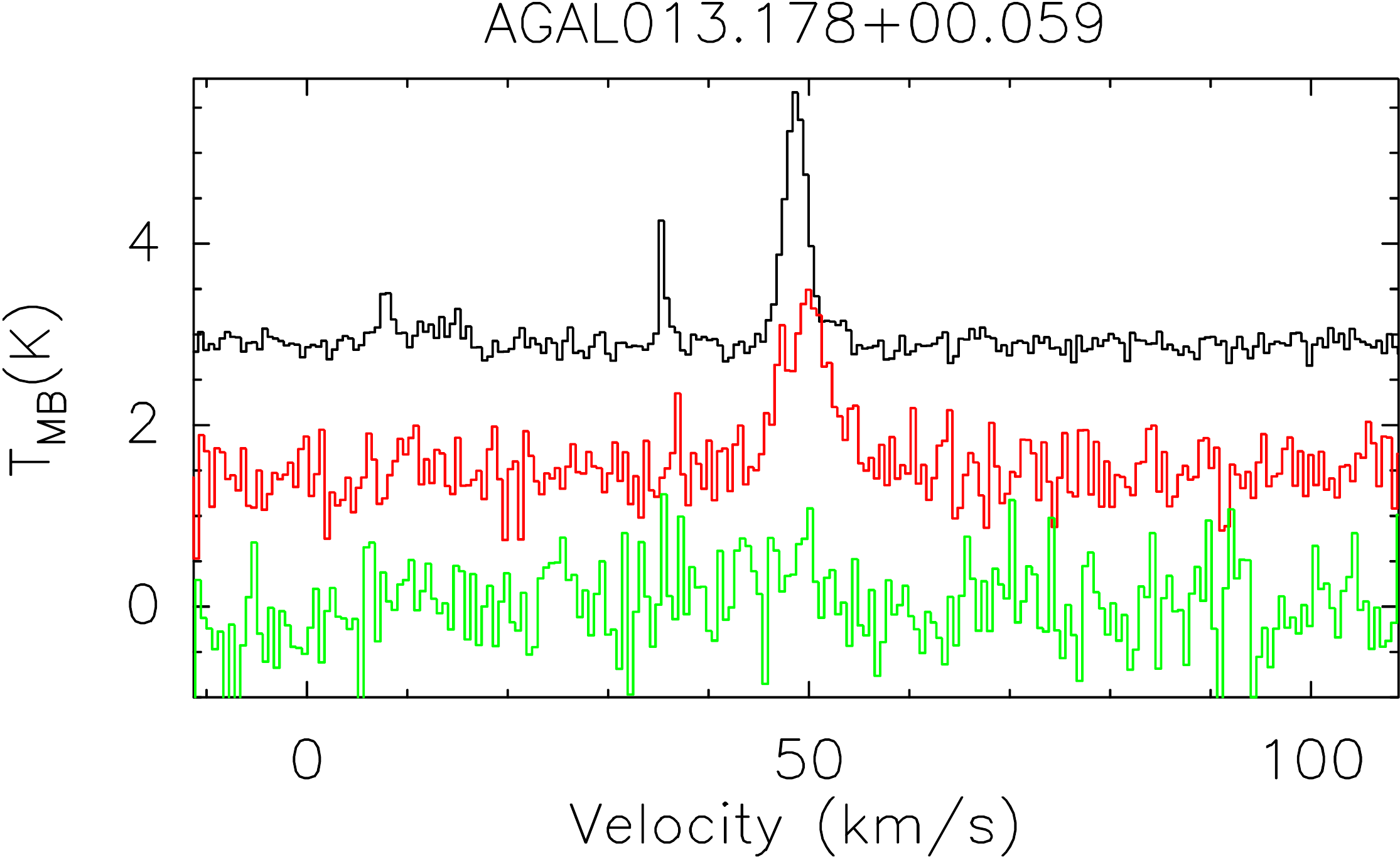} 
\includegraphics[angle=-90,width=0.3\textwidth]{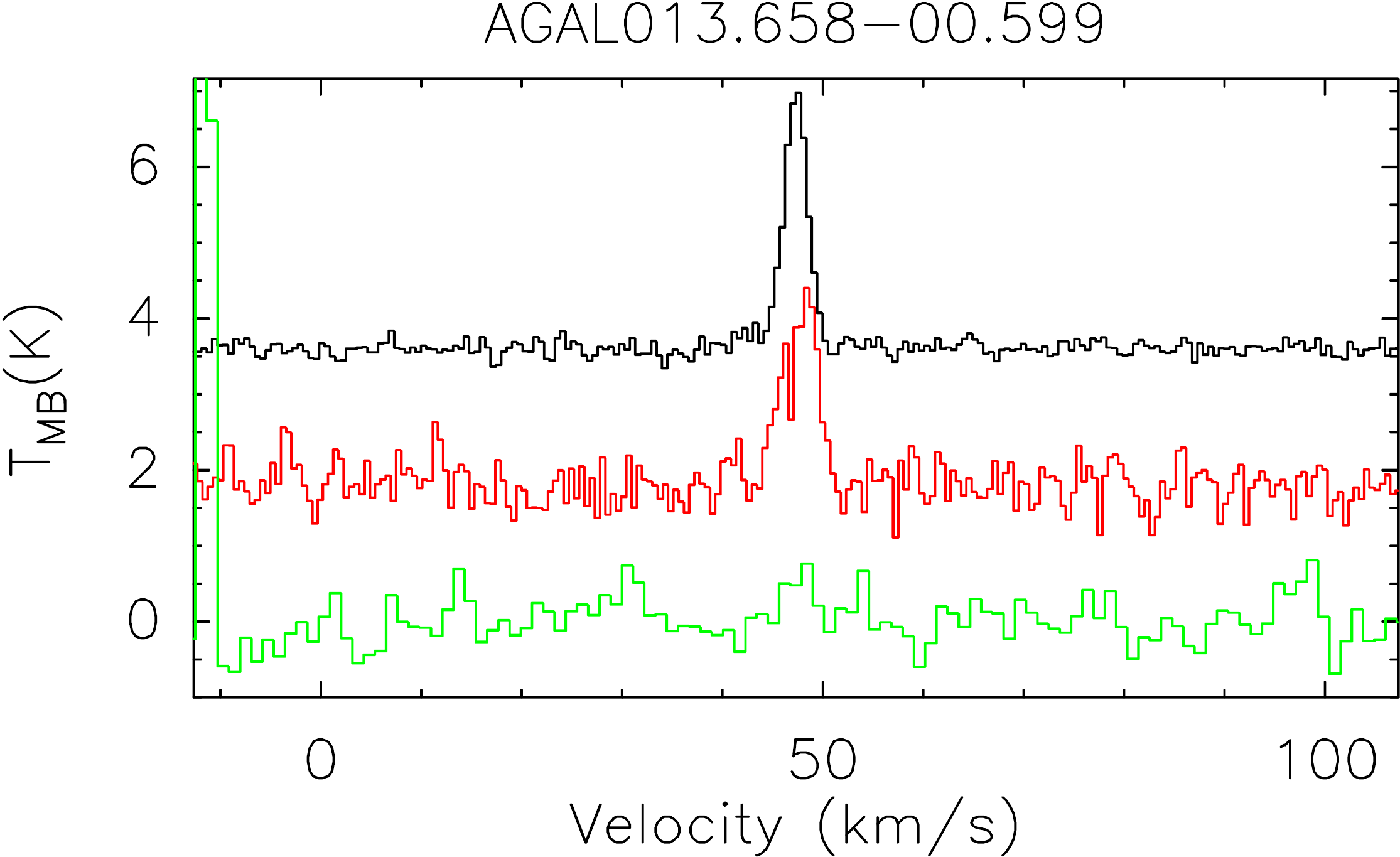}
\includegraphics[angle=-90,width=0.3\textwidth]{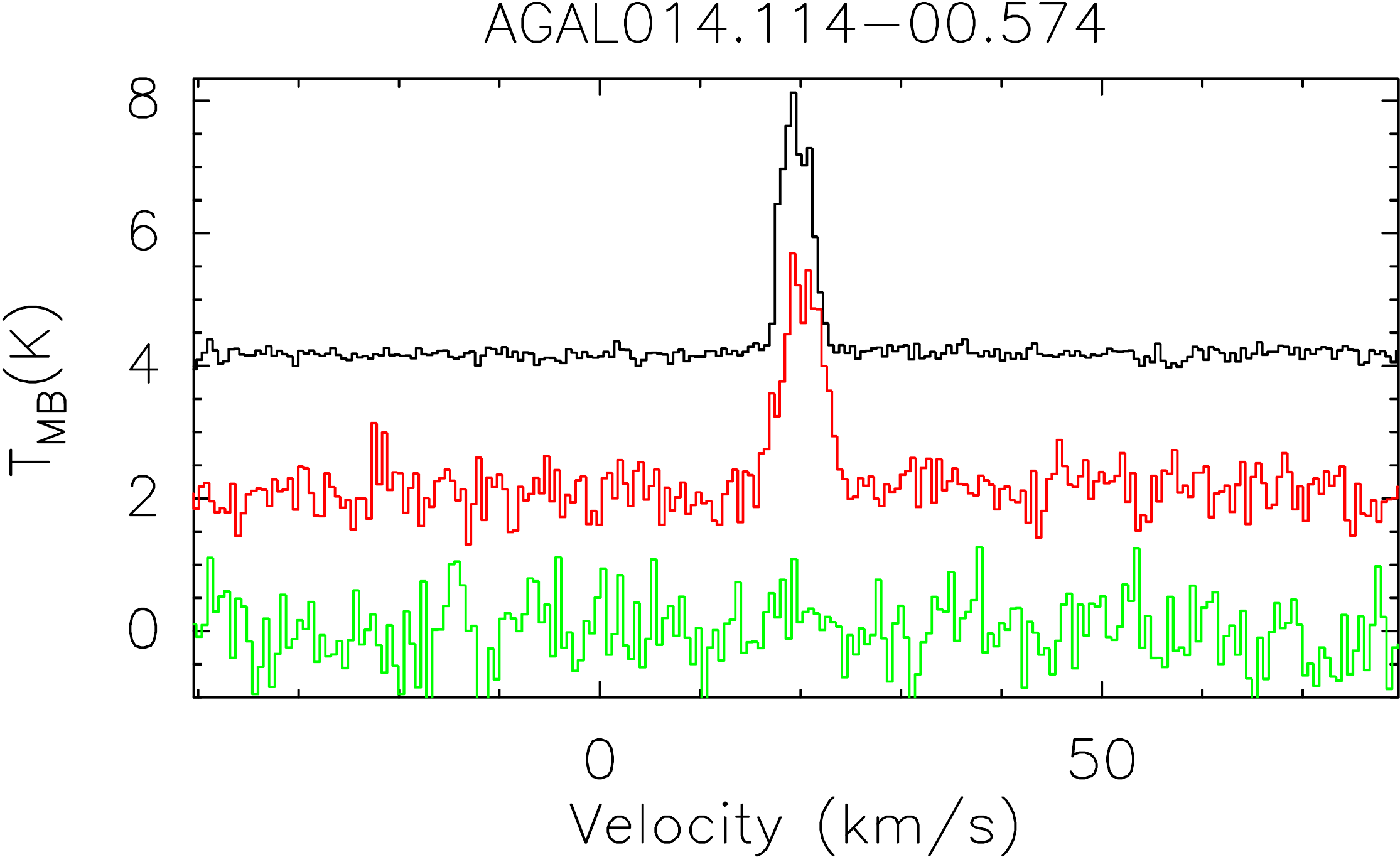} \\
\includegraphics[angle=-90,width=0.3\textwidth]{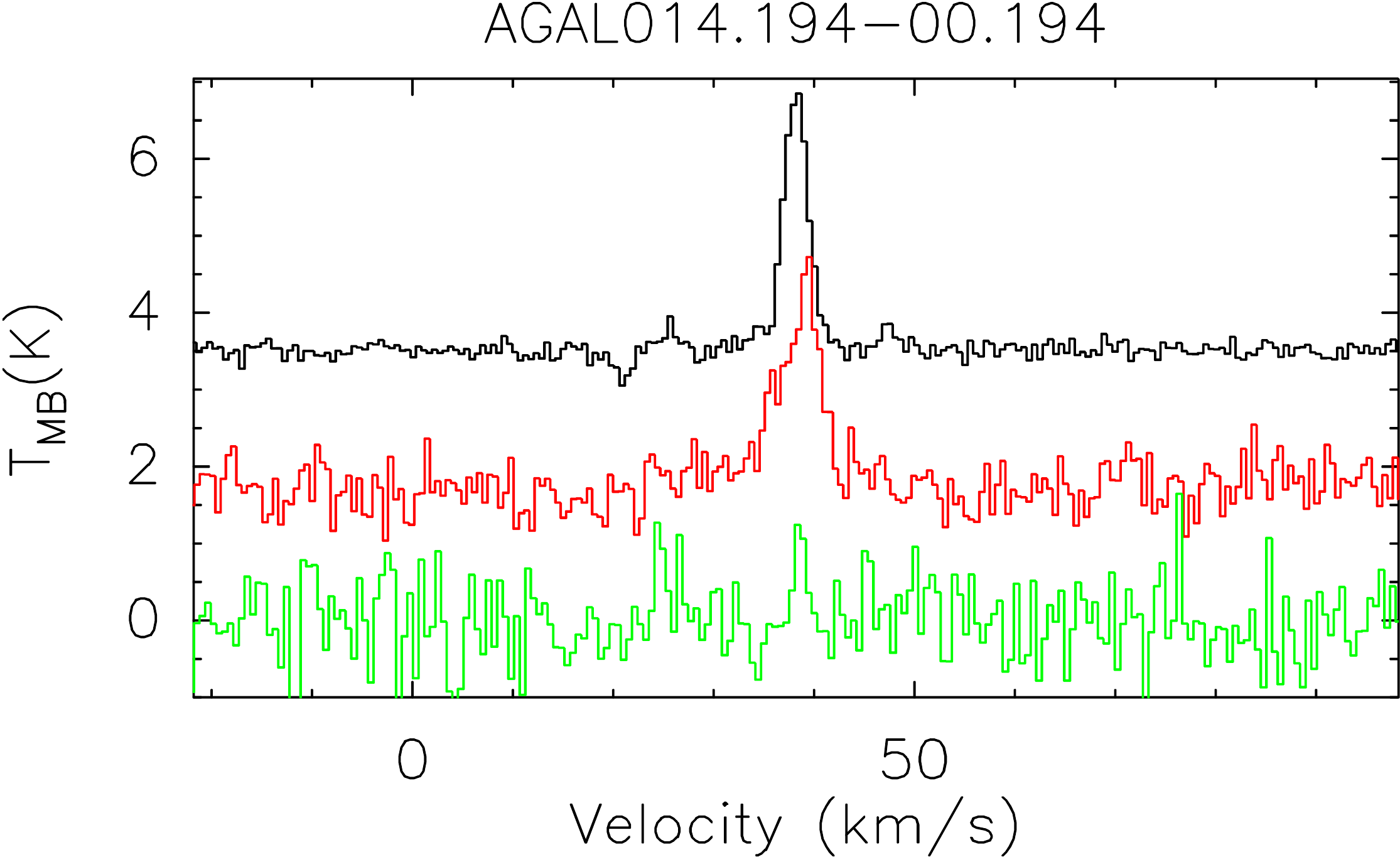}
\includegraphics[angle=-90,width=0.3\textwidth]{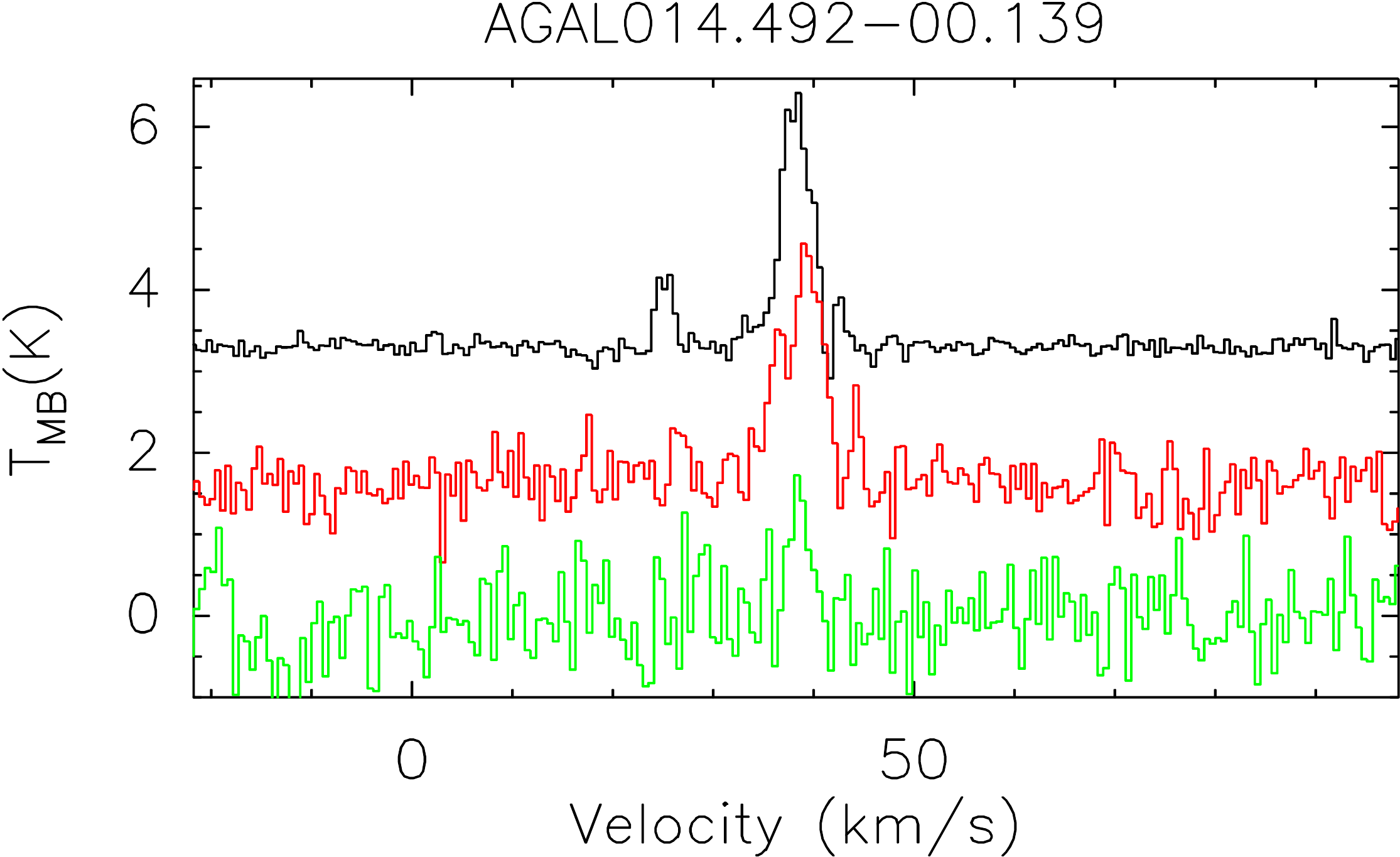} 
\includegraphics[angle=-90,width=0.3\textwidth]{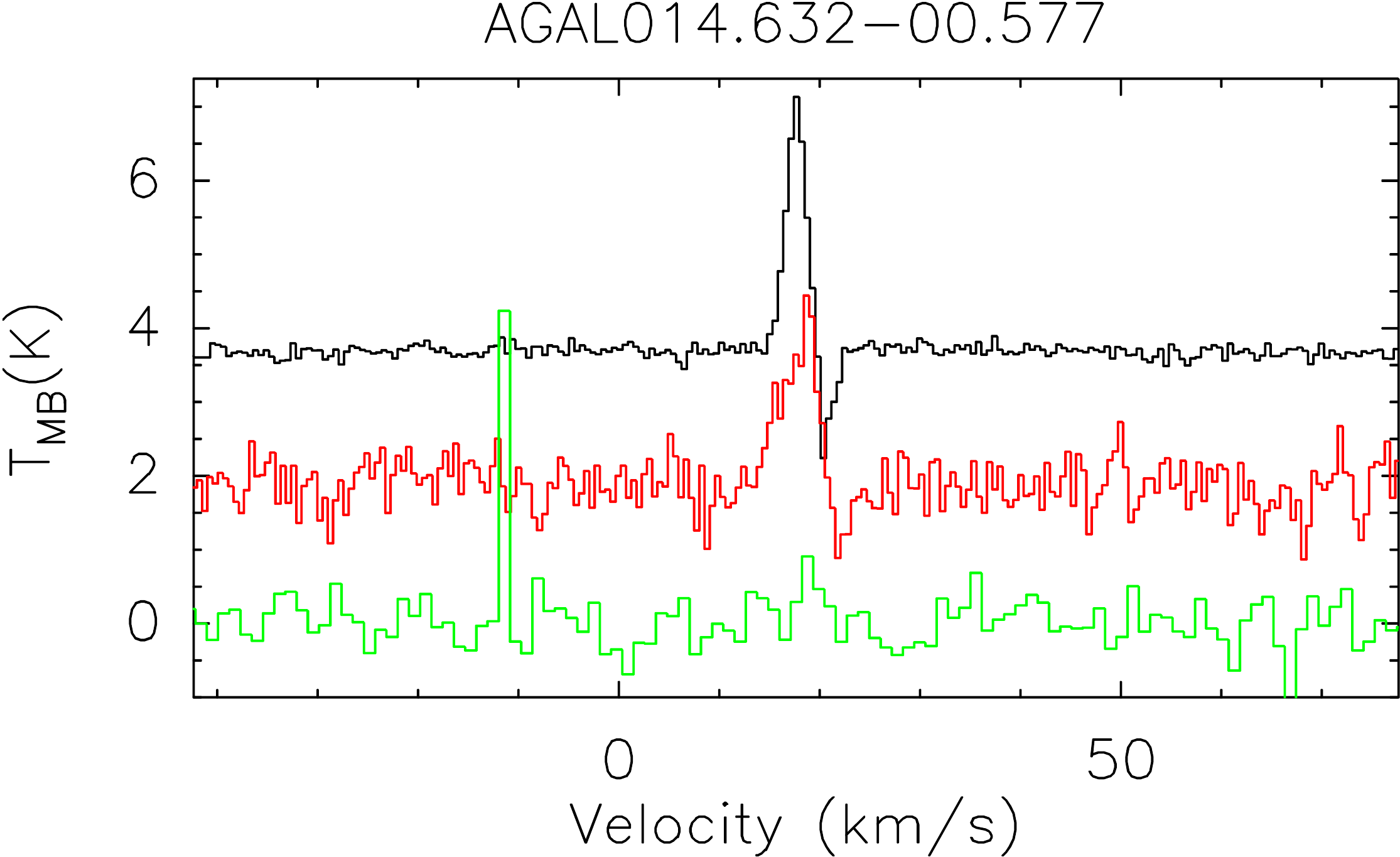} \\
\includegraphics[angle=-90,width=0.3\textwidth]{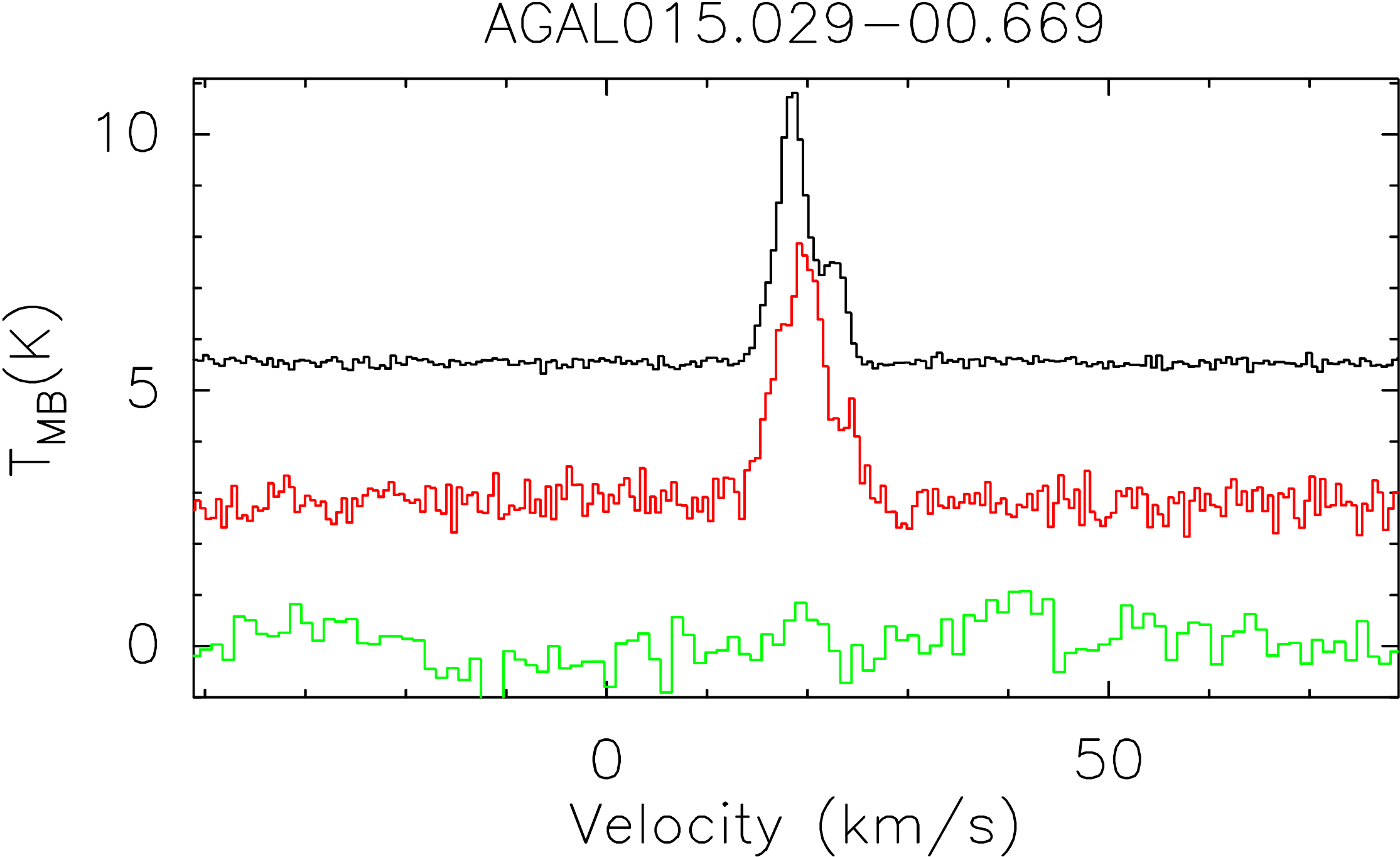}
\includegraphics[angle=-90,width=0.3\textwidth]{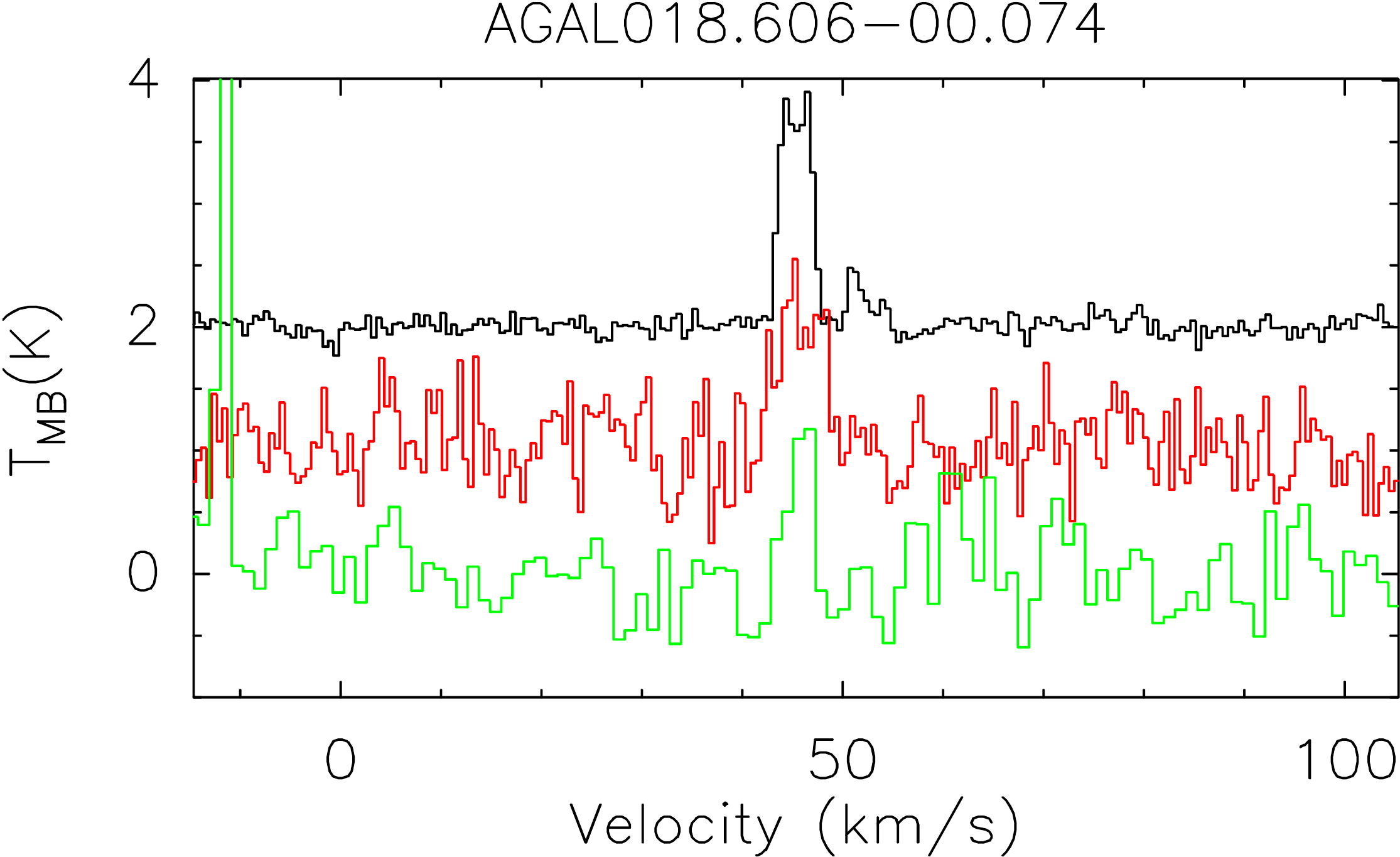} 
\includegraphics[angle=-90,width=0.3\textwidth]{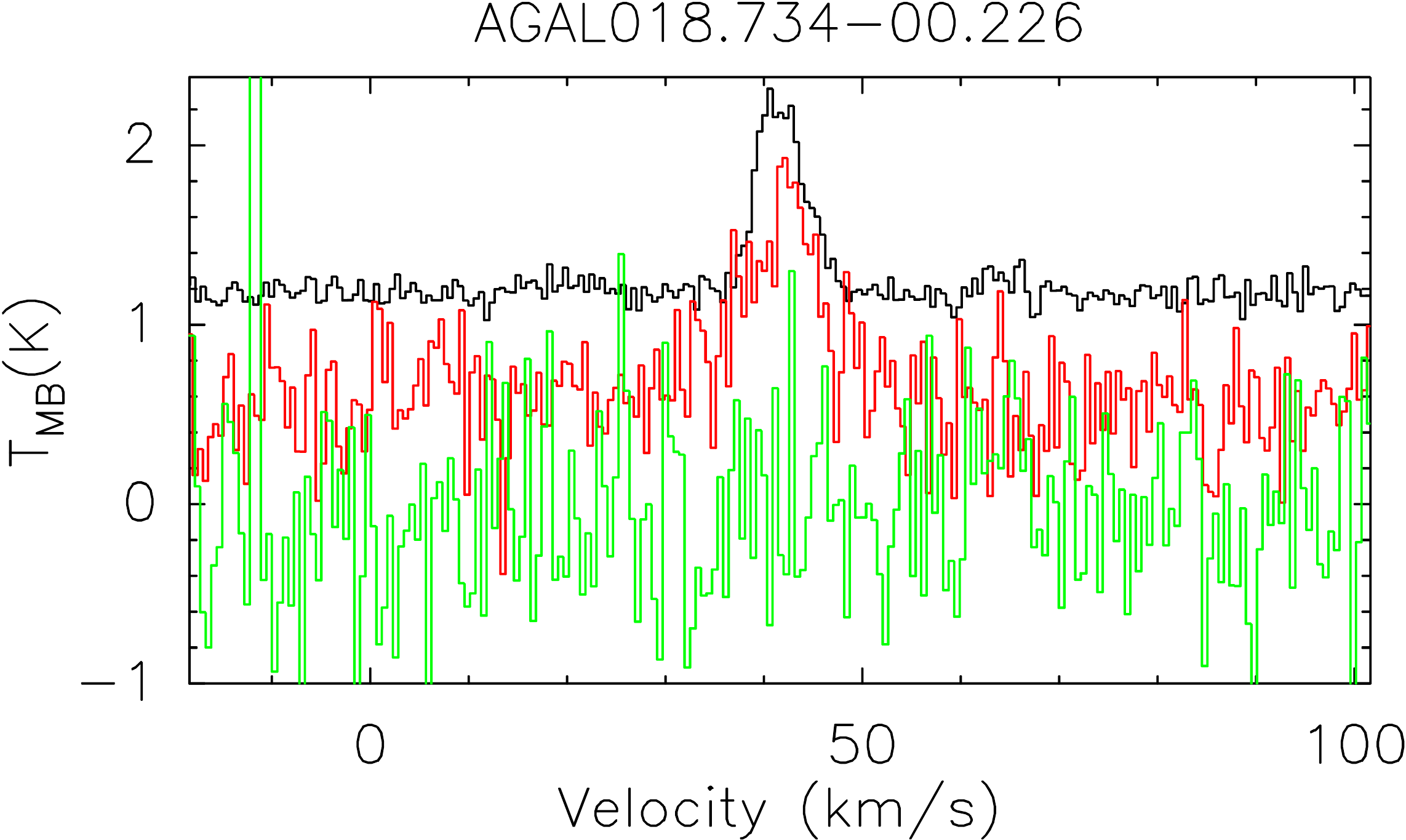} \\
\includegraphics[angle=-90,width=0.3\textwidth]{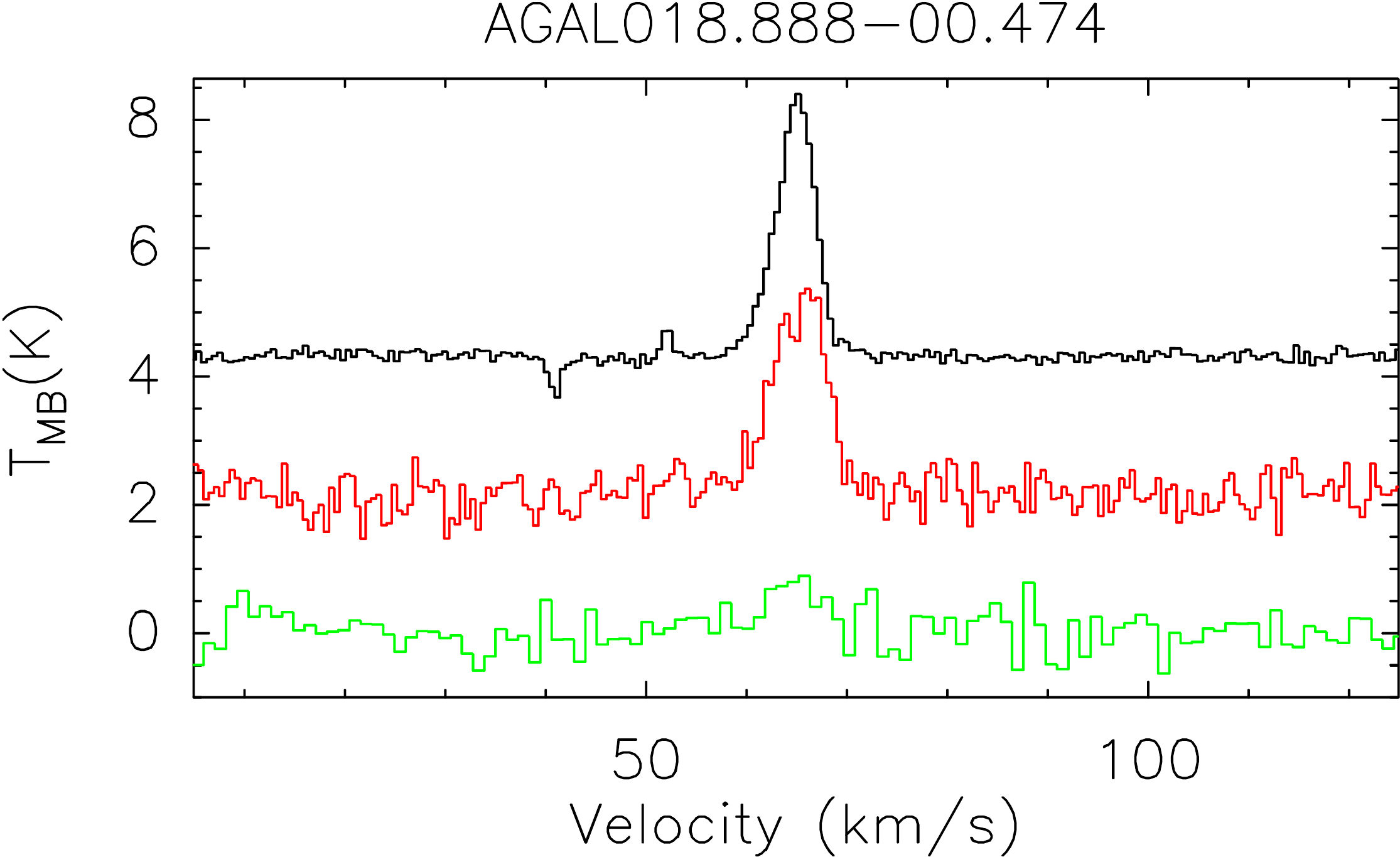}
\includegraphics[angle=-90,width=0.3\textwidth]{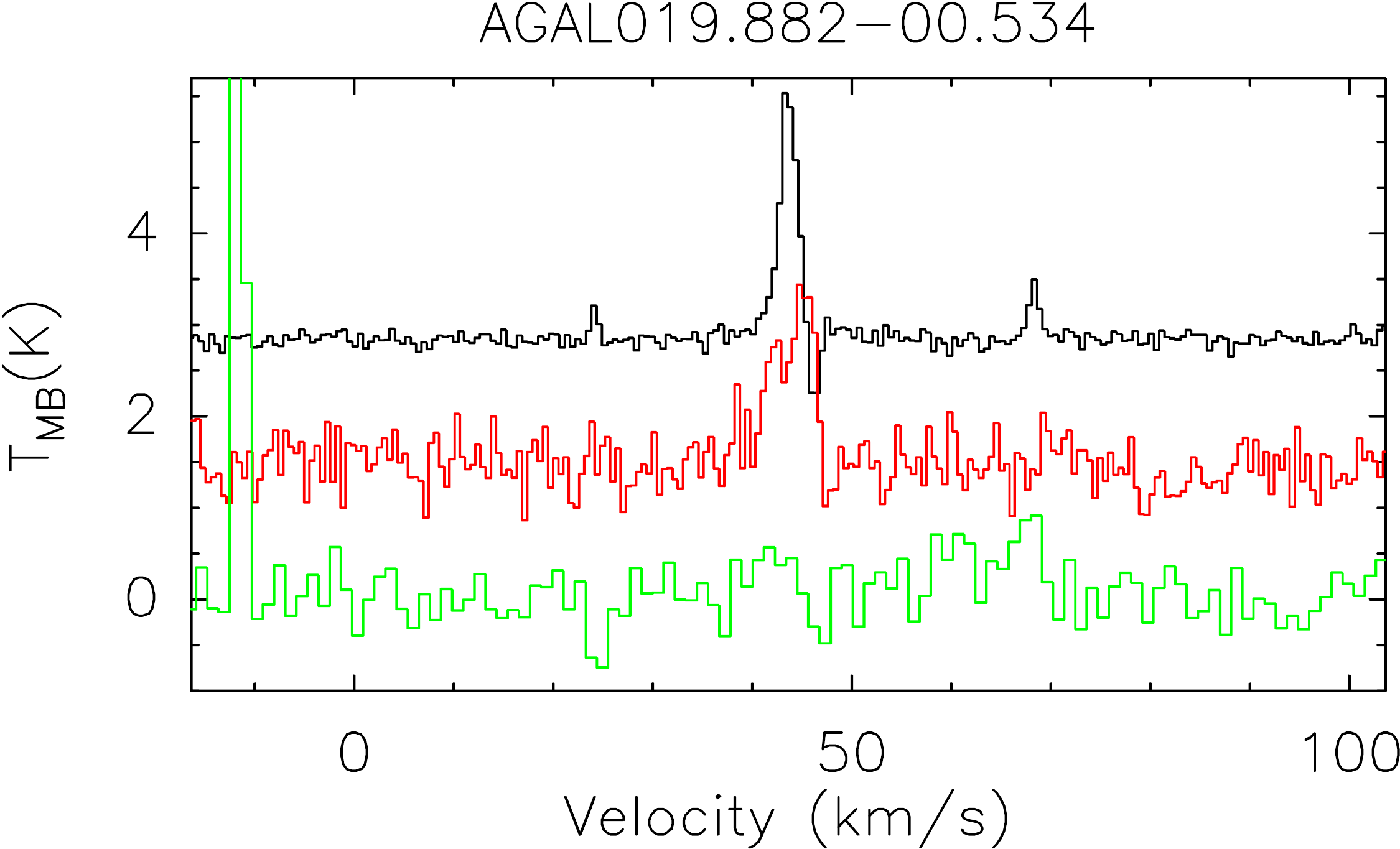}
\includegraphics[angle=-90,width=0.3\textwidth]{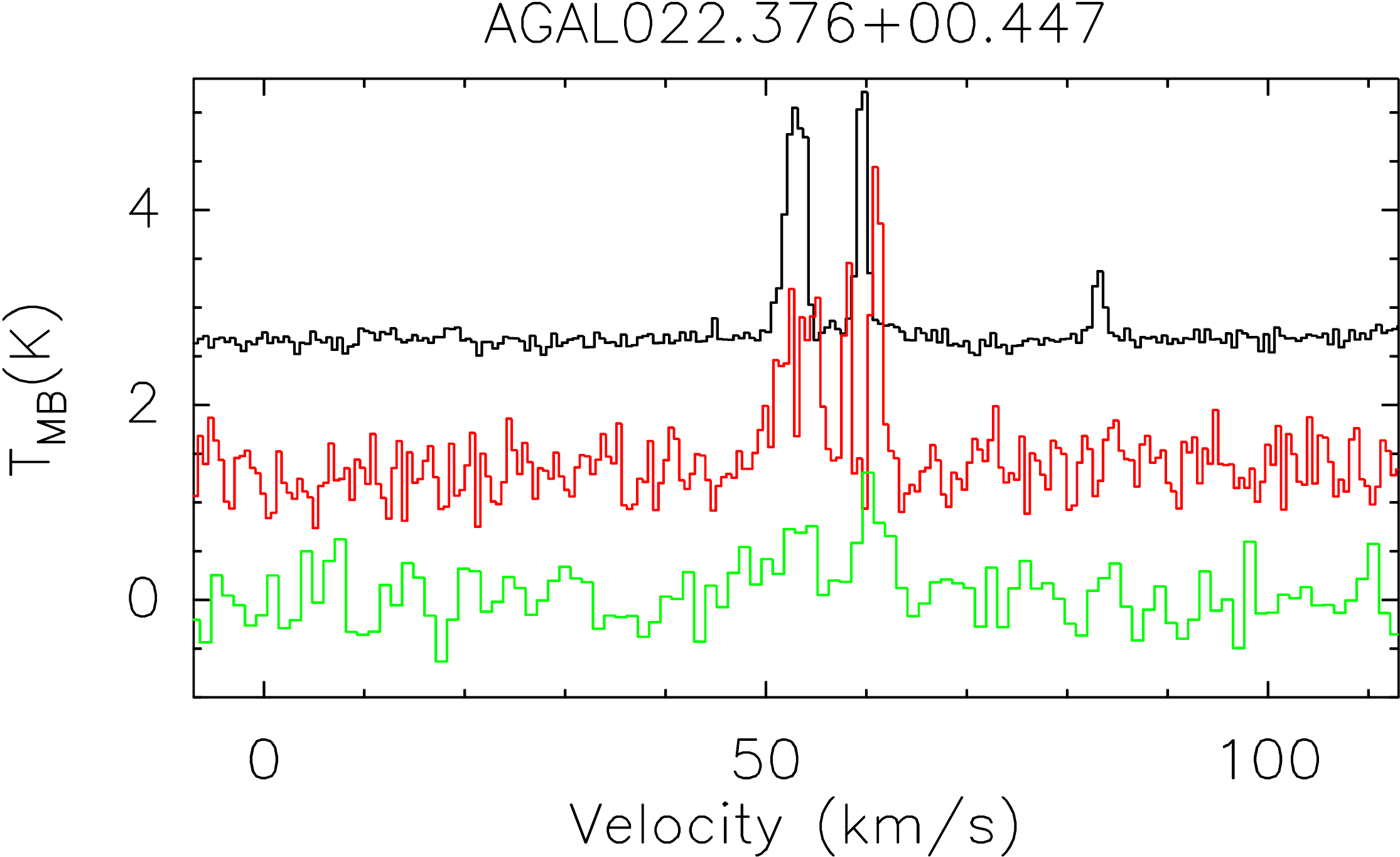}  \hfill
\caption{$^{13}$C$^{18}$O$(1-0)$ (x10, green), C$^{17}$O$(1-0)$ (x3, red) and C$^{18}$O$(1-0)$ (black). The spectra are displaced for clarity.} \label{fig:spectra_10_A}
\end{figure*} 

\begin{figure*} 
\ContinuedFloat
\centering 
\includegraphics[angle=-90,width=0.3\textwidth]{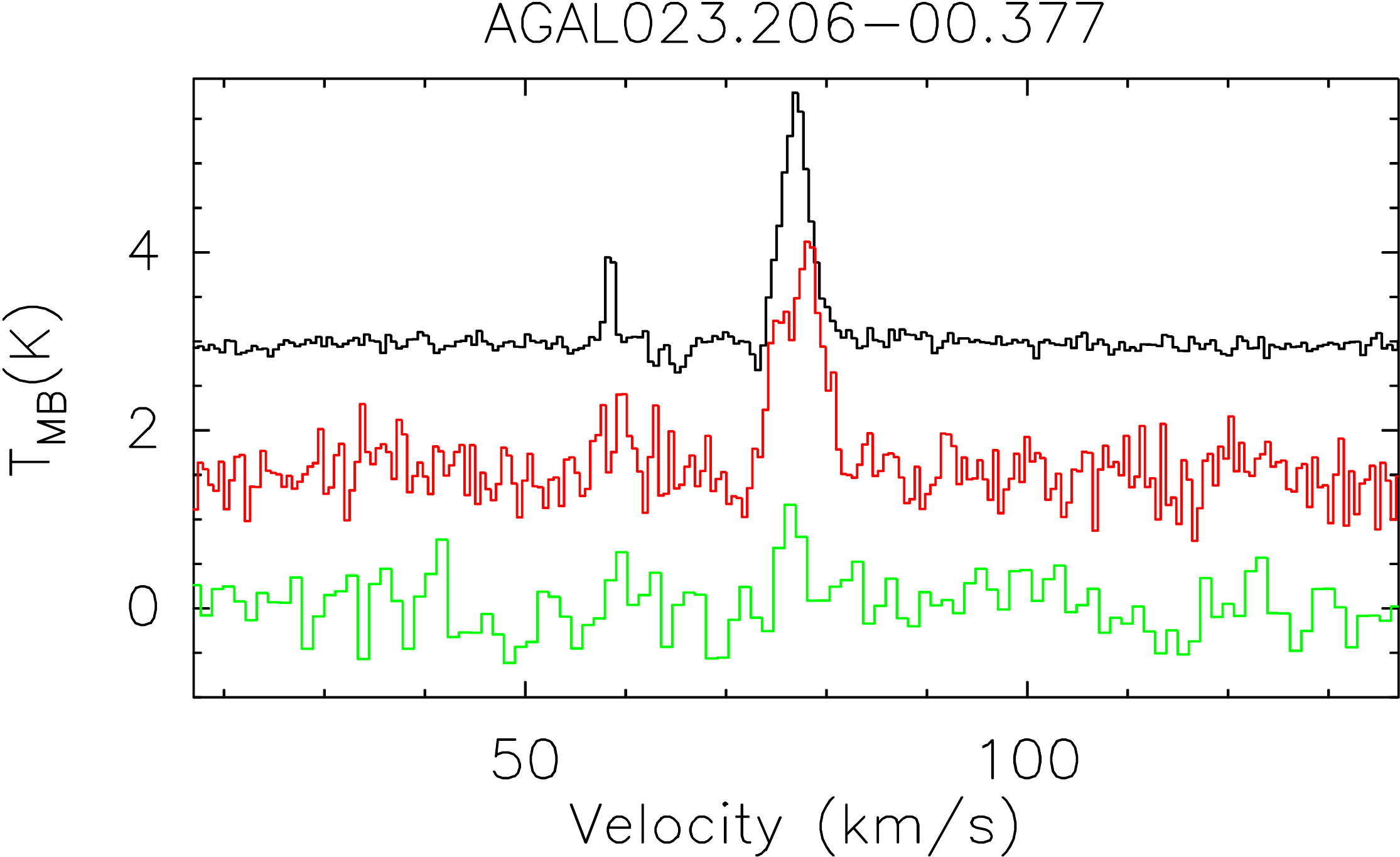}
\includegraphics[angle=-90,width=0.3\textwidth]{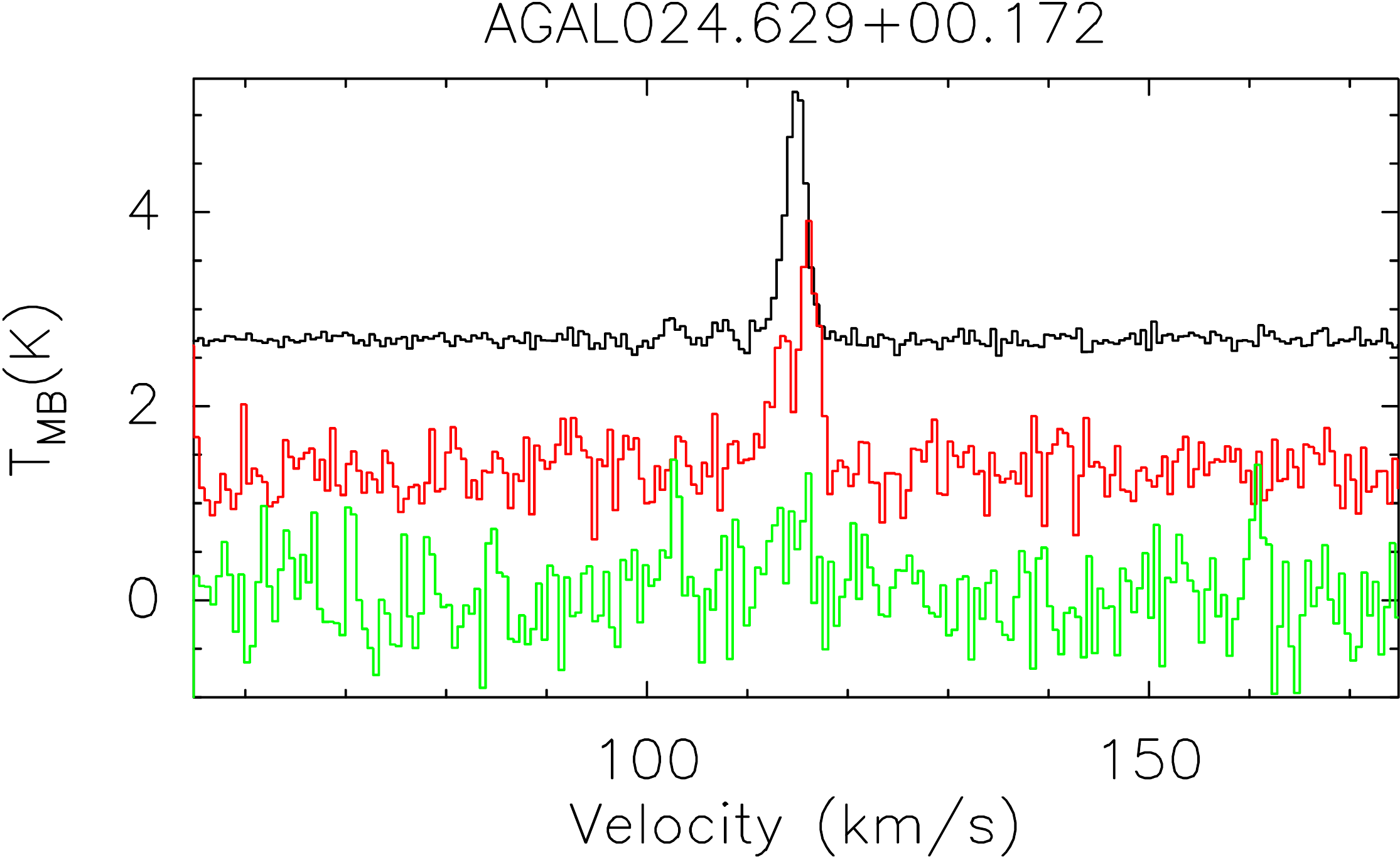}
\includegraphics[angle=-90,width=0.3\textwidth]{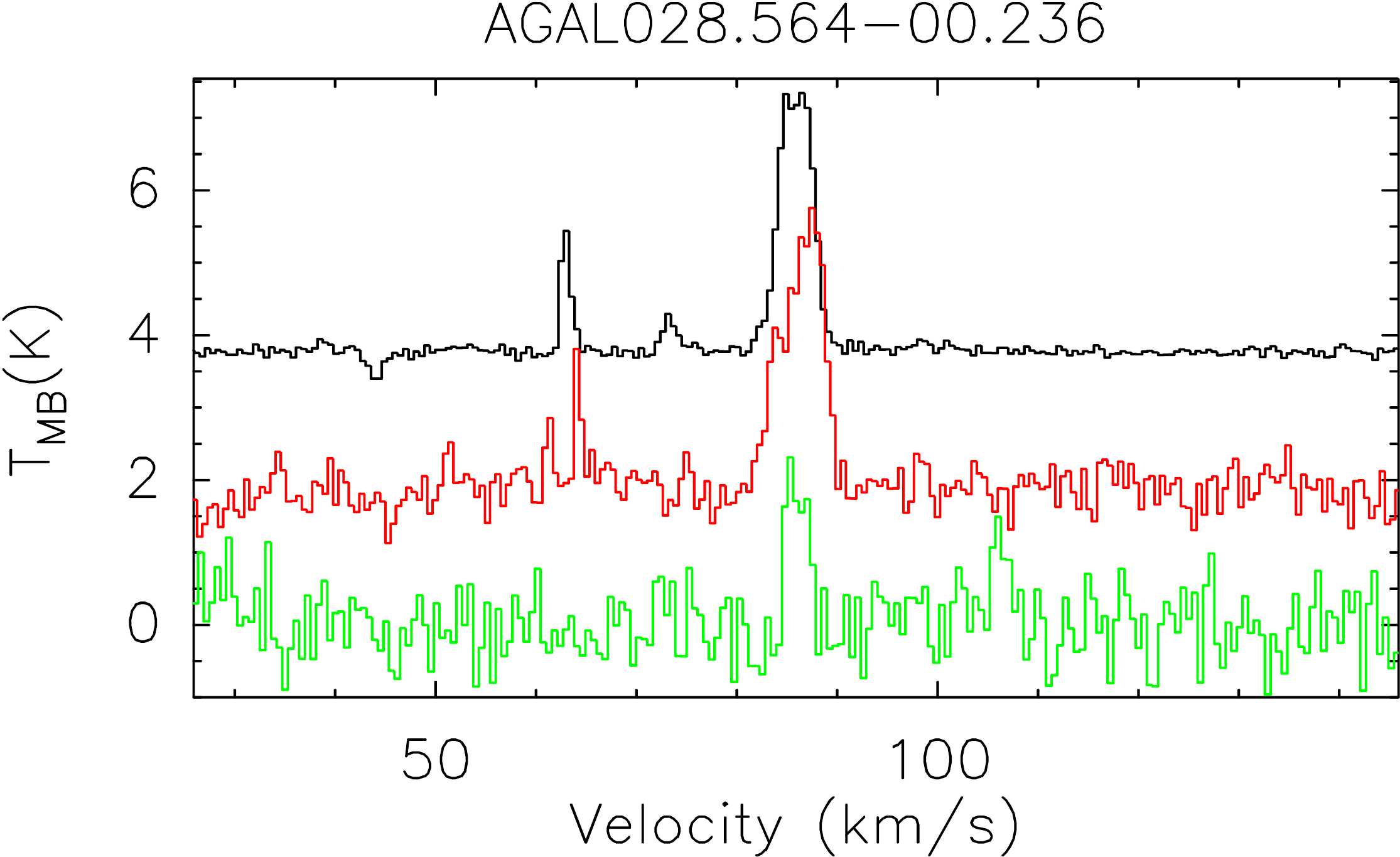} \\
\includegraphics[angle=-90,width=0.3\textwidth]{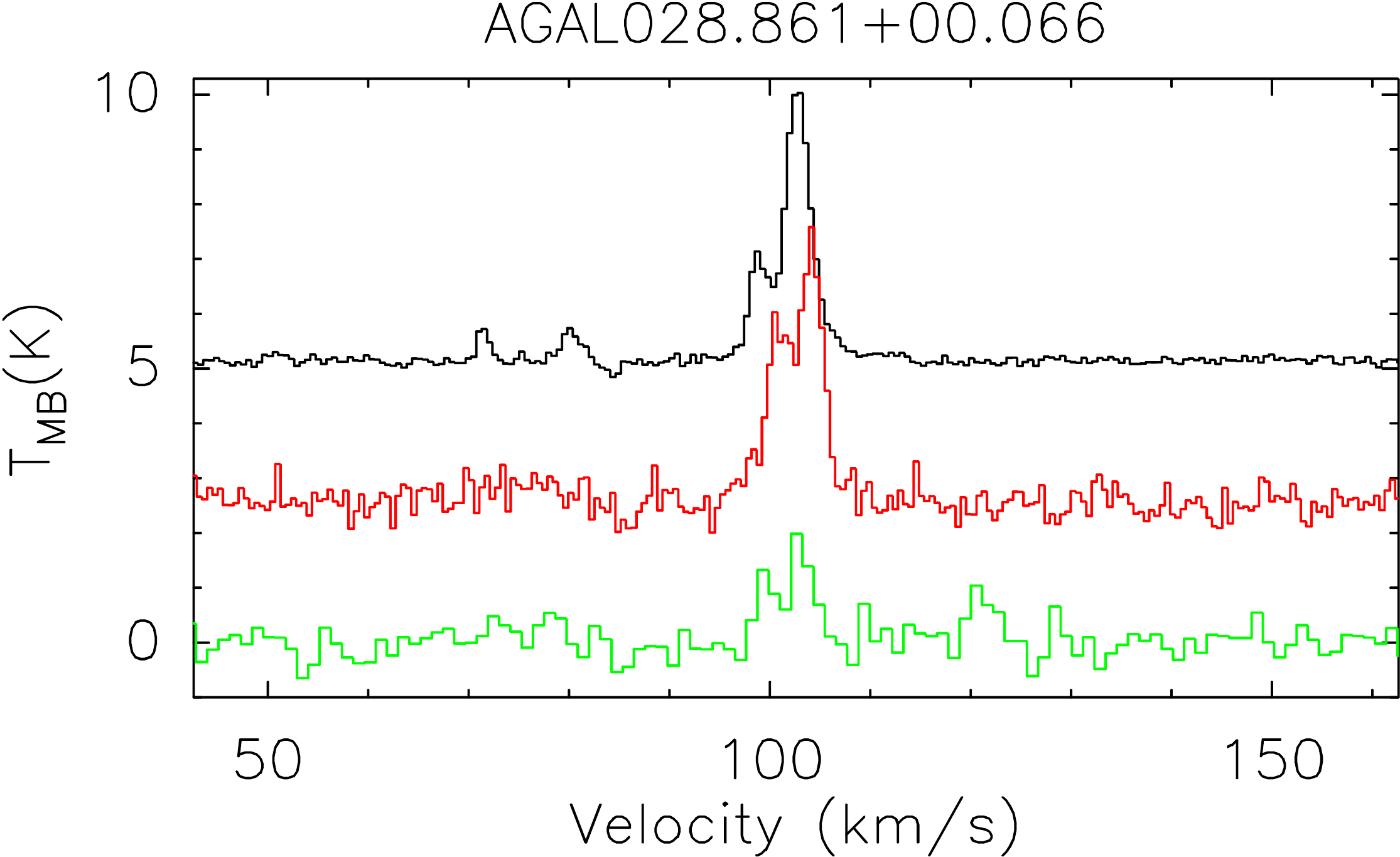} 
\includegraphics[angle=-90,width=0.3\textwidth]{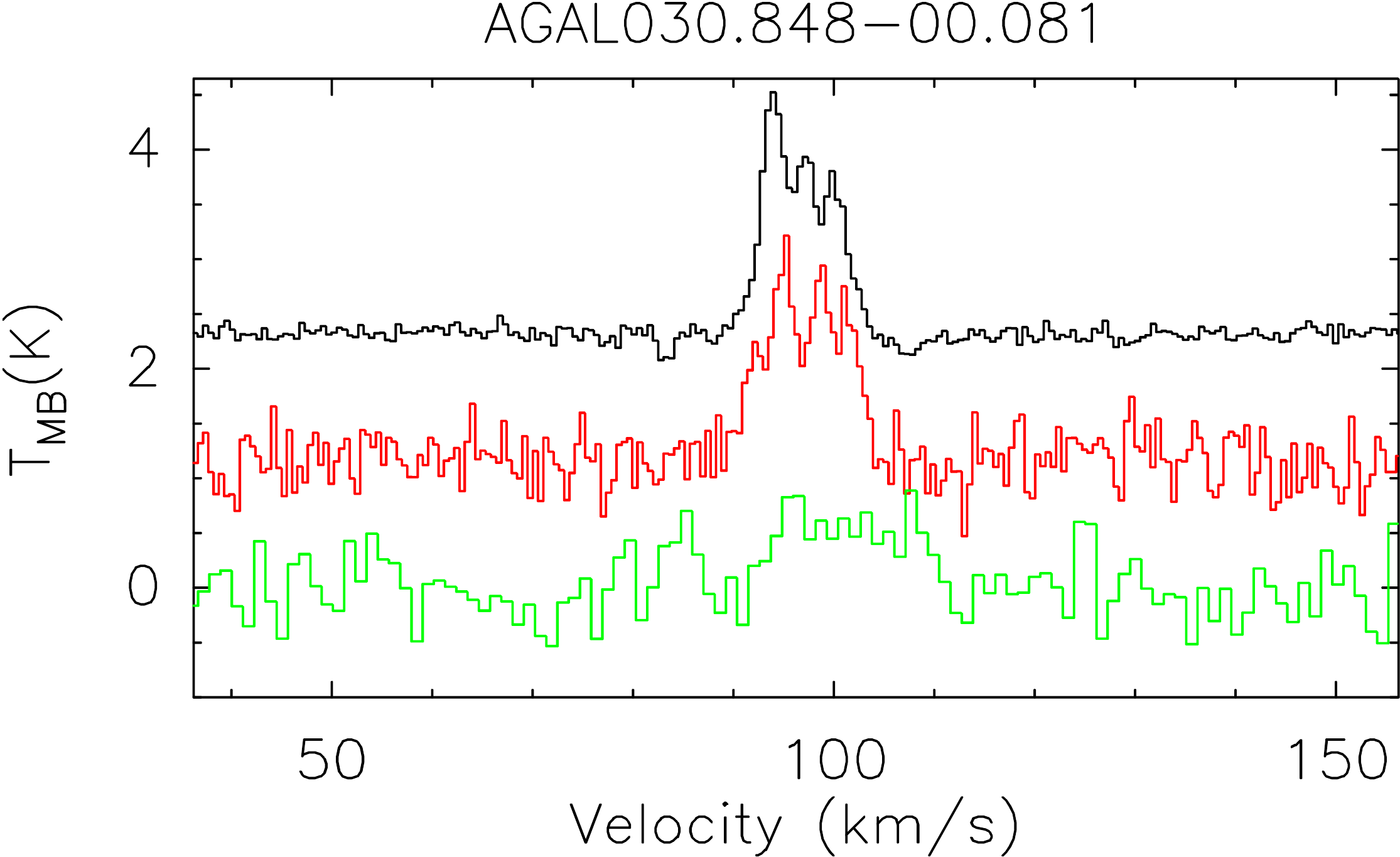}
\includegraphics[angle=-90,width=0.3\textwidth]{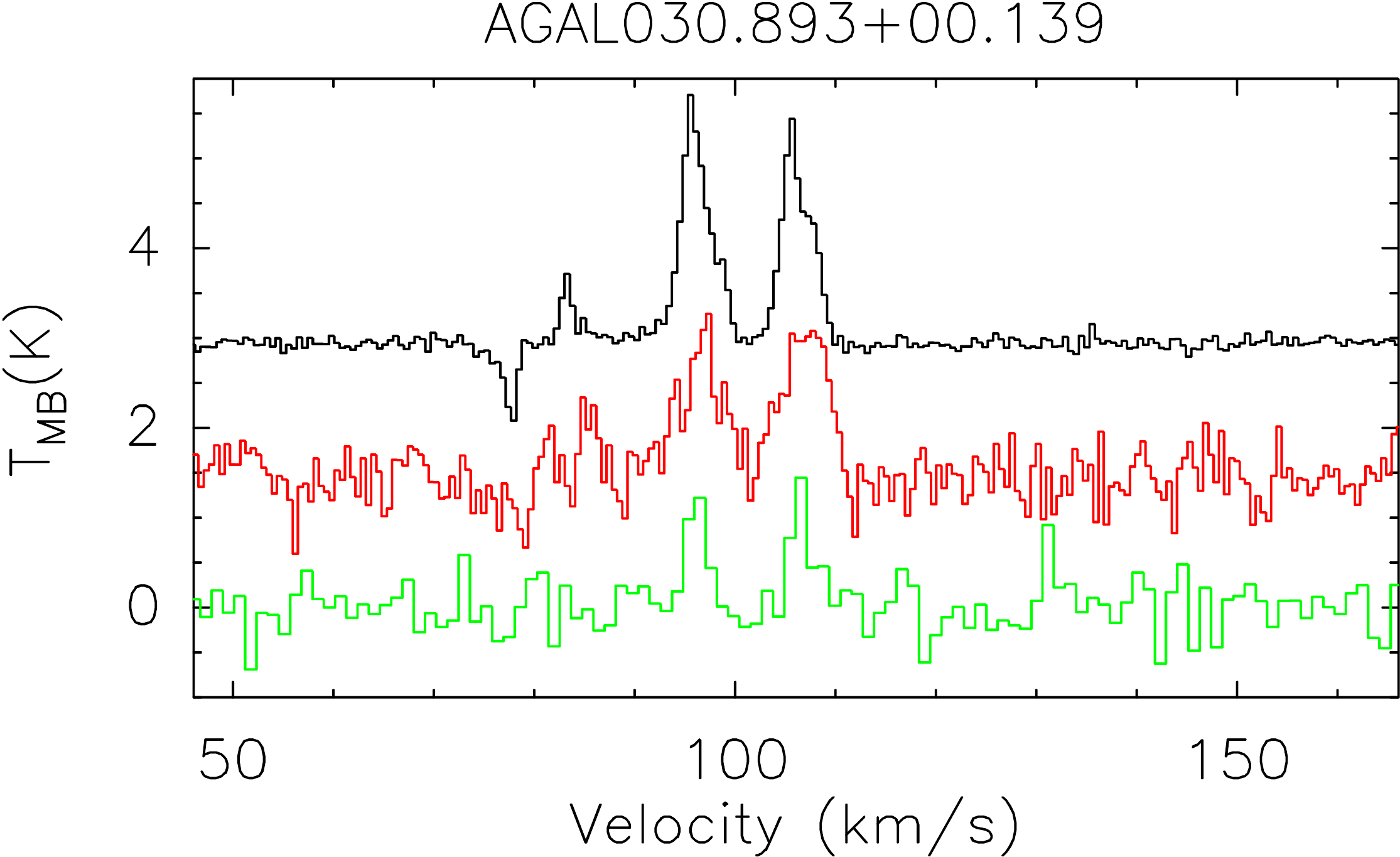} \\
\includegraphics[angle=-90,width=0.3\textwidth]{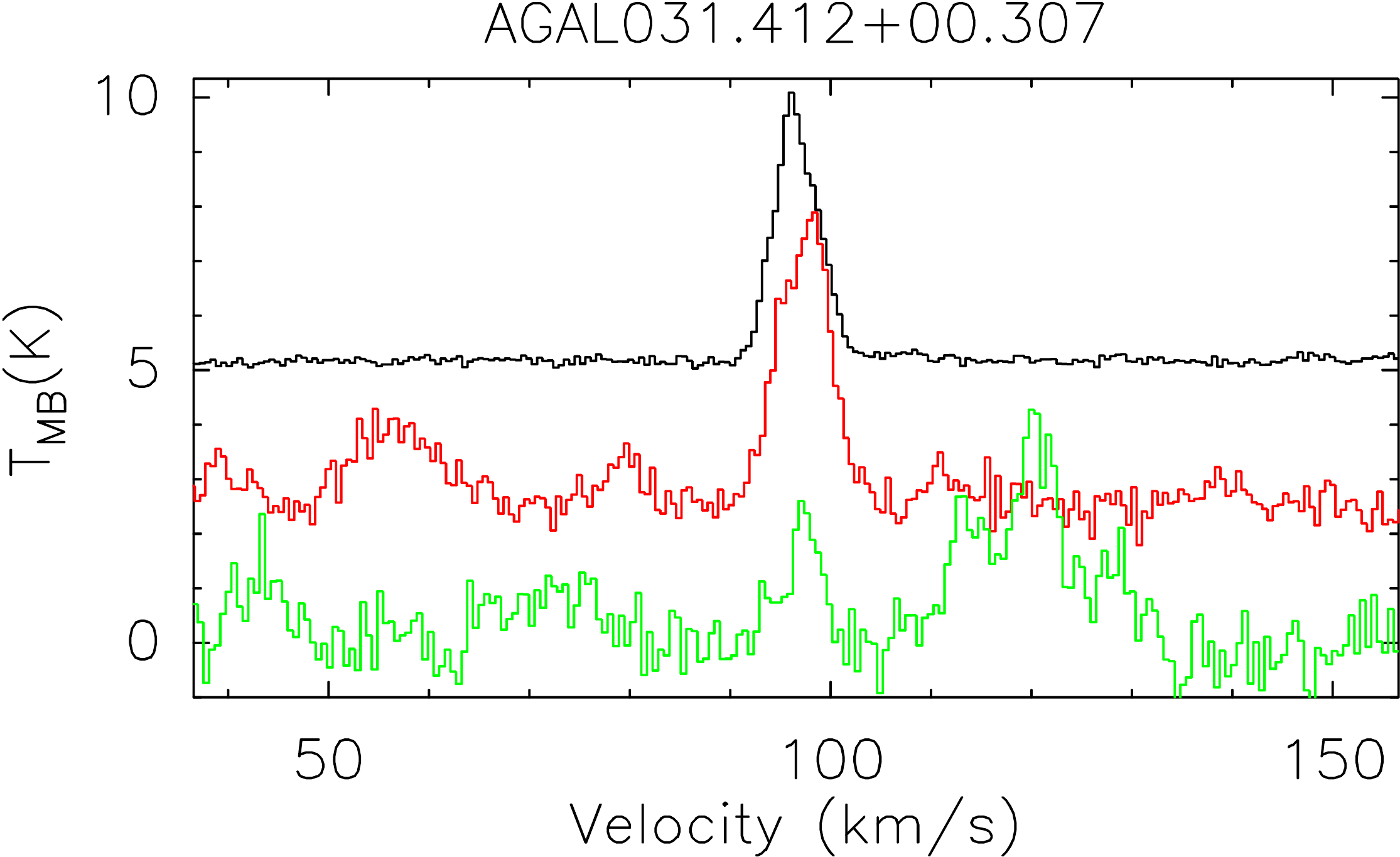}
\includegraphics[angle=-90,width=0.3\textwidth]{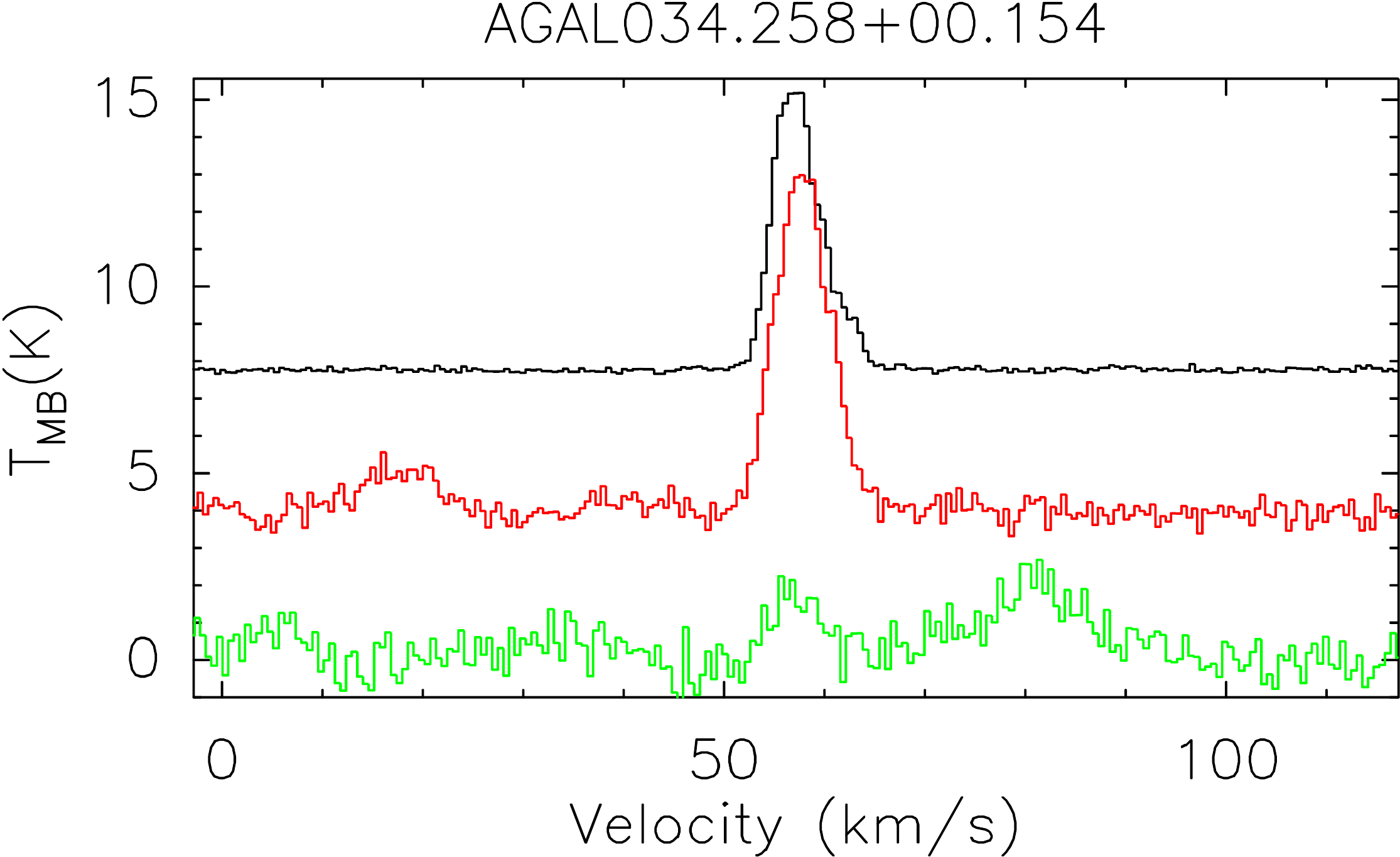} 
\includegraphics[angle=-90,width=0.3\textwidth]{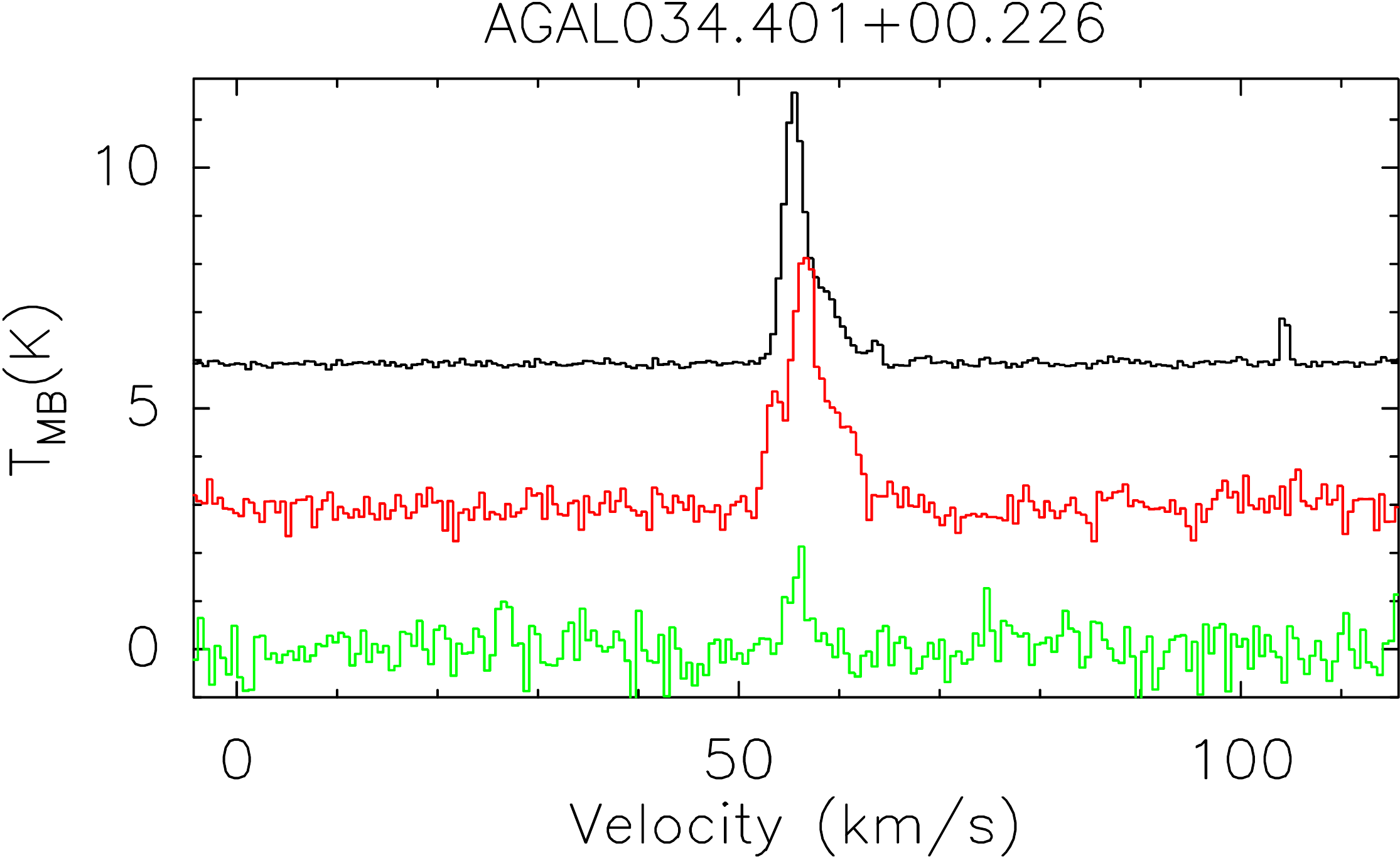} \\
\includegraphics[angle=-90,width=0.3\textwidth]{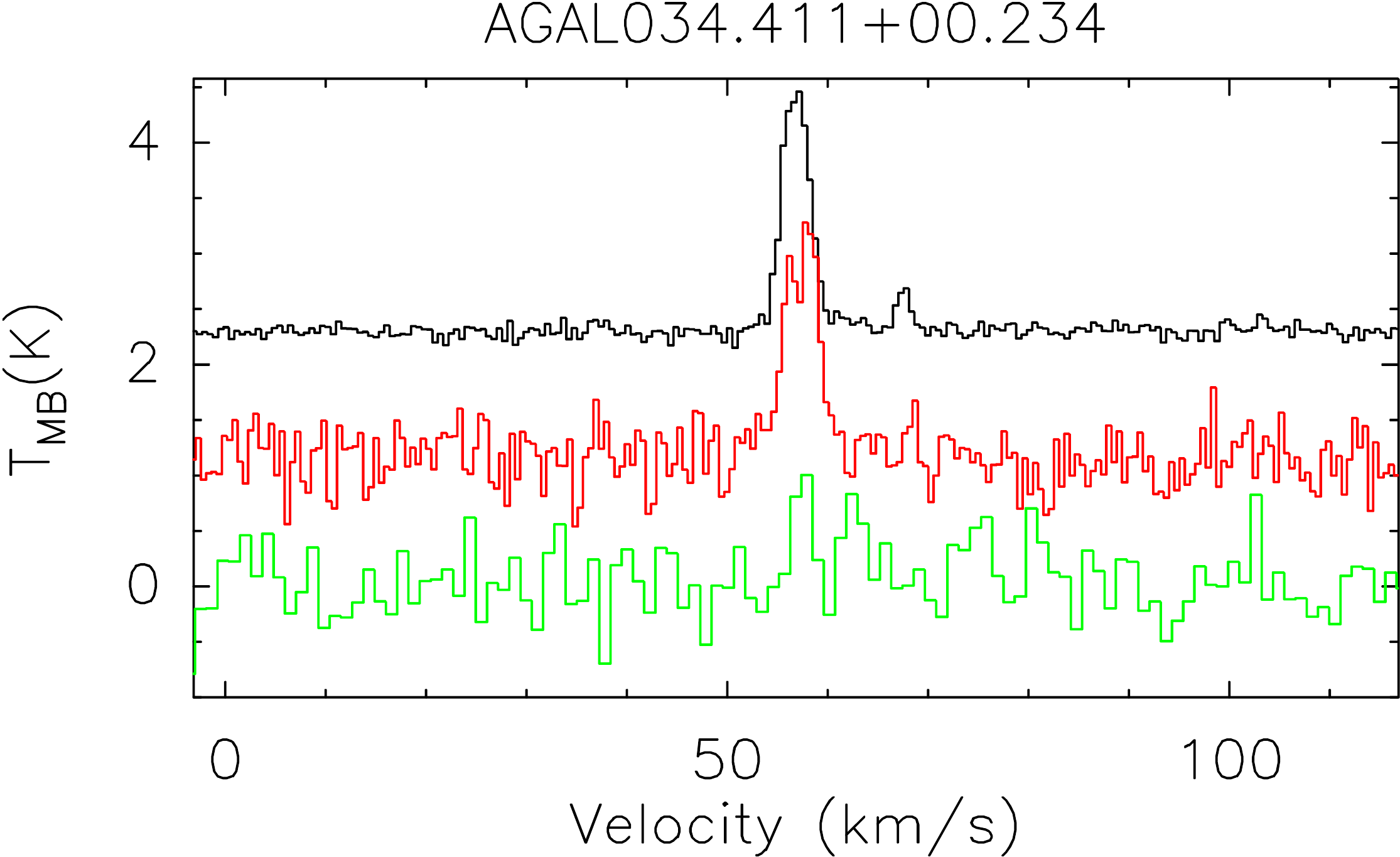}
\includegraphics[angle=-90,width=0.3\textwidth]{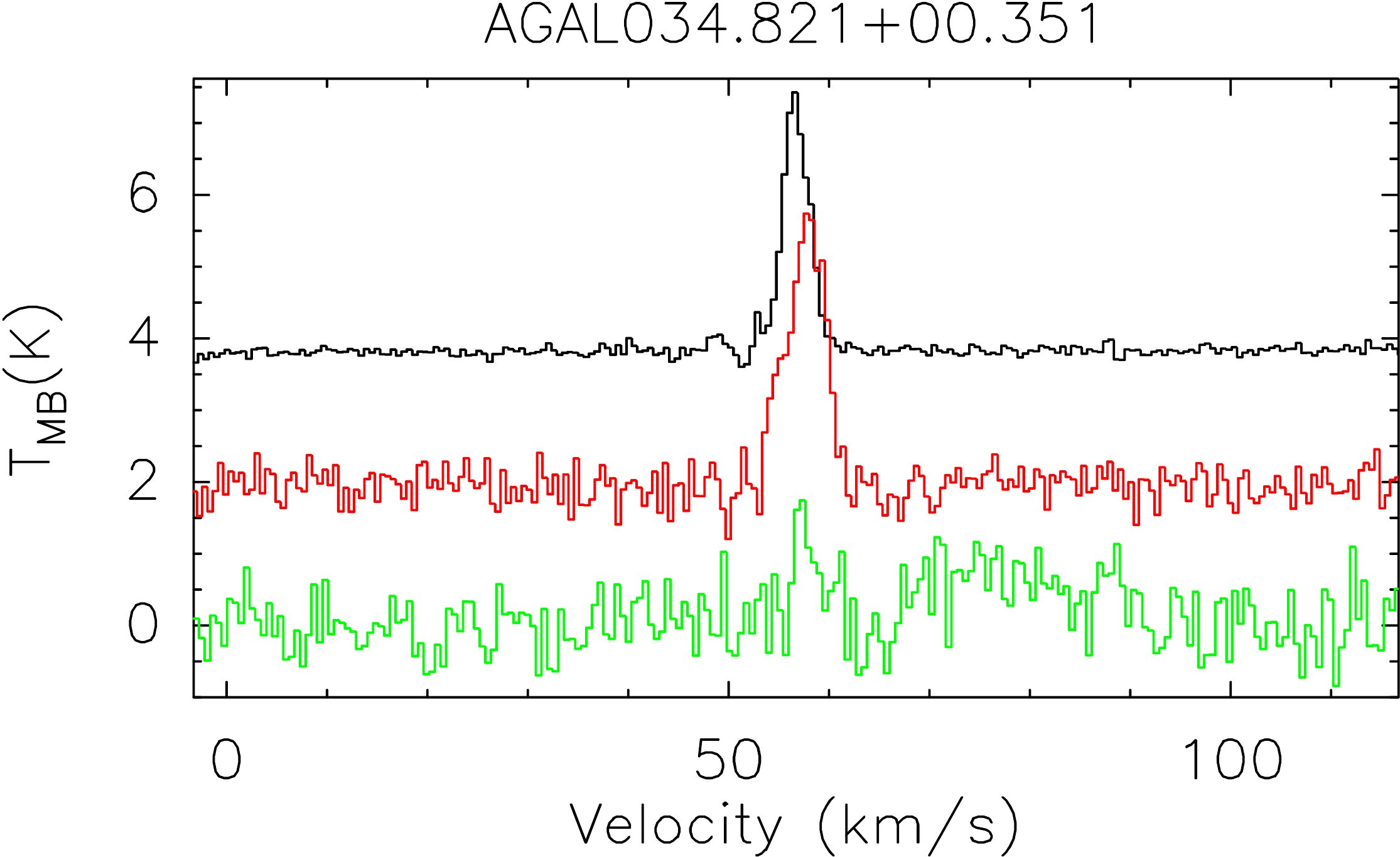} 
\includegraphics[angle=-90,width=0.3\textwidth]{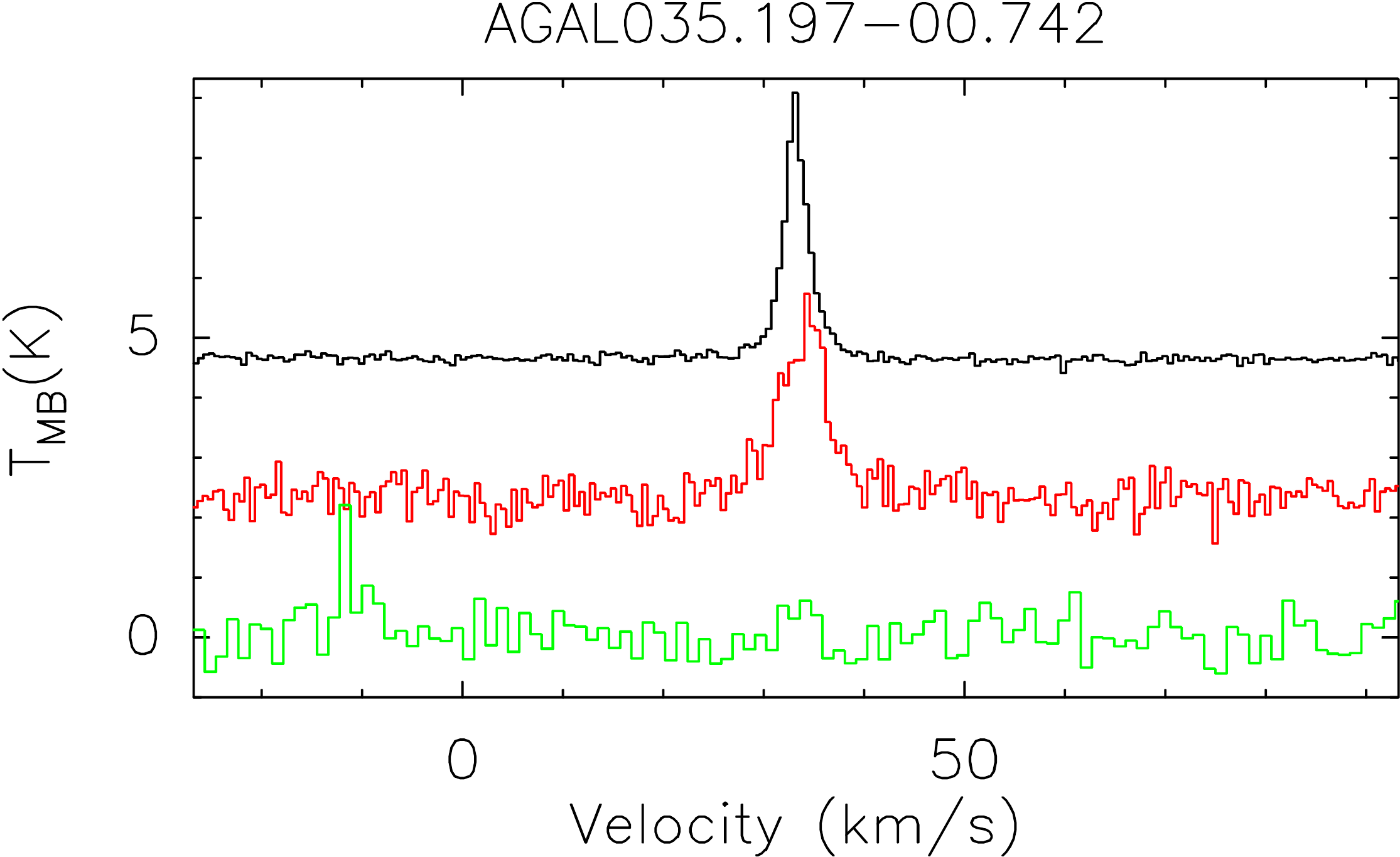} \\
\includegraphics[angle=-90,width=0.3\textwidth]{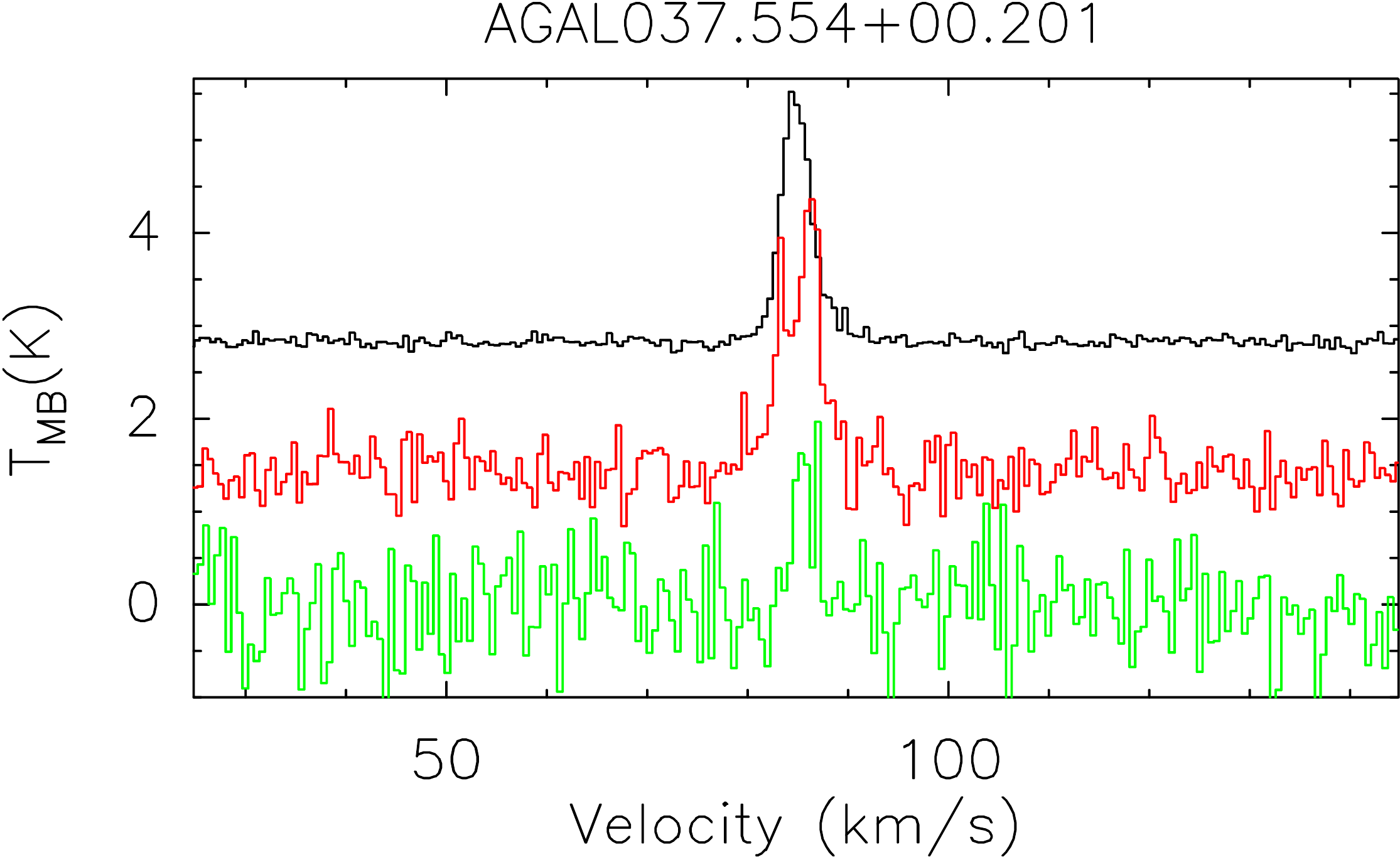}
\includegraphics[angle=-90,width=0.3\textwidth]{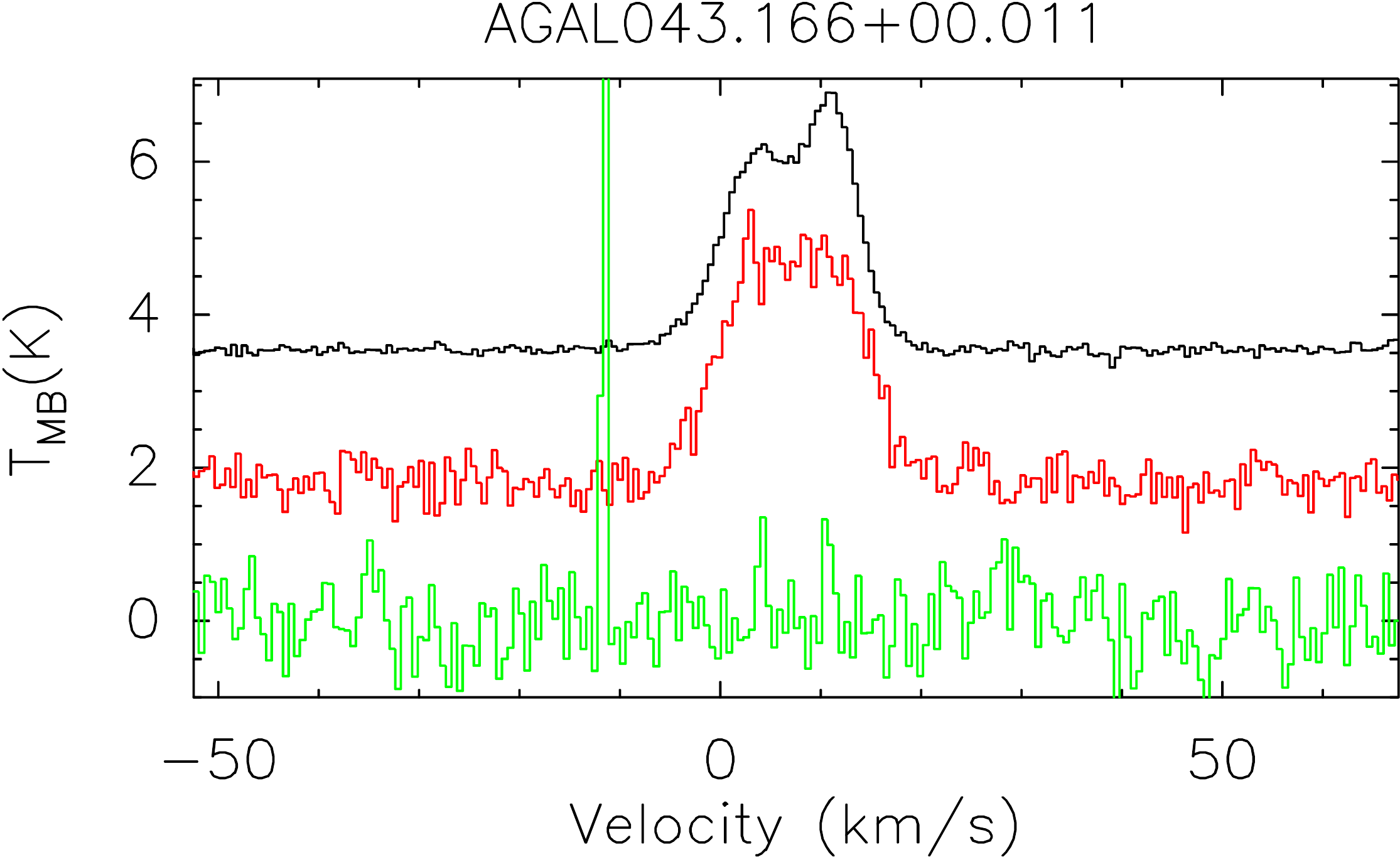}
\includegraphics[angle=-90,width=0.3\textwidth]{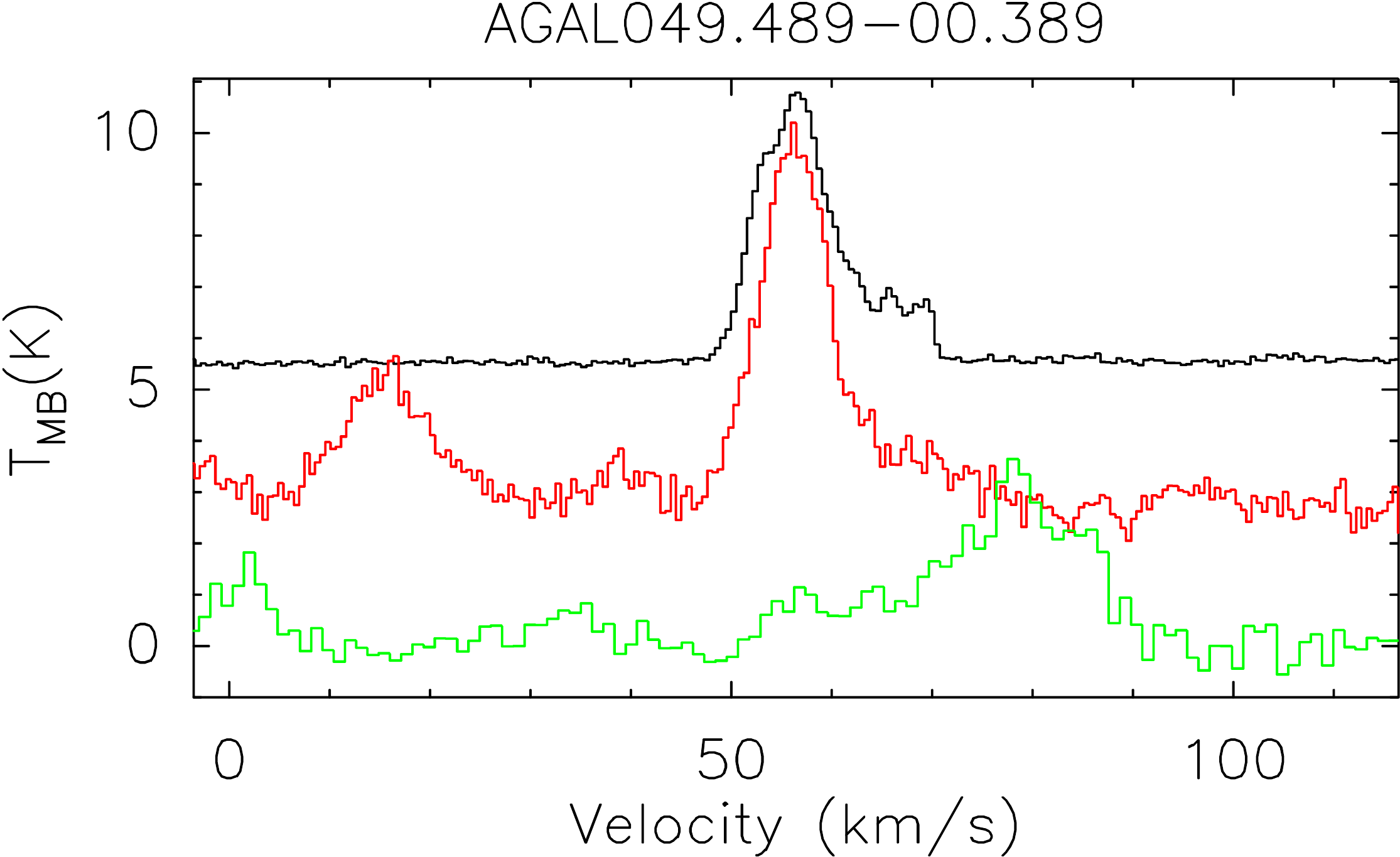} \\
\includegraphics[angle=-90,width=0.3\textwidth]{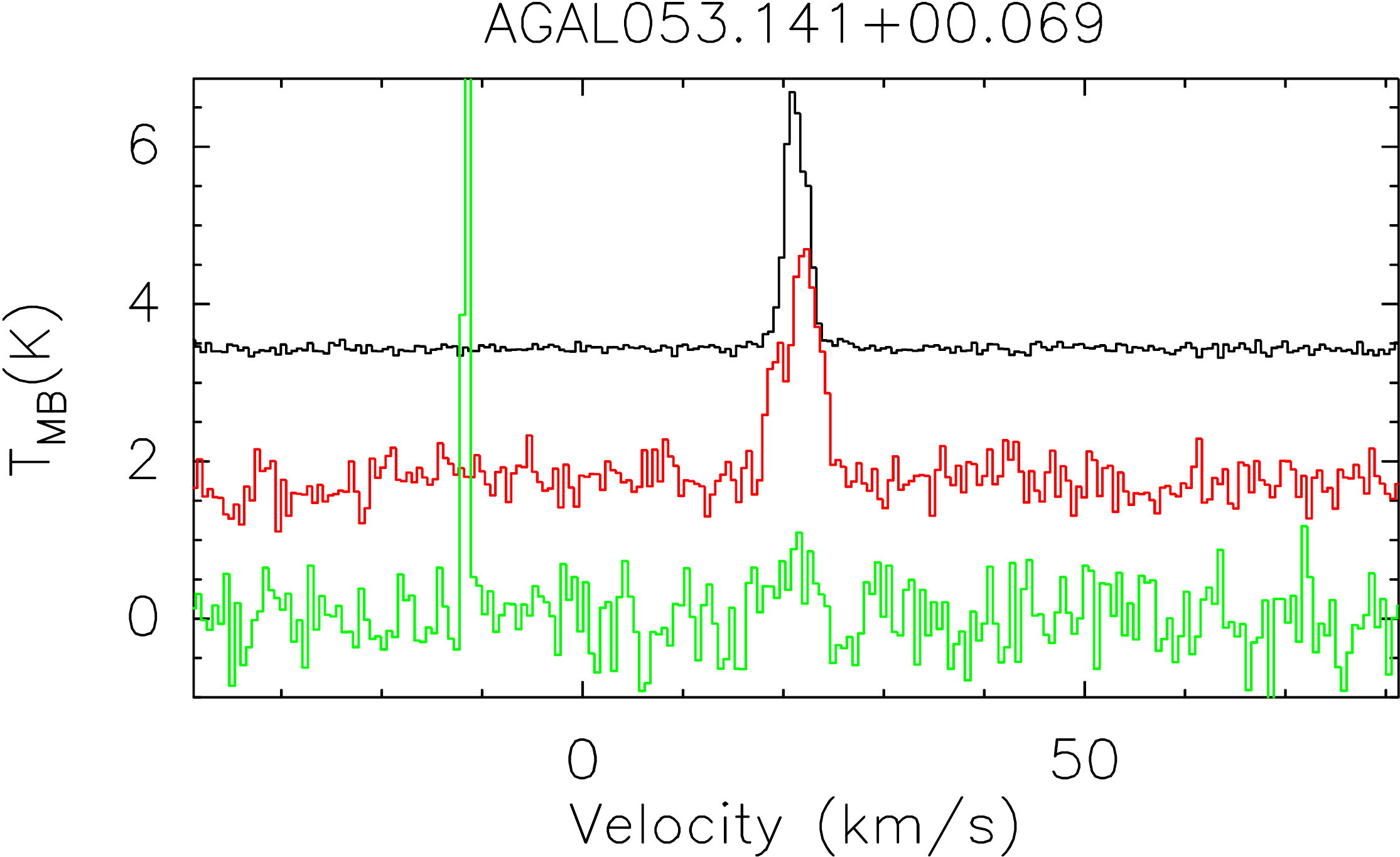}
\includegraphics[angle=-90,width=0.3\textwidth]{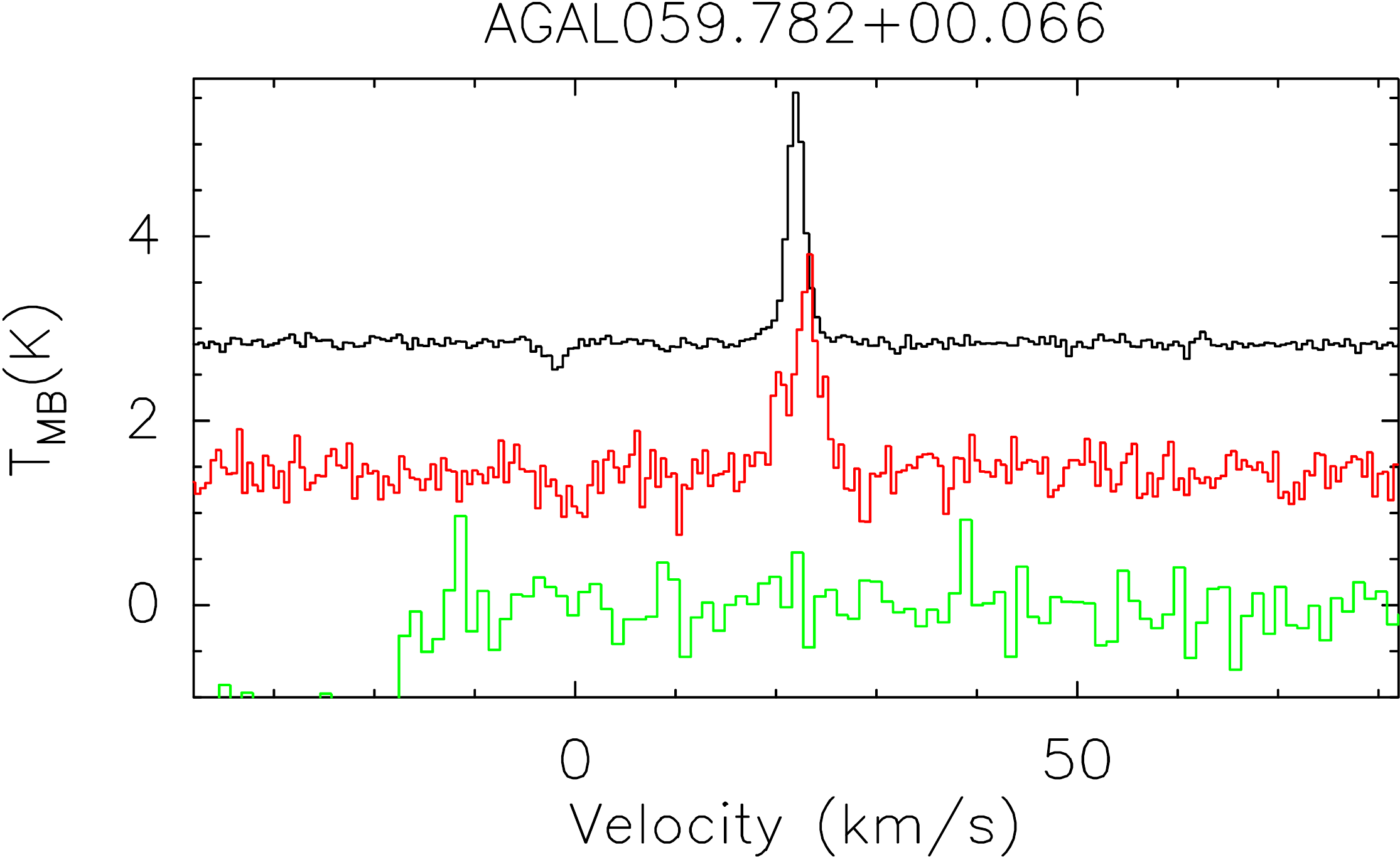} \hfill
\caption{Continued.} 
\end{figure*} 
}
\onlfig{2}{
\begin{figure*} 
\centering 
\includegraphics[angle=-90,width=0.3\textwidth]{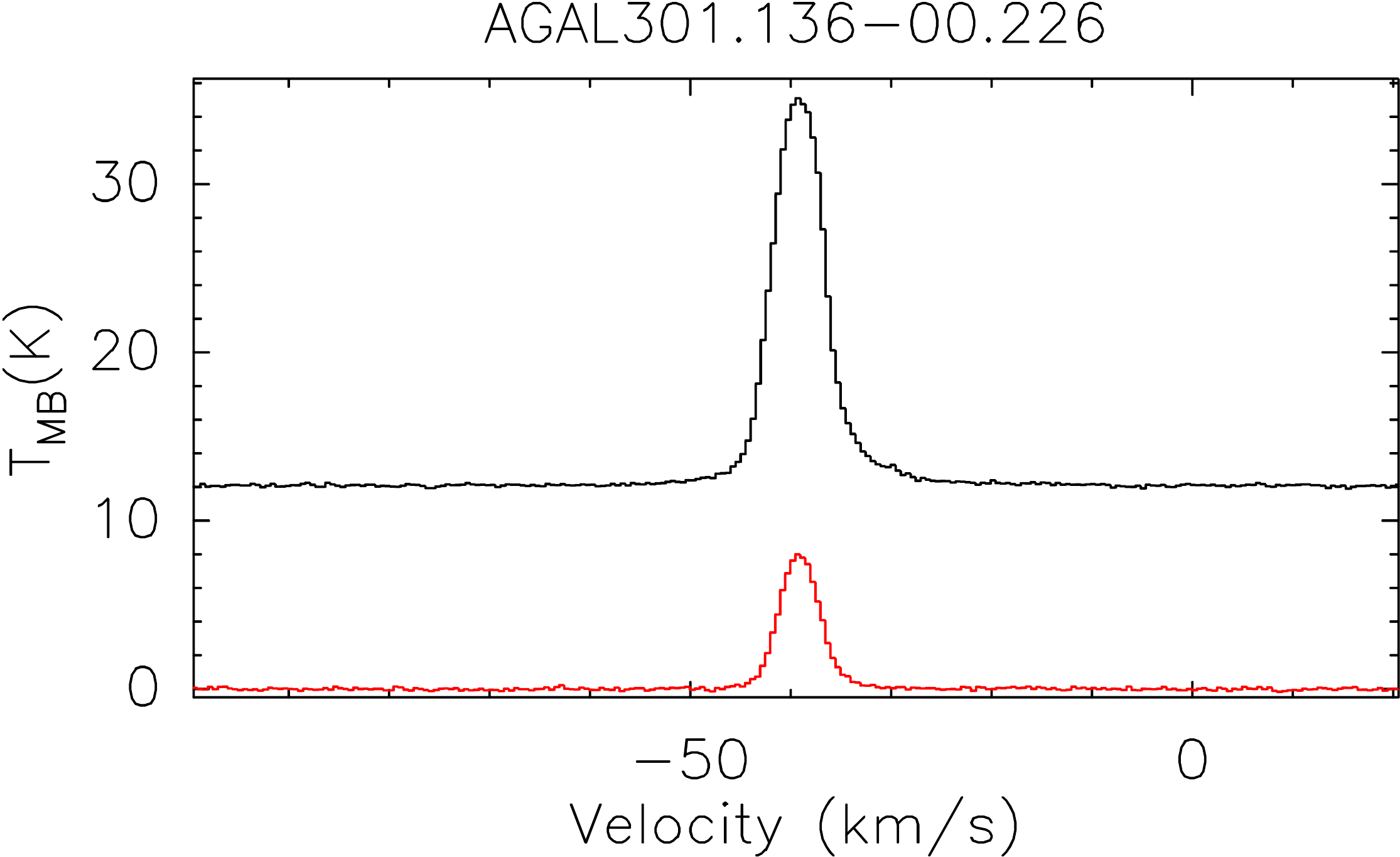} 
\includegraphics[angle=-90,width=0.3\textwidth]{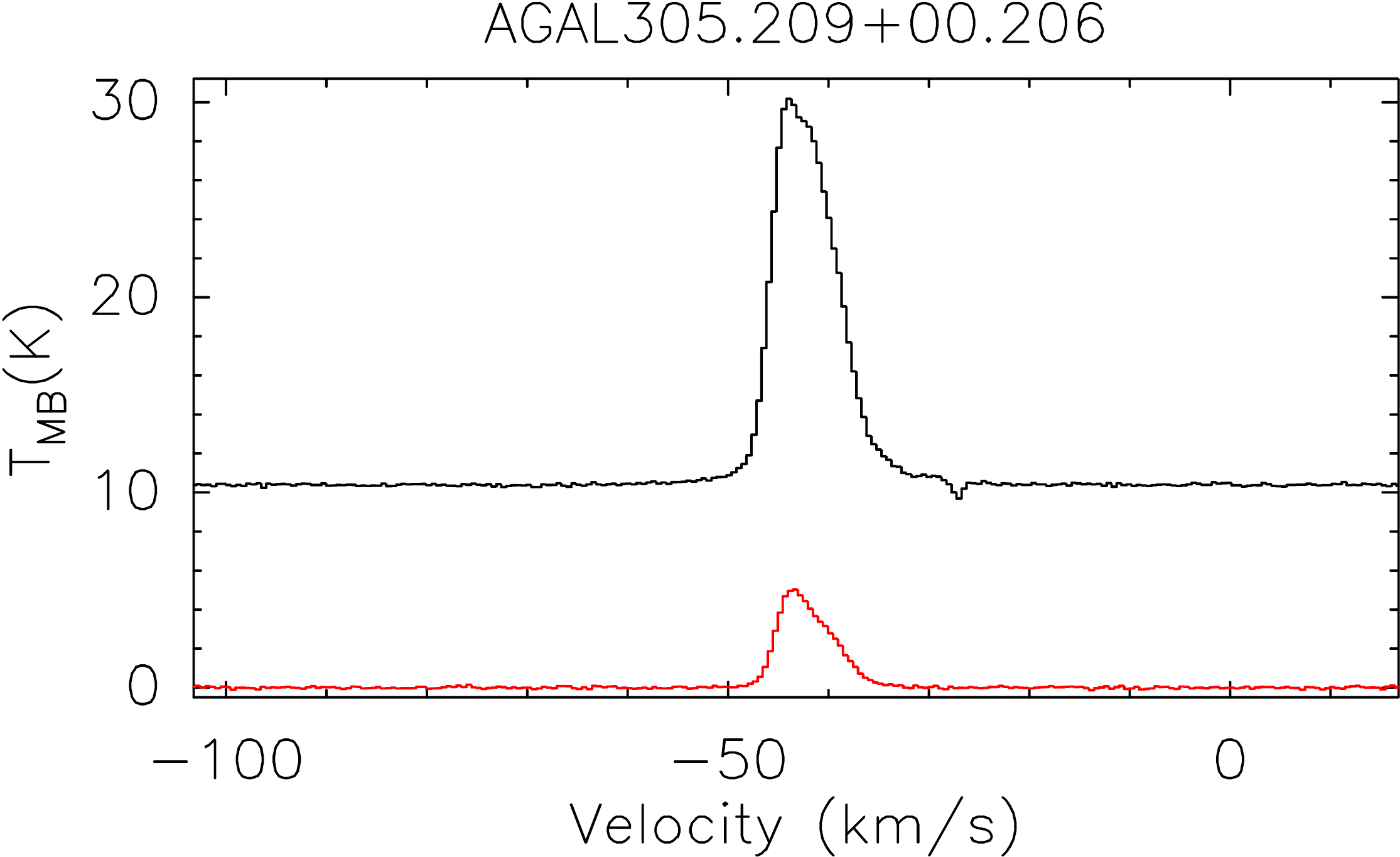} 
\includegraphics[angle=-90,width=0.3\textwidth]{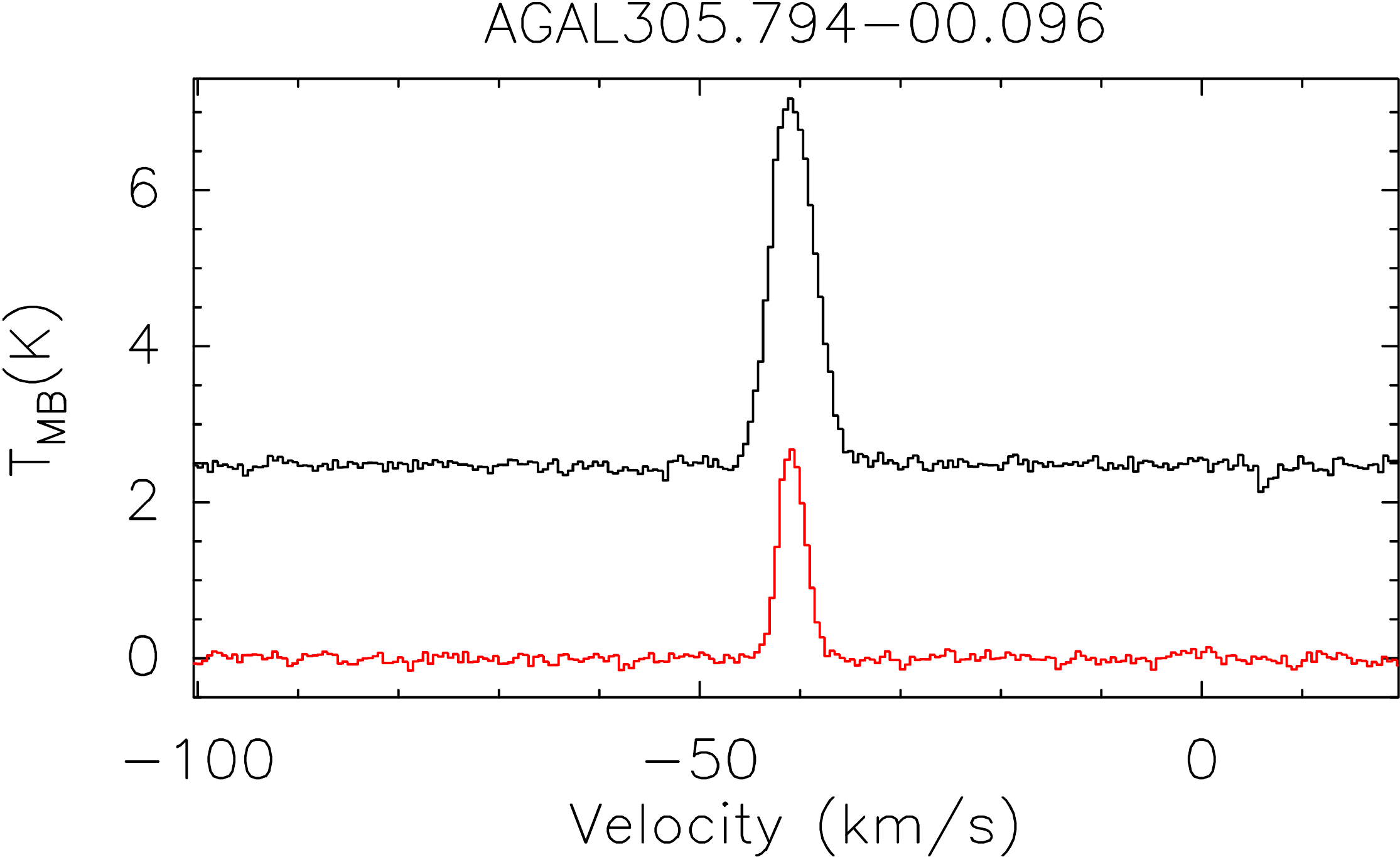} \\ 
\includegraphics[angle=-90,width=0.3\textwidth]{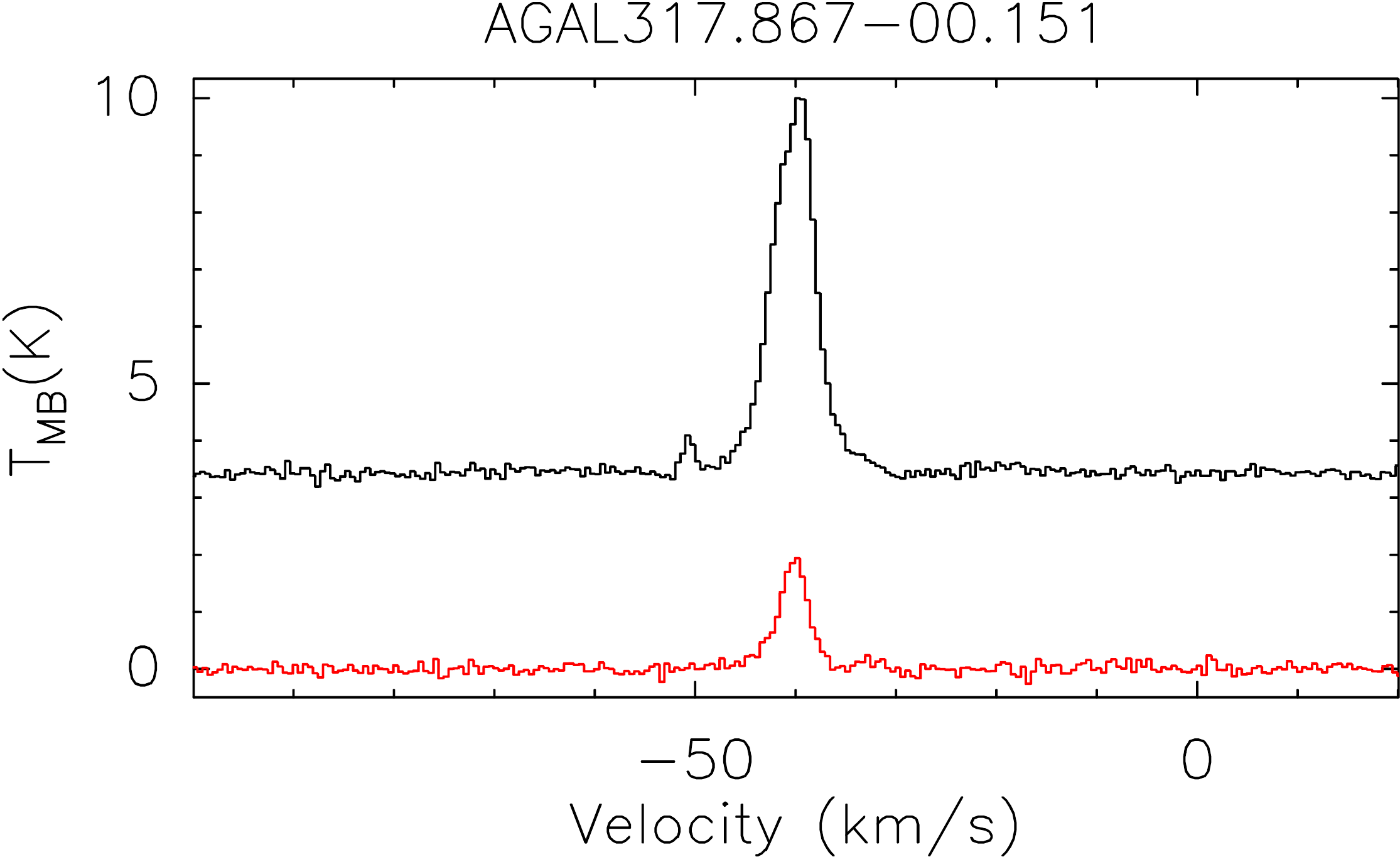} 
\includegraphics[angle=-90,width=0.3\textwidth]{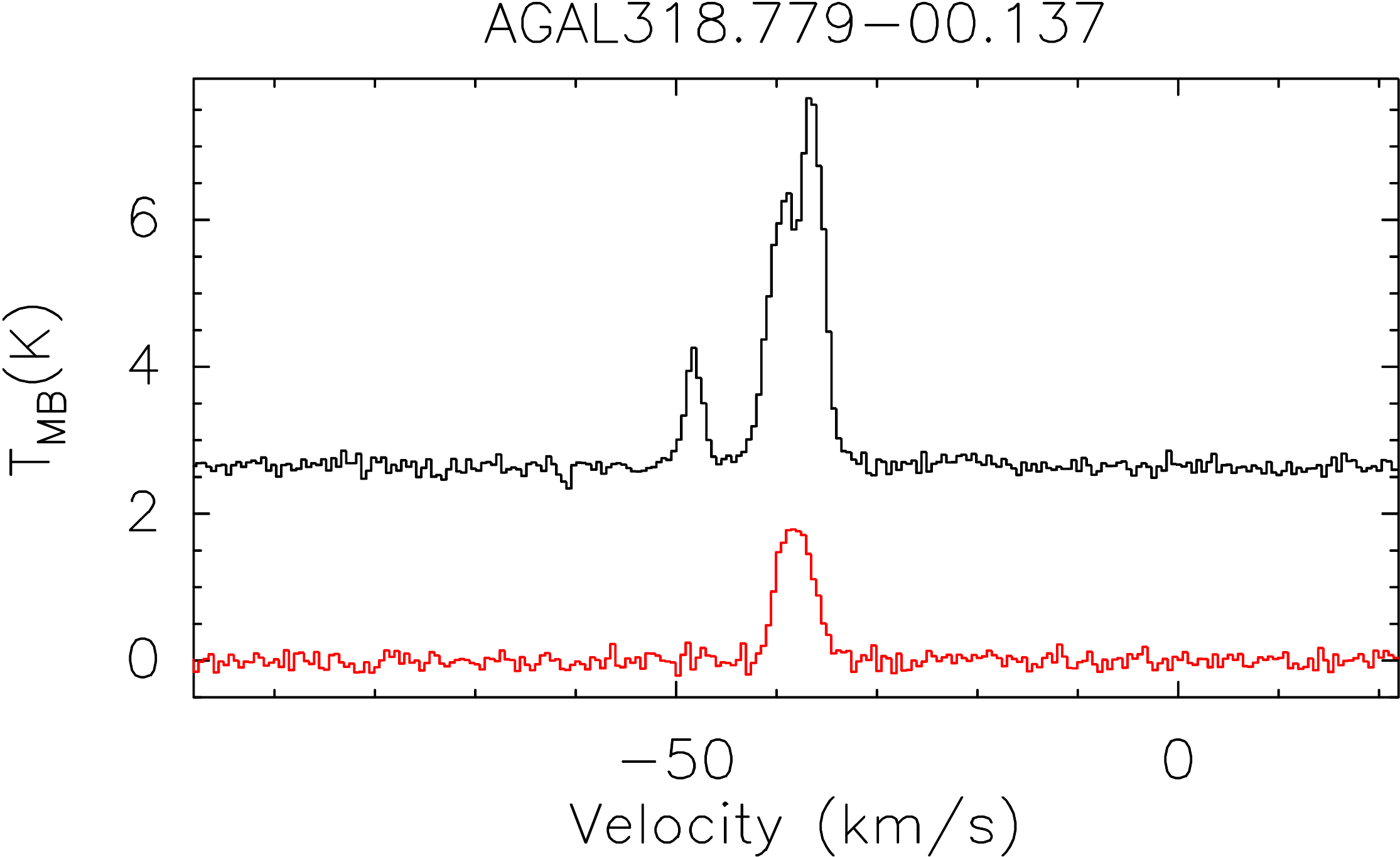} 
\includegraphics[angle=-90,width=0.3\textwidth]{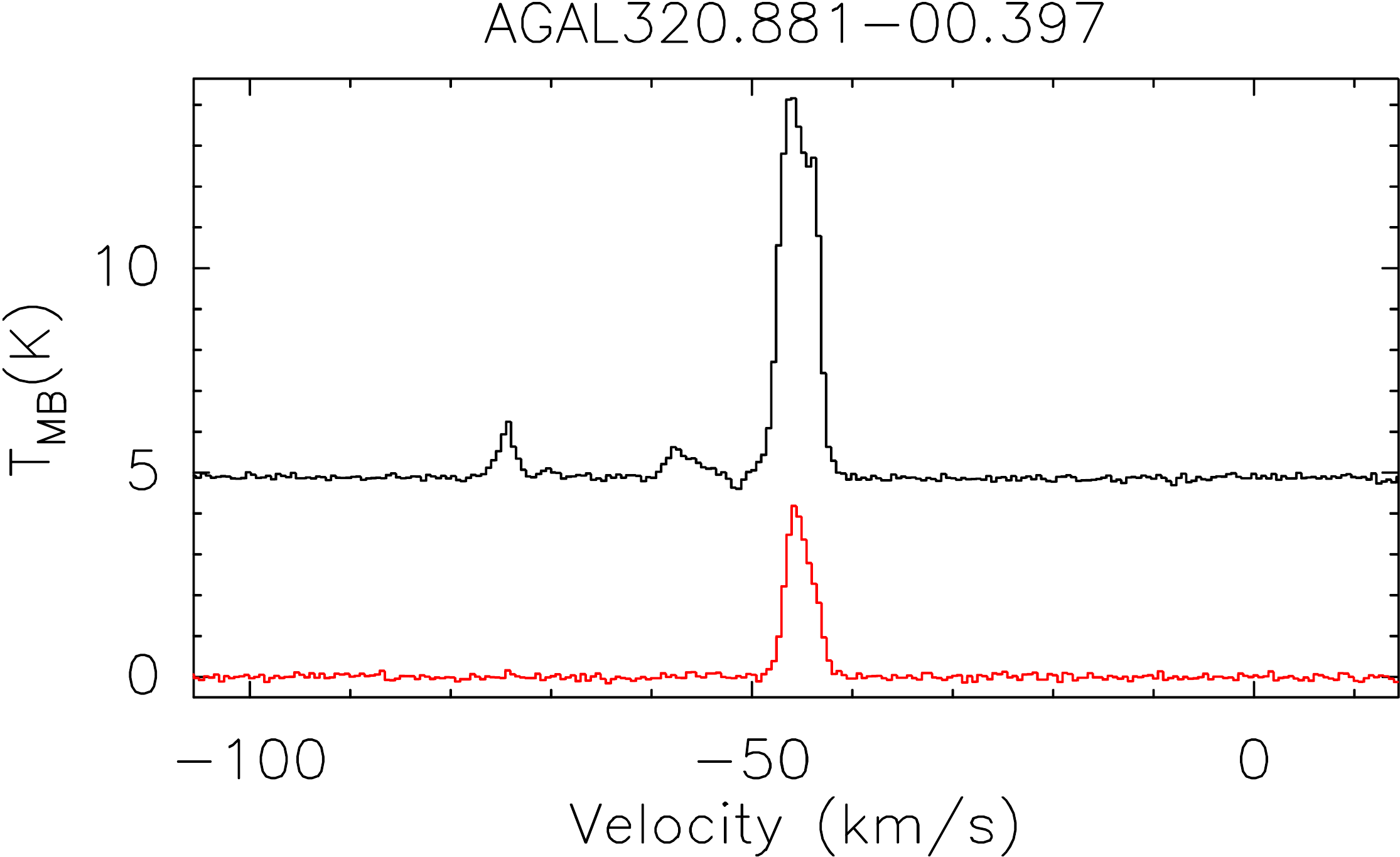} \\ 
\includegraphics[angle=-90,width=0.3\textwidth]{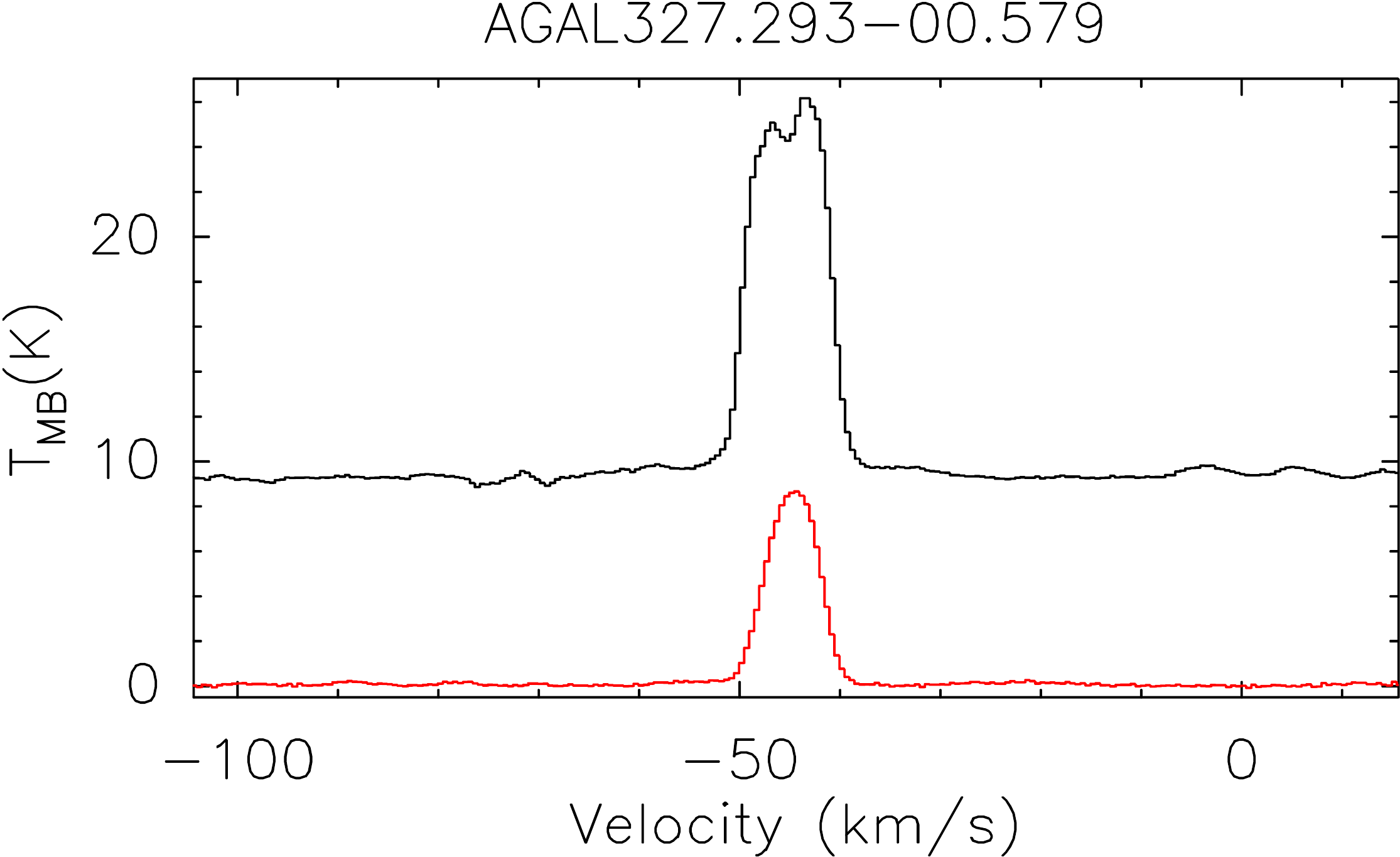} 
\includegraphics[angle=-90,width=0.3\textwidth]{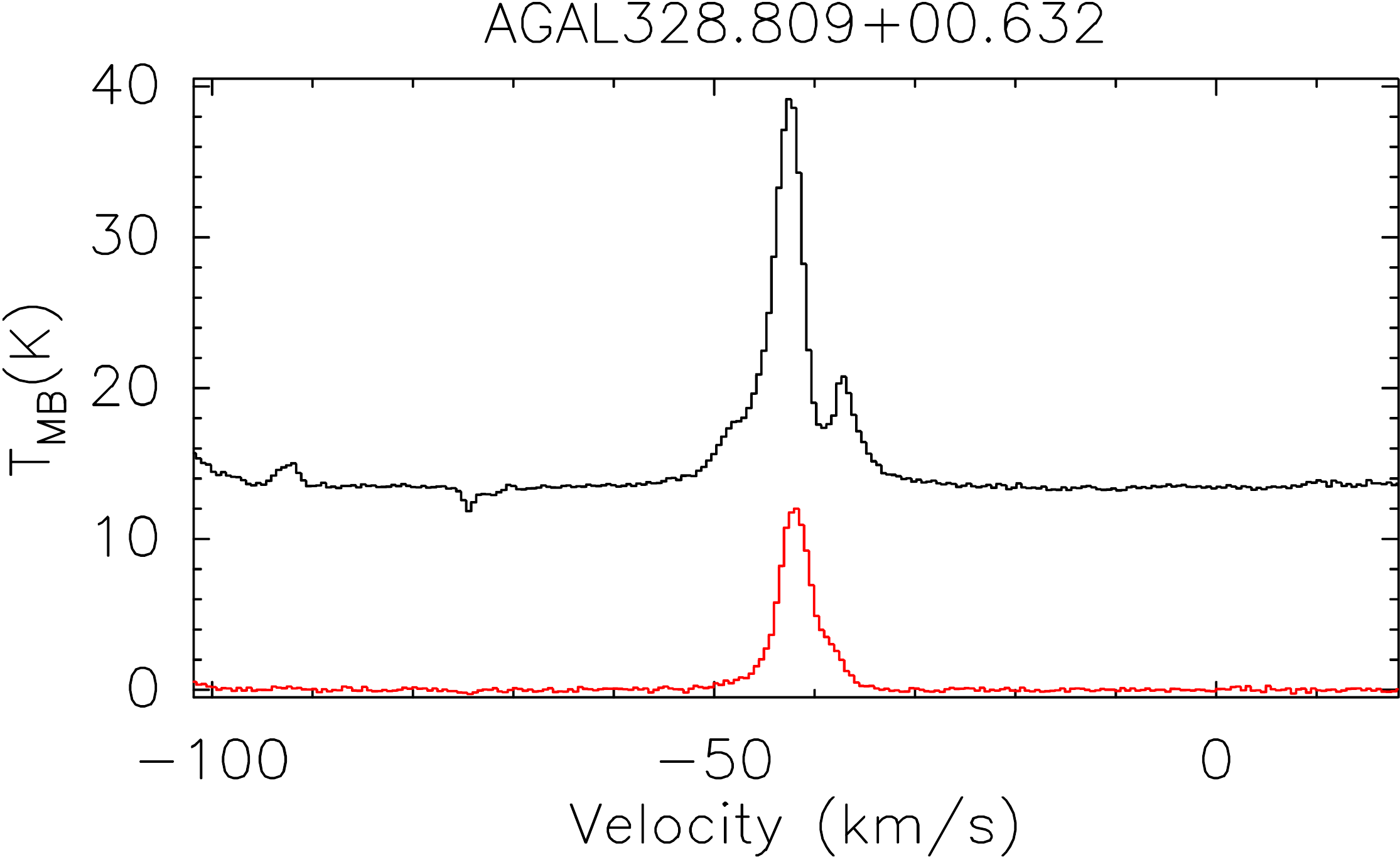} 
\includegraphics[angle=-90,width=0.3\textwidth]{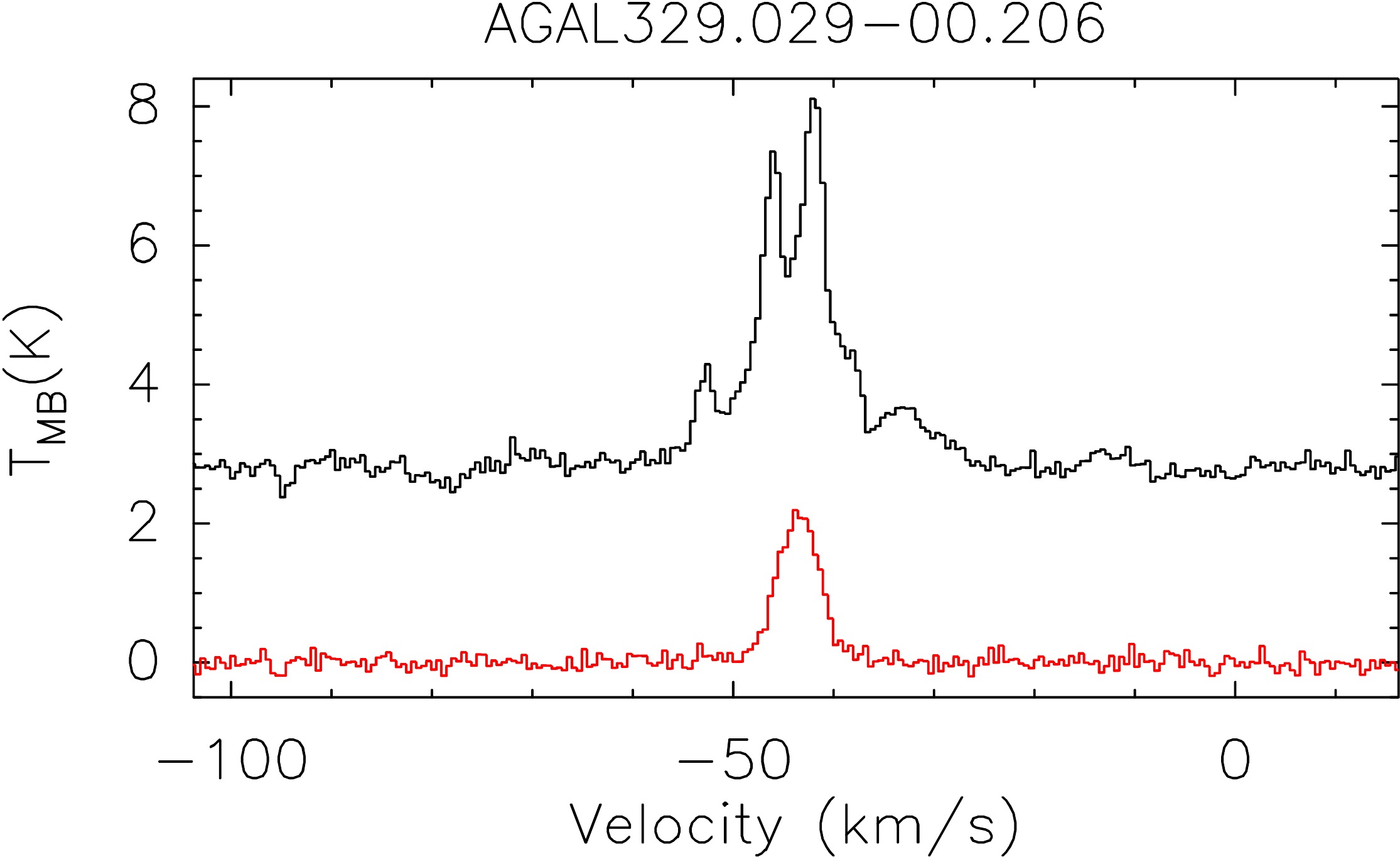} \\ 
\includegraphics[angle=-90,width=0.3\textwidth]{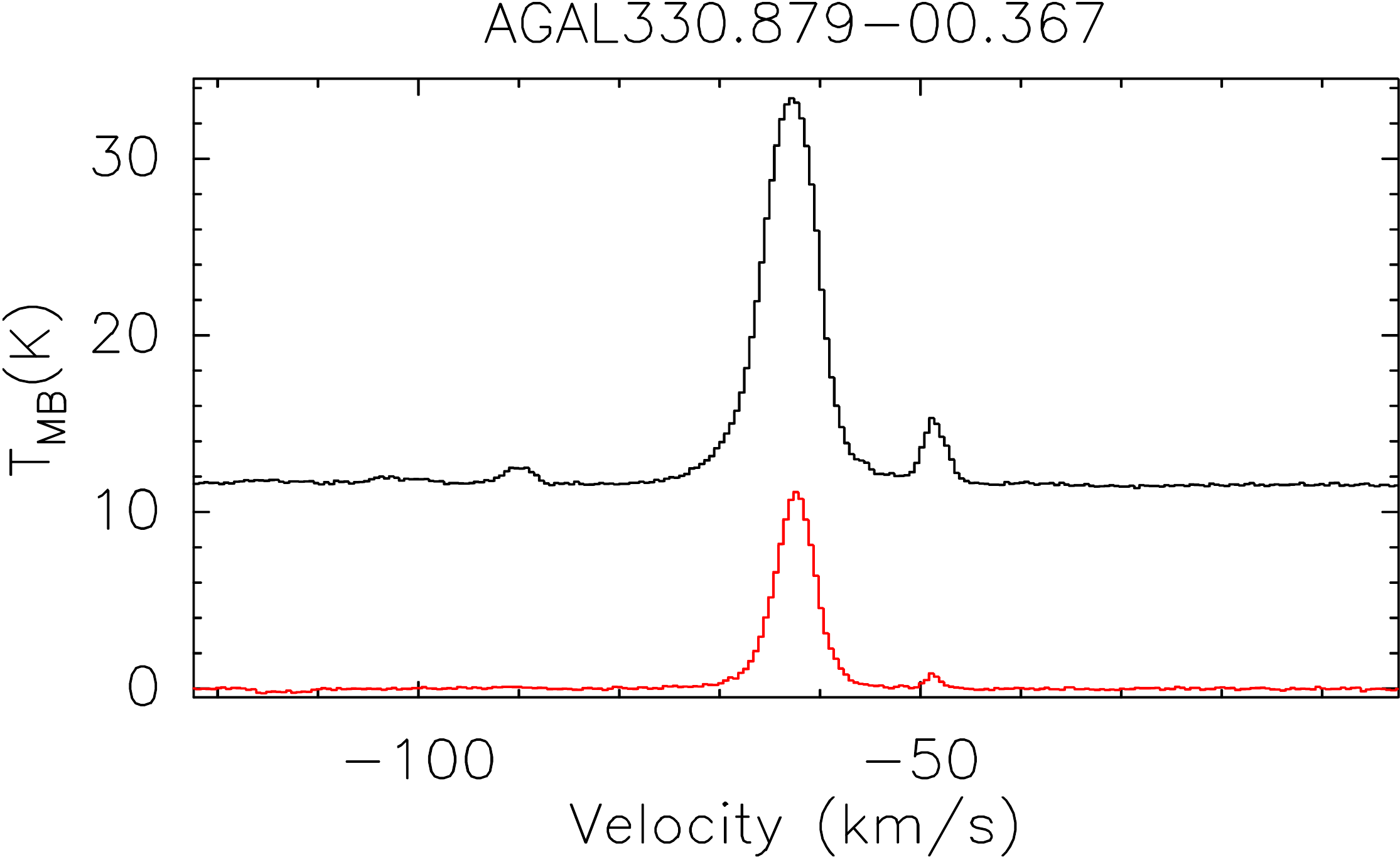} 
\includegraphics[angle=-90,width=0.3\textwidth]{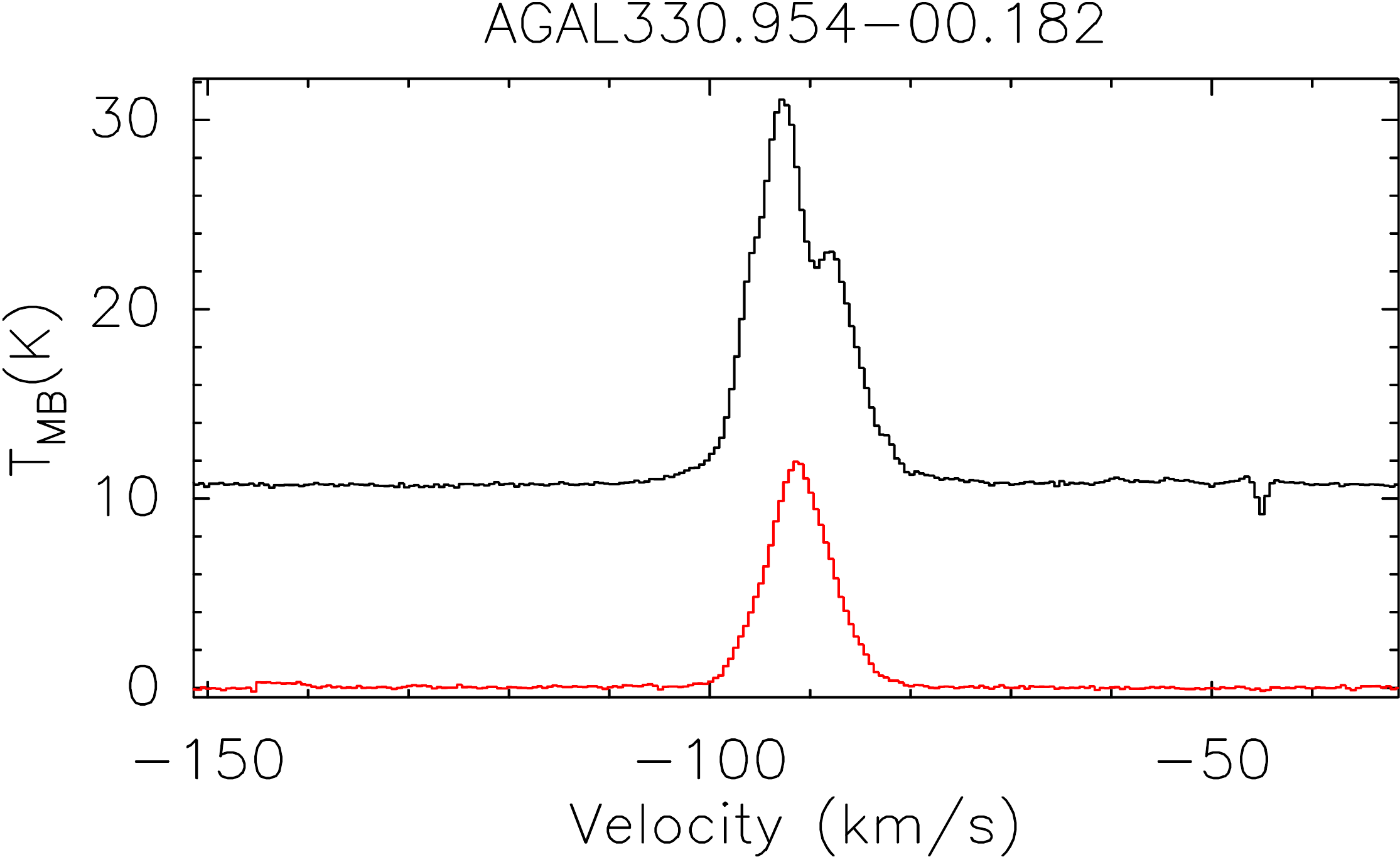} 
\includegraphics[angle=-90,width=0.3\textwidth]{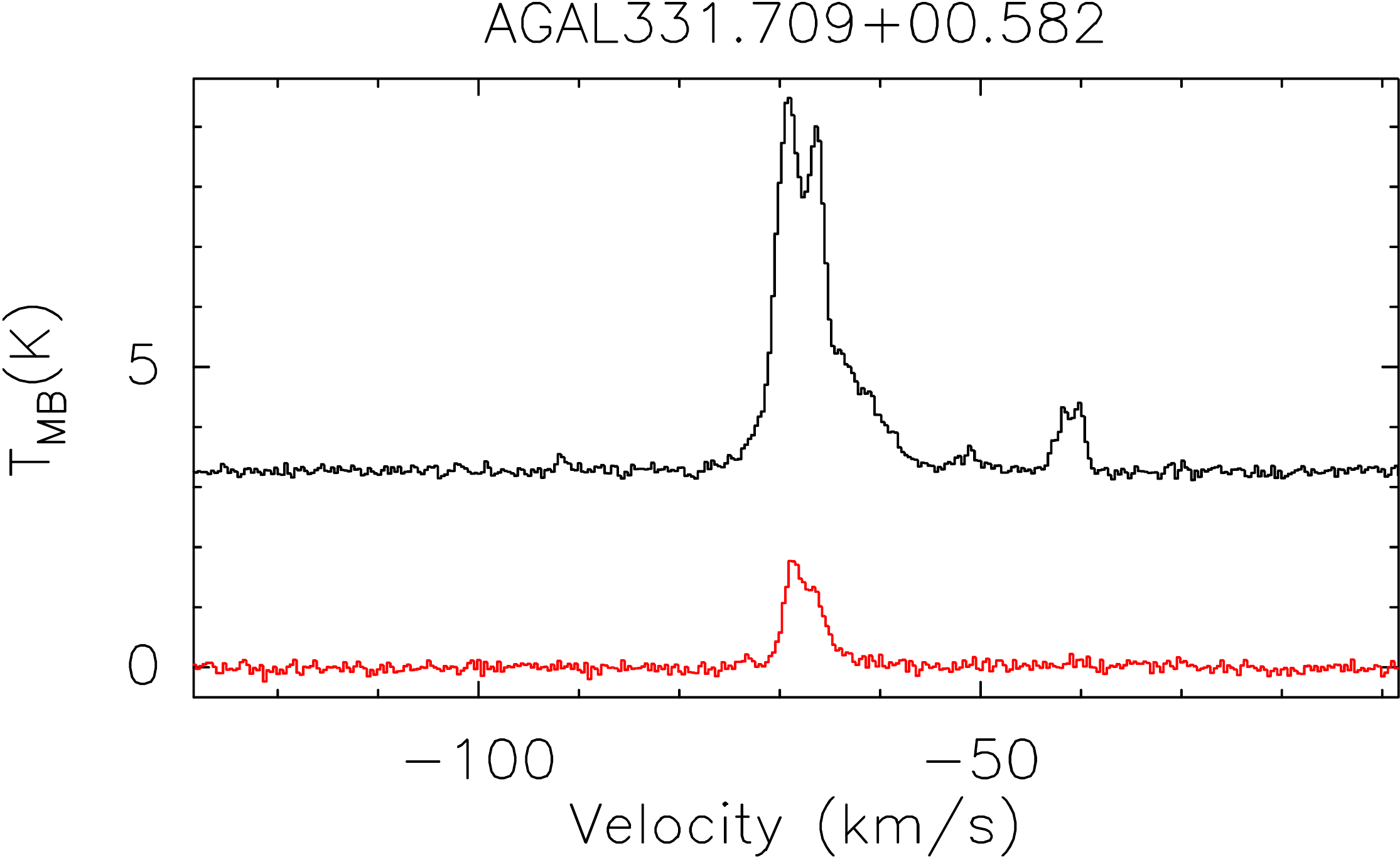} \\ 
\includegraphics[angle=-90,width=0.3\textwidth]{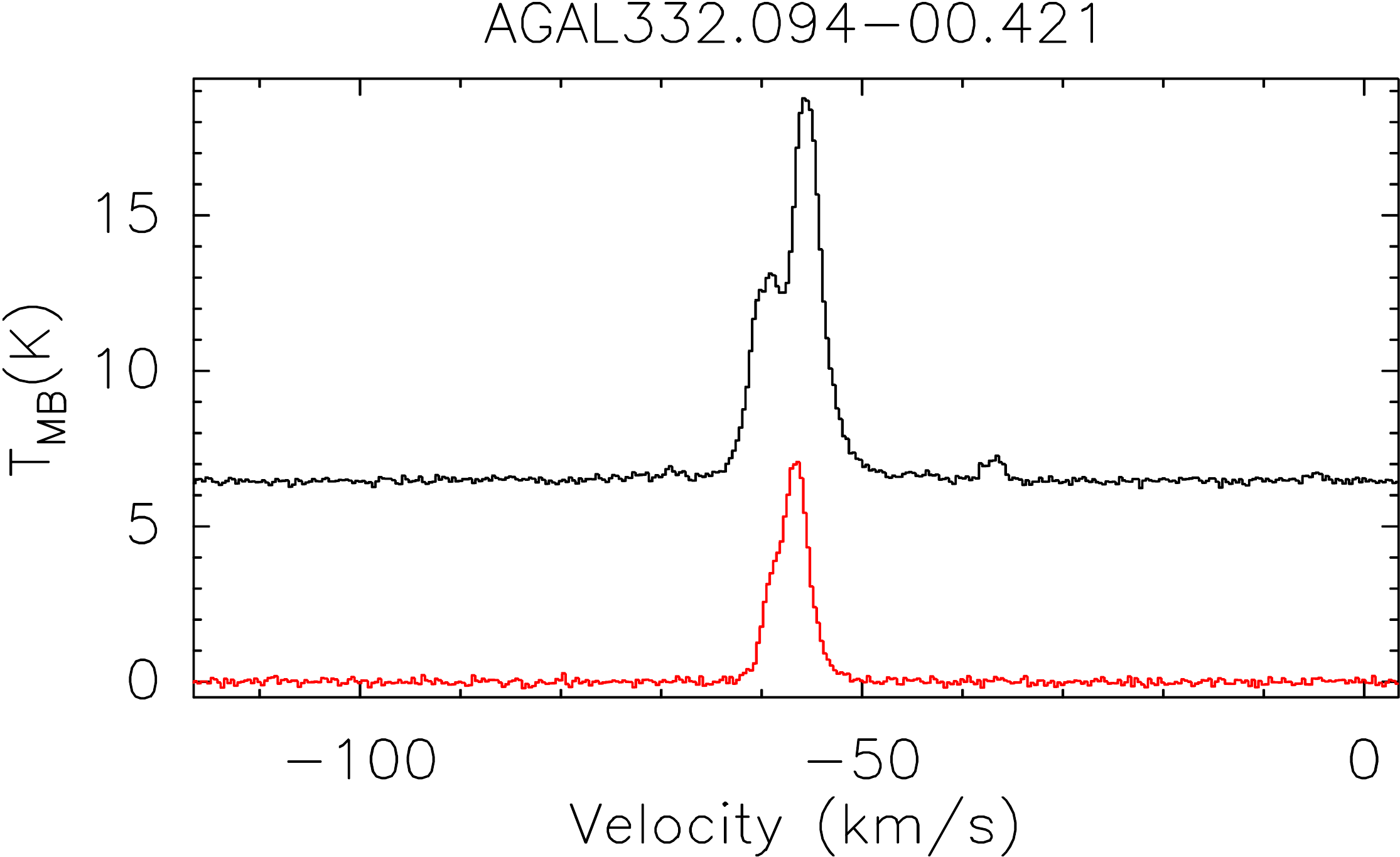} 
\includegraphics[angle=-90,width=0.3\textwidth]{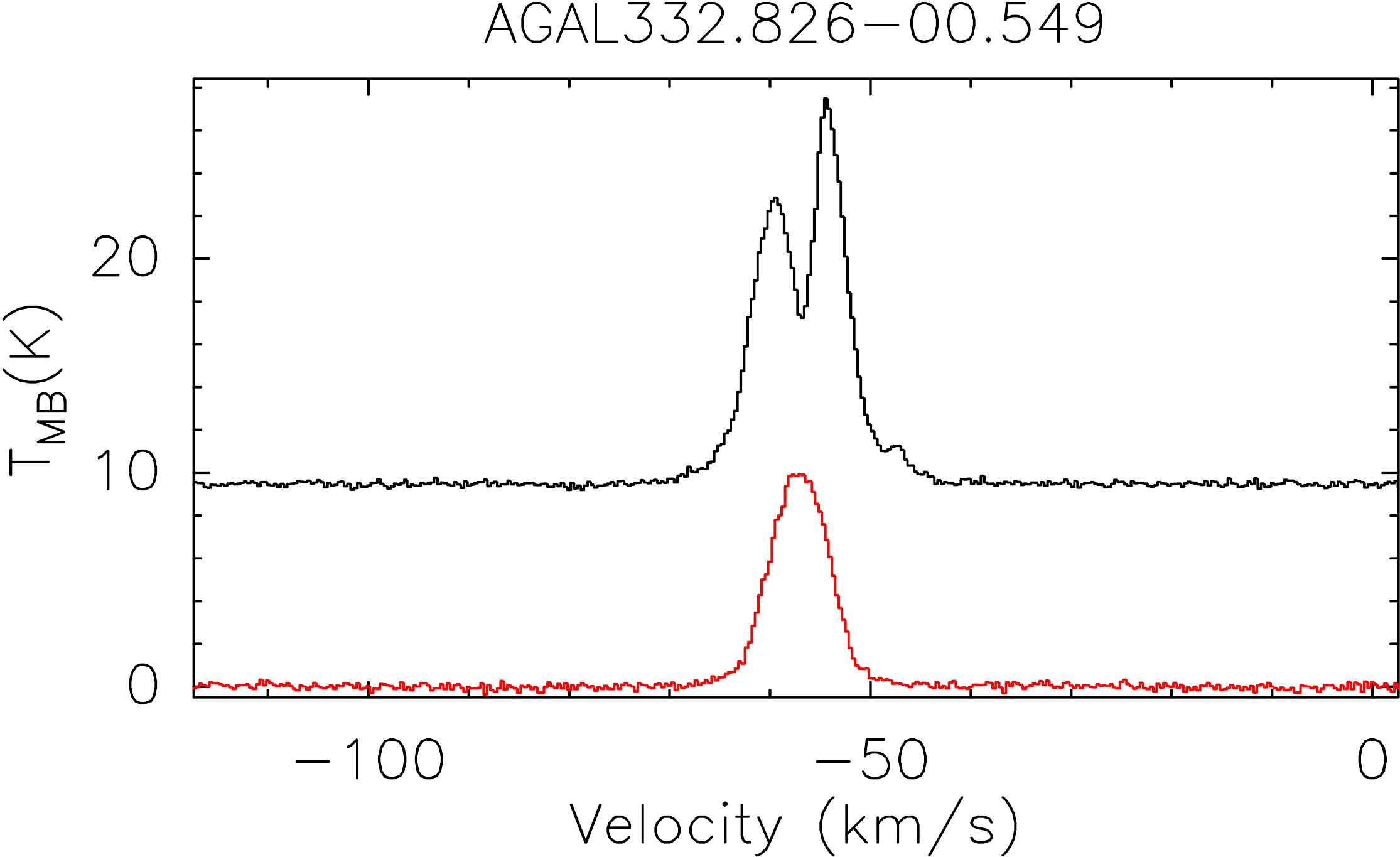} 
\includegraphics[angle=-90,width=0.3\textwidth]{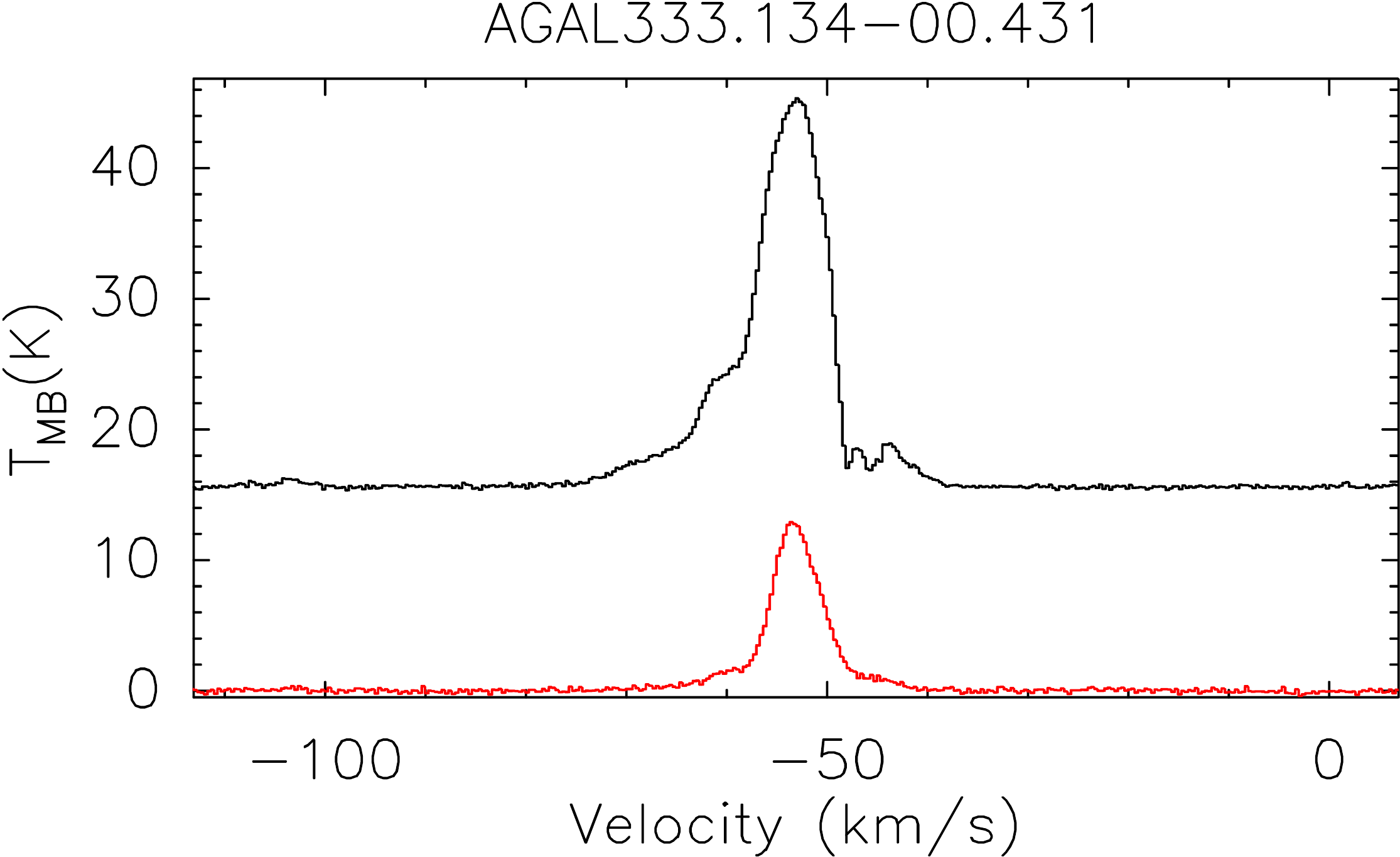} \\ 
\includegraphics[angle=-90,width=0.3\textwidth]{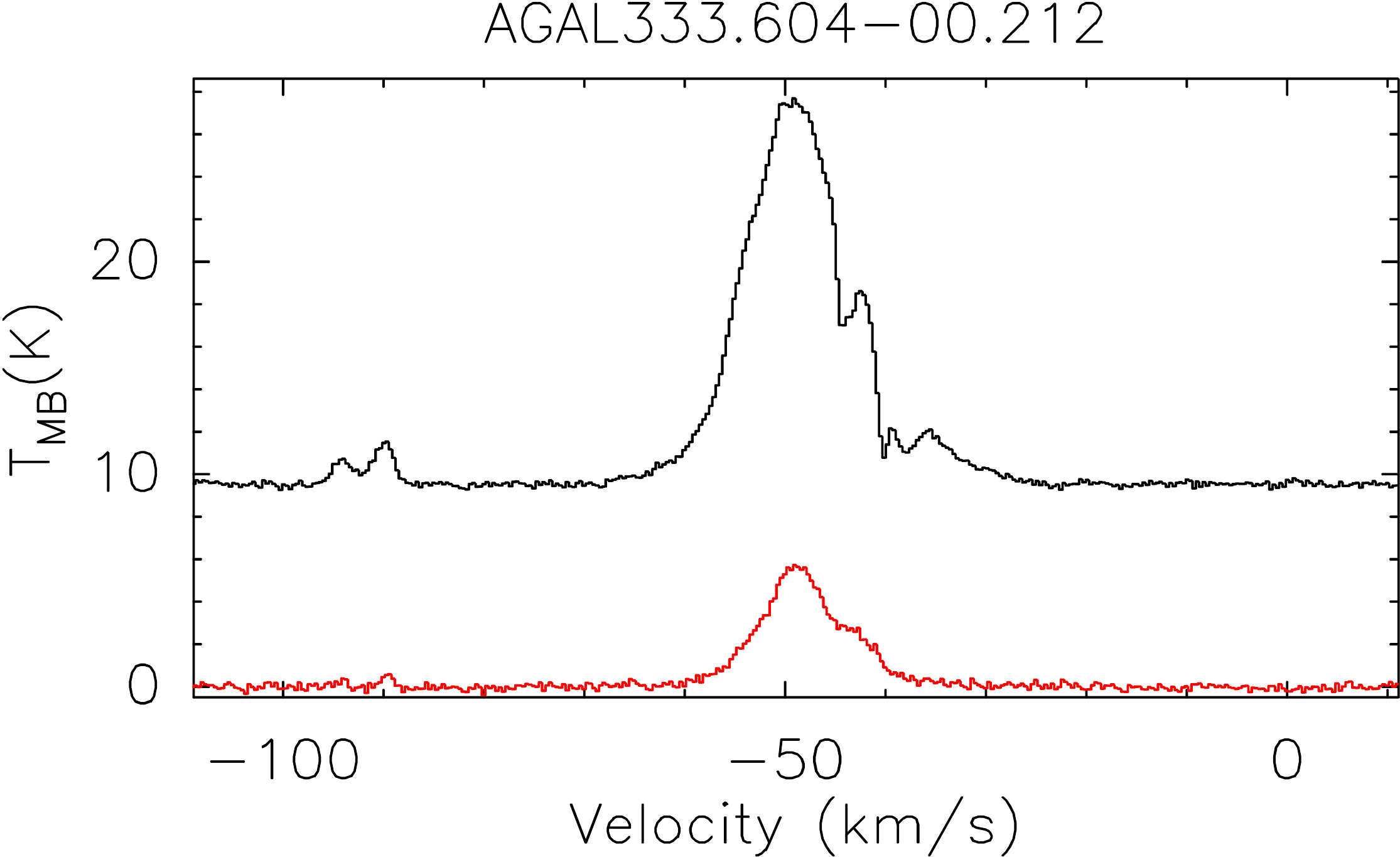} 
\includegraphics[angle=-90,width=0.3\textwidth]{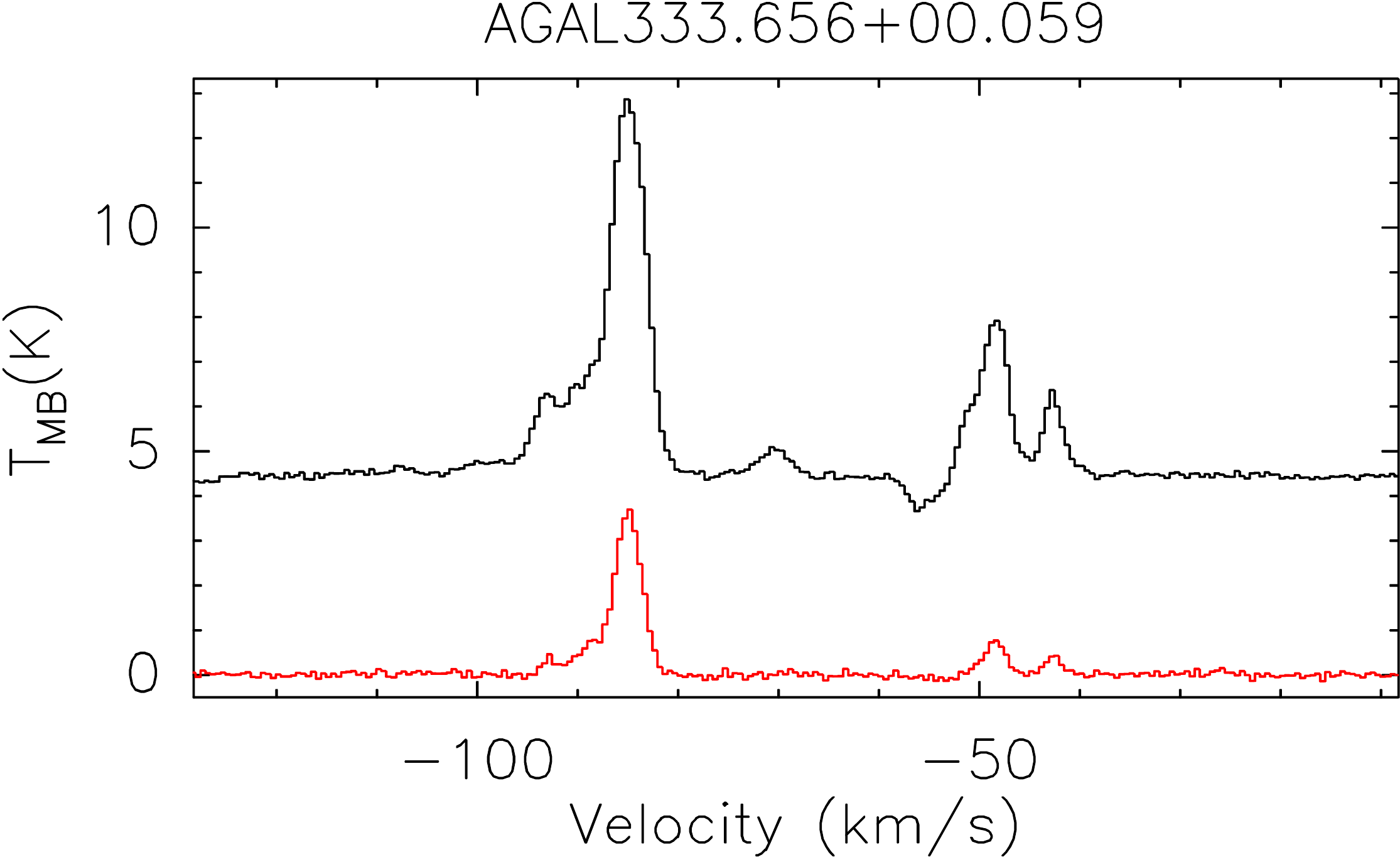} 
\includegraphics[angle=-90,width=0.3\textwidth]{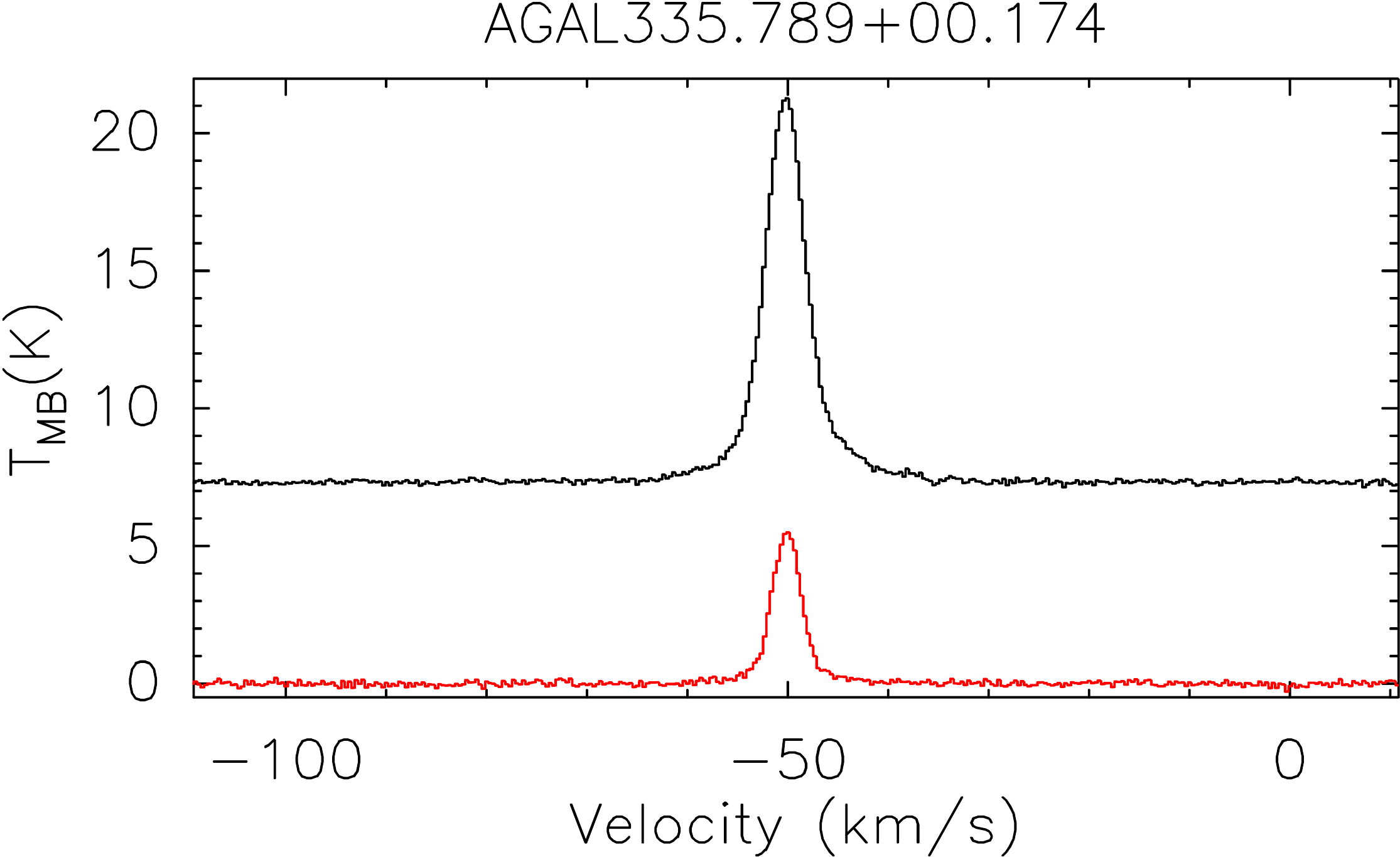} \hfill 
\caption{C$^{18}$O$(2-1)$ (red) and $^{13}$CO$(2-1)$ (black). The spectra are displaced for clarity.} \label{fig:spectra_21_B}
\end{figure*} 

\begin{figure*} 
\ContinuedFloat
\centering 
\includegraphics[angle=-90,width=0.3\textwidth]{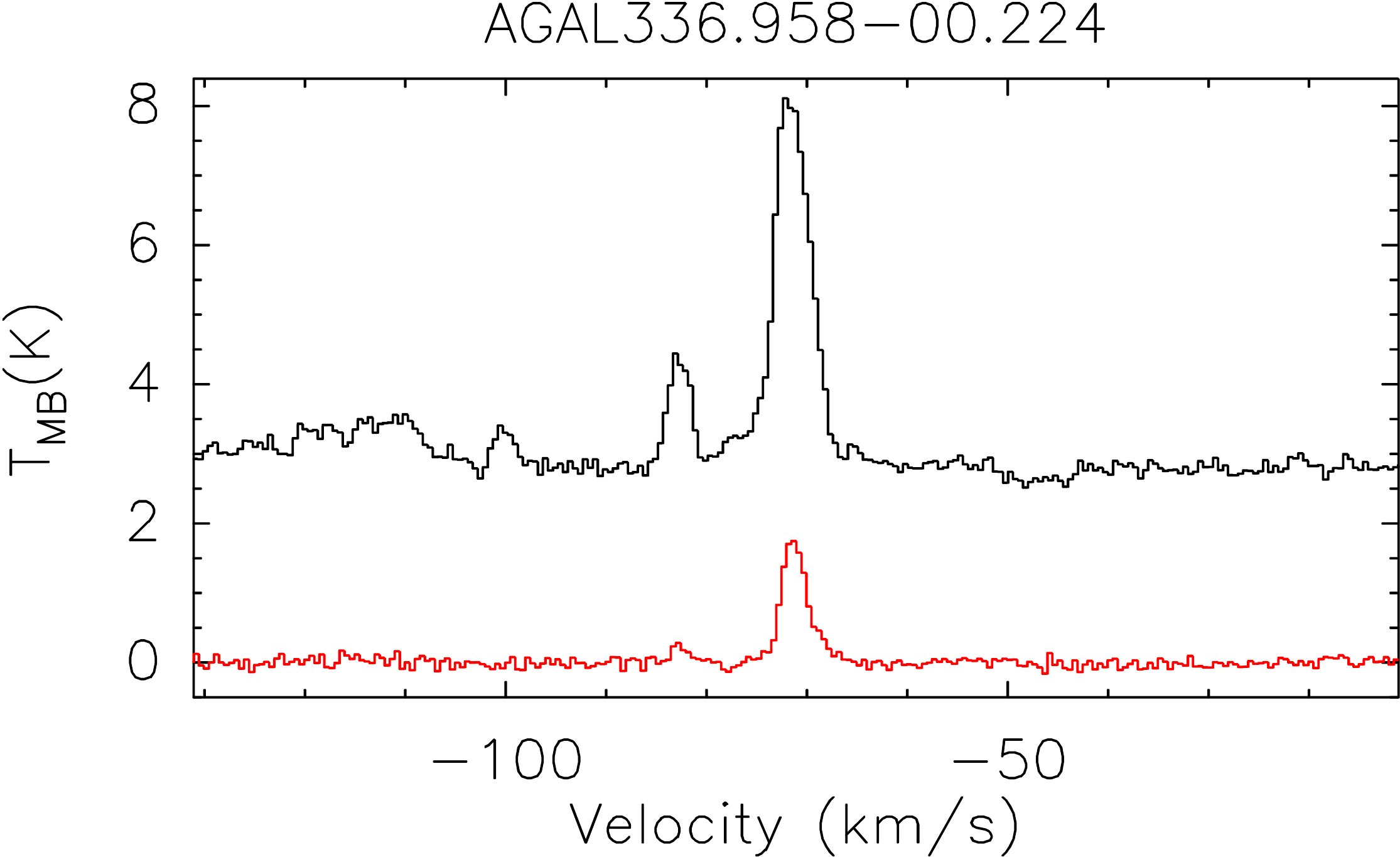} 
\includegraphics[angle=-90,width=0.3\textwidth]{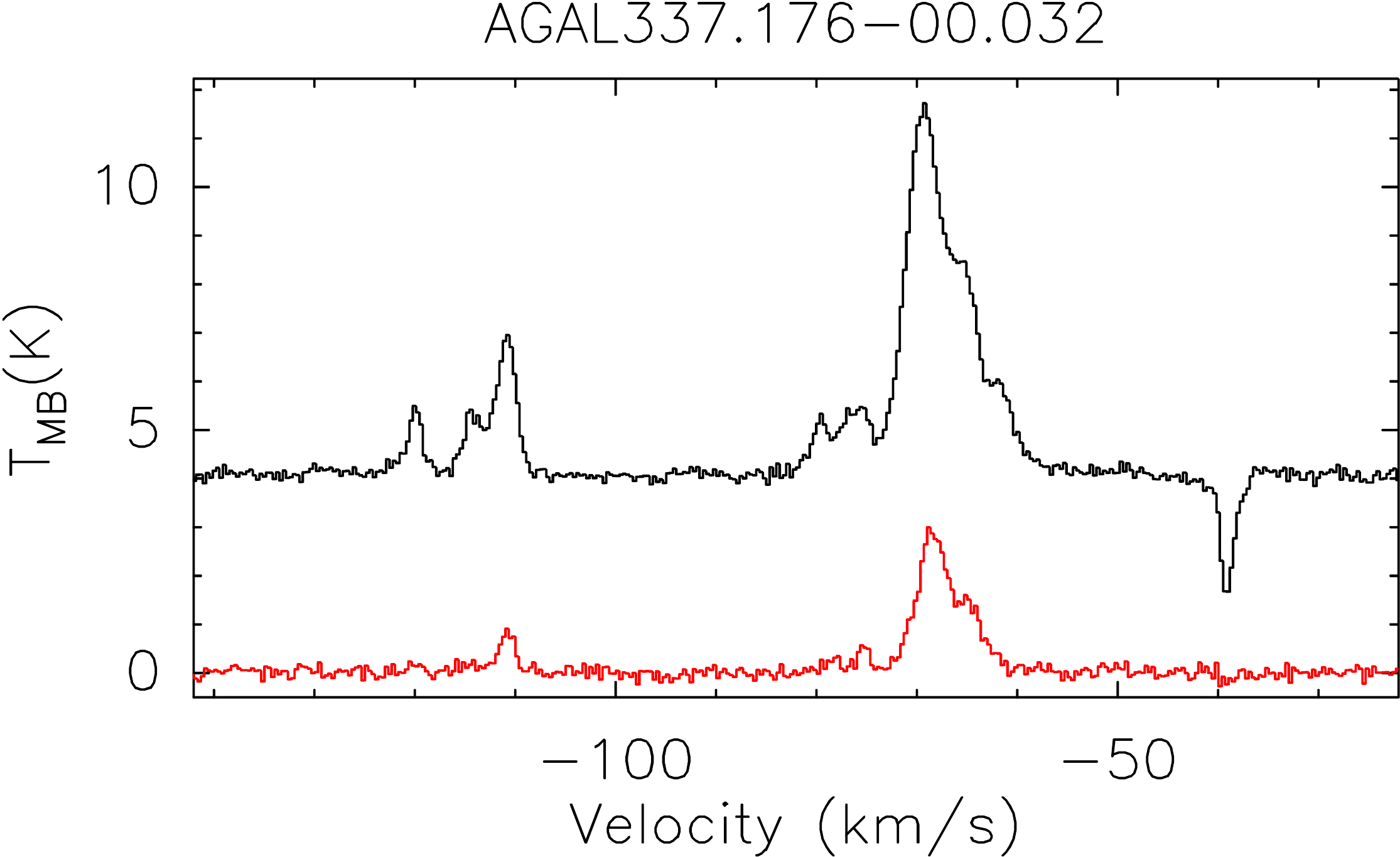} 
\includegraphics[angle=-90,width=0.3\textwidth]{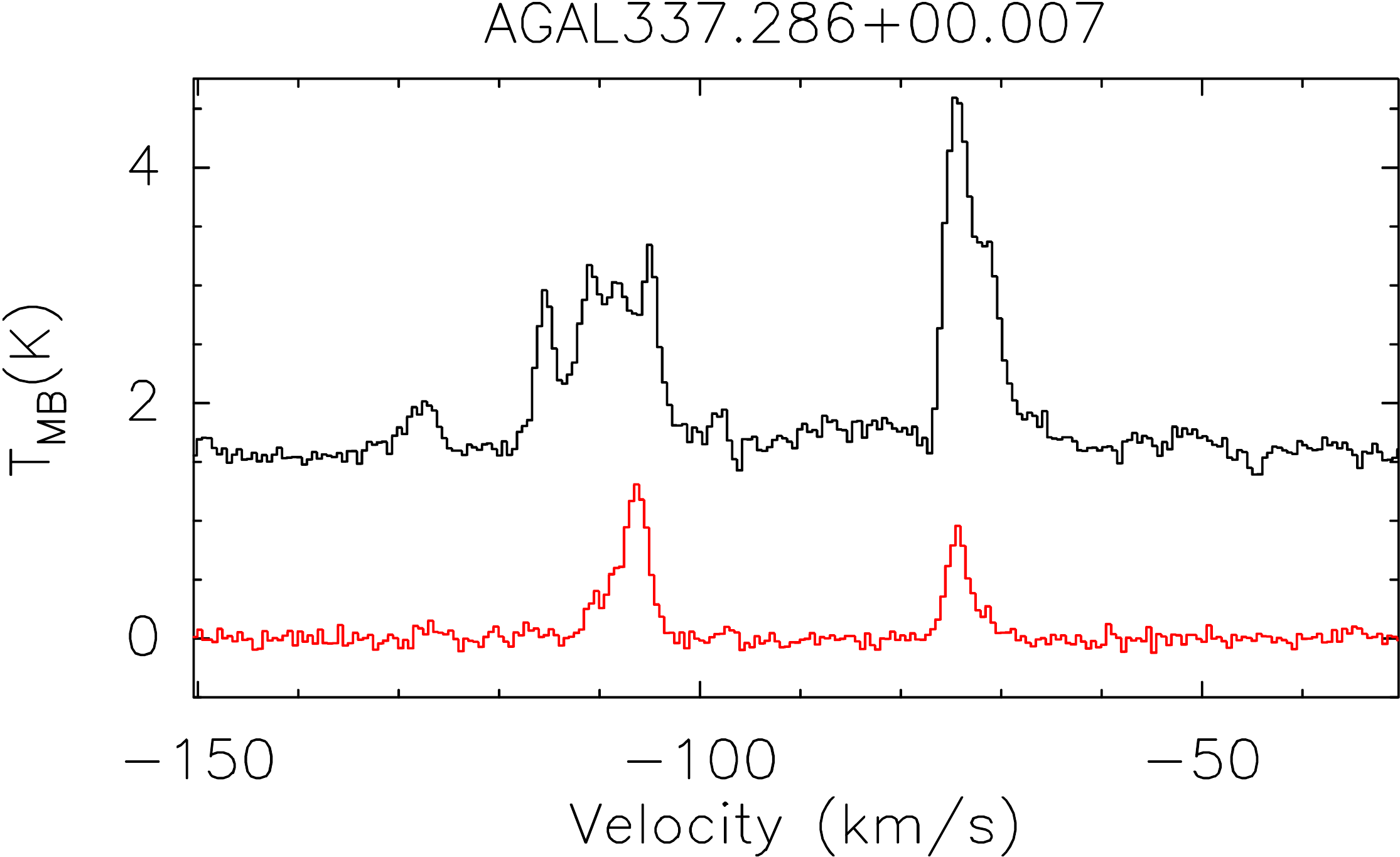} \\ 
\includegraphics[angle=-90,width=0.3\textwidth]{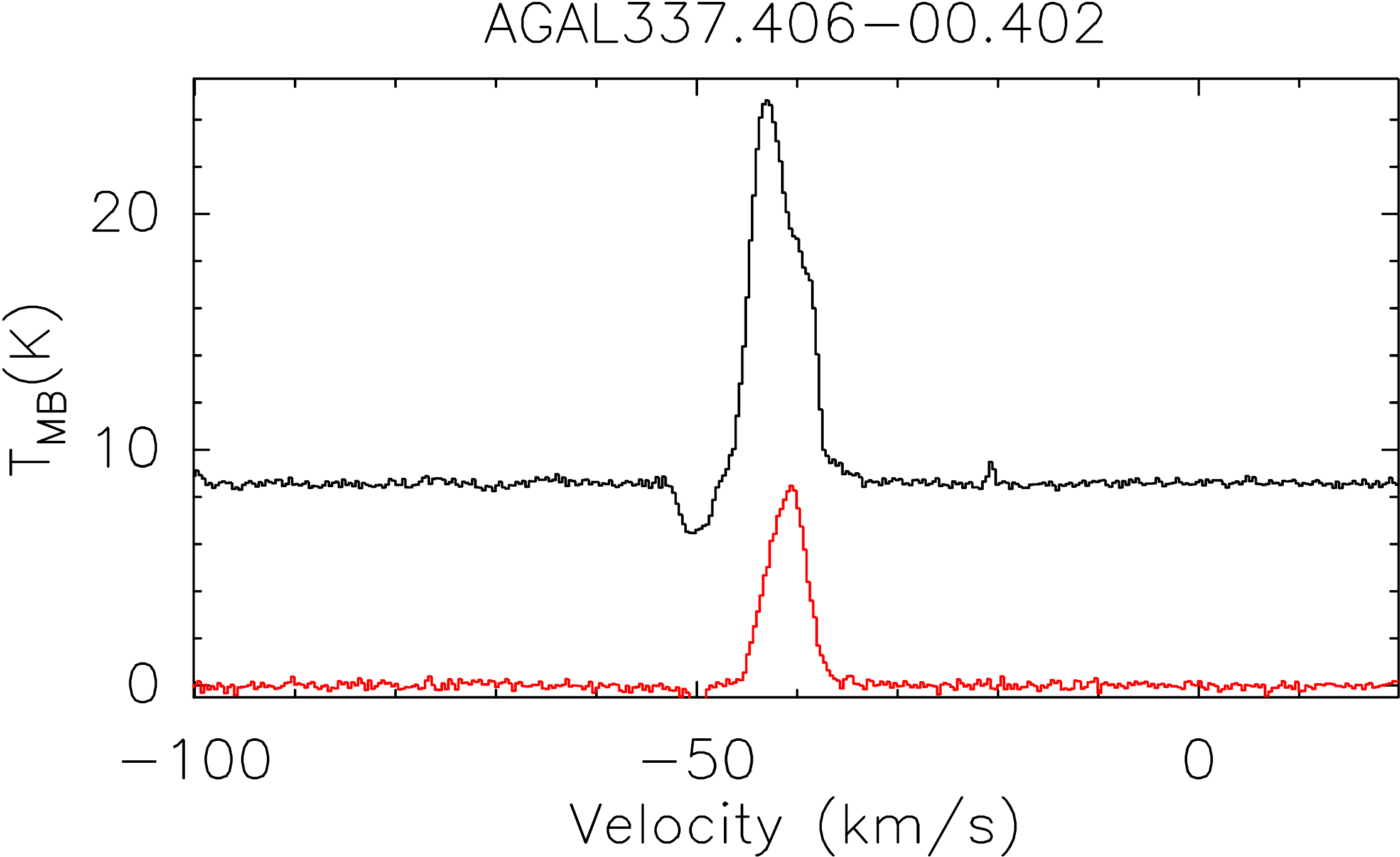} 
\includegraphics[angle=-90,width=0.3\textwidth]{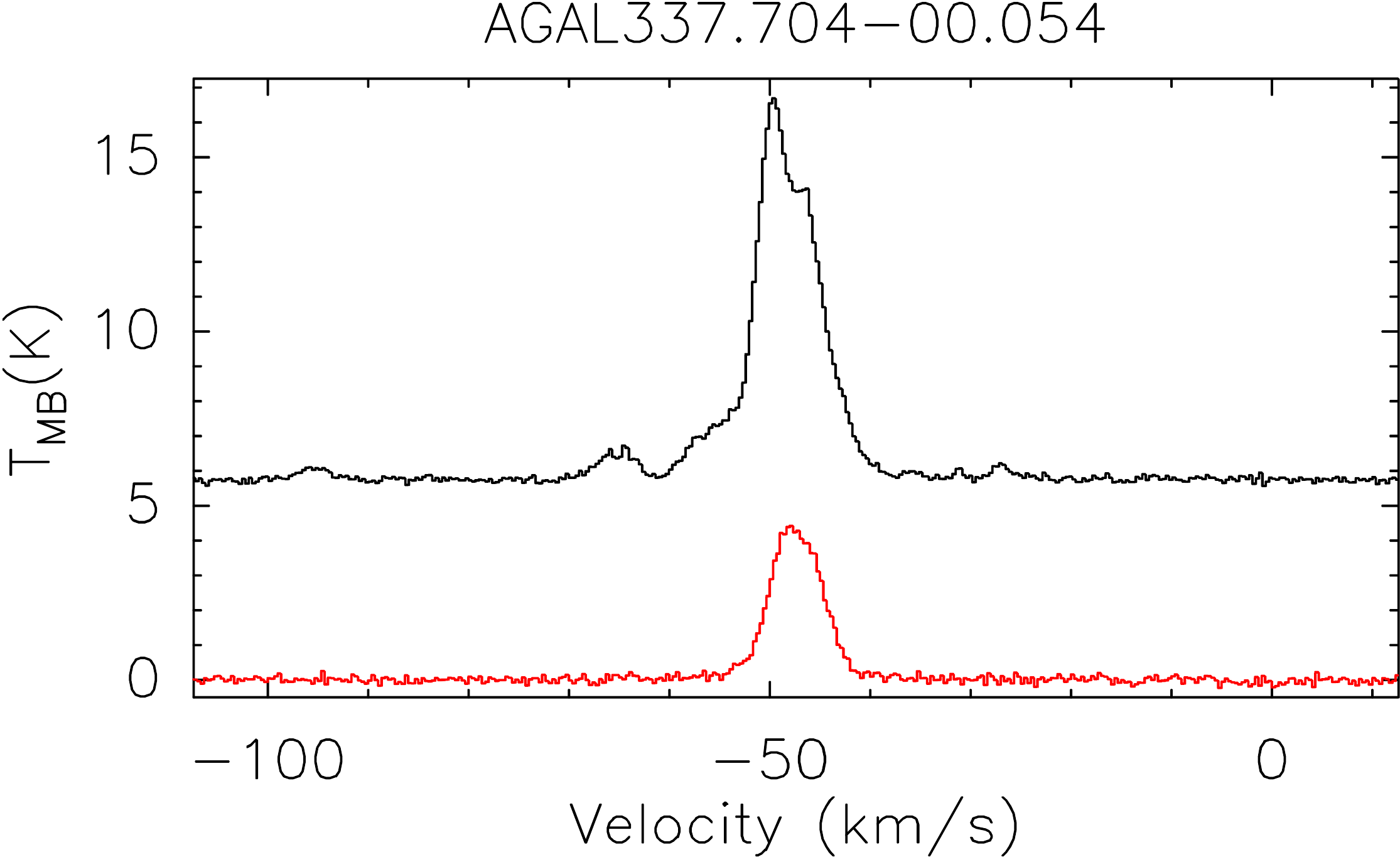} 
\includegraphics[angle=-90,width=0.3\textwidth]{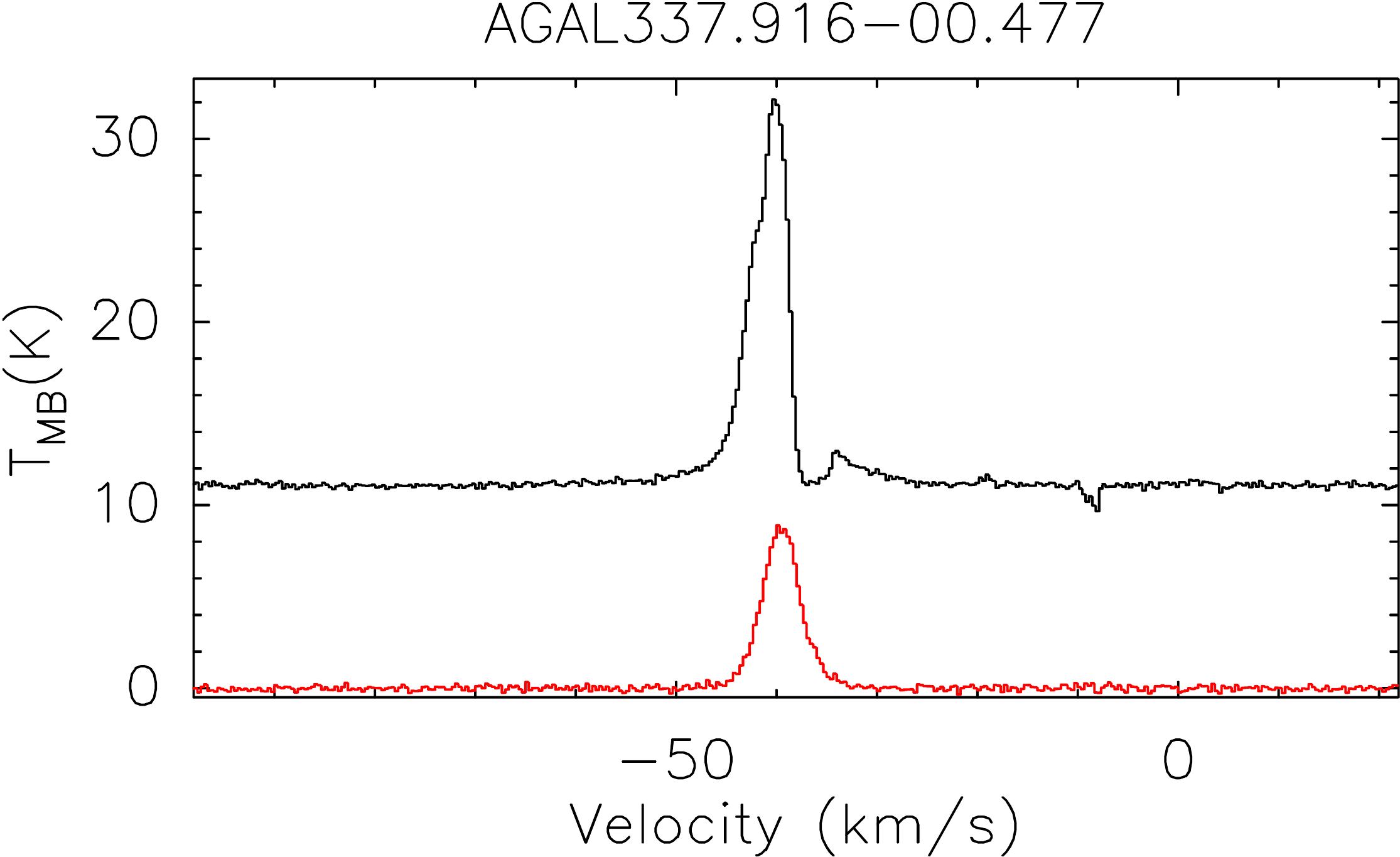} \\ 
\includegraphics[angle=-90,width=0.3\textwidth]{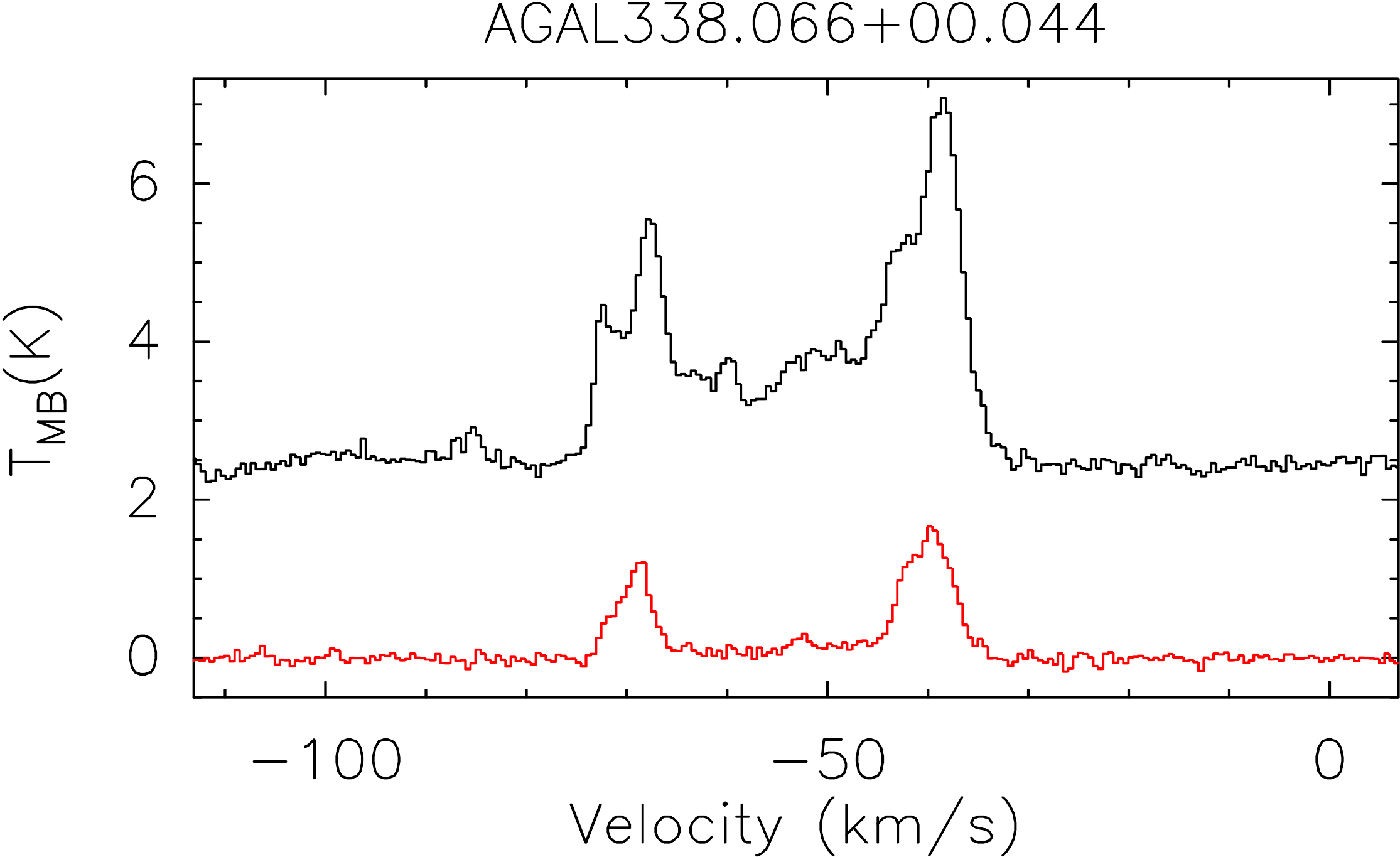} 
\includegraphics[angle=-90,width=0.3\textwidth]{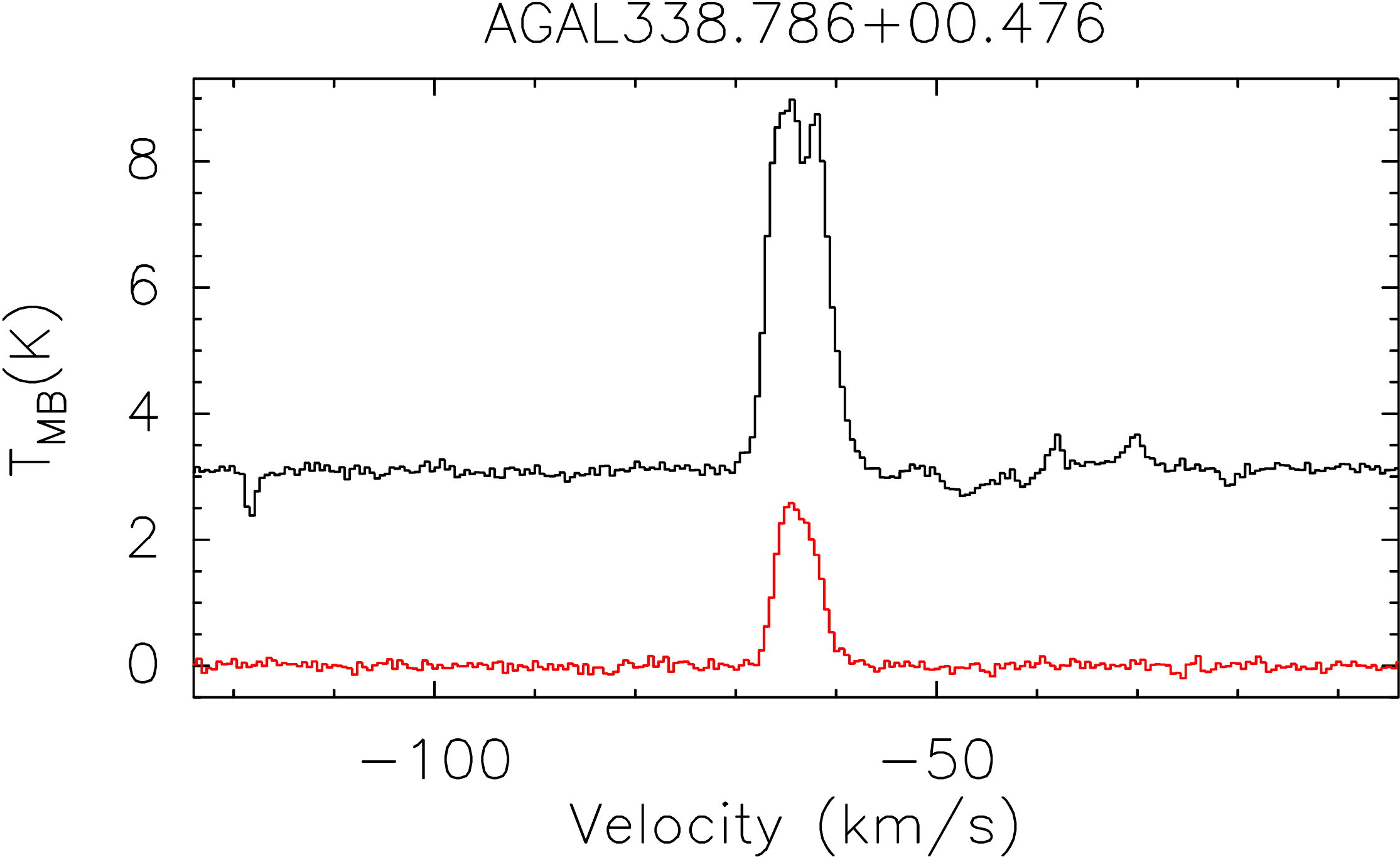} 
\includegraphics[angle=-90,width=0.3\textwidth]{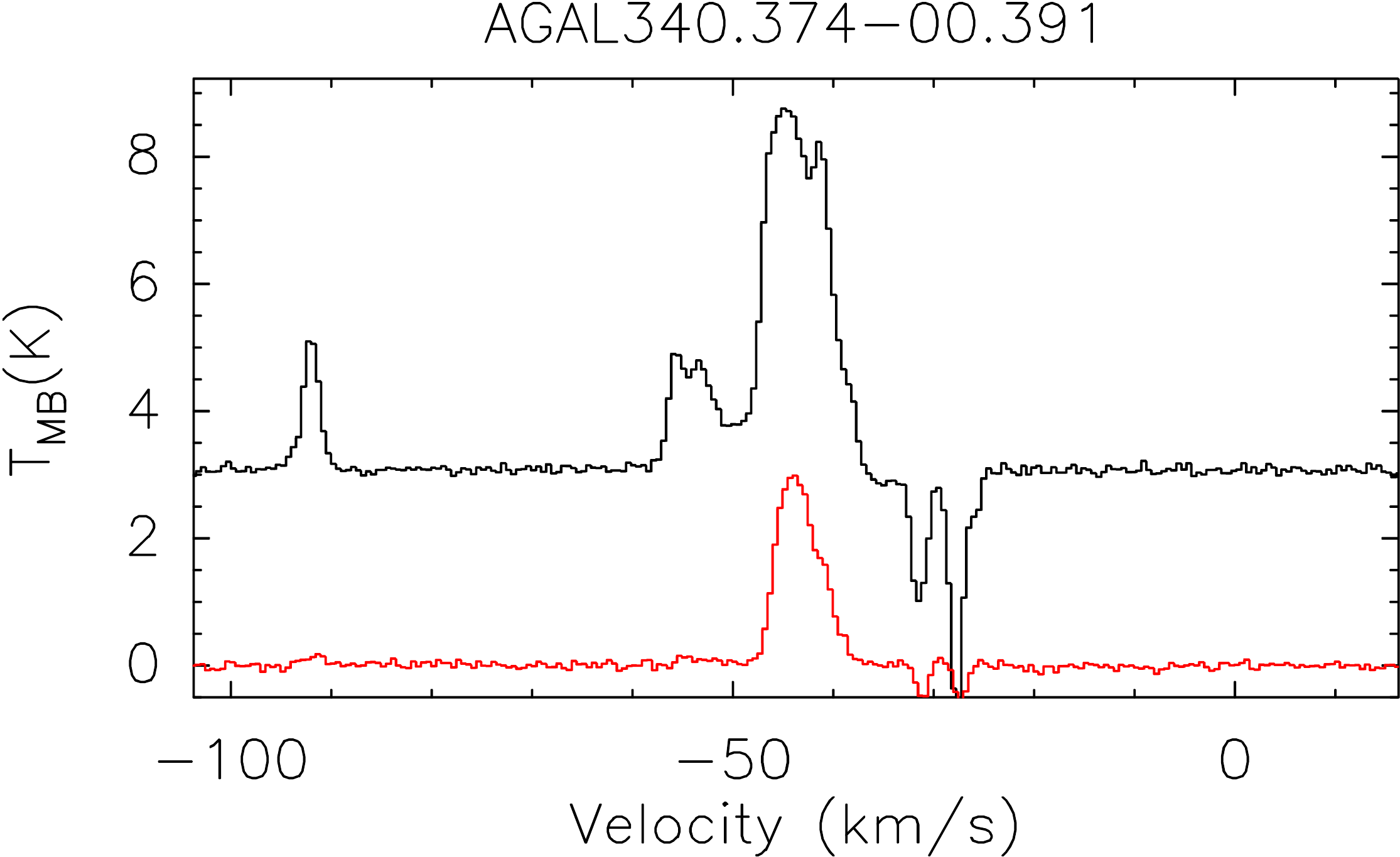} \\ 
\includegraphics[angle=-90,width=0.3\textwidth]{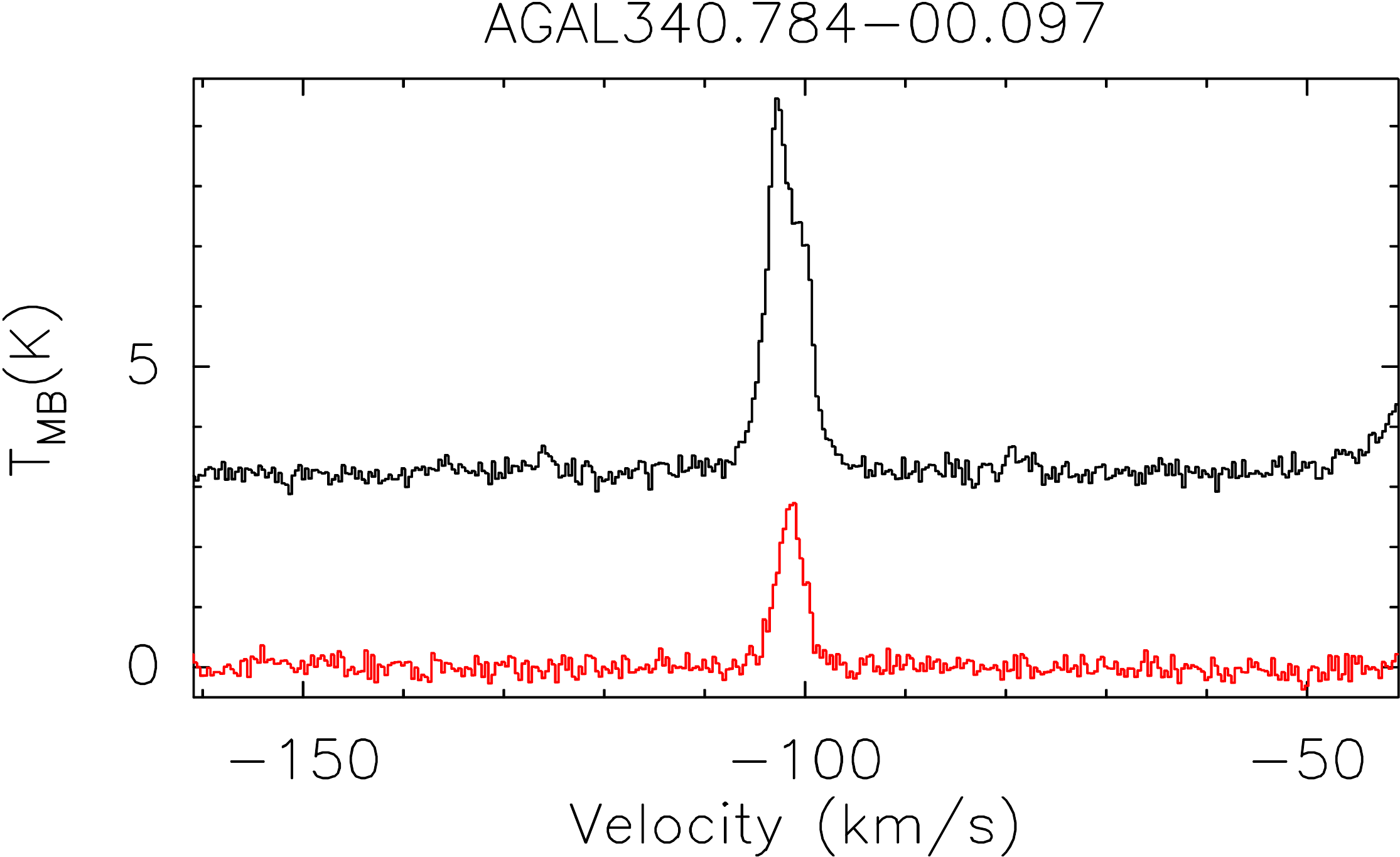} 
\includegraphics[angle=-90,width=0.3\textwidth]{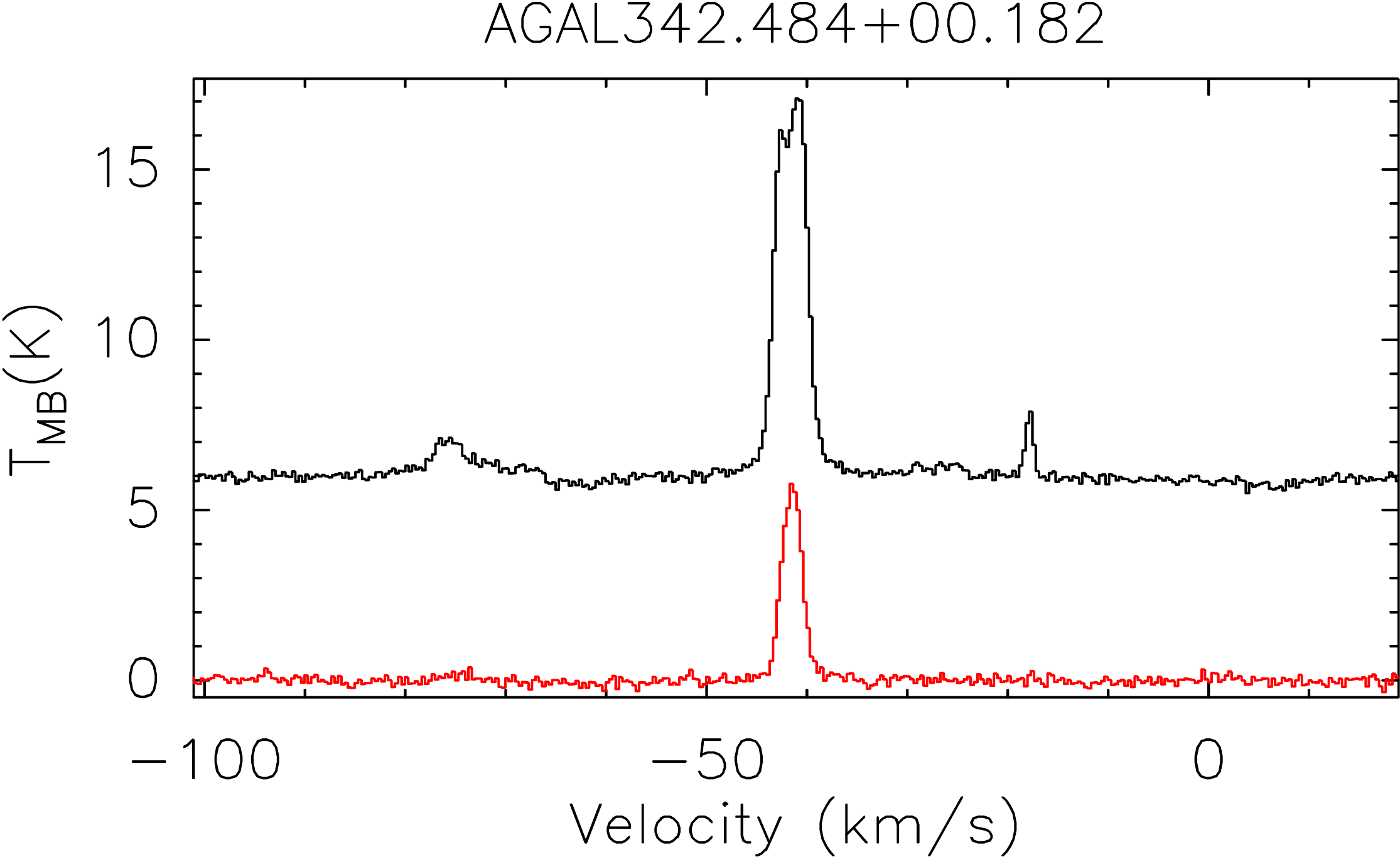} 
\includegraphics[angle=-90,width=0.3\textwidth]{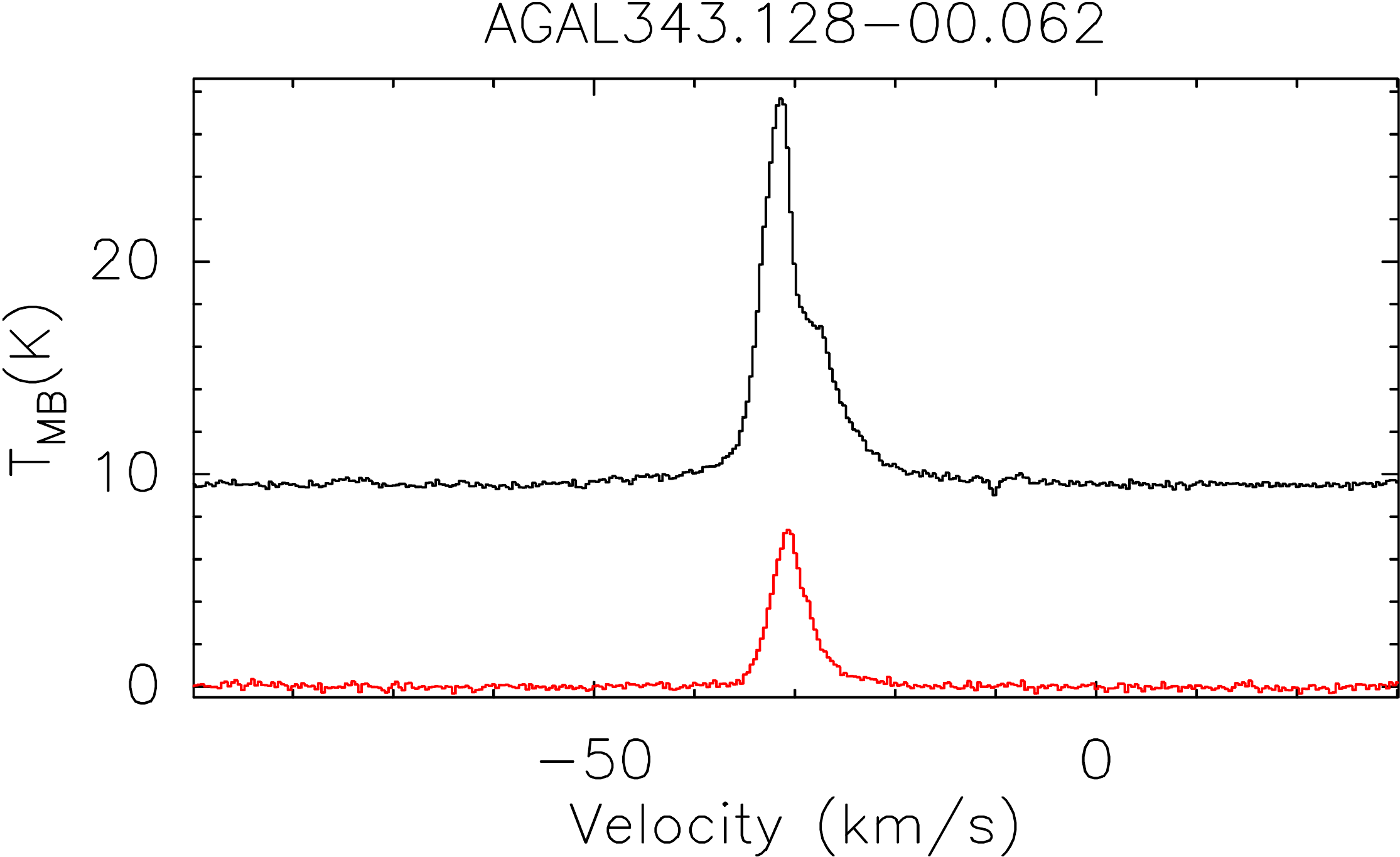} \\ 
\includegraphics[angle=-90,width=0.3\textwidth]{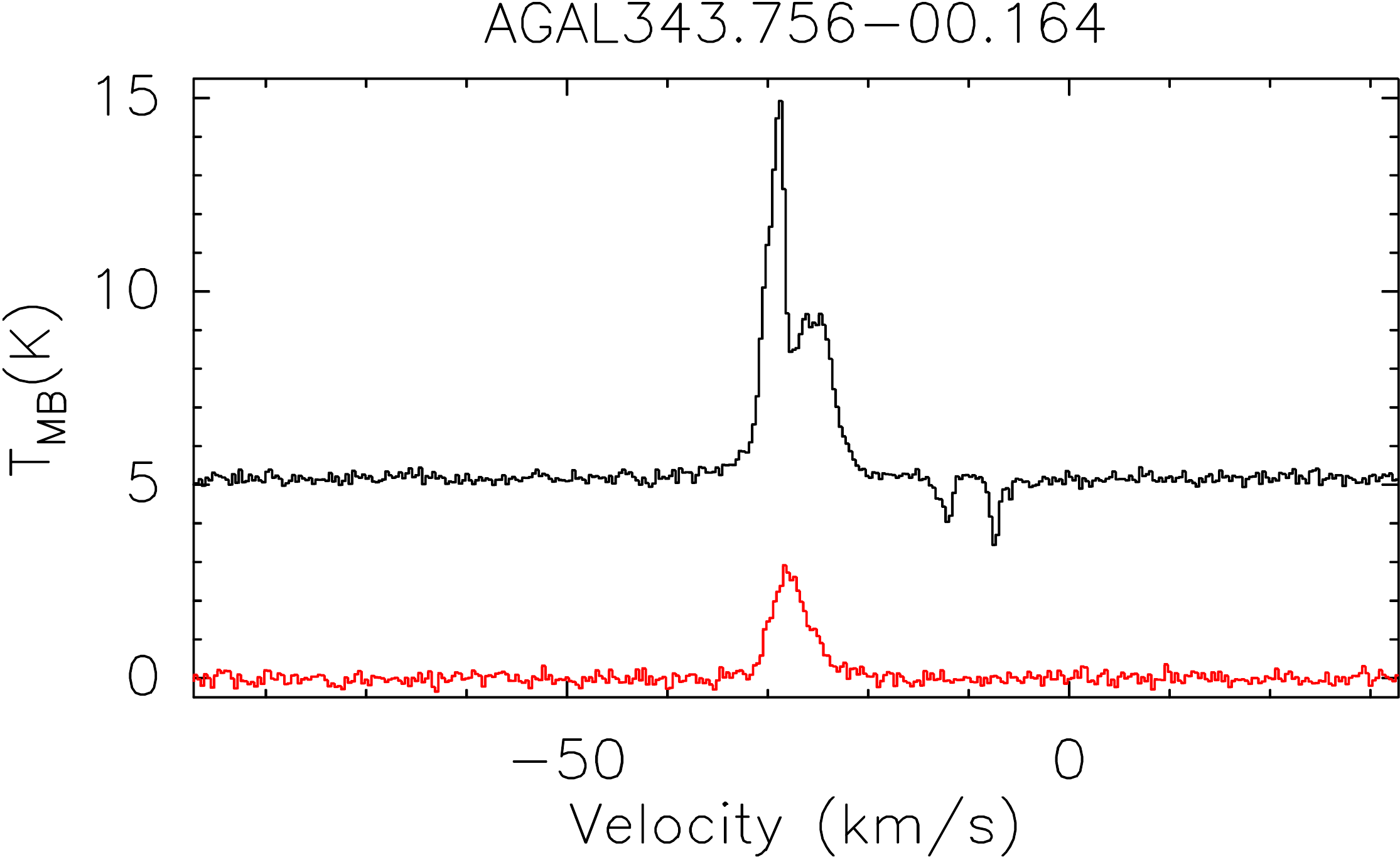} 
\includegraphics[angle=-90,width=0.3\textwidth]{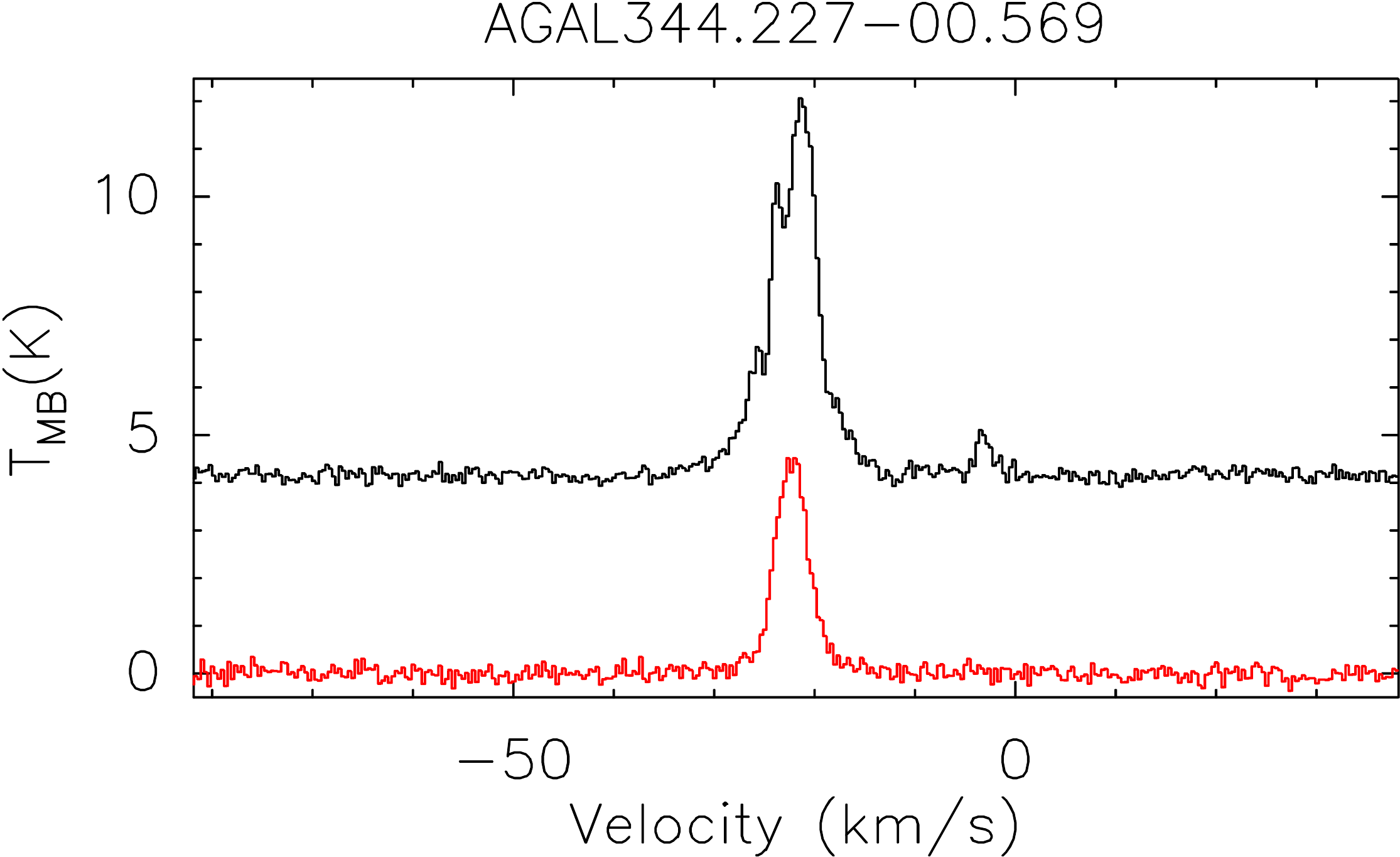} 
\includegraphics[angle=-90,width=0.3\textwidth]{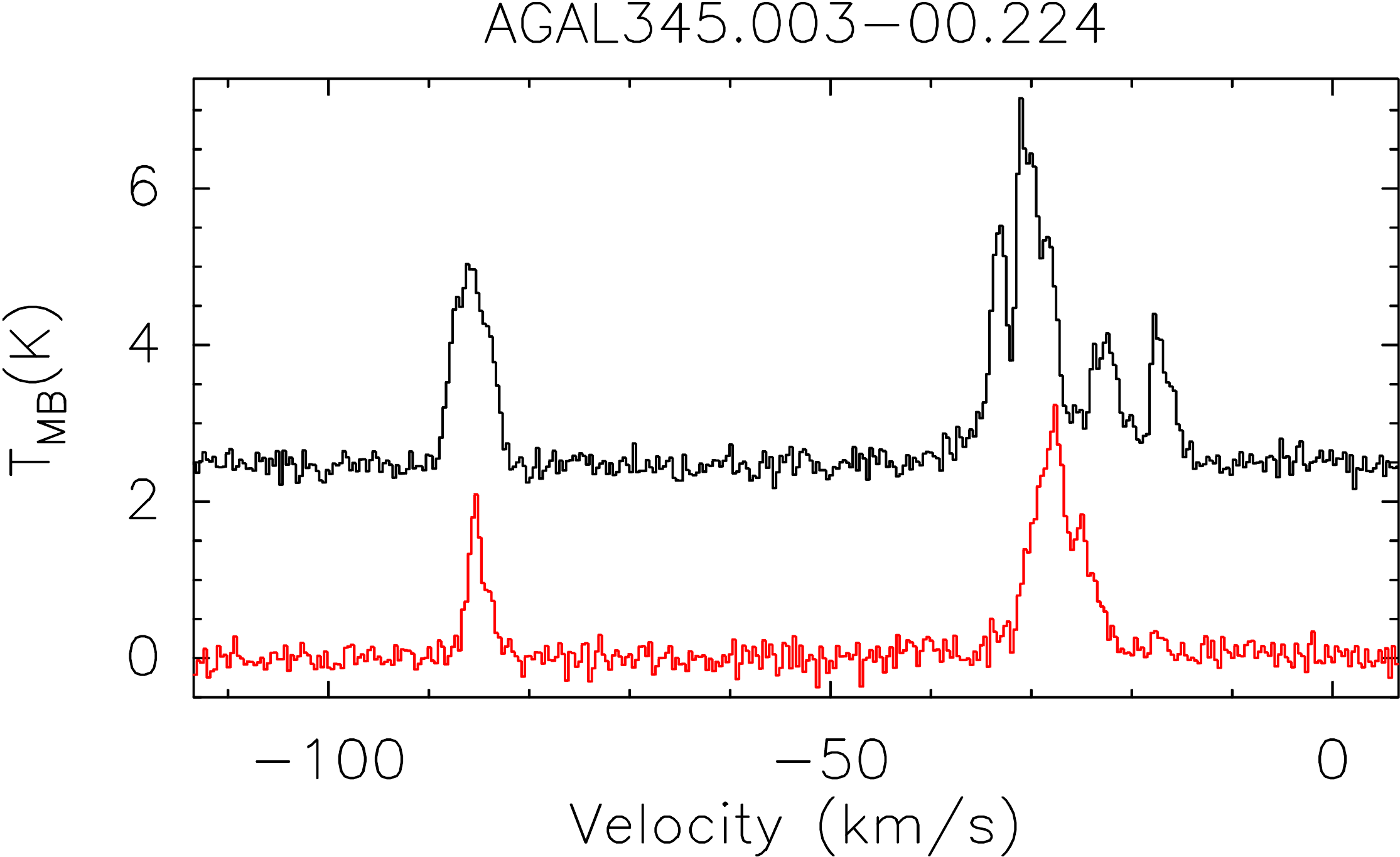} \\ 
\includegraphics[angle=-90,width=0.3\textwidth]{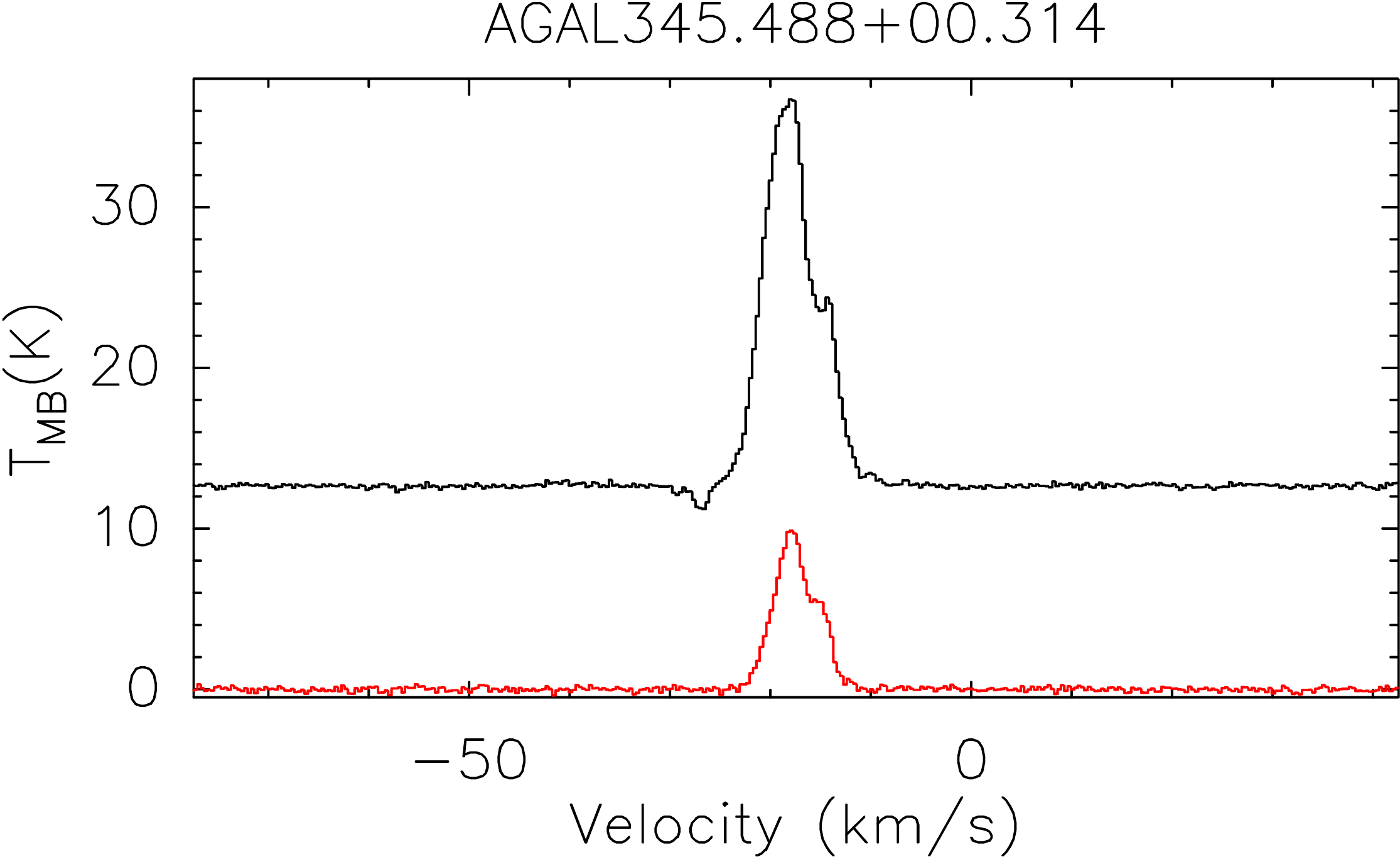} 
\includegraphics[angle=-90,width=0.3\textwidth]{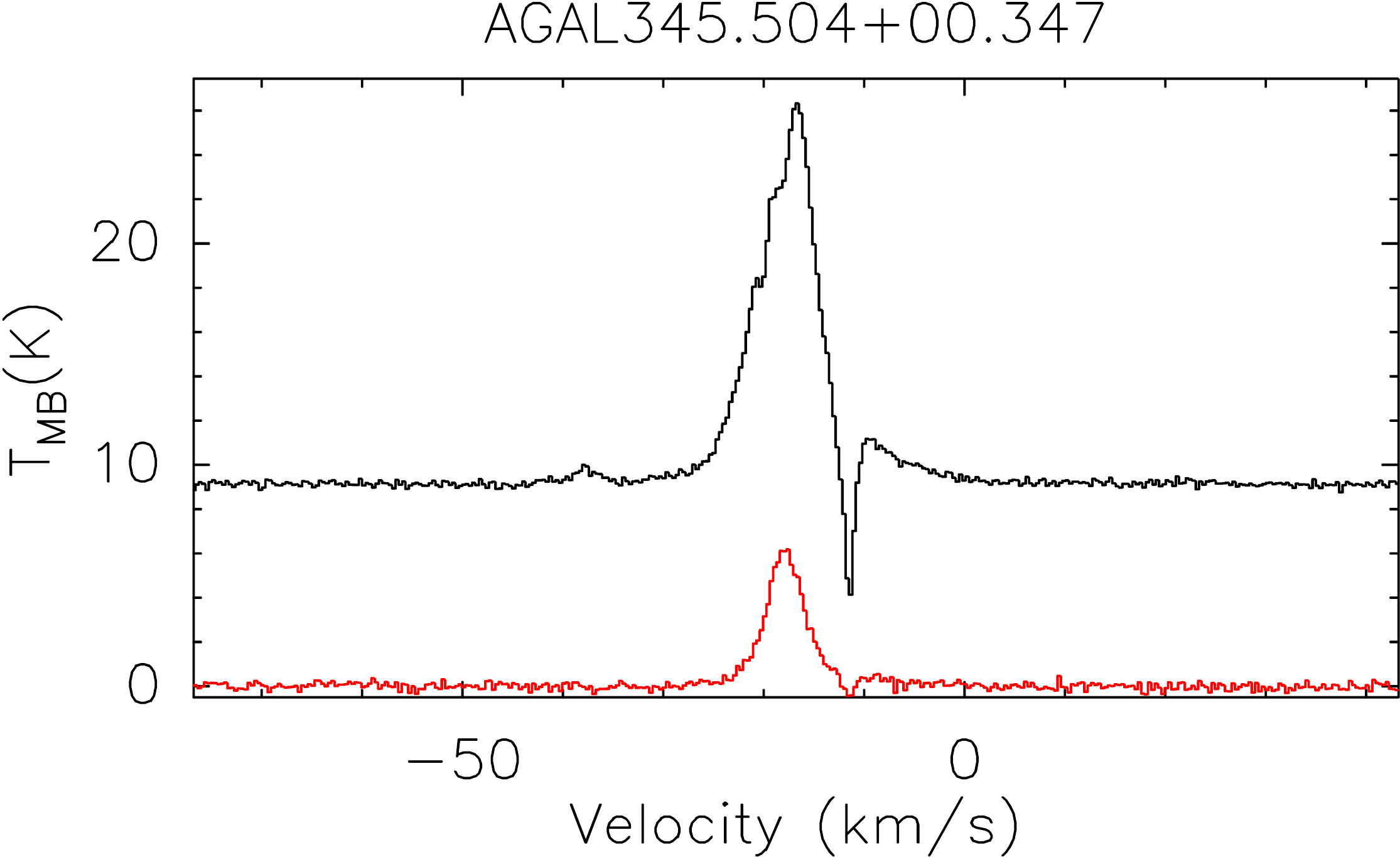} 
\includegraphics[angle=-90,width=0.3\textwidth]{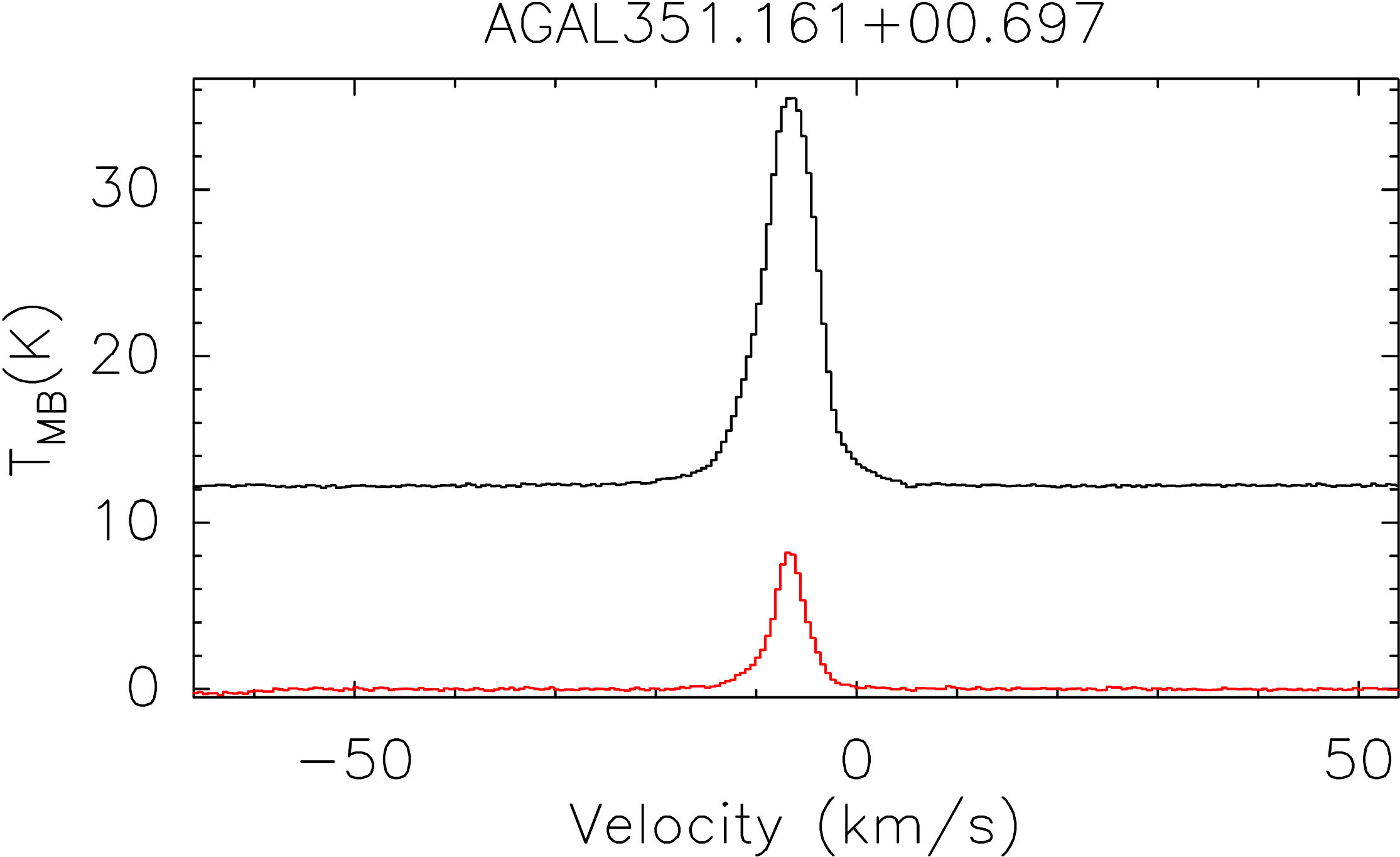} \hfill 
\caption{Continued.} 
\end{figure*} 

\begin{figure*} 
\ContinuedFloat
\centering 
\includegraphics[angle=-90,width=0.3\textwidth]{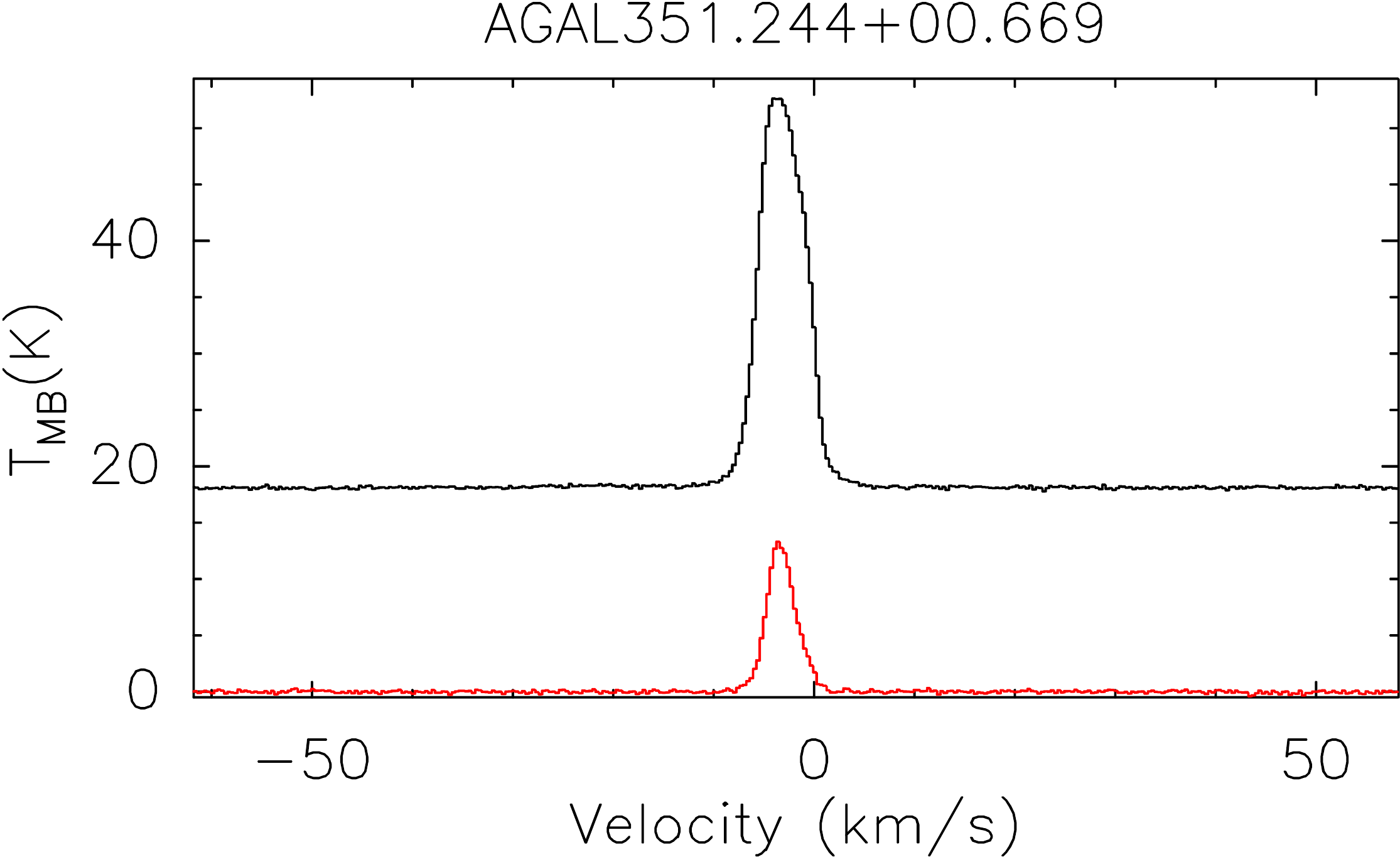} 
\includegraphics[angle=-90,width=0.3\textwidth]{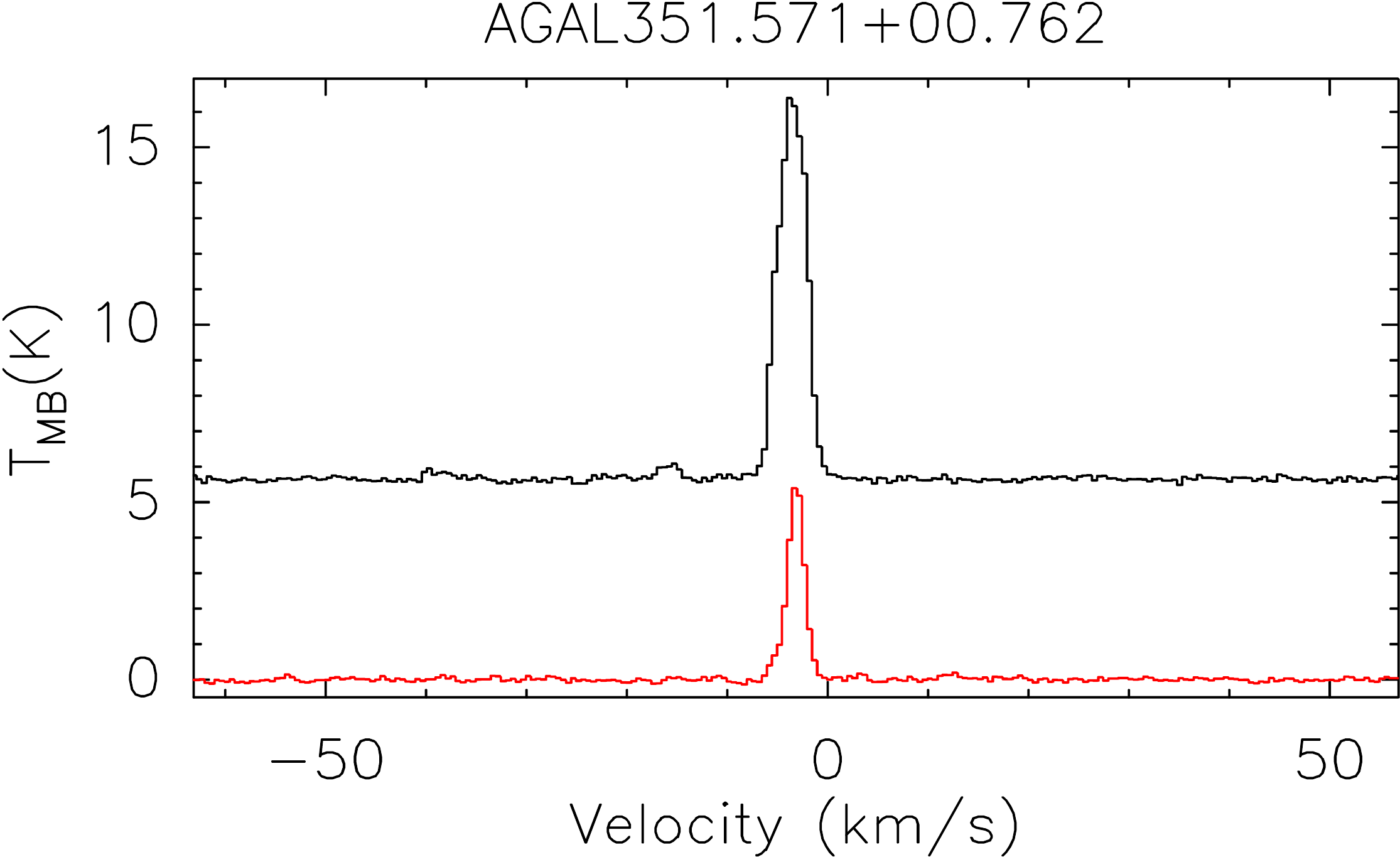} 
\includegraphics[angle=-90,width=0.3\textwidth]{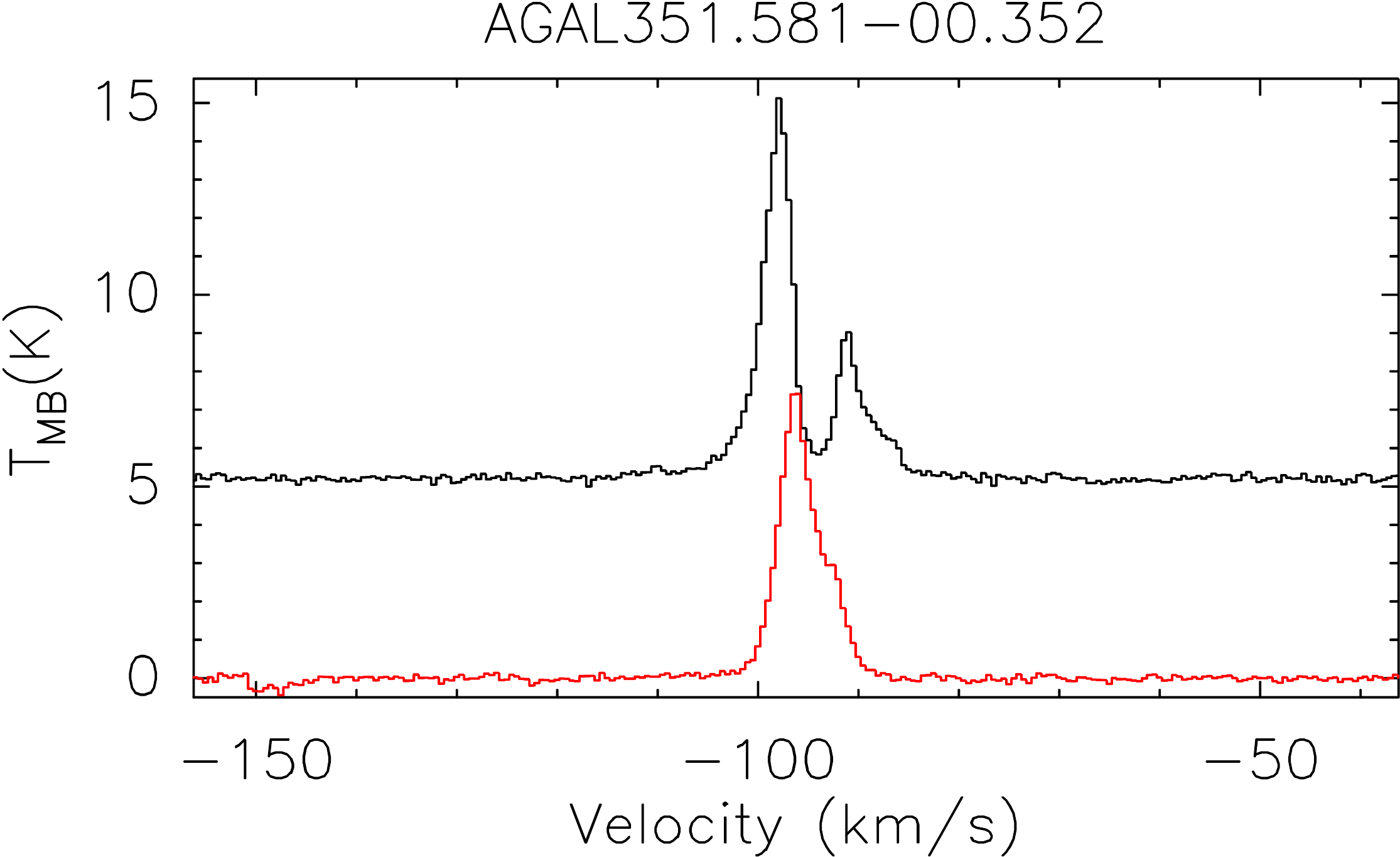} \\
\includegraphics[angle=-90,width=0.3\textwidth]{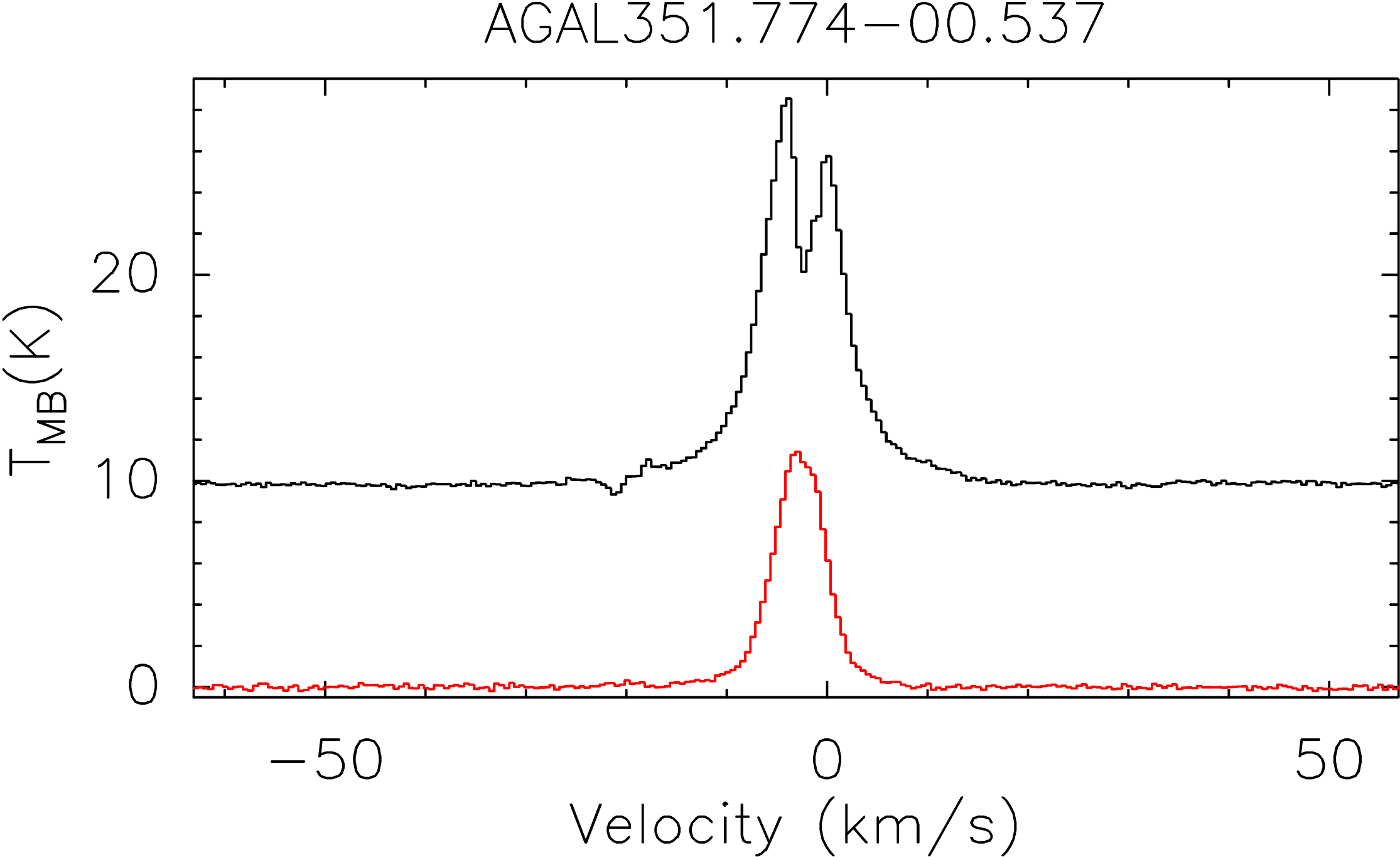} 
\includegraphics[angle=-90,width=0.3\textwidth]{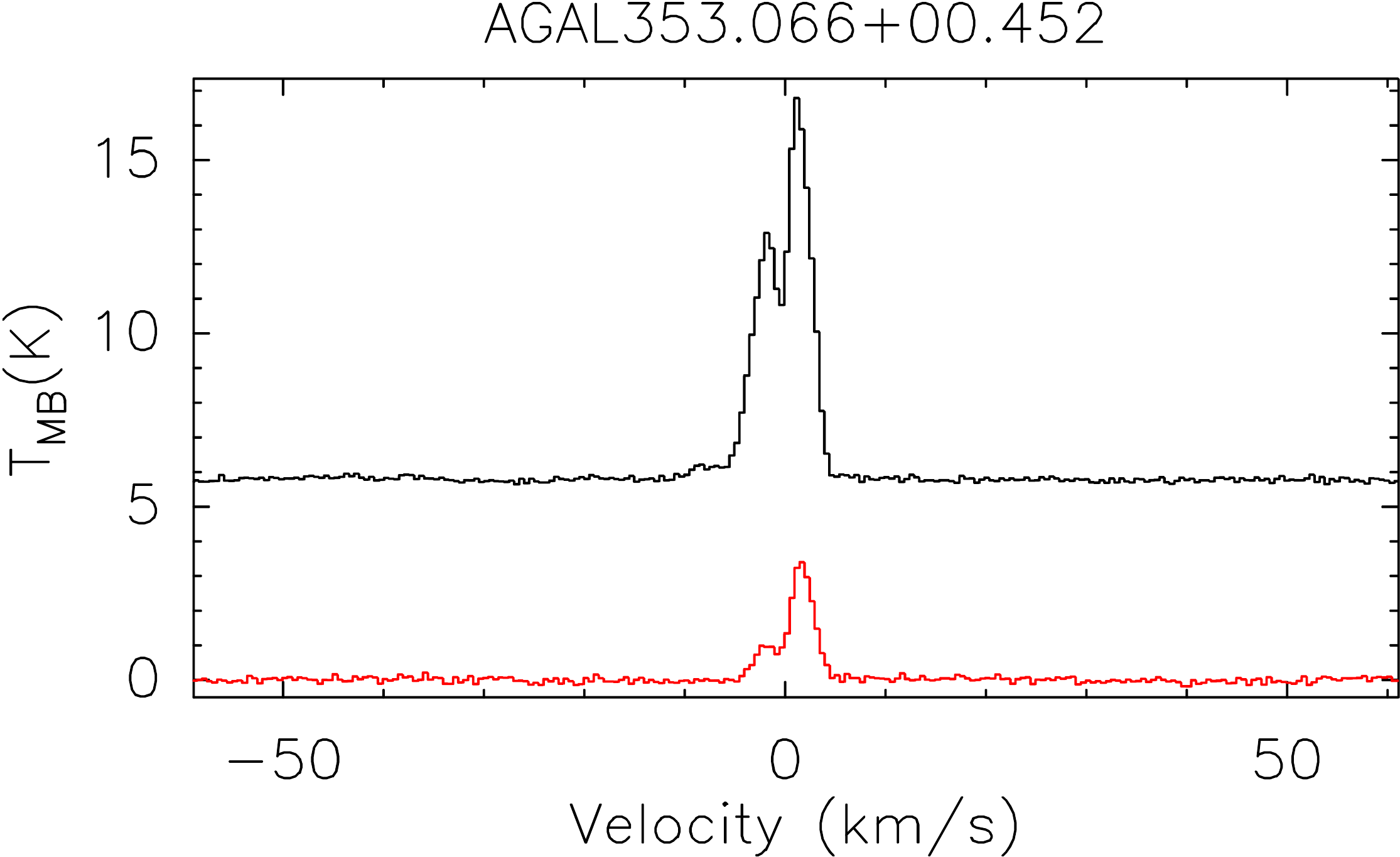}  
\includegraphics[angle=-90,width=0.3\textwidth]{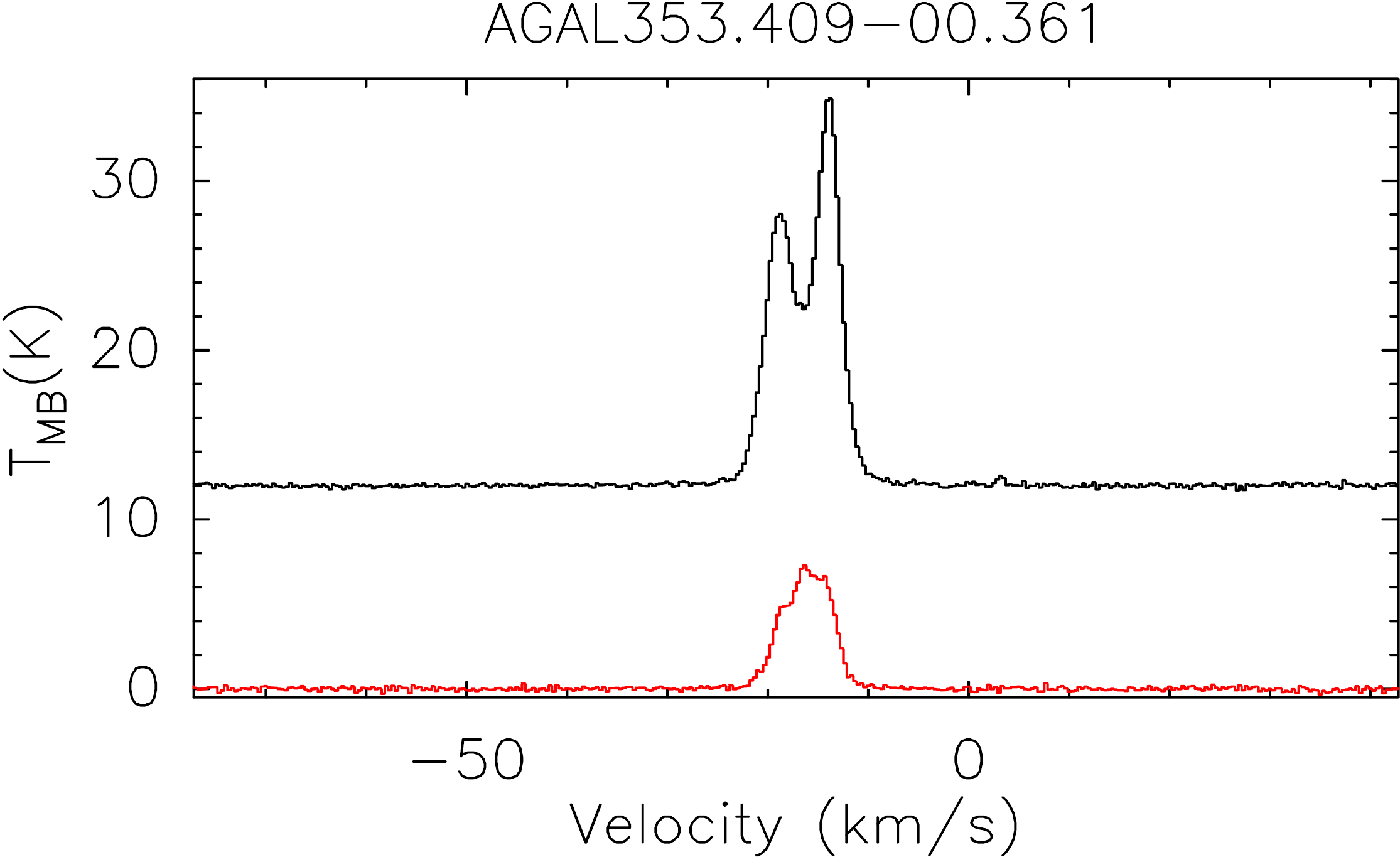} \\
\includegraphics[angle=-90,width=0.3\textwidth]{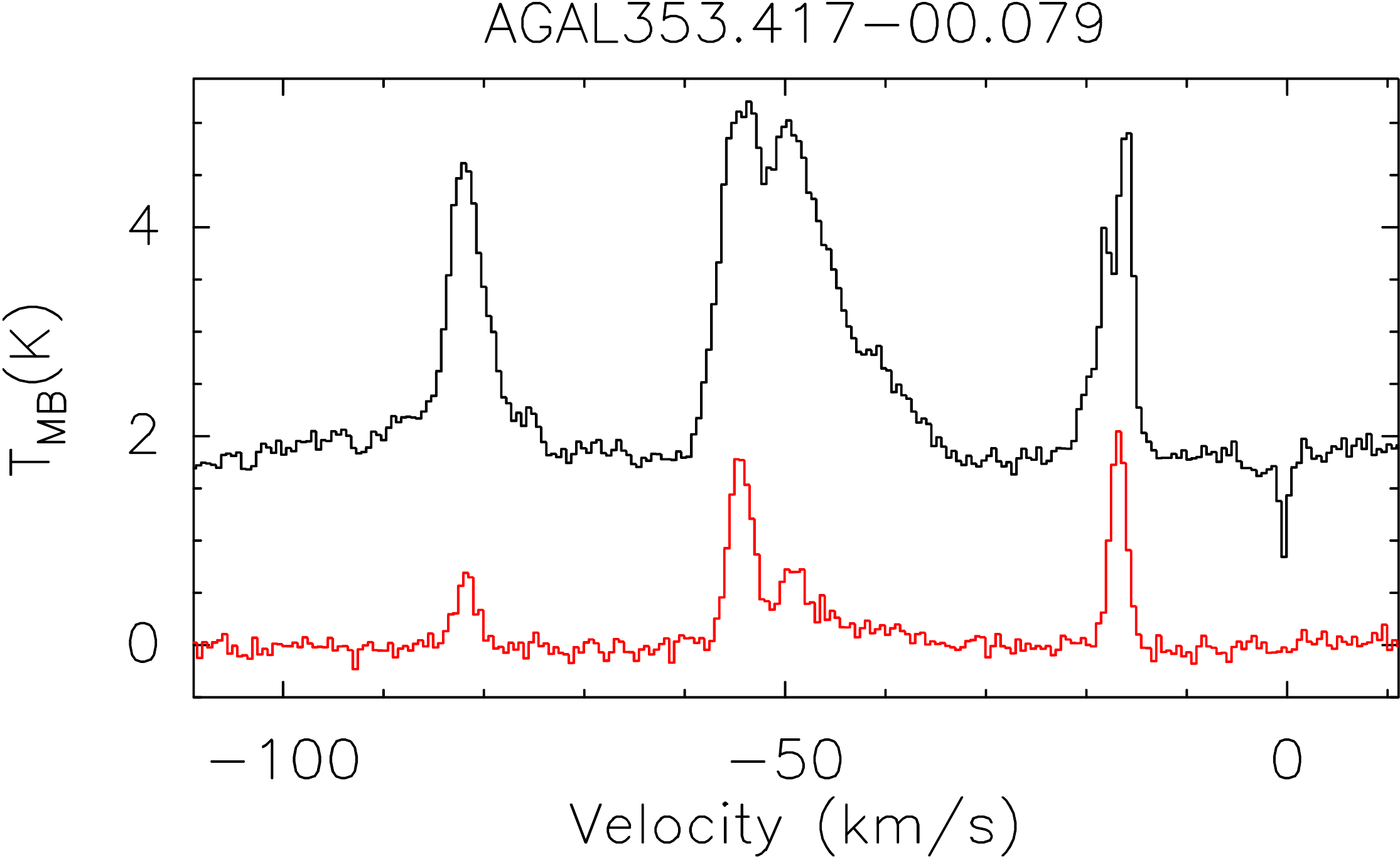} 
\includegraphics[angle=-90,width=0.3\textwidth]{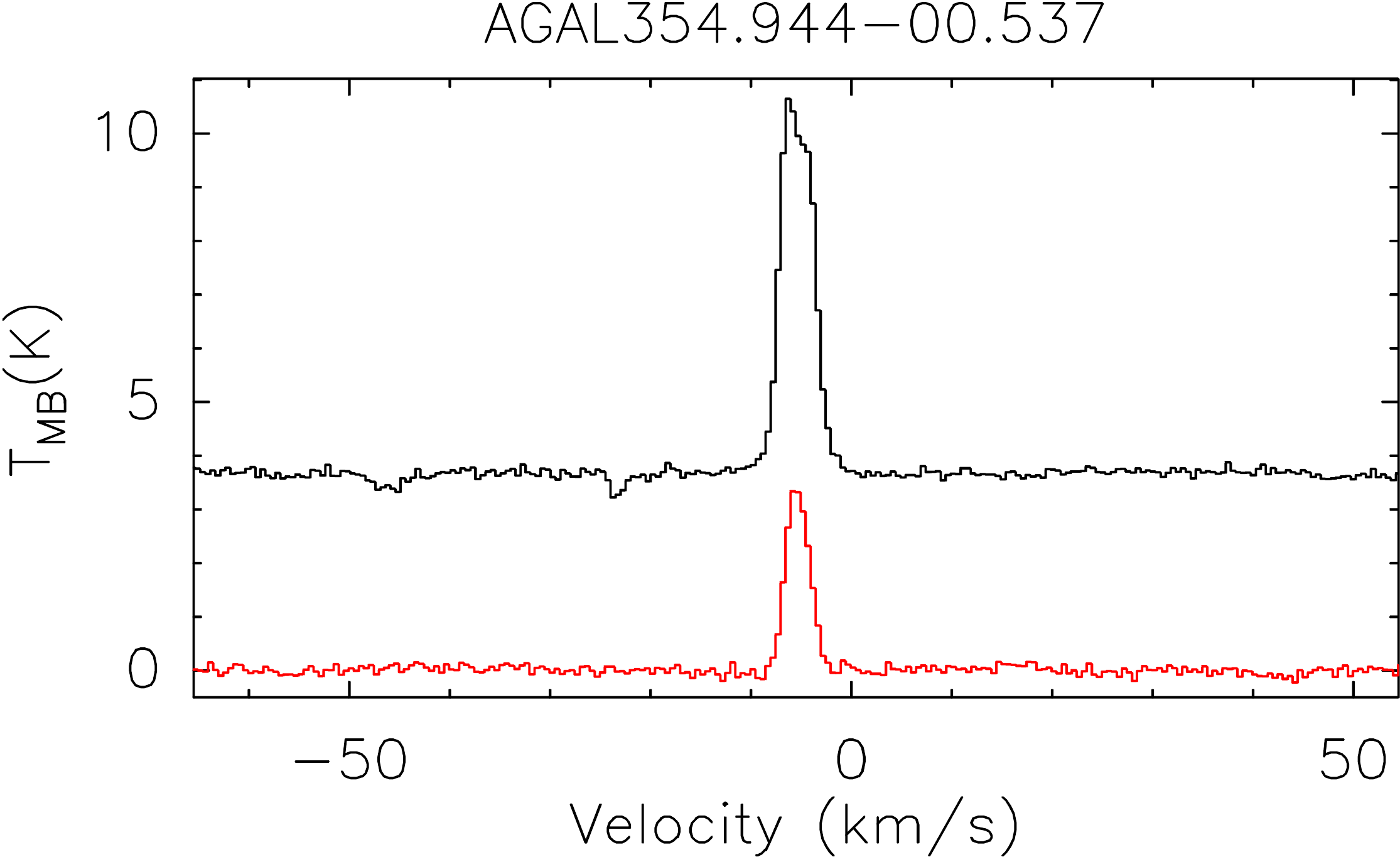} \hfill 
\caption{Continued.} 
\end{figure*} 
}
\onlfig{3}{
\begin{figure*} 
\centering 
\includegraphics[angle=-90,width=0.3\textwidth]{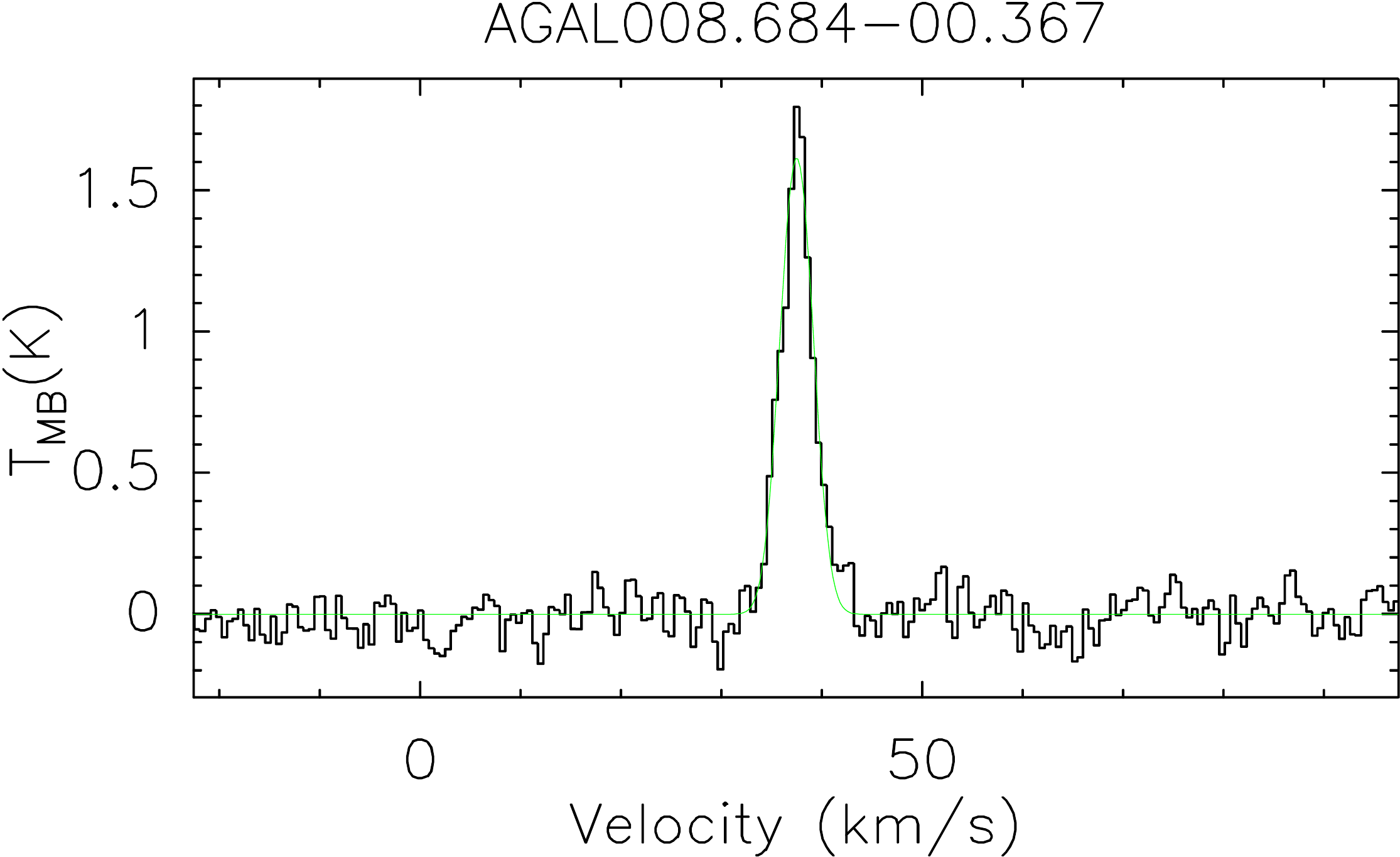}
\includegraphics[angle=-90,width=0.3\textwidth]{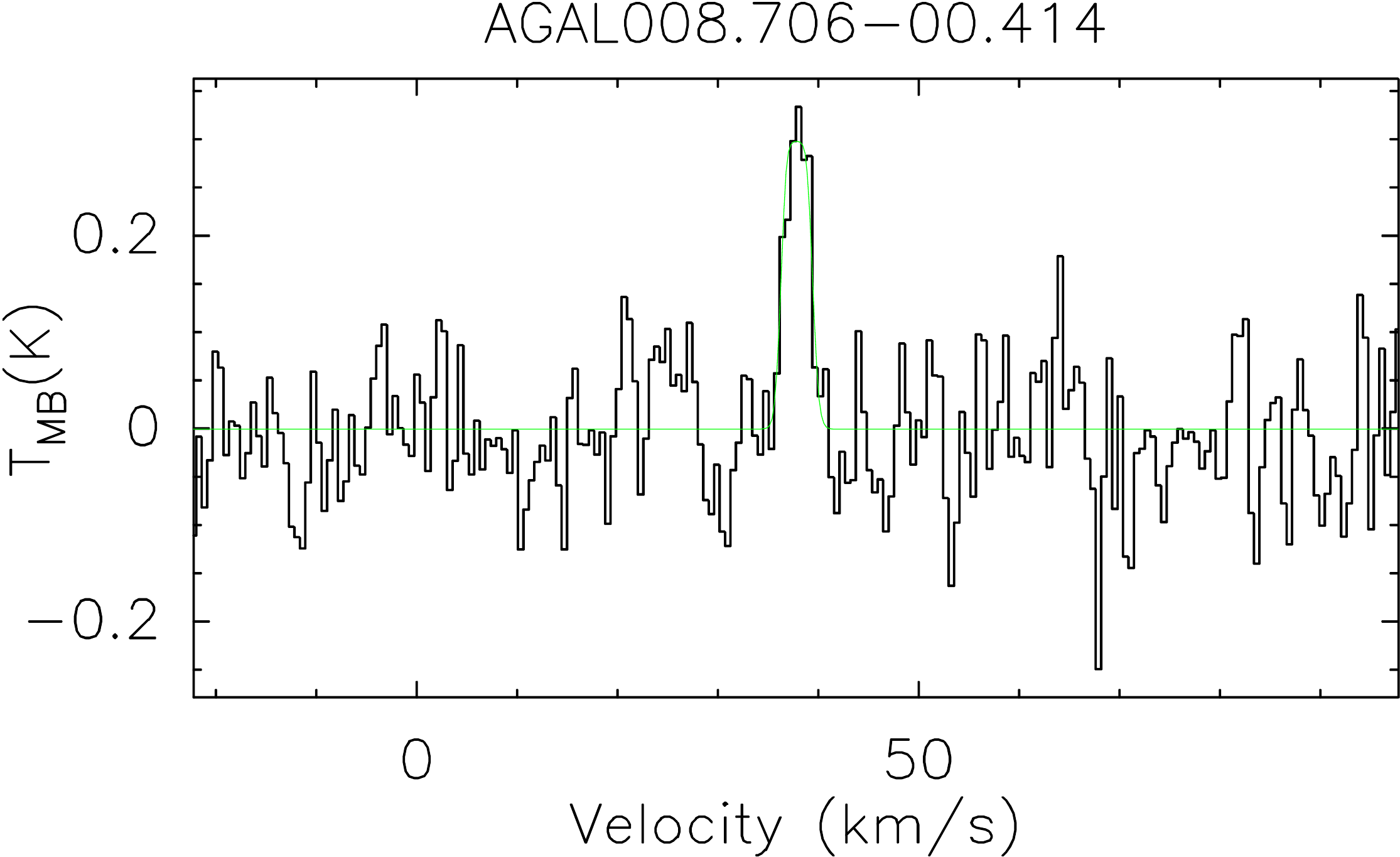}
\includegraphics[angle=-90,width=0.3\textwidth]{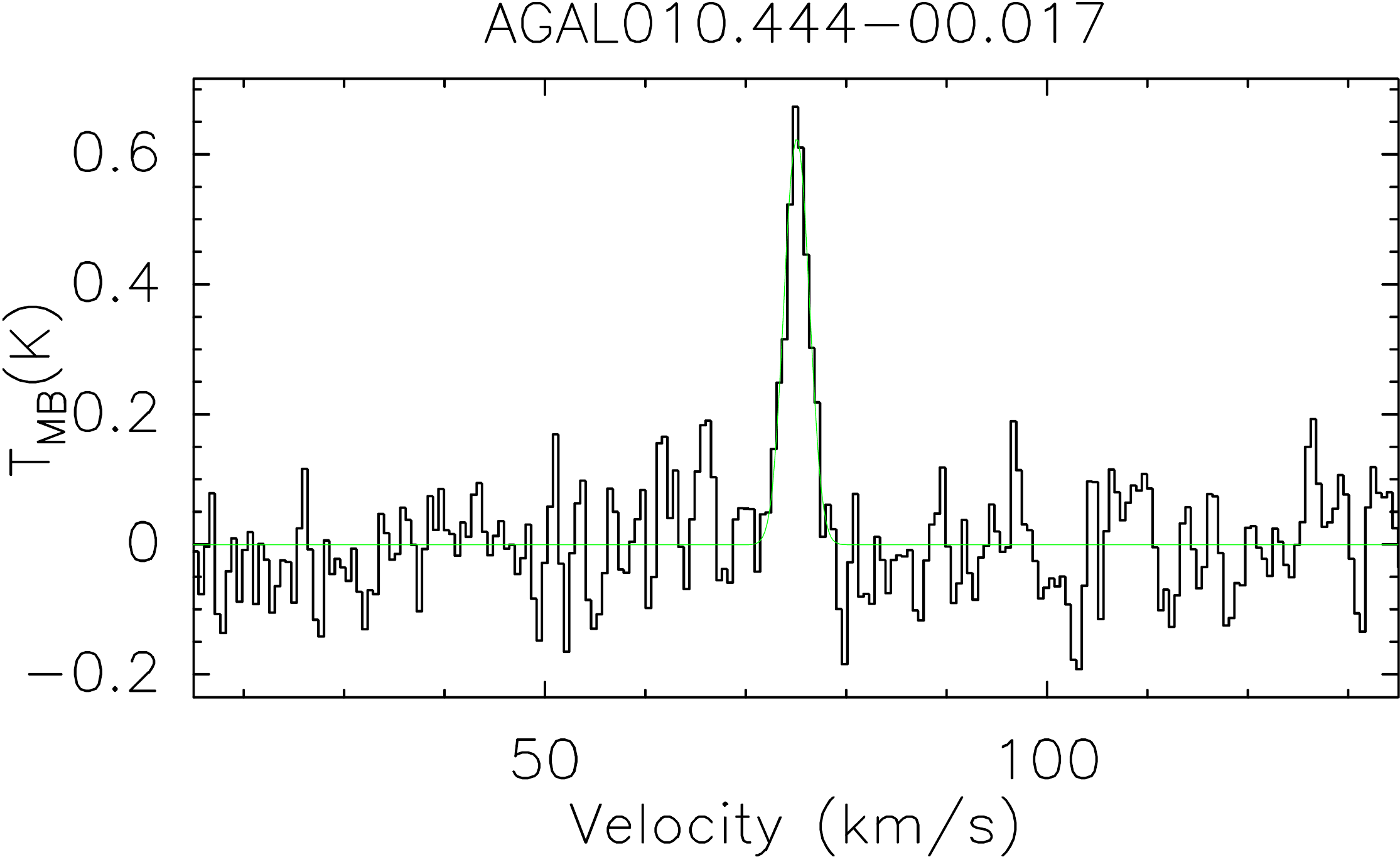} \\
\includegraphics[angle=-90,width=0.3\textwidth]{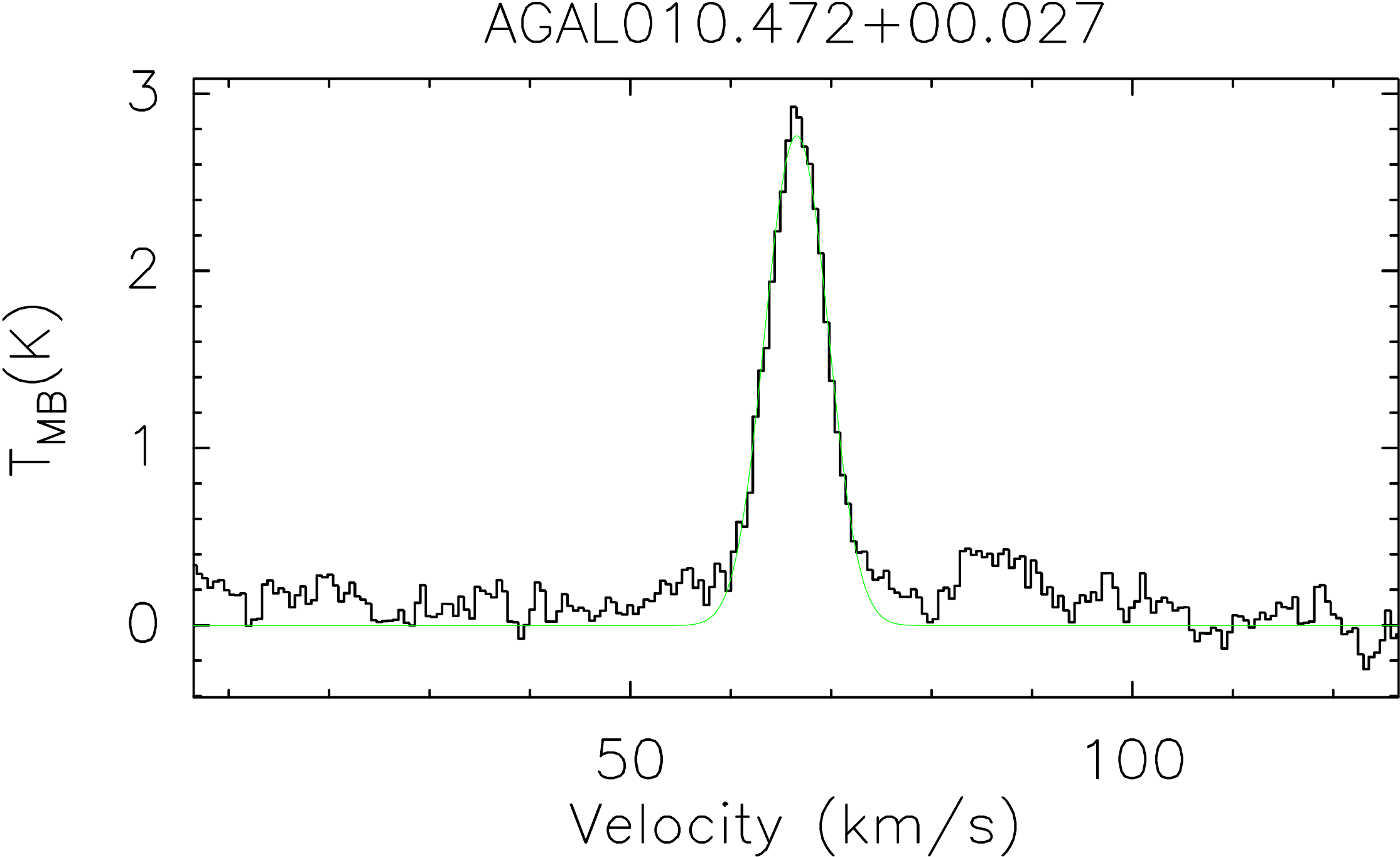}
\includegraphics[angle=-90,width=0.3\textwidth]{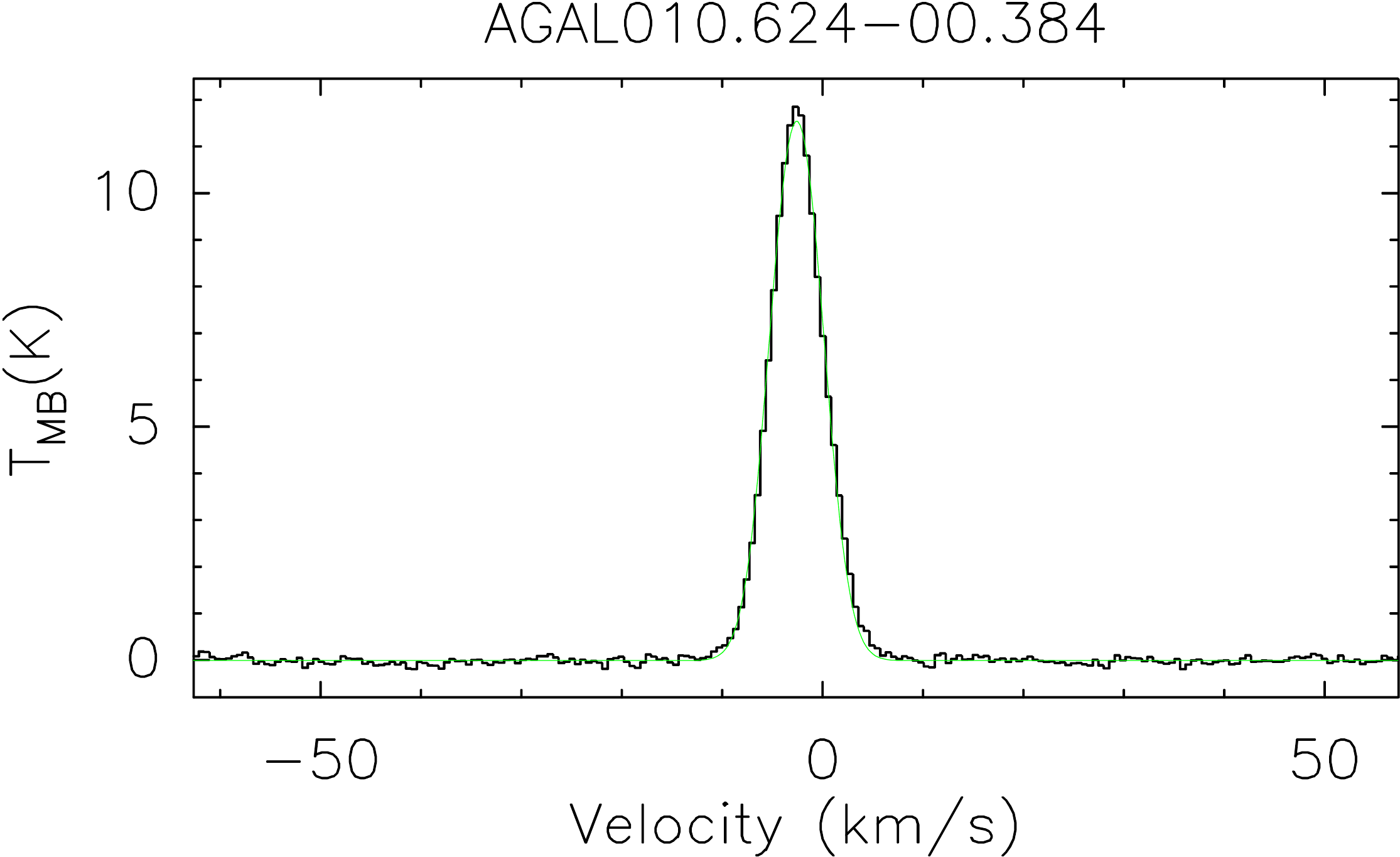}
\includegraphics[angle=-90,width=0.3\textwidth]{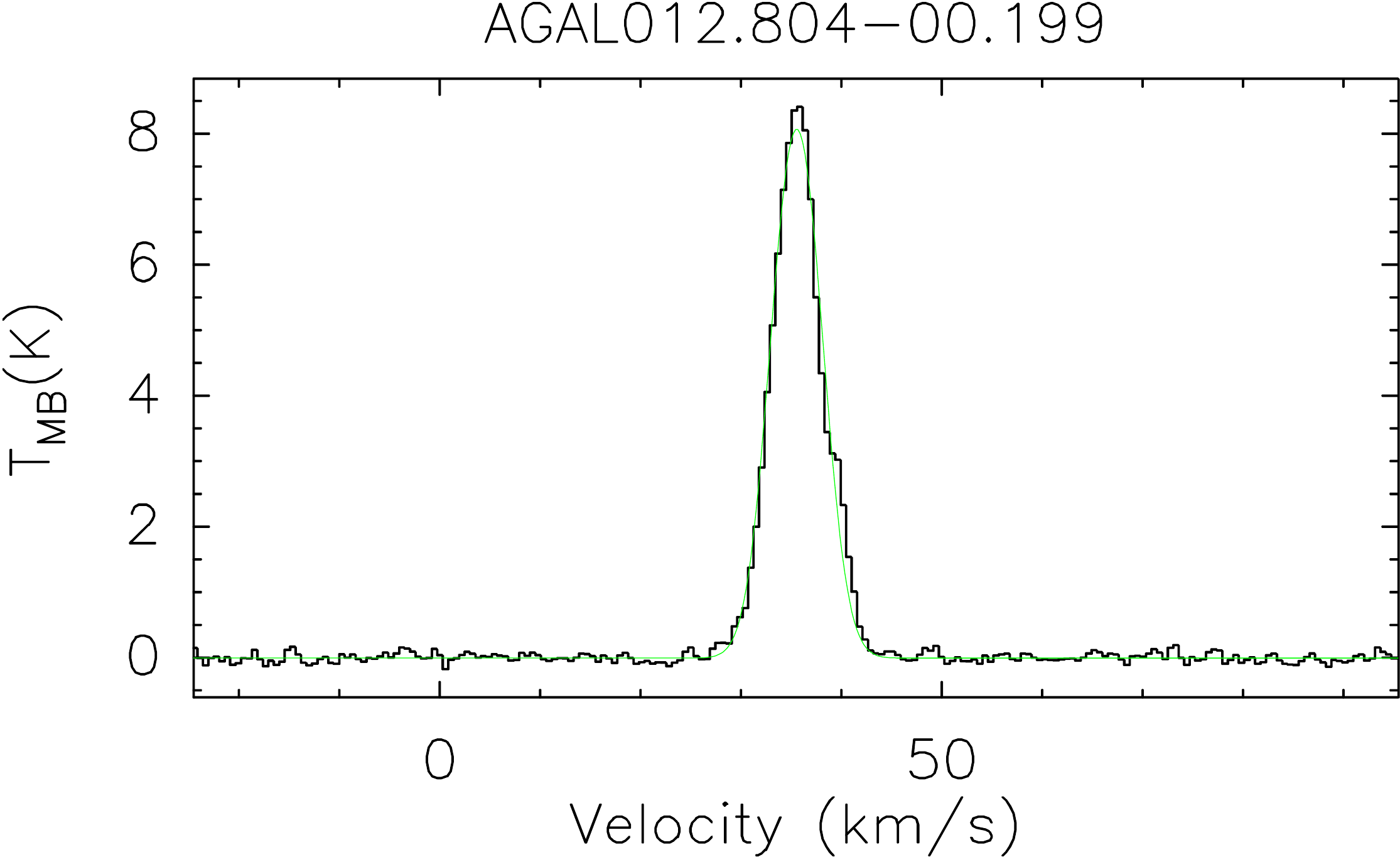} \\
\includegraphics[angle=-90,width=0.3\textwidth]{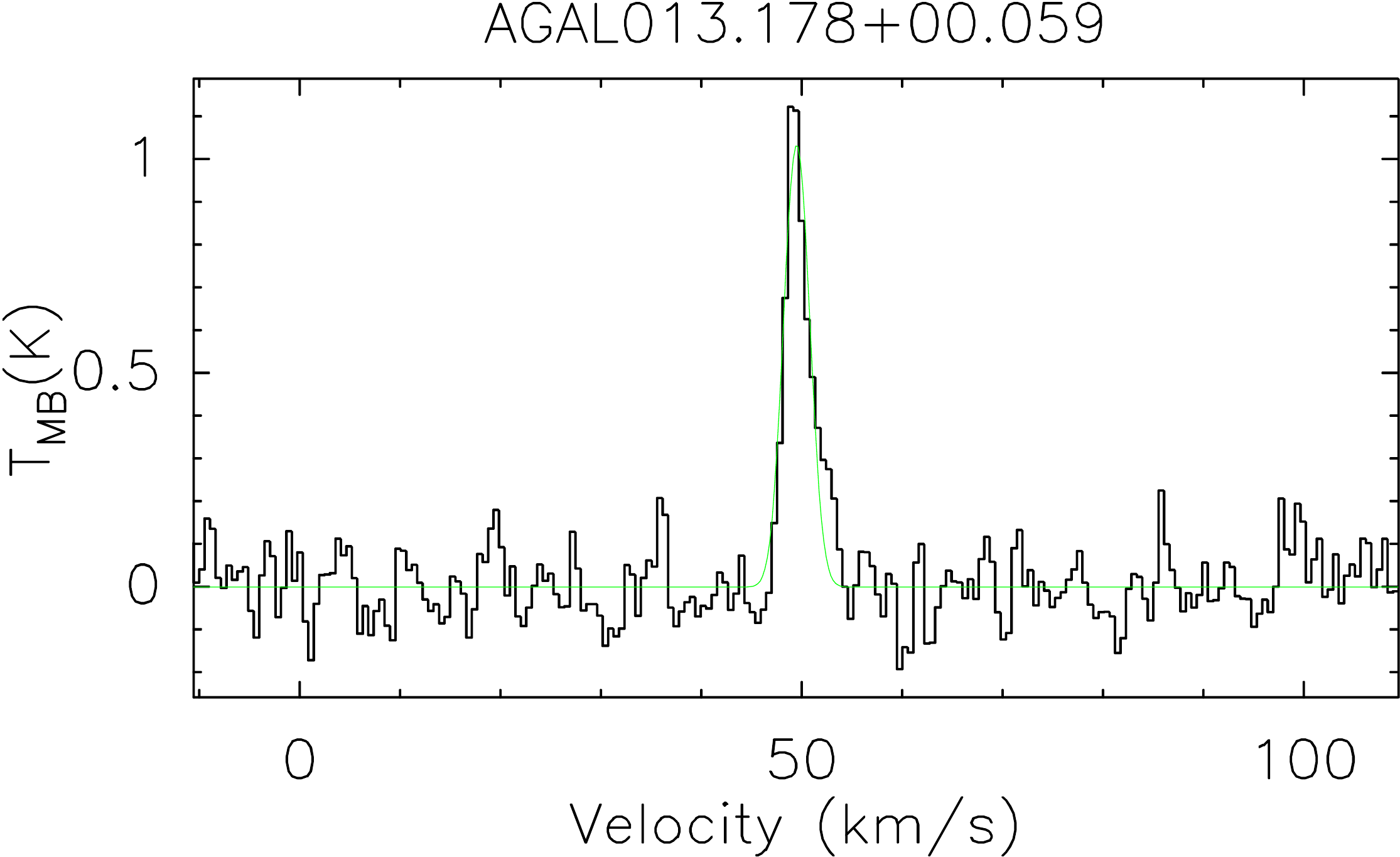}
\includegraphics[angle=-90,width=0.3\textwidth]{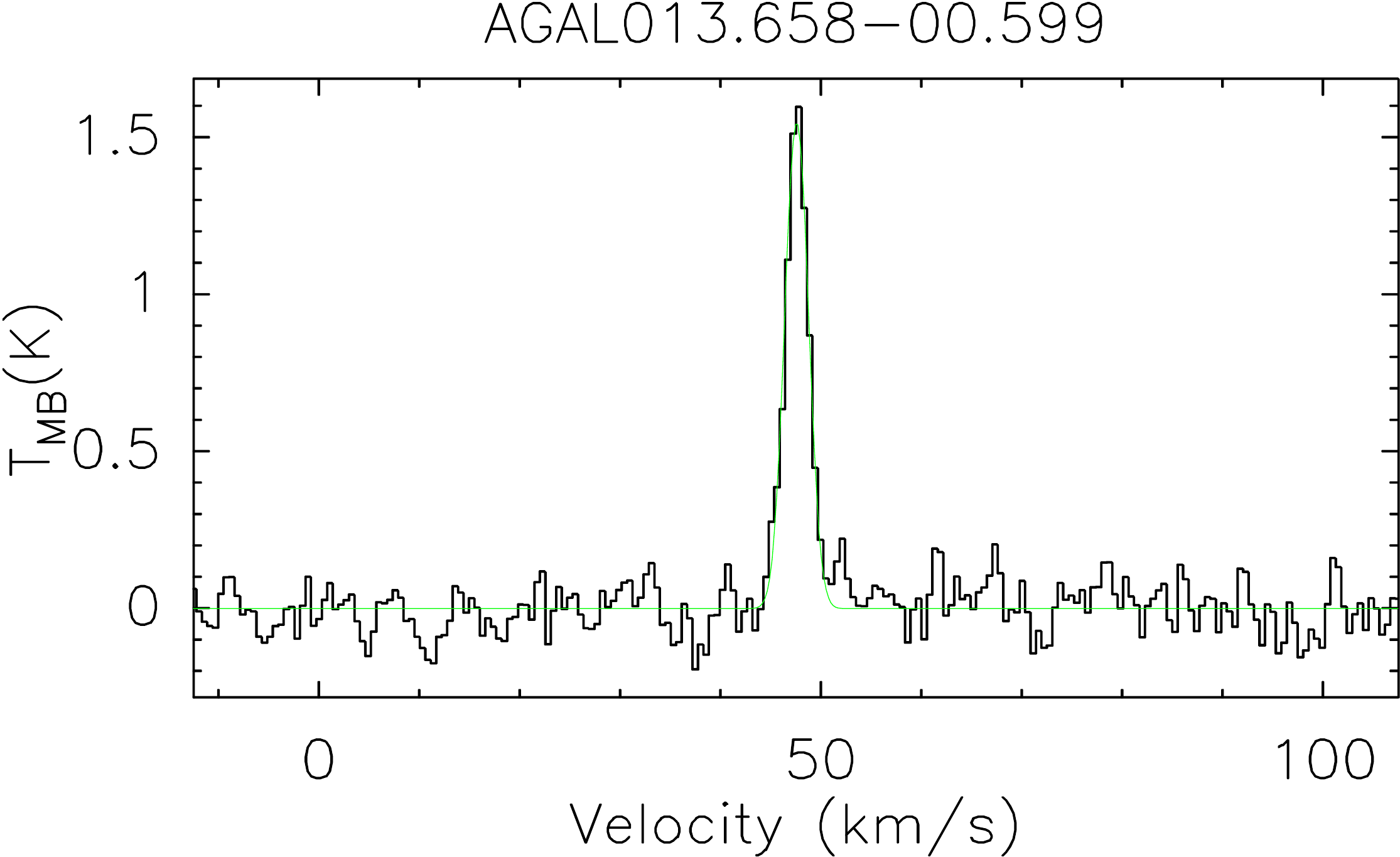}
\includegraphics[angle=-90,width=0.3\textwidth]{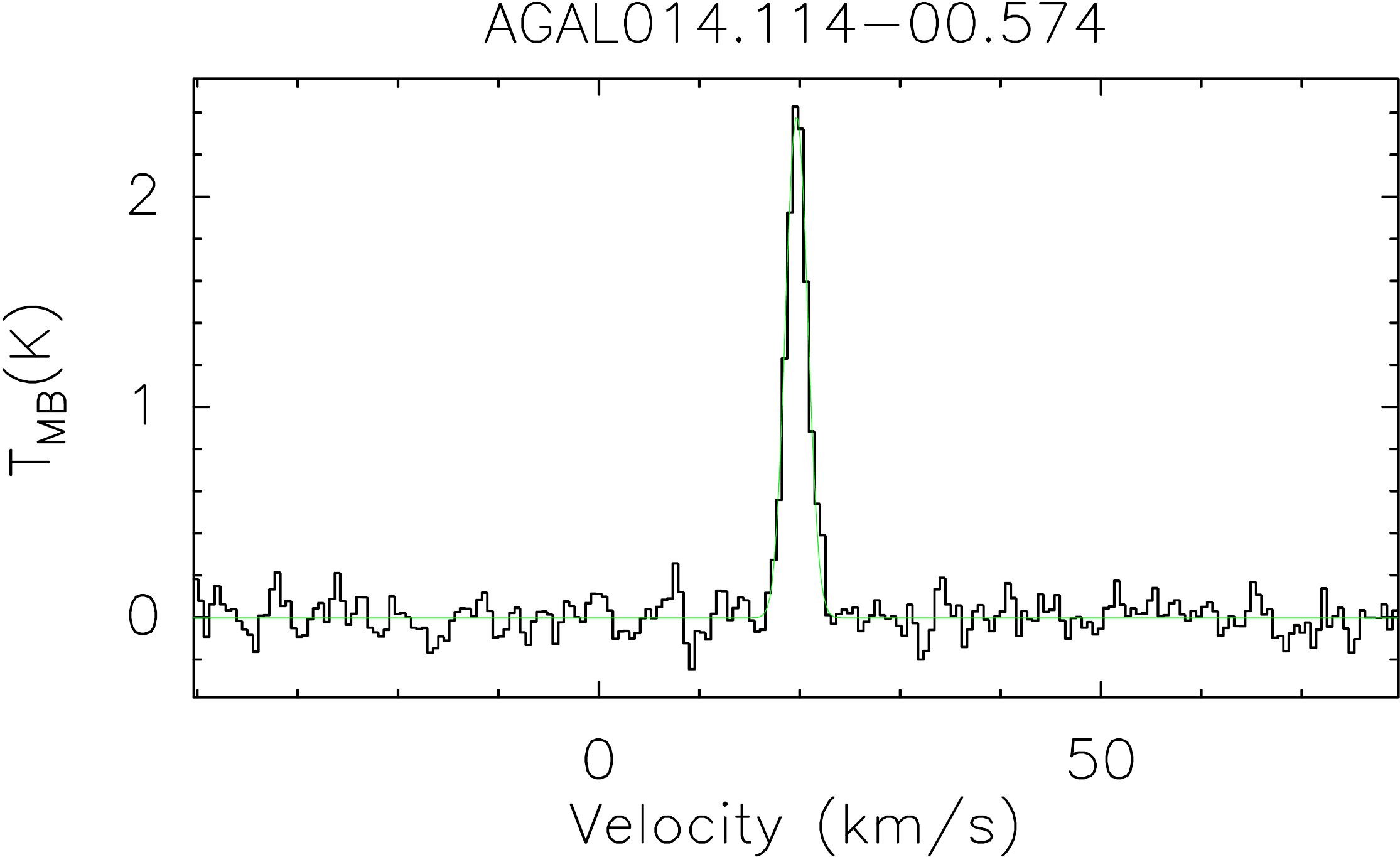} \\
\includegraphics[angle=-90,width=0.3\textwidth]{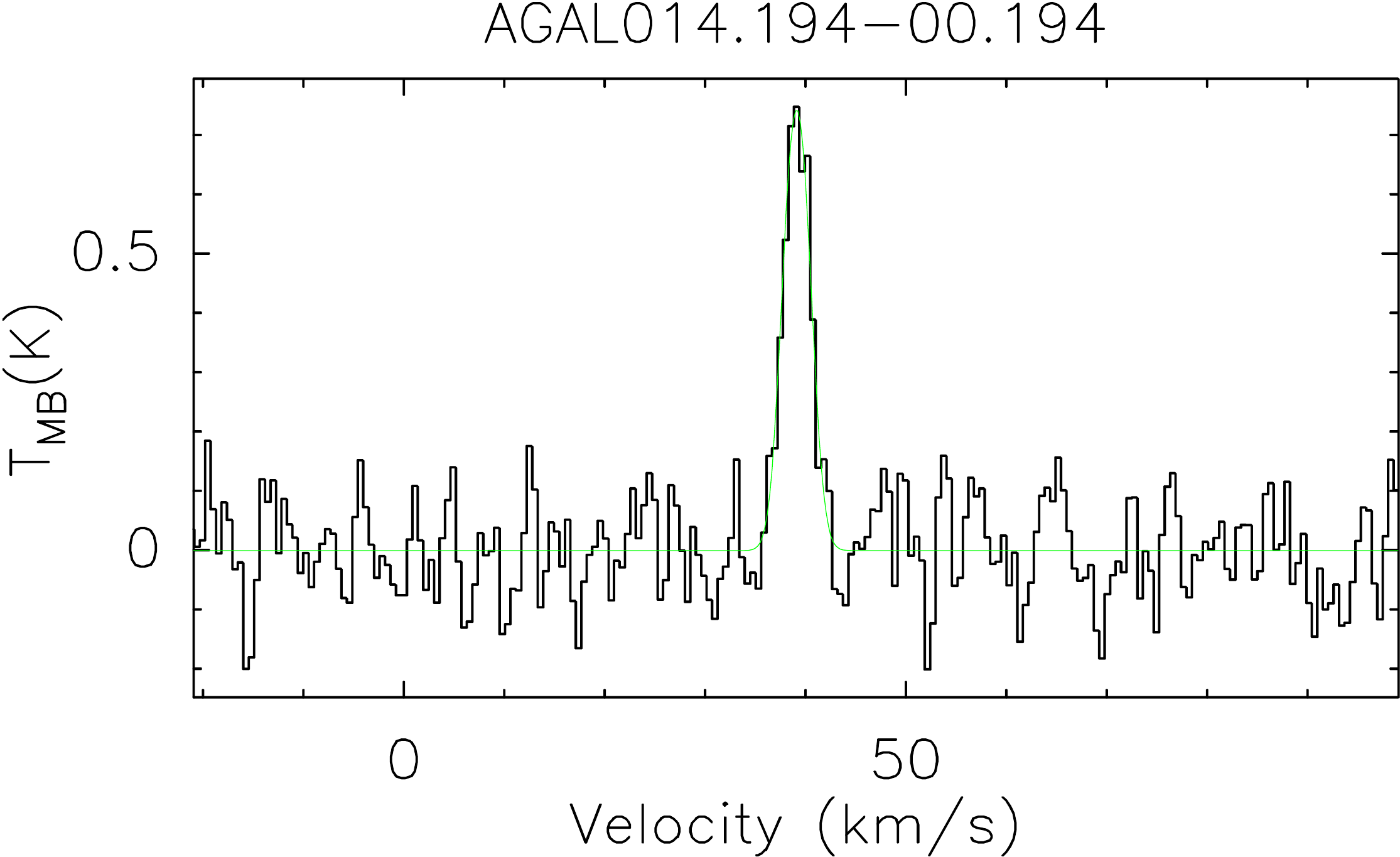}
\includegraphics[angle=-90,width=0.3\textwidth]{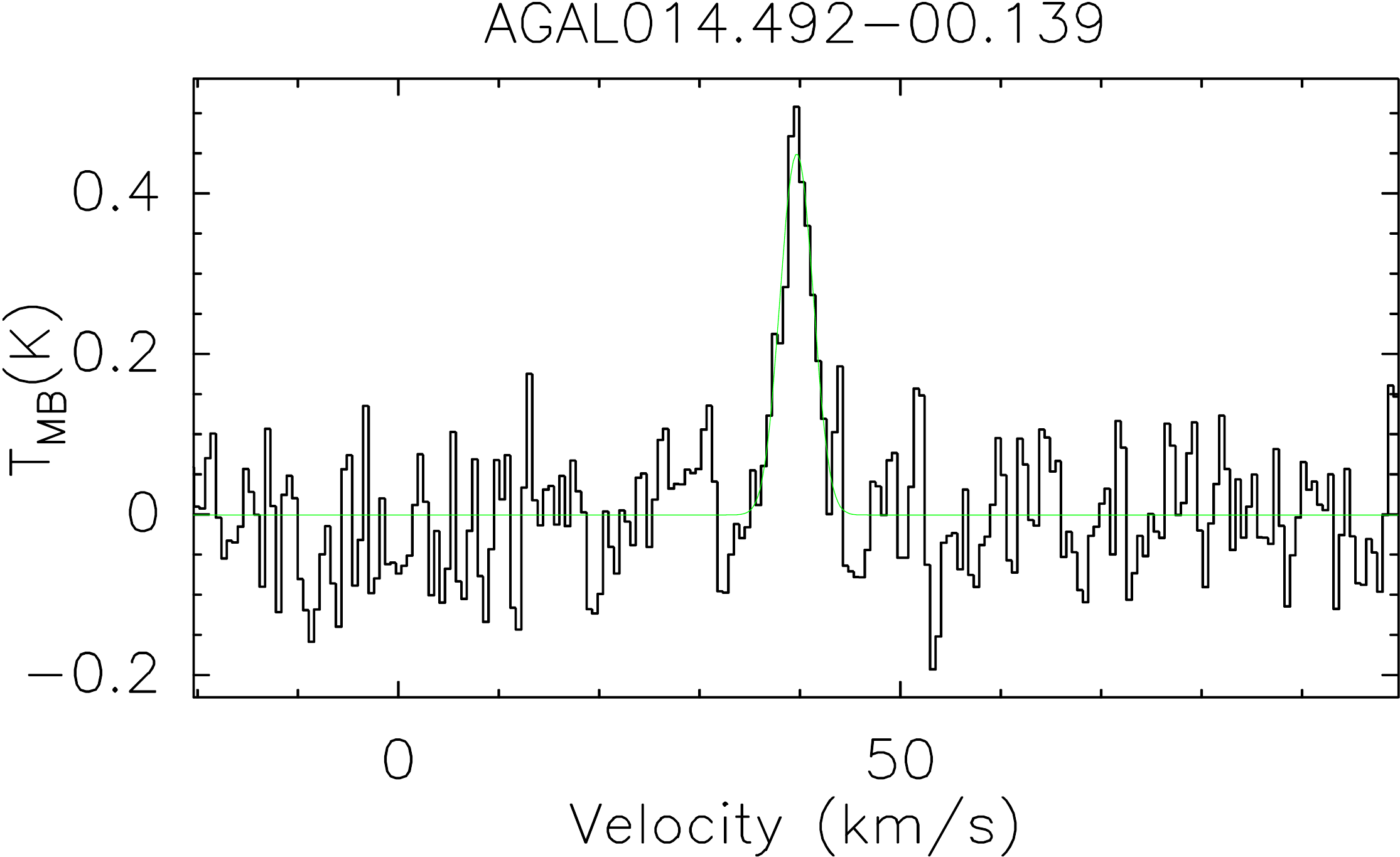}
\includegraphics[angle=-90,width=0.3\textwidth]{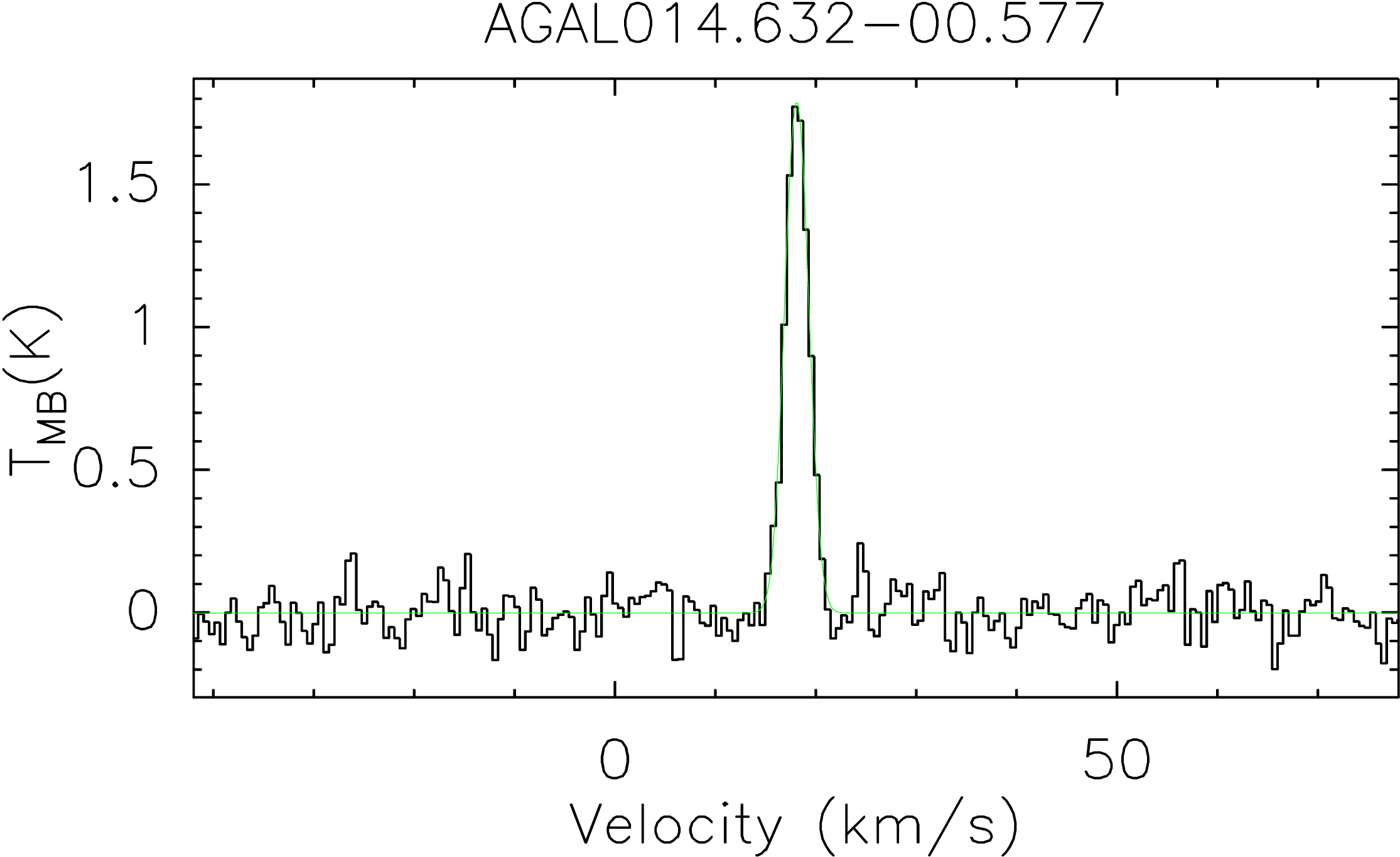} \\
\includegraphics[angle=-90,width=0.3\textwidth]{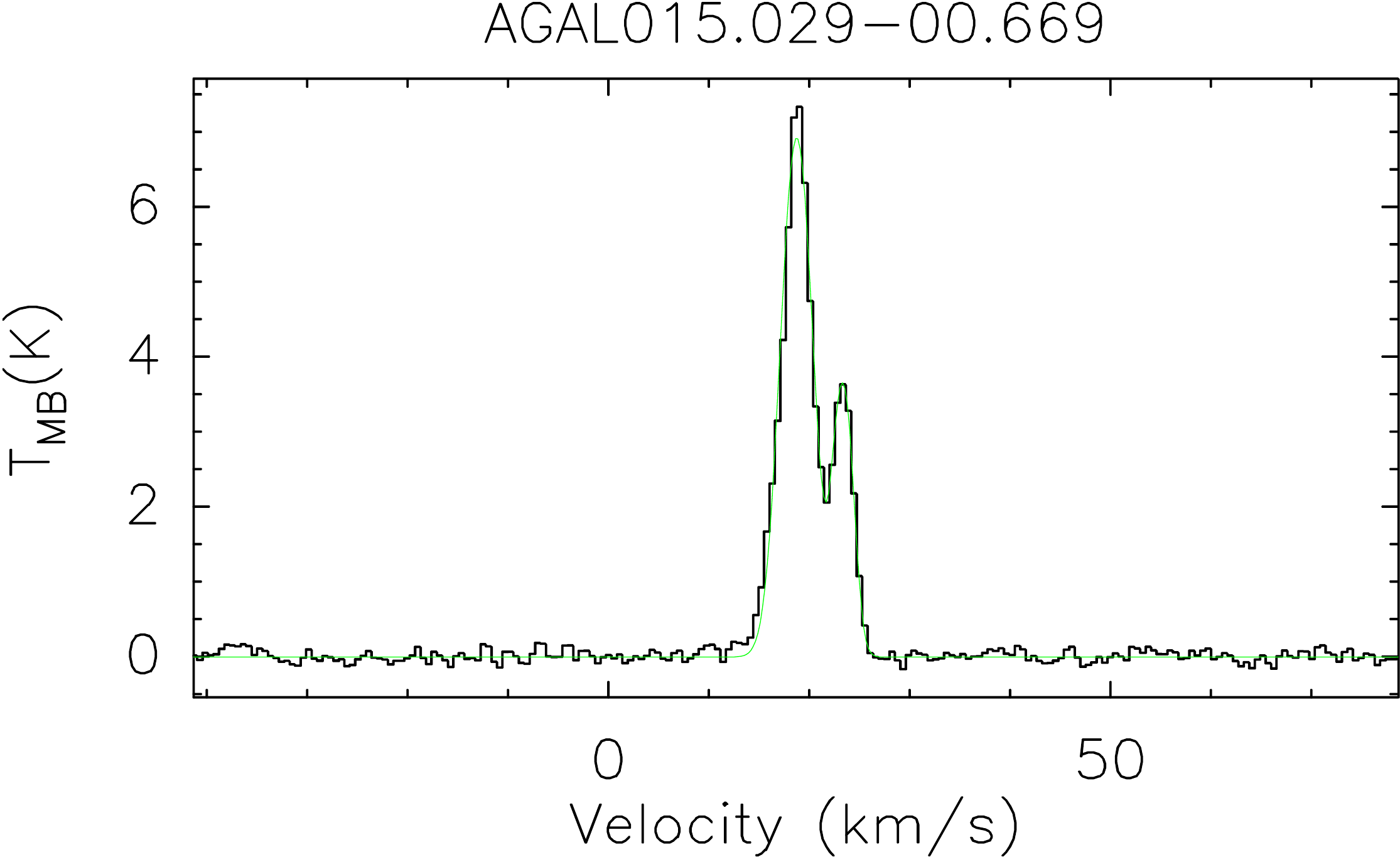}
\includegraphics[angle=-90,width=0.3\textwidth]{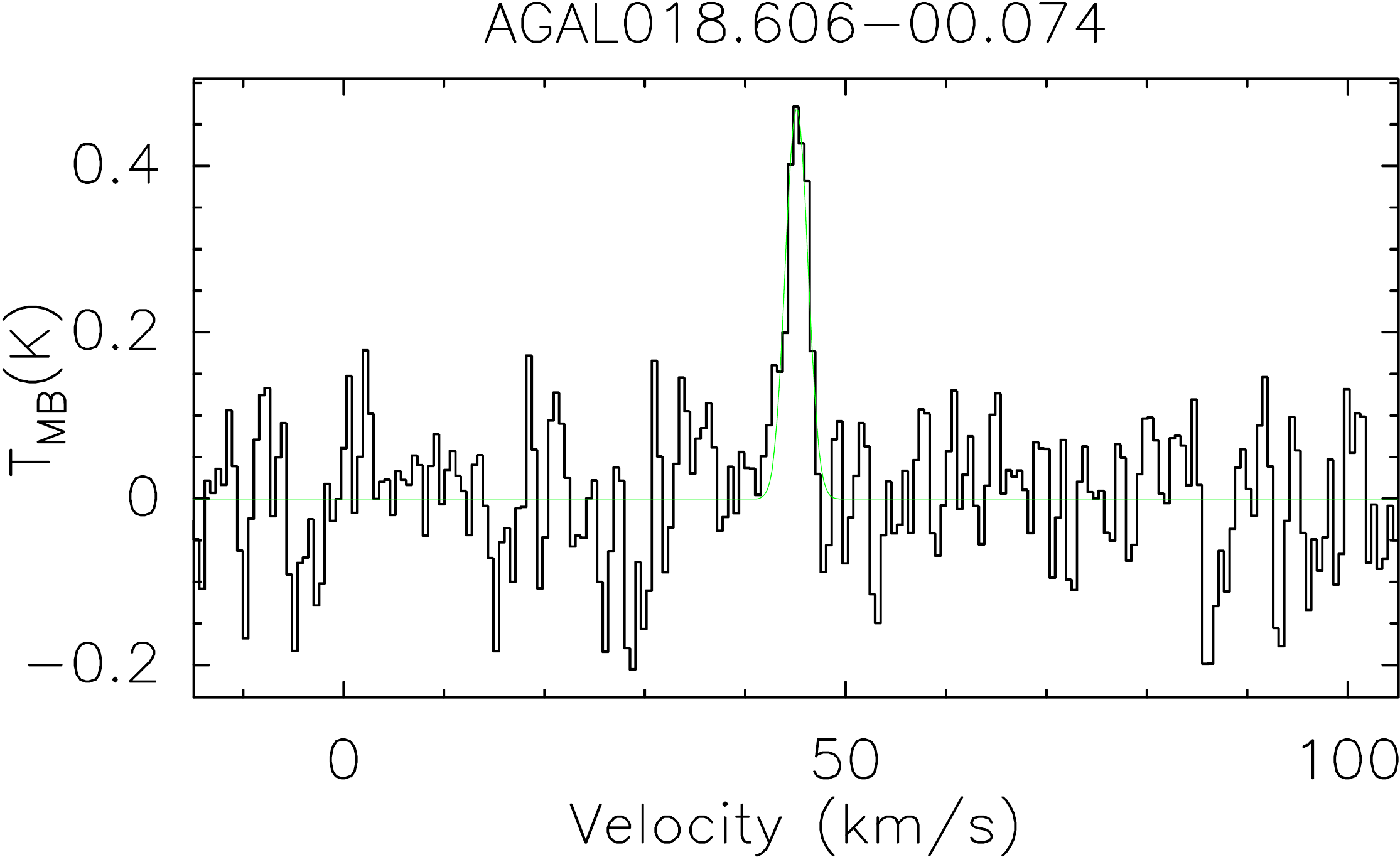}
\includegraphics[angle=-90,width=0.3\textwidth]{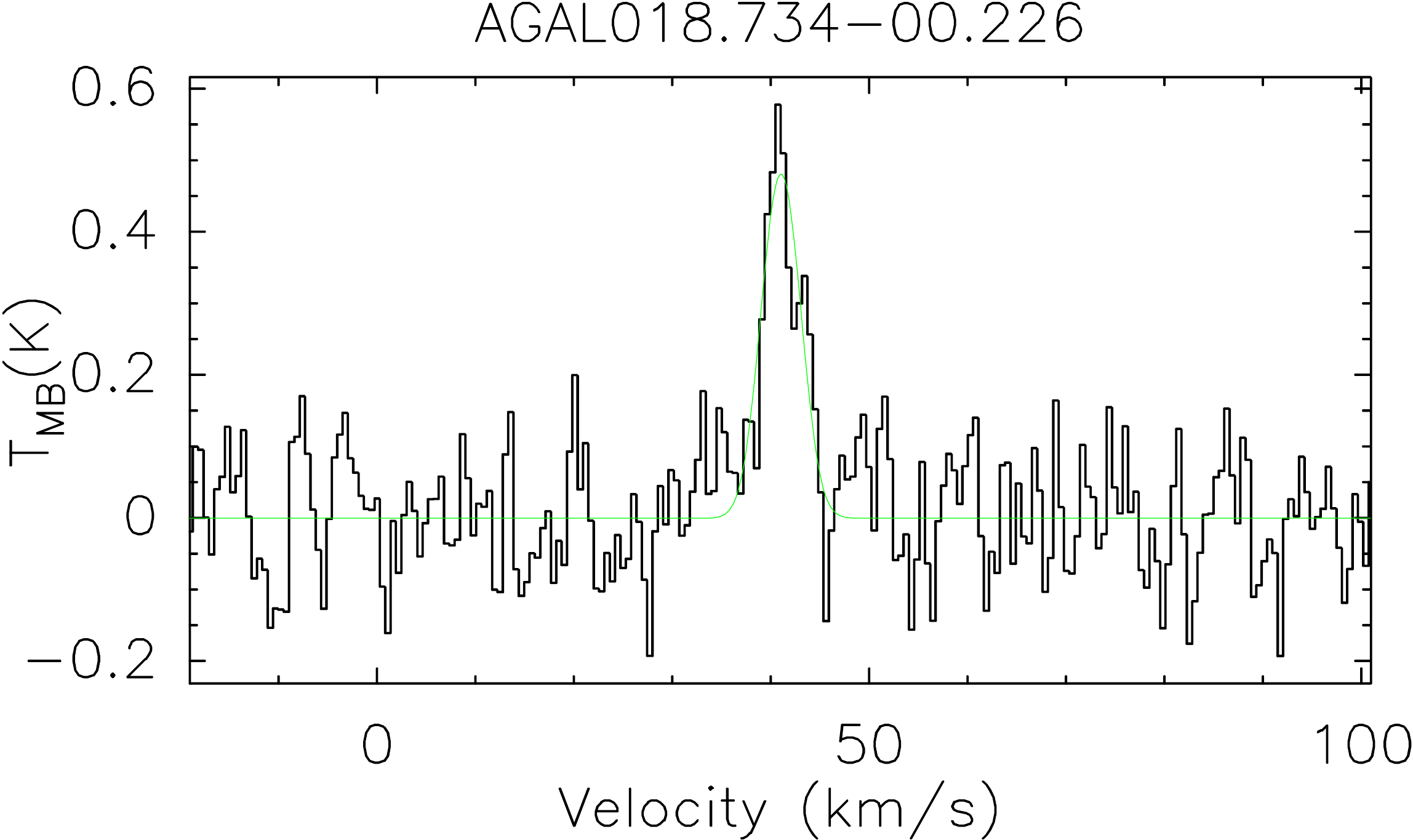} \\
\includegraphics[angle=-90,width=0.3\textwidth]{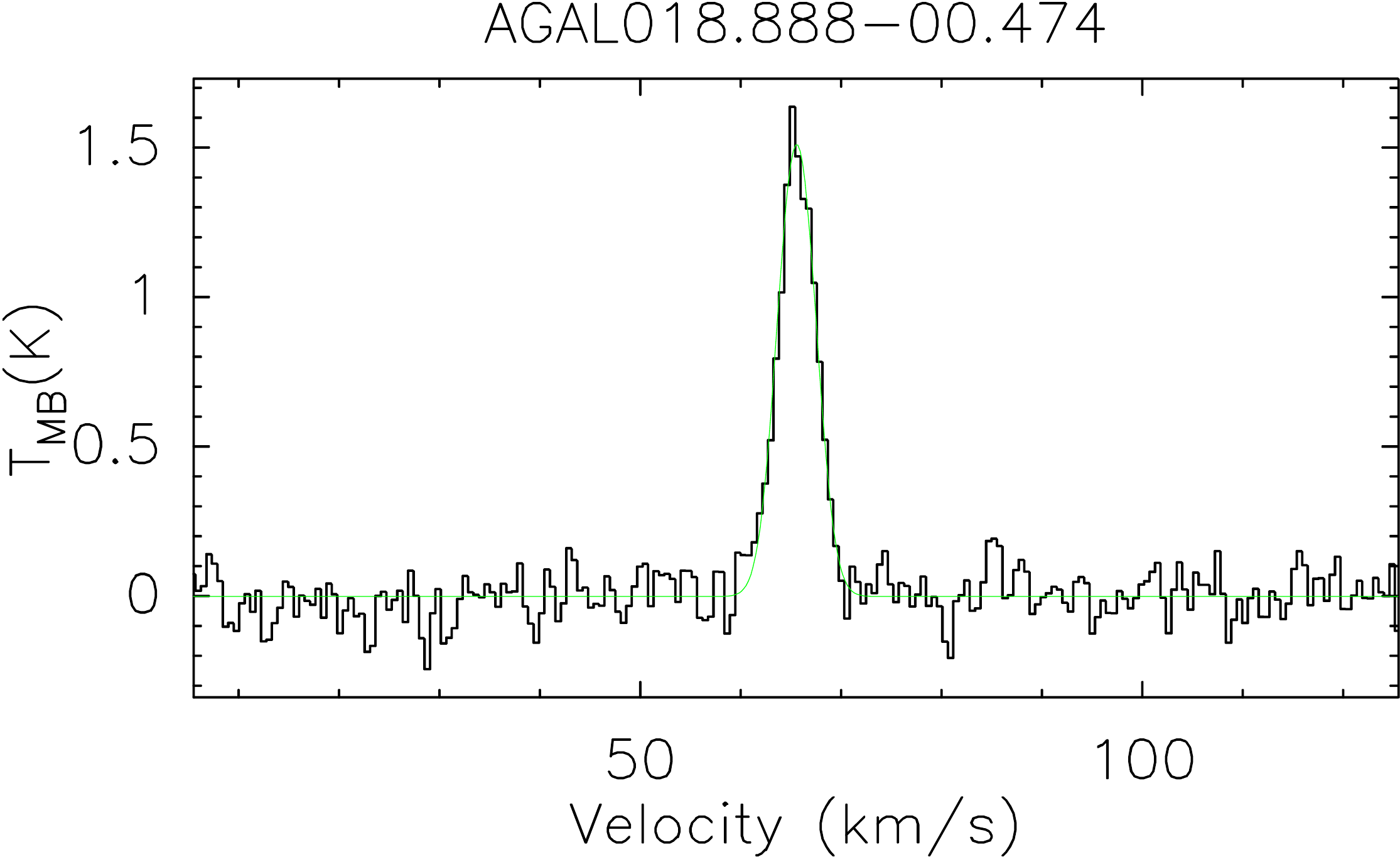}
\includegraphics[angle=-90,width=0.3\textwidth]{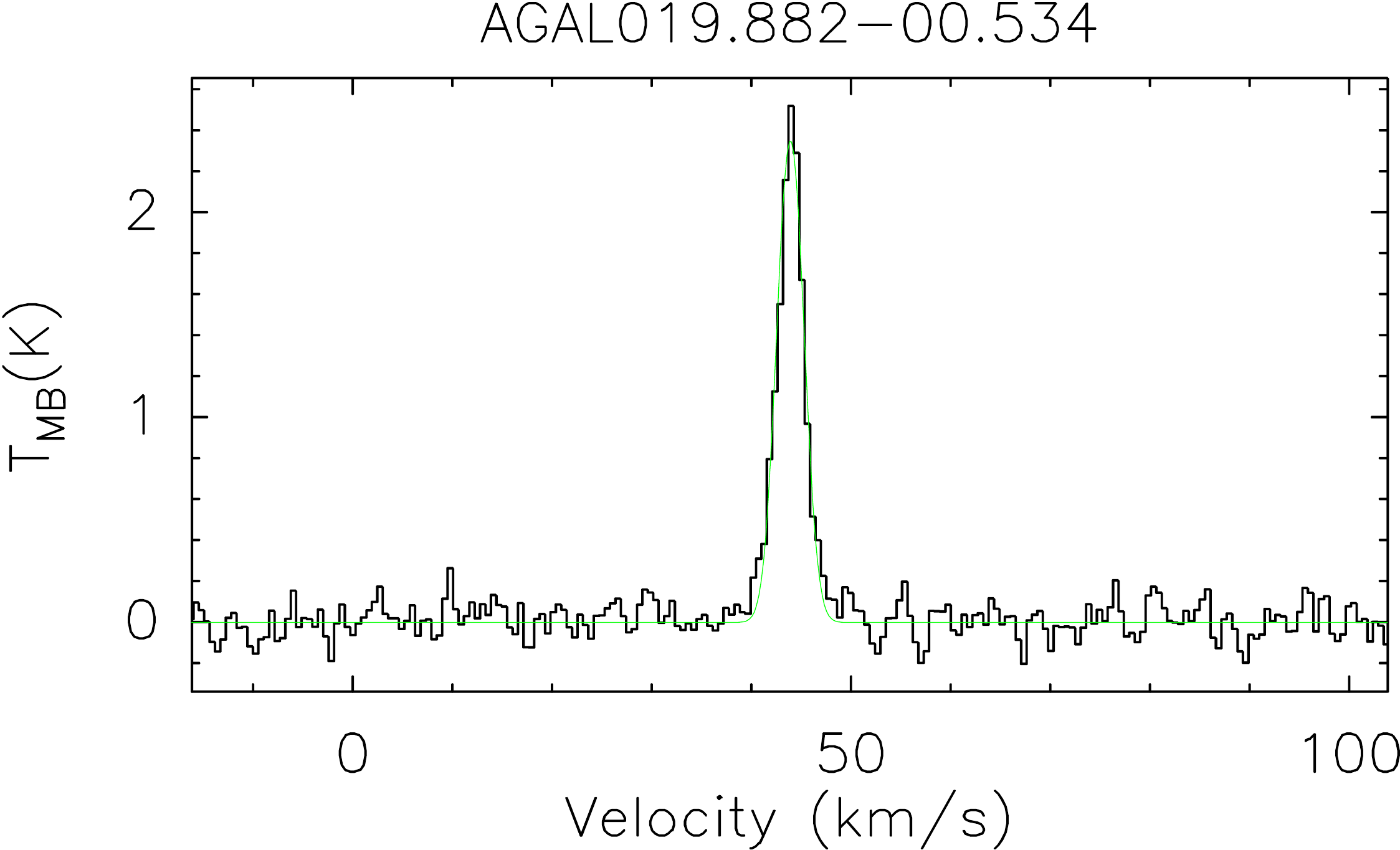}
\includegraphics[angle=-90,width=0.3\textwidth]{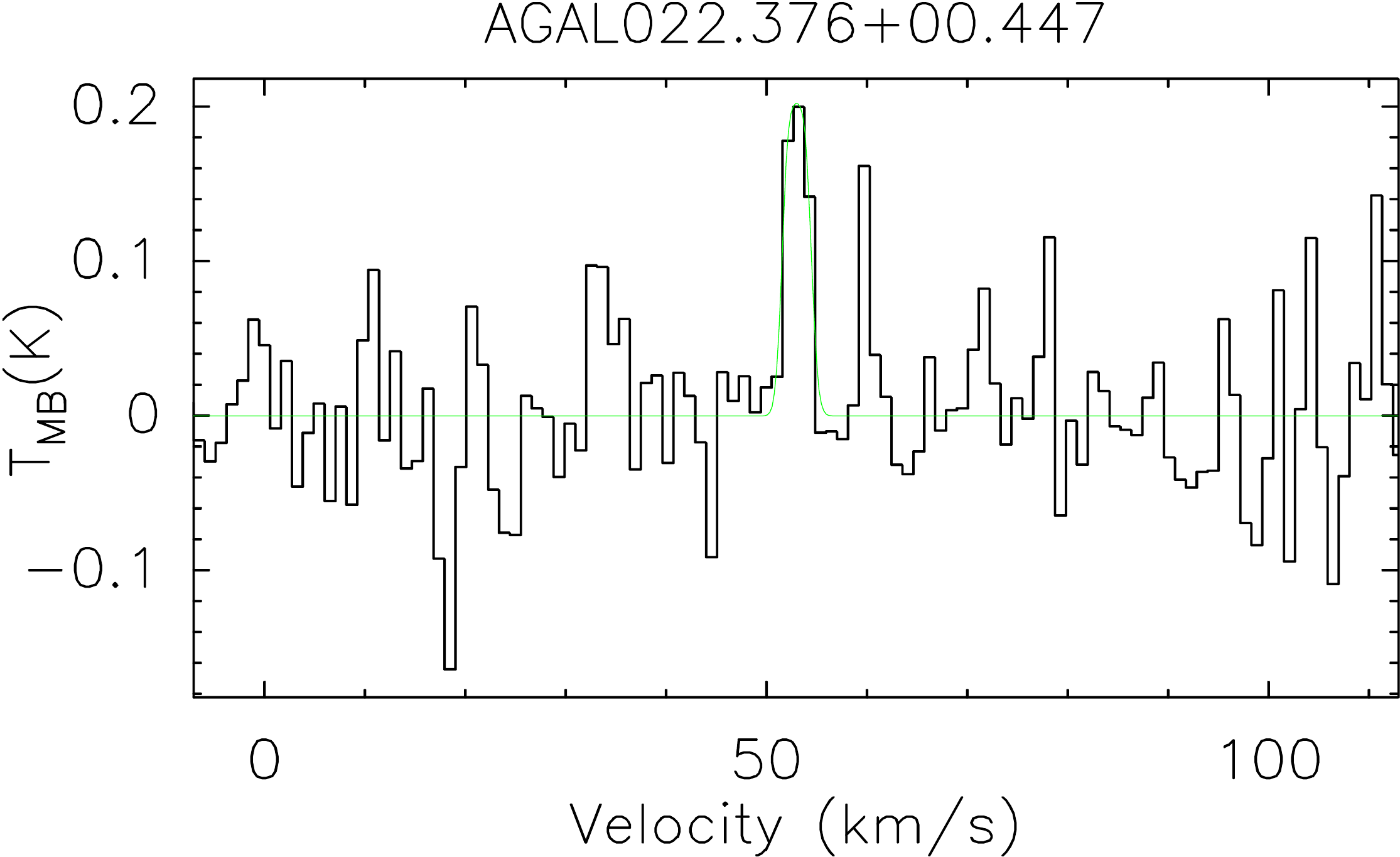} \hfill
\caption{C$^{17}$O$(3-2)$ for subsample S1. The fit is shown in green.} \label{fig:spectra_32_A}
\end{figure*} 

\begin{figure*} 
\ContinuedFloat
\centering 
\includegraphics[angle=-90,width=0.3\textwidth]{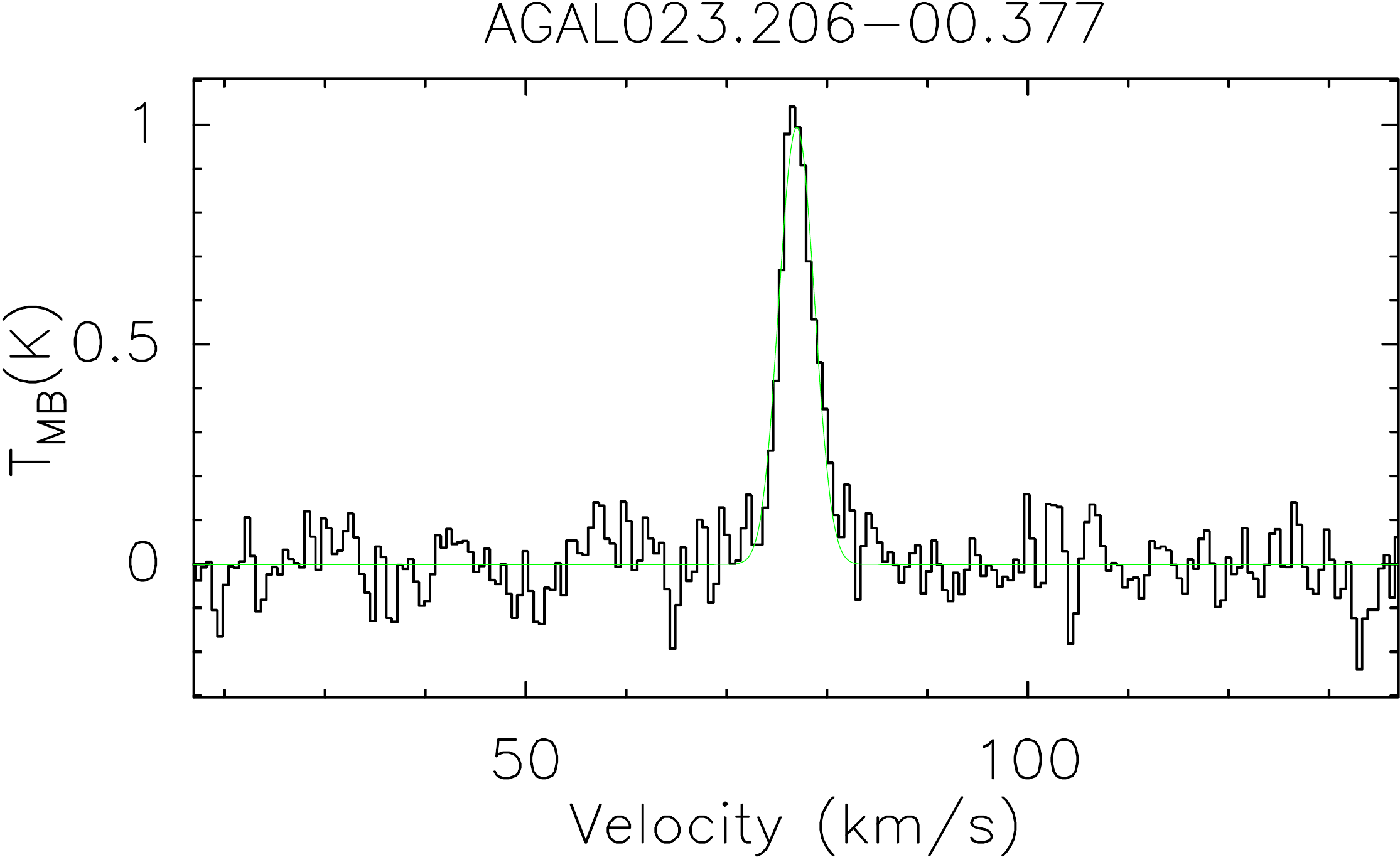}
\includegraphics[angle=-90,width=0.3\textwidth]{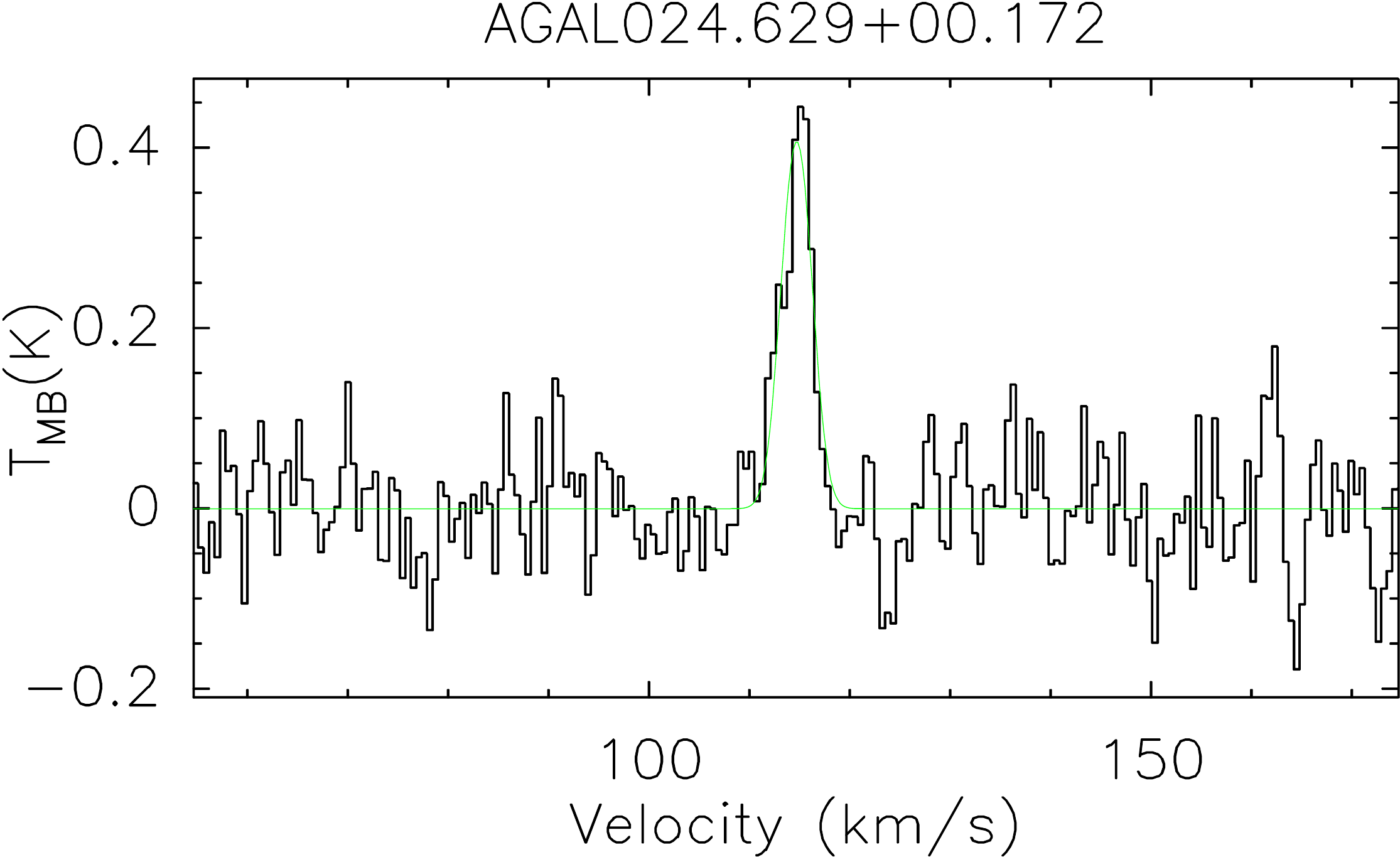}
\includegraphics[angle=-90,width=0.3\textwidth]{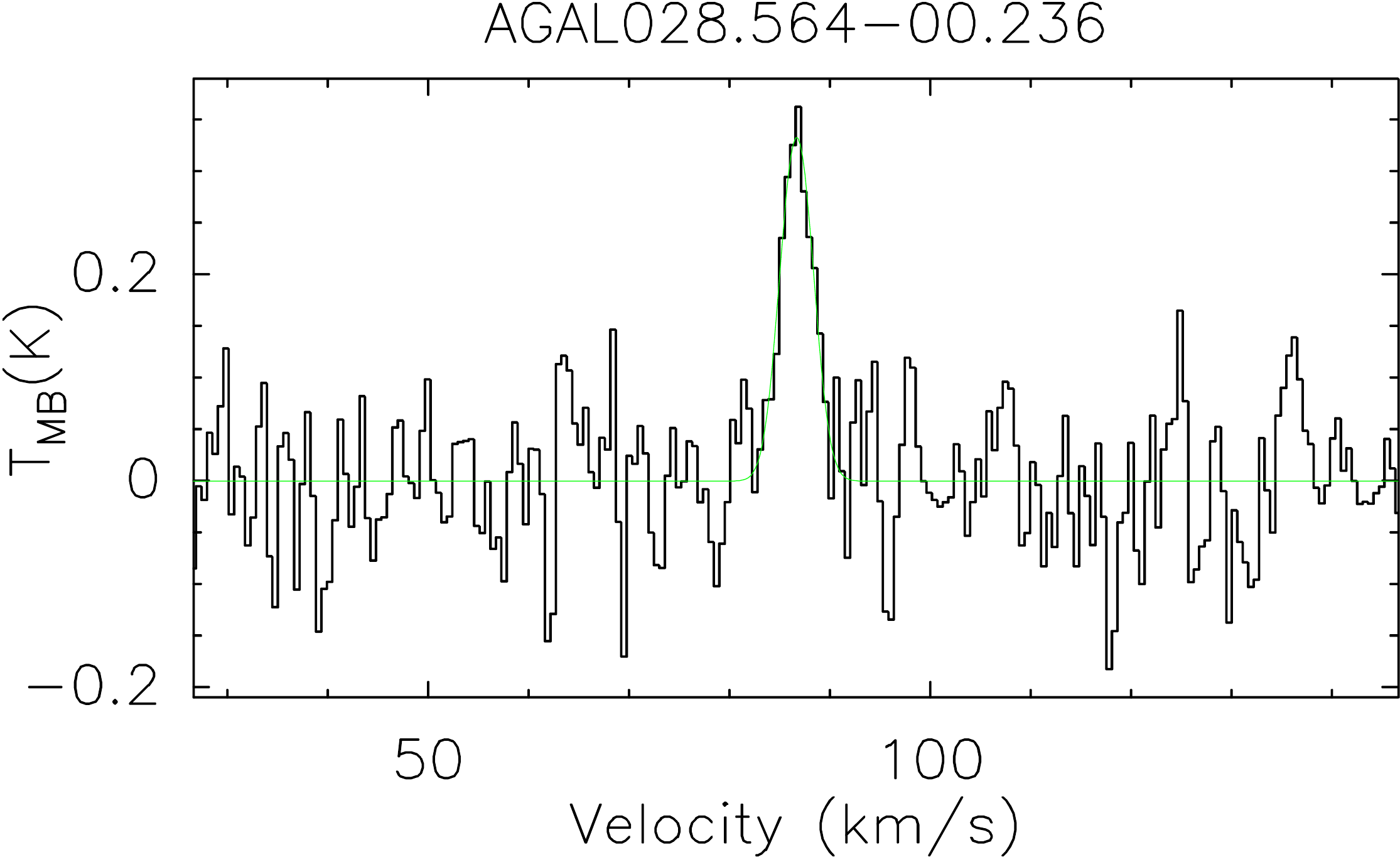} \\
\includegraphics[angle=-90,width=0.3\textwidth]{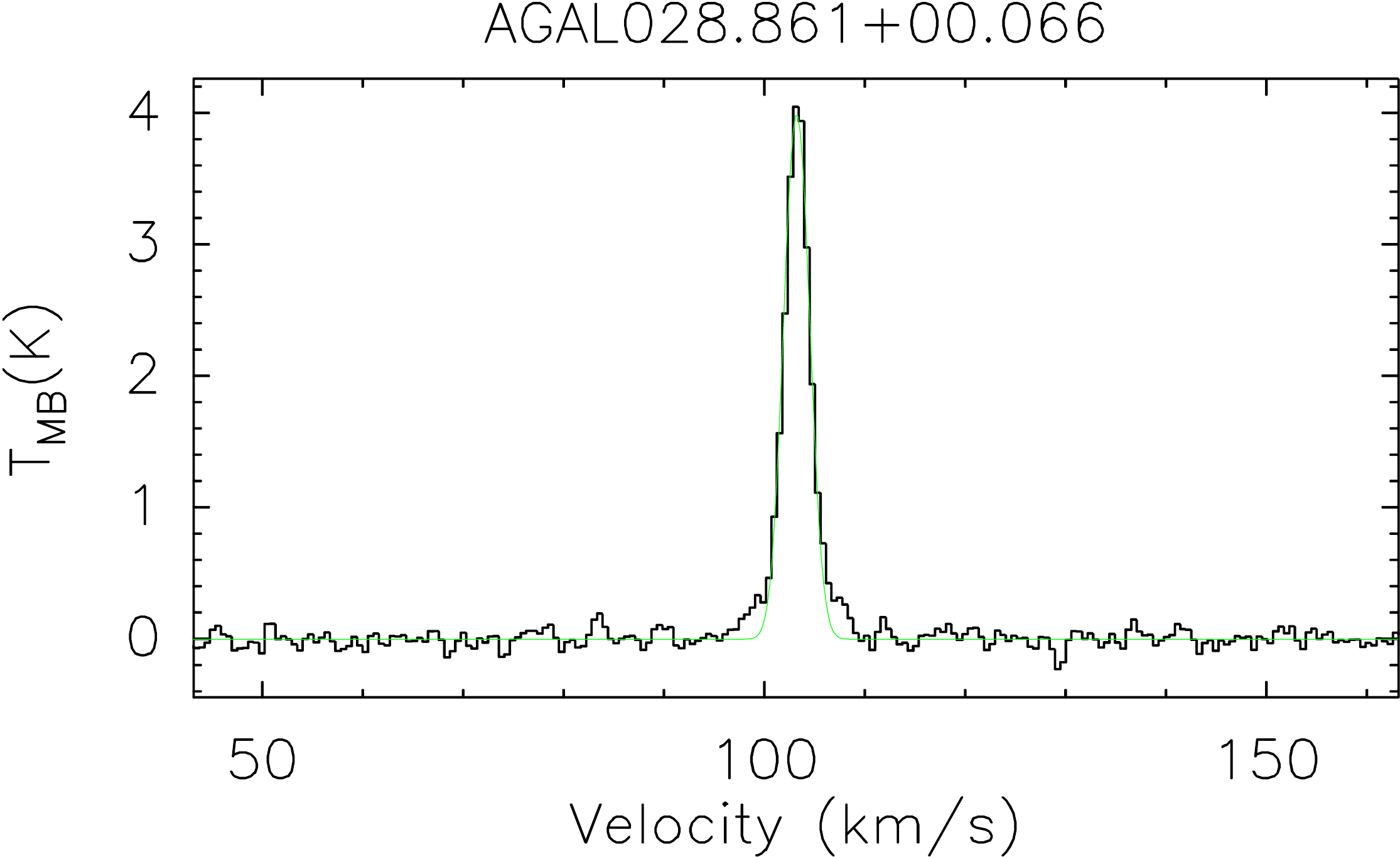}
\includegraphics[angle=-90,width=0.3\textwidth]{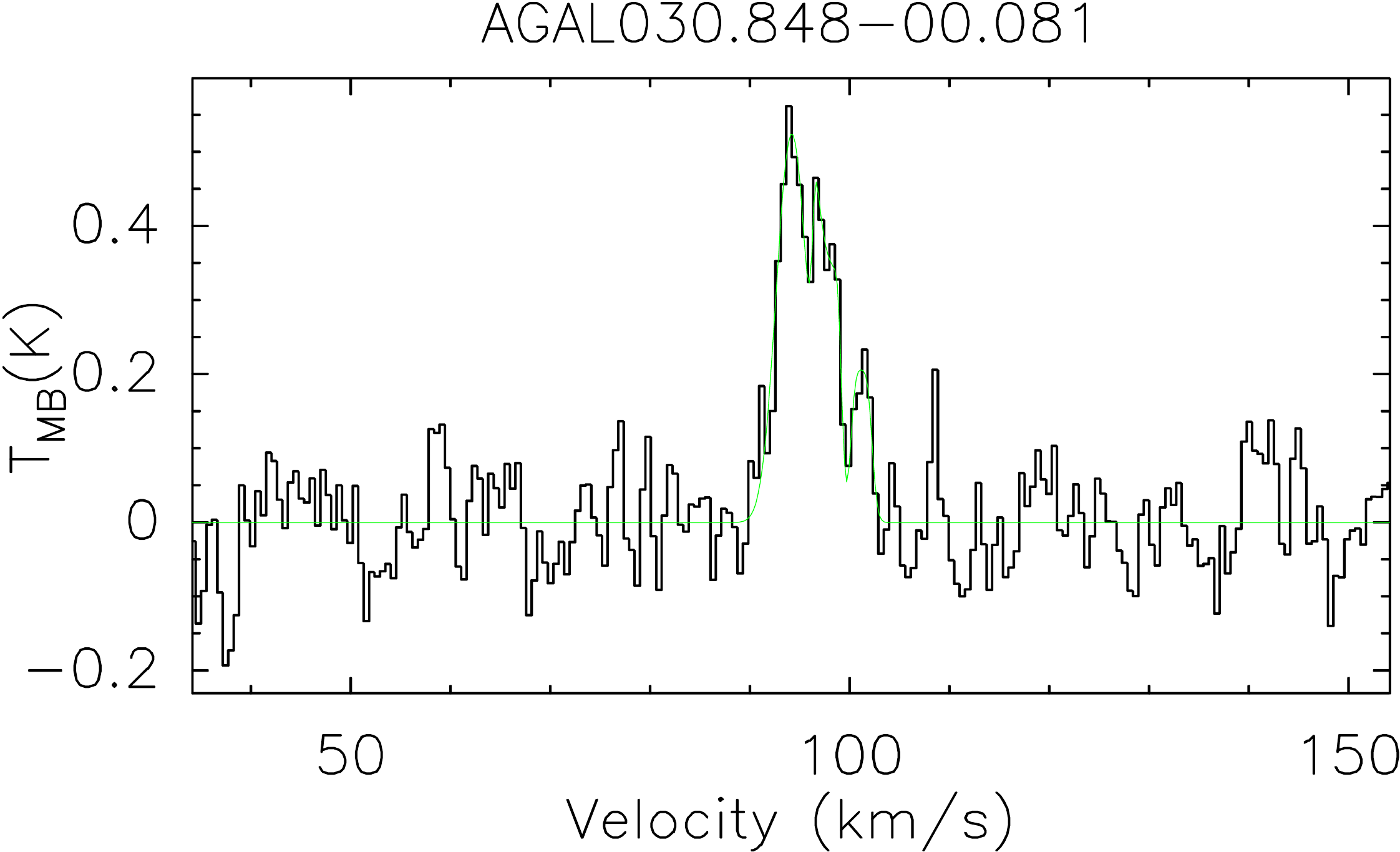}
\includegraphics[angle=-90,width=0.3\textwidth]{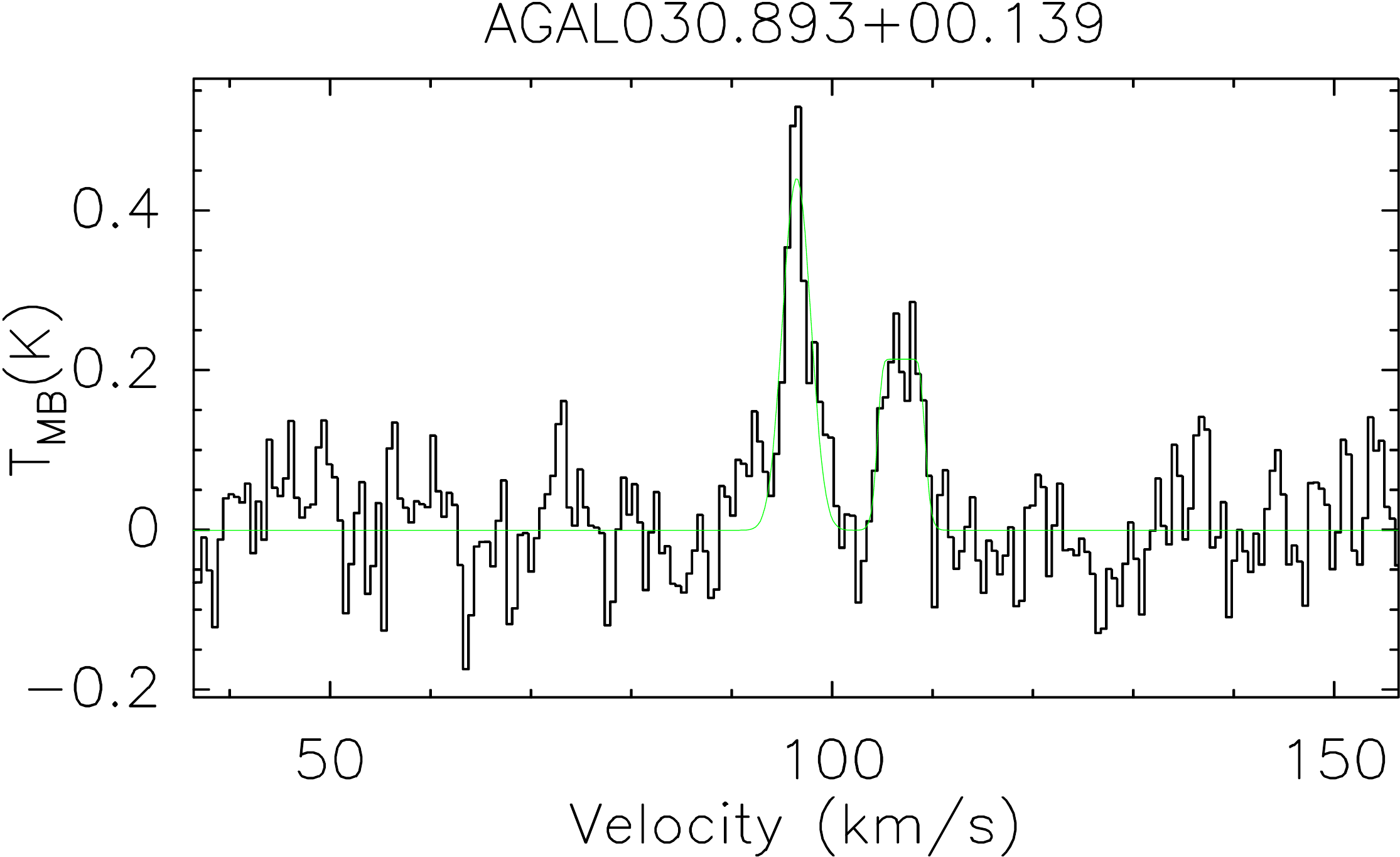} \\
\includegraphics[angle=-90,width=0.3\textwidth]{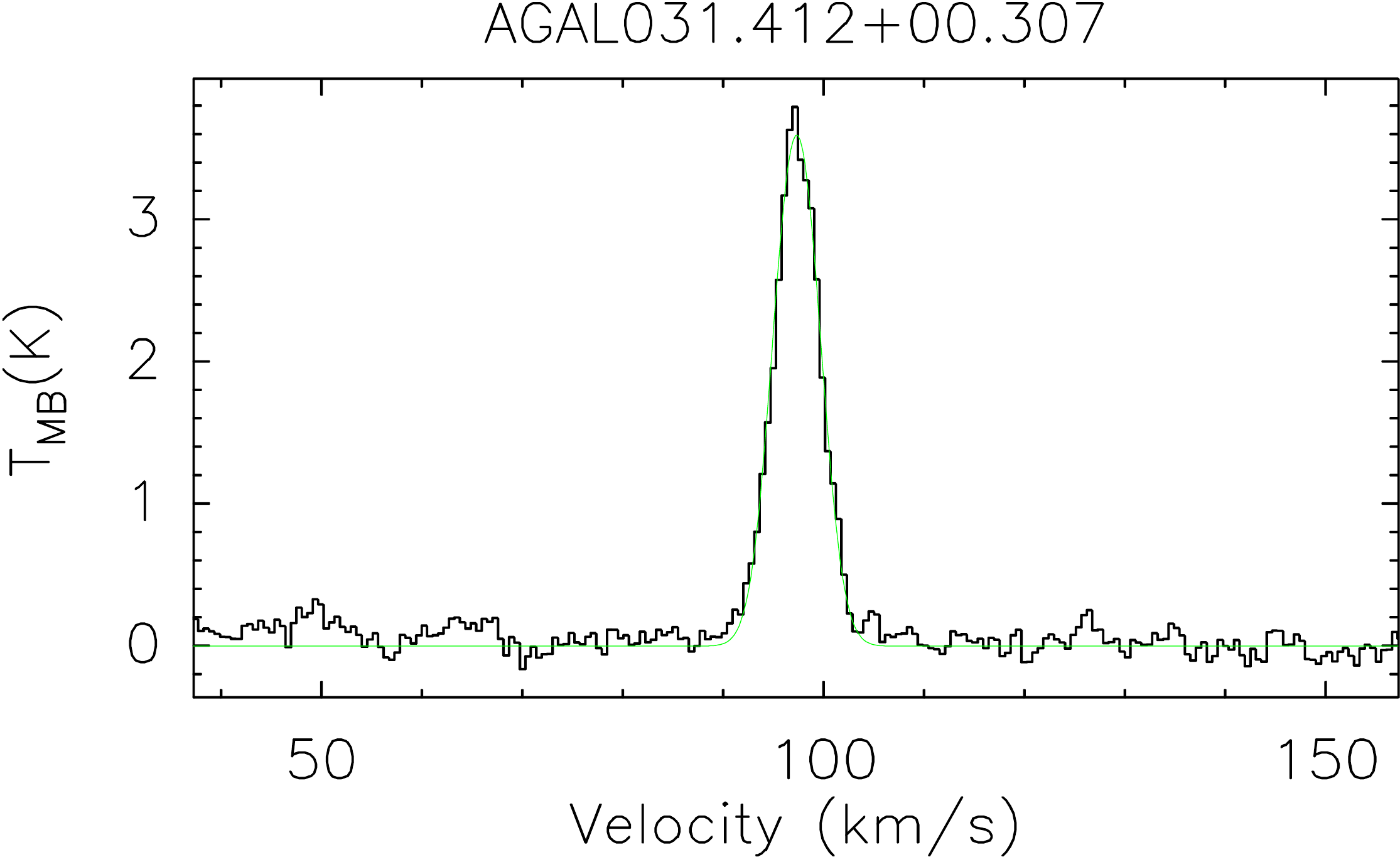}
\includegraphics[angle=-90,width=0.3\textwidth]{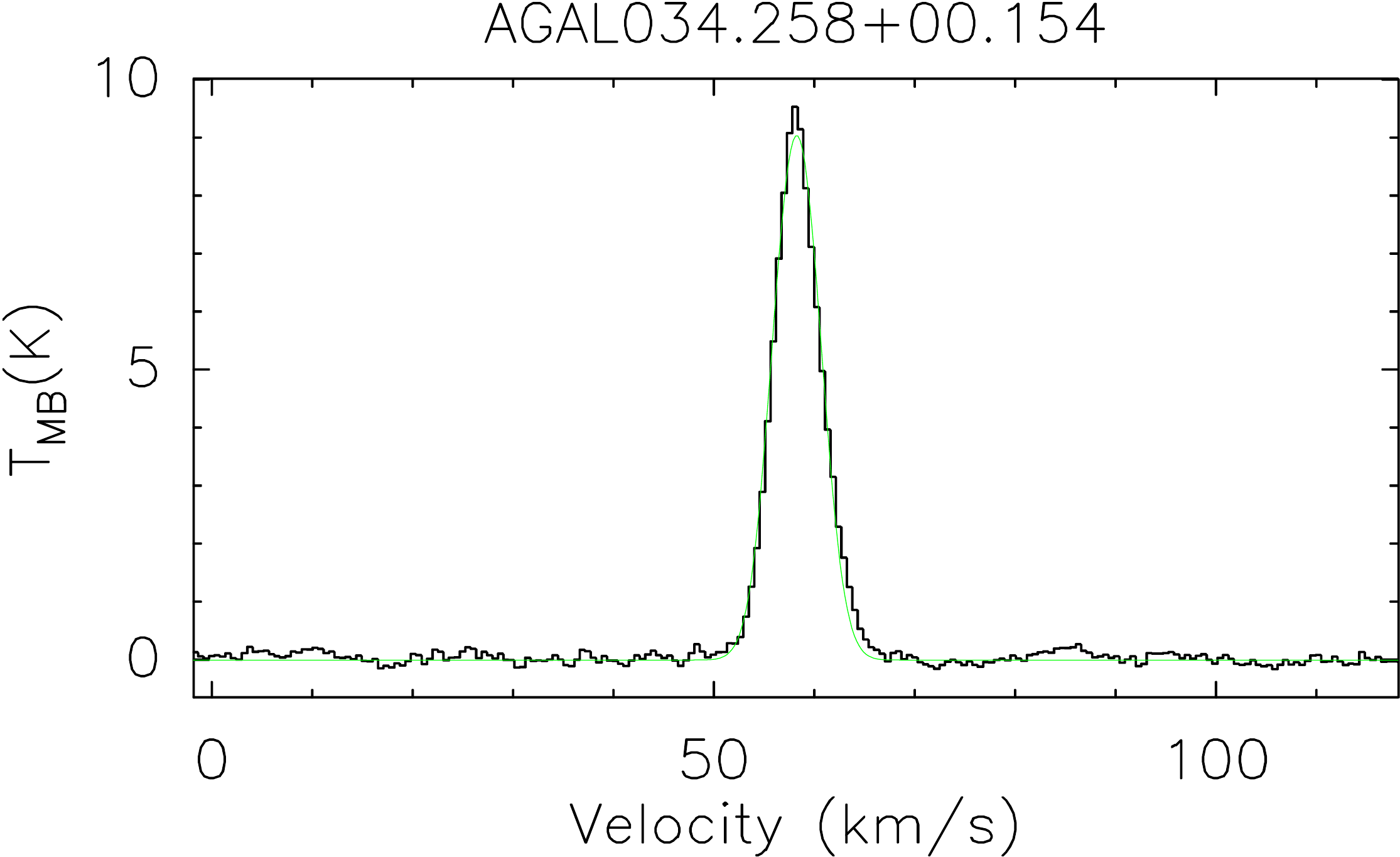}
\includegraphics[angle=-90,width=0.3\textwidth]{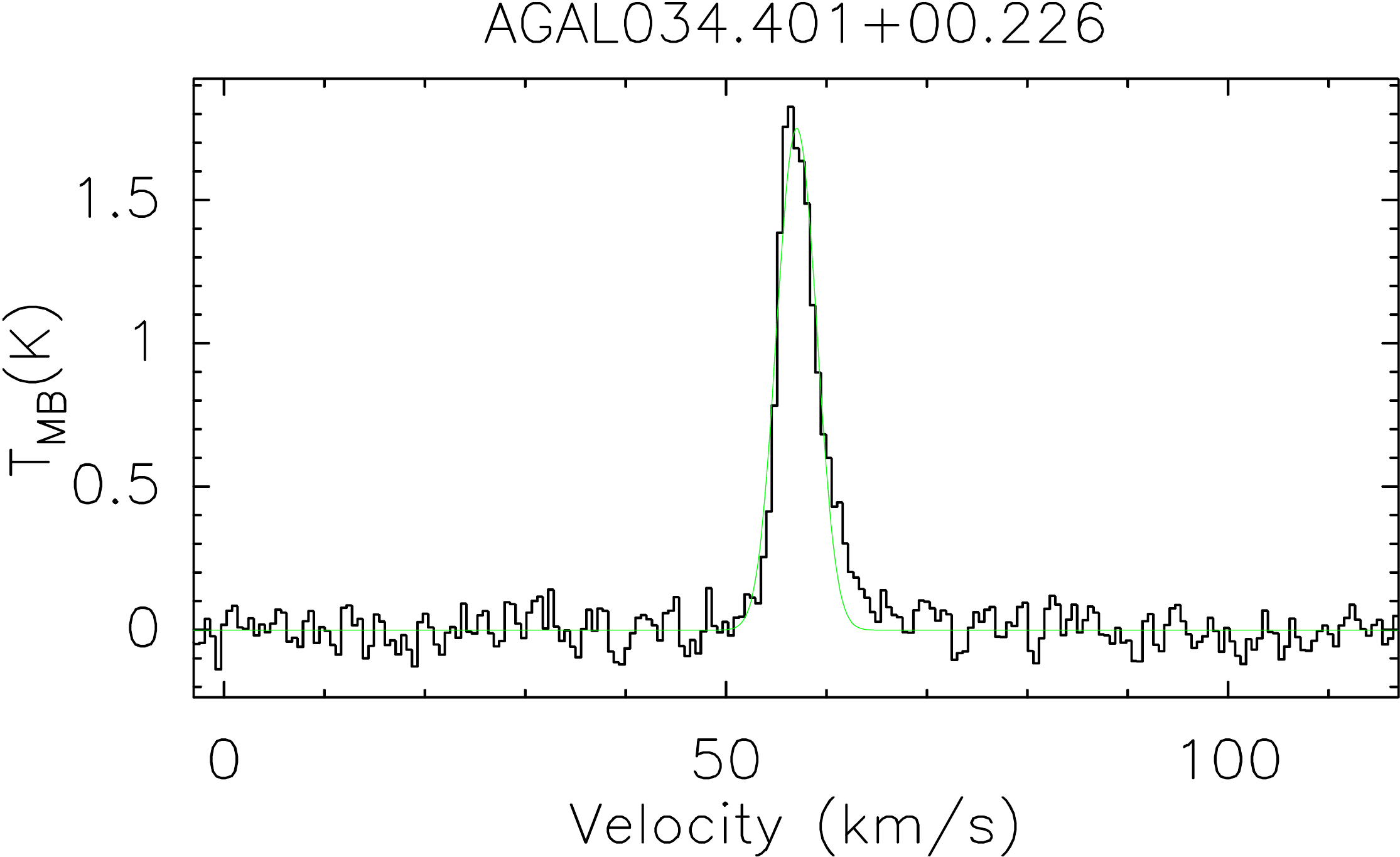} \\
\includegraphics[angle=-90,width=0.3\textwidth]{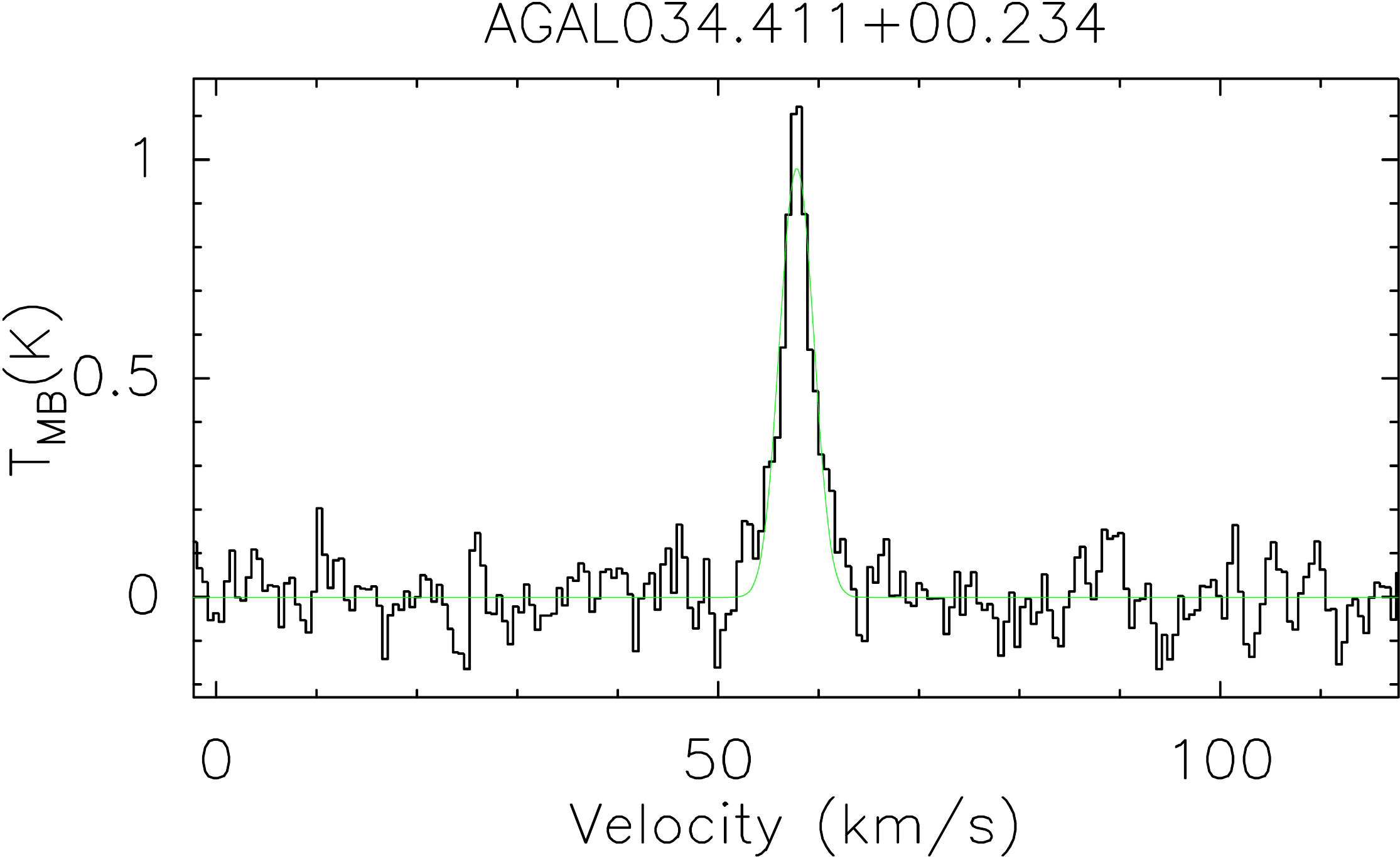}
\includegraphics[angle=-90,width=0.3\textwidth]{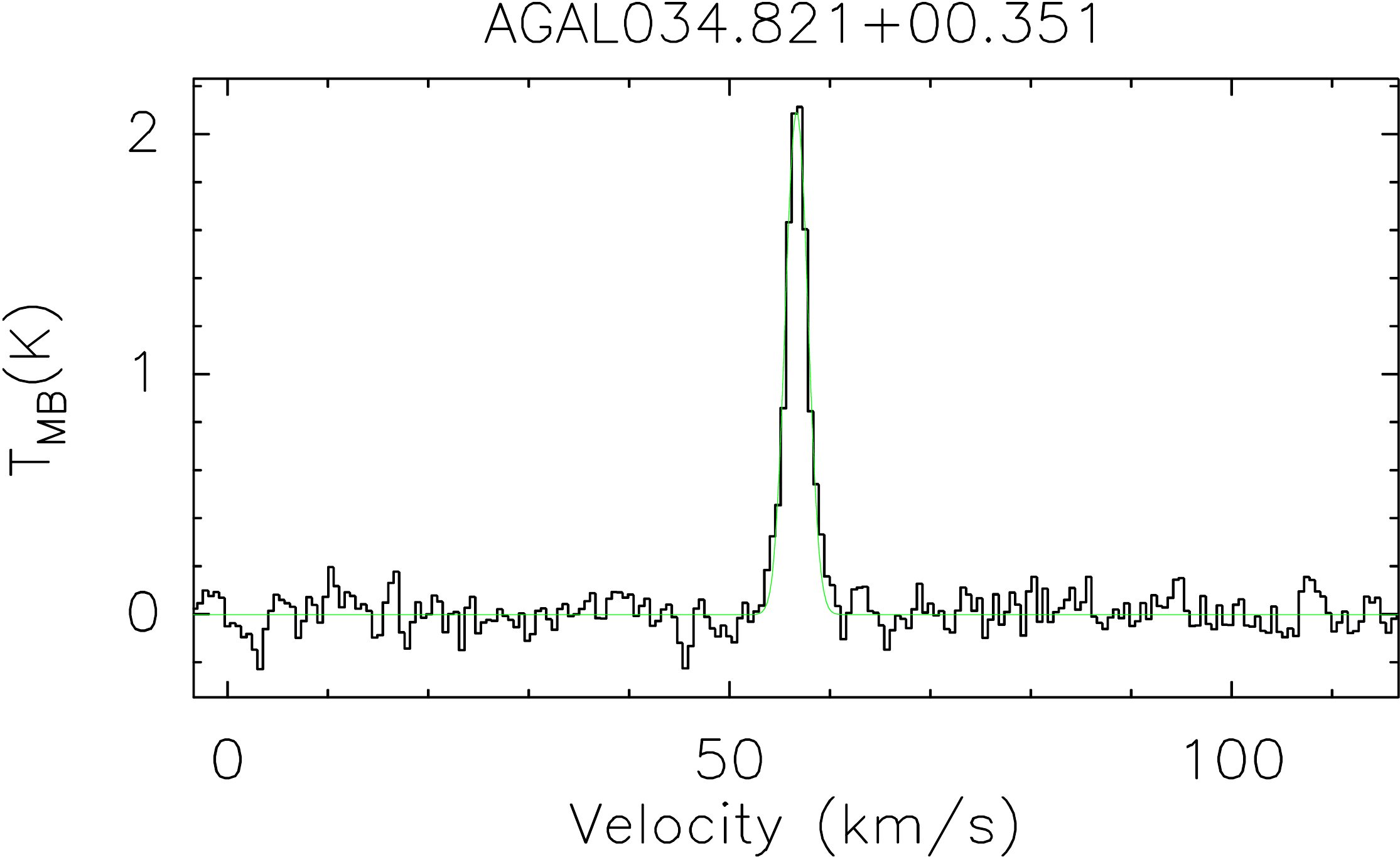}
\includegraphics[angle=-90,width=0.3\textwidth]{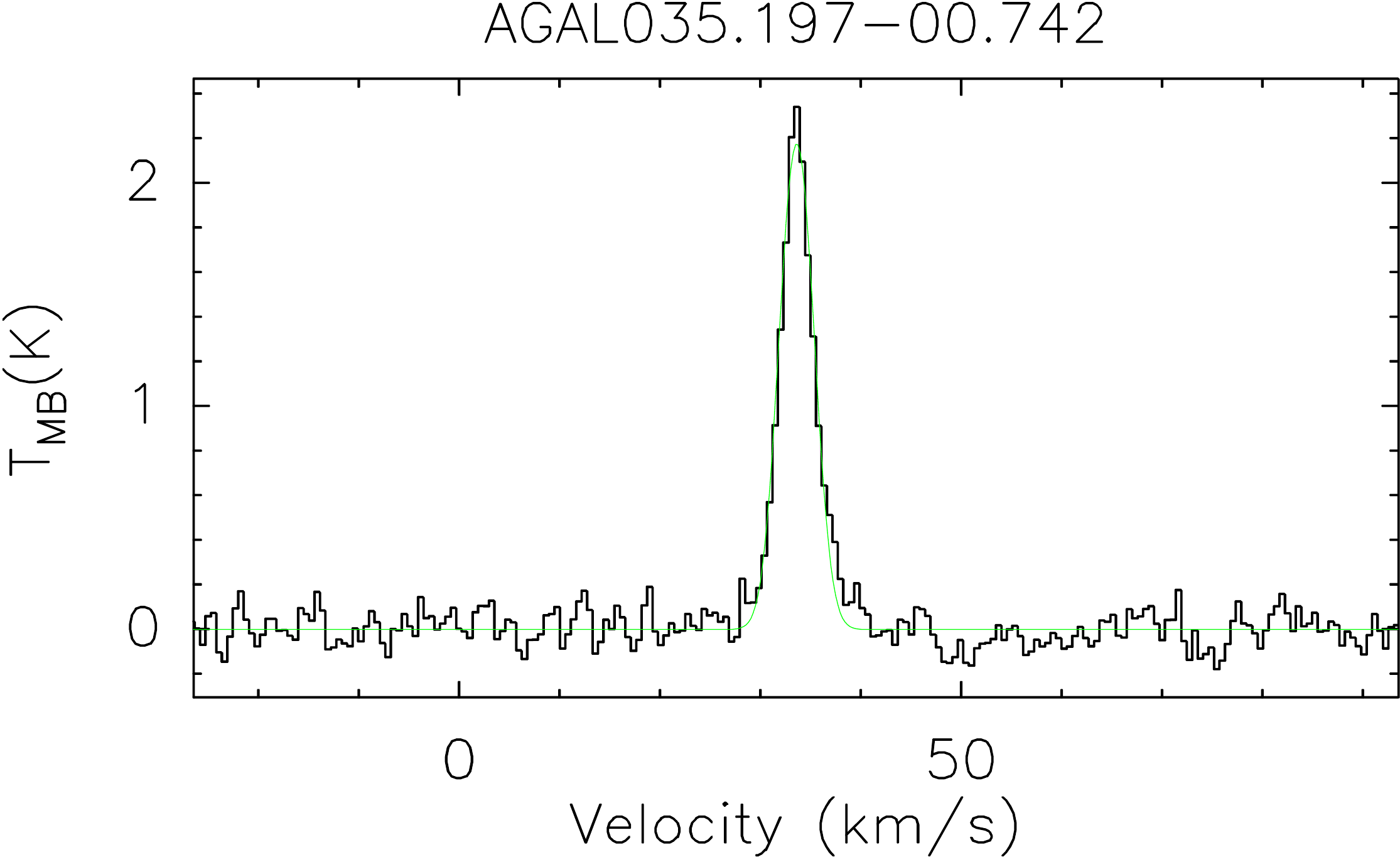} \\
\includegraphics[angle=-90,width=0.3\textwidth]{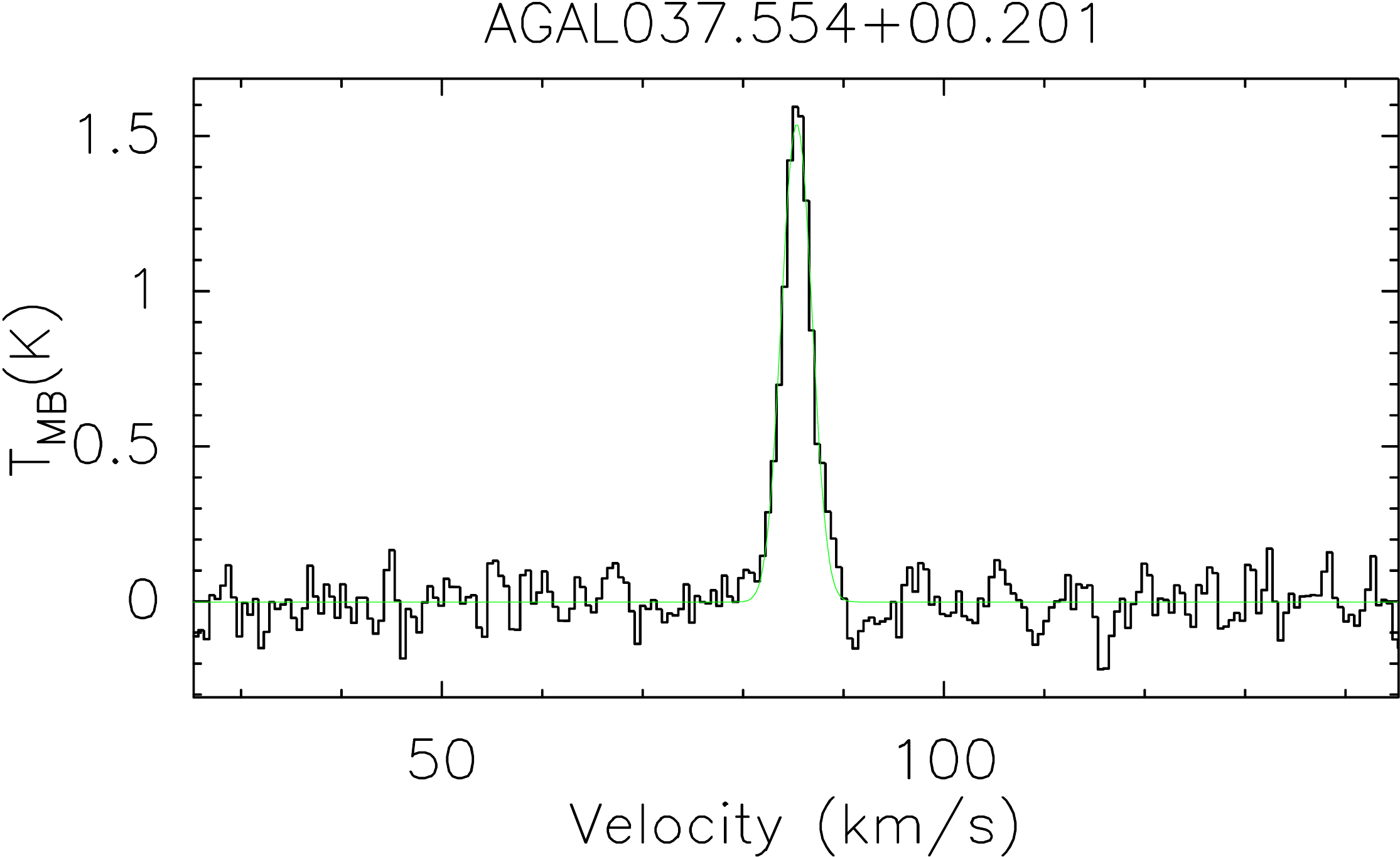}
\includegraphics[angle=-90,width=0.3\textwidth]{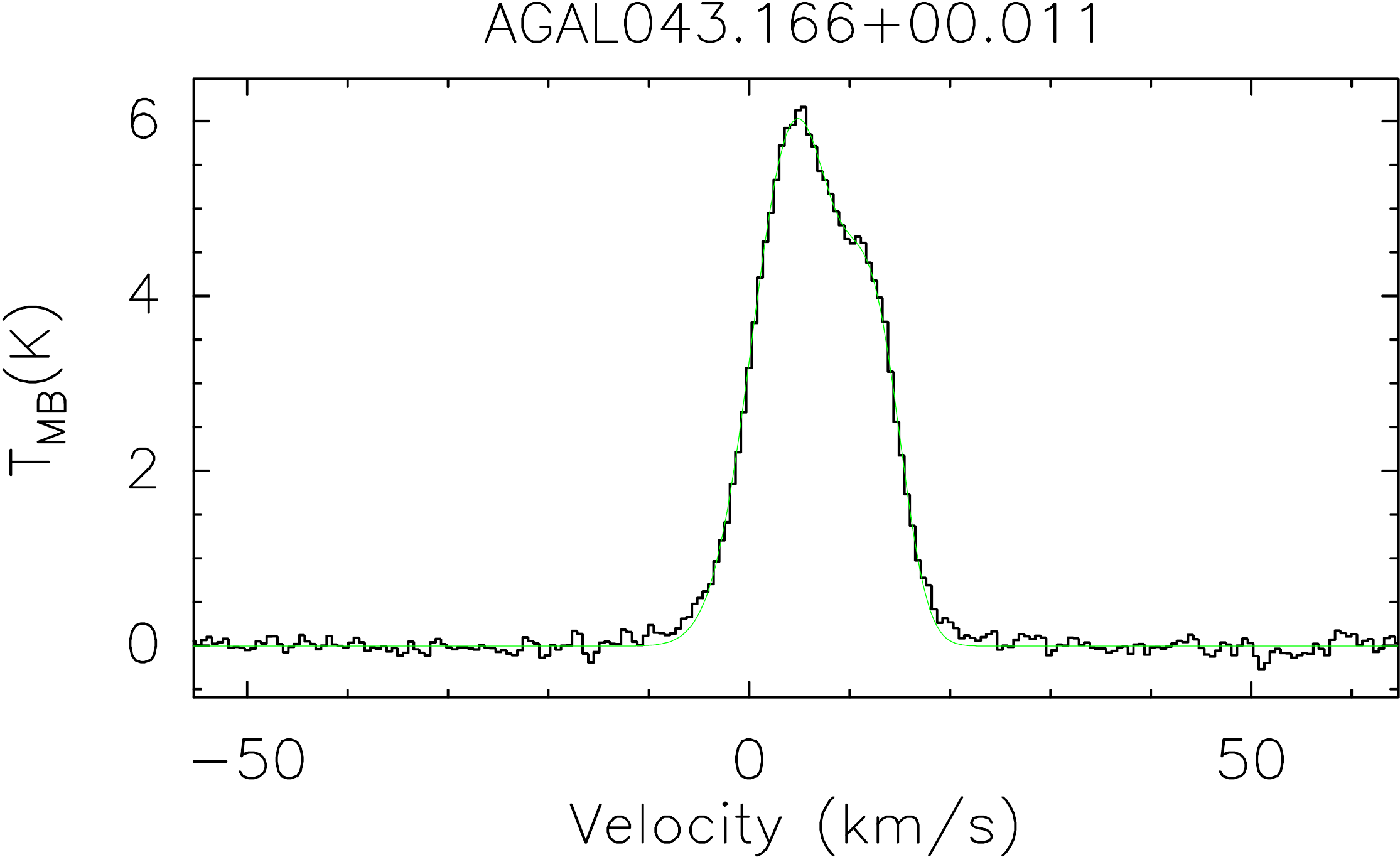}
\includegraphics[angle=-90,width=0.3\textwidth]{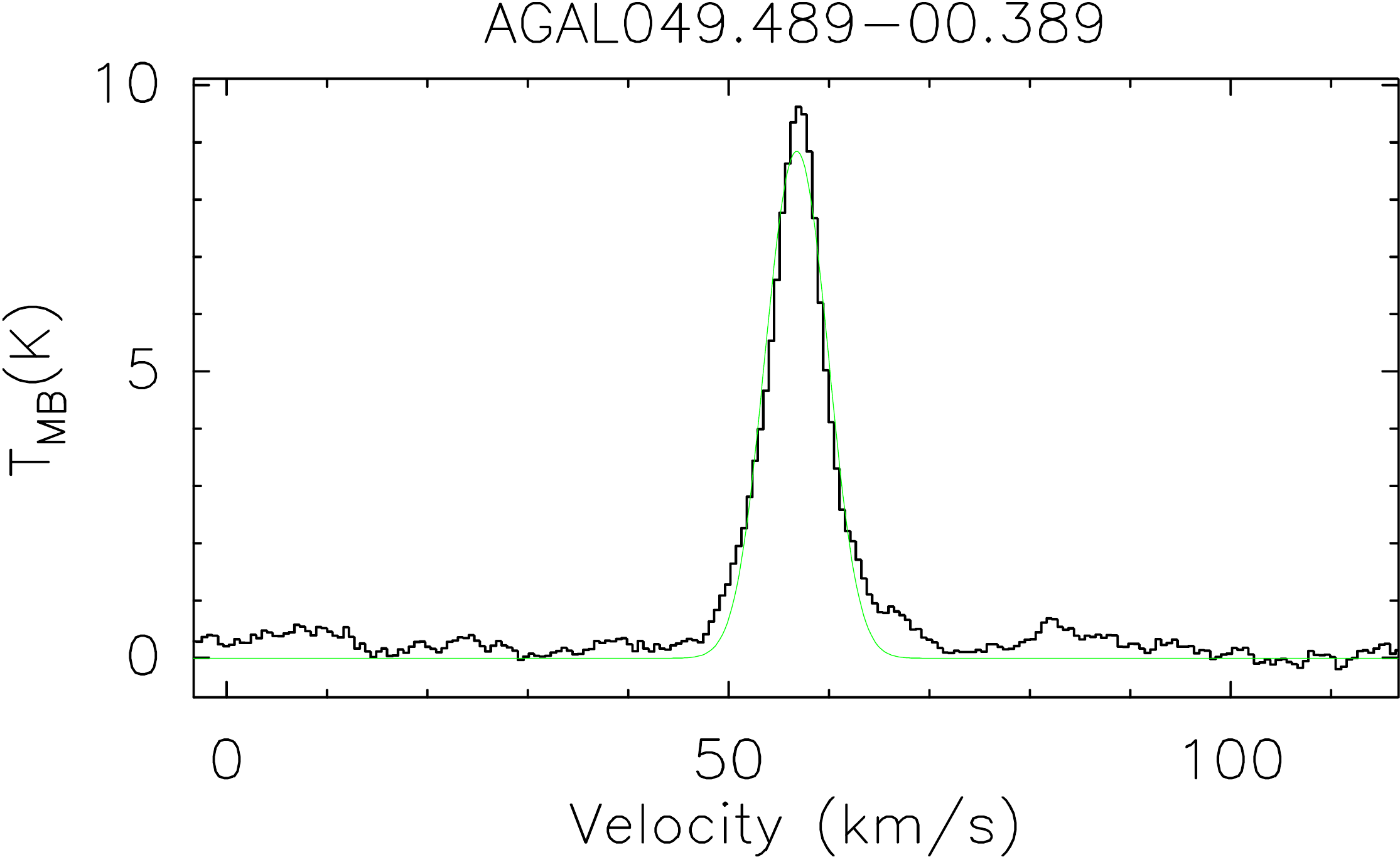} \\
\includegraphics[angle=-90,width=0.3\textwidth]{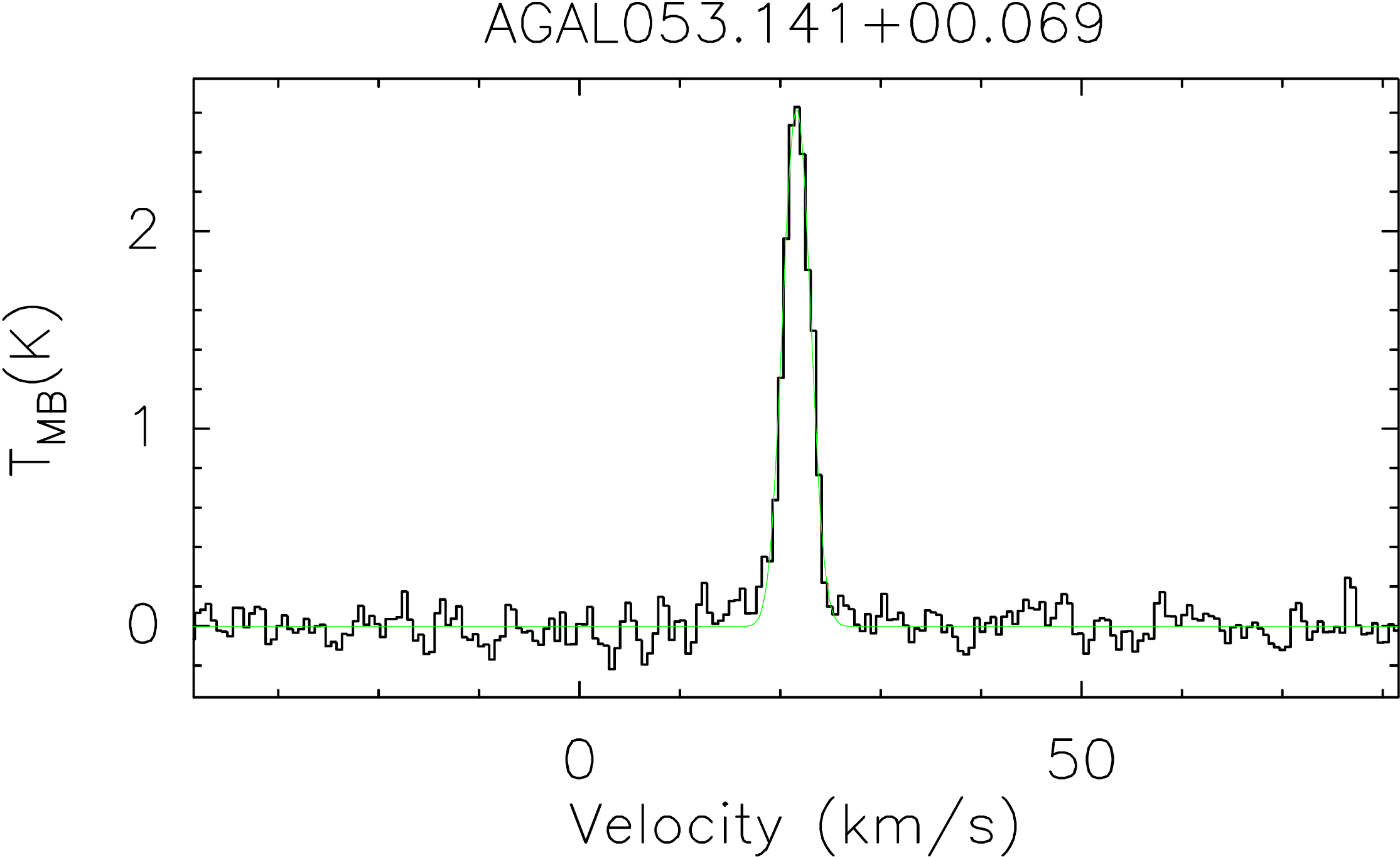}
\includegraphics[angle=-90,width=0.3\textwidth]{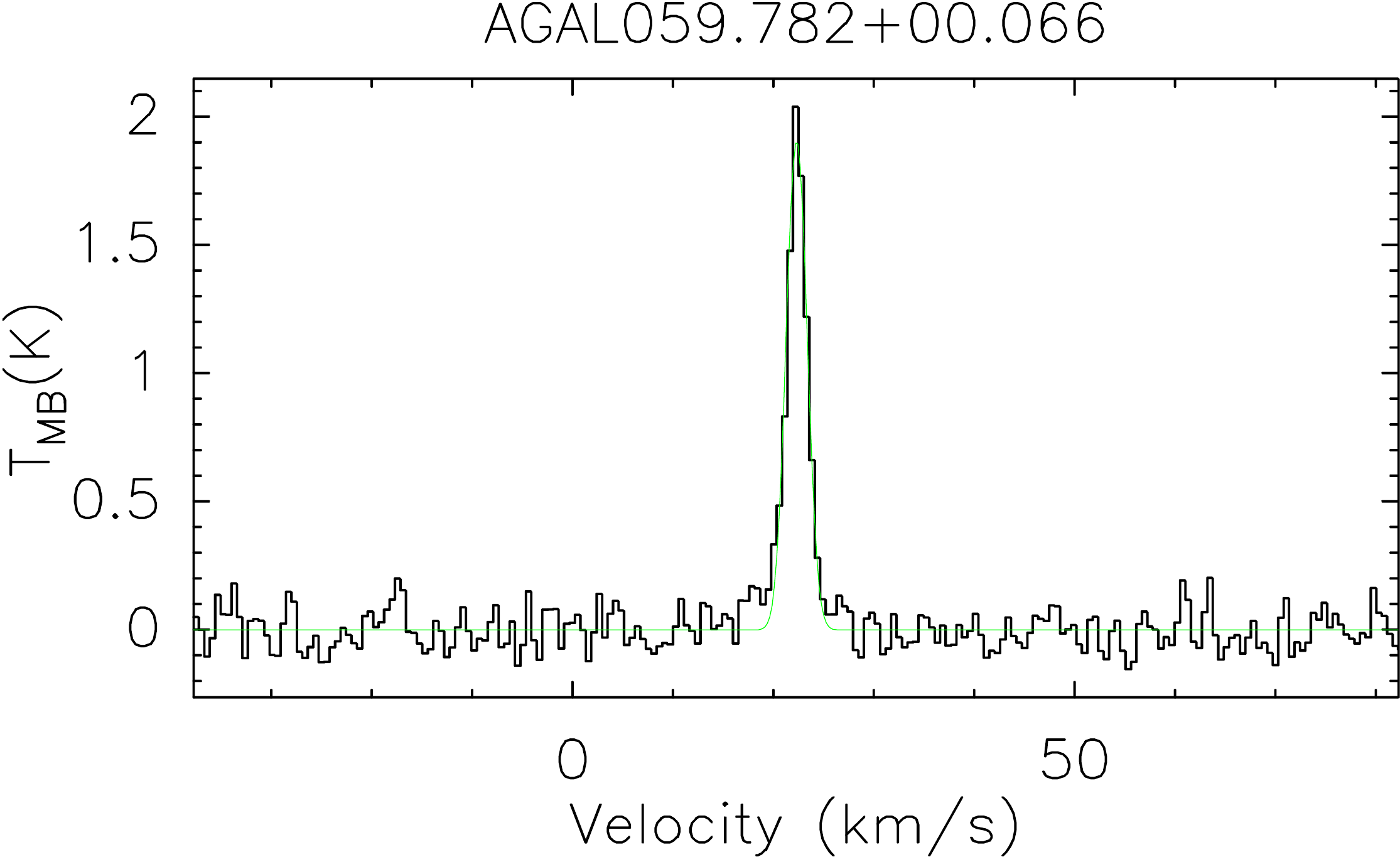} \hfill 
\caption{Continued.} 
\end{figure*} 
}
\onlfig{4}{
\begin{figure*} 
\centering 
\includegraphics[angle=-90,width=0.3\textwidth]{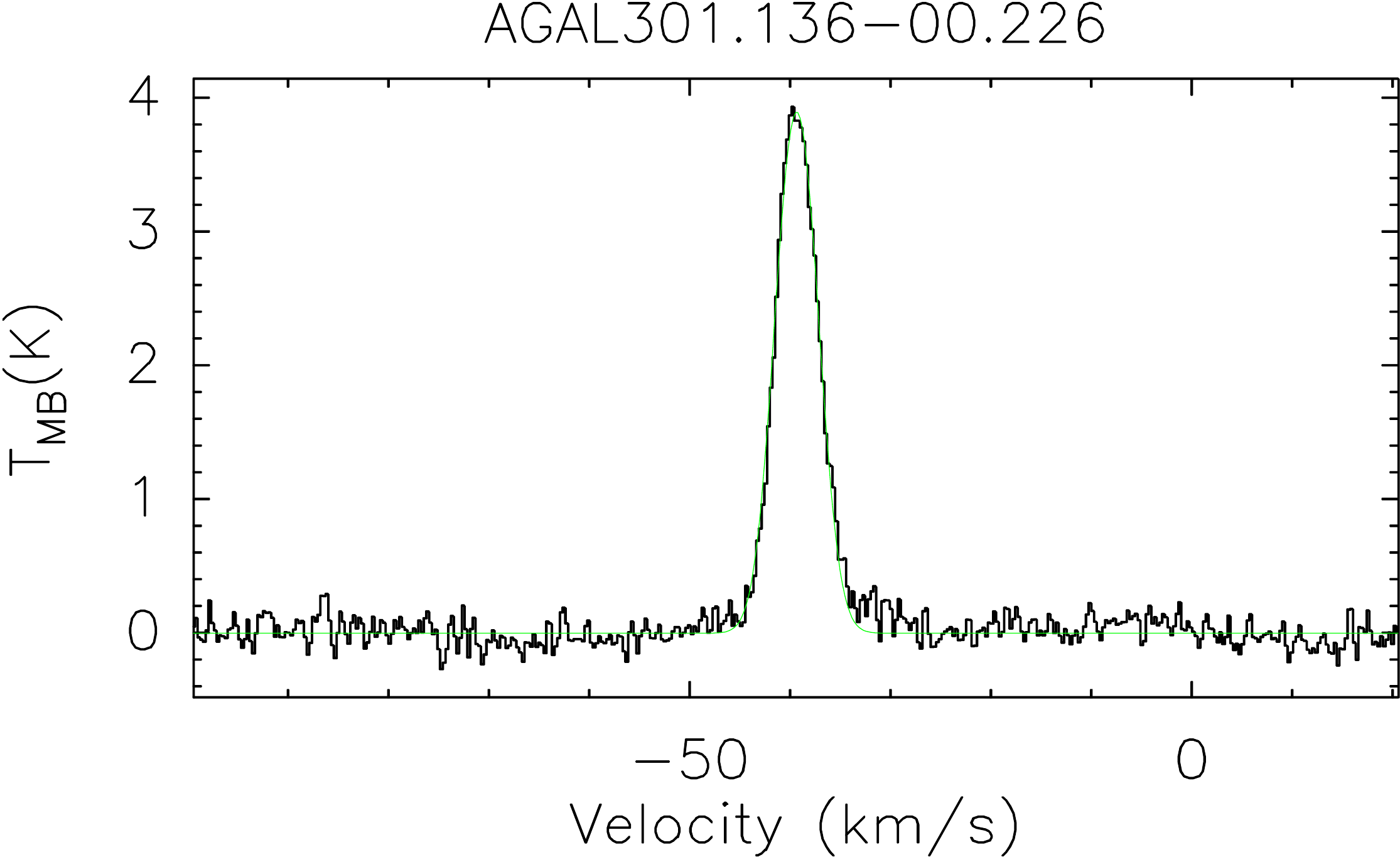}
\includegraphics[angle=-90,width=0.3\textwidth]{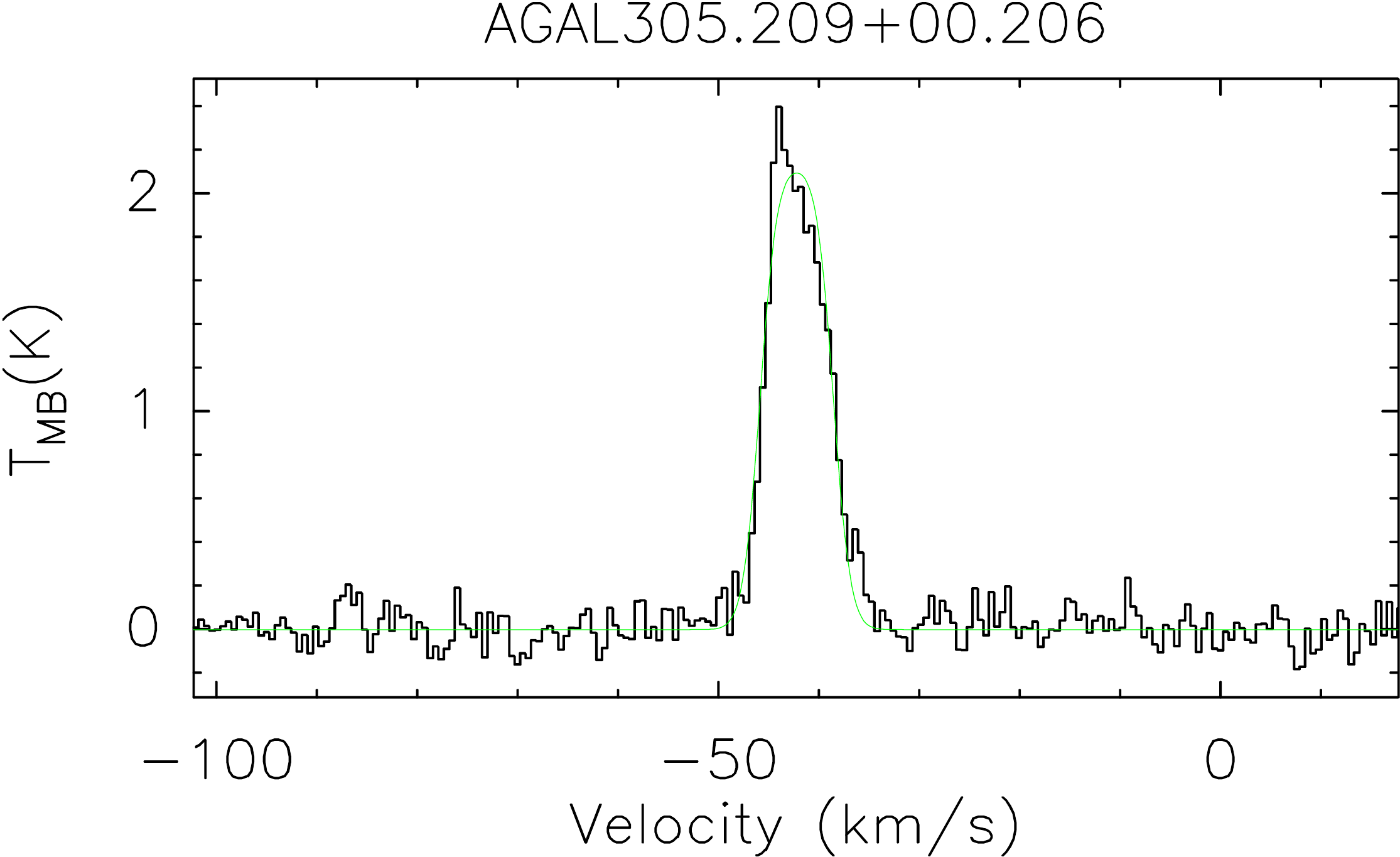}
\includegraphics[angle=-90,width=0.3\textwidth]{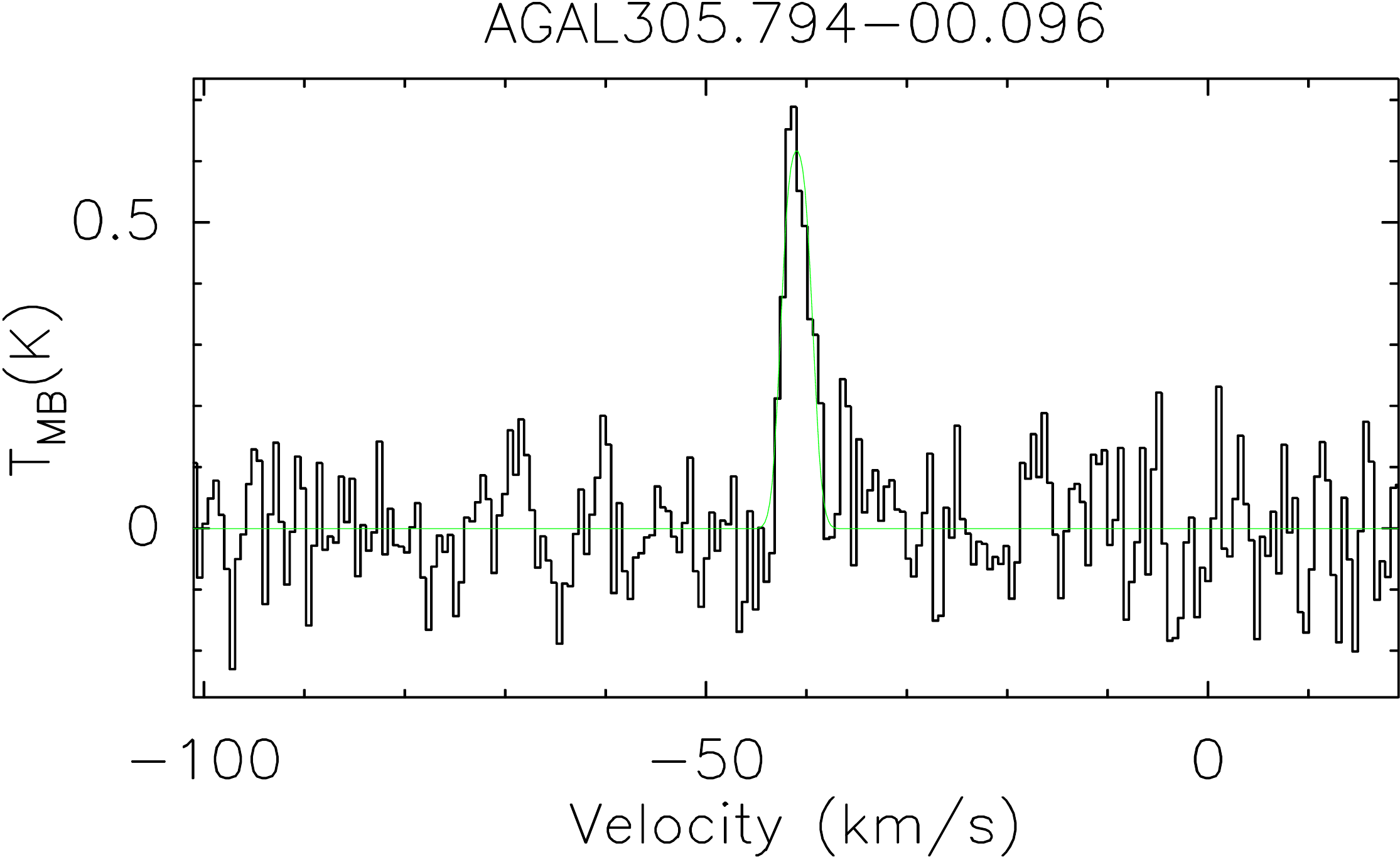} \\ 
\includegraphics[angle=-90,width=0.3\textwidth]{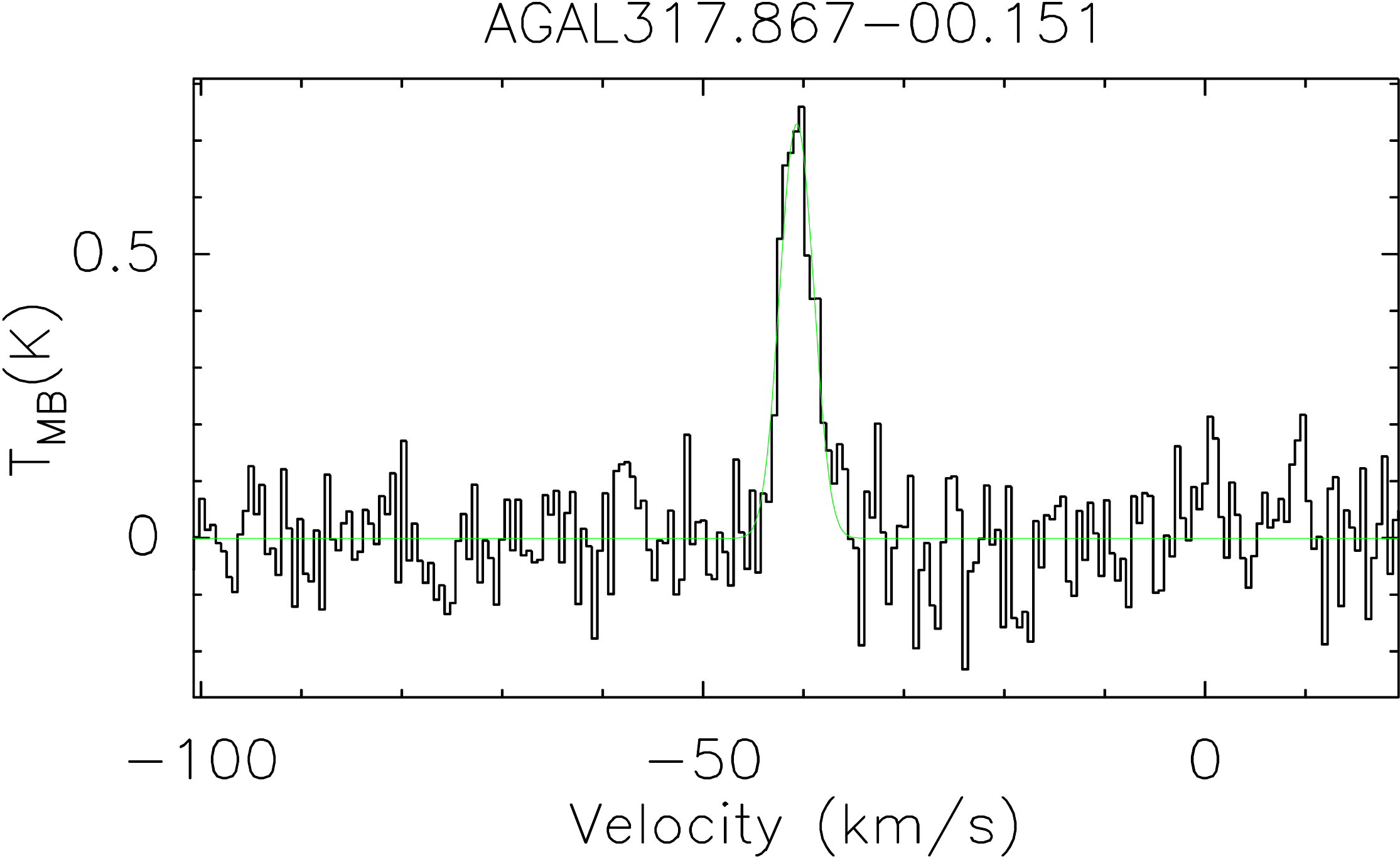} 
\includegraphics[angle=-90,width=0.3\textwidth]{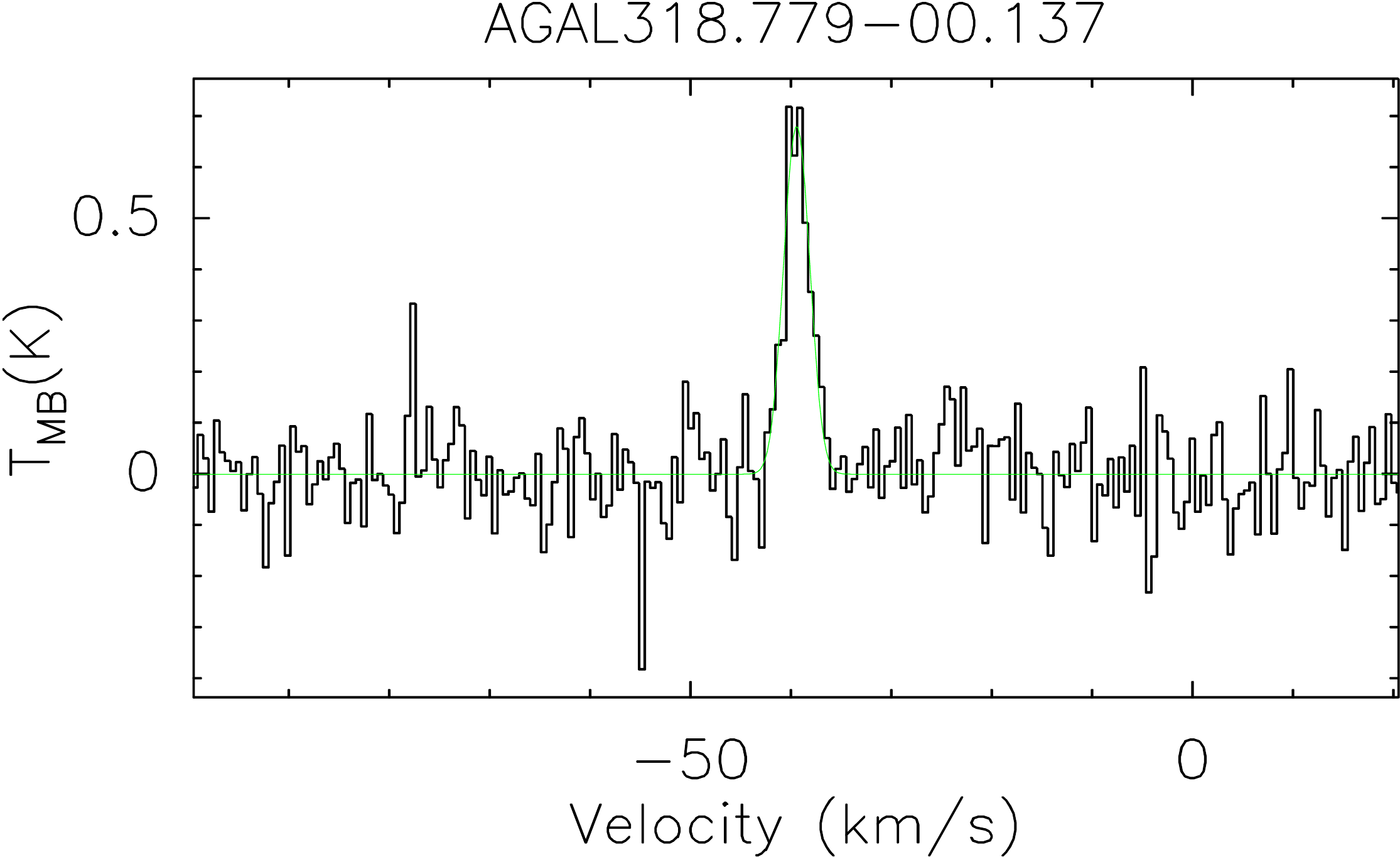}
\includegraphics[angle=-90,width=0.3\textwidth]{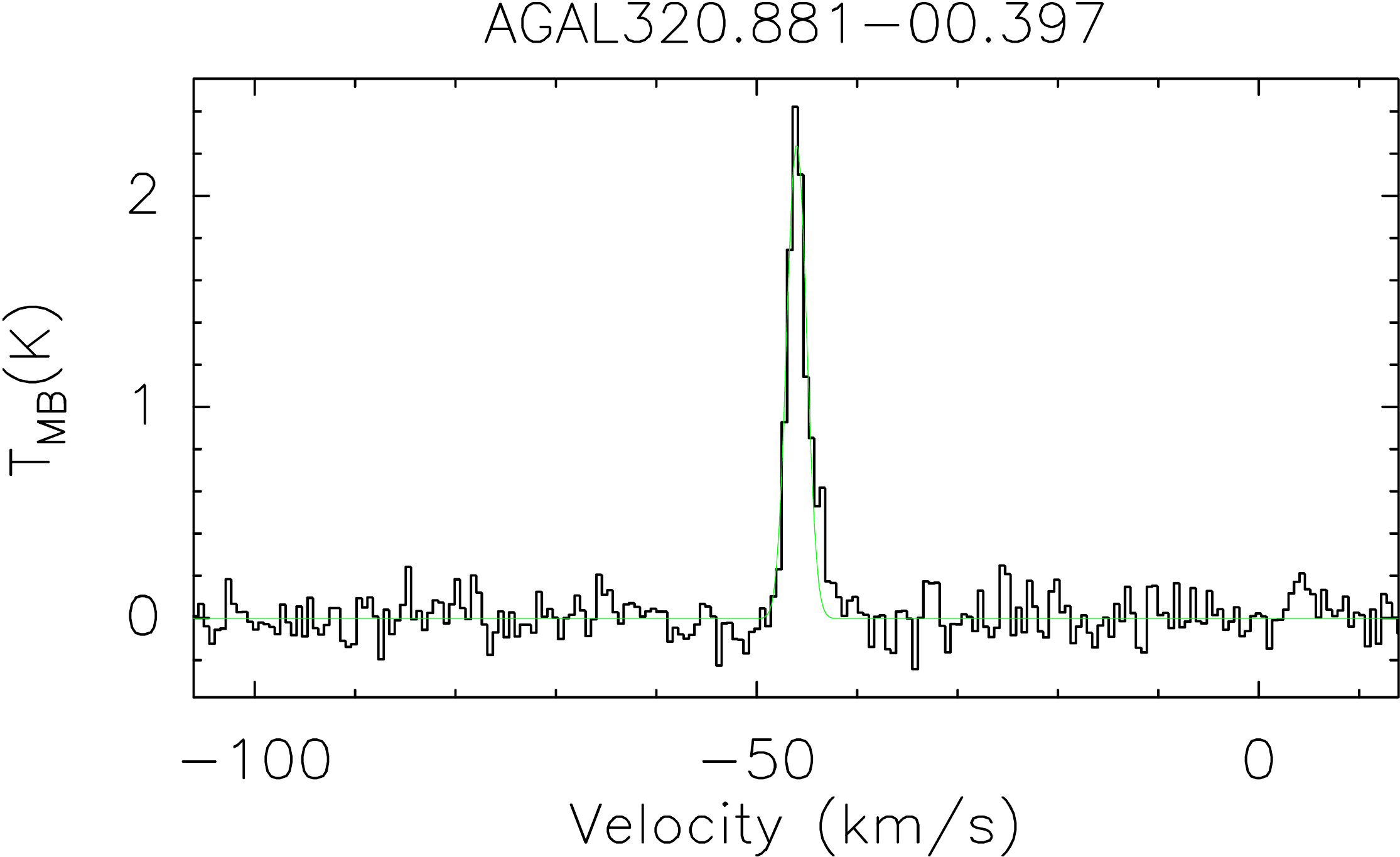} \\ 
\includegraphics[angle=-90,width=0.3\textwidth]{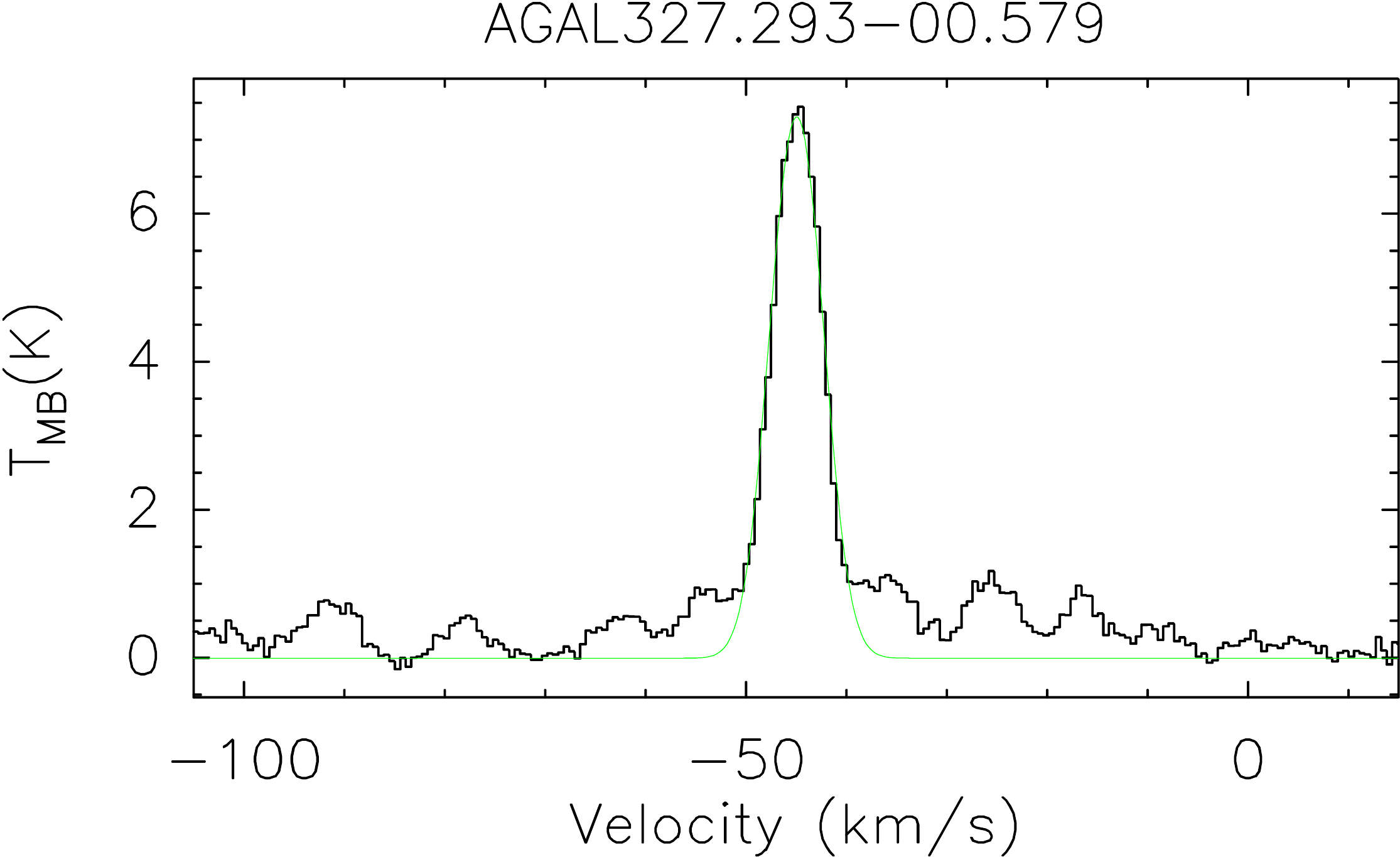} 
\includegraphics[angle=-90,width=0.3\textwidth]{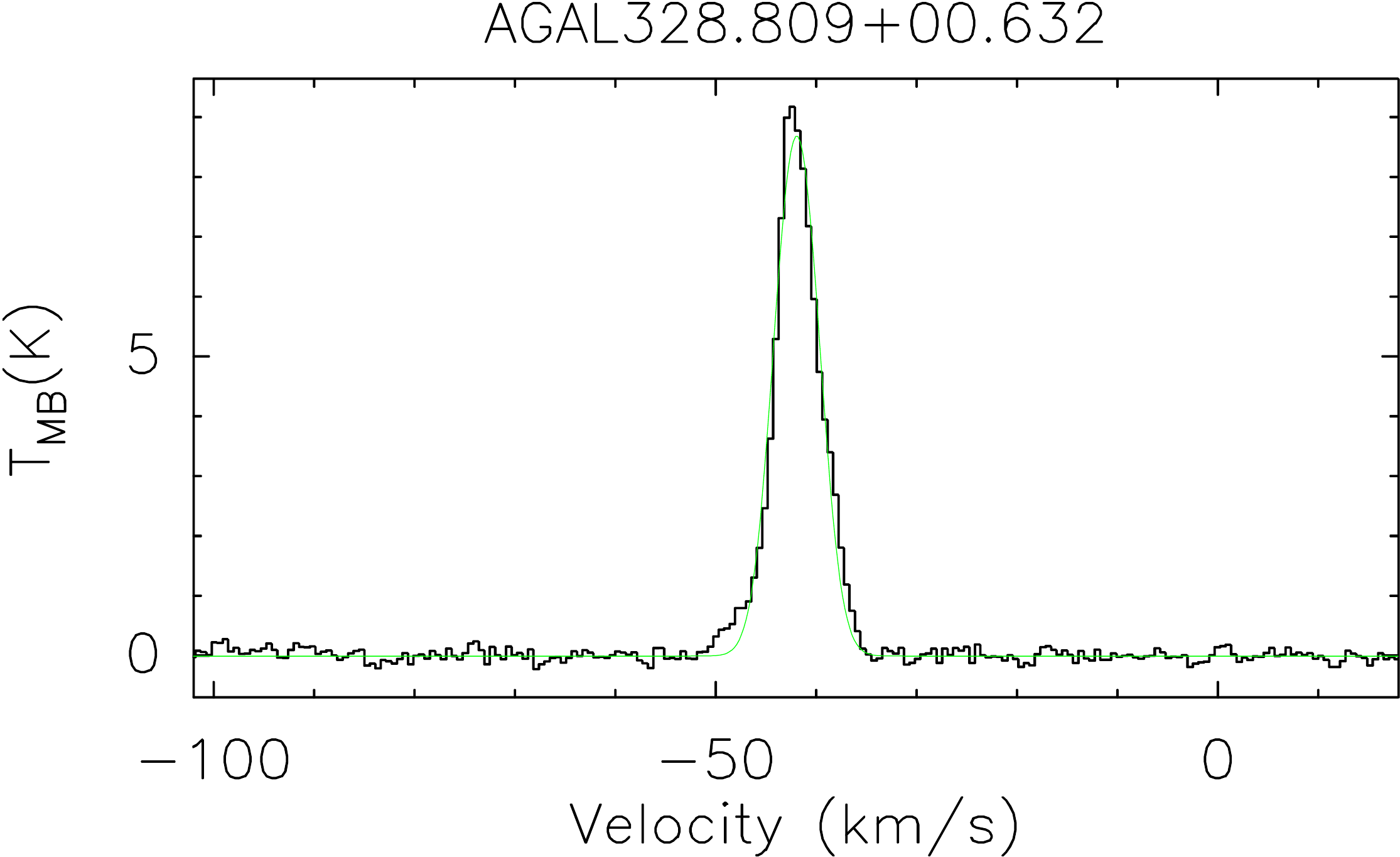} 
\includegraphics[angle=-90,width=0.3\textwidth]{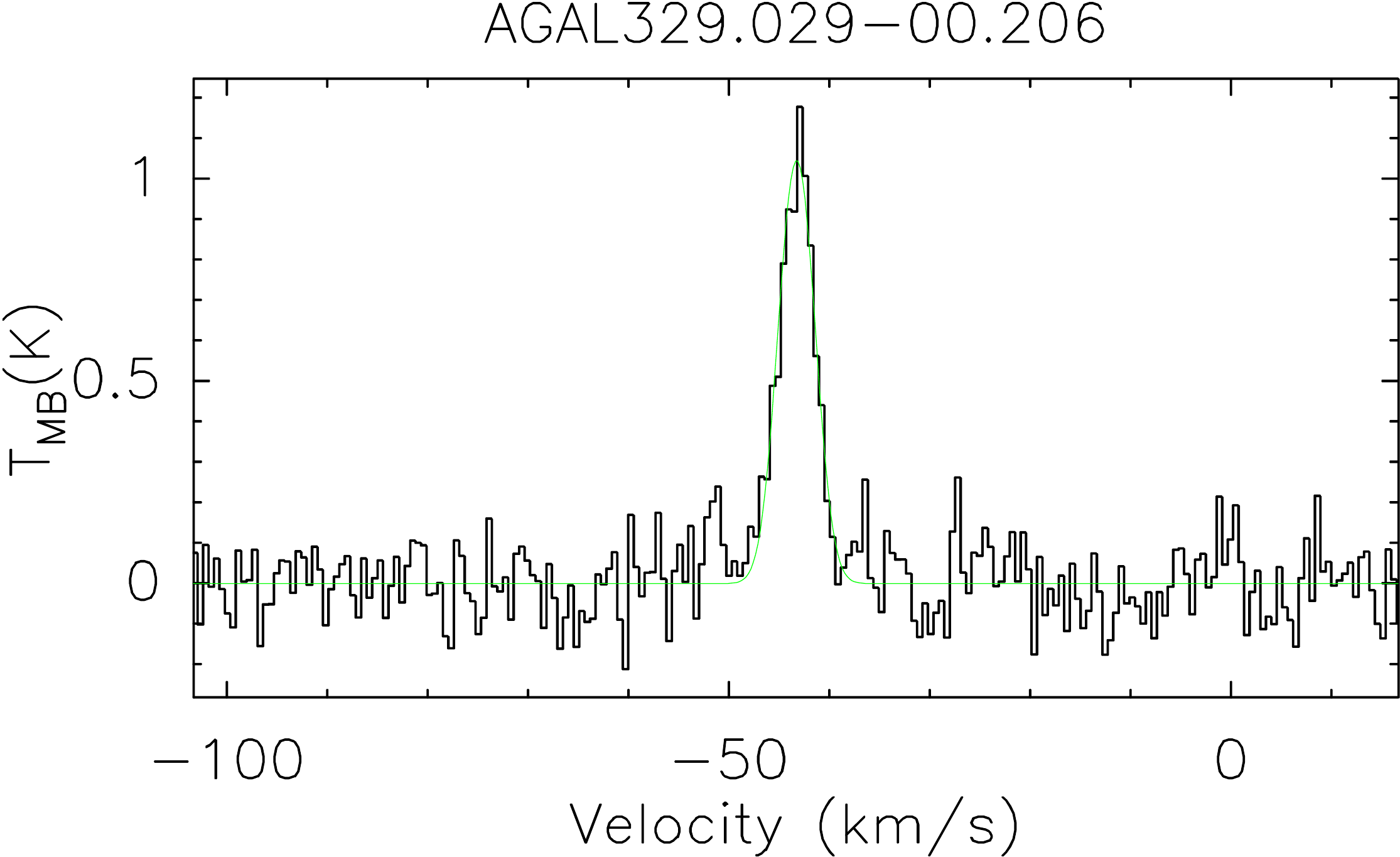} \\ 
\includegraphics[angle=-90,width=0.3\textwidth]{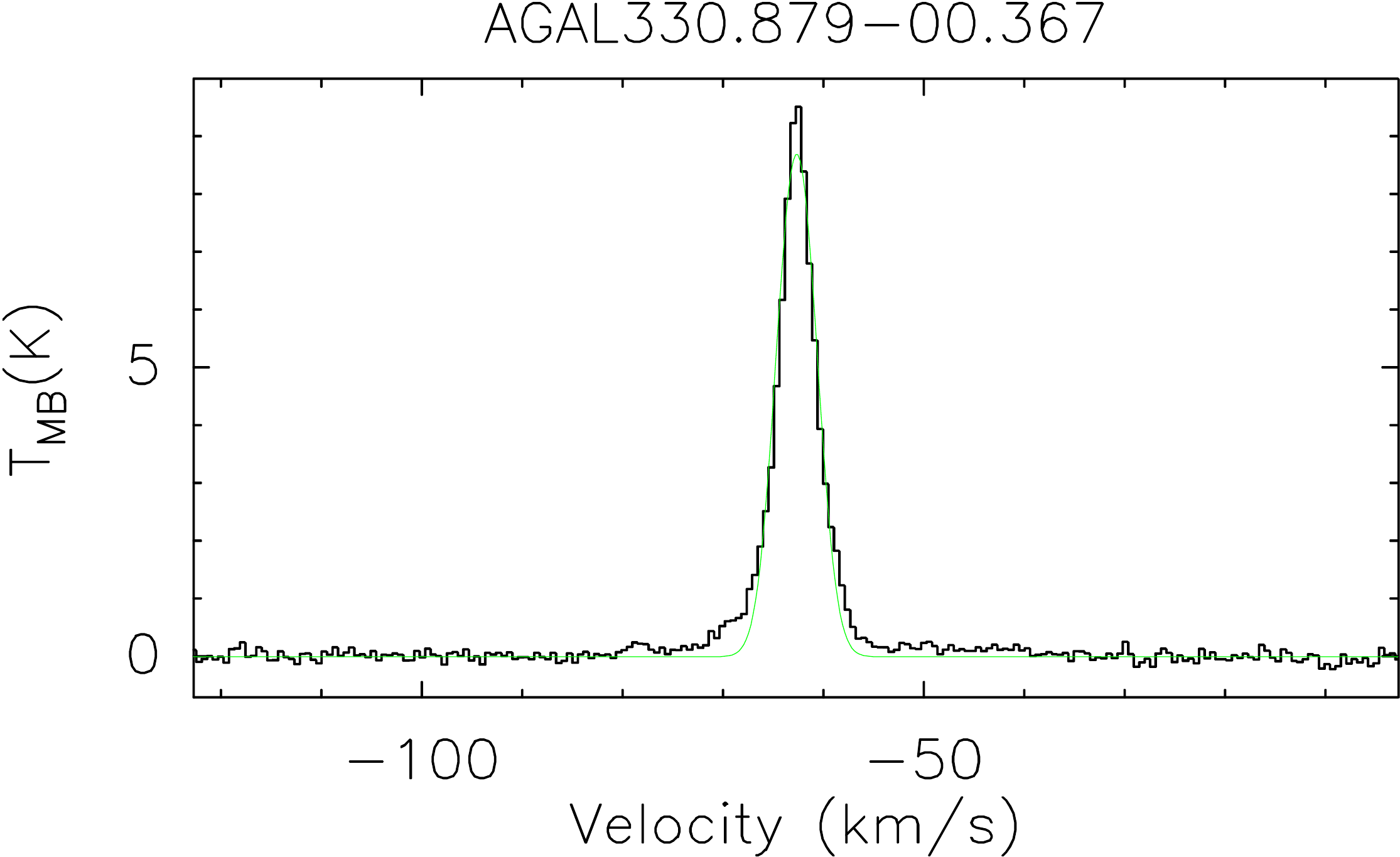} 
\includegraphics[angle=-90,width=0.3\textwidth]{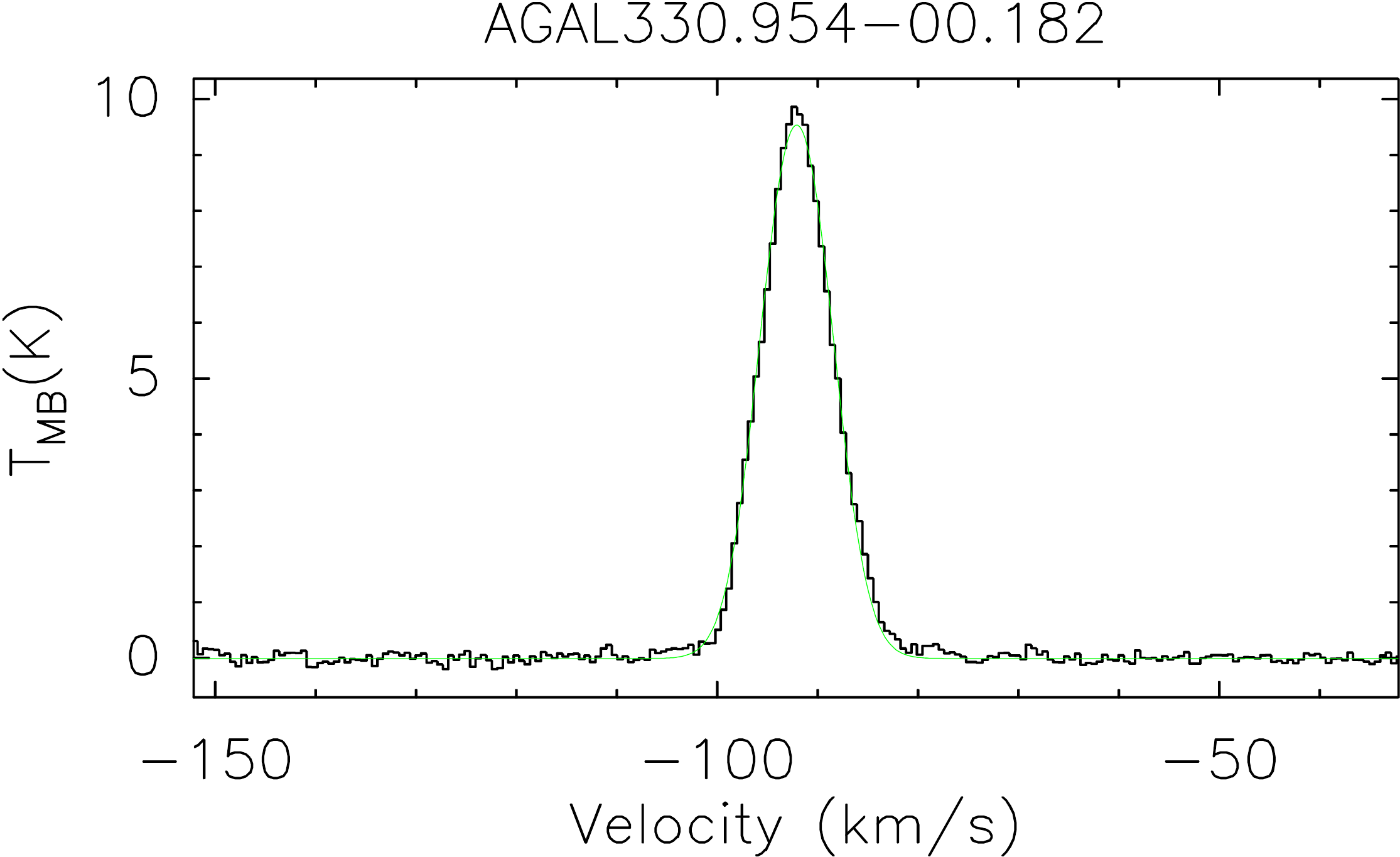} 
\includegraphics[angle=-90,width=0.3\textwidth]{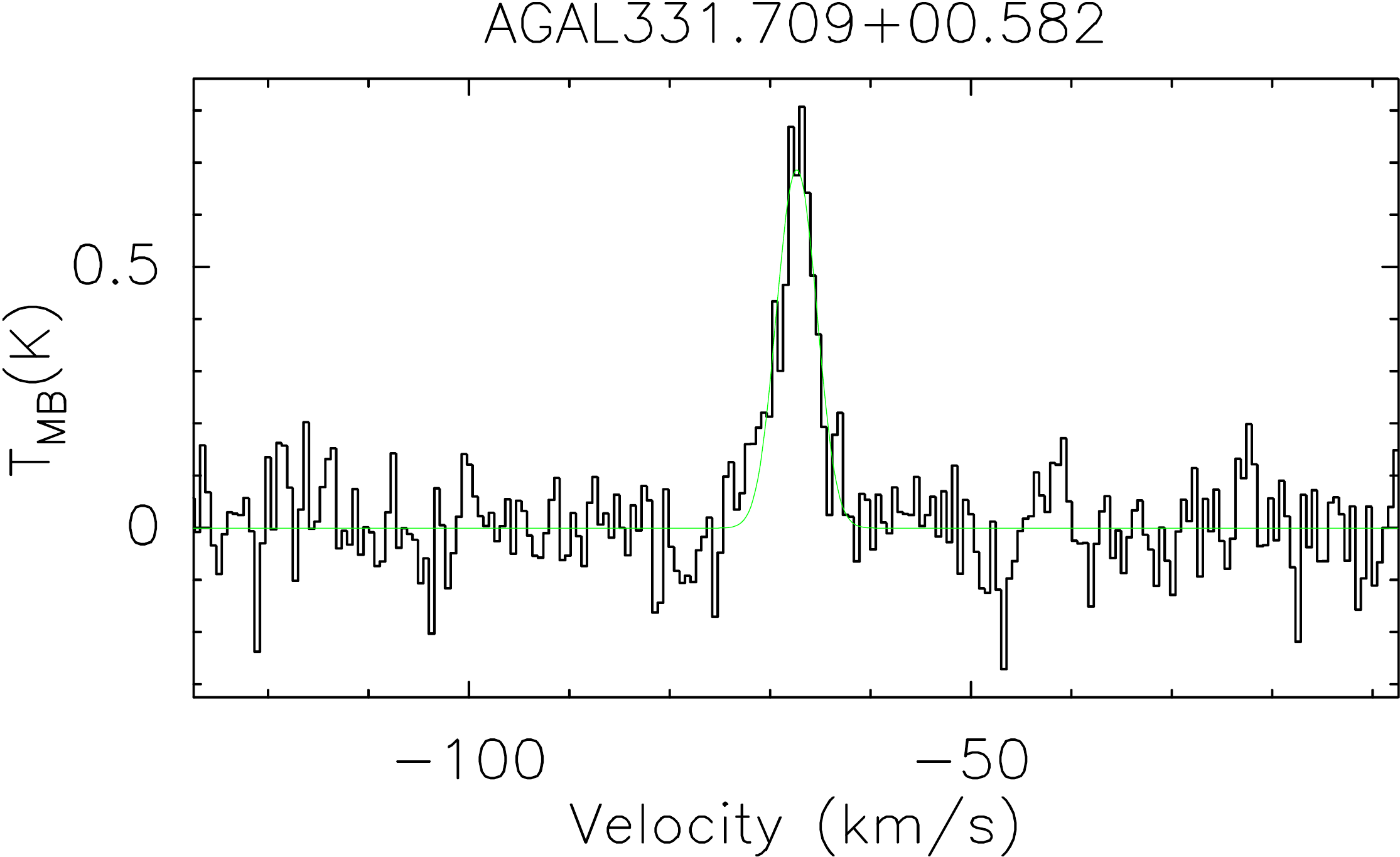} \\ 
\includegraphics[angle=-90,width=0.3\textwidth]{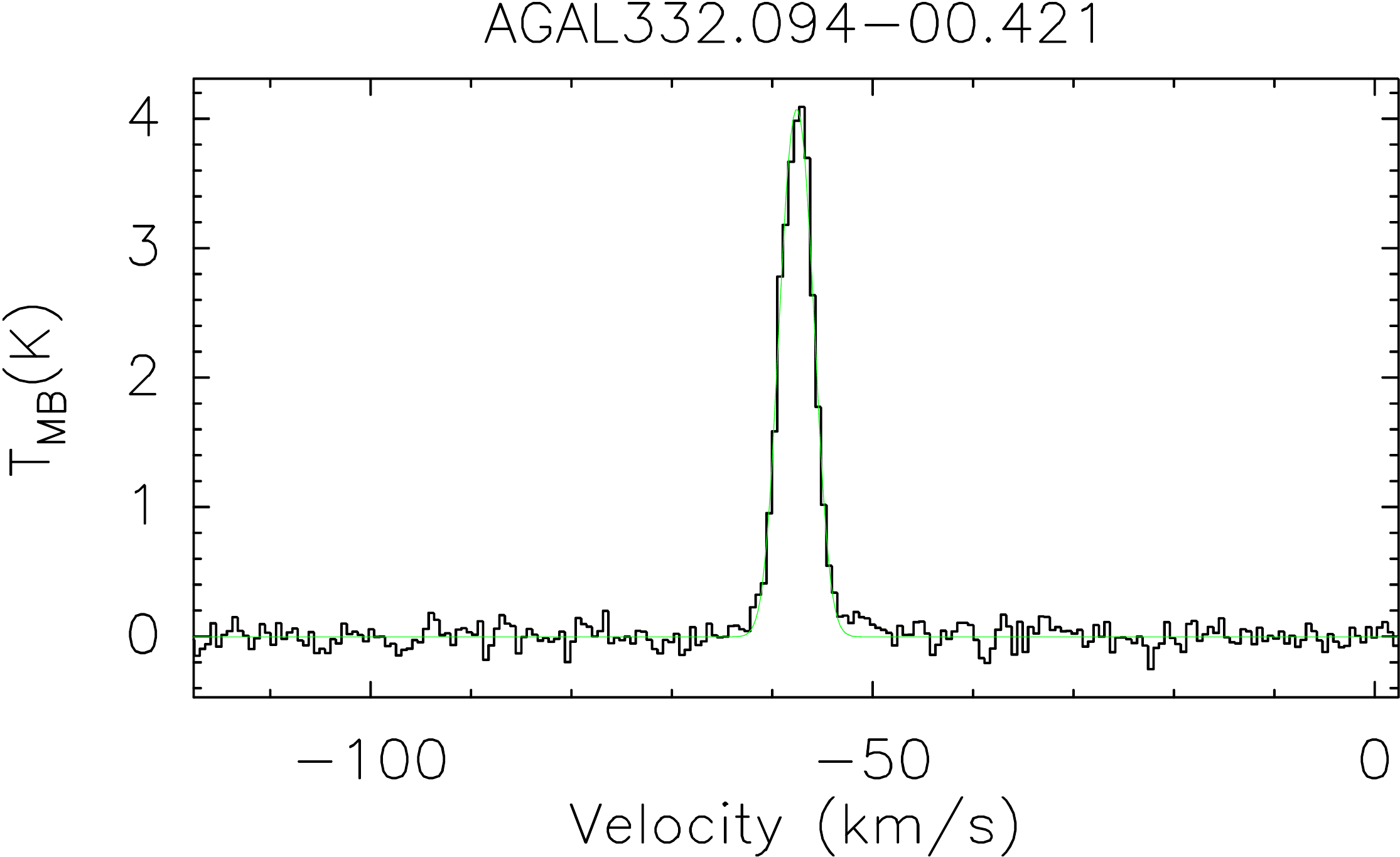} 
\includegraphics[angle=-90,width=0.3\textwidth]{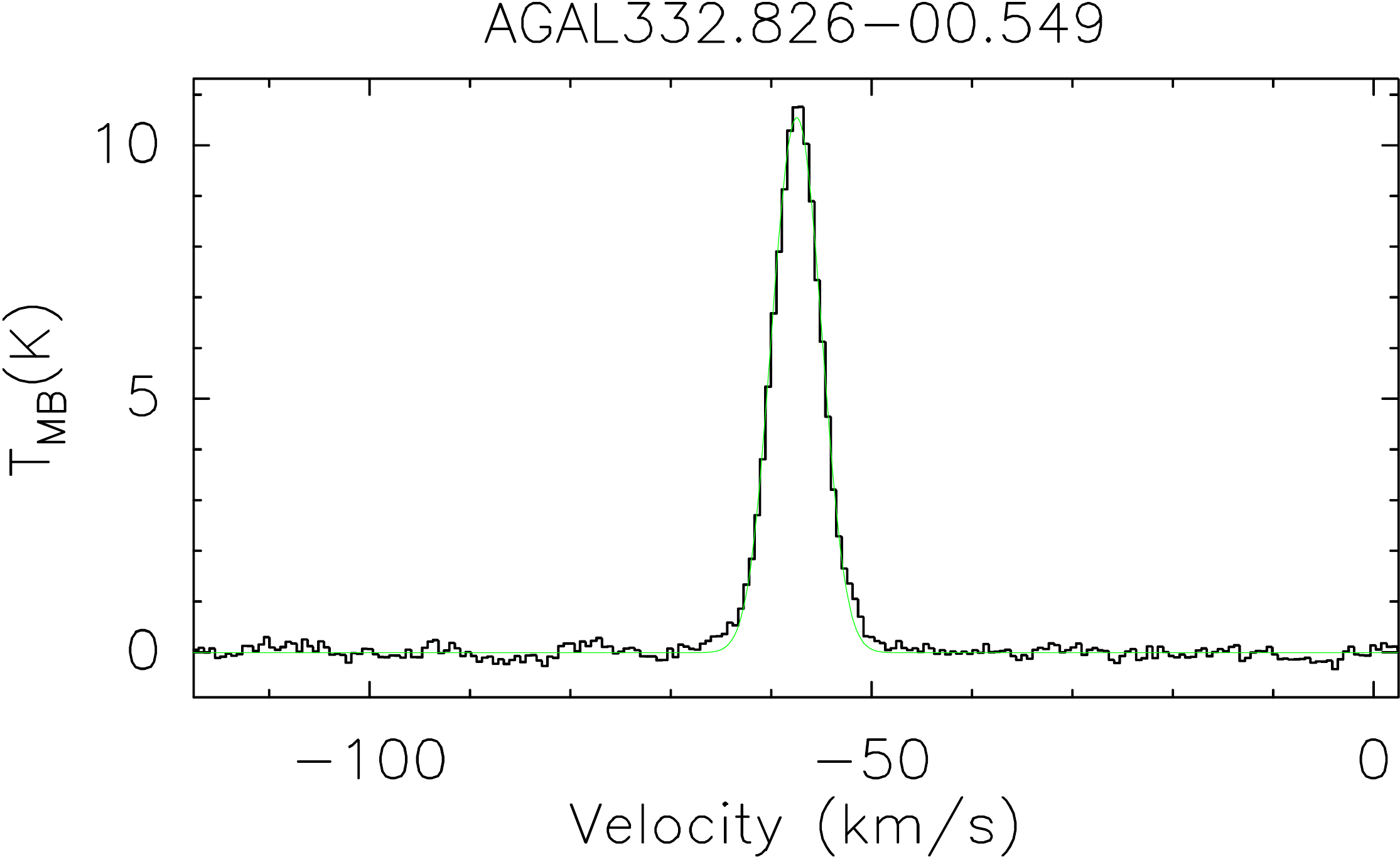} 
\includegraphics[angle=-90,width=0.3\textwidth]{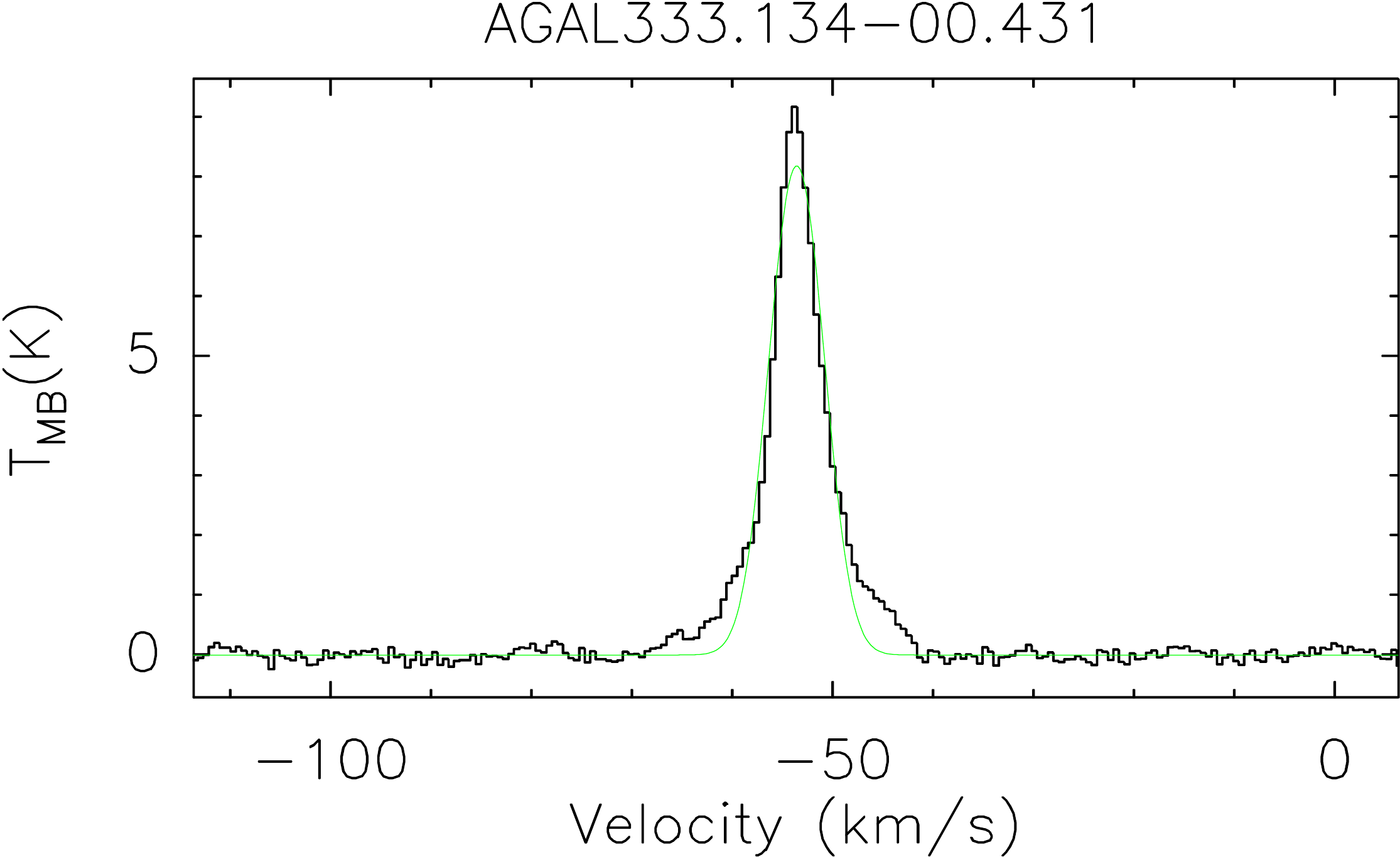} \\ 
\includegraphics[angle=-90,width=0.3\textwidth]{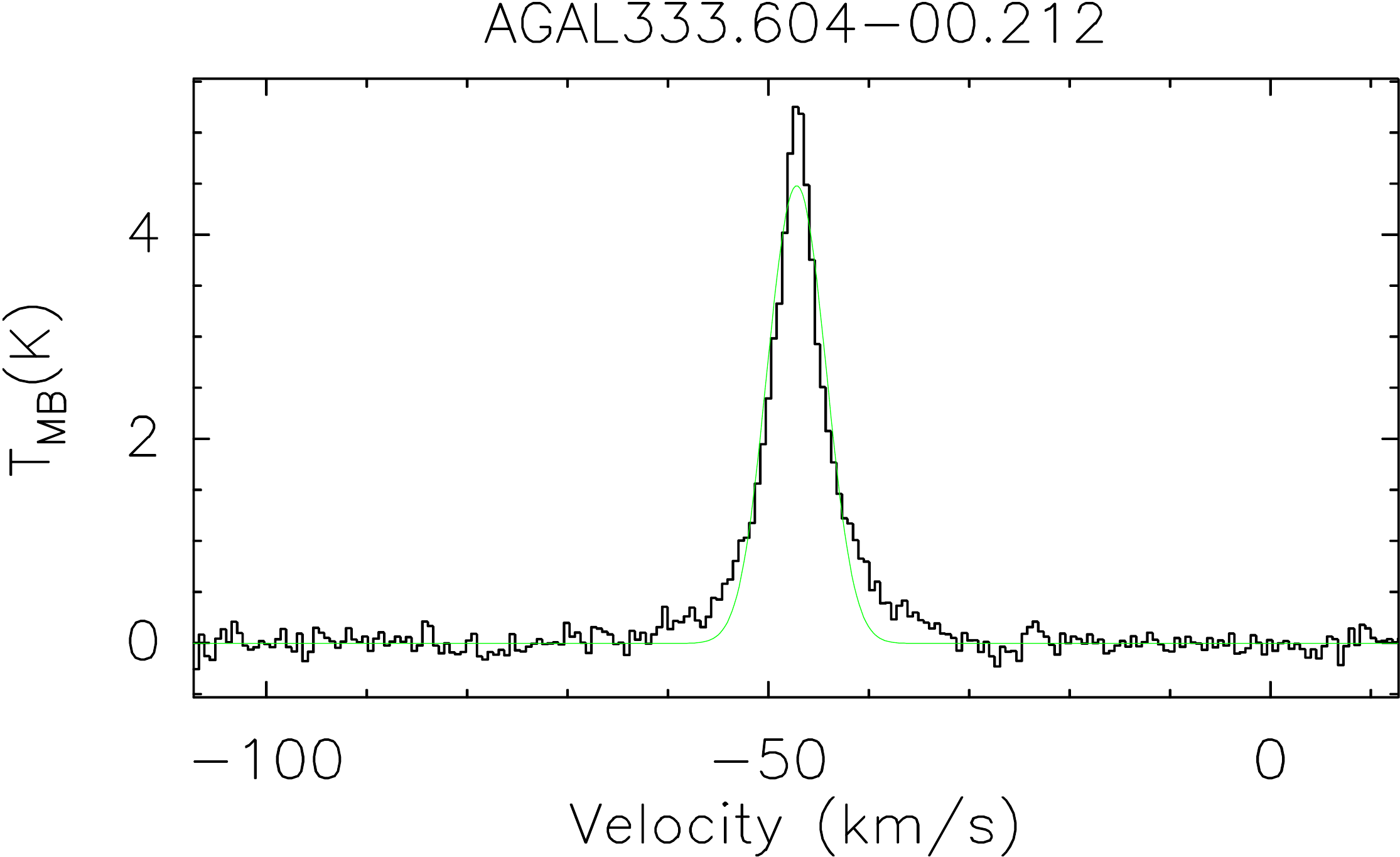} 
\includegraphics[angle=-90,width=0.3\textwidth]{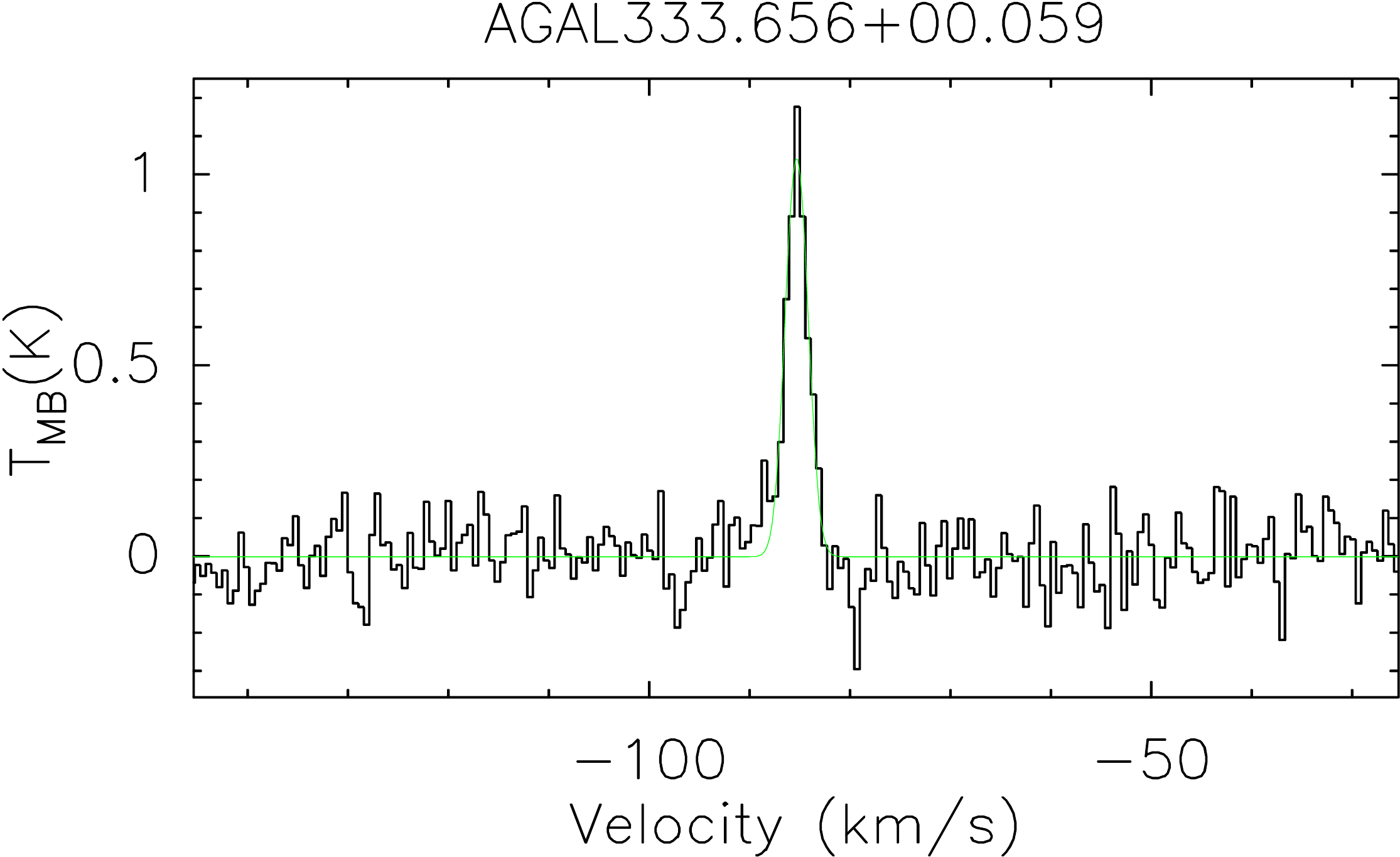} 
\includegraphics[angle=-90,width=0.3\textwidth]{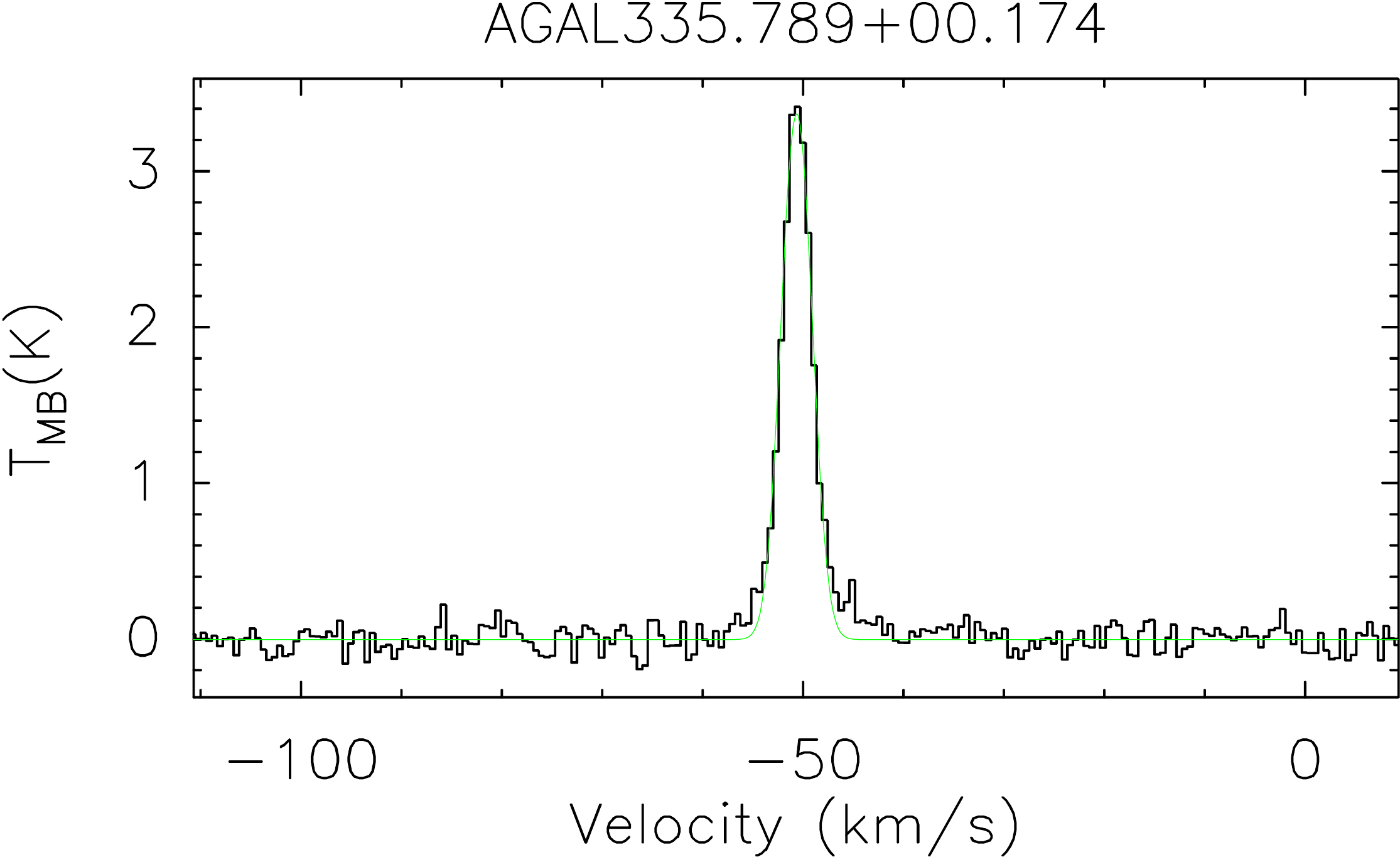} \hfill 
\caption{C$^{17}$O$(3-2)$ for subsample S2. The fit is shown in green.} \label{fig:spectra_32_B}
\end{figure*} 

\begin{figure*} 
\ContinuedFloat
\centering 
\includegraphics[angle=-90,width=0.3\textwidth]{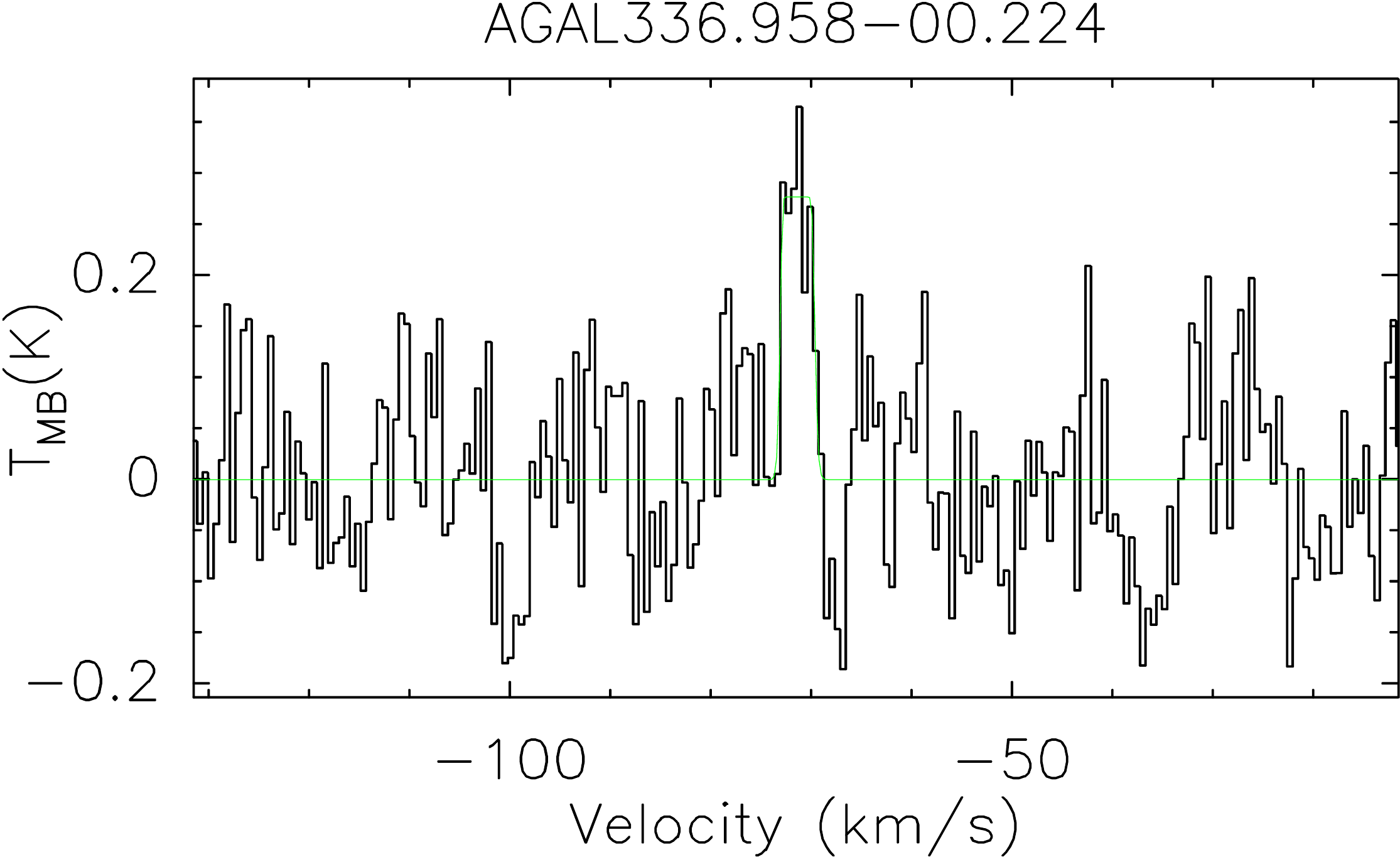} 
\includegraphics[angle=-90,width=0.3\textwidth]{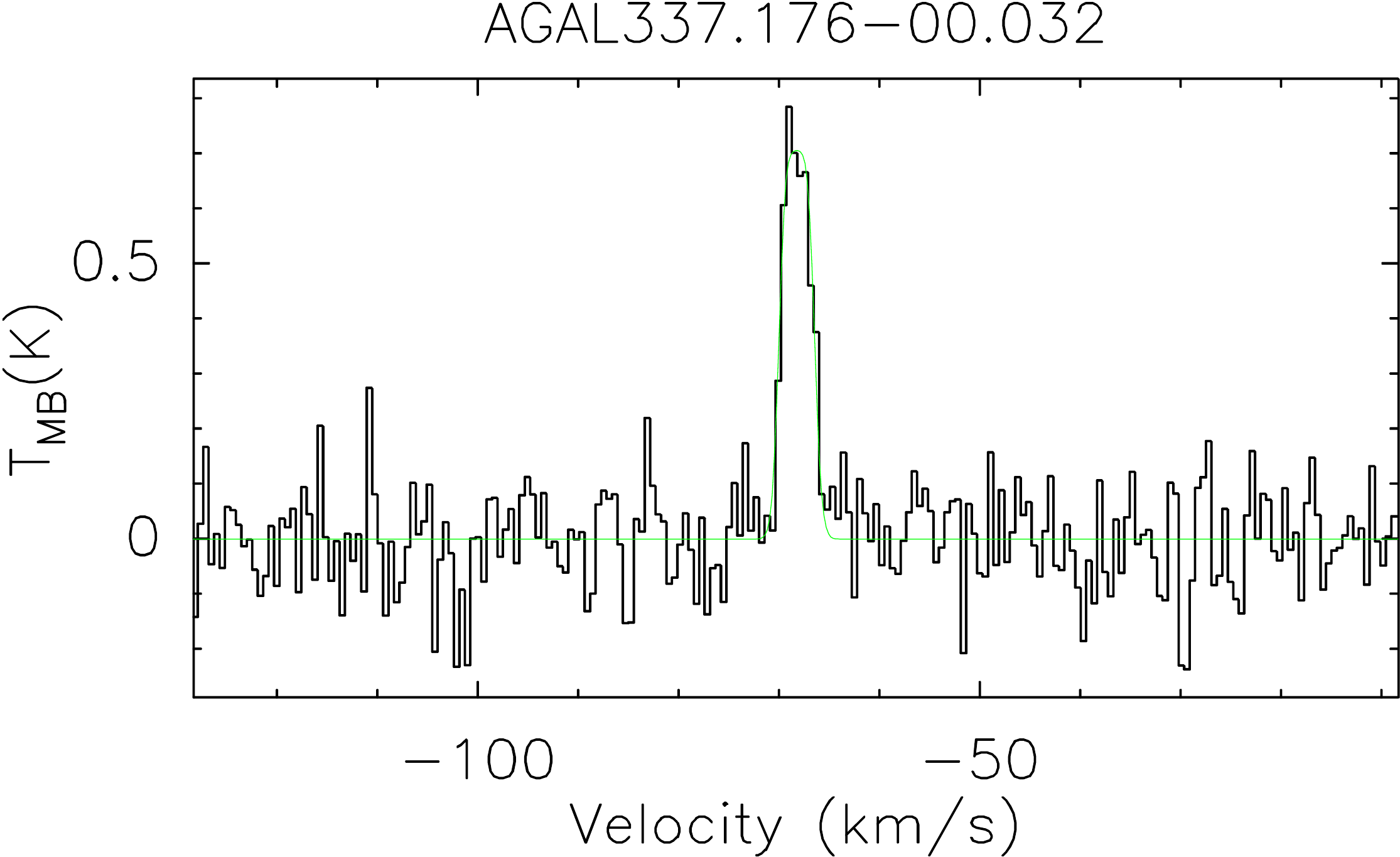} 
\includegraphics[angle=-90,width=0.3\textwidth]{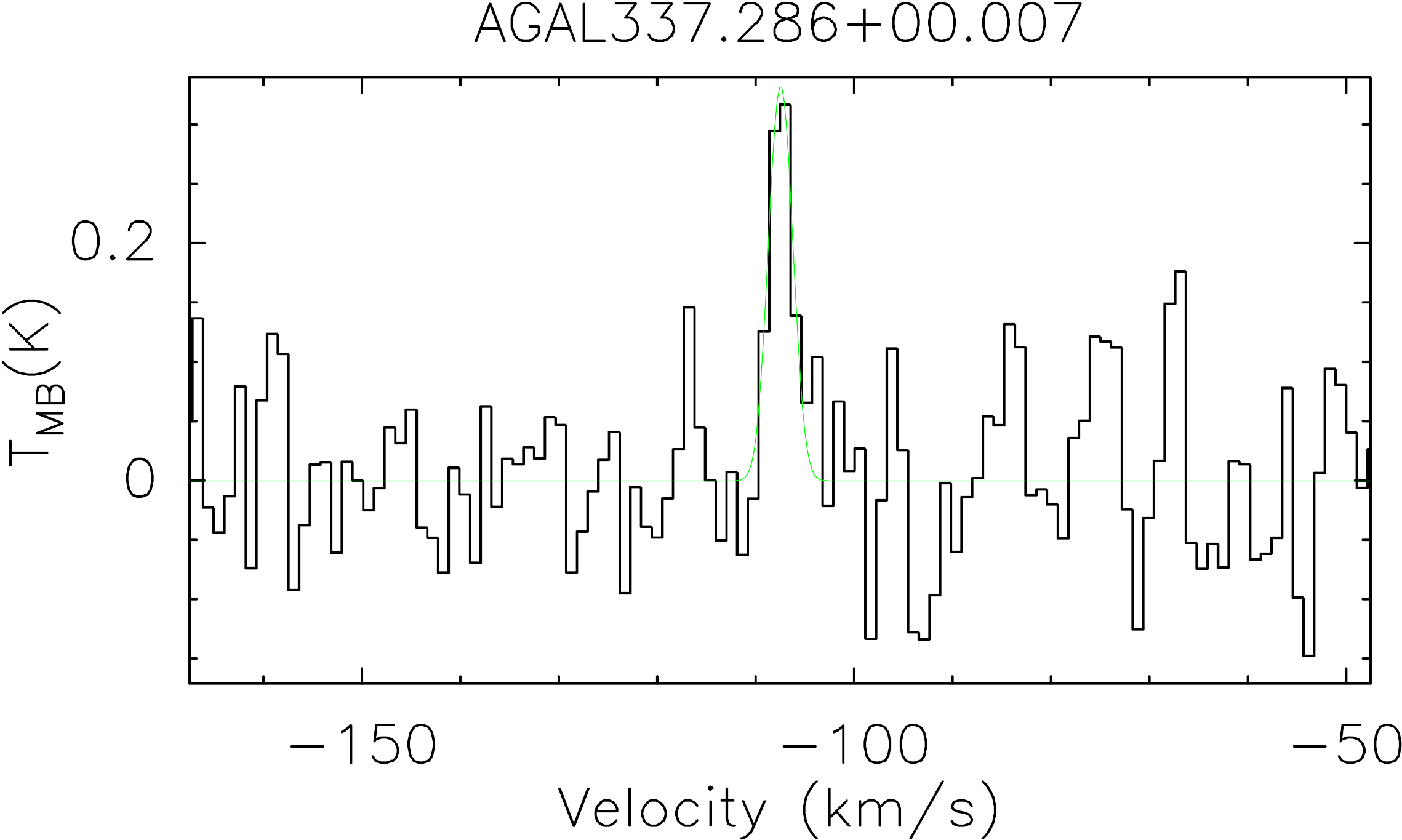} \\ 
\includegraphics[angle=-90,width=0.3\textwidth]{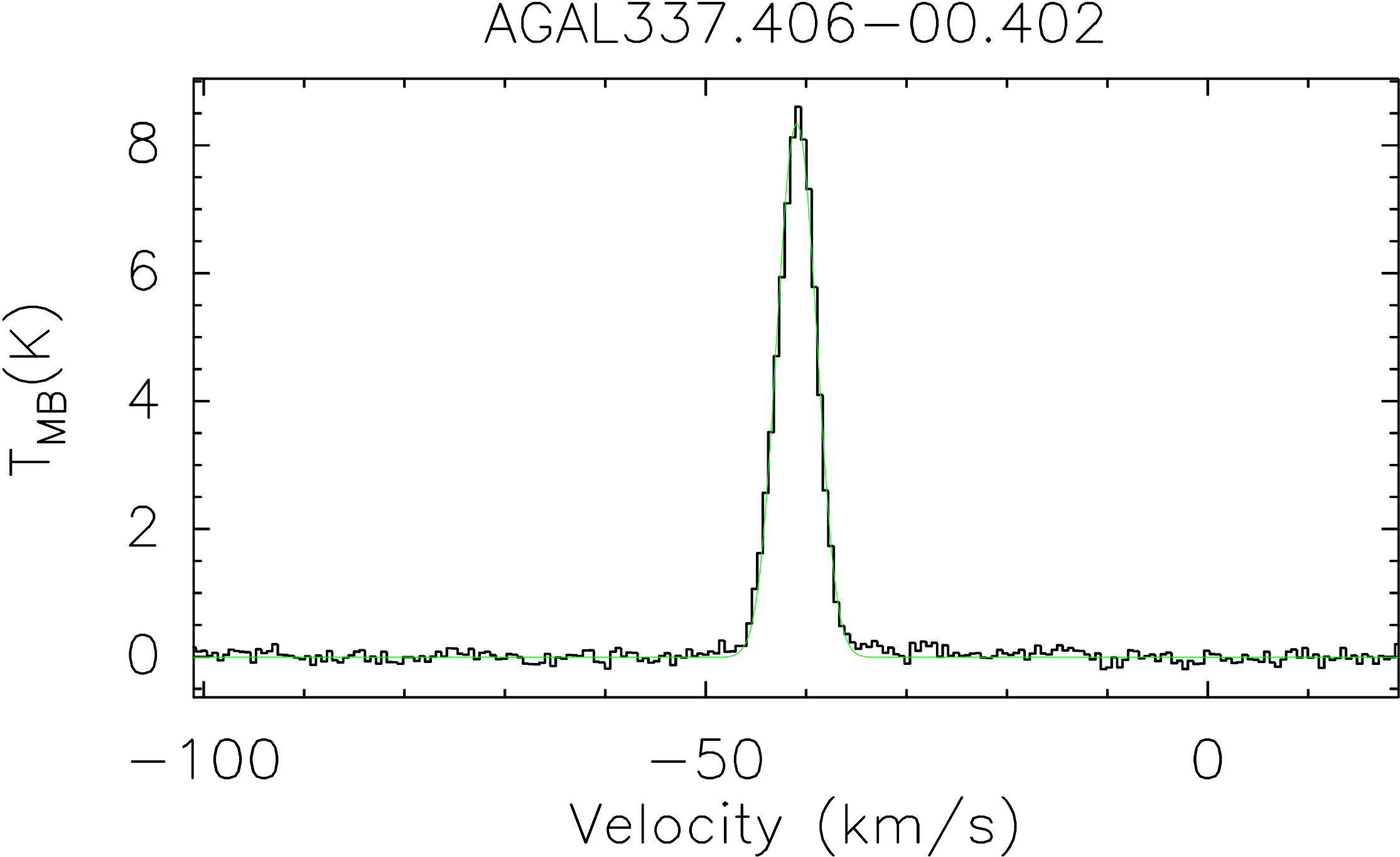} 
\includegraphics[angle=-90,width=0.3\textwidth]{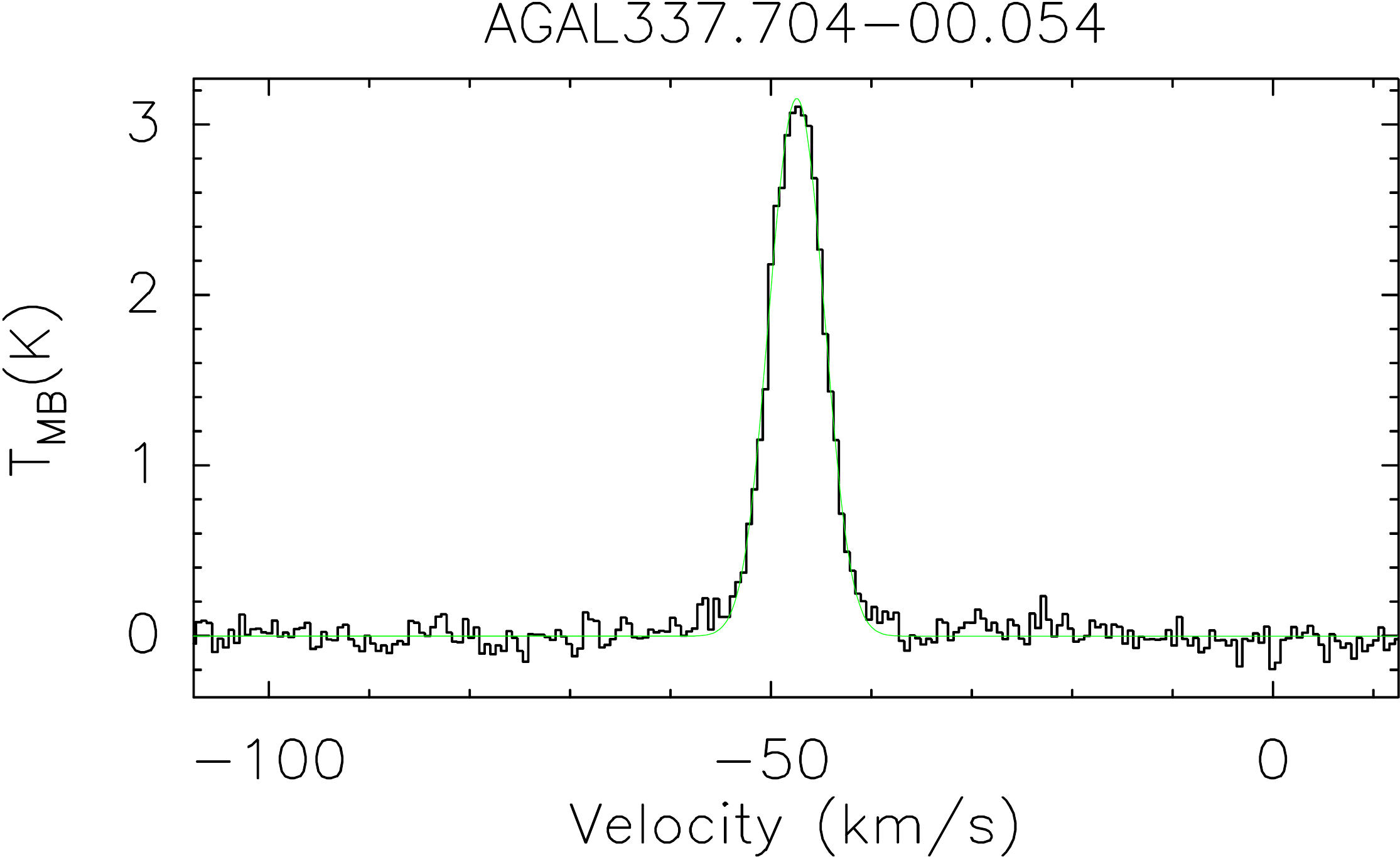} 
\includegraphics[angle=-90,width=0.3\textwidth]{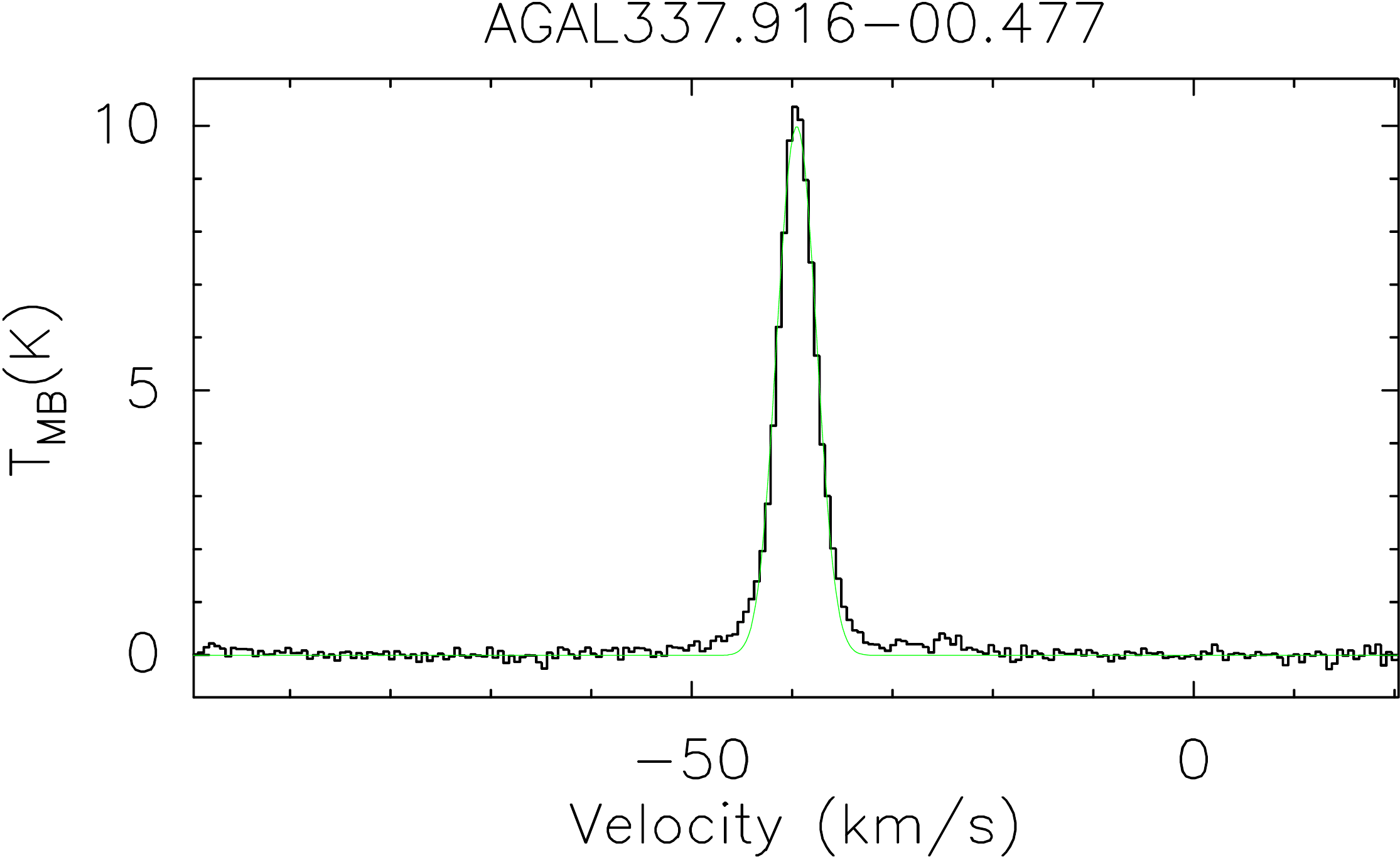} \\ 
\includegraphics[angle=-90,width=0.3\textwidth]{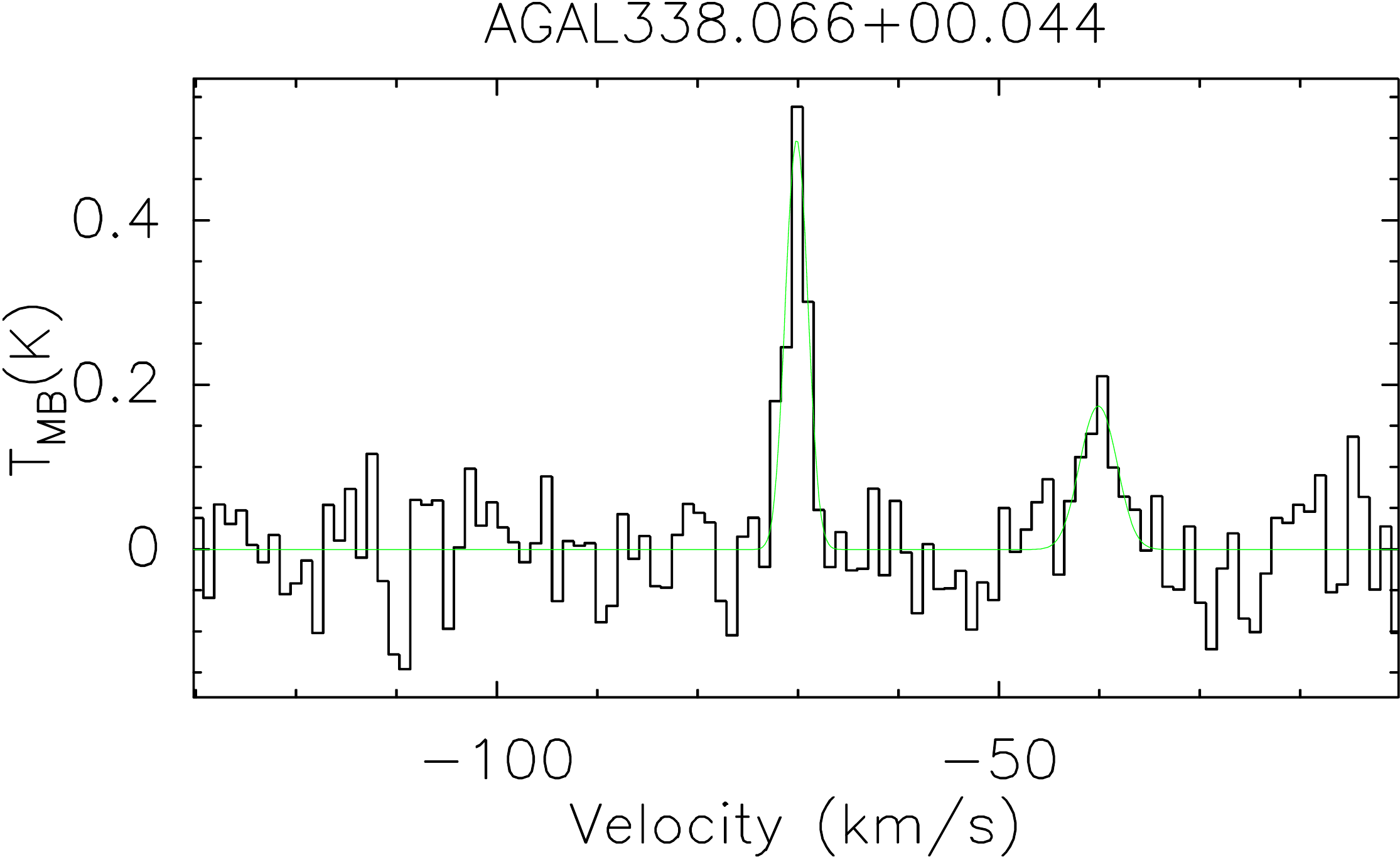} 
\includegraphics[angle=-90,width=0.3\textwidth]{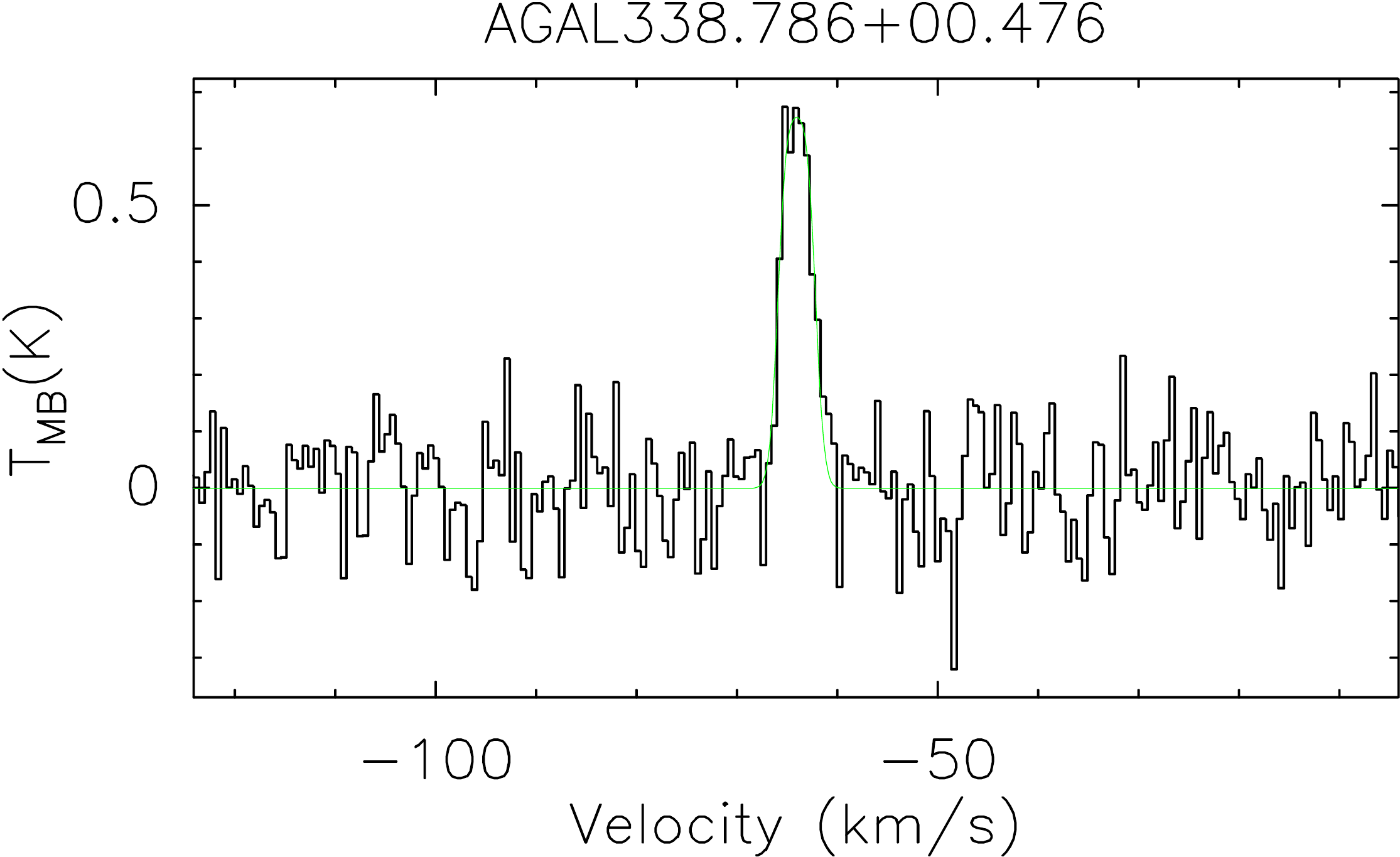} 
\includegraphics[angle=-90,width=0.3\textwidth]{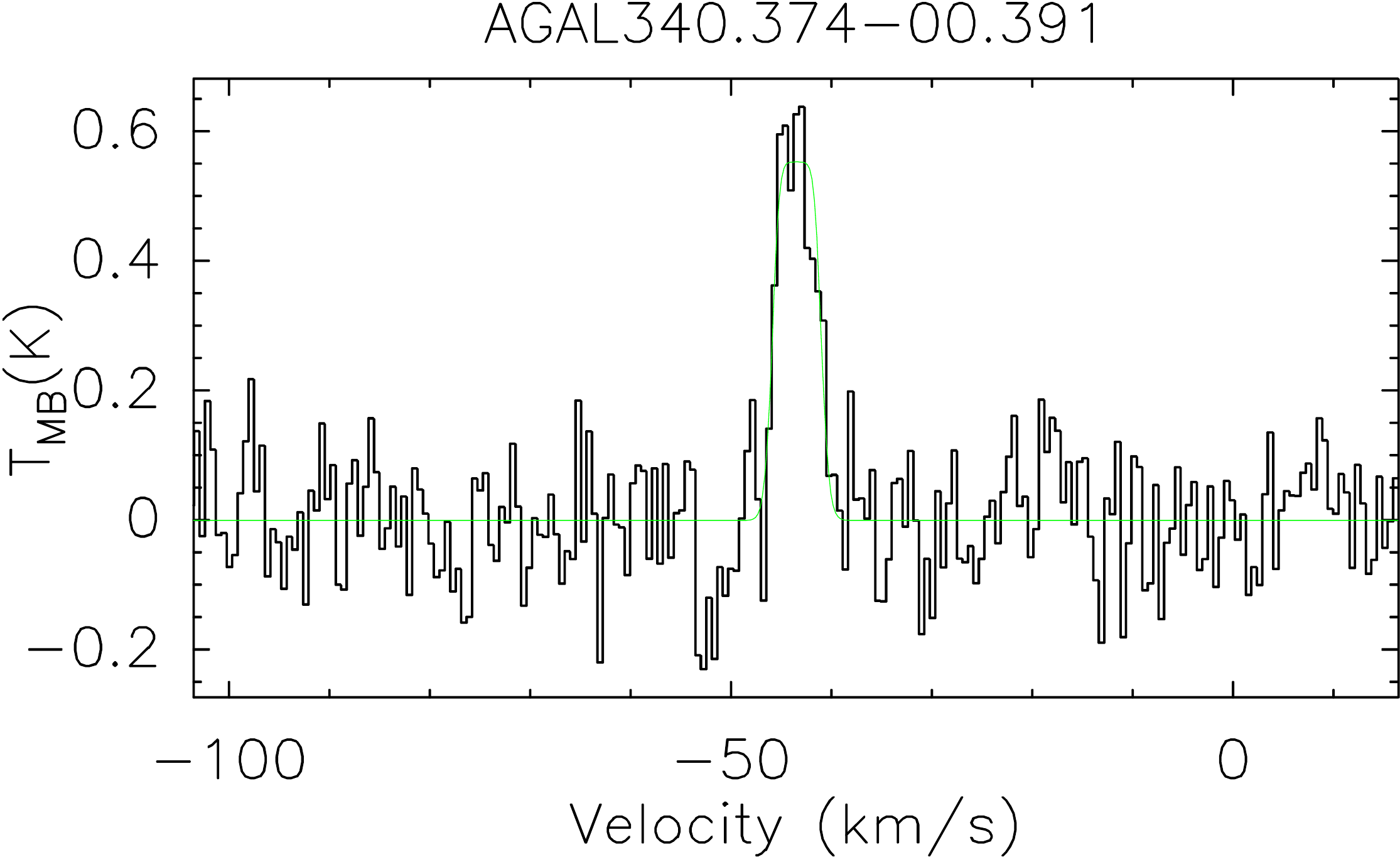} \\
\includegraphics[angle=-90,width=0.3\textwidth]{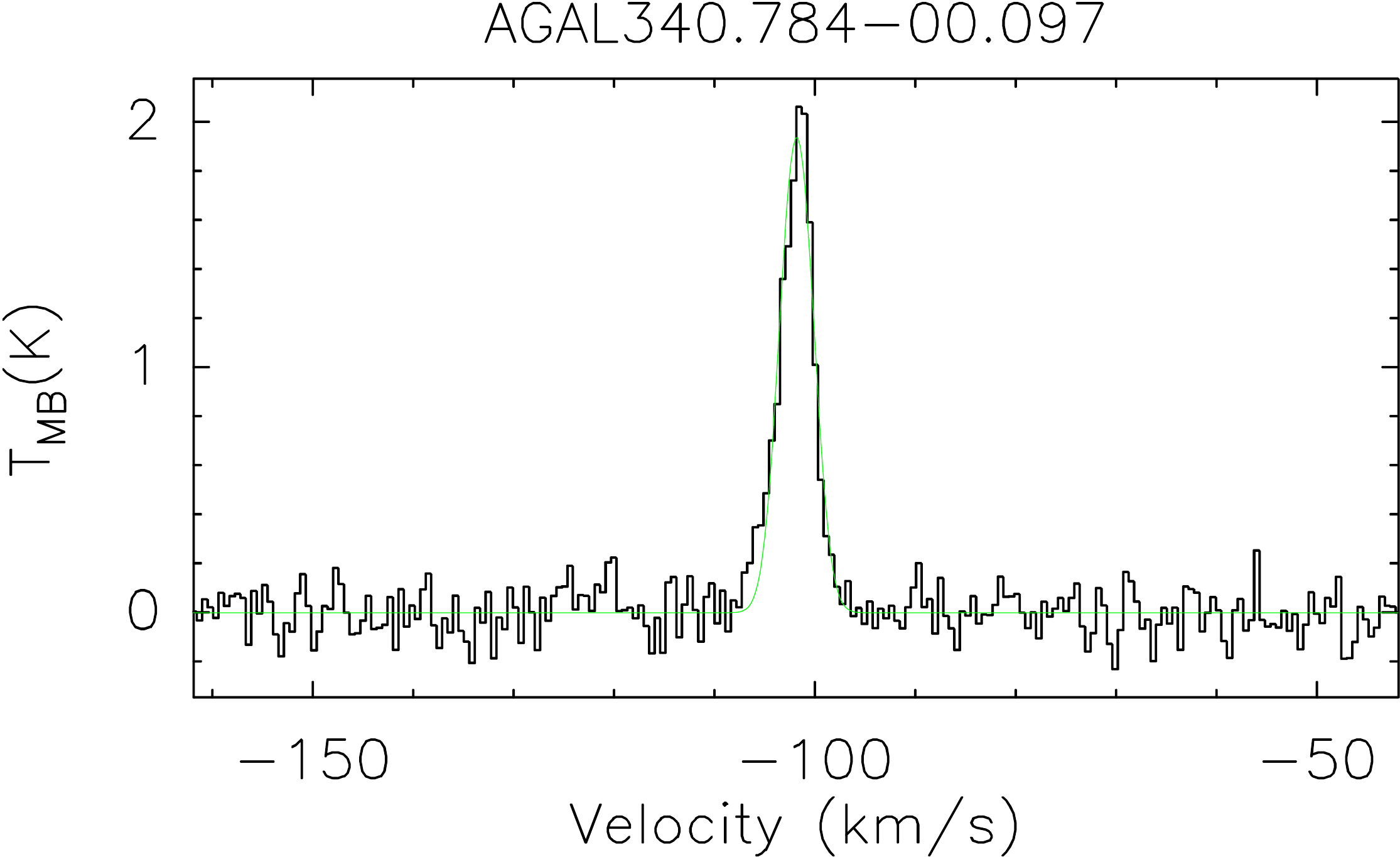} 
\includegraphics[angle=-90,width=0.3\textwidth]{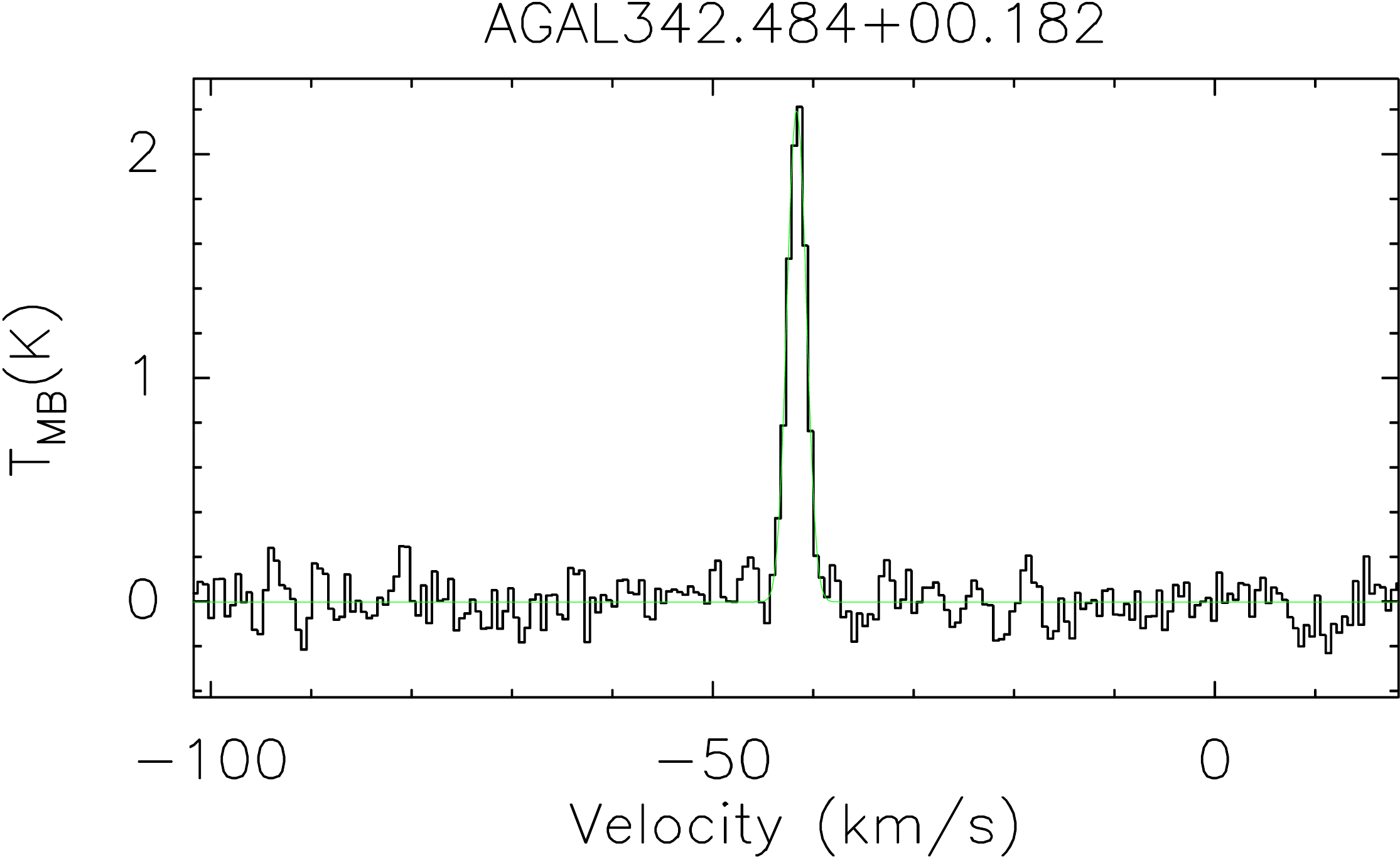} 
\includegraphics[angle=-90,width=0.3\textwidth]{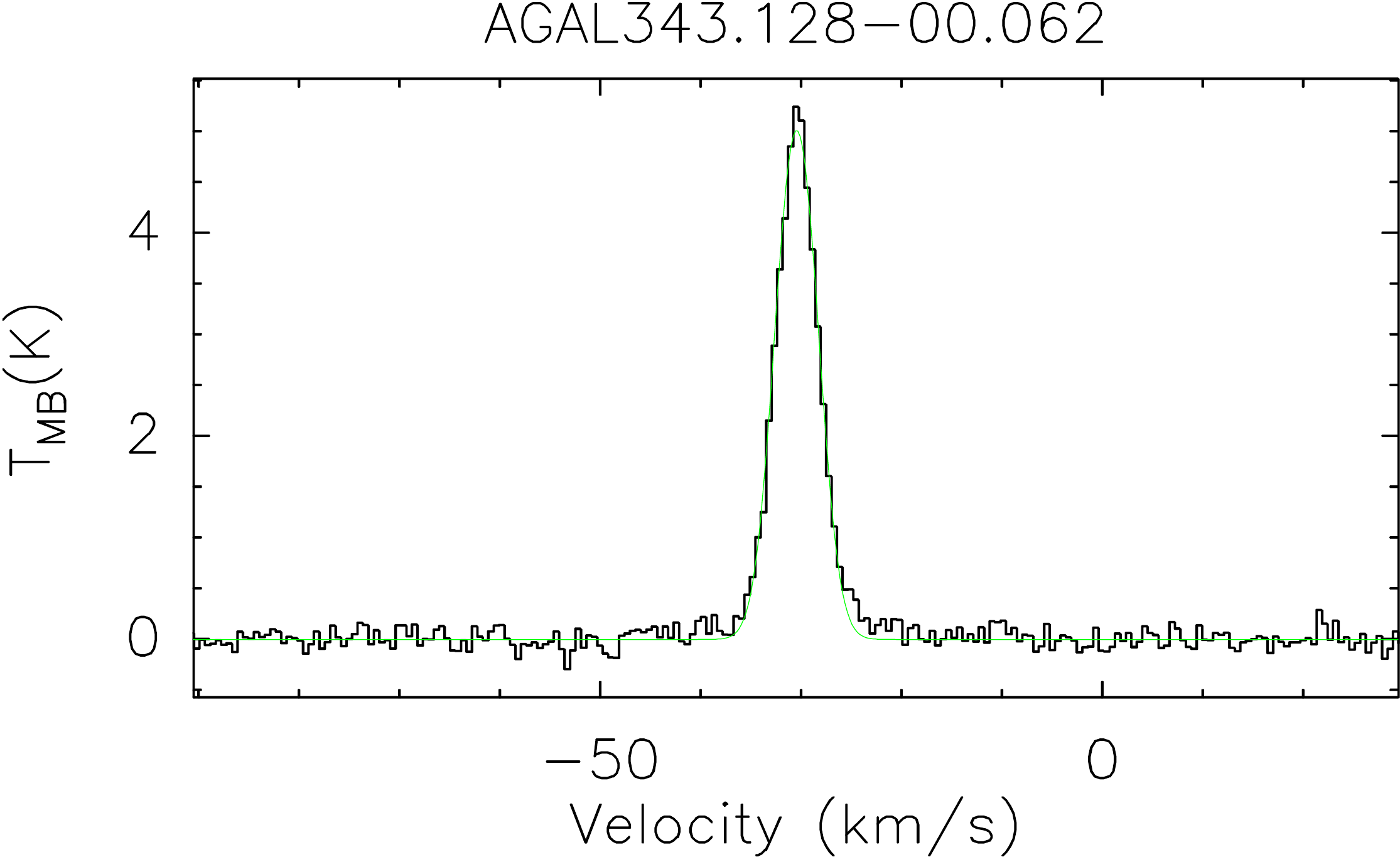} \\ 
\includegraphics[angle=-90,width=0.3\textwidth]{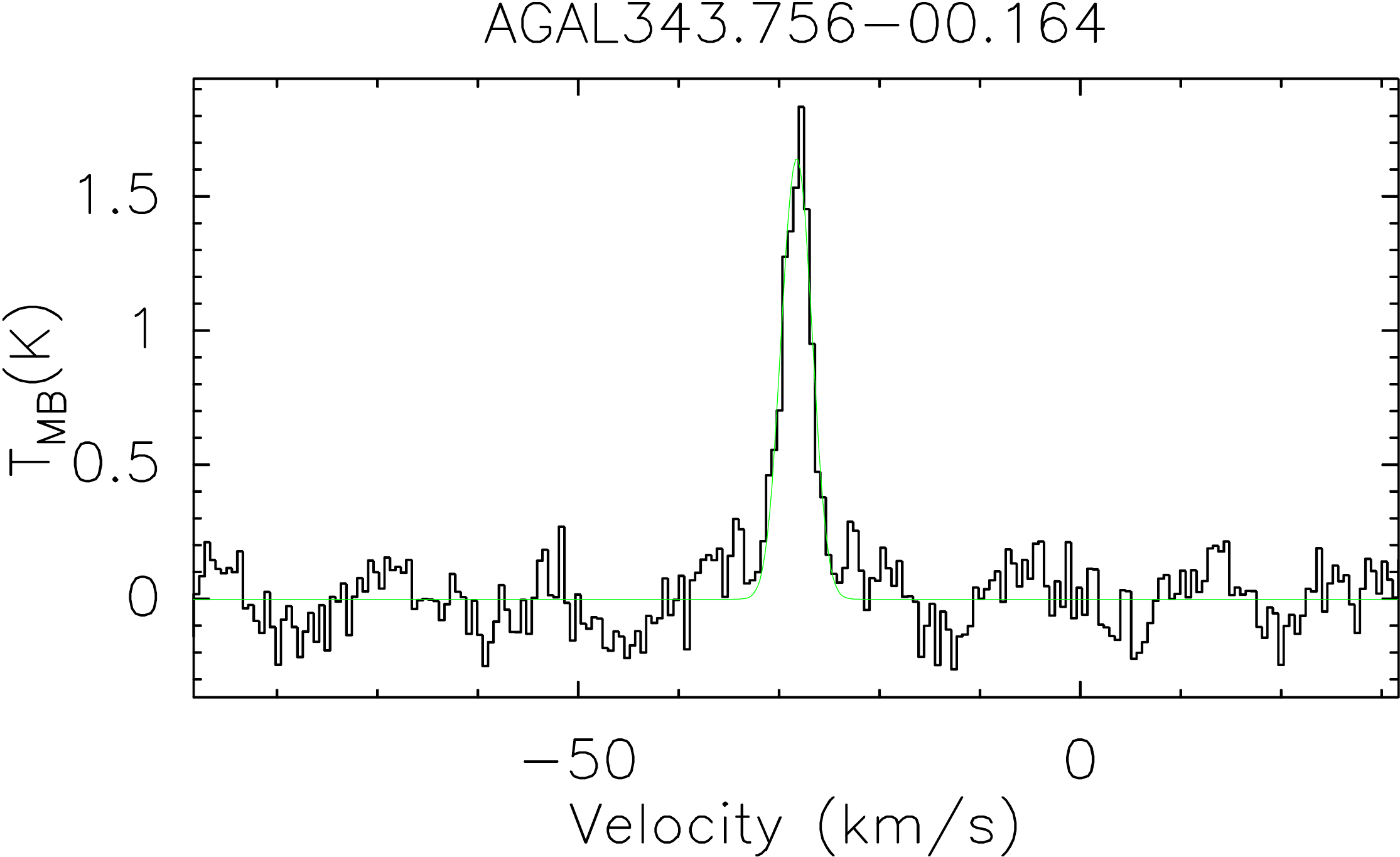} 
\includegraphics[angle=-90,width=0.3\textwidth]{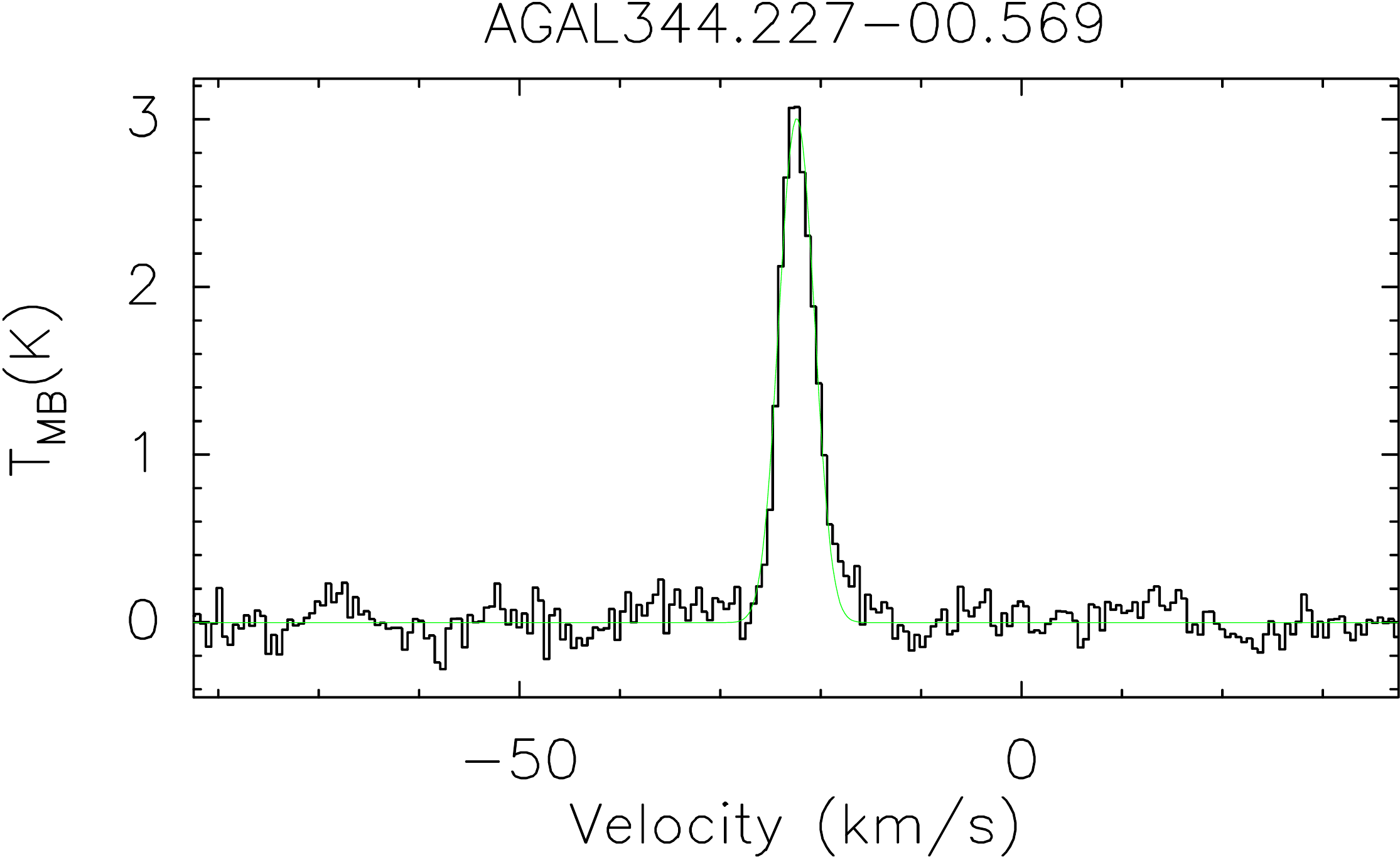} 
\includegraphics[angle=-90,width=0.3\textwidth]{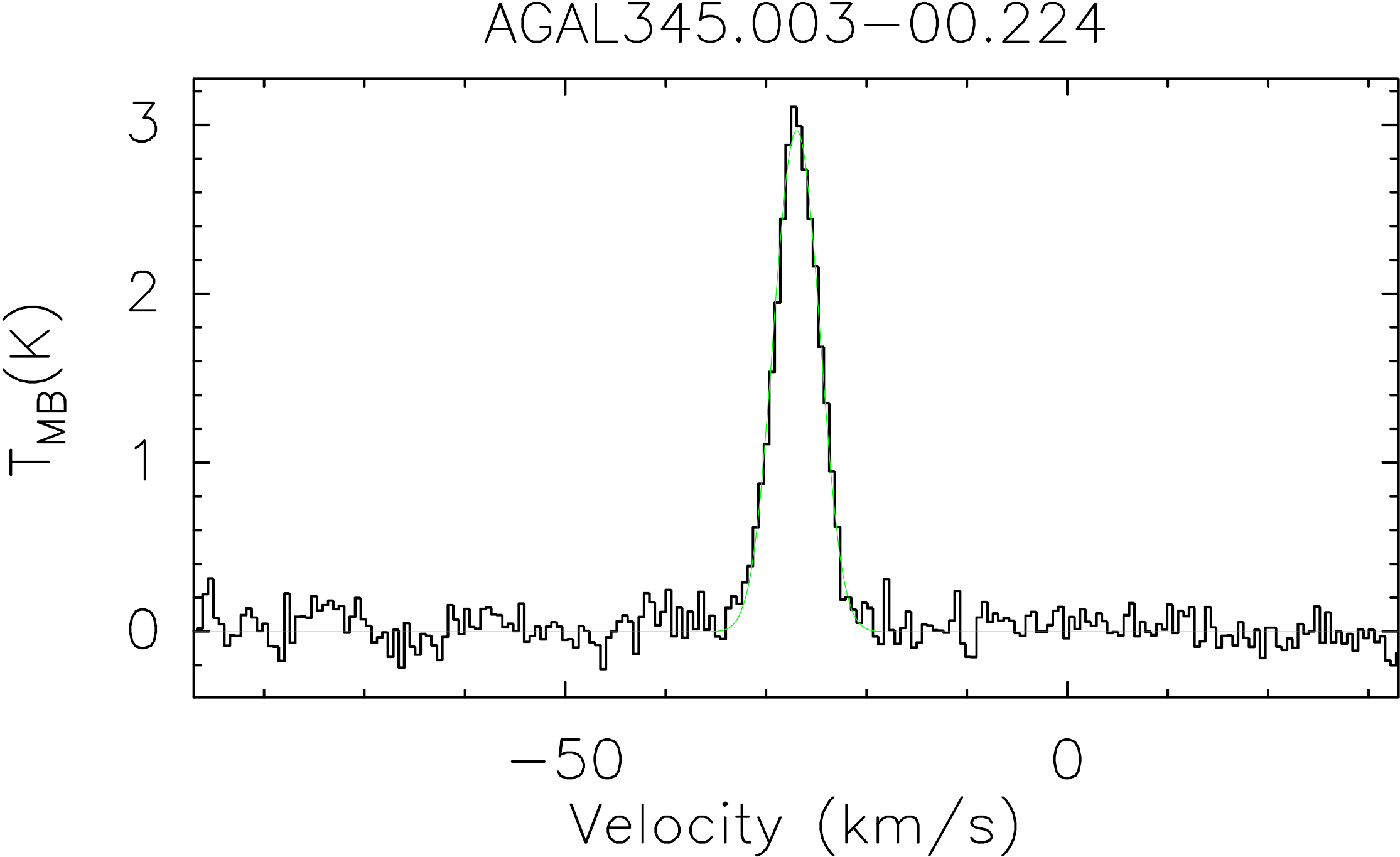} \\ 
\includegraphics[angle=-90,width=0.3\textwidth]{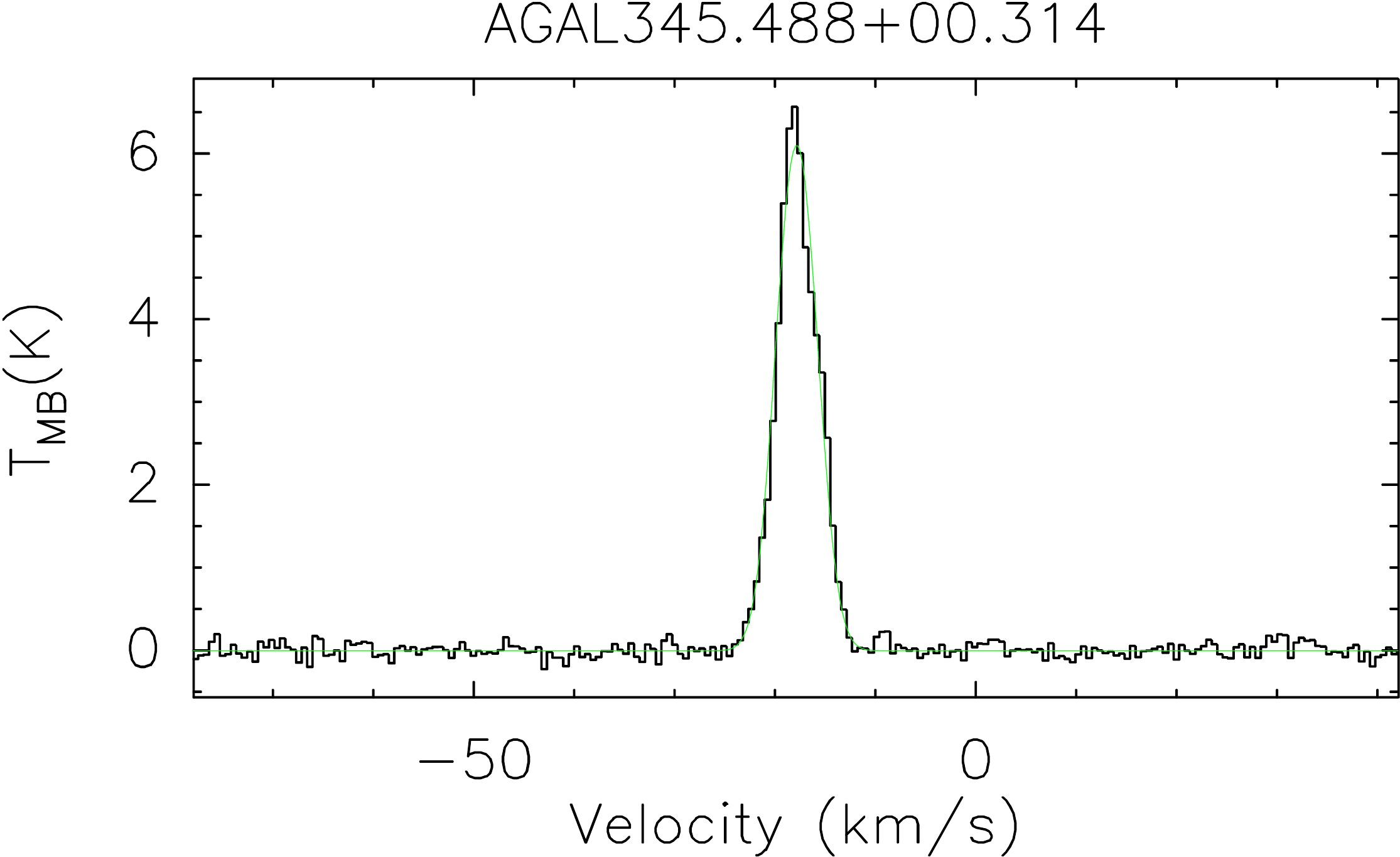} 
\includegraphics[angle=-90,width=0.3\textwidth]{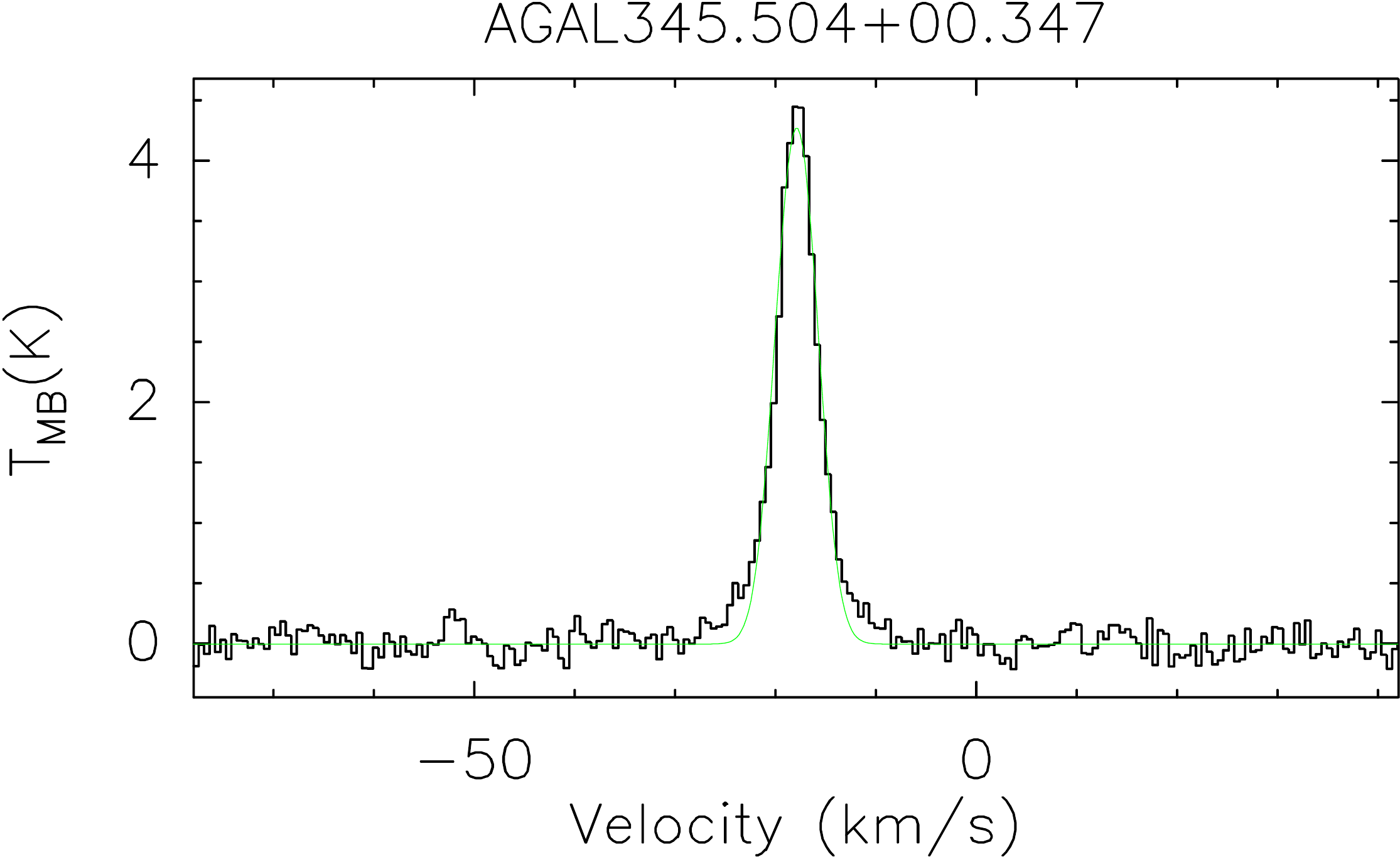} 
\includegraphics[angle=-90,width=0.3\textwidth]{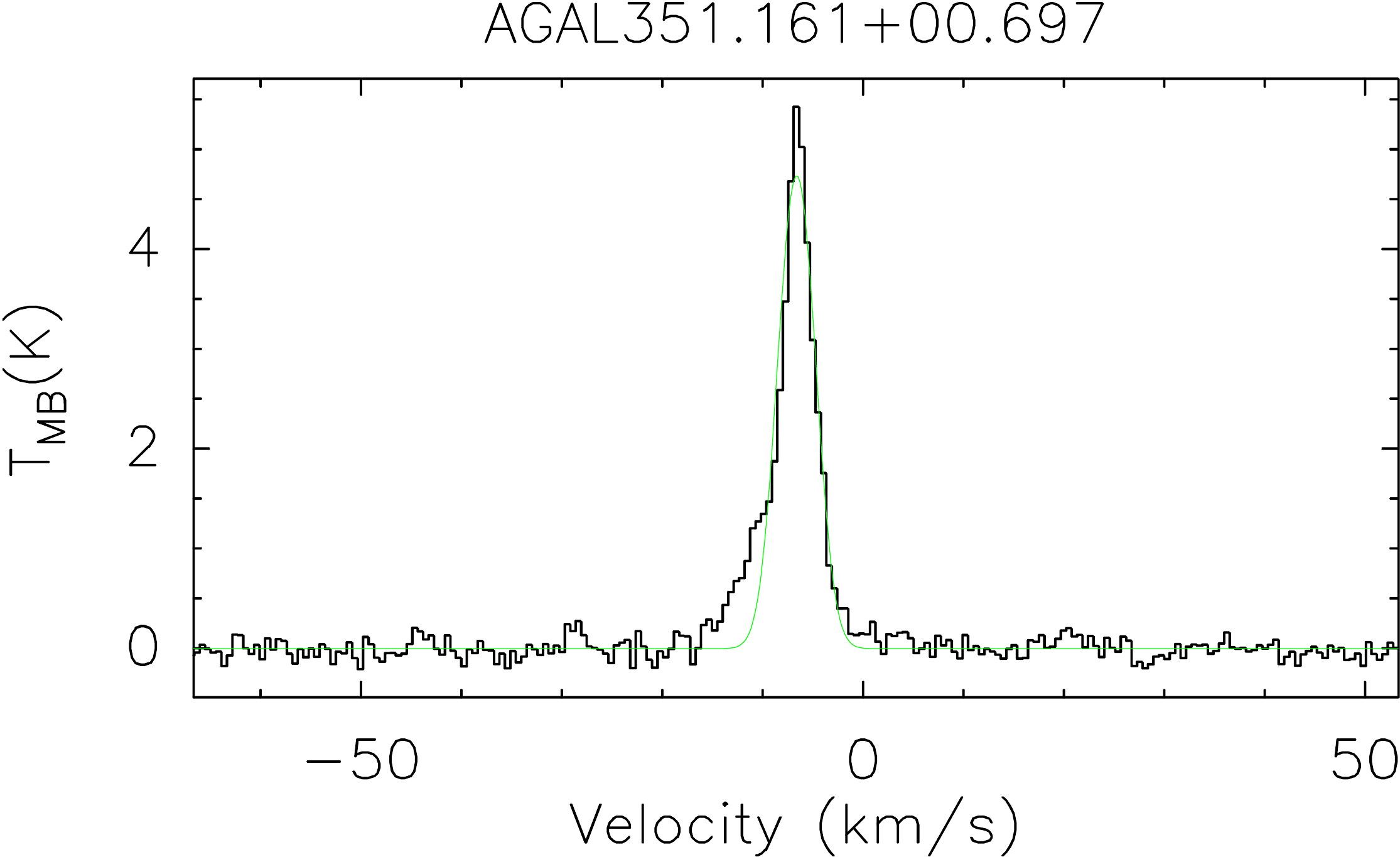} \hfill 
\caption{Continued.} 
\end{figure*} 

\begin{figure*} 
\ContinuedFloat
\centering 
\includegraphics[angle=-90,width=0.3\textwidth]{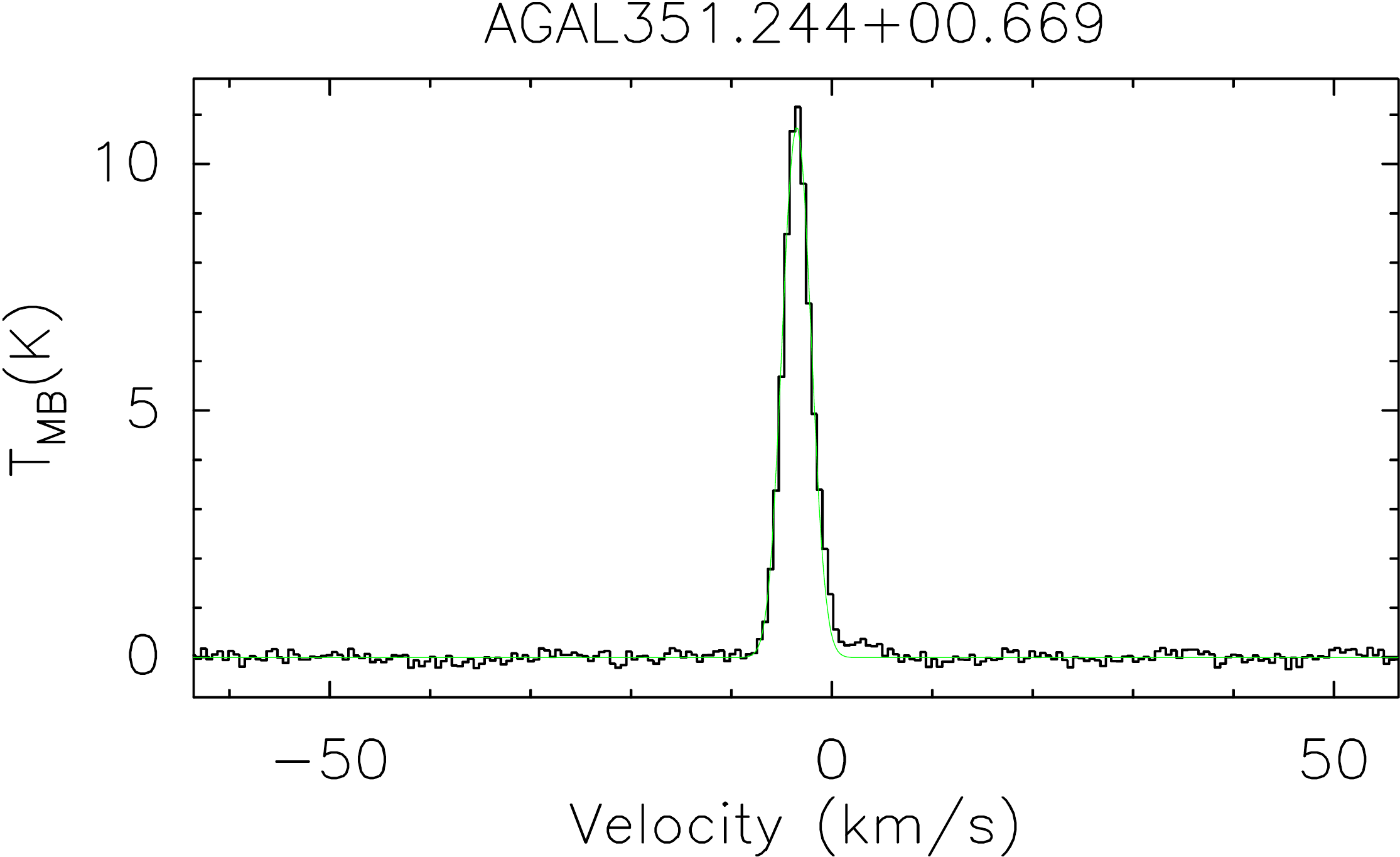} 
\includegraphics[angle=-90,width=0.3\textwidth]{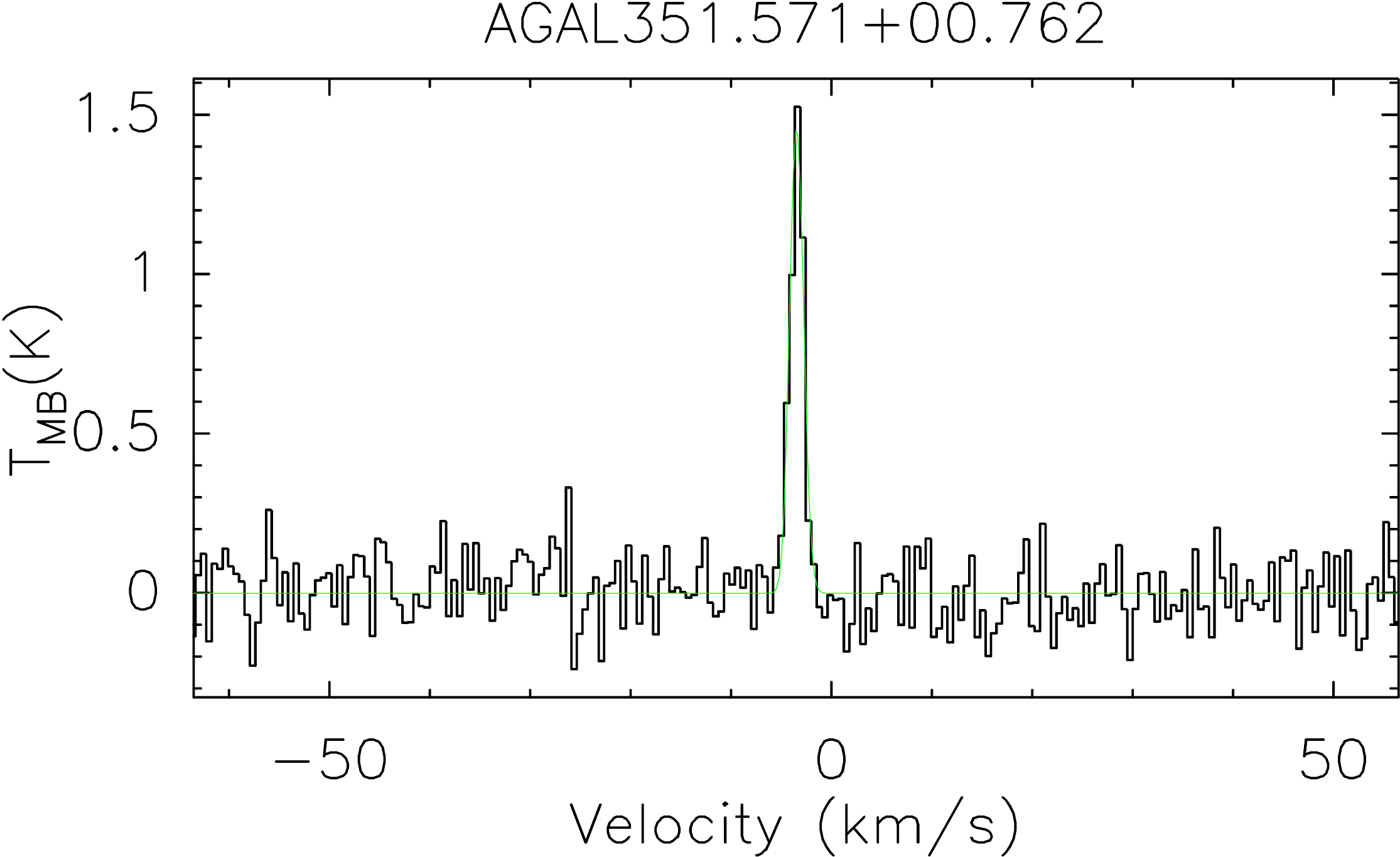} 
\includegraphics[angle=-90,width=0.3\textwidth]{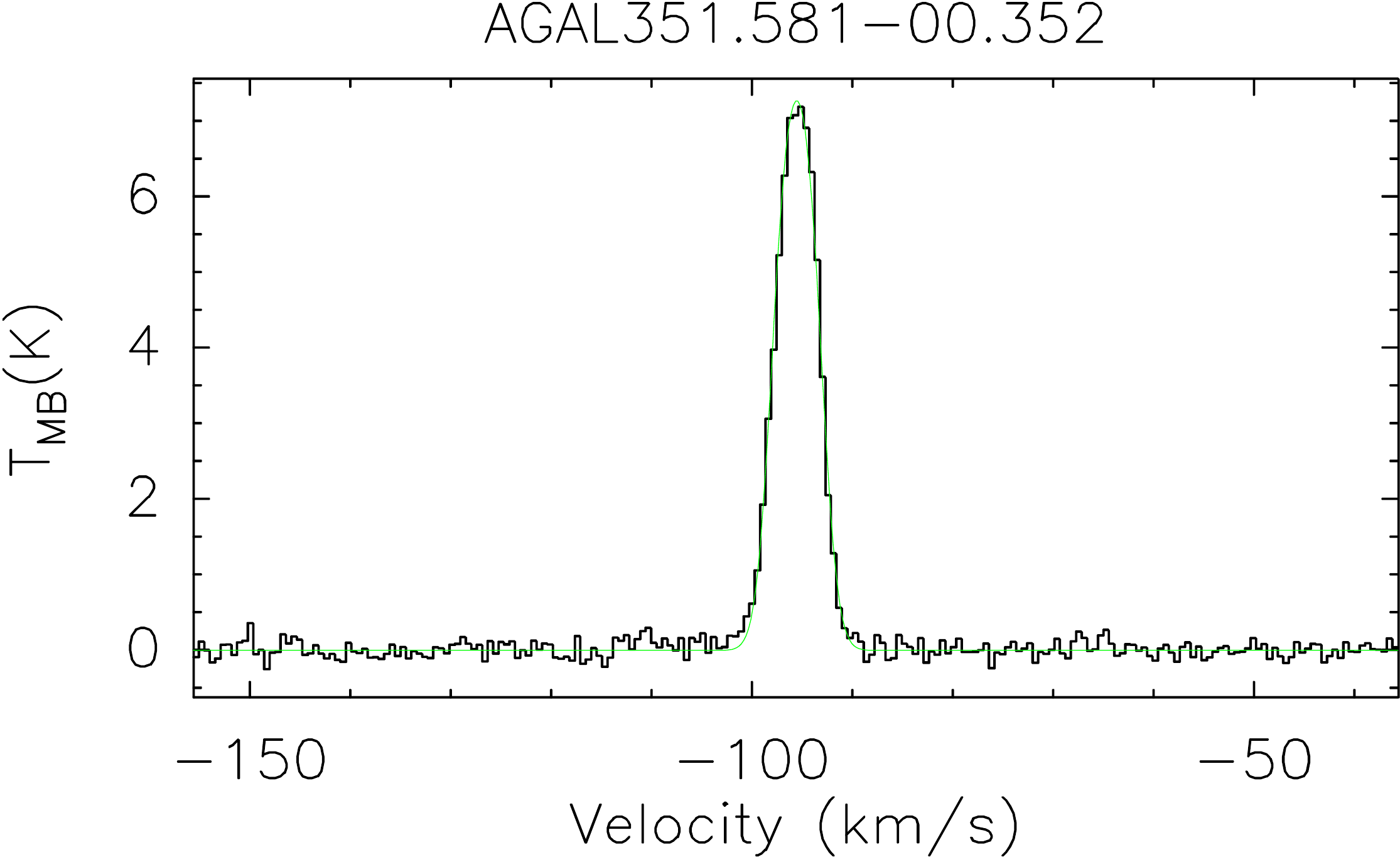} \\ 
\includegraphics[angle=-90,width=0.3\textwidth]{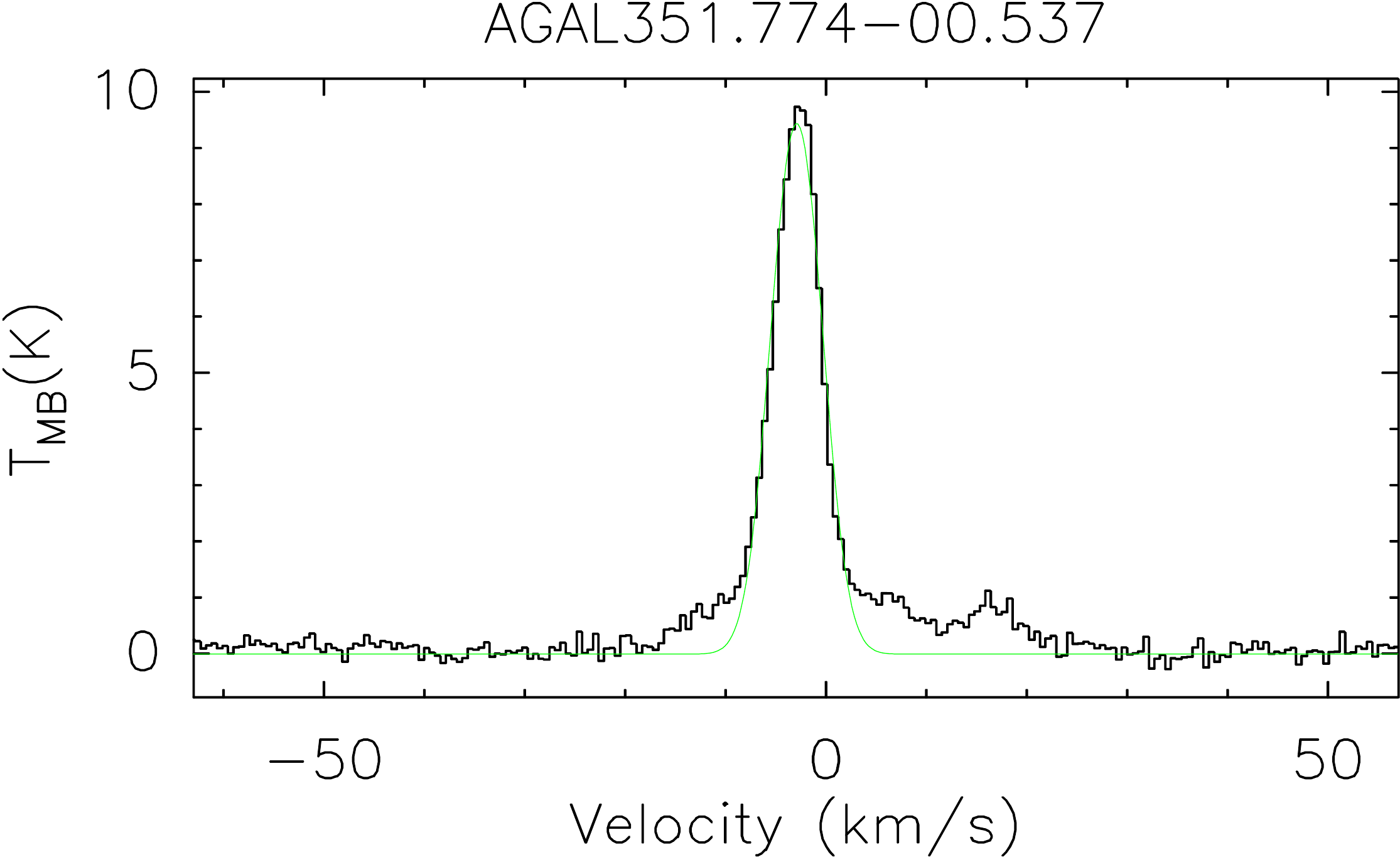} 
\includegraphics[angle=-90,width=0.3\textwidth]{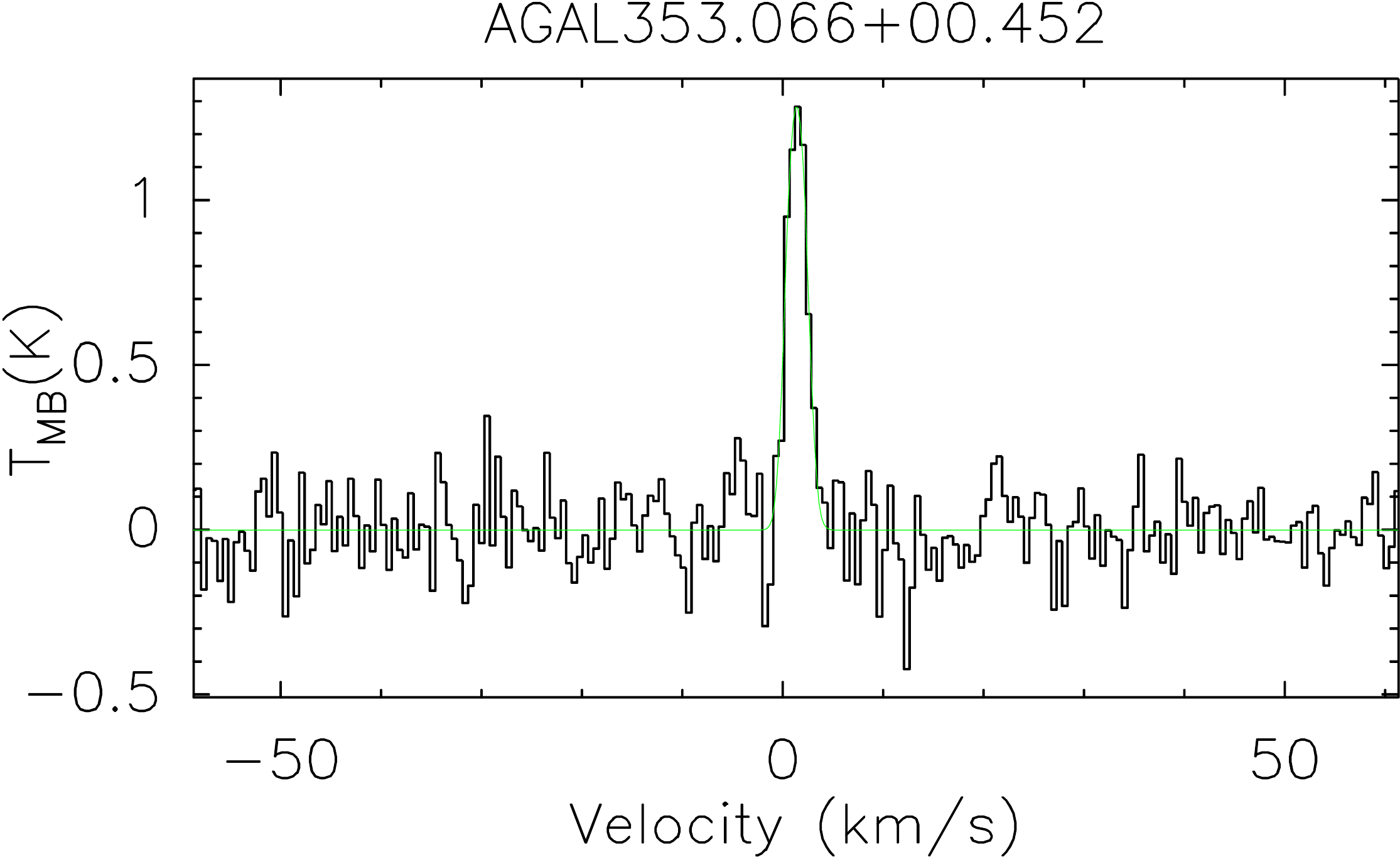} 
\includegraphics[angle=-90,width=0.3\textwidth]{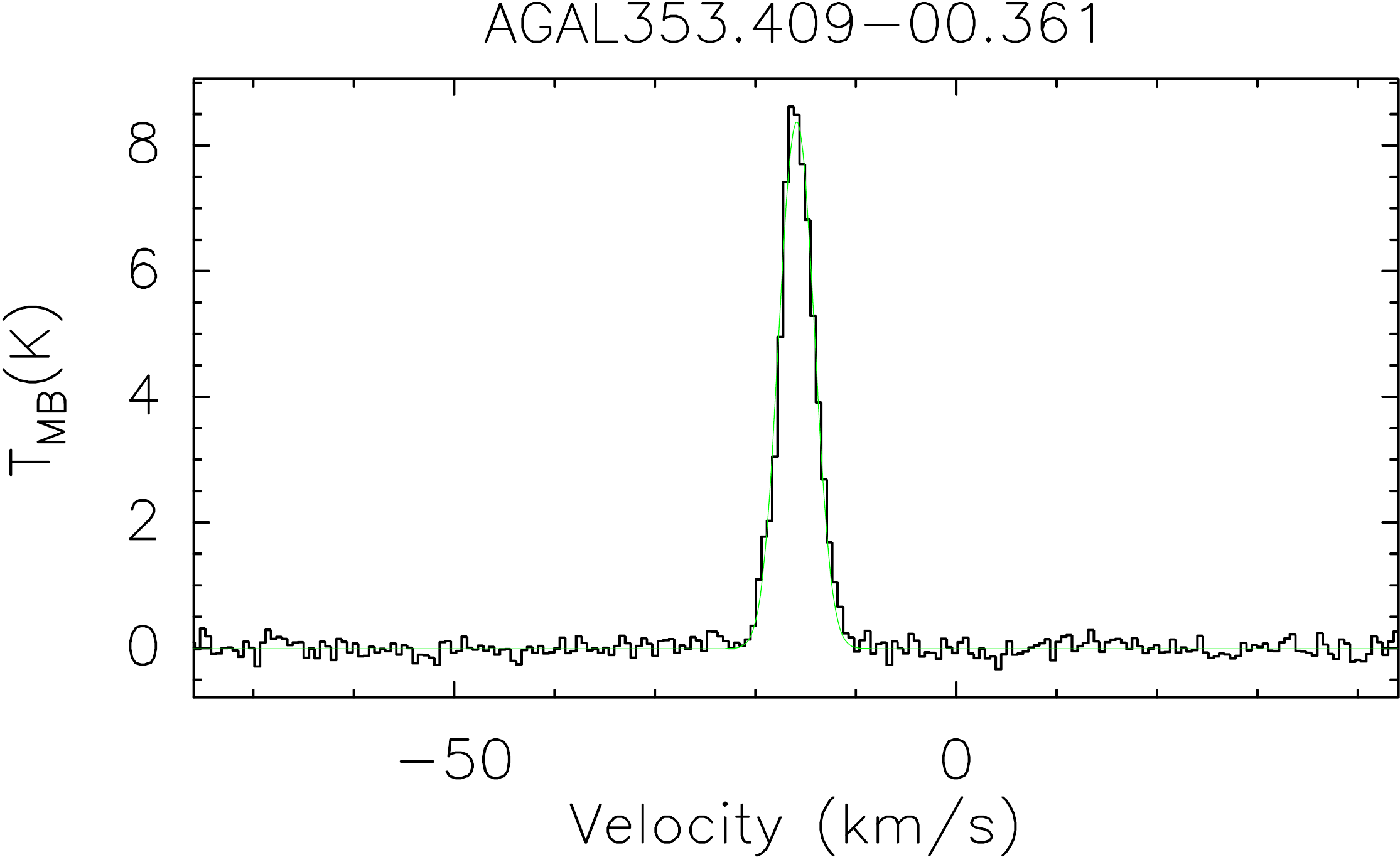} \\  
\includegraphics[angle=-90,width=0.3\textwidth]{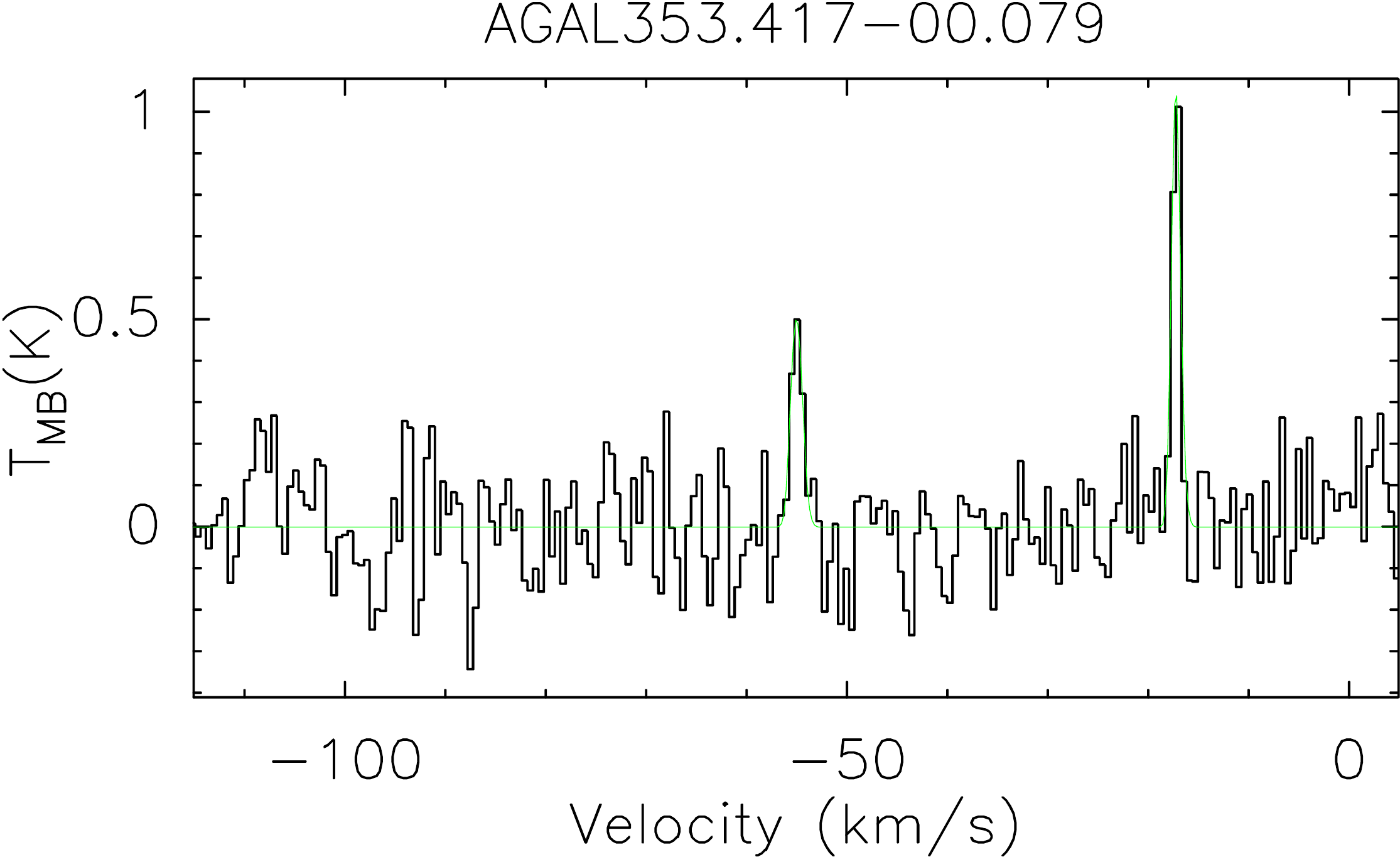} 
\includegraphics[angle=-90,width=0.3\textwidth]{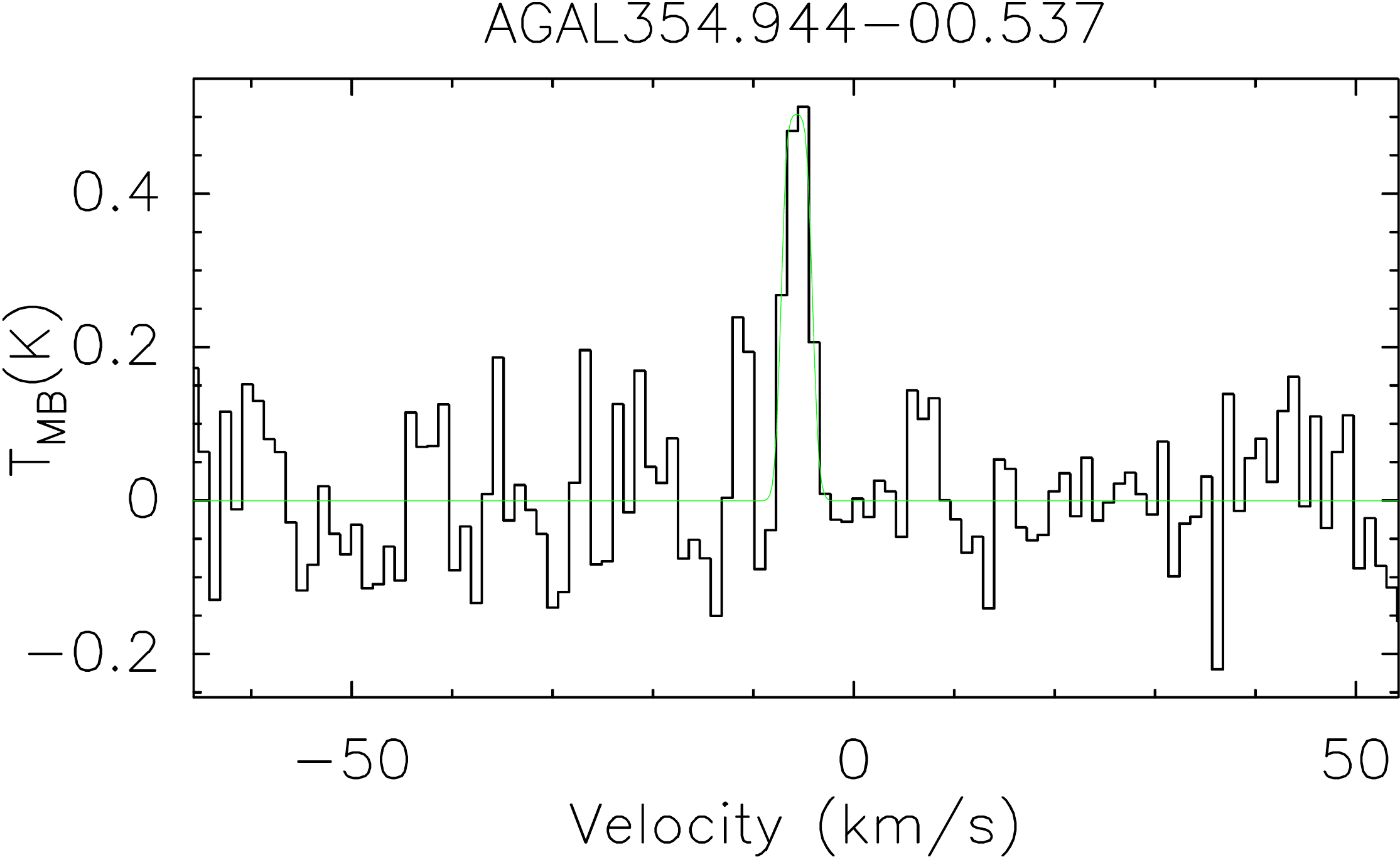} \hfill
\caption{Continued.} 
\end{figure*} 
}
\onlfig{5}{
\begin{figure*} 
\centering 
\includegraphics[angle=-90,width=0.3\textwidth]{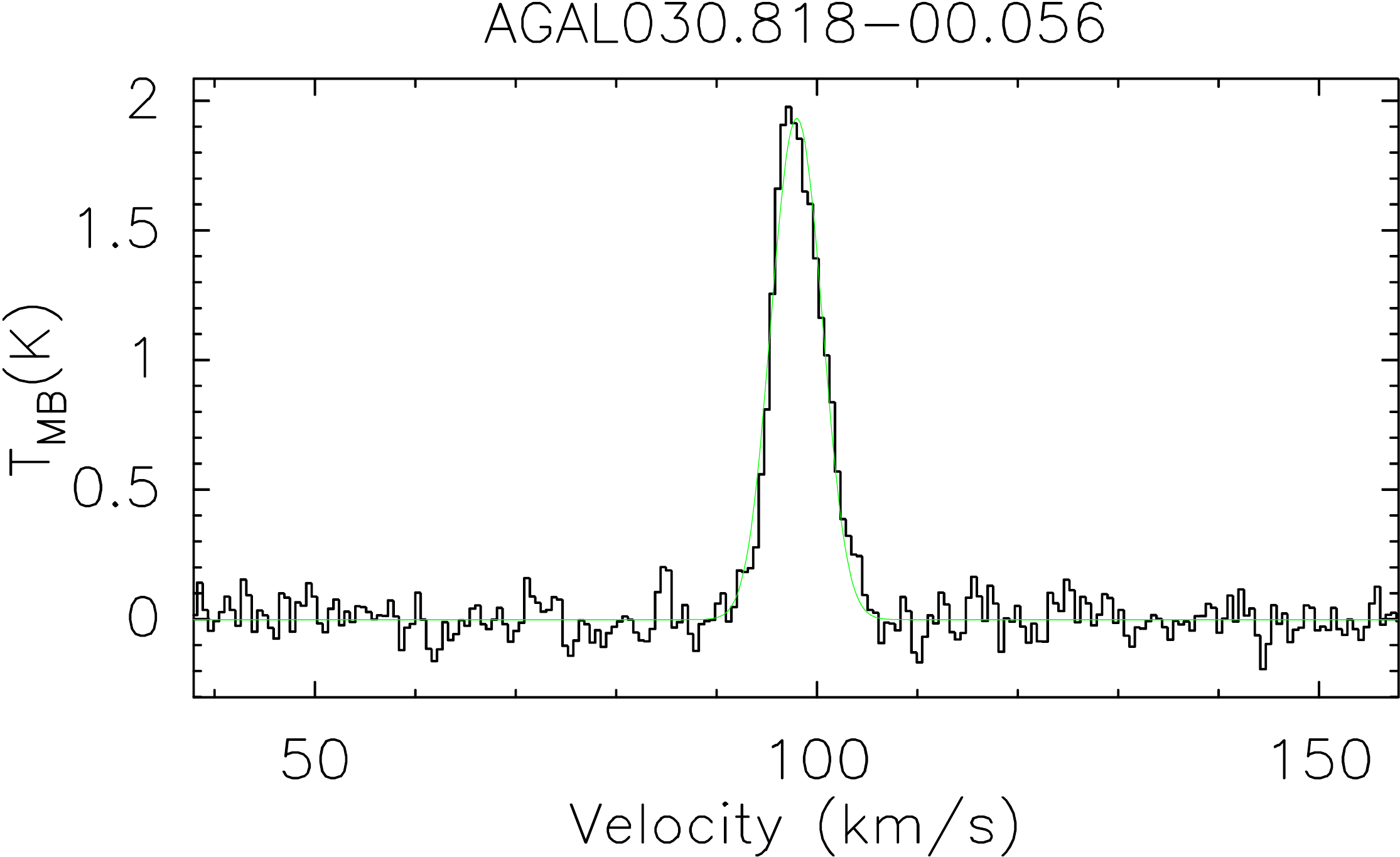} 
\includegraphics[angle=-90,width=0.3\textwidth]{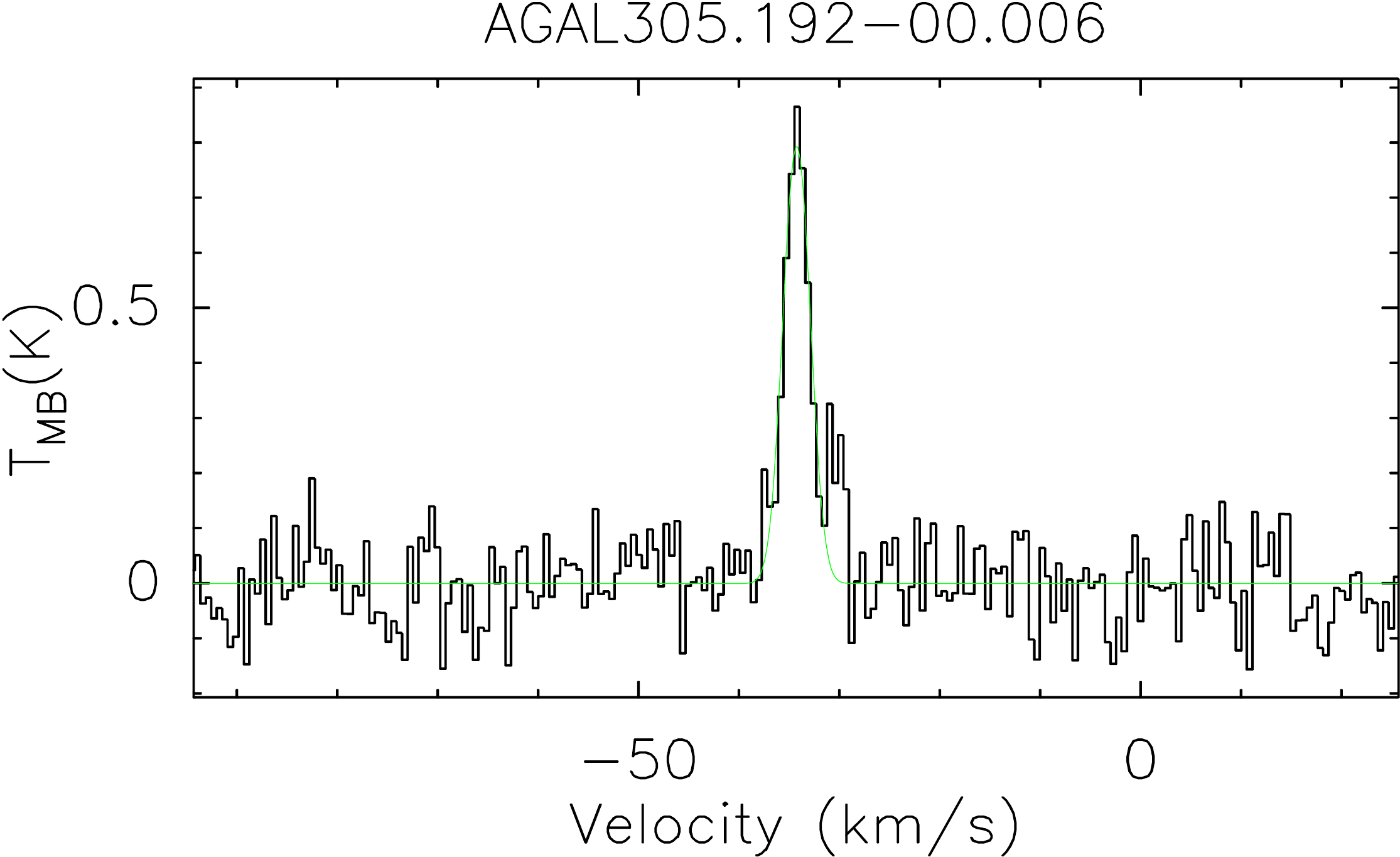} 
\includegraphics[angle=-90,width=0.3\textwidth]{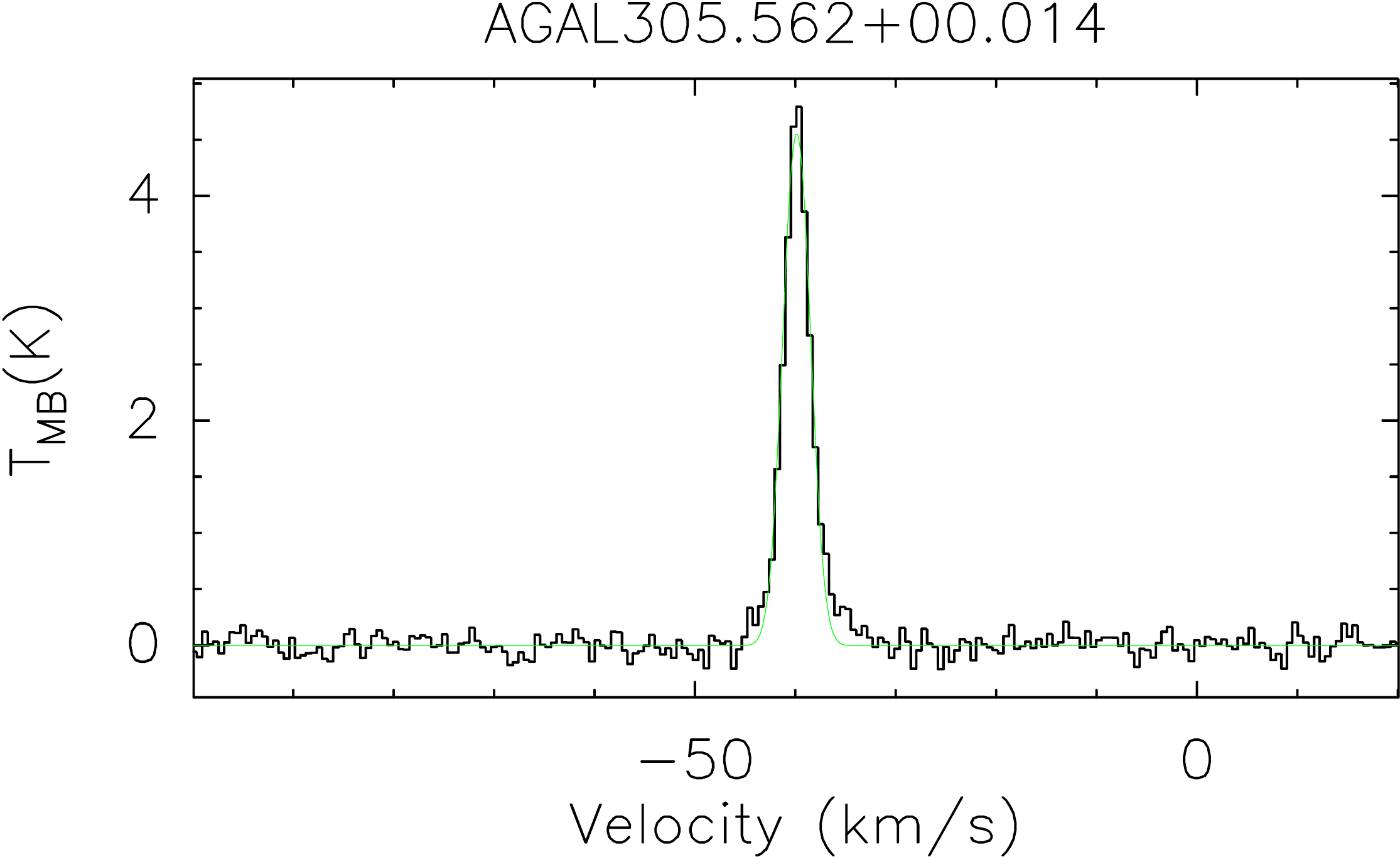} \\ 
\includegraphics[angle=-90,width=0.3\textwidth]{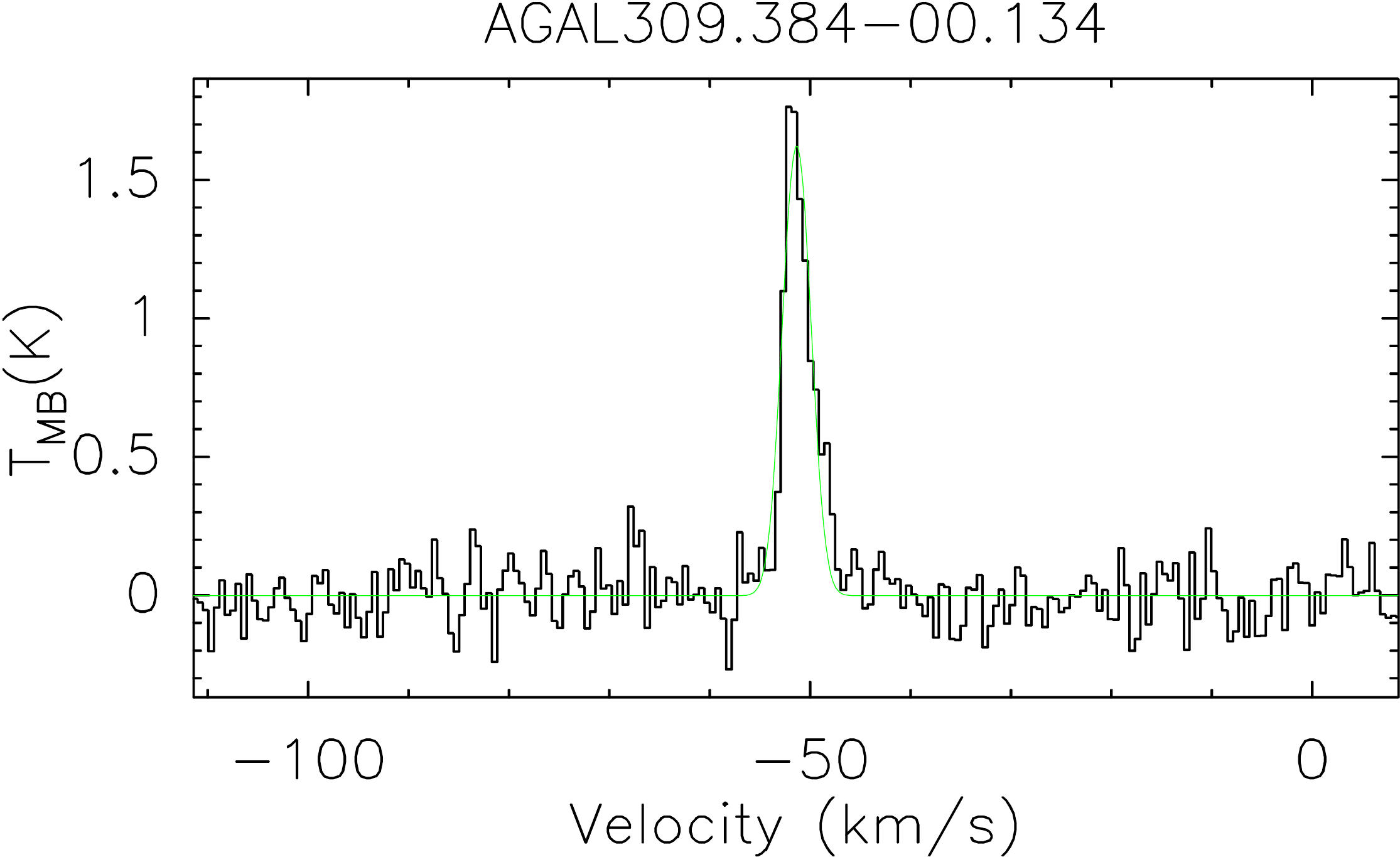} 
\includegraphics[angle=-90,width=0.3\textwidth]{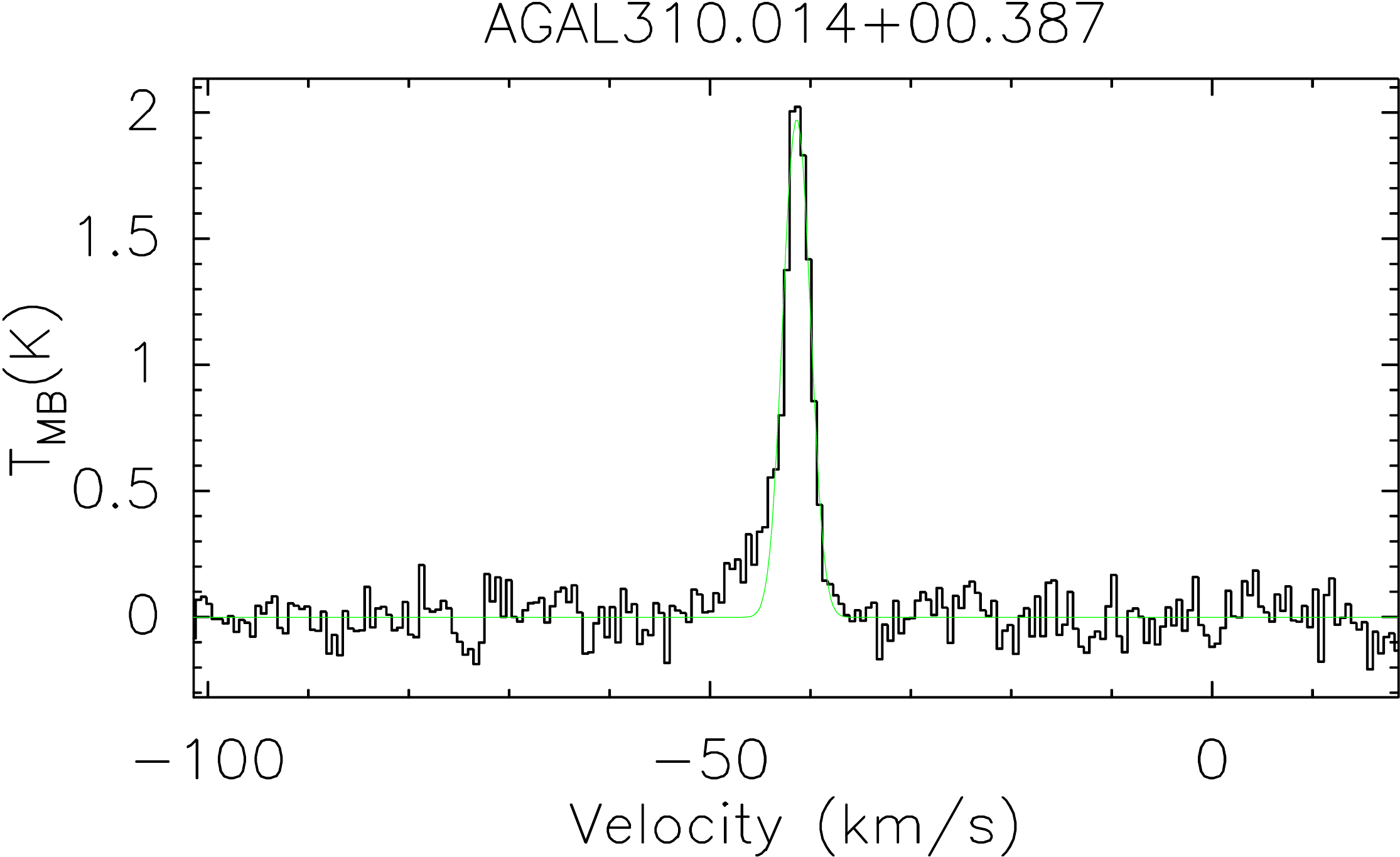} 
\includegraphics[angle=-90,width=0.3\textwidth]{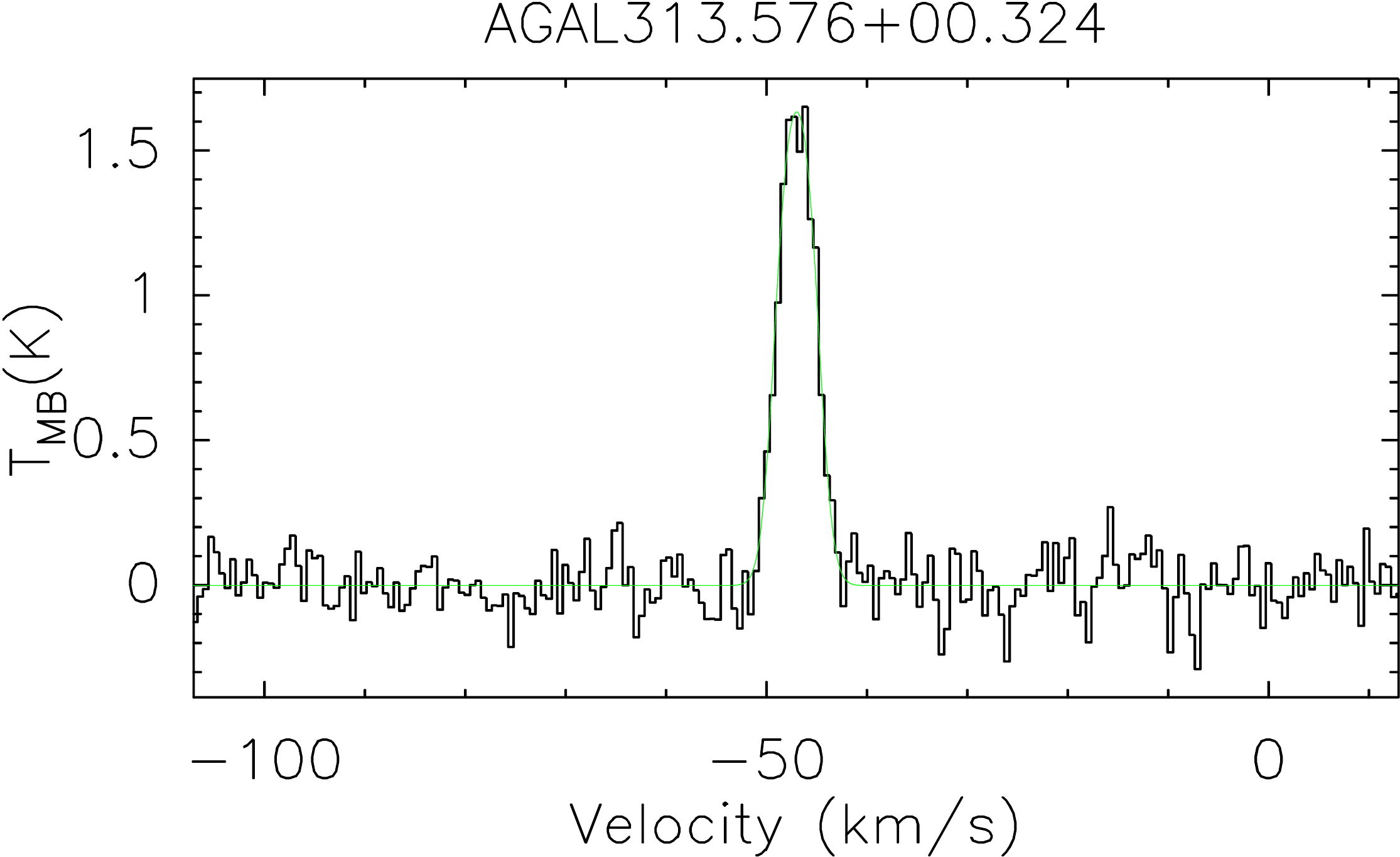} \\ 
\includegraphics[angle=-90,width=0.3\textwidth]{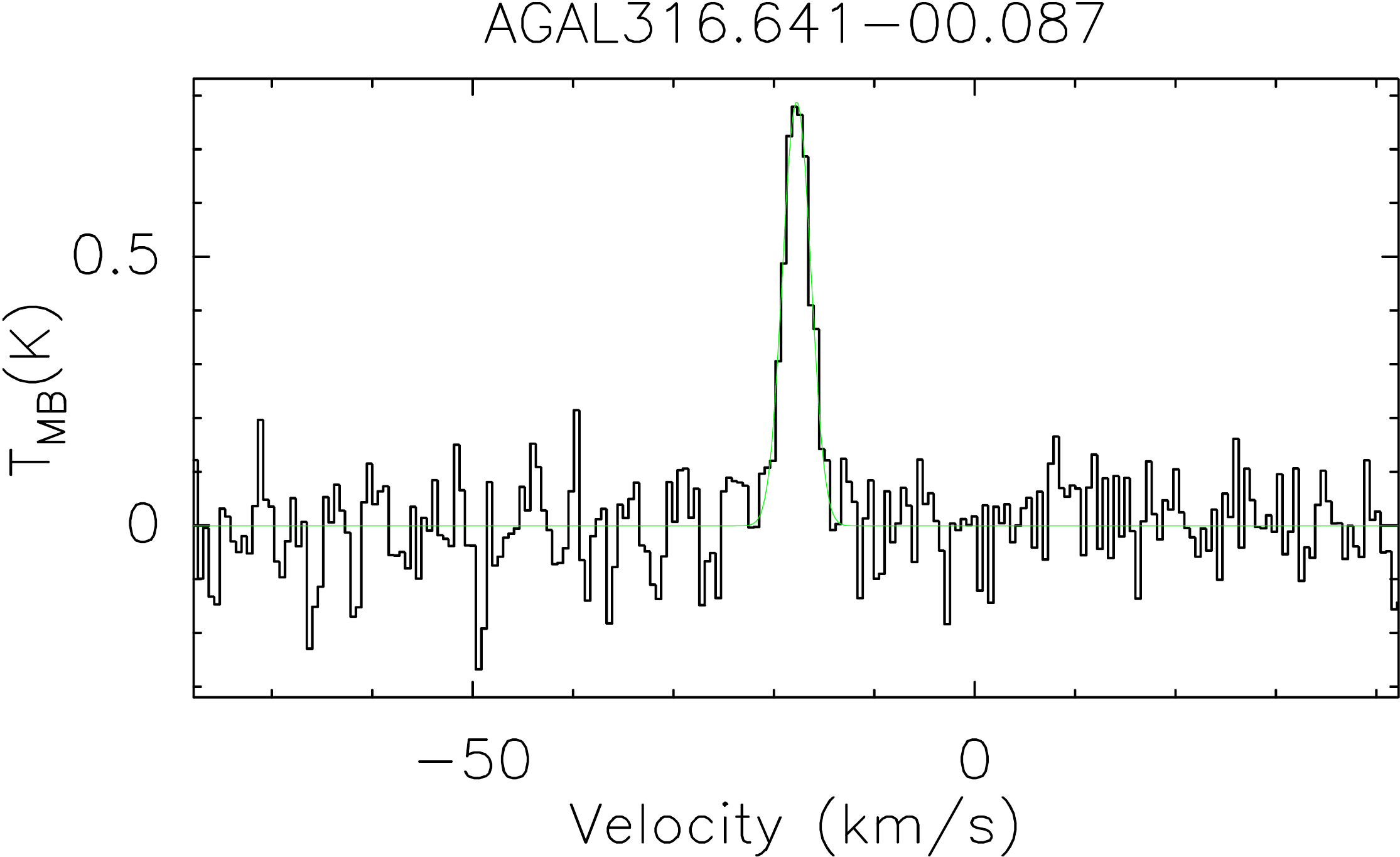} 
\includegraphics[angle=-90,width=0.3\textwidth]{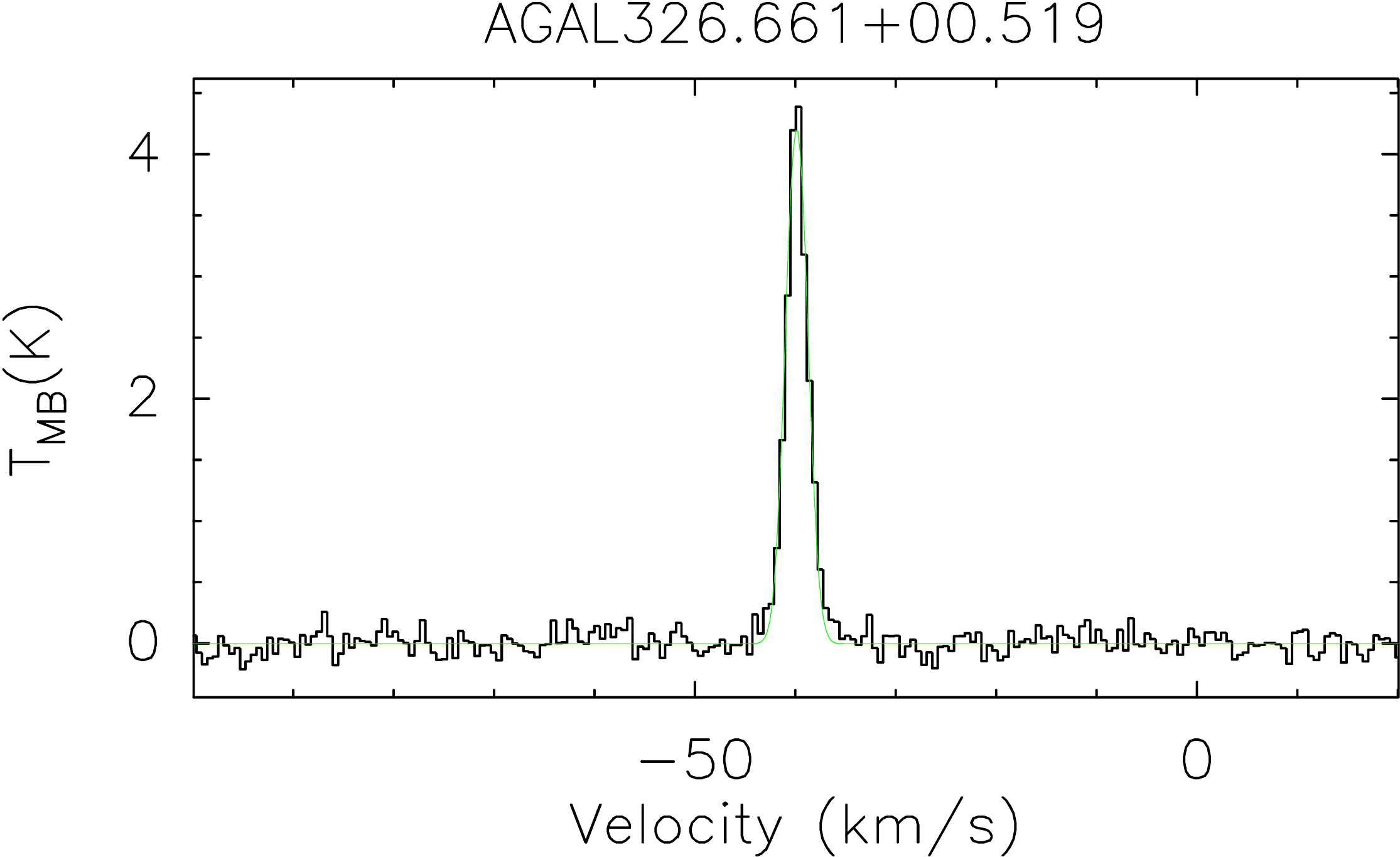} 
\includegraphics[angle=-90,width=0.3\textwidth]{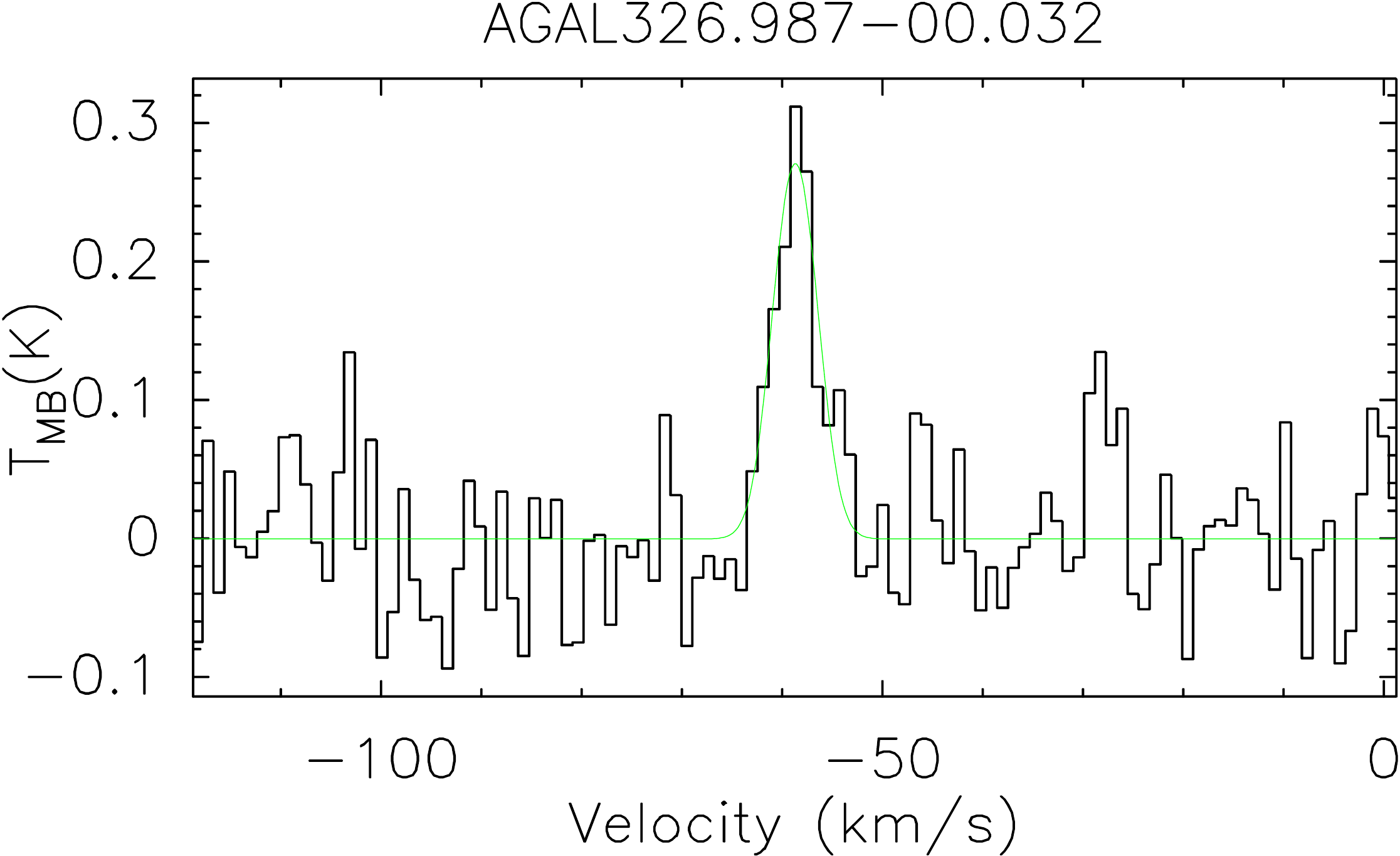} \\ 
\includegraphics[angle=-90,width=0.3\textwidth]{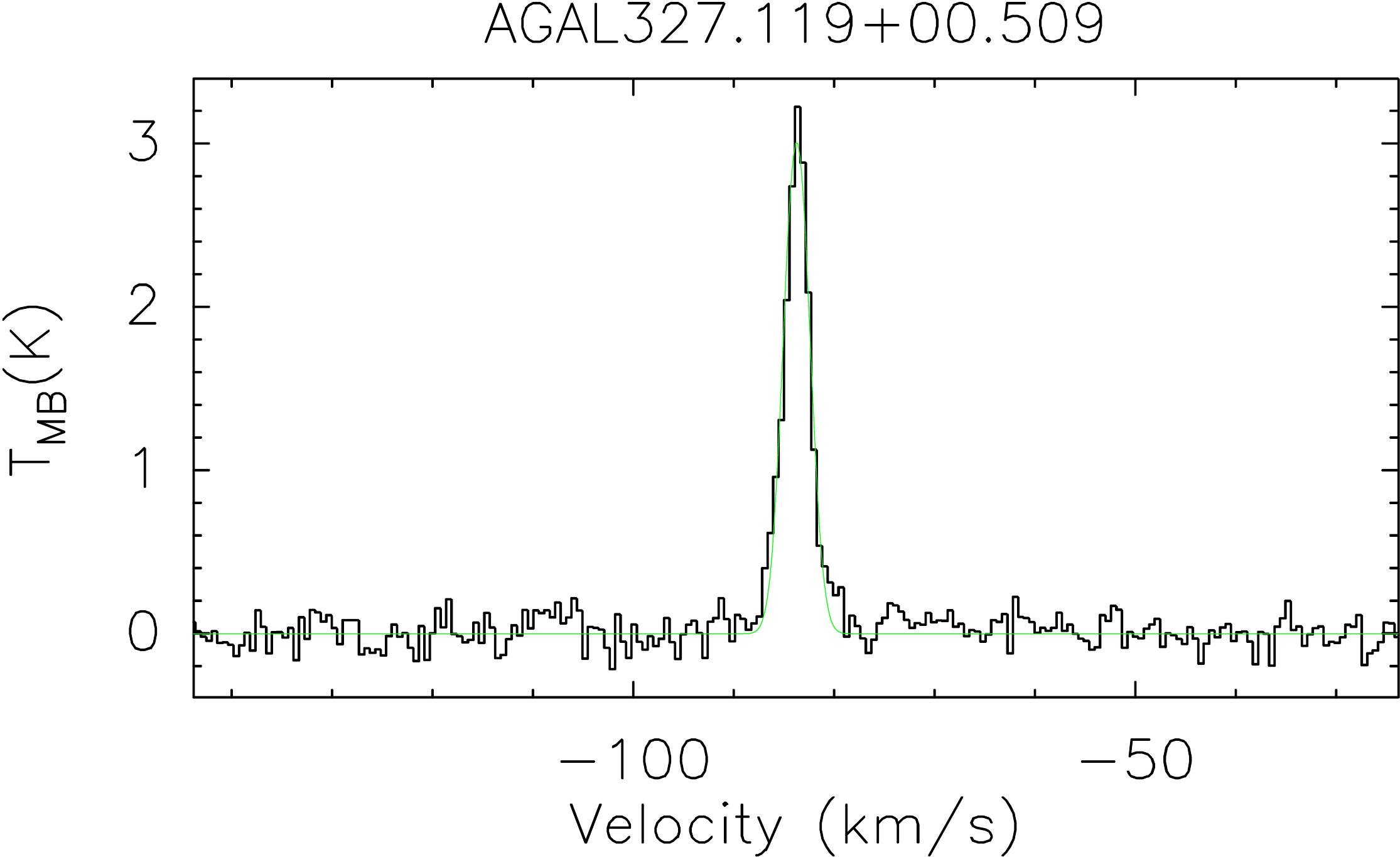} 
\includegraphics[angle=-90,width=0.3\textwidth]{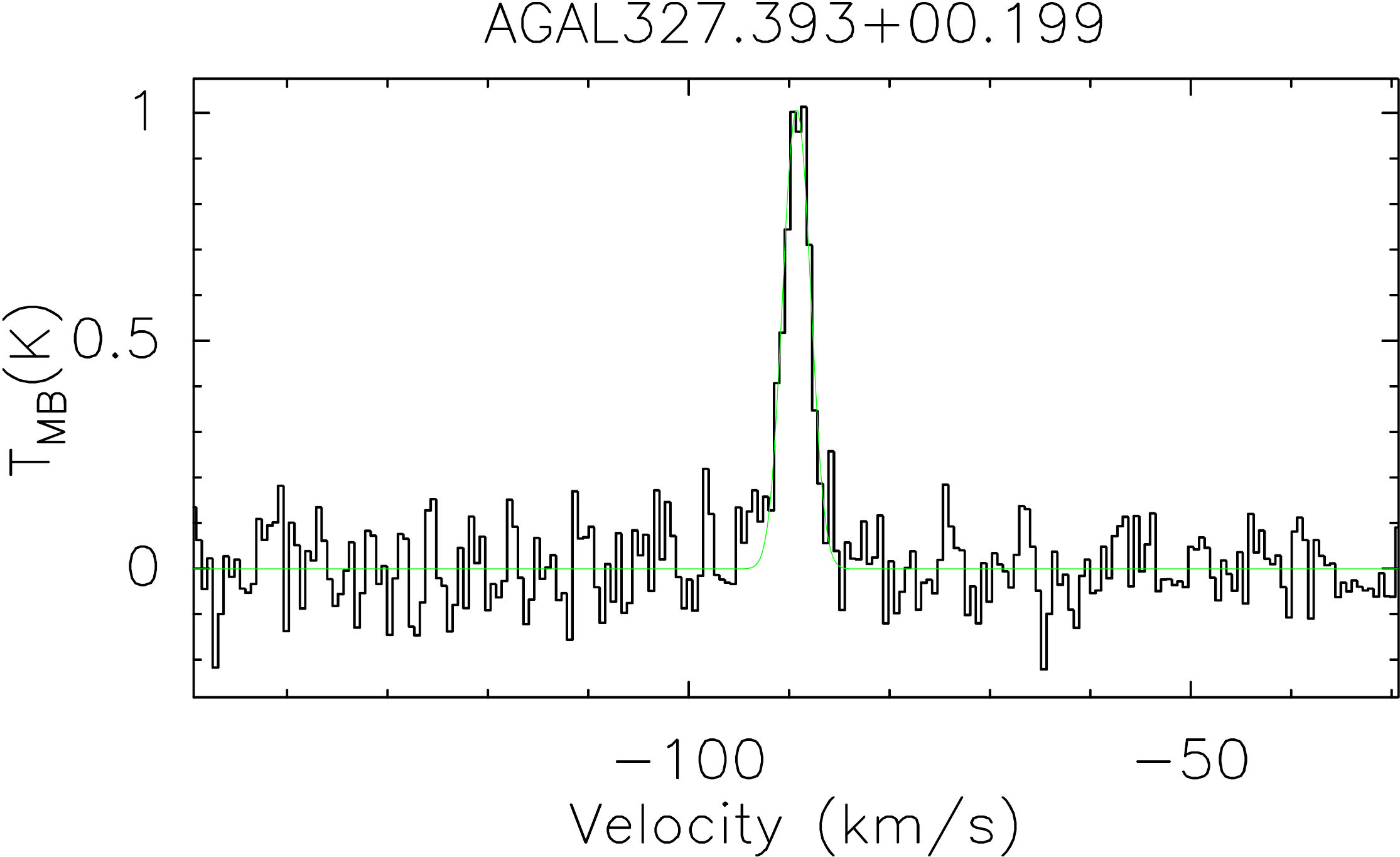} 
\includegraphics[angle=-90,width=0.3\textwidth]{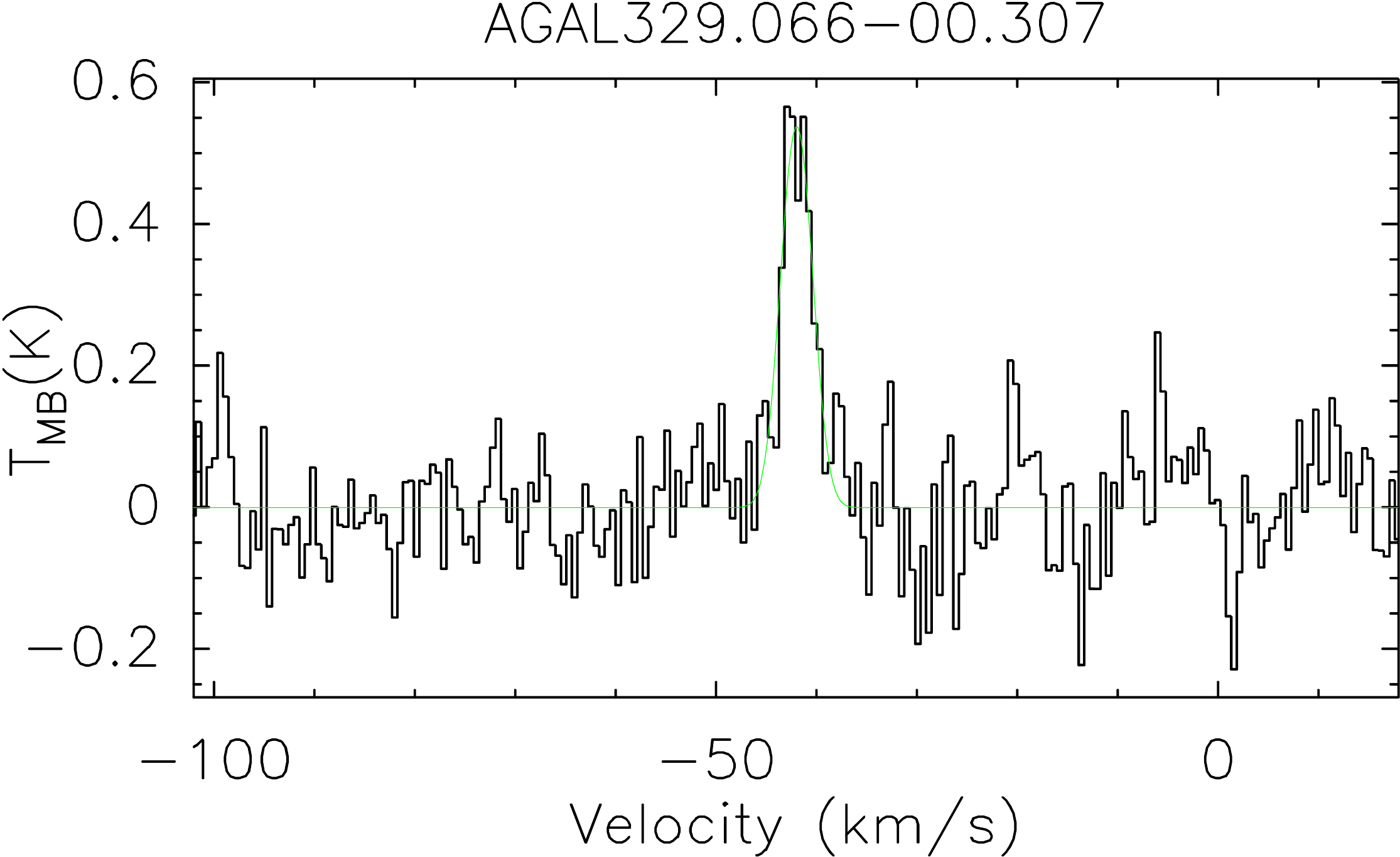} \\ 
\includegraphics[angle=-90,width=0.3\textwidth]{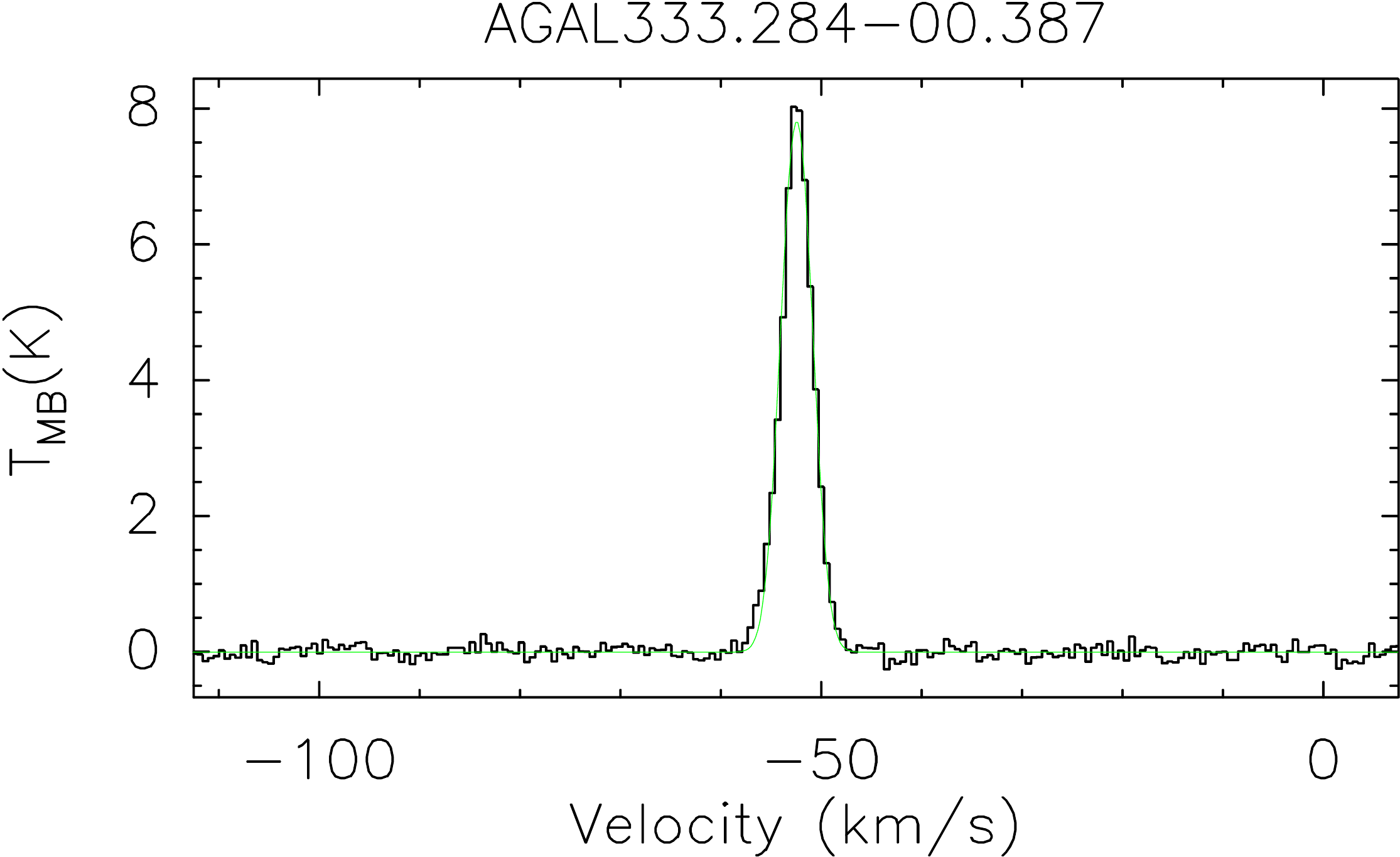} 
\includegraphics[angle=-90,width=0.3\textwidth]{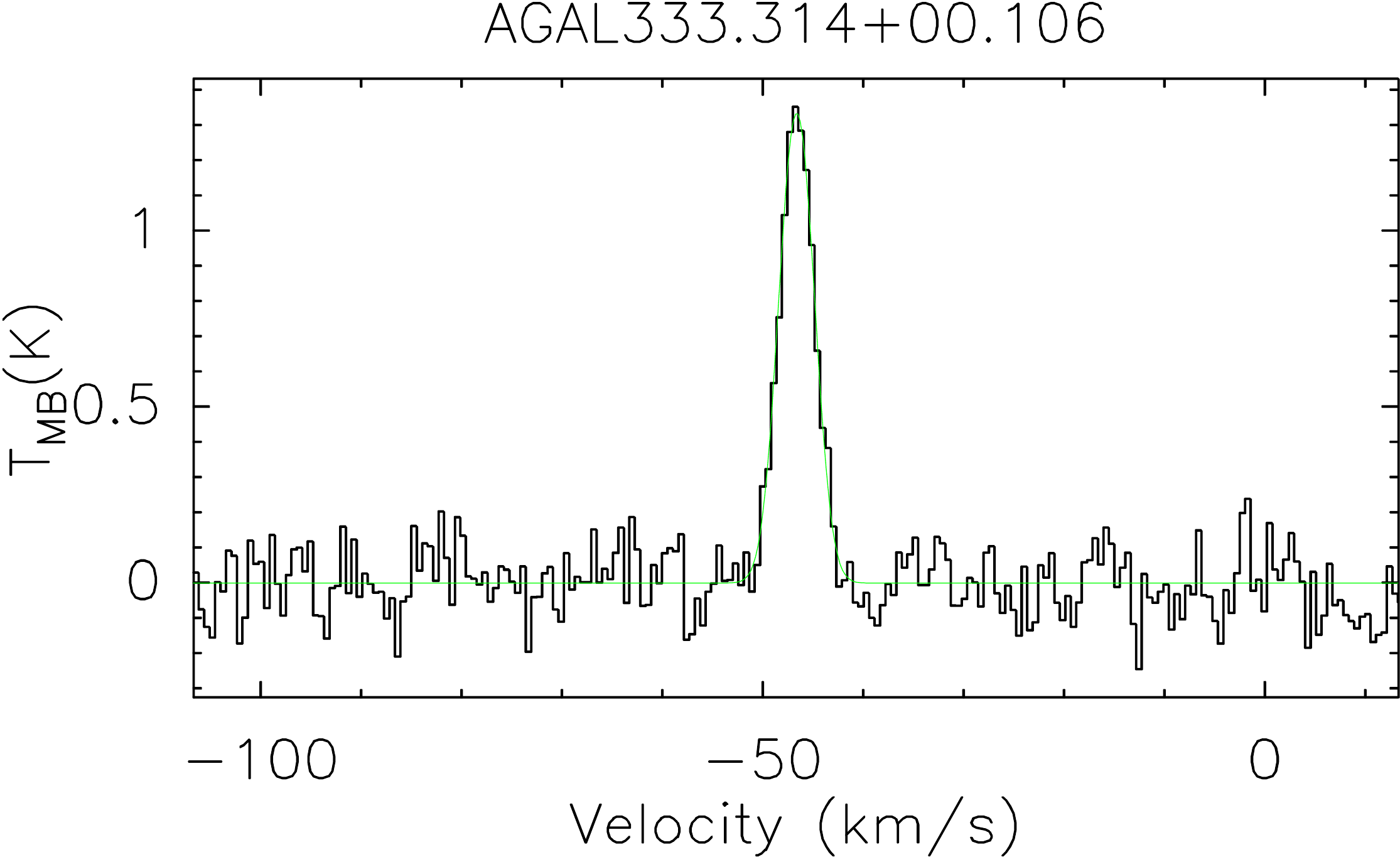} 
\includegraphics[angle=-90,width=0.3\textwidth]{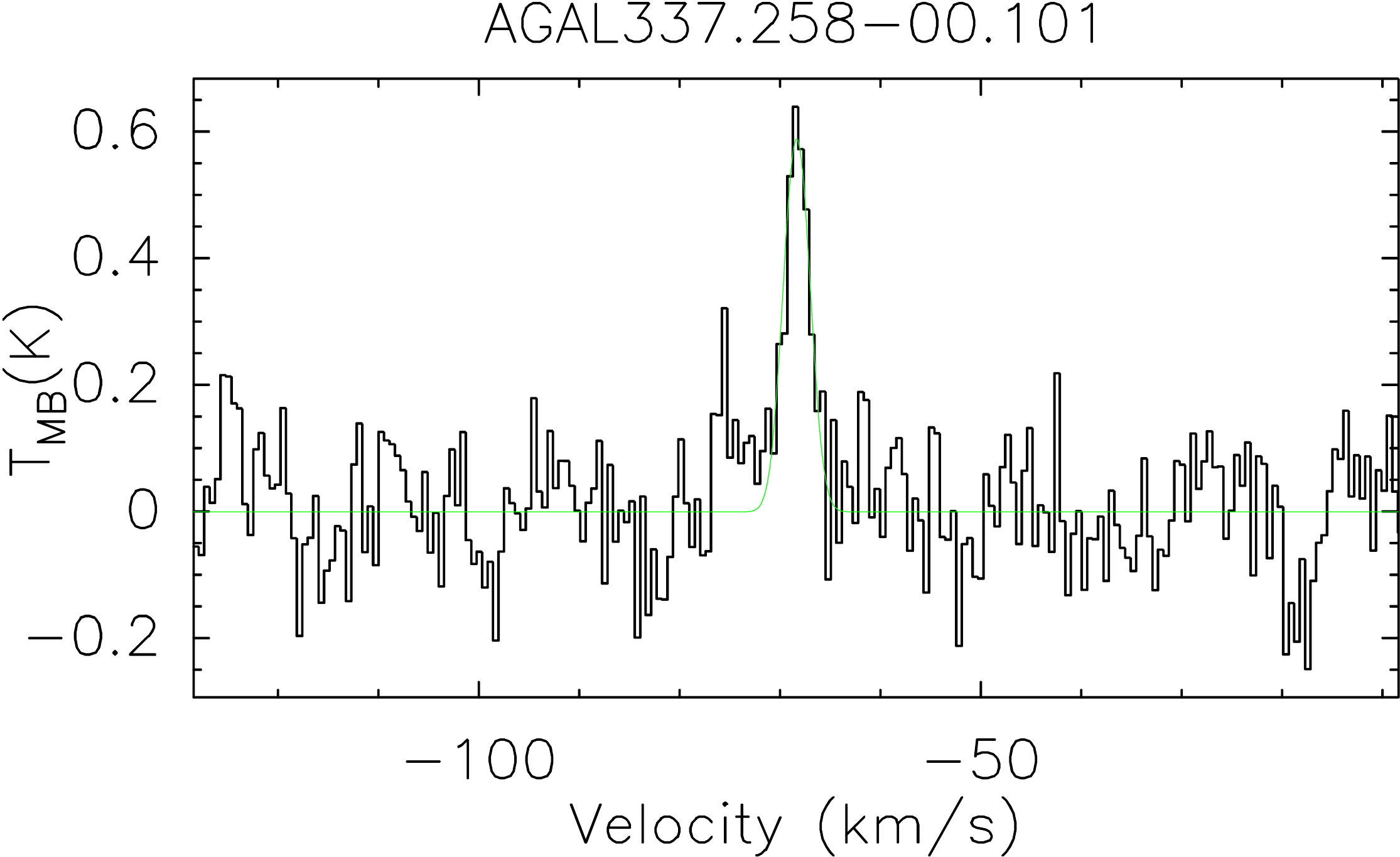} \\ 
\includegraphics[angle=-90,width=0.3\textwidth]{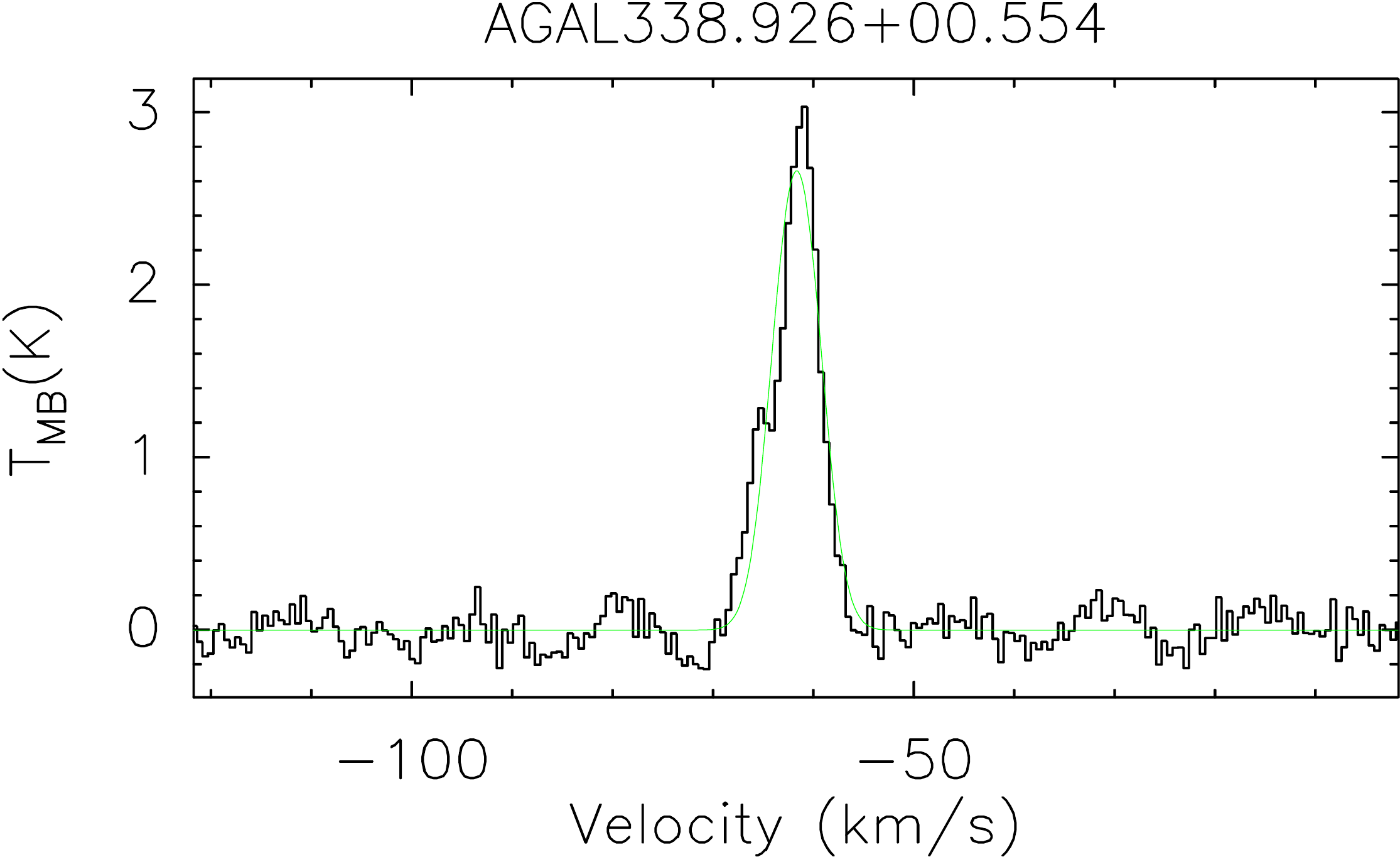} 
\includegraphics[angle=-90,width=0.3\textwidth]{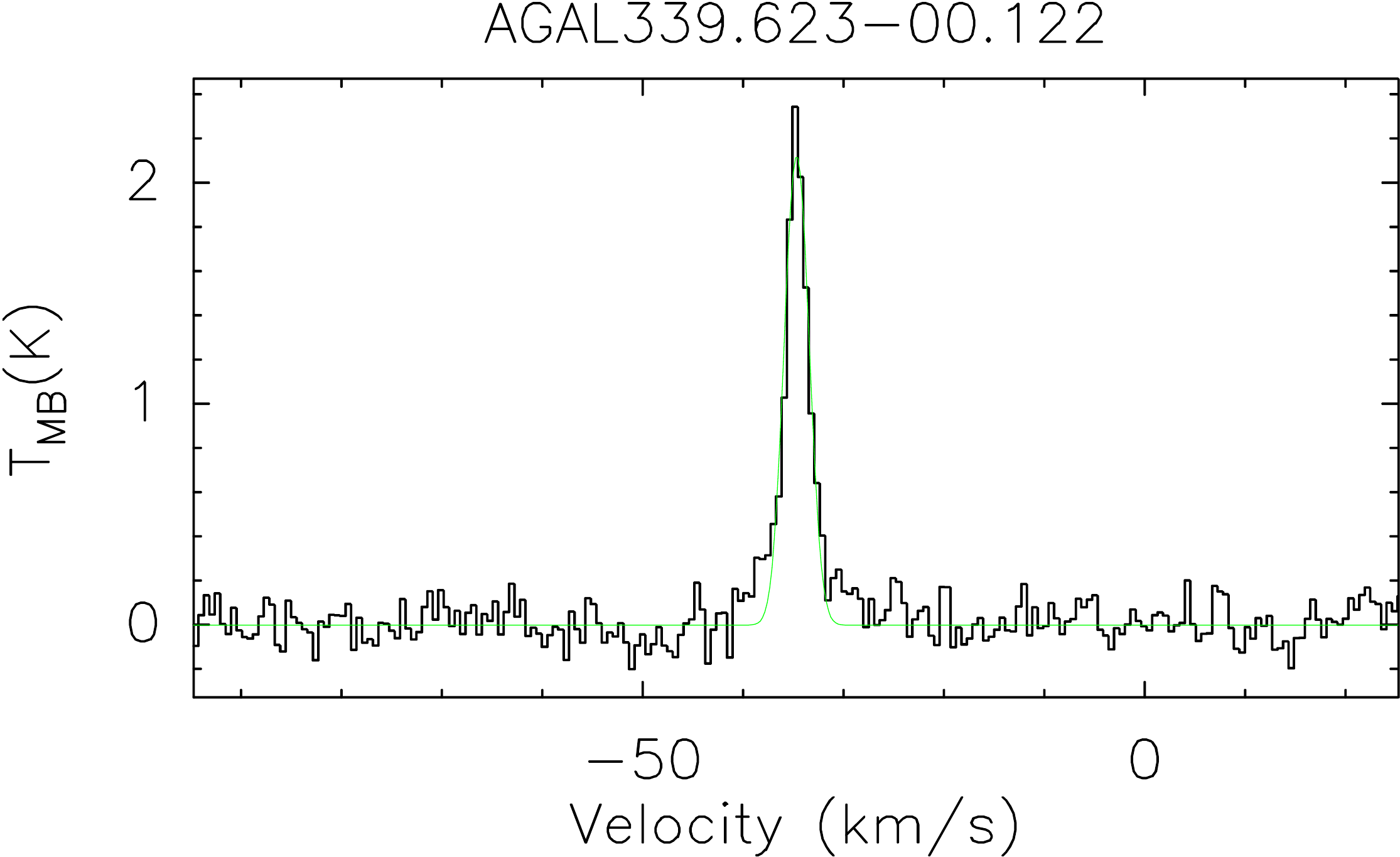} 
\includegraphics[angle=-90,width=0.3\textwidth]{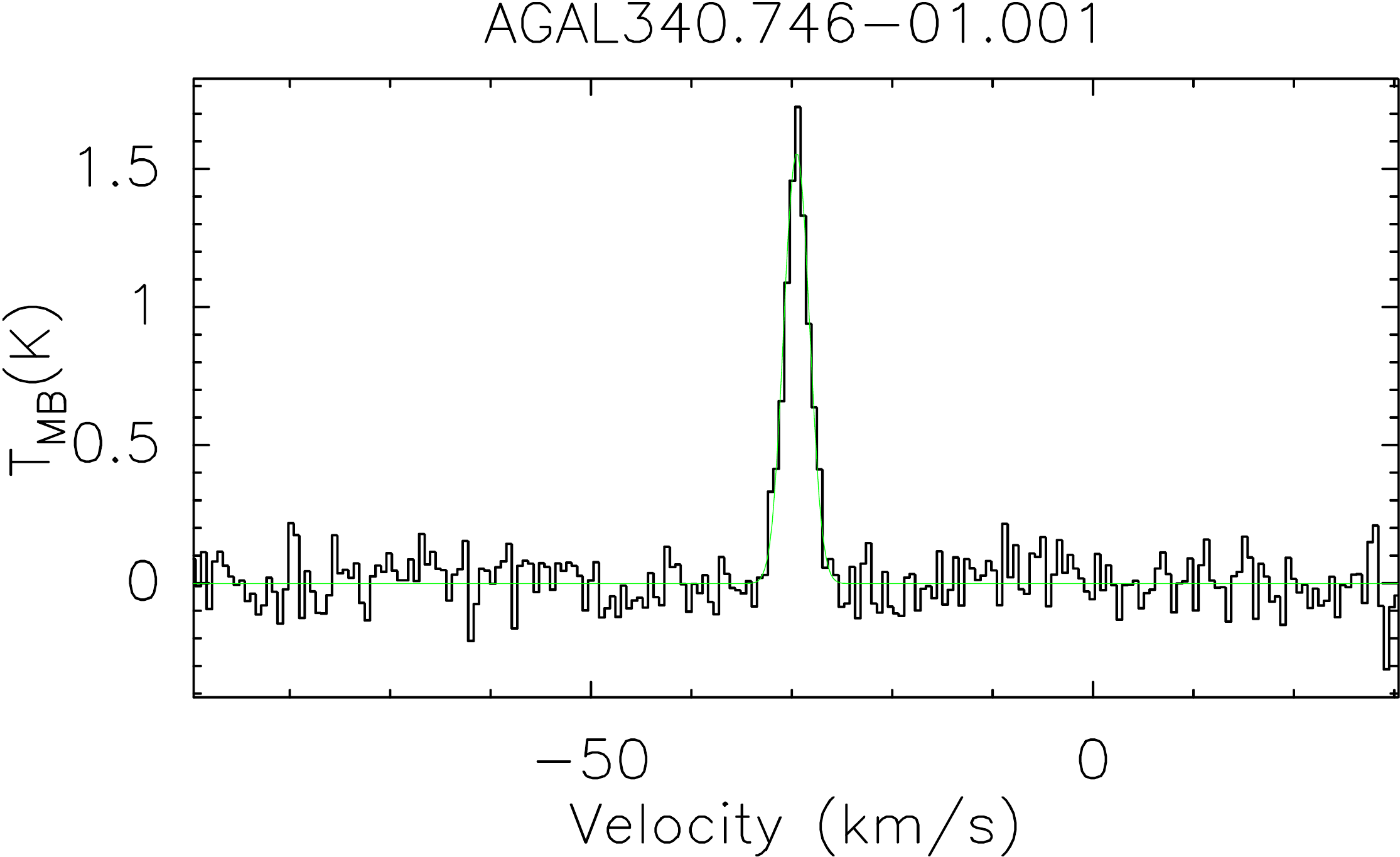} \hfill 
\caption{C$^{17}$O$(3-2)$ for subsample S3. The fit is shown in green.} \label{fig:spectra_32_C}
\end{figure*} 

\begin{figure*} 
\ContinuedFloat
\centering 
\includegraphics[angle=-90,width=0.3\textwidth]{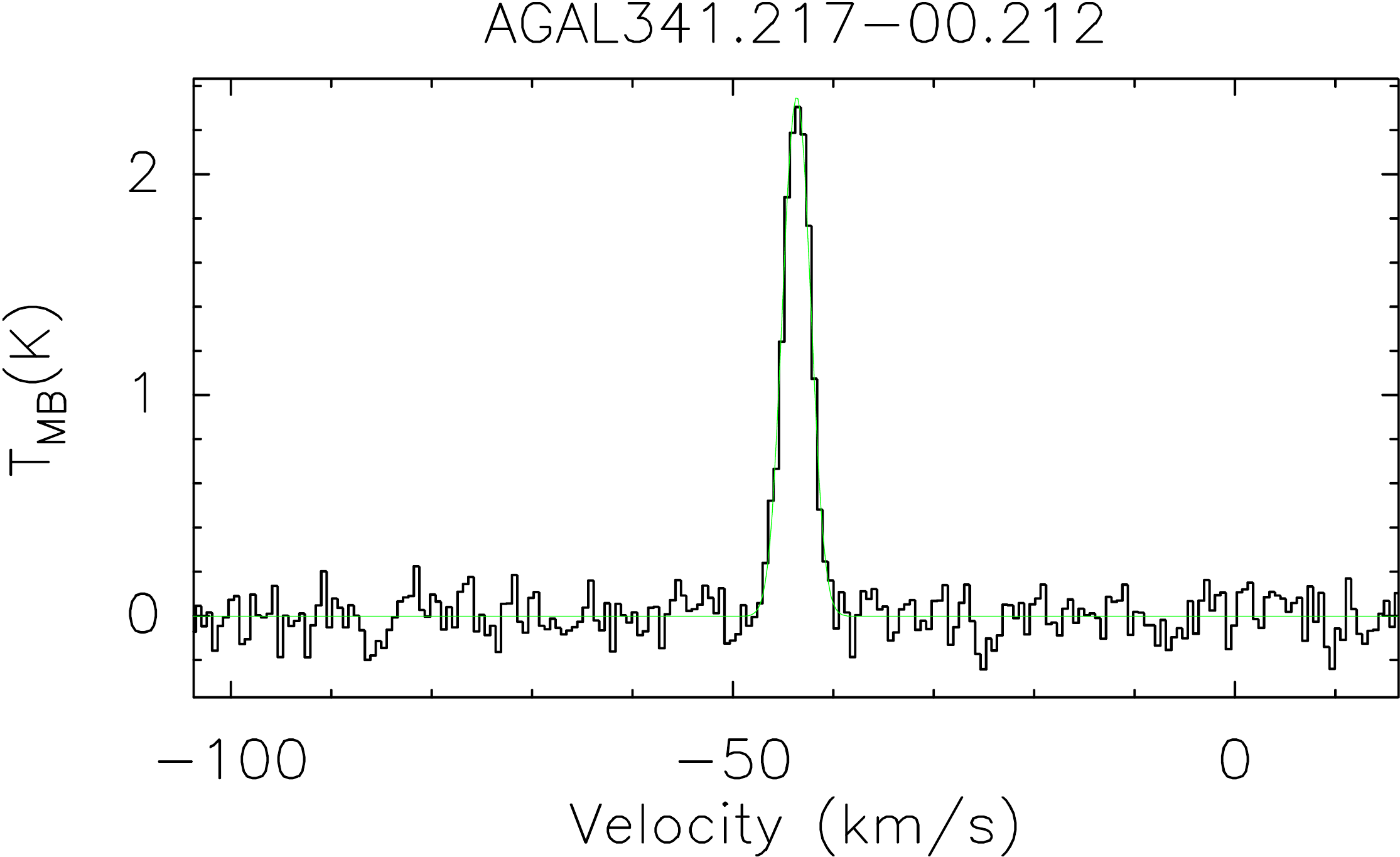} 
\includegraphics[angle=-90,width=0.3\textwidth]{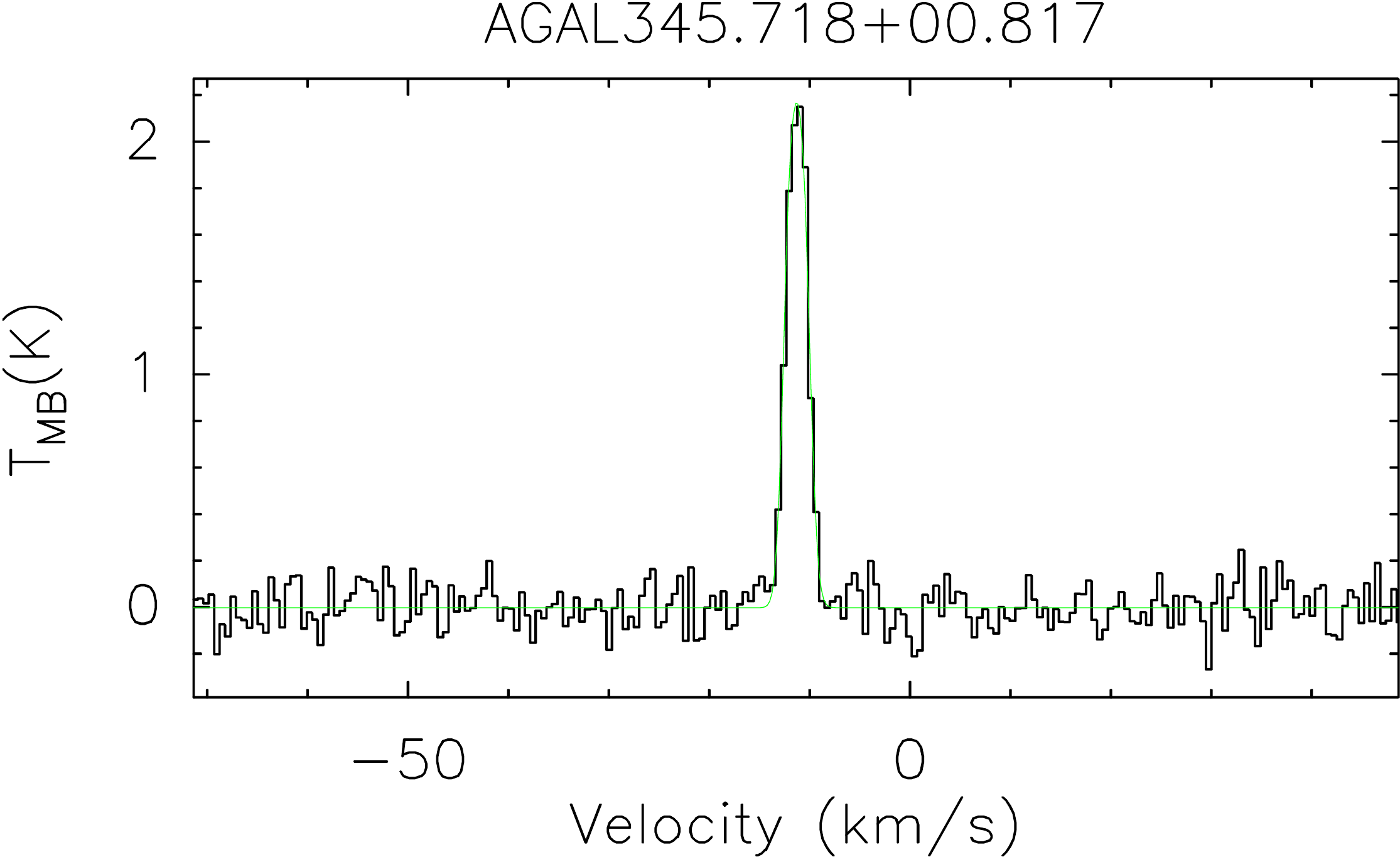} 
\includegraphics[angle=-90,width=0.3\textwidth]{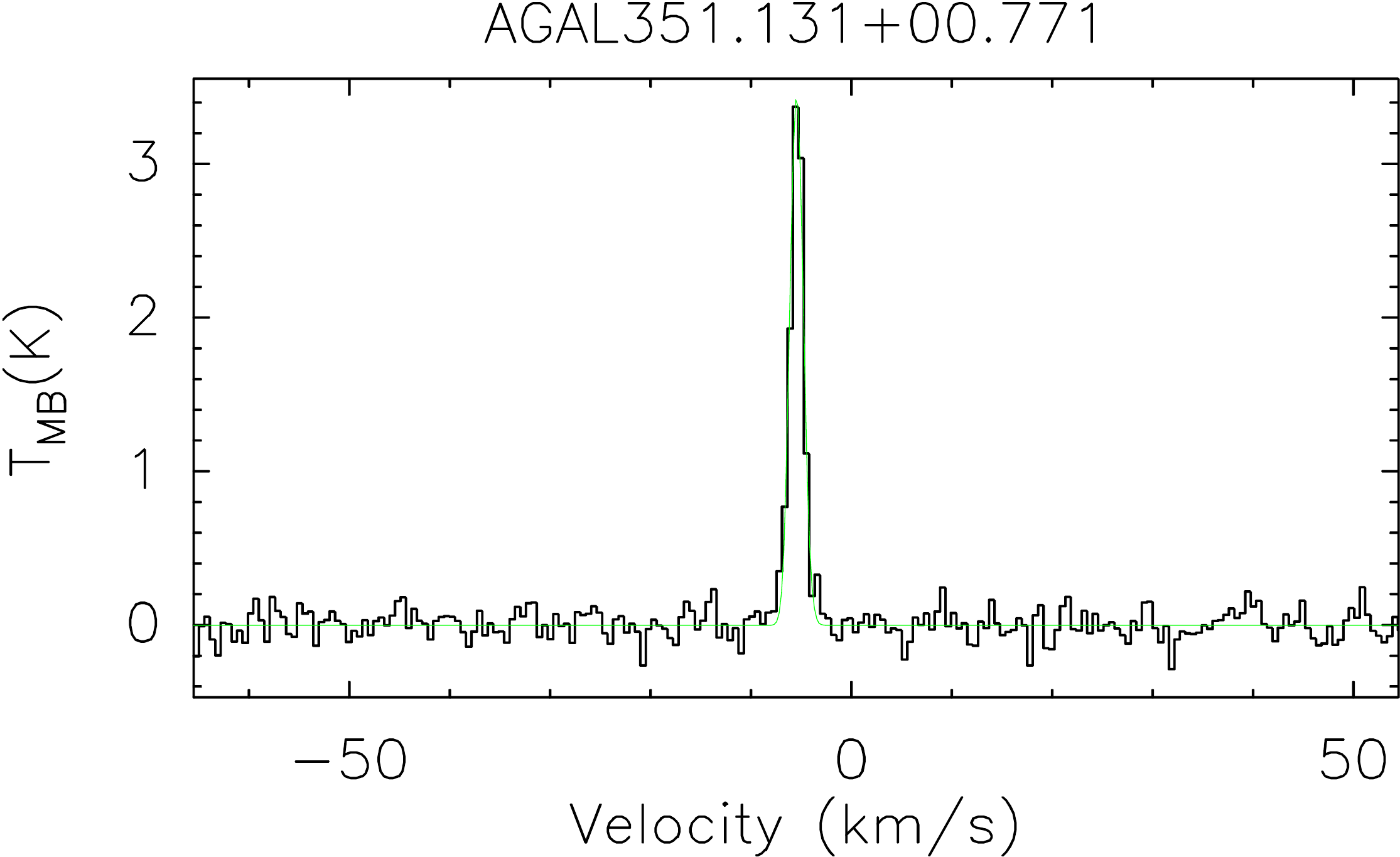} \\ 
\includegraphics[angle=-90,width=0.3\textwidth]{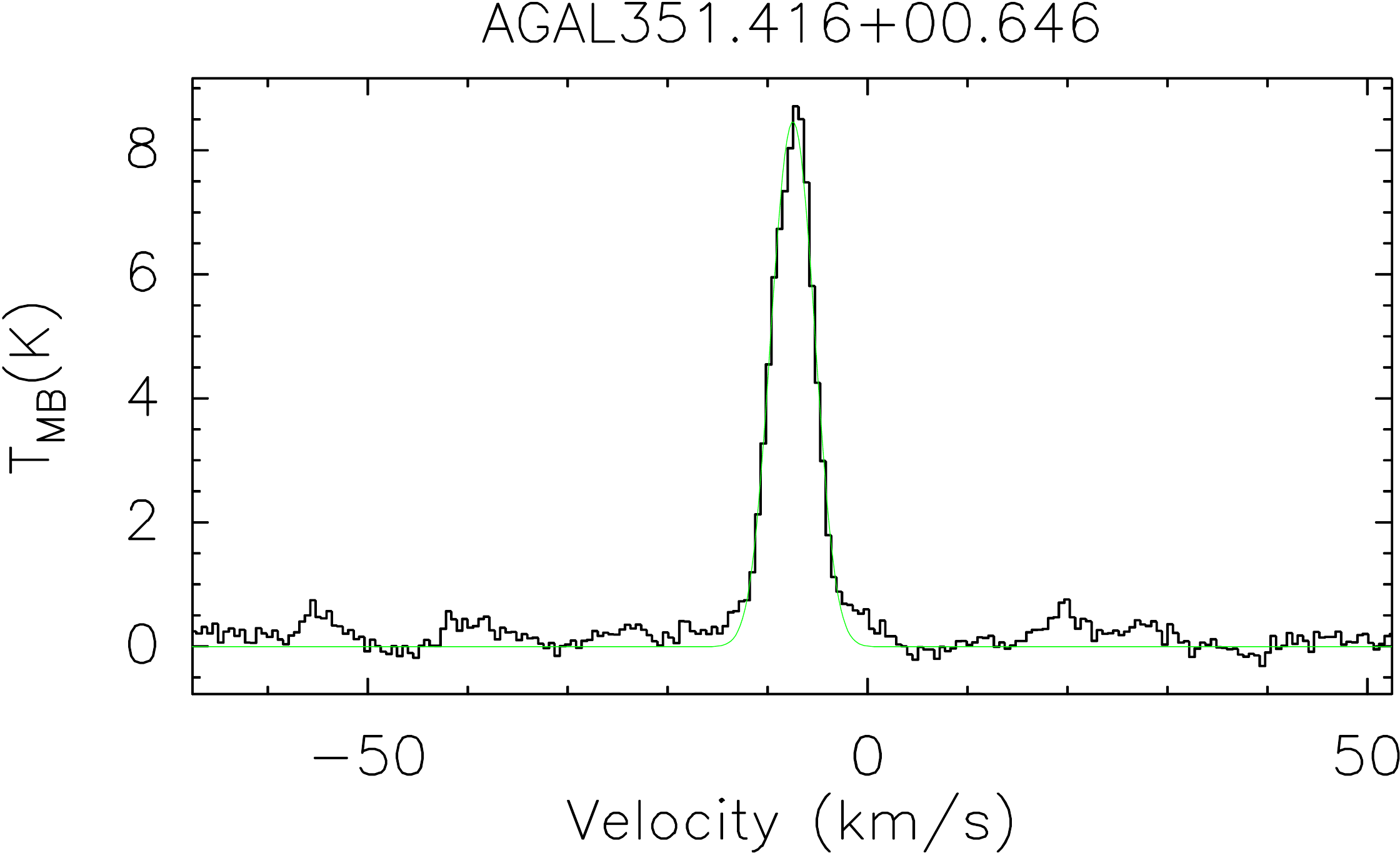} 
\includegraphics[angle=-90,width=0.3\textwidth]{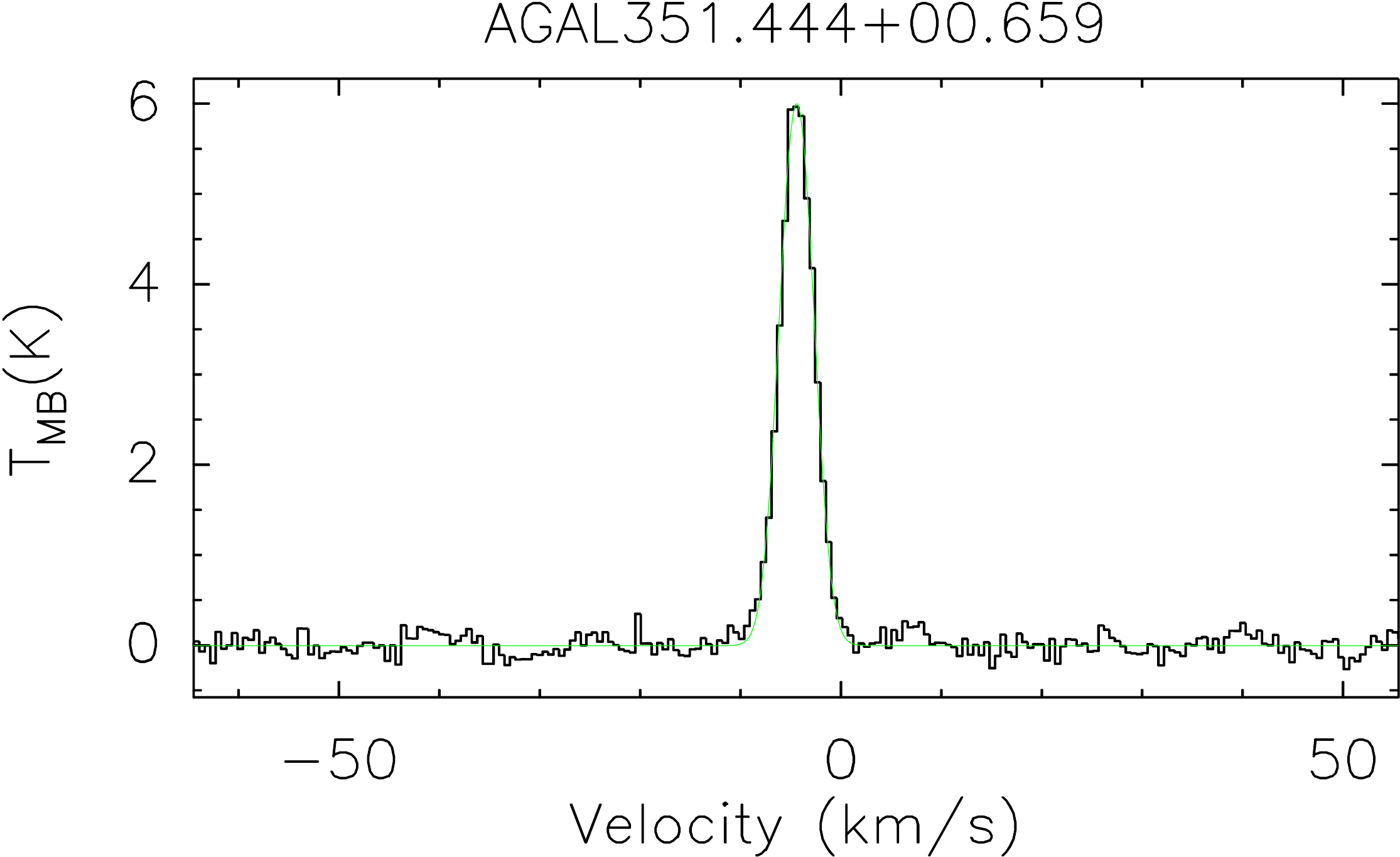} \hfill 
\caption{Continued.} 
\end{figure*} 
\clearpage
}

\onlfig{1}{
\section{RATRAN figures}

\begin{figure*} 
\centering 
\hspace*{0.02\textwidth}
\includegraphics[angle=-90,width=0.63\textwidth]{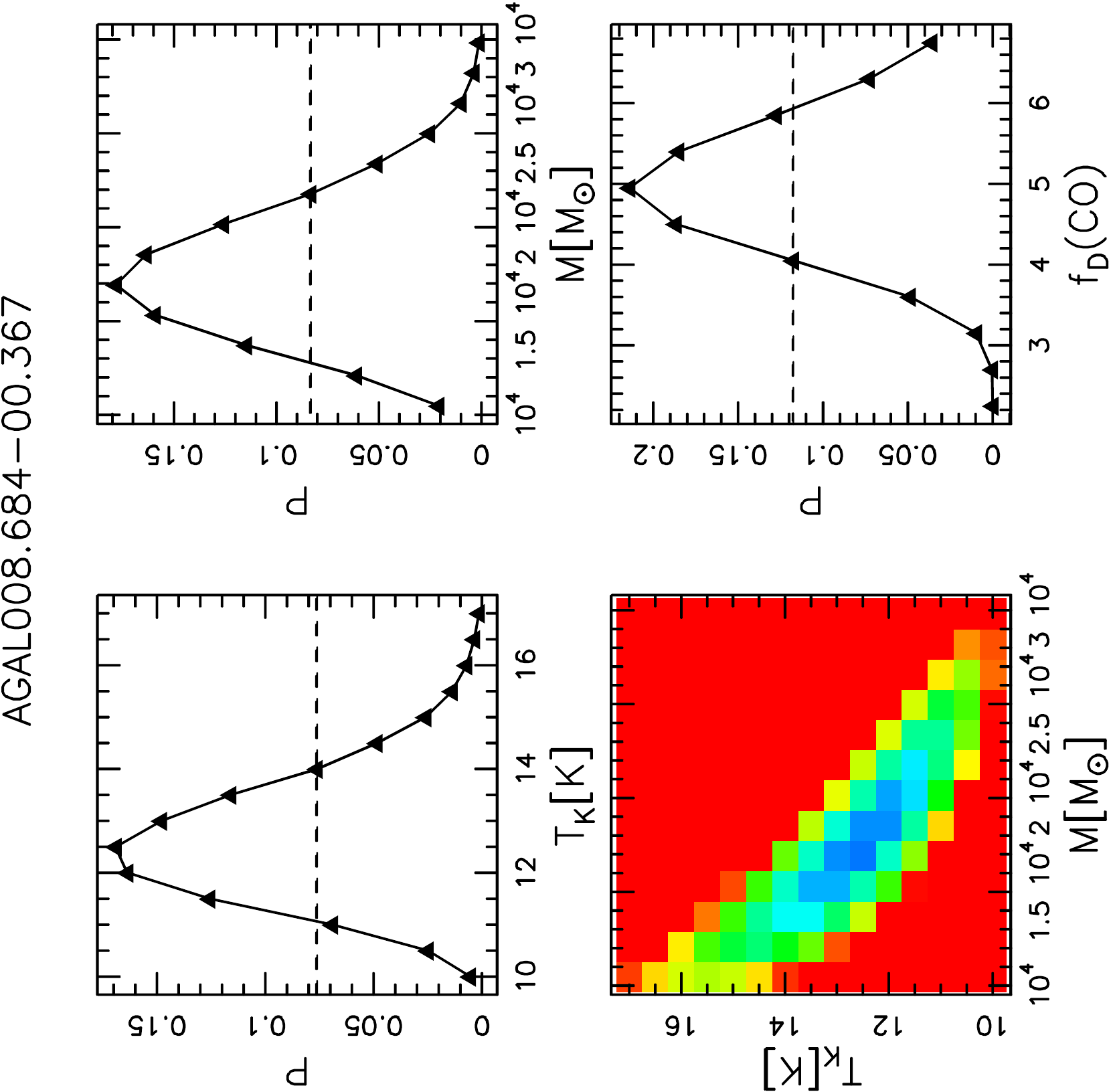}\\ 
\hspace*{0.02\textwidth}\\
\vspace*{0.4cm}
\includegraphics[angle=-90,width=0.63\textwidth]{fig_clean/{AG10.44-0.0-gc}.pdf}
\caption{RATRAN results for individual sources, using models with a constant abundance profile. The panels show: \textbf{(top left)} marginal probability distribution of the temperature/luminosity, depending on the group of the source, \textbf{(top right)} marginal probability distribution of the mass, \textbf{(bottom left)} joint probability distribution of mass and $T$ or $L$ (depending on whether model is centrally heated or isothermal), \textbf{(bottom right)} marginal probability distribution of the depletion factor.} \label{fig:ind_sou_ratran}
\end{figure*} 
 
\begin{figure*} 
\ContinuedFloat
\centering 
\hspace*{0.02\textwidth}
\includegraphics[angle=-90,width=0.63\textwidth]{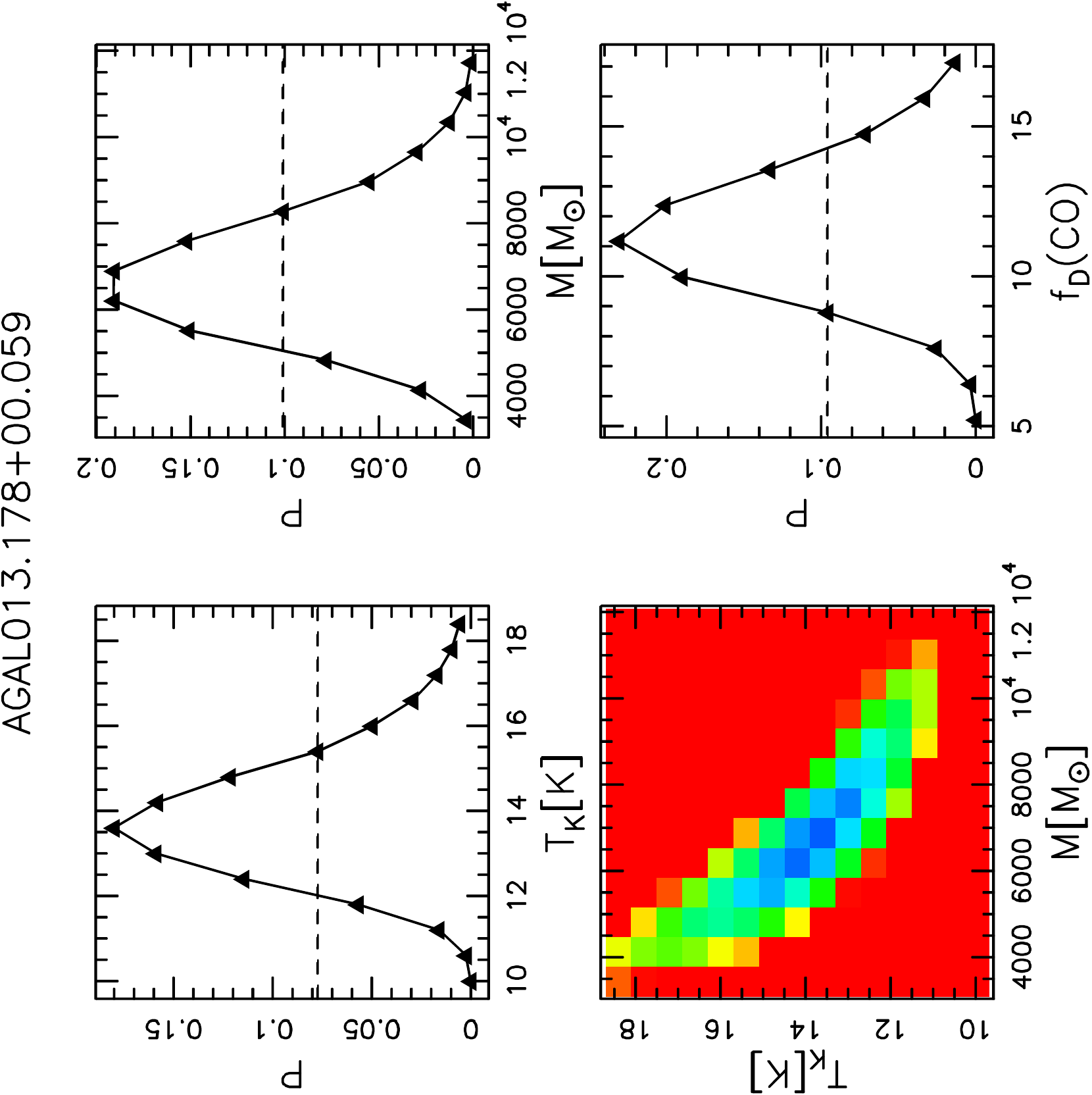} \\
\vspace*{0.4cm}
\includegraphics[angle=-90,width=0.6\textwidth]{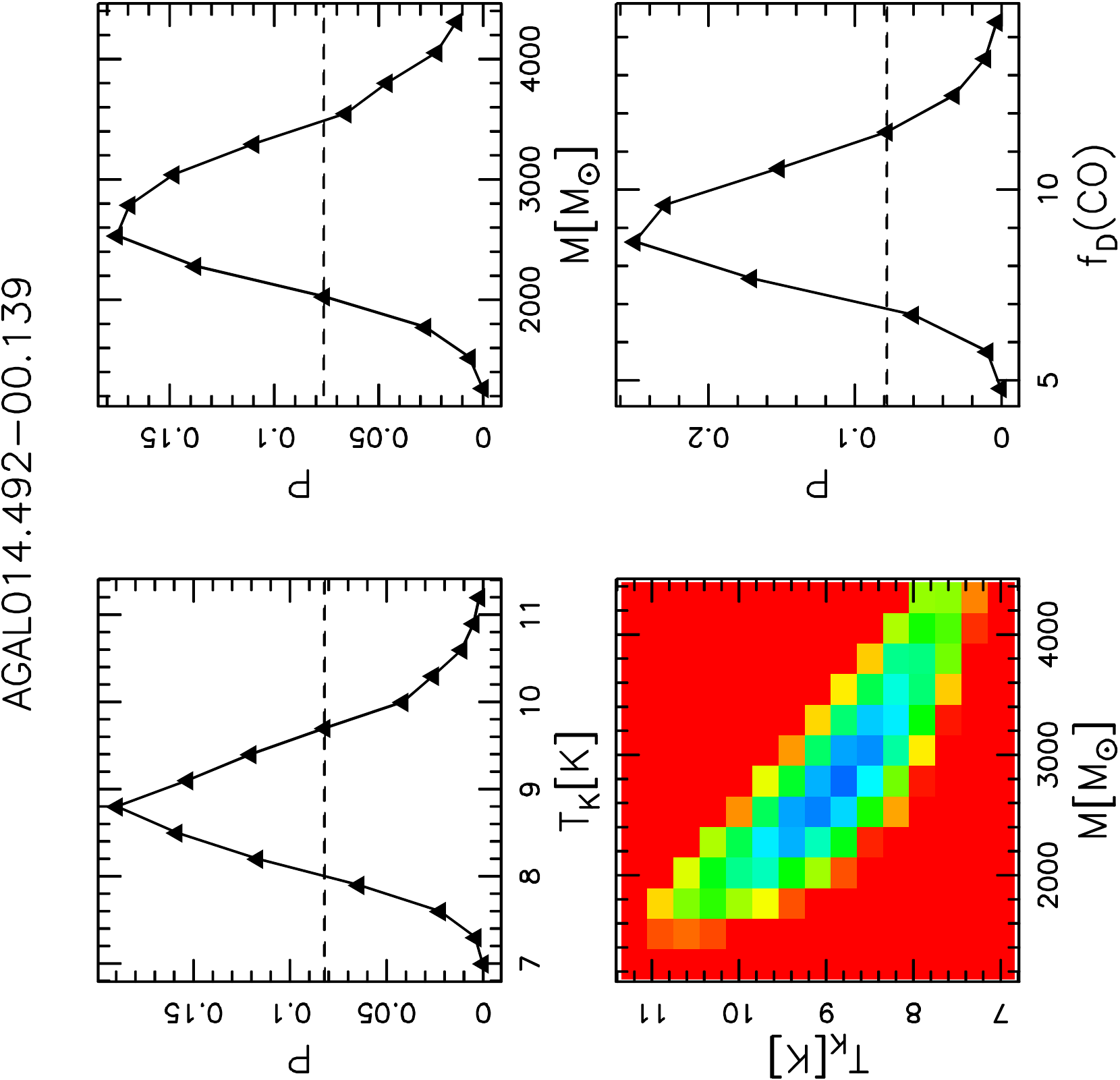} 
\caption{Continued.}
\end{figure*} 

\begin{figure*} 
\ContinuedFloat
\centering 
\includegraphics[angle=-90,width=0.6\textwidth]{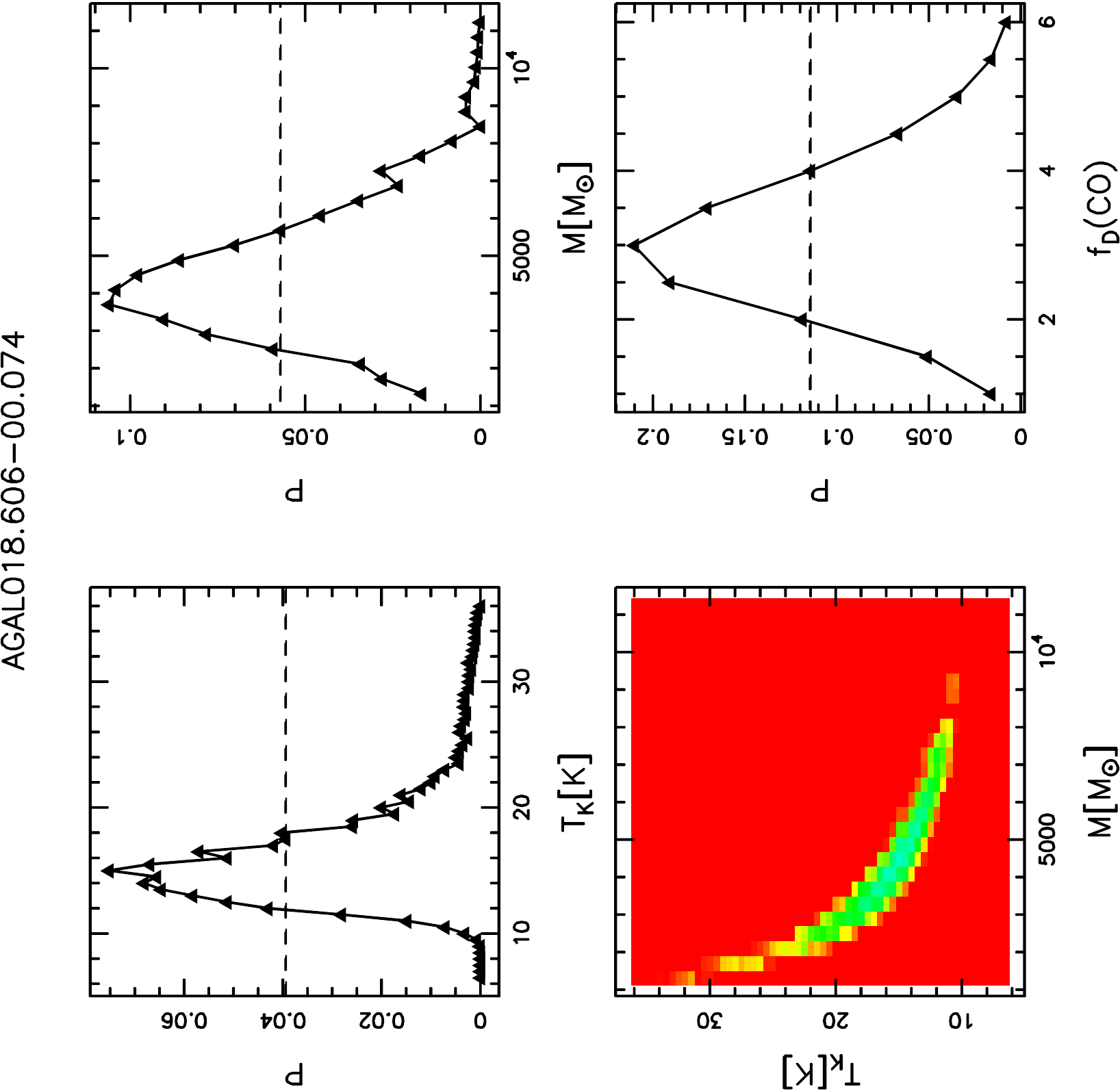} 
\vspace*{0.4cm}
\includegraphics[angle=-90,width=0.6\textwidth]{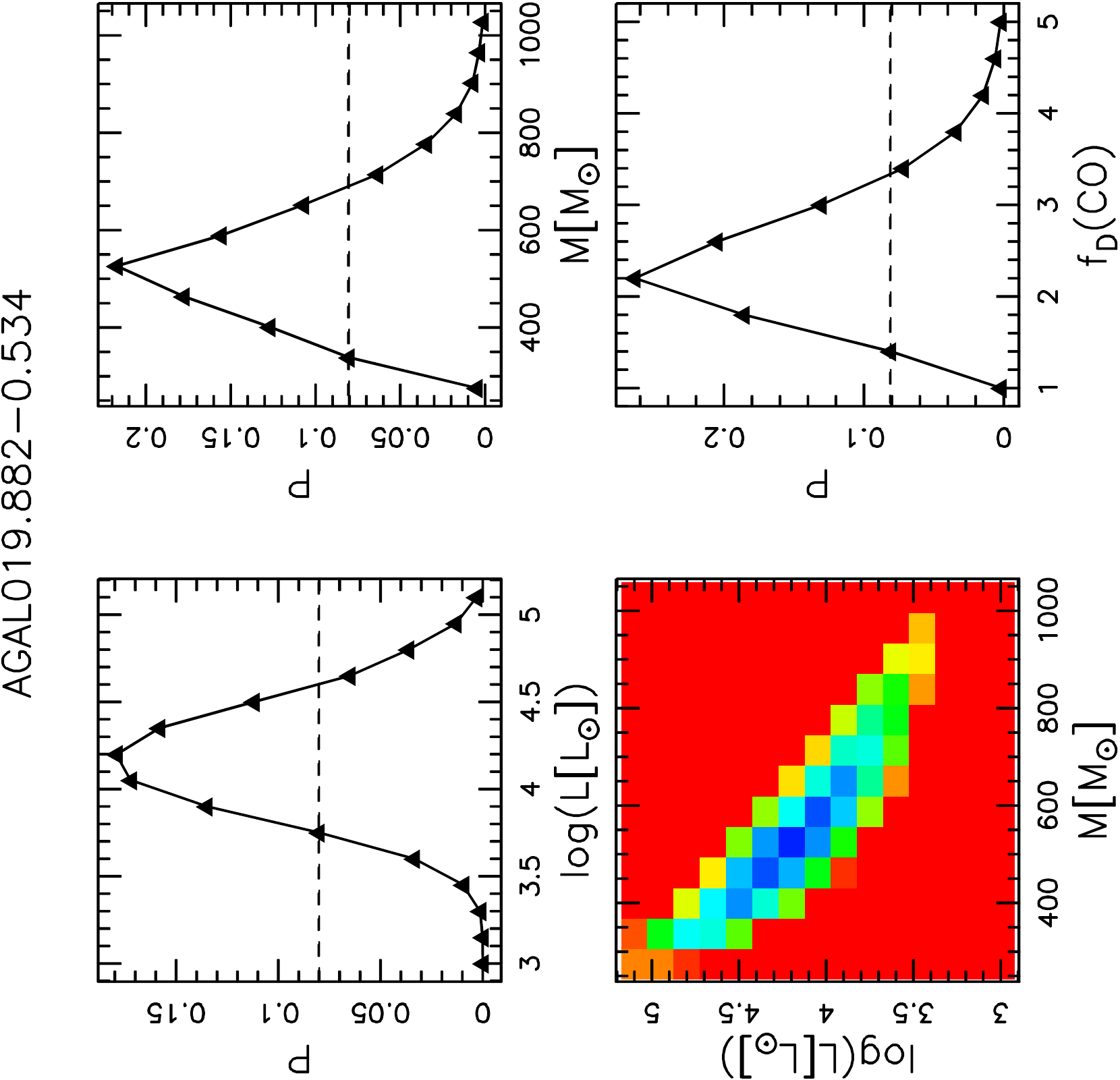} 
\caption{Continued.}
\end{figure*} 

\begin{figure*} 
\ContinuedFloat
\centering 
\includegraphics[angle=-90,width=0.6\textwidth]{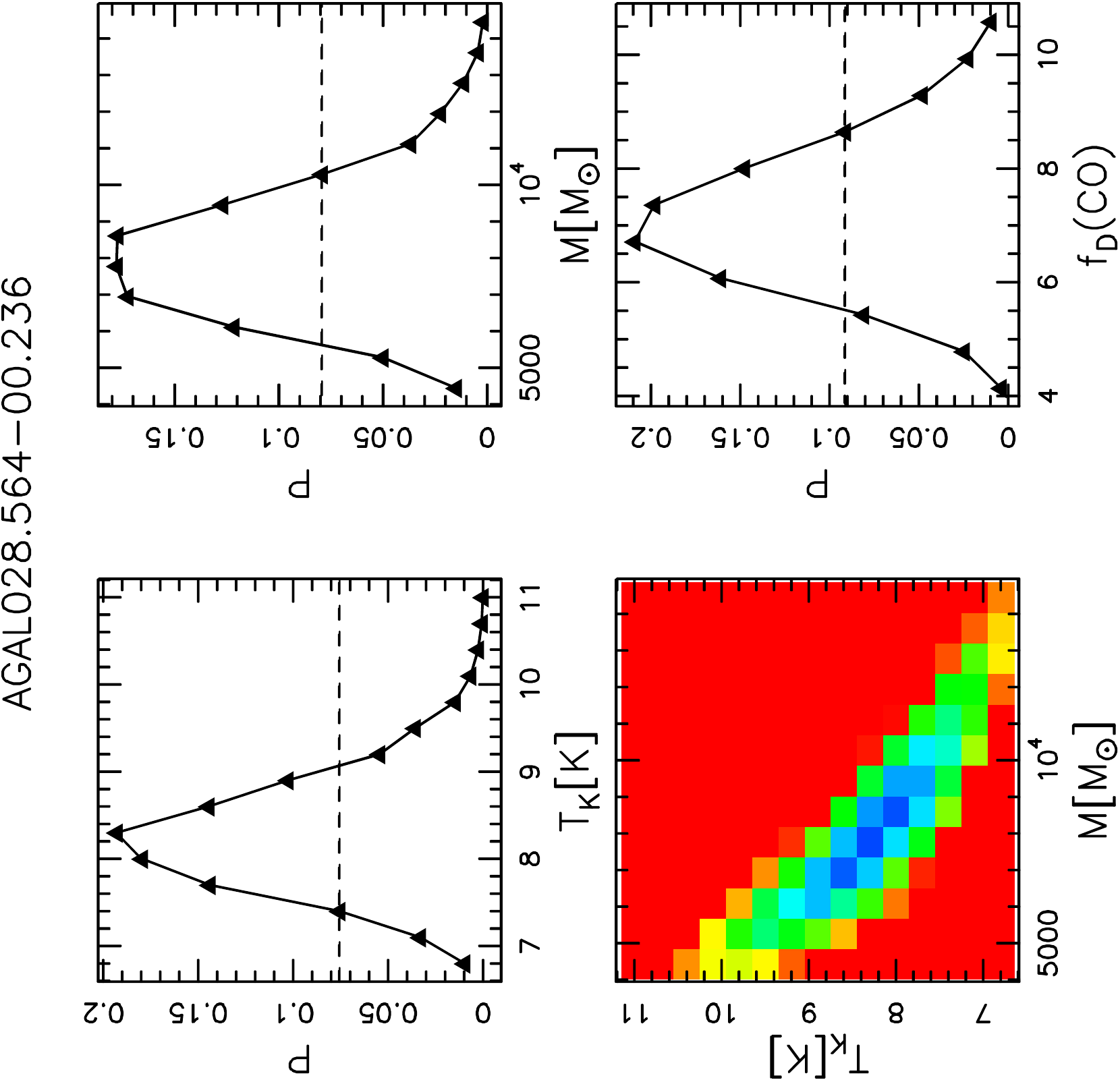} 
\vspace*{0.4cm}
\includegraphics[angle=-90,width=0.6\textwidth]{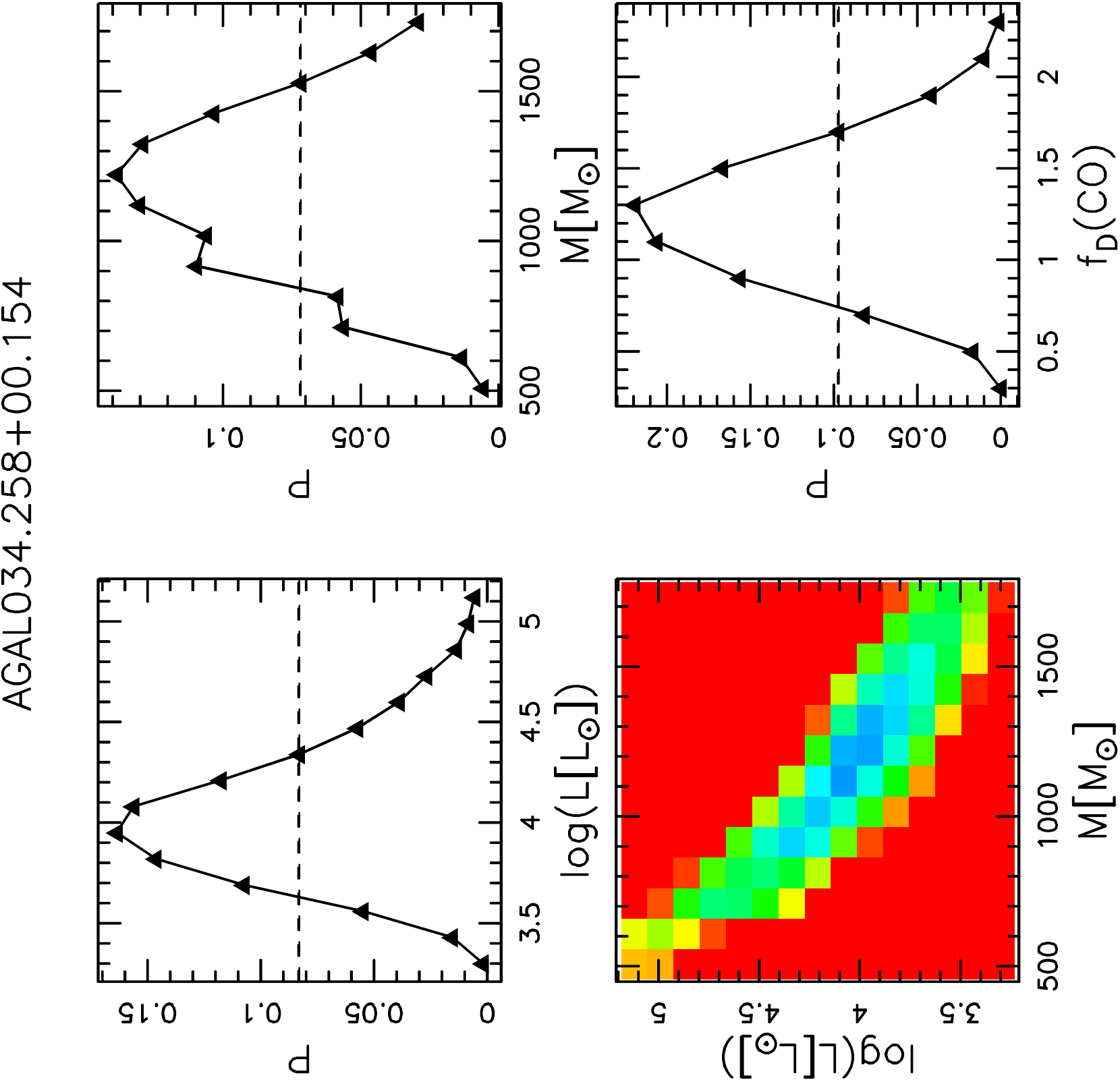} 
\caption{Continued.}
\end{figure*} 

\begin{figure*} 
\ContinuedFloat
\centering 
\includegraphics[angle=-90,width=0.6\textwidth]{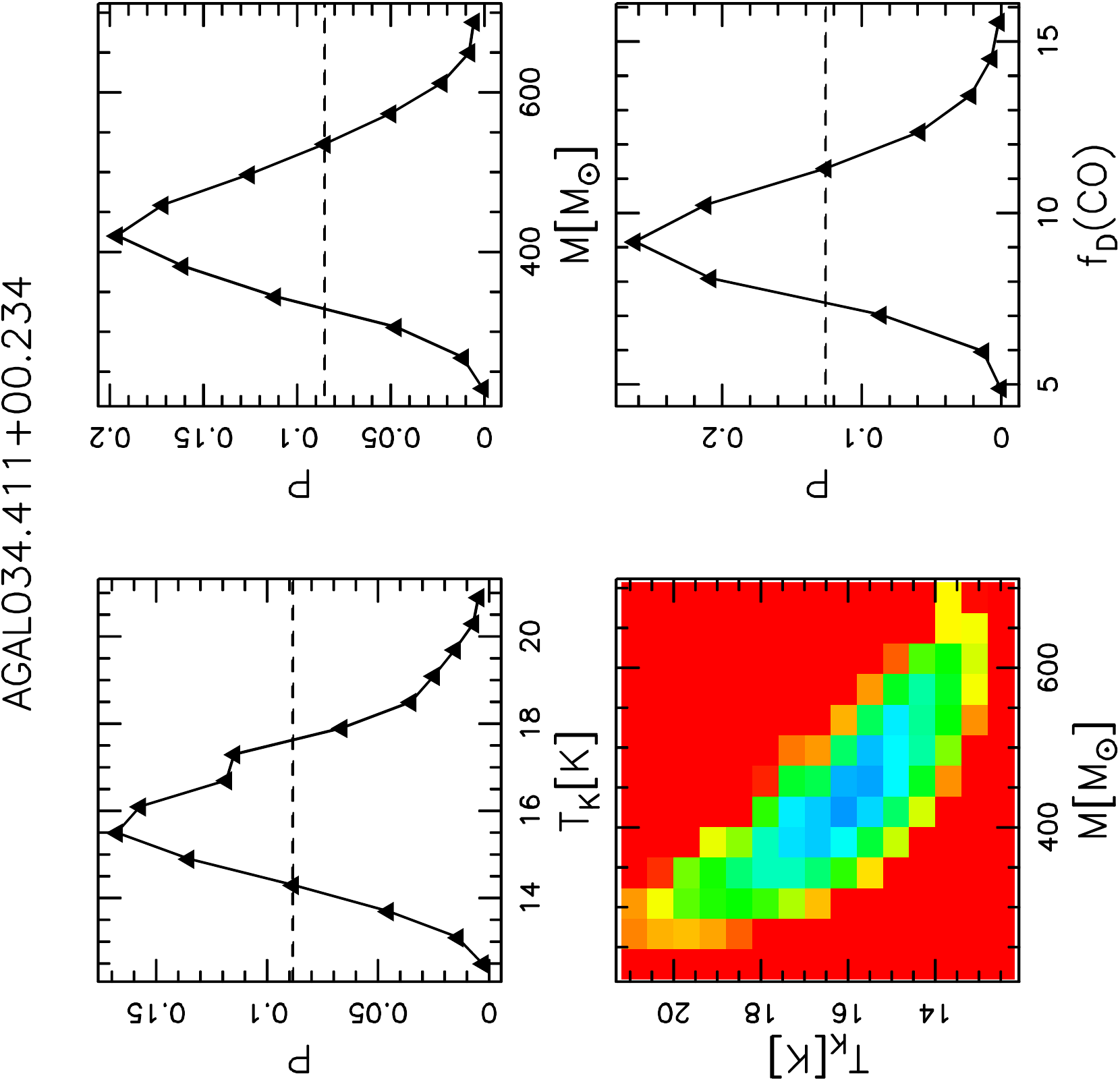} 
\vspace*{0.4cm}
\includegraphics[angle=-90,width=0.6\textwidth]{fig_clean/{AG49.49-0.39-gc}.pdf} 
\caption{Continued.}
\end{figure*} 
}
\onlfig{2}{
\begin{figure*} 
\centering 

\includegraphics[angle=-90,width=0.8\textwidth]{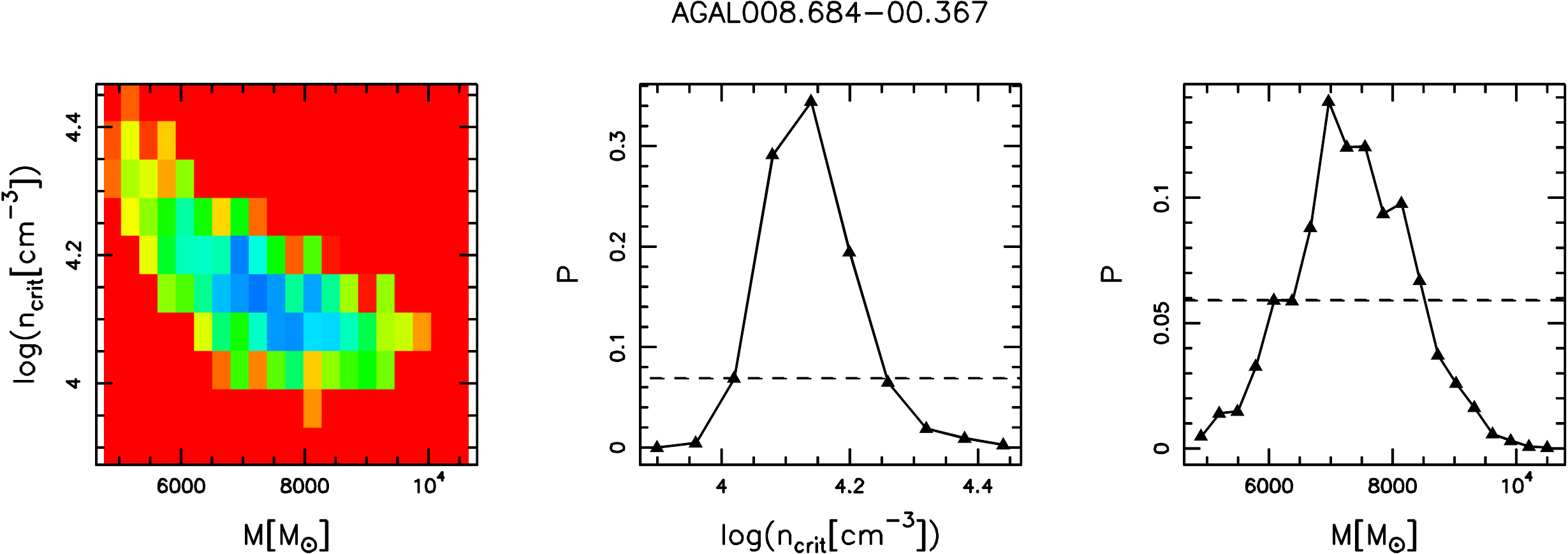} \hfill \\
\vspace*{0.5cm}
\includegraphics[angle=-90,width=0.8\textwidth]{fig_clean/{drop_AG10}.pdf} \hfill \\
\vspace*{0.5cm}
\includegraphics[angle=-90,width=0.8\textwidth]{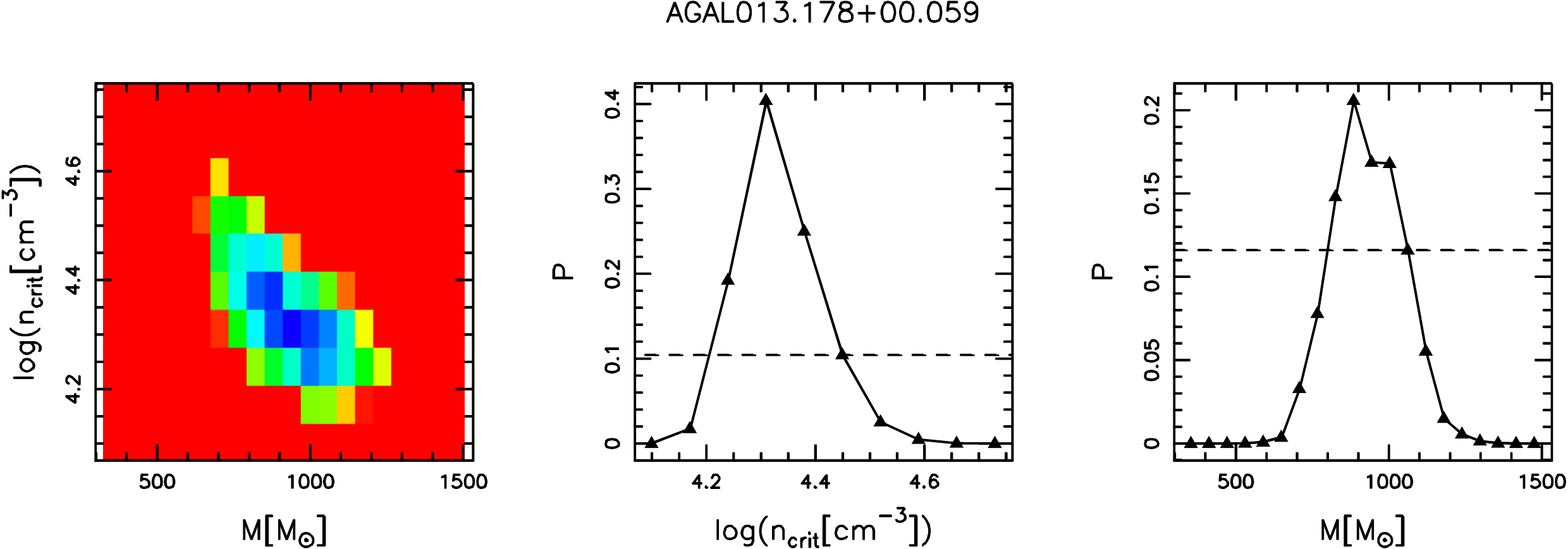} \hfill \\
\vspace*{0.5cm}
\includegraphics[angle=-90,width=0.8\textwidth]{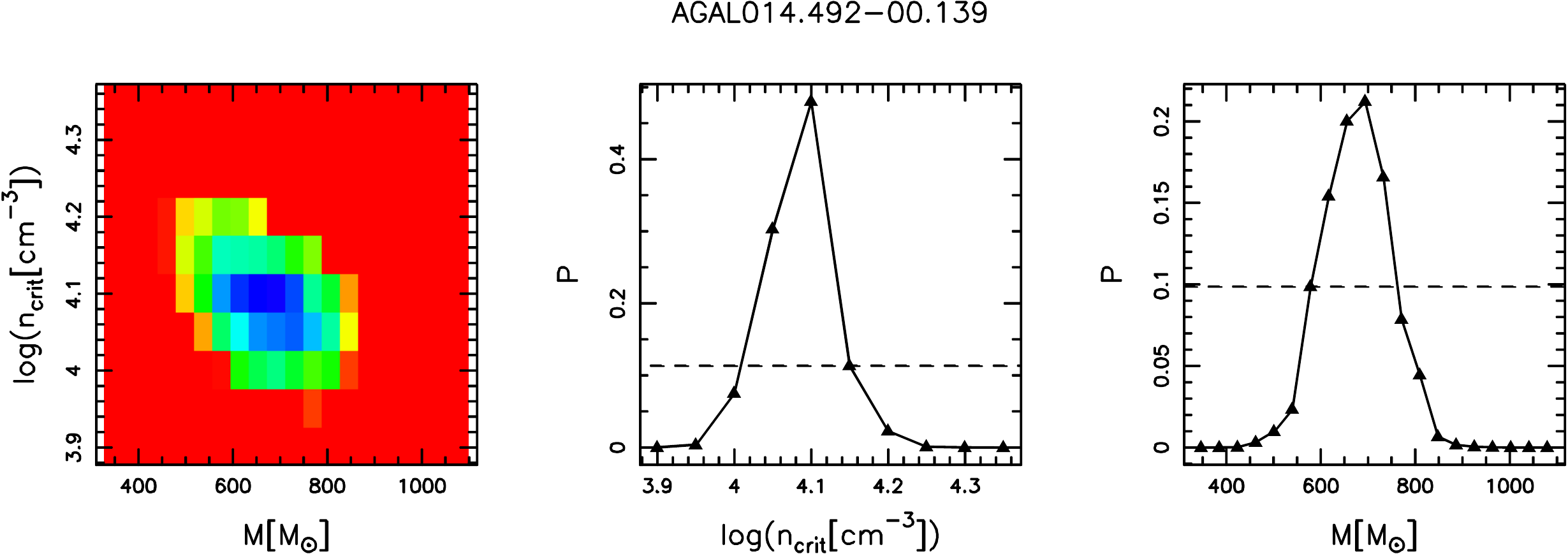} \hfill

\caption{RATRAN results for individual sources, using models with a drop profile. The panels show: \textbf{(left)} joint probability distribution of mass and critical density of molecular hydrogen above which all CO is locked onto dust grains, \textbf{(centre)} marginal probability distribution of critical density, \textbf{(right)} marginal probability distribution of mass.} \label{fig:RATRAN_ind_drop}
\end{figure*} 

\begin{figure*} 
\centering 
\ContinuedFloat

\includegraphics[angle=-90,width=0.8\textwidth]{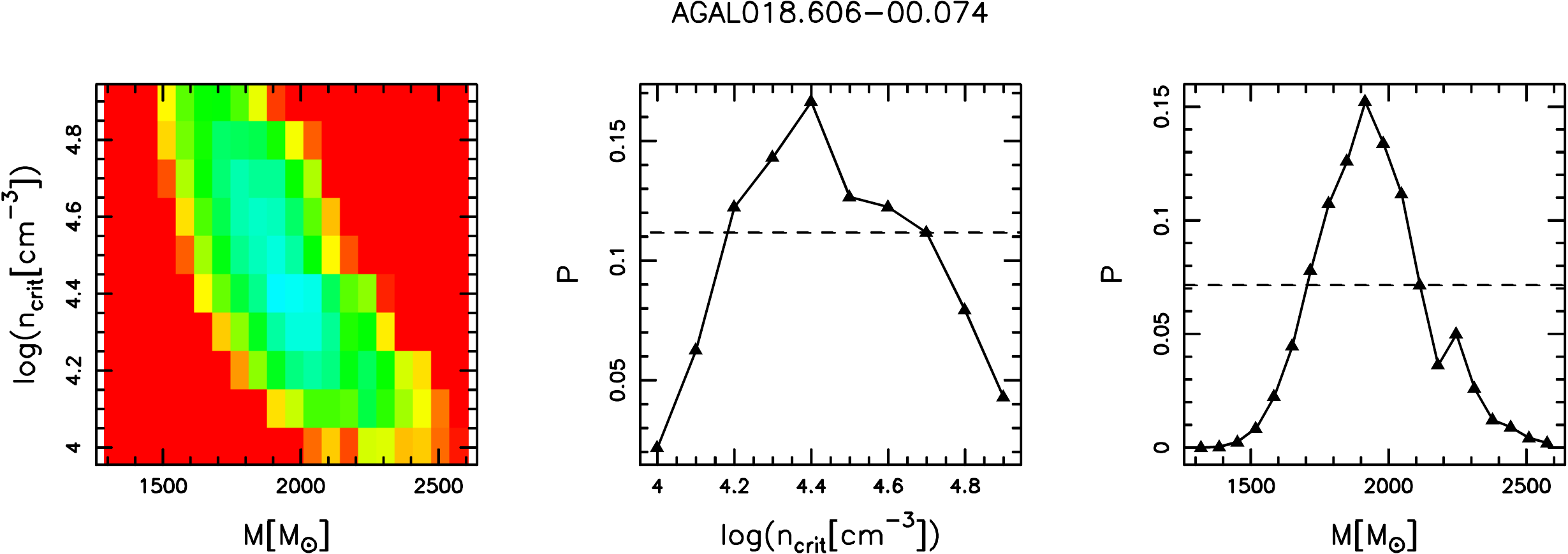} \hfill \\
\vspace*{0.5cm}
\includegraphics[angle=-90,width=0.8\textwidth]{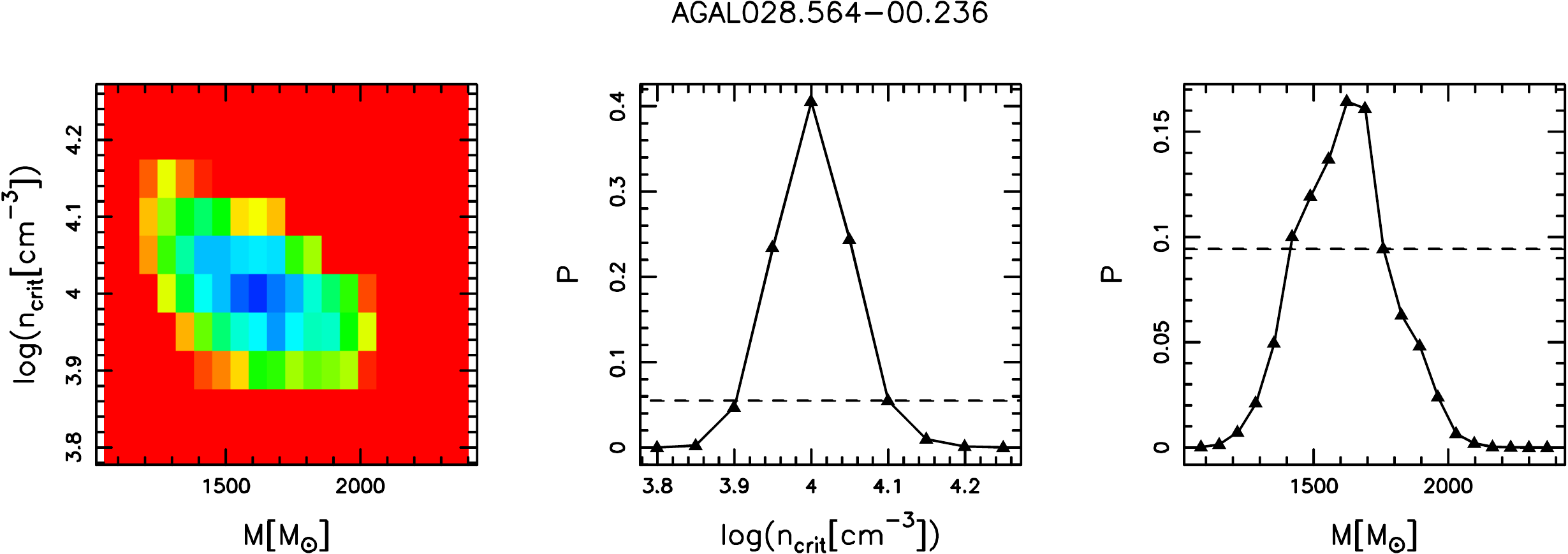} \hfill \\
\vspace*{0.5cm}
\includegraphics[angle=-90,width=0.8\textwidth]{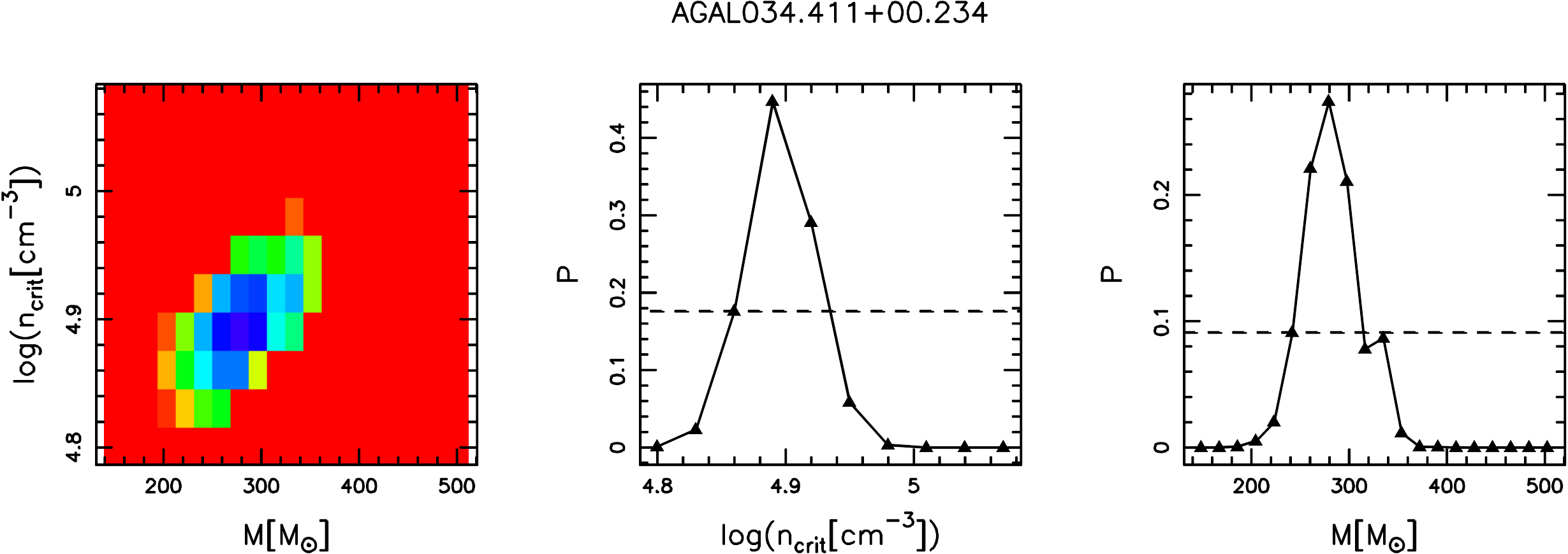} \hfill

\caption{Continued.}
\end{figure*} 
\clearpage
}
\end{document}